\documentclass[11pt,twoside]{article}
\usepackage{amssymb}
\usepackage{amsmath}
\usepackage[mathscr]{eucal}
\usepackage{amsfonts}
\usepackage{latexsym}
\usepackage{amsxtra}
\usepackage{amsbsy}
\usepackage{amsthm}
\usepackage{amscd}
\usepackage{amsopn}
\usepackage{amstext}
\newcounter{z0}
\newcounter{z1}
\newtheorem{ay}{Lemma}[section]

\newtheorem{by}{Proposition}[section]

\newtheorem{aaaaa}{Definition}[section]
\newtheorem{bbbbb}{Proposition}[section]
\newtheorem{ccccc}{Lemma}[section]
\newtheorem{ddddd}{Theorem}[section]
\newtheorem{fffff}{Corollary}[section]

\theoremstyle{definition}

\theoremstyle{definition}

\theoremstyle{definition}
\newtheorem{eeeee}{Remark}[section]
\newcommand{\me}{\mathrm{e}}
\newcommand{\mi}{\mathrm{i}}
\newcommand{\md}{\mathrm{d}}
\begin{document}
\baselineskip=12pt
\frenchspacing
\title{Long-Time Asymptotics of Solutions to the Cauchy Problem for the
Defocusing Non-Linear Schr\"{o}dinger Equation with Finite-Density Initial
Data.~I.~Solitonless Sector}
\author{A.~H.~Vartanian\thanks{
\texttt{E-mail: vartaniana@arthur.winthrop.edu}. Current address: Department
of Mathematics, Duke University, Durham, North Carolina 27708, U.~S.~A.,
\texttt{e-mail: arthur@math.duke.edu}} \\
Department of Mathematics \\
Winthrop University \\
Rock Hill, South Carolina 29733 \\
U.~S.~A.}
\date{3 December 2001}
\maketitle
\begin{abstract}
\noindent
The methodology of the Riemann-Hilbert (RH) factorisation approach for
Lax-pair isos\-p\-e\-c\-t\-r\-a\-l deformations is used to derive, in
the solitonless sector, the leading-order asymptotics as $t \! \to \!
\pm \infty$ $(x/t \! \sim \! \mathcal{O}(1))$ of solutions to the
Cauchy problem for the defocusing non-linear Schr\"{o}dinger equation
(D${}_{f}$NLSE), $\mi \partial_{t}u \! + \! \partial_{x}^{2}u\! - \!
2(\vert u \vert^{2} \! - \! 1)u \! = \! 0$, with (finite-density) initial
data $u(x,0) \! =_{x \, \to \, \pm \infty} \! \exp (\tfrac{\mi (1 \mp 1)
\theta}{2})(1 \! + \! o(1))$, $\theta \! \in \! [0,2 \pi)$. A limiting
case of these asymptotics related to the RH problem for the Painlev\'{e}
II equation, or one of its special reductions, is also identified.

\vspace{1.35cm}

37K15: (Secondary) 30E20, 30E25, 81U40

\vspace{0.50cm}
{\bf Abbreviated Title.} Asymptotics of the Defocusing NLSE

\vspace{0.50cm}

a\-n\-n-H\-i\-l\-b\-e\-r\-t problems, singular integral equations
\end{abstract}
\clearpage
\section{Introduction}
In the optical fibre literature, the mathematical model, in normalised and
dimensionless form, describing dark soliton pulse propagation (which consists
of a rapid dip in the intensity of a broad pulse of a continuous wave
background) in polarisation preserving, single-mode optical fibres in the
picosecond time scale is the following non-linear Schr\"{o}dinger equation
(NLSE) \cite{a1,a2,a3}, $\mi \partial_{z}q \! + \! \partial_{\tau}^{2}q
\! - \! 2q \vert q \vert^{2} \! = \! 0$, where $q \! = \! q(\tau,z)$ is the
slowly varying amplitude of the complex field envelope, $z$ is the distance
along the fibre length, and $\tau$ is the retarded time measured in a
reference frame moving along the fibre at the group velocity, with
non-vanishing boundary conditions $q(\tau,z) \! =_{\tau \to \pm \infty} \!
\varrho \me^{\mi (\varphi_{\infty}^{\pm}-2 \varrho^{2}z)}$, where $\varrho$
$(> \! 0)$ is the so-called density, and $\varphi_{\infty}^{\pm}$ $(\in \!
[0,2 \pi))$ are the asymptotic phases. Mapping, isomorphically, the
physical variables onto the mathematical variables, $(\tau,z) \! \mapsto
\! (\widetilde{x},\widetilde{t})$, setting $q(\widetilde{x},\widetilde{t})
\! := \! \widehat{q}(\widetilde{x},\widetilde{t}) \me^{-2 \mi \varrho^{
2} \widetilde{t}}$, scaling according to the rule $\widetilde{t} \!
\to \! \varrho^{-2} t$, $\widetilde{x} \! \to \! \varrho^{-1} x$, and
$\widehat{q} \! \to \! \varrho u(x,t) \me^{\mi \varphi_{\infty}^{+}}$,
and defining $\theta \! := \! \varphi_{\infty}^{-} \! - \! \varphi_{
\infty}^{+}$, one arrives at considering solutions of the following
non-linear evolution equation (NLEE), hereafter referred to as the
defocusing non-linear Schr\"{o}dinger equation (D${}_{f}$NLSE),
with finite-density initial data,
\begin{equation}
\begin{split}
\mi \partial_{t}u \! + \! \partial_{x}^{2}u \! - \! 2(\vert u \vert^{2}
\! - \! 1)u \! = \! 0, \qquad (x,t) \! \in \! \mathbb{R} \! \times \!
\mathbb{R}, \\
u(x,0) \! := \! u_{o}(x) \! \underset{x \, \to \, \pm \infty}{=} \! \exp
(\tfrac{\mi (1 \mp 1) \theta}{2})(1 \! + \! o(1)),
\end{split}
\end{equation}
where $u_{o}(x) \! \in \! \mathbf{C}^{\infty}(\mathbb{R})$, $\theta
\! \in \! [0,2 \pi)$ (see Lemma~2.2), and the $o(1)$ term is to be
understood in the sense that, $\forall \, (k,l) \! \in \! \mathbb{Z}_{
\geqslant 0} \! \times \! \mathbb{Z}_{\geqslant 0}$, $\vert x \vert^{
k}(\tfrac{\md}{\md x})^{l}(u_{o}(x) \! - \! \exp (\tfrac{\mi (1 \mp 1)
\theta}{2})) \! =_{x \to \pm \infty} \! 0$. Only for initial data
satisfying $\vert x \vert^{k}(\tfrac{\md}{\md x})^{l}(u_{o}(x) \! - \!
\exp (\tfrac{\mi (1 \mp 1) \theta}{2})) \! =_{x \to \pm \infty} \! 0$,
$(k,l) \! \in \! \mathbb{Z}_{\geqslant 0} \! \times \! \mathbb{Z}_{
\geqslant 0}$, is it true that the closure of the set of reflectionless
(soliton) potentials of the D${}_{f}$NLSE in the topology of uniform
convergence of functions on compact sets of $\mathbb{R}$ (denoted by
$\mathcal{B}$) remains an invariant set of this model for $t \! \not= \!
0$ (a solution of the D${}_{f}$NLSE with finite density initial data, in
the above-defined sense, remains in $\mathcal{B} \, \, \forall \, t \! \in
\! \mathbb{R}$, and not just for $t \! = \! 0)$ \cite{a4}.

It is instructive to study the asymptotics of solutions to the Cauchy
problem for the D${}_{f}$NLSE for finite-density initial data having a
decomposition of the form $u_{o}(x) \! := \! u_{\mathrm{sol}}(x) \! + \!
u_{\mathrm{rad}}(x)$, $x \! \in \! \mathbb{R}$, where $u_{o}(x)$ satisfies
the conditions stated heretofore, $u_{\mathrm{sol}}(x)$ is responsible for
``generating'' the multi- or $N$-dark soliton solution, and $u_{\mathrm{rad}}
(x)$ is the ``small'' non-dark-soliton component manifesting as the
asymptotically decaying dispersive component of the solution. In fact, this
is the principal objective of the present series of works devoted to the
asymptotic analysis of solutions to the D${}_{f}$NLSE for finite-density
initial data; in particular, in this work, the case $u_{\mathrm{sol}}(x)
\! \equiv \! 0$ and $u_{\mathrm{rad}}(x) \! \not\equiv \! 0$ is treated,
and the case $(u_{\mathrm{sol}}(x),u_{\mathrm{rad}}(x)) \! \not\equiv \!
(0,0)$ is presently under study. Another objective of this series of works,
which will be pursued elsewhere, is to use the results obtained herein to
derive an explicit asymptotic expression for the transfer matrix for an
$N$-dark soliton \textsf{X} junction \cite{a5}.

It is well-known that, within the framework of the inverse scattering
method (ISM) \cite{a6,a7,a8}, the D${}_{f}$NLSE is a completely integrable
NLEE with an explicit representation as an infinite-dimensional Hamiltonian
system \cite{a9}. Even though the analysis of NLS-like NLEEs with rapidly
decaying, e.g., Schwartz class, initial data on $\mathbb{R}$ has received
the vast majority of the attention in the context of direct and inverse
spectral treatments, there have been a handful of, in some cases seminal,
works devoted exclusively to the direct and inverse scattering analysis of
completely integrable NLEEs belonging to the ZS-AKNS class with non-vanishing
values of the initial data \cite{a10,a11,a12}. As shown in Part~1 of
\cite{a9}, a two-sheeted Riemann surface plays a central role in the
direct/inverse spectral formulation associated with the D${}_{f}$NLSE for
finite-density initial data. Other interesting classes of finite-density
(or non-vanishing)-type initial data for completely integrable NLEEs,
e.g., NLS, derivative and modified NLS, KdV, and sinh/e-Gordon, have
also been considered \cite{a13,a14,a15,a16,a17,a18}. To the best of the
author's knowledge as at the time of the presents, the first to consider
the asymptotics of solutions to the D${}_{f}$NLSE for finite-density
initial data were Its \emph{et al.} \cite{a19,a20}.

In the framework of the ISM, the asymptotic analysis of solutions to the
Cauchy problem for the D${}_{f}$NLSE with finite-density initial data is
divided into two steps: (1) the analysis of the solitonless (continuum)
component of the solution; and (2) the inclusion of the $N$-dark soliton
component via the application of a ``dressing'' procedure to the solitonless
background/component \cite{a21,a22}. In this work, stage~(1) of the
above-mentioned two-step asymptotic paradigm, which is the more
technical of the two, is carried out systematically using the methodology
of the Riemann-Hilbert (RH) factorisation \cite{a23} approach to the
ISM \cite{a6,a8,a24,a25,a26}.

This paper is organized as follows. In Section~2, starting {}from the
Lax-pair isospectral deformation condition associated with the D${}_{
f}$NLSE, all necessary formulae {}from the direct and inverse scattering
analyses associated with the solution of the Cauchy problem for the D${
}_{f}$NLSE with finite-density initial data are derived, the corresponding
(matrix) Riemann-Hilbert problem (RHP) is formulated, and the particular
case of this RHP studied asymptotically in this work is stated. In Section~3,
a self-contained synopsis of the Beals-Coifman \cite{a24} construction for
the solution of a matrix RHP on an oriented contour is given, a detailed
account of the Deift-Zhou \cite{a27} non-linear steepest descent method
for the asymptotic analysis of the RHP stated in Section~2 is presented,
and the results of this paper are summarised in Theorems~3.1--3.3. In
Section~4, as $t \! \to \! +\infty$ $(x/t \! \sim \! \mathcal{O}(1))$, the
RHP is reformulated as an auxiliary RHP on an augmented contour which is
then dissected to produce an equivalent RHP on a truncated contour. In
Section~5, it is shown that, to leading order as $t \! \to \! +\infty$
$(x/t \! \sim \! \mathcal{O}(1))$, modulo terms that are $\mathcal{O}
(t^{-1/2} \ln t)$, the solution of the equivalent RHP on the truncated contour
``tends to'' the solution of an explicitly solvable model RHP on a contour
which consists of the disjoint union of two rotated crosses. In Section~6,
as $t \! \to \! +\infty$ $(x/t \! \sim \! \mathcal{O}(1))$, the model RHP
on the disjoint union of the two rotated crosses is reformulated as an
asymptotic system of linear singular integral equations which are then
solved explicitly to yield the asymptotics of solutions (and related
integrals of solutions) to the Cauchy problem for the D${}_{f}$NLSE. In
Section~7, the above asymptotic paradigm is succinctly reworked for the
case when $t \! \to \! -\infty$ $(x/t \! \sim \! \mathcal{O}(1))$. The
paper concludes with an Appendix.
\section{The Direct/Inverse Scattering Analysis and the
Ri\-em\-an\-n-Hi\-lb\-ert Problem}
The necessary facts {}from the direct/inverse scattering analysis of
the Lax pair (see Proposition~2.1) associated with the D${}_{f}$NLSE
for finite-density initial data are derived, the corresponding RHP is
formulated, and the particular case of this RHP which is analysed
asymptotically as $t \! \to \! \pm \infty$ $(x/t \! \sim \! \mathcal{O}
(1))$ in this work is stated. Before proceeding, however, the
notation/nomenclature used throughout this work is summarised.
\begin{center}
\underline{\textsc{Notational Conventions}}
\end{center}
\begin{enumerate}
\item[(1)] $\mathrm{I} \! = \!
\left(
\begin{smallmatrix}
1 & 0 \\
0 & 1
\end{smallmatrix}
\right)$ is the $2 \! \times \! 2$ identity matrix, $\sigma_{1} \! = \!
\left(
\begin{smallmatrix}
0 & 1 \\
1 & 0
\end{smallmatrix}
\right)$, $\sigma_{2} \! = \!
\left(
\begin{smallmatrix}
0 & -\mi \\
\mi & 0
\end{smallmatrix}
\right)$, and $\sigma_{3} \! = \!
\left(
\begin{smallmatrix}
1 & 0 \\
0 & -1
\end{smallmatrix}
\right)$ are the Pauli matrices, $\sigma_{+} \! = \!
\left(
\begin{smallmatrix}
0 & 1 \\
0 & 0
\end{smallmatrix}
\right)$ and $\sigma_{-} \! = \!
\left(
\begin{smallmatrix}
0 & 0 \\
1 & 0
\end{smallmatrix}
\right)$ are, respectively, the raising and lowering matrices, $\mathrm{sgn}
(z) \! := \! +1$ if $z \! > \! 0$, $0$ if $z \! = \! 0$, and $-1$ if $z \!
< \! 0$, $\mathbb{R}_{\pm} \! := \! \{\mathstrut x; \, \pm x \! > \! 0\}$,
and $\pm \mi \! := \! \exp (\pm \mi \pi/2)$;
\item[(2)] for a scalar $\varpi$ and a $2 \! \times \! 2$ matrix
$\Upsilon$, $\varpi^{\mathrm{ad}(\sigma_{3})} \Upsilon \! := \!
\varpi^{\sigma_{3}} \Upsilon \varpi^{-\sigma_{3}}$;
\item[(3)] for each segment of an oriented contour $\mathcal{D}$, according
to the given orientation, the ``+'' side is to the left and the ``-'' side
is to the right as one traverses the contour in the direction of orientation,
i.e., for a matrix $\mathcal{A}_{ij}(\cdot)$, $i,j \! \in \! \{1,2\}$,
$(\mathcal{A}_{ij}(\cdot))_{\pm}$ denote the non-tangential limits
$(\mathcal{A}_{ij}(z))_{\pm} \! := \! \lim_{\genfrac{}{}{0pt}{2}{z^{\prime}
\, \to \, z}{z^{\prime} \, \in \, \pm \, \mathrm{side} \, \mathrm{of} \,
\mathcal{D}}} \mathcal{A}_{ij}(z^{\prime})$;
\item[(4)] for a matrix $\mathcal{A}_{ij}(\cdot)$, $i,j \! \in \! \{1,2\}$,
to have boundary values in the $\mathcal{L}^{2}$ sense on an oriented
contour $\mathcal{D}$, it is meant that $\lim_{\genfrac{}{}{0pt}{2}
{z^{\prime} \, \to \, z}{z^{\prime} \, \in \, \pm \, \mathrm{side} \,
\mathrm{of} \, \mathcal{D}}} \int_{\mathcal{D}} \vert \mathcal{A}(z^{
\prime}) \! - \! (\mathcal{A}(z))_{\pm} \vert^{2} \, \vert \md z \vert \!
= \! 0$, where $\vert \mathcal{A}(\cdot) \vert$ denotes the Hilbert-Schmidt
norm, $\vert \mathcal{A}(\cdot) \vert \! := \! (\sum_{i,j=1}^{2} \overline{
\mathcal{A}_{ij}(\cdot)} \, \mathcal{A}_{ij}(\cdot))^{1/2}$, with $\overline{
(\bullet)}$ denoting complex conjugation of $(\bullet)$, i.e., if, say,
$\mathcal{D} \! = \! \mathbb{R}$ oriented {}from $+\infty$ to $-\infty$,
then $\mathcal{A}(\cdot)$ has $\mathcal{L}^{2}$ boundary values on
$\mathcal{D}$ means that $\lim_{\varepsilon \downarrow 0} \int_{\mathbb{R}}
\vert \mathcal{A}(x \! \mp \! \mi \varepsilon) \! - \! (\mathcal{A}(x))_{
\pm} \vert^{2} \, \md x \! = \! 0$;
\item[(5)] for $1 \! \leqslant \! p \! < \! \infty$ and $\mathcal{D}$ some
point set,
\begin{equation*}
\mathcal{L}^{p}_{\mathrm{M}_{2}(\mathbb{C})}(\mathcal{D}) \! := \! \{
\mathstrut f \colon \mathcal{D} \! \to \! \mathrm{M}(2,\mathbb{C}); \, \vert
\vert f(\cdot) \vert \vert_{\mathcal{L}^{p}_{\mathrm{M}_{2}(\mathbb{C})}
(\mathcal{D})} \! := \! (\smallint\nolimits_{\mathcal{D}} \vert f(z) \vert^{
p} \, \vert \md z \vert)^{1/p} \! < \! \infty\},
\end{equation*}
and, for $p \! = \! \infty$,
\begin{equation*}
\mathcal{L}^{\infty}_{\mathrm{M}_{2}(\mathbb{C})}(\mathcal{D}) \! := \!
\{\mathstrut g \colon \mathcal{D} \! \to \! \mathrm{M}(2,\mathbb{C}); \,
\vert \vert g(\cdot) \vert \vert_{\mathcal{L}^{\infty}_{\mathrm{M}_{2}
(\mathbb{C})}(\mathcal{D})} \! := \! \max_{i,j \in \{1,2\}} \sup_{z \in
\mathcal{D}} \vert g_{ij}(z) \vert \! < \! \infty\};
\end{equation*}
\item[(6)] for $D$ an unbounded domain of $\mathbb{R}$, $\mathcal{S}_{
\mathbb{C}}(D)$ (respectively~$\mathcal{S}_{\mathrm{M}_{2}(\mathbb{
C})}(D))$ denotes the Schwartz space on $D$, namely, the space of all
infinitely continuously differentiable (smooth) $\mathbb{C}$-valued
(respectively~$\mathrm{M}(2,\mathbb{C})$-valued) functions which together
with all their derivatives tend to zero faster than any positive power
of $\vert \bullet \vert^{-1}$ as $\vert \bullet \vert \! \to \! \infty$,
that is, $\mathcal{S}_{\mathbb{C}}(D) \! := \! \mathbf{C}^{\infty}(D)
\cap \{ \mathstrut f \colon D \! \to \! \mathbb{C}; \, \vert \vert f
(\cdot) \vert \vert_{k,l} \! := \! \sup_{x \in \mathbb{R}} \vert x^{k}
(\tfrac{\md}{\md x})^{l}f(x) \vert \! < \! \infty \, \forall \, (k,l) \!
\in \! \mathbb{Z}_{\geqslant 0} \times \mathbb{Z}_{\geqslant 0}\}$ and
$\mathcal{S}_{\mathrm{M}_{2}(\mathbb{C})}(D) \! := \! \{ \mathstrut F
\colon D \! \to \! \mathrm{M}(2,\mathbb{C}); \, F_{ij}(\cdot) \! \in
\! \mathbf{C}^{\infty}(D), \, i,j \! \in \! \{1,2\}\} \cap \{ \mathstrut
G \colon D \! \to \! \mathrm{M}(2,\mathbb{C}); \, \vert \vert G_{ij}
(\cdot) \vert \vert_{k,l} \! := \! \sup_{x \in \mathbb{R}} \vert x^{k}
(\tfrac{\md}{\md x})^{l}G_{ij}(x) \vert \! < \! \infty \, \forall \, (k,
l) \! \in \! \mathbb{Z}_{\geqslant 0} \! \times \! \mathbb{Z}_{\geqslant
0}, \, i,j \! \in \! \{1,2\}\}$, and $\mathbf{C}^{\infty}_{0}(\ast) \! :=
\! \cap_{k=0}^{\infty} \mathbf{C}^{k}_{0}(\ast)$;
\item[(7)] for $D$ an unbounded domain of $\mathbb{R}$, $\mathcal{S}_{
\mathbb{C}}^{1}(D) \! := \! \mathcal{S}_{\mathbb{C}}(D) \cap \{\mathstrut
h(z); \, \vert \vert h(\cdot) \vert \vert_{\mathcal{L}^{\infty}(D)} \!
:= \! \sup_{z \in D} \vert h(z) \vert \linebreak[4]
< \! 1\}$;
\item[(8)] $\vert \vert \mathscr{F}(\cdot) \vert \vert_{\cap_{p \in
\mathfrak{J}} \mathcal{L}^{p}_{\mathrm{M}_{2}(\mathbb{C})}(\ast)} \! := \!
\sum_{p \in \mathfrak{J}} \vert \vert \mathscr{F}(\cdot) \vert \vert_{
\mathcal{L}^{p}_{\mathrm{M}_{2}(\mathbb{C})}(\ast)}$, where $\mathfrak{J}$
is a finite index set;
\item[(9)] for $(\mu,\widetilde{\nu}) \! \in \! \mathbb{R} \! \times \!
\mathbb{R}$, the function $(\bullet \! - \! \mu)^{\mi \widetilde{\nu}}
\colon \mathbb{C} \setminus (-\infty,\mu) \! \to \! \mathbb{C} \colon
\bullet \! \mapsto \! \me^{\mi \widetilde{\nu} \ln (\bullet-\mu)}$, with
the branch cut taken along $(-\infty,\mu)$ and the principal branch of the
logarithm chosen, $\ln (\bullet \! - \! \mu) \! := \! \ln \! \vert \!
\bullet - \mu \vert \! + \! \mi \arg (\bullet \! - \! \mu)$, $\arg (\bullet
\! - \! \mu) \! \in \! (-\pi,\pi)$;
\item[(10)] a contour, $\mathcal{D}$, say, which is the finite union of
piecewise smooth simple closed curves, is said to be \emph{orientable} if
its complement, $\mathbb{C} \setminus \mathcal{D}$, can always be divided
into two, possibly disconnected, disjoint open sets $\mho^{+}$ and $\mho^{
-}$, either of which has finitely many components, such that $\mathcal{D}$
admits an orientation so that it can either be viewed as a positively
oriented boundary $\mathcal{D}^{+}$ for $\mho^{+}$ or as a negatively
oriented boundary $\mathcal{D}^{-}$ for $\mho^{-}$ \cite{a28}, i.e., the
(possibly disconnected ) components of $\mathbb{C} \setminus \mathcal{D}$
can be coloured by $+$ or $-$ in such a way that the $+$ regions do not
share boundary with the $-$ regions, except, possibly, at finitely many
points \cite{a29}.
\end{enumerate}
\setcounter{equation}{1}
\begin{bbbbb}[$\cite{a9,a10,a30}$]
The necessary and sufficient condition for the compatibility of the
following linear system (Lax-pair), for arbitrary $\zeta \! \in \!
\mathbb{C}$,
\begin{equation}
\partial_{x} \varPsi (x,t;\zeta) \! = \! \mathcal{U}(x,t;\zeta) \varPsi
(x,t;\zeta), \qquad \partial_{t} \varPsi (x,t;\zeta) \! = \! \mathcal{V}
(x,t;\zeta) \varPsi (x,t;\zeta),
\end{equation}
where
\begin{align}
\mathcal{U}(x,t;\zeta) &= \! -\mi \lambda (\zeta) \sigma_{3} \! + \!
\begin{pmatrix}
0 & u \\
\overline{u} & 0
\end{pmatrix}, \nonumber \\
\mathcal{V}(x,t;\zeta) &= \! -2 \mi (\lambda(\zeta))^{2} \sigma_{3} \!
+ \! 2 \lambda (\zeta) \!
\begin{pmatrix}
0 & u \\
\overline{u} & 0
\end{pmatrix} \! - \! \mi
\begin{pmatrix}
u \overline{u}-1 & \partial_{x}u \\
\partial_{x} \overline{u} & u \overline{u}-1
\end{pmatrix} \! \sigma_{3}, \nonumber
\end{align}
and $\lambda (\zeta) \! := \! \tfrac{1}{2}(\zeta \! + \! \tfrac{1}
{\zeta})$, with $\mathrm{tr}(\mathcal{U}(x,t;\zeta)) \! = \! \mathrm{tr}
(\mathcal{V}(x,t;\zeta)) \! = \! 0$, is that $u \! = \! u(x,t)$ satisfies
the {\rm D${}_{f}$NLSE}.
\end{bbbbb}

\emph{Proof.} Invoking the isospectral deformation condition, $\partial_{
\ast} \zeta \! = \! 0$, $\ast \! \in \! \{x,t\}$, one shows that the
D${}_{f}$NLSE is the Frobenius compatibility, or zero-curvature, condition
for system~(2), $\partial_{t} \mathcal{U}(x,t;\zeta) \! - \! \partial_{x}
\mathcal{V}(x,t;\zeta) \! + \! [\mathcal{U}(x,t;\zeta),\mathcal{V}(x,t;
\zeta)] \! = \!
\left(
\begin{smallmatrix}
0 & 0 \\
0 & 0
\end{smallmatrix}
\right)$, $\zeta \! \in \! \mathbb{C}$, where $[\mathfrak{A},\mathfrak{B}]
\! := \! \mathfrak{A} \mathfrak{B} \! - \! \mathfrak{B} \mathfrak{A}$ is
the matrix commutator. \hfill $\square$
\begin{bbbbb}
Let $u(x,t)$ be a solution of the {\rm D${}_{f}$NLSE} and $\varPsi (x,t;
\zeta)$ the corresponding solution of system~{\rm (2)}. Then $\Psi (x,t;
\zeta) \! := \! \varPsi (x,t;\zeta) \mathcal{Q}(\zeta)$, with $\mathcal{Q}
(\zeta) \! \in \! \mathrm{M}(2,\mathbb{C})$, is also a solution of
system~{\rm (2)}.
\end{bbbbb}

\emph{Proof.} Let $u(x,t)$ be a solution of the D${}_{f}$NLSE and $\varPsi
(x,t;\zeta)$ the corresponding solution of system~(2). Multiply system~(2)
on the right by $\mathcal{Q} \! \in \! \mathrm{M}(2,\mathbb{C})$ and define
$\Psi (x,t;\zeta)$ as in the Proposition. \hfill $\square$

As a consequence of Proposition~2.2, $\Psi (x,t;\zeta)$ becomes the principal
object of study. The ISM analysis for the D${}_{f}$NLSE is based on the
direct scattering problem for the (self-adjoint) operator
(cf.~Proposition~2.1) $\mathcal{O}^{\mathcal{D}} \! := \! \mi \sigma_{3}
\partial_{x} \! - \!
\left(
\begin{smallmatrix}
0 & \mi u_{o}(x) \\
\overline{\mi u_{o}(x)} & 0
\end{smallmatrix}
\right) \! - \! \mathrm{diag} \! \left(\tfrac{1}{2}(\zeta \! + \! \tfrac{
1}{\zeta}) \right)$, where $u(x,0) \! := \! u_{o}(x)$ satisfies $u_{o}(x)
\! =_{x \to \pm \infty} \! u_{o}(\pm \infty)(1 \! + \! o(1))$, with $u_{o}
(\pm \infty) \! := \! \exp (\tfrac{\mi (1 \mp 1) \theta}{2})$, $\theta \!
\in \! [0,2 \pi)$ (see Lemma~2.2), $u_{o}(x) \! \in \! \mathbf{C}^{\infty}
(\mathbb{R})$, and $u_{o}(x) \! - \! u_{o}(\pm \infty) \! \in \! \mathcal{
S}_{\mathbb{C}}(\mathbb{R}_{\pm})$.
\begin{bbbbb}
Let $u(x,t)$ be a solution of the {\rm D${}_{f}$NLSE} and $\Psi
(x,t;\zeta)$ the corresponding solution of system~{\rm (2)} defined
in Proposition~{\rm 2.2}. Then $\Psi (x,t;\zeta)$ satisfies the
symmetry reductions $\sigma_{1} \overline{\Psi (x,t;\overline{
\zeta})} \, \sigma_{1} \! = \! \Psi (x,t;\zeta) \mathrm{M}_{1}(\zeta)$
and $\Psi (x,t;\tfrac{1}{\zeta}) \! = \! \Psi (x,t;\zeta) \mathrm{M}_{
2}(\zeta)$, where $\mathrm{M}_{i}(\zeta) \! \in \! \mathrm{GL}
(2,\mathbb{C})$, $i \! \in \! \{1,2\}$.
\end{bbbbb}

\emph{Proof.} For the $\zeta \! \to \! \overline{\zeta}$
(respectively~$\zeta \! \to \! \tfrac{1}{\zeta})$ involution, one shows
that $\partial_{x}(\sigma_{1} \overline{\Psi (x,t;\overline{\zeta})} \,
\sigma_{1}) \! = \! \mathcal{U}(x,t;\zeta) \sigma_{1} \overline{\Psi (x,
t;\overline{\zeta})} \, \sigma_{1}$ and $\partial_{t}(\sigma_{1} \overline{
\Psi (x,t;\overline{\zeta})} \, \sigma_{1}) \! = \! \mathcal{V}(x,t;\zeta)
\sigma_{1} \overline{\Psi (x,t;\overline{\zeta})} \, \sigma_{1}$
(respectively~$\partial_{x} \Psi (x,\linebreak[4]
t;\tfrac{1}{\zeta}) \! = \! \mathcal{U}(x,t;\zeta) \Psi (x,t;\tfrac{1}{
\zeta})$ and $\partial_{t} \Psi (x,t;\tfrac{1}{\zeta}) \! = \! \mathcal{
V}(x,t;\zeta) \Psi (x,t;\tfrac{1}{\zeta}))$; hence, $\exists \, \mathrm{
M}_{1}(\zeta) \! \in \! \mathrm{GL}(2,\mathbb{C})$
(respectively~$\mathrm{M}_{2}(\zeta) \! \in \! \mathrm{GL}(2,\mathbb{C}))$
such that $\sigma_{1} \overline{\Psi (x,t;\overline{\zeta})} \, \sigma_{1}
\! = \! \Psi (x,t;\zeta) \mathrm{M}_{1}(\zeta)$ (respectively~$\Psi (x,t;
\tfrac{1}{\zeta}) \linebreak[4]
= \! \Psi (x,t;\zeta) \mathrm{M}_{2}(\zeta))$ solves system~(2). \hfill
$\square$
\begin{aaaaa}
Let $u(x,t)$ be a solution of the {\rm D${}_{f}$NLSE} with $u(x,0) \! := \!
u_{o}(x) \! =_{x \to \pm \infty} \! u_{o}(\pm \infty)(1 \! + \! o(1))$,
where $u_{o}(\pm \infty) \! := \! \exp (\tfrac{\mi (1 \mp 1) \theta}{2})$,
$\theta \! \in \! [0,2 \pi)$ (see Lemma~{\rm 2.2)}, $u_{o}(x) \! \in \!
\mathbf{C}^{\infty}(\mathbb{R})$, and $u_{o}(x) \! - \! u_{o}(\pm \infty) \!
\in \! \mathcal{S}_{\mathbb{C}}(\mathbb{R}_{\pm})$. Define the $\mathrm{M}
(2,\mathbb{C})$-valued functions $\Psi^{\pm}(x,0;\zeta)$ as the (Jost)
solutions of the first equation of system~{\rm (2)}, $\mathcal{O}^{\mathcal{
D}} \Psi^{\pm}(x,0;\zeta) \! = \!
\left(
\begin{smallmatrix}
0 & 0 \\
0 & 0
\end{smallmatrix}
\right)$, with the following asymptotics,
\begin{equation*}
\Psi^{\pm}(x,0;\zeta) \! \underset{x \, \to \, \pm \infty}{=} \! \left(
\me^{\frac{\mi (1 \mp 1) \theta}{4} \sigma_{3}} \!
\left(
\begin{smallmatrix}
1 & -\mi \zeta^{-1} \\
\mi \zeta^{-1} & 1
\end{smallmatrix}
\right)
\! + \! o(1) \right) \! \me^{-\mi k(\zeta)x \sigma_{3}},
\end{equation*}
where $k(\zeta) \! := \! \tfrac{1}{2}(\zeta \! - \! \tfrac{1}{\zeta})$.
\end{aaaaa}
\begin{fffff}
$\sigma_{1} \overline{\Psi (x,t;\overline{\zeta})} \, \sigma_{1} \! =
\! \Psi (x,t;\zeta)$ and $\Psi (x,t;\tfrac{1}{\zeta}) \! = \! \zeta
\Psi (x,t;\zeta) \sigma_{2}$.
\end{fffff}

\emph{Proof.} Since, {}from Definition~2.1, $\Psi^{\pm}(x,0;\zeta)$
satisfy $\mathcal{O}^{\mathcal{D}} \Psi^{\pm}(x,0;\zeta) \! = \! \left(
\begin{smallmatrix}
0 & 0 \\
0 & 0
\end{smallmatrix}
\right)$, and, as a consequence of Proposition~2.3, $\sigma_{1} \overline{
\Psi^{\pm}(x,0;\overline{\zeta})} \, \sigma_{1} \! = \! \Psi^{\pm}(x,0;
\zeta) \mathrm{M}_{1}(\zeta)$ (respectively~$\Psi^{\pm}(x,0;\tfrac{1}{
\zeta}) \! = \! \Psi^{\pm}(x,0;\zeta) \mathrm{M}_{2}(\zeta))$, one uses
the asymptotics for $\Psi^{\pm}(x,0;\zeta)$ given in Definition~2.1
and the fact that $\overline{k(\overline{\zeta})} \! = \! k(\zeta)$
(respectively~$k(\tfrac{1}{\zeta}) \! = \! -k(\zeta))$ to deduce that
$\mathrm{M}_{1}(\zeta) \! = \! \mathrm{I}$ (respectively~$\mathrm{M}_{2}
(\zeta) \! = \! \zeta \sigma_{2})$. \hfill $\square$
\begin{bbbbb}
Set $\Psi^{\pm}(x,0;\zeta) \! := \!
\left(
\begin{smallmatrix}
\Psi^{\pm}_{11}(\zeta) & \Psi^{\pm}_{12}(\zeta) \\
\Psi^{\pm}_{21}(\zeta) & \Psi^{\pm}_{22}(\zeta)
\end{smallmatrix}
\right)$. Then
$\left(
\begin{smallmatrix}
\Psi_{12}^{+}(\zeta) \\
\Psi^{+}_{22}(\zeta)
\end{smallmatrix}
\right)$ and
$\left(
\begin{smallmatrix}
\Psi_{11}^{-}(\zeta) \\
\Psi^{-}_{21}(\zeta)
\end{smallmatrix}
\right)$ have analytic continuation to $\mathbb{C}_{+}$
(respectively~$\left(
\begin{smallmatrix}
\Psi_{11}^{+}(\zeta) \\
\Psi^{+}_{21}(\zeta)
\end{smallmatrix}
\right)$ and~$\left(
\begin{smallmatrix}
\Psi_{12}^{-}(\zeta) \\
\Psi^{-}_{22}(\zeta)
\end{smallmatrix}
\right)$ have analytic continuation to $\mathbb{C}_{-})$, the monodromy
(scattering) matrix, $\mathrm{T}(\zeta)$, is defined by $\Psi^{-}(x,0;
\zeta) \! := \! \Psi^{+}(x,0;\zeta) \mathrm{T}(\zeta)$, $\Im (\zeta) \!
= \! 0$, where $\mathrm{T}(\zeta) \! = \! \left(
\begin{smallmatrix}
a(\zeta) & \overline{b(\overline{\zeta})} \\
b(\zeta) & \overline{a(\overline{\zeta})}
\end{smallmatrix}
\right)$, with $a(\zeta) \! = \! (1 \! - \! \zeta^{-2})^{-1}(\Psi_{22}^{+}
(\zeta) \Psi_{11}^{-}(\zeta) \! - \! \Psi_{12}^{+}(\zeta) \Psi_{21}^{-}
(\zeta))$, $b(\zeta) \! = \! (1 \! - \! \zeta^{-2})^{-1}(\overline{\Psi_{
22}^{+}(\overline{\zeta})} \, \Psi_{21}^{-}(\zeta) \! - \! \overline{\Psi
_{12}^{+}(\overline{\zeta})} \, \Psi_{11}^{-}(\zeta))$, $\vert a(\zeta)
\vert^{2} \! - \! \vert b(\zeta) \vert^{2} \! = \! 1$, $a(\tfrac{1}{\zeta}
) \! = \! \overline{a(\overline{\zeta})}$, $b(\tfrac{1}{\zeta}) \! = \!
-\overline{b(\overline{\zeta})}$, and $\det (\Psi^{\pm}(x,0;\zeta)) \vert
_{\zeta = \pm 1} \! = \! 0$.
\end{bbbbb}

\emph{Proof.} The analytic continuation of the respective columns
of $\Psi^{\pm}(x,0;\zeta)$ to $\mathbb{C}_{\pm}$ follows {}from
Definition~2.1. Introduce the monodromy matrix according to $\Psi^{+}
(x,0;\zeta) \mathrm{T}(\zeta) \! = \! \Psi^{-}(x,0;\zeta)$, $\Im (\zeta)
\! = \! 0$, where $\mathrm{T}(\zeta) \! = \!
\left(
\begin{smallmatrix}
a(\zeta) & \widetilde{b}(\zeta) \\
b(\zeta) & \widetilde{a}(\zeta)
\end{smallmatrix}
\right)$. {}From the $\sigma_{1}$ symmetry reduction $\sigma_{1}
\overline{\Psi^{\pm}(x,0;\overline{\zeta})}\\
\cdot \sigma_{1} \! = \! \Psi^{\pm}(x,0;\zeta)$, it follows
that $\widetilde{a}(\zeta) \! = \! \overline{a(\overline{\zeta})
}$ and $\widetilde{b}(\zeta) \! = \! \overline{b(\overline{\zeta})}$;
hence, the expression for $\mathrm{T}(\zeta)$ given in the Proposition.
Since $(\Psi^{+}(x,0;\zeta))^{-1} \Psi^{-}(x,0;\zeta) \! = \! \mathrm{
T}(\zeta) \! = \!
\left(
\begin{smallmatrix}
a(\zeta) & \overline{b(\overline{\zeta})} \\
b(\zeta) & \overline{a(\overline{\zeta})}
\end{smallmatrix}
\right)$ and $\det (\Psi^{+}(x,0;\zeta)) \! = \! \det (\Psi^{-}(x,0;\zeta))
\! = \! 1 \! - \! \zeta^{-2}$, namely, $\det (\mathrm{T}(\zeta)) \! = \!
1$, one deduces the expressions for $a(\zeta)$ and $b(\zeta)$ given in
the Proposition, and, using the unimodularity of $\mathrm{T}(\zeta)$, one
deduces that $a(\zeta) \overline{a(\overline{\zeta})} \! - \! b(\zeta)
\overline{b(\overline{\zeta})} \! = \! 1$, $\Im (\zeta) \! = \! 0$. Using
the $\sigma_{2}$ symmetry reduction, $\Psi^{\pm}(x,0;\tfrac{1}{\zeta})
\! = \! \zeta \Psi^{\pm}(x,0;\zeta) \sigma_{2}$, and the expression for
$\mathrm{T}(\zeta)$ given in the Proposition, one shows that $a(\tfrac{
1}{\zeta}) \! = \! \overline{a(\overline{\zeta})}$ and $b(\tfrac{1}{\zeta})
\! = \! - \overline{b(\overline{\zeta})}$: finally, since $\det (\Psi^{
\pm}(x,0;\zeta)) \! = \! 1 \! - \! \zeta^{-2}$, setting $\zeta \! = \!
\pm 1$, the degeneracy of $\Psi^{\pm}(x,0;\zeta)$ at $\zeta \! = \! \pm
1$ follows. \hfill $\square$
\begin{fffff}
Let the reflection coefficient associated with the direct scattering
problem for the operator $\mathcal{O}^{\mathcal{D}}$ be defined by $r(\zeta)
\! := \! \tfrac{b(\zeta)}{a(\zeta)}$. Then $r(\tfrac{1}{\zeta}) \! = \!
-\overline{r(\overline{\zeta})}$.
\end{fffff}

\emph{Proof.} The relation $r(\tfrac{1}{\zeta}) \! = \! -\overline{r
(\overline{\zeta})}$ is an immediate consequence of the definition of
$r(\zeta)$ given in the Corollary and the properties of $a(\zeta)$ and
$b(\zeta)$ given in Proposition~2.4. \hfill $\square$
\begin{eeeee}
Note that, {}from Proposition~2.4, even though $a(\zeta)$
(respectively~$a^{\ast}(\zeta) \! := \! \overline{a(\overline{\zeta})})$
has an analytic continuation off $\Im (\zeta) \! = \! 0$ to $\mathbb{C}_{
+}$ (respectively~$\mathbb{C}_{-})$ and is continuous on $\overline{\mathbb{
C}}_{+}$ (respectively~$\overline{\mathbb{C}}_{-})$, in general, $b(\zeta)$
does not have an analytic continuation off $\Im (\zeta) \! = \! 0$: in this
work, $b(\zeta)$ has an analytic continuation to (compact subsets of)
$\{\mathstrut \zeta; \, \vert \zeta \vert \! \leqslant \! 1\}$; in
particular, to rays of the form $r_{n} \me^{\pm \mi \phi_{n}}$, $n \! \in \!
\{1,2,\ldots,N\}$, where $(r_{n},\phi_{n}) \! \in \! [0,1] \times (0,\pi)$.
\end{eeeee}
\begin{ccccc}
Let $u(x,t)$ be the solution of the Cauchy problem for the {\rm D${}_{f}$NLSE}
with finite-density initial data and $\Psi^{\pm}(x,0;\zeta)$ the corresponding
(Jost) solutions of $\mathcal{O}^{\mathcal{D}} \Psi^{\pm}(x,0;\zeta) \! = \!
\left(
\begin{smallmatrix}
0 & 0 \\
0 & 0
\end{smallmatrix}
\right)$ given in Definition~{\rm 2.1}. Then $\Psi^{\pm}(x,0;\zeta)$
have the following asymptotics:
\begin{align}
\Psi^{-}(x,0;\zeta) \! &\underset{\zeta \, \to \, \infty}{=} \! \me^{\frac{
\mi \theta}{2} \sigma_{3}} \! \left( \mathrm{I} \! + \! \tfrac{1}{\zeta}
\left(
\begin{smallmatrix}
\mi \int_{-\infty}^{x}(\vert u_{o}(\xi) \vert^{2}-1) \, \md \xi &
-\mi u_{o}(x) \me^{-\mi \theta} \\
\mi \overline{u_{o}(x)} \, \me^{\mi \theta} & -\mi \int_{-\infty}^{x}
(\vert u_{o}(\xi) \vert^{2}-1) \, \md \xi
\end{smallmatrix}
\right) \! + \! \mathcal{O}(\zeta^{-2}) \right) \! \me^{-\mi k(\zeta)x
\sigma_{3}}, \nonumber \\
\Psi^{+}(x,0;\zeta) \! &\underset{\zeta \, \to \, \infty}{=} \! \left(
\mathrm{I} \! + \! \tfrac{1}{\zeta}
\left(
\begin{smallmatrix}
\mi \int_{+\infty}^{x}(\vert u_{o}(\xi) \vert^{2}-1) \, \md \xi &
-\mi u_{o}(x) \\
\mi \overline{u_{o}(x)} & -\mi \int_{+\infty}^{x}(\vert u_{o}(\xi)
\vert^{2}-1) \, \md \xi
\end{smallmatrix}
\right) \! + \! \mathcal{O}(\zeta^{-2}) \right) \! \me^{-\mi k(\zeta)x
\sigma_{3}}, \nonumber \\
\Psi^{-}(x,0;\zeta) &\underset{\zeta \, \to \, 0}{=} \! \left( \tfrac{
1}{\zeta} \sigma_{2} \me^{-\frac{\mi \theta}{2} \sigma_{3}} \! + \!
\mathcal{O}(1) \right) \! \me^{-\mi k(\zeta)x \sigma_{3}}, \qquad
\Psi^{+}(x,0;\zeta) \! \underset{\zeta \, \to \, 0}{=} \! \left( \tfrac{1}{
\zeta} \sigma_{2} \! + \! \mathcal{O}(1) \right) \! \me^{-\mi k(\zeta)
x \sigma_{3}}. \nonumber
\end{align}
\end{ccccc}

\emph{Proof.} {}From the asymptotics of the (Jost) solutions of
$\mathcal{O}^{\mathcal{D}} \Psi^{\pm}(x,0;\zeta) \! = \! \left(
\begin{smallmatrix}
0 & 0 \\
0 & 0
\end{smallmatrix}
\right)$ given in Definition~2.1, the $\sigma_{1}$ symmetry reduction
(Corollary~2.1), and a Volterra-type integral representation, one arrives
at $\Psi^{\pm}(x,0;\zeta) \! = \!
\left(
\begin{smallmatrix}
\Psi^{\pm}_{11}(x,0;\zeta) & \Psi^{\pm}_{12}(x,0;\zeta) \\
\Psi^{\pm}_{21}(x,0;\zeta) & \Psi^{\pm}_{22}(x,0;\zeta)
\end{smallmatrix}
\right)$, with
\begin{align}
\Psi^{-}_{11}(x,0;\zeta) &= \overline{\Psi^{-}_{22}(x,0;\overline{
\zeta})} = \exp (\tfrac{\mi \theta}{2}) \exp \! \left( \mi \left(
\int_{-\infty}^{x} \mathscr{S}(\xi;\zeta) \, \md \xi \! - \! k(\zeta)
x \right) \right), \nonumber \\
\Psi^{-}_{21}(x,0;\zeta) &= \overline{\Psi^{-}_{12}(x,0;\overline{\zeta}
)} = \exp (-\tfrac{\mi \theta}{2}) \mathcal{A}(x;\zeta) \exp \! \left(
\mi \left( \int_{-\infty}^{x} \mathscr{S}(\xi;\zeta) \, \md \xi \! - \!
k(\zeta)x \right) \right), \nonumber \\
\Psi^{+}_{12}(x,0;\zeta) &= \overline{\Psi^{+}_{21}(x,0;\overline{\zeta}
)} = \mathcal{B}(x;\zeta) \exp \! \left( \mi \left( \int_{+\infty}^{x}
\mathscr{T}(\xi;\zeta) \, \md \xi \! + \! k(\zeta)x \right) \right),
\nonumber \\
\Psi^{+}_{22}(x,0;\zeta) &= \overline{\Psi^{+}_{11}(x,0;\overline{\zeta}
)} = \exp \! \left( \mi \left( \int_{+\infty}^{x} \mathscr{T}(\xi;\zeta)
\, \md \xi \! + \! k(\zeta)x \right) \right), \nonumber
\end{align}
where, in the neighbourhood of the singular points, namely, the origin
$(\zeta \! = \! 0)$ and the point at infinity $(\zeta \! = \! \infty)$,
$\mathcal{A}(x;\zeta)$, $\mathcal{B}(x;\zeta)$, $\mathscr{S}(x;\zeta)$
and $\mathscr{T}(x;\zeta)$ have the asymptotic expansions given below.
Since $\Psi^{\pm}(x,0;\zeta)$ are the solutions of $\mathcal{O}^{\mathcal{
D}} \Psi^{\pm}(x,0;\zeta) \! = \! \left(
\begin{smallmatrix}
0 & 0 \\
0 & 0
\end{smallmatrix}
\right)$ with the asymptotics given in Definition~2.1, using the matrix
representations for $\Psi^{\pm}(x,0;\zeta)$ given above, one arrives at
the following system of equations (as well as their complex conjugates):
\begin{gather}
\mathcal{A}(x;\zeta) \! \left( -\tfrac{\mi}{2}(\zeta \! + \! \tfrac{1}{
\zeta}) \! + \! u_{o}(x) \me^{-\mi \theta} \mathcal{A}(x;\zeta) \right)
\! + \! \partial_{x} \mathcal{A}(x;\zeta) \! = \! \tfrac{\mi}{2}(\zeta
\! + \! \tfrac{1}{\zeta}) \mathcal{A}(x;\zeta) \! + \! \overline{u_{o}
(x)} \, \me^{\mi \theta}, \nonumber \\
\mathcal{B}(x;\zeta) \! \left( \tfrac{\mi}{2}(\zeta \! + \! \tfrac{1}{
\zeta}) \! + \! \overline{u_{o}(x)} \, \mathcal{B}(x;\zeta) \right) \!
+ \! \partial_{x} \mathcal{B}(x;\zeta) \! = \! -\tfrac{\mi}{2}(\zeta \!
+ \! \tfrac{1}{\zeta}) \mathcal{B}(x;\zeta) \! + \! u_{o}(x), \tag{L2.1a}
\\
\mi \mathscr{S}(x;\zeta) \! = \! -\tfrac{\mi}{\zeta} \! + \! u_{o}(x)
\me^{-\mi \theta} \mathcal{A}(x;\zeta), \qquad \mi \mathscr{T}(x;\zeta)
\! = \! \tfrac{\mi}{\zeta} \! + \! \overline{u_{o}(x)} \, \mathcal{B}(x;
\zeta). \nonumber
\end{gather}
In the neighbourhood of $\zeta \! = \! \infty$, one has the following
(formal) asymptotic expansions (see Theorem~2.9 in \cite{a29}: Note,
$v(x)$ in Eq.~(2.10) of \cite{a29} should be changed to $v(z))$,
$\mathcal{A}(x;\zeta) \! = \! \sum_{n=1}^{\infty} \widetilde{I}_{n}^{a}
[u,\overline{u}]\zeta^{-n} \! + \! \mathcal{O}(\vert \zeta \vert^{-\infty})$,
$\mathcal{B}(x;\zeta) \! = \! \sum_{n=1}^{\infty} \widetilde{I}_{n}^{b}[u,
\overline{u}] \zeta^{-n} \! + \! \mathcal{O}(\vert \zeta \vert^{-\infty})$,
$\mathscr{S}(x;\zeta) \! = \! \sum_{n=1}^{\infty} \widetilde{I}_{n}^{s}[u,
\overline{u}] \zeta^{-n} \! + \! \mathcal{O}(\vert \zeta \vert^{-\infty})$,
and $\mathscr{T}(x;\zeta) \! = \! \sum_{n=1}^{\infty} \widetilde{I}_{n}^{
t}[u,\overline{u}] \zeta^{-n} \! + \! \mathcal{O}(\vert \zeta \vert^{-
\infty})$, where $\widetilde{I}_{n}^{\star}[u,\overline{u}] \! := \!
\star^{\infty}(x)$, $\star \! \in \! \{a,b,s,t\}$, are functionals of
$u_{o}(x)$ and $\overline{u_{o}(x)}$, and, in the neighbourhood of $\zeta
\! = \! 0$, one has the following (formal) asymptotic expansions, $\mathcal{
A}(x;\zeta) \! = \! \sum_{n=-1}^{\infty} \widehat{I}_{n}^{a}[u,\overline{u}]
\zeta^{n} \! + \! \mathcal{O}(\vert \zeta \vert^{+\infty})$, $\widehat{I}_{
-1}^{a}[u,\overline{u}] \! \not\equiv \! 0$, $\mathcal{B}(x;\zeta) \! = \!
\sum_{n=-1}^{\infty} \widehat{I}_{n}^{b}[u,\overline{u}] \zeta^{n} \! + \!
\mathcal{O}(\vert \zeta \vert^{+\infty})$, $\widehat{I}_{-1}^{b}[u,
\overline{u}] \! \not\equiv \! 0$, $\mathscr{S}(x;\zeta) \! = \! \sum_{n=-
1}^{\infty} \widehat{I}_{n}^{s}[u,\overline{u}] \zeta^{n} \! + \! \mathcal{
O}(\vert \zeta \vert^{+\infty})$, and $\mathscr{T}(x;\zeta) \! = \! \sum_{n=
-1}^{\infty} \widehat{I}_{n}^{t}[u,\overline{u}] \zeta^{n} \! + \! \mathcal{
O}(\vert \zeta \vert^{+\infty})$, where $\widehat{I}_{n}^{\star}[u,\overline{
u}] \! := \! \star^{0}(x)$ are functionals of $u_{o}(x)$ and $\overline{u_{
o}(x)}$. Substituting the above asymptotic expansions as $\zeta \! \to \!
\infty$ into system~(L2.1a), one arrives at (for the first
few---leading---terms), with $d_{x}^{n} \! := \! (\tfrac{\md}{\md x})^{n}$,
\begin{align*}
\mathcal{O}(1): -\mi a_{1}^{\infty}(x) &= \overline{u_{o}(x)} \, \me^{
\mi \theta} \Rightarrow a_{1}^{\infty}(x) \! = \! \mi \overline{u_{o}
(x)} \, \me^{\mi \theta}, \quad \mi b_{1}^{\infty}(x) \! = \! u_{o}(x)
\Rightarrow b_{1}^{\infty}(x) \! = \! -\mi u_{o}(x), \nonumber \\
\mathcal{O}(\tfrac{1}{\zeta}): -\mi a_{2}^{\infty}(x) &+ \! d_{x}a_{
1}^{\infty}(x) \! = \! 0 \Rightarrow a_{2}^{\infty}(x) \! = \! \me^{
\mi \theta} d_{x} \overline{u_{o}(x)}, \nonumber \\
\mi b_{2}^{\infty}(x) &+ \! d_{x}b_{1}^{\infty}(x) \! = \! 0
\Rightarrow b_{2}^{\infty}(x) \! = \! d_{x}u_{o}(x), \nonumber \\
\mi s_{1}^{\infty}(x) &= \! -\mi \! + \! u_{o}(x) \me^{-\mi \theta} a_{
1}^{\infty}(x) \Rightarrow s_{1}^{\infty}(x) \! = \! -1 \! + \! \vert
u_{o}(x) \vert^{2}, \nonumber \\
\mi t_{1}^{\infty}(x) &= \! \mi \! + \! u_{o}(x)b_{1}^{\infty}(x)
\Rightarrow t_{1}^{\infty}(x) \! = \! 1 \! - \! \vert u_{o}(x) \vert^{
2}, \\
\mathcal{O}(\tfrac{1}{\zeta^{2}}): -\mi a_{3}^{\infty}(x) &- \! \mi a_{
1}^{\infty}(x) \! + \! u_{o}(x) \me^{-\mi \theta}(a_{1}^{\infty}(x))^{2}
\! + \! d_{x}a_{2}^{\infty}(x) \! = \! 0 \Rightarrow \nonumber \\
a_{3}^{\infty}(x) &= \! -\mi \me^{\mi \theta} \! \left(\overline{u_{o}
(x)} \! - \! \vert u_{o}(x) \vert^{2} \, \overline{u_{o}(x)} + \! d_{
x}^{2}u_{o}(x) \right), \nonumber \\
\mi b_{3}^{\infty}(x) &+ \! \mi b_{1}^{\infty}(x) \! + \! \overline{
u_{o}(x)}(b_{1}^{\infty}(x))^{2} \! + \! d_{x}b_{2}^{\infty}(x) \! =
\! 0 \Rightarrow \nonumber \\
b_{3}^{\infty}(x) &= \! \mi \! \left( u_{o}(x) \! - \! \vert u_{o}(x)
\vert^{2}u_{o}(x) \! + \! d_{x}^{2}u_{o}(x) \right), \nonumber \\
\mi s_{2}^{\infty}(x) &= \! u_{o}(x) \me^{-\mi \theta}a_{2}^{\infty}(x)
\Rightarrow s_{2}^{\infty}(x) \! = \! -\mi u_{o}(x) d_{x} \overline{u_{
o}(x)}, \nonumber \\
\mi t_{2}^{\infty}(x) &= \overline{u_{o}(x)}b_{2}^{\infty} \Rightarrow
t_{2}^{\infty}(x) \! = \! -\mi \overline{u_{o}(x)} d_{x}u_{o}(x); \nonumber
\end{align*}
hence, {}from these $\zeta \! \to \! \infty$ results and the Volterra-type
integral representation for $\Psi^{\pm}(x,0;\zeta)$ given at the beginning
of the Lemma, one shows that
\begin{align}
\Psi^{-}(x,0;\zeta) \! \underset{\zeta \, \to \, \infty}{=}& \! \me^{\frac{
\mi \theta}{2} \sigma_{3}} \left( \!
\left(
\begin{smallmatrix}
\me^{\frac{\mi}{\zeta} \int_{-\infty}^{x}(\vert u_{o}(\xi) \vert^{2}-1)
\, \md \xi} & \frac{1}{\zeta}(-\mi u_{o}(x)) \me^{-\mi (\theta +\frac{1}
{\zeta} \int_{-\infty}^{x}(\vert u_{o}(\xi) \vert^{2}-1) \, \md \xi)} \\
\frac{1}{\zeta}(\mi \overline{u_{o}(x)}) \me^{\mi (\theta +\frac{1}
{\zeta} \int_{-\infty}^{x}(\vert u_{o}(\xi) \vert^{2}-1) \, \md \xi)} &
\me^{-\frac{\mi}{\zeta} \int_{-\infty}^{x}(\vert u_{o}(\xi) \vert^{2}-1)
\, \md \xi}
\end{smallmatrix}
\right) \right. \nonumber \\
 +& \left. \mathcal{O}(\zeta^{-2}) \right) \! \me^{-\mi k(\zeta)x
\sigma_{3}}, \nonumber \\
\Psi^{+}(x,0;\zeta) \! \underset{\zeta \, \to \, \infty}{=}& \! \left( \!
\left(
\begin{smallmatrix}
\me^{\frac{\mi}{\zeta} \int_{+\infty}^{x}(\vert u_{o}(\xi) \vert^{2}-
1) \, \md \xi} & \frac{1}{\zeta}(-\mi u_{o}(x)) \me^{-\frac{\mi}{\zeta}
\int_{+\infty}^{x}(\vert u_{o}(\xi) \vert^{2}-1) \, \md \xi} \\
\frac{1}{\zeta}(\mi \overline{u_{o}(x)}) \me^{\frac{\mi}{\zeta} \int_{
+\infty}^{x}(\vert u_{o}(\xi) \vert^{2}-1) \, \md \xi} & \me^{-\frac{\mi}
{\zeta} \int_{+\infty}^{x}(\vert u_{o}(\xi) \vert^{2}-1) \, \md \xi}
\end{smallmatrix}
\right) \right. \nonumber \\
 +& \left. \mathcal{O}(\zeta^{-2}) \right) \! \me^{-\mi k(\zeta)x
\sigma_{3}}: \nonumber
\end{align}
finally, expanding the exponentials in power series in $\zeta^{-1}$, one
obtains the $\zeta \! \to \! \infty$ asymptotics stated in the Lemma.
Similarly, substituting the asymptotic expansions as $\zeta \! \to \! 0$
into system~(L2.1a), one arrives at (for the first few---leading---terms),
\begin{align}
\mathcal{O}(\tfrac{1}{\zeta^{2}}): -\mi &a_{-1}^{0}(x) \! + \! (a_{-1}^{
0}(x))^{2}u_{o}(x) \me^{-\mi \theta} \! = \! 0 \Rightarrow a_{-1}^{0}(x)
\! = \! \mi \me^{\mi \theta}(u_{o}(x))^{-1}, \nonumber \\
\mi &b_{-1}^{0}(x) \! + \! (b_{-1}^{0}(x))^{2} \, \overline{u_{o}(x)} \!
= \! 0 \Rightarrow b_{-1}^{0}(x) \! = \! -\mi (\overline{u_{o}(x)})^{-1},
\nonumber \\
\mathcal{O}(\tfrac{1}{\zeta}): -\mi &a_{0}^{0}(x) \! + \! 2u_{o}(x) \me^{
-\mi \theta}a_{-1}^{0}(x)a_{0}^{0}(x) \! + \! d_{x}a_{-1}^{0}(x) \! = \!
0 \Rightarrow a_{0}^{0}(x) \! = \! -\me^{\mi \theta} d_{x}((u_{o}(x))^{-
1}), \nonumber \\
\mi &b_{0}^{0}(x) \! + \! 2 \overline{u_{o}(x)} \, b_{-1}^{0}(x)b_{0}^{0}
(x) \! + \! d_{x}b_{-1}^{0}(x) \! = \! 0 \Rightarrow b_{0}^{0}(x) \! = \!
-d_{x}((\overline{u_{o}(x)})^{-1}), \nonumber \\
\mi &s_{-1}^{0}(x) \! = \! -\mi \! + \! u_{o}(x) \me^{-\mi \theta}a_{-1}
^{0}(x) \Rightarrow s_{-1}^{0}(x) \! = \! 0, \nonumber \\
\mi &t_{-1}^{0}(x) \! = \! \mi \! + \! \overline{u_{o}(x)} \, b_{-1}^{0}
(x) \Rightarrow t_{-1}^{0}(x) \! = \! 0, \nonumber \\
\mathcal{O}(1): \! \quad \mi &s_{0}^{0}(x) \! = \! u_{o}(x) \me^{-\mi
\theta}a_{0}^{0}(x) \Rightarrow s_{0}^{0}(x) \! = \! -\mi d_{x} \ln
(u_{o}(x)), \nonumber \\
\mi &t_{0}^{0}(x) \! = \! \overline{u_{o}(x)} \, b_{0}^{0}(x)
\Rightarrow t_{0}^{0}(x) \! = \! -\mi d_{x} \ln (\overline{u_{o}
(x)}), \nonumber
\end{align}
and the expressions for $a_{1}^{0}(x)$ and $b_{1}^{0}(x)$, resulting
{}from the $\mathcal{O}(1)$ terms, have not been written as they will
not actually be used; hence, {}from these $\zeta \! \to \! 0$ results, and
the Volterra-type integral representation for $\Psi^{\pm}(x,0;\zeta)$
given at the beginning of the Lemma, one shows that
\begin{align}
\Psi^{-}(x,0;\zeta) &\underset{\zeta \, \to \, 0}{=} \! \me^{\frac{
\mi \theta}{2} \sigma_{3}} \!
\left(
\begin{smallmatrix}
\mathcal{O}(1) & -\tfrac{1}{\zeta} \tfrac{\mi \me^{-\mi \theta}}{
\overline{u_{o}(x)}} \me^{\int_{-\infty}^{x} d_{\xi} \ln (\overline{
u_{o}(\xi)}) \, \md \xi}+\mathcal{O}(1) \\
\tfrac{1}{\zeta} \tfrac{\mi \me^{\mi \theta}}{u_{o}(x)} \me^{\int_{
-\infty}^{x} d_{\xi} \ln (u_{o}(\xi)) \, \md \xi}+\mathcal{O}(1) &
\mathcal{O}(1)
\end{smallmatrix}
\right) \! \me^{-\mi k(\zeta)x \sigma_{3}}, \nonumber \\
\Psi^{+}(x,0;\zeta) &\underset{\zeta \, \to \, 0}{=} \!
\left(
\begin{smallmatrix}
\mathcal{O}(1) & -\tfrac{1}{\zeta} \tfrac{\mi}{\overline{u_{o}(x)}}
\me^{\int_{+\infty}^{x} d_{\xi} \ln (\overline{u_{o}(\xi)}) \, \md
\xi}+\mathcal{O}(1) \\
\tfrac{1}{\zeta} \tfrac{\mi}{u_{o}(x)} \me^{\int_{+\infty}^{x}d_{\xi}
\ln (u_{o}(\xi)) \, \md \xi}+\mathcal{O}(1) & \mathcal{O}(1)
\end{smallmatrix}
\right) \! \me^{-\mi k(\zeta)x \sigma_{3}}: \nonumber
\end{align}
finally, using the fact that $\int_{\pm \infty}^{x}d_{\xi} \ln (u_{o}(\xi))
\, \md \xi \! = \! \ln (u_{o}(x)) \! - \! \ln (u_{o}(\pm \infty)) \! = \!
\ln (u_{o}(x)) \! - \! \tfrac{\mi (1 \mp 1) \theta}{2}$ (having taken the
principal branch for $\ln (\cdot))$, one obtains the $\zeta \! \to \! 0$
asymptotics stated in the Lemma. \hfill $\square$
\begin{fffff}
The following asymptotics are valid:
\begin{gather}
a(\zeta) \! \underset{\zeta \, \to \, \infty}{=} \! \me^{\frac{\mi
\theta}{2}} \left( 1 \! + \! \left( \mi \int\nolimits_{-\infty}^{+
\infty}(\vert u_{o}(\xi) \vert^{2} \! - \! 1) \, \md \xi \right)
\zeta^{-1} \! + \! \mathcal{O}(\zeta^{-2}) \right), \quad a(\zeta) \!
\underset{\zeta \, \to \, 0}{=} \! \me^{-\frac{\mi \theta}{2}}(1 \! +
\! \mathcal{O}(\zeta)), \nonumber \\
r(\zeta) \! \underset{\zeta \, \to \, \infty}{=} \! \mathcal{O}(\zeta^{
-1}), \qquad r(\zeta) \! \underset{\zeta \, \to \, 0}{=} \! \mathcal{O}
(\zeta). \nonumber
\end{gather}
\end{fffff}

\emph{Proof.} {}From Proposition~2.4,
\begin{align}
a(\zeta) &= \! \tfrac{\zeta^{2}}{\zeta^{2}-1}(\Psi^{+}_{22}(x,0;\zeta)
\Psi^{-}_{11}(x,0;\zeta) \! - \! \Psi^{+}_{12}(x,0;\zeta) \Psi^{-}_{21}
(x,0;\zeta)), \nonumber \\
b(\zeta) &= \! \tfrac{\zeta^{2}}{\zeta^{2}-1}(\overline{\Psi^{+}_{22}(x,
0;\overline{\zeta})} \, \Psi^{-}_{21}(x,0;\zeta) \! - \! \overline{\Psi^{
+}_{12}(x,0;\overline{\zeta})} \, \Psi^{-}_{11}(x,0;\zeta)). \nonumber
\end{align}
Using the $\zeta \! \to \! \infty$ and $\zeta \! \to \! 0$ asymptotics for
$\Psi^{\pm}_{ij}(x,0;\zeta)$, $i,j \! \in \! \{1,2\}$, given in Lemma~2.1,
one obtains the asymptotics for $a(\zeta)$ stated in the Corollary,
and also $b(\zeta) \! =_{\zeta \to \infty} \! \mathcal{O}(\zeta^{-1})$
and $b(\zeta) \! =_{\zeta \, \to \, 0} \! \mathcal{O}(\zeta)$; hence,
since $r(\zeta) \! := \! \tfrac{b(\zeta)}{a(\zeta)}$, one arrives at the
asymptotics for $r(\zeta)$ stated in the Corollary (in particular,
$r(0) \! = \! 0)$. \hfill $\square$
\begin{eeeee}
Although the technical details of this argument are not presented here, using
the fact that $u_{o}(x) \! \in \! \mathbf{C}^{\infty}(\mathbb{R})$ and $u_{o}
(x) \! - \! u_{o}(\pm \infty) \! \in \! \mathcal{S}_{\mathbb{C}}(\mathbb{R}_{
\pm})$, one shows that, using the Volterra-type integral representation for
the elements of $\Psi^{\pm}(x,0;\zeta)$ given in Lemma~2.1, based on a
successive approximations (Neumann series-type) argument, $r(\zeta) \!
\in \! \mathcal{S}_{\mathbb{C}}(\mathbb{R})$ (see the second article in
\cite{a12} for complete details; see, also, Part~1 of \cite{a9}).
\end{eeeee}
\begin{ccccc}
Let $u(x,t)$ be the solution of the Cauchy problem for the {\rm D${}_{f}$NLSE}
with finite-density initial data, and $\Psi^{\pm}(x,0;\zeta)$ the (Jost)
solutions of $\mathcal{O}^{\mathcal{D}} \Psi^{\pm}(x,0;\zeta) \! = \!
\left(
\begin{smallmatrix}
0 & 0 \\
0 & 0
\end{smallmatrix}
\right)$ with the asymptotics stated in Definition~{\rm 2.1} and satisfying
the symmetry reductions of Corollary~{\rm 2.1}. In the absence of spectral
singularities, denote the discrete and continuous spectra of the direct
scattering problem for the operator $\mathcal{O}^{\mathcal{D}}$ by $\sigma_{
d}$ and $\sigma_{c}$, respectively, with $\sigma_{\mathcal{O}^{\mathcal{D}}}
\! := \! \mathrm{spec}(\mathcal{O}^{\mathcal{D}})$ partitioned such that
$\sigma_{\mathcal{O}^{\mathcal{D}}} \! = \! \sigma_{d} \cup \sigma_{c}$, and
$\sigma_{d} \cap \sigma_{c} \! = \! \emptyset$. Then, for $r(\zeta) \! \in
\! \mathcal{S}_{\mathbb{C}}(\mathbb{R})$, $\sigma_{d} \! = \! \Delta_{a}
\cup \overline{\Delta_{a}}$, with $\Delta_{a} \! := \! \{\mathstrut
\varsigma_{n}; \, a(\zeta) \vert_{\zeta = \varsigma_{n}} \! = \! 0, \,
\varsigma_{n} \! = \! \me^{\mi \phi_{n}}, \, \phi_{n} \! \in \! (0,\pi), \,
n \! \in \! \{1,2,\ldots,N\}\}$, $\Delta_{a} \cap \overline{\Delta_{a}} \! =
\! \emptyset$, and $\mathrm{card}(\sigma_{d}) \! = \! 2N \! < \! \infty$,
and $\sigma_{c} \! = \! \{ \mathstrut \zeta; \, \Im(\zeta) \! = \! 0\}$,
with $\mathrm{card}(\sigma_{c}) \! = \! \infty$. Furthermore,
\begin{equation*}
a(\zeta) \! = \! \me^{\frac{\mi \theta}{2}} \prod_{n=1}^{N} \dfrac{(\zeta
\! - \! \varsigma_{n})}{(\zeta \! - \! \overline{\varsigma_{n}})} \exp \!
\left(-\int\nolimits_{-\infty}^{+\infty} \dfrac{\ln (1 \! - \! \vert r(\mu)
\vert^{2})}{(\mu \! - \! \zeta)} \, \dfrac{\md \mu}{2 \pi \mi} \right), \quad
\zeta \! \in \! \mathbb{C}_{+},
\end{equation*}
and
\begin{equation*}
0 \! \leqslant \theta \! = \! -2 \sum_{n=1}^{N} \phi_{n} \! - \!
\int\nolimits_{-\infty}^{+\infty} \dfrac{\ln (1 \! - \! \vert r(\mu)
\vert^{2})}{\mu} \, \dfrac{\md \mu}{2 \pi} < \! 2 \pi.
\end{equation*}
\end{ccccc}

\emph{Proof.} {}From classical complex analysis, a function $f(z)$ analytic
and of finite order $\eta$ for $z \! \in \! \mathbb{C}_{+}$ admits the
inner-outer factorisation
\begin{align*}
f(z) &= \exp \! \left( \mi \left( \sum_{k=0}^{q}b_{k}z^{k} \right) \!
+ \! \int\nolimits_{-1}^{1} \dfrac{\ln (\vert f(\mu) \vert)}{(\mu \! - \!
z)} \, \dfrac{\md \mu}{\pi \mi} \! + \! \int\nolimits_{-1}^{1} \dfrac{
\md \tau (\mu)}{(\mu \! - \! z)} \, \dfrac{1}{\pi \mi} \right) \prod_{
\vert z_{n} \vert <1} \dfrac{(z \! - \! z_{n})}{(z \! - \! \overline{
z_{n}})} \nonumber \\
 &\times \exp \! \left( z^{q+1} \! \left( \int\nolimits_{\vert \mu \vert
\geqslant 1} \dfrac{\ln (\vert f(\mu) \vert)}{\mu^{q+1}(\mu \! - \! z)} \,
\dfrac{\md \mu}{\pi \mi} \! + \! \int\nolimits_{\vert \mu \vert \geqslant
1} \dfrac{\md \tau (\mu)}{\mu^{q+1}(\mu \! - \! z)} \, \dfrac{1}{\pi
\mi} \right) \right) \prod_{\vert z_{n} \vert \geqslant 1}D_{q}(z,z_{n}),
\end{align*}
where $q \! = \! [\eta]$, with $[\bullet]$ denoting the greatest integer
less than or equal to $\bullet$, $b_{k} \! \in \! \mathbb{R}_{\geqslant 0}$,
$\tau (\cdot)$ is a singular boundary function, $z_{n} \! := \! r_{n} \me^{
\mi \theta_{z}^{n}}$, with $(r_{n},\theta_{z}^{n}) \! \in \! \mathbb{R}_{+}
\! \times \! (0,\pi)$, are the zeros of $f(z)$, $D_{q}(u,v) \! := \! \tfrac{
E(u/v,q)}{E(u/\overline{v},q)}$ is the canonical Nevanlinna factor, with
\begin{equation*}
E(w,q) \! = \!
\begin{cases}
1 \! - \! w, &\text{$q=0$}, \\
(1 \! - \! w) \exp (w \! + \! \tfrac{w^{2}}{2} \! + \! \cdots \! +
\tfrac{w^{q}}{q}), &\text{$q \! \in \! \mathbb{N}$},
\end{cases}
\end{equation*}
the canonical Weierstrass factor, and all integrals and finite products
converge absolutely (hence converge), $\sum_{r_{n} \leqslant 1}r_{n}
\sin \theta_{z}^{n} \! < \! \infty$, $\sum_{r_{n}>1}r_{n}^{-\mu-\epsilon}
\sin \theta_{z}^{n} \! < \! \infty$, $\int_{-\infty}^{+\infty} \tfrac{\vert
\ln \vert f(\xi) \vert \vert}{1+\vert \xi \vert^{1+\mu+\epsilon}} \, \md
\xi \! < \! \infty$, and $\int_{-\infty}^{+\infty} \tfrac{\vert \md \tau
(\xi) \vert}{1+\vert \xi \vert^{1+\mu +\epsilon}} \! < \! \infty$, where
$\mu \! := \! \max (\eta,1)$, and $\epsilon$ is an arbitrary positive (real)
number. Specialise the above to $a(\zeta)$. Let $\varsigma_{n}$ be the
(necessarily simple: see below) zeros of $a(\zeta)$: since $u_{o}(x) \! \in
\! \mathbf{C}^{\infty}(\mathbb{R})$, $u_{o}(x) \! - \! u_{o}(\pm \infty) \!
\in \! \mathcal{S}_{\mathbb{C}}(\mathbb{R}_{\pm})$, and there are no spectral
singularities, thus only a finite number of simple poles (see the paragraph
preceding Section~3 in \cite{a28}), it follows {}from an argument in
\cite{a24} that $\mathrm{card} \{\mathstrut \varsigma_{n}; \, a(\zeta)
\vert_{\zeta =\varsigma_{n}} \! = \! 0\} \! := \! N \! < \! \infty$. Since
$a(\zeta)$ has an analytic continuation to $\mathbb{C}_{+}$, and, {}from
Proposition~2.4, $a(\tfrac{1}{\zeta}) \! = \! \overline{a(\overline{
\zeta})}$, it follows that $\{\varsigma_{n}\}_{n=1}^{N}$ are distributed
along $\mathbb{C}_{+} \cap (\{\mathstrut z; \, \vert z \vert \! = \! 1\}
\setminus \{\pm 1\})$; hence, $\sigma_{d}$, the discrete spectrum of
$\mathcal{O}^{\mathcal{D}}$, is given by $\sigma_{d} \! = \! \Delta_{a}
\cup \overline{\Delta_{a}}$, where $\Delta_{a}$ is defined in the Lemma.
Since $\Delta_{a} \cap \overline{\Delta_{a}} \! = \! \emptyset$, it follows
({}from the Inclusion-Exclusion Principle) that $\mathrm{card}(\Delta_{a}
\cup \overline{\Delta_{a}}) \! = \! 2N$. {}From \cite{a24}, the continuous
spectrum of $\mathcal{O}^{\mathcal{D}}$ is given by $\sigma_{c} \! = \! \{
\mathstrut \zeta; \, \Re (-\mi \lambda (\zeta)) \! = \! \Re (\mi \lambda
(\zeta))\} \! = \! \{ \mathstrut \zeta; \, \Im (\zeta) \! = \! 0\}$ (oriented
{}from $-\infty$ to $+\infty)$. As there are no spectral singularities
\cite{a28}, $\tau (\cdot) \! \equiv \! 0$; hence, {}from the above argument,
the $\zeta \! \to \! \infty$ asymptotics for $a(\zeta)$ given in
Corollary~2.3 (in which case $\eta \! = \! 0$, that is, $q \! = \! [0] \! =
\! 0)$, the unimodularity relation $\vert a(\zeta) \vert^{2} \! - \! \vert
b(\zeta) \vert^{2} \! = \! 0$, $\Im (\zeta) \! = \! 0$, given in
Proposition~2.4, and the definition of the reflection coefficient,
$r(\zeta)$, given in Corollary~2.2, it follows that, for $b_{0} \! := \!
(\tfrac{1}{2} \theta \! + \! 2 \sum_{n=1}^{N} \phi_{n} \! + \! \int_{\vert
\mu \vert \geqslant 1} \tfrac{\ln (1-\vert r(\mu) \vert^{2})}{\mu} \,
\tfrac{\md \mu}{2 \pi}) \, \mathrm{mod}(2 \pi)$, $a(\zeta)$ has the
simplified representation $a(\zeta) \! = \! \me^{\frac{\mi \theta}{2}}
\prod_{\varsigma_{n} \in \Delta_{a}} \! \tfrac{(\zeta -\varsigma_{n})}{
(\zeta -\overline{\varsigma_{n}})} \exp \! \left(\! -\int_{\sigma_{c}}
\tfrac{\ln (1-\vert r(\mu) \vert^{2})}{(\mu - \zeta)} \, \tfrac{\md \mu}{2
\pi \mi} \right)$, $\zeta \! \in \! \mathbb{C}_{+}$. Writing $a(\zeta) \!
= \! \me^{\frac{\mi \theta}{2}} \! \prod_{n=1}^{N} \! \tfrac{(\zeta -
\varsigma_{n})}{(\zeta -\overline{\varsigma_{n}})} \exp \! \left(\! -\int_{
-\infty}^{+\infty} \tfrac{\ln (1-\vert r(\mu) \vert^{2})}{(\mu -\zeta)} \,
\tfrac{\md \mu}{2 \pi \mi} \right)$, $\zeta \! \in \! \mathbb{C}_{+}$, and
using the $\zeta \! \to \! 0$ asymptotics for $a(\zeta)$ given in
Corollary~2.3, one obtains the expression for $\theta$ given in the Lemma;
moreover, with this $\theta$, one also shows that $a(\zeta)$ satisfies the
involution $a(\tfrac{1}{\zeta}) \! = \! \overline{a(\overline{\zeta})}$,
and, since $r(\zeta) \! \in \! \mathcal{S}_{\mathbb{C}}(\mathbb{R})$,
$\int_{-\infty}^{+\infty} \tfrac{\ln (1-\vert r(\xi) \vert^{2})}{1+\vert
\xi \vert^{2+\epsilon}} \, \md \xi \! < \! \infty$. \hfill $\square$
\begin{ccccc}
For $r(\zeta) \! \in \! \mathcal{S}_{\mathbb{C}}^{1}(\mathbb{R})$,
\begin{equation*}
a(s \! + \! \mi \varepsilon) \! \underset{\varepsilon \downarrow 0}{=}
\! \dfrac{(-s)^{N} \me^{\mi (\frac{\theta}{2}+\sum_{n=1}^{N} \phi_{
n})} \exp (-\mathrm{P.V.} \int_{\mathbb{R} \setminus \{s\}} \frac{\ln
(1-\vert r(\mu) \vert^{2})}{(\mu-s)} \, \frac{\md \mu}{2 \pi \mi})}{(1 \!
 - \! \vert r(s) \vert^{2})^{\varkappa_{\mathrm{sgn}(s)}}}(1 \! + \!
o(1)), \quad s \! \in \! \{\pm 1\},
\end{equation*}
where $\mathrm{P.V.} \int$ denotes the principal value integral,
and $\varkappa_{\pm}$ are real, possibly zero, constants.
\end{ccccc}

\emph{Proof.} This result will be proved via two independent approaches:
(1) using the representation for $a(\zeta)$ given in Lemma~2.2
(\textsc{Method}~(i)); and (2) analysing the expression for $a(\zeta)$
given in Proposition~2.4 (\textsc{Method}~(ii)).

\textsc{Method}~(i). One starts with the representation for
$a(\zeta)$ given in Lemma~2.2, namely, $a(\zeta) \! = \! \me^{\frac{
\mi \theta}{2}} \prod_{n=1}^{N} \tfrac{(\zeta -\varsigma_{n})}{(\zeta
-\overline{\varsigma_{n}})} \exp \! \left(\! -\int_{-\infty}^{+\infty}
\tfrac{\ln (1-\vert r(\mu) \vert^{2})}{(\mu -\zeta)} \, \tfrac{\md \mu}
{2 \pi \mi} \right)$, $\zeta \! \in \! \mathbb{C}_{+}$, where $\theta \!
\in \! [0,2 \pi)$ is given in Lemma~2.2, $\varsigma_{n} \! = \! \me^{
\mi \phi_{n}}$, $\phi_{n} \! \in \! (0,\pi)$, $n \! \in \! \{1,2,\ldots,N\}$,
and $r(\zeta) \! \in \! \mathcal{S}_{\mathbb{C}}(\mathbb{R})$.
The fact that $\vert \vert r(\cdot) \vert \vert_{\mathcal{L}^{\infty}
(\mathbb{R})} \! := \! \sup_{z \in \mathbb{R}} \vert r(z) \vert \! < \!
1$ is essential for the proof: this will be proved in Lemma~2.4 below.
To study the behaviour of $a(\zeta)$ as $\zeta \! \to \! \pm 1$ {}from
above $(\mathbb{C}_{+})$, set $\zeta \! = \! \pm 1 \! + \! \epsilon
\me^{\mi \beta}$, $\beta \! \in \! (0,\pi)$, and consider the limit as
$\epsilon \! \downarrow \! 0$. Set $\mathbb{U}_{\varepsilon}(\pm 1) \!
:= \! \{ \mathstrut z; \, \vert z \! \mp \! 1 \vert \! < \! \varepsilon\}$,
where $\varepsilon$ is an arbitrarily fixed, sufficiently small positive
real number. Write $I \! = \! \int_{-\infty}^{+\infty} \tfrac{\ln (1-
\vert r(\mu) \vert^{2})}{(\mu -\zeta)} \, \tfrac{\md \mu}{2 \pi \mi}
\! = \! I_{o}^{r} \! + \! (\int_{\mathbb{U}_{\varepsilon}(-1)} \! + \!
\int_{\mathbb{U}_{\varepsilon}(+1)}) \tfrac{\ln (1-\vert r(\mu) \vert^{
2})}{(\mu -\zeta )} \, \tfrac{\md \mu}{2 \pi \mi}$, where $I_{o}^{r}
\! := \! \int_{\mathbb{R} \setminus \mathbb{U}_{\varepsilon}(\pm 1)} \!
\tfrac{\ln (1-\vert r(\mu) \vert^{2})}{(\mu -\zeta)} \, \tfrac{\md \mu}{
2 \pi \mi}$. Since $r(\zeta) \! \in \! \mathcal{S}_{\mathbb{C}}(\mathbb{
R})$, one Taylor expands $(1 \! - \! \vert r(\zeta) \vert^{2})$ to show
that, for $\zeta \! \in \! \mathbb{U}_{\varepsilon}(\pm 1)$, $(1 \! - \!
\vert r(\zeta) \vert^{2}) \! = \! 1 \! - \! r_{o}(\pm 1) \! + \! \widehat{
r}_{1}(\pm 1)(\zeta \! \mp \! 1) \! + \! \tfrac{\widehat{r}_{2}(\pm 1)}{
2!}(\zeta \! \mp \! 1)^{2} \! + \! \mathcal{O}((\zeta \! \mp \! 1)^{3})$,
where $r_{o}(\pm 1) \! := \! \vert r(\pm 1) \vert^{2}$ $(\not= \! 1)$,
and $\widehat{r}_{j}(\pm 1)$, $j \! \in \! \{1,2\}$, are some $\mathbb{
C}$-valued constants: now, using the fact that $\vert \vert r(\cdot)
\vert \vert_{\mathcal{L}^{\infty}(\mathbb{R})} \! < \! 1$, and the
expansion $\ln (1 \! + \! x) \! =_{\vert x \vert < 1} \! x \! + \!
\mathcal{O}(x^{2})$, one shows that $I \! = \! I_{o}^{r} \! + \! \sum_{l
\in \{\pm 1\}}(\int_{\mathbb{U}_{\varepsilon}(l)} \! \tfrac{\ln (1-r_{o}
(l))}{(\mu -\zeta)} \, \tfrac{\md \mu}{2 \pi \mi} \! + \! \int_{\mathbb{
U}_{\varepsilon}(l)} \! \tfrac{(\widetilde{r}_{1}(l)(\mu-l)+\mathcal{O}
((\mu-l)^{2}))}{(\mu -\zeta)} \, \tfrac{\md \mu}{2 \pi \mi})$, where
$\widetilde{r}_{j}(l)$, $j \! \in \! \{1,2\}$, $l \! \in \! \{\pm 1\}$,
are some $\mathbb{C}$-valued constants. Using the distributional identities
$(x \! - (x_{o} \! \pm \! \mi 0))^{-1} \! = \! \mathrm{P.V.}(x \! - \! x_{
o})^{-1} \! \pm \! \mi \pi \delta (x \! - \! x_{o})$, where $\mathrm{P.V.}$
denotes the principal value integral and $\delta (x \! - \! x_{o})$ is the
Dirac delta function, and $\int_{x_{1}}^{x_{2}} f(x) \delta (x \! - \! x_{
o}) \, \md x \! = \!
\begin{cases}
f(x_{o}), &\text{$x_{o} \! \in \! (x_{1},x_{2})$}, \\
0, &\text{$x_{o} \! \in \! \mathbb{R} \setminus (x_{1},x_{2})$},
\end{cases}$ one shows that, with the choice $\vert \epsilon \cos \beta
\vert \! < \! \varepsilon$ $(\forall \, \beta \! \in \! (0,\pi))$, and
taking the principal branch, for $\vert r(\pm 1) \vert \! \not= \! 1$,
$I \! =_{\varepsilon \downarrow 0} \! \mathrm{P.V.} \int_{\mathbb{R}
\setminus \{\pm 1\}} \! \tfrac{\ln (1-\vert r(\mu) \vert^{2})}{(\mu \mp
1)} \, \tfrac{\md \mu}{2 \pi \mi} \! + \! \sum_{l \in \{\pm 1\}} \!
\varkappa_{\mathrm{sgn}(l)} \ln (1 \! - \! \vert r(l) \vert^{2}) \! +
\! o(1)$, with $\varkappa_{\mathrm{sgn}(l)}$, $l \! \in \! \{\pm 1\}$,
some $\mathbb{R}$-valued, possibly zero, constants. One also shows
that $\prod_{n=1}^{N} \tfrac{(\pm 1+\mi \varepsilon -\zeta_{n})}{(\pm
1+\mi \varepsilon -\overline{\zeta_{n}})} \! =_{\varepsilon \downarrow
0} \! (\mp 1)^{N} \exp (\mi \sum_{n=1}^{N} \phi_{n})(1 \! + \! o(1))$;
hence, {}from the above estimates, it follows that $a(\pm 1 \! + \! \mi
\varepsilon) \! =_{\varepsilon \downarrow 0} \! \tfrac{\exp (\frac{\mi
\theta}{2}) \exp (-\mathrm{P.V.} \int_{\mathbb{R} \setminus \{\pm
1\}} \tfrac{\ln (1-\vert r(\mu) \vert^{2})}{(\mu \mp 1)} \, \tfrac{\md \mu}
{2 \pi \mi})(\mp 1)^{N} \exp (\mi \sum_{n=1}^{N} \phi_{n})}{(1-\vert
r(\pm 1) \vert^{2})^{\varkappa_{\pm}}}(1 \! + \! o(1))$.

\textsc{Method}~(ii). {}From Proposition~2.4, $a(\zeta) \!
= \! \tfrac{\zeta^{2}}{\zeta^{2}-1} \! \left( \Psi^{+}_{22}(\zeta) \Psi^{
-}_{11}(\zeta) \! - \! \Psi^{+}_{12}(\zeta) \Psi^{-}_{21}(\zeta) \right)$,
$\zeta \! \in \! \mathbb{C}_{+}$, where, {}from Lemma~2.1,
\begin{align}
\Psi^{-}(\zeta) \! := \! \Psi^{-}(x,0;\zeta) &= \!
\left(
\begin{smallmatrix}
\me^{\frac{\mi \theta}{2}} \me^{\mi (\int_{-\infty}^{x} \mathscr{S}(\xi;
\zeta) \, \md \xi -k(\zeta)x)} & \me^{\frac{\mi \theta}{2}} \overline{
\mathcal{A}(x;\overline{\zeta})} \, \me^{-\mi (\int_{-\infty}^{x}
\overline{\mathscr{S}(\xi;\overline{\zeta})} \, \md \xi - k(\zeta)x)} \\
\me^{-\frac{\mi \theta}{2}} \mathcal{A}(x;\zeta) \me^{\mi (\int_{-\infty}
^{x} \mathscr{S}(\xi;\zeta) \, \md \xi -k(\zeta)x)} & \me^{-\frac{\mi
\theta}{2}} \me^{-\mi (\int_{-\infty}^{x} \overline{\mathscr{S}(\xi;
\overline{\zeta})} \, \md \xi - k(\zeta)x)}
\end{smallmatrix}
\right), \nonumber \\
\Psi^{+}(\zeta) \! := \! \Psi^{+}(x,0;\zeta) &= \!
\left(
\begin{smallmatrix}
\me^{-\mi (\int_{+\infty}^{x} \overline{\mathscr{T}(\xi;\overline{\zeta})}
\, \md \xi +k(\zeta)x)} & \mathcal{B}(x;\zeta) \me^{\mi (\int_{+\infty}^{
x} \mathscr{T}(\xi;\zeta) \, \md \xi +k(\zeta)x)} \\
\overline{\mathcal{B}(x;\overline{\zeta})} \, \me^{-\mi (\int_{+\infty}^{
x} \overline{\mathscr{T}(\xi;\overline{\zeta})} \, \md \xi +k(\zeta)x)} &
\me^{\mi (\int_{+\infty}^{x} \mathscr{T}(\xi;\zeta) \, \md \xi +k(\zeta)x)}
\end{smallmatrix}
\right), \nonumber
\end{align}
with $\mathcal{A}(x;\zeta)$, $\mathcal{B}(x;\zeta)$, $\mathscr{S}(x;\zeta)$,
and $\mathscr{T}(x;\zeta)$ satisfying system~(L2.1a), and $\theta$ given
in Lemma~2.2. Near $\pm 1$, let the small parameters be defined by
$\mu_{\pm} \! := \! \zeta \! \mp \! 1$: since $r(\zeta) \! \in \! \mathcal{
S}_{\mathbb{C}}(\mathbb{R})$, it follows {}from a result of Zhou \cite{a29}
that $\mathcal{A}(x;\mu_{\pm}) \! = \! \sum_{n=0}^{\infty}a_{n}^{\pm}[u,
\overline{u}] \mu_{\pm}^{n} \! + \! \mathcal{O}(\vert \mu_{\pm} \vert^{+
\infty})$ and $\mathcal{B}(x;\mu_{\pm}) \! = \! \sum_{n=0}^{\infty}b_{n}^{
\pm}[u,\overline{u}] \mu_{\pm}^{n} \! + \! \mathcal{O}(\vert \mu_{\pm}
\vert^{+\infty})$, where $\star^{\pm}_{n}[u,\overline{u}] \! := \! \star_{
n}^{\pm}$, with $\star \! \in \! \{a,b\}$, are functionals of $u_{o}(x)$
and $\overline{u_{o}(x)}$. Substituting these asymptotic expansions into
the first two equations of system~(L2.1a) and using the geometric
progression $(1 \! \pm \! z)^{-1} \! = \! \sum_{n=0}^{\infty}(\mp 1)^{n}
z^{n}$, $\vert z \vert \! < \! 1$, one shows that, for $\zeta \! \approx
\! \pm 1$,
\begin{gather}
\mp \mi (2 \! + \! \mu_{\pm}^{2} \! \mp \! \mu_{\pm}^{3} \! + \! \cdots)
(a_{0}^{\pm} \! + \! a_{1}^{\pm} \mu_{\pm} \! + \! a_{2}^{\pm} \mu_{\pm}
^{2} \! + \! a_{3}^{\pm} \mu_{\pm}^{3} \! + \! \cdots) \! + \! d_{x}(a_{
0}^{\pm} \! + \! a_{1}^{\pm} \mu_{\pm} \! + \! a_{2}^{\pm} \mu_{\pm}^{2}
\! + \! a_{3}^{\pm} \mu_{\pm}^{3} \! + \! \cdots) \nonumber \\
+u_{o}(x) \me^{-\mi \theta}(a_{0}^{\pm} \! + \! a_{1}^{\pm} \mu_{\pm} \!
+ \! a_{2}^{\pm} \mu_{\pm}^{2} \! + \! a_{3}^{\pm} \mu_{\pm}^{3} \! + \!
\cdots)(a_{0}^{\pm} \! + \! a_{1}^{\pm} \mu_{\pm} \! + \! a_{2}^{\pm}
\mu_{\pm}^{2} \! + \! a_{3}^{\pm} \mu_{\pm}^{3} \! + \! \cdots) \! =
\overline{u_{o}(x)} \, \me^{\mi \theta}, \tag{L2.2a} \\
\pm \mi (2 \! + \! \mu_{\pm}^{2} \! \mp \! \mu_{\pm}^{3} \! + \! \cdots)
(b_{0}^{\pm} \! + \! b_{1}^{\pm} \mu_{\pm} \! + \! b_{2}^{\pm} \mu_{\pm}
^{2} \! + \! b_{3}^{\pm} \mu_{\pm}^{3} \! + \! \cdots) \! + \! d_{x}(b_{
0}^{\pm} \! + \! b_{1}^{\pm} \mu_{\pm} \! + \! b_{2}^{\pm} \mu_{\pm}^{2}
\! + \! b_{3}^{\pm} \mu_{\pm}^{3} \! + \! \cdots) \nonumber \\
+ \, \overline{u_{o}(x)}(b_{0}^{\pm} \! + \! b_{1}^{\pm} \mu_{\pm} \! + \!
b_{2}^{\pm} \mu_{\pm}^{2} \! + \! b_{3}^{\pm} \mu_{\pm}^{3} \! + \! \cdots)
(b_{0}^{\pm} \! + \! b_{1}^{\pm} \mu_{\pm} \! + \! b_{2}^{\pm} \mu_{\pm}^{
2} \! + \! b_{3}^{\pm} \mu_{\pm}^{3} \! + \! \cdots) \! = \! u_{o}(x).
\tag{L2.2b}
\end{gather}
{}From Eqs.~(L2.2a) and~(L2.2b), one has, {}from the $\mathcal{O}(1)$
terms,
\begin{gather}
-2 \mi a_{0}^{+} \! + \! d_{x}a_{0}^{+} \! + \! u_{o}(x) \me^{-\mi \theta}
(a_{0}^{+})^{2} \! = \overline{u_{o}(x)} \, \me^{\mi \theta}, \quad 2 \mi
b_{0}^{+} \! + \! d_{x}b_{0}^{+} \! + \! \overline{u_{o}(x)}(b_{0}^{+})^{
2} \! = \! u_{o}(x), \tag{L2.2c} \\
2 \mi a_{0}^{-} \! + \! d_{x}a_{0}^{-} \! + \! u_{o}(x) \me^{-\mi \theta}
(a_{0}^{-})^{2} \! = \overline{u_{o}(x)} \, \me^{\mi \theta}, \quad -2 \mi
b_{0}^{-} \! + \! d_{x}b_{0}^{-} \! + \! \overline{u_{o}(x)}(b_{0}^{-})^{
2} \! = \! u_{o}(x). \tag{L2.2d}
\end{gather}
Eqs.~(L2.2c) and~(L2.2d) are non-linear ordinary differential equations
(ODEs) of the Riccati type for $a_{0}^{\pm}$ and $b_{0}^{\pm}$: for the
purposes of this proof, their explicit solutions are not necessary (they
can also be transformed into linear 2nd-order non-constant coefficient
ODEs). {}From Eqs.~(L2.2c) and~(L2.2d), one shows that $a_{0}^{\pm}
b_{0}^{\pm} \! = \! \me^{\mi \theta}$; hence, with the representations $a_{
0}^{\pm} \! := \! \vert a_{0}^{\pm}(x) \vert \exp (\mi \phi_{\pm}^{a}(x))$
and $b_{0}^{\pm} \! := \! \vert b_{0}^{\pm}(x) \vert \exp (\mi \phi_{\pm}^{
b}(x))$, with $\vert a_{0}^{\pm}(x) \vert \colon \mathbb{R} \! \to \!
\mathbb{R}_{+}$ (respectively~$\vert b_{0}^{\pm}(x) \vert \colon \mathbb{
R} \! \to \! \mathbb{R}_{+})$ and $\phi_{\pm}^{a}(x) \colon \mathbb{R} \!
\to \! \mathbb{R} \setminus \{0\}$ (respectively~$\phi_{\pm}^{b}(x) \colon
\mathbb{R} \! \to \! \mathbb{R} \setminus \{0\})$, it follows that $\vert
a_{0}^{\pm}(x) \vert \vert b_{0}^{\pm}(x) \vert \! = \! 1$ and $\phi_{\pm}
^{a}(x) \! + \! \phi_{\pm}^{b}(x) \! = \! \theta \, \mathrm{mod}(2 \pi)$:
one also requires that (zero-integral conditions) $\int_{\mathbb{R}}
\vert u_{o}(\xi) \vert (\vert a_{0}^{\pm}(\xi) \vert \! + \! \vert a_{0}^{\pm}
(\xi) \vert^{-1}) \cos (\arg (u_{o}(\xi)) \! - \! \theta \! + \! \phi_{\pm}^{a}
(\xi)) \, \md \xi \! = \! 0$ and $\int_{\mathbb{R}} \vert u_{o}(\xi) \vert
(\vert a_{0}^{\pm}(\xi) \vert \! - \! \vert a_{0}^{\pm}(\xi) \vert^{-1}) \sin
(\arg (u_{o}(\xi)) \! - \! \theta \! + \! \phi_{\pm}^{a}(\xi)) \, \md \xi \!
= \! 0$. {}From Eqs.~(L2.2a) and~(L2.2b), one has, {}from the $\mathcal{O}
(\mu_{\pm})$ terms,
\begin{gather}
d_{x}a_{1}^{+} \! + \! a_{1}^{+}(2u_{o}(x) \me^{-\mi \theta}a_{
0}^{+} \! - \! 2 \mi) \! = \! 0, \quad d_{x}b_{1}^{+} \! + \! b_{1}^{+}
(2 \overline{u_{o}(x)} \, b_{0}^{+} \! + \! 2 \mi) \! = \! 0, \tag{L2.2e} \\
d_{x}a_{1}^{-} \! + \! a_{1}^{-}(2u_{o}(x) \me^{-\mi \theta}a_{
0}^{-} \! + \! 2 \mi) \! = \! 0, \quad d_{x}b_{1}^{-} \! + \! b_{1}^{-}
(2 \overline{u_{o}(x)} \, b_{0}^{-} \! - \! 2 \mi) \! = \! 0.
\tag{L2.2f}
\end{gather}
Eqs.~(L2.2e) and~(L2.2f) are linear 1st-order ODEs for $a_{1}^{\pm}$ and
$b_{1}^{\pm}$ which can be solved explicitly to yield $a_{1}^{\pm} \! =
\! c_{\pm} \exp (- \! \smallint_{-\infty}^{x}(2u_{o}(\xi) \me^{-\mi \theta}
a_{0}^{\pm} \! \mp \! 2 \mi) \md \xi)$ and $b_{1}^{\pm} \! = \! d_{\pm} \exp
(- \! \smallint_{+\infty}^{x}(2 \overline{u_{o}(\xi)} \, b_{0}^{\pm} \! \pm
\! 2 \mi) \md \xi)$, with $(c_{\pm},d_{\pm}) \! \in \! \mathbb{C} \times
\mathbb{C}$. Once again, since $r(\zeta) \! \in \! \mathcal{S}_{\mathbb{C}}
(\mathbb{R})$, it follows {}from a result of Zhou \cite{a29} that, near
$\pm 1$, with $\mu_{\pm} \! := \! \zeta \! \mp \! 1$, $\mathscr{S}(x;
\mu_{\pm}) \! = \! \sum_{n=0}^{\infty}s_{n}^{\pm}[u,\overline{u}]
\mu_{\pm}^{n} \! + \! \mathcal{O}(\vert \mu_{\pm} \vert^{+\infty})$
and $\mathscr{T}(x;\mu_{\pm}) \! = \! \sum_{n=0}^{\infty}t_{n}^{\pm}[u,
\overline{u}] \mu_{\pm}^{n} \! + \! \mathcal{O}(\vert \mu_{\pm} \vert^{
+\infty})$, where $\star_{n}^{\pm}[u,\overline{u}] \! := \! \star_{n}^{
\pm}$, with $\star \! \in \! \{s,t\}$, are functionals of $u_{o}(x)$ and
$\overline{u_{o}(x)}$. Substituting these asymptotic expansions into the
third equation of system~(L2.1a) and using the geometric progression $(1
\! \pm \! z)^{-1} \! = \! \sum_{n=0}^{\infty}(\mp 1)^{n}z^{n}$, $\vert z
\vert \! < \! 1$, one shows that, for $\zeta \! \approx \! \pm 1$,
\begin{gather}
\mi (s_{0}^{\pm} \! + \! s_{1}^{\pm} \mu_{\pm} \! + \! s_{2}^{\pm} \mu_{
\pm}^{2} \! + \! \cdots) \! = \! \mp \mi (1 \! \mp \! \mu_{\pm} \! + \!
\mu_{\pm}^{2} \! + \! \cdots) \! + \! u_{o}(x) \me^{-\mi \theta}(a_{0}^{
\pm} \! + \! a_{1}^{\pm} \mu_{\pm} \! + \! \cdots), \tag{L2.2g} \\
\mi (t_{0}^{\pm} \! + \! t_{1}^{\pm} \mu_{\pm} \! + \! t_{2}^{\pm} \mu_{
\pm}^{2} \! + \! \cdots) \! = \! \pm \mi (1 \! \mp \! \mu_{\pm} \! + \!
\mu_{\pm}^{2} \! + \! \cdots) \! + \! \overline{u_{o}(x)} \, (b_{0}^{\pm}
\! + \! b_{1}^{\pm} \mu_{\pm} \! + \! \cdots). \tag{L2.2h}
\end{gather}
Solving, algebraically, Eqs.~(L2.2g) and~(L2.2h) for the first two
non-zero terms, one arrives at $s_{0}^{\pm} \! = \! \mp 1 \! - \! \mi
u_{o}(x) \me^{-\mi \theta}a_{0}^{\pm}$, $t_{0}^{\pm} \! = \! \pm 1 \! -
\! \mi \overline{u_{o}(x)} \, b_{0}^{\pm}$, $s_{1}^{\pm} \! = \! 1 \! -
\! \mi u_{o}(x) \me^{-\mi \theta}a_{1}^{\pm}$, and $t_{1}^{\pm} \! = \!
-1 \! - \! \mi \overline{u_{o}(x)} \, b_{1}^{\pm}$, with $a_{0}^{\pm}$
and $b_{0}^{\pm}$ (respectively~$a_{1}^{\pm}$ and $b_{1}^{\pm})$ as
stated (respectively~given) above. Substituting the above asymptotic
expansions into the matrix representations for $\Psi^{\pm}(x,0;\zeta)$
given at the beginning of \textsc{Method}~(ii), isolating the elements
$\Psi^{+}_{22}(\zeta)$, $\Psi^{-}_{11}(\zeta)$, $\Psi^{+}_{12}(\zeta)$,
and $\Psi^{-}_{21}(\zeta)$, recalling that $a_{0}^{\pm}b_{0}^{\pm} \!
= \! \me^{\mi \theta}$, choosing $c_{\pm}$ and $d_{\pm}$ judiciously,
along with the convergence conditions $\int_{-\infty}^{+\infty} \vert u_{
o}(\xi) \vert \vert a_{0}^{\pm}(\xi) \vert \cos (\arg (u_{o}(\xi)) \! - \!
\theta \! + \! \phi_{\pm}^{a}(\xi)) \, \md \xi \! < \! \infty$ and $\int_{
-\infty}^{+\infty}(\vert u_{o}(\xi) \vert \vert a_{0}^{\pm}(\xi) \vert \sin
(\arg (u_{o}(\xi)) \! - \! \theta \! + \! \phi_{\pm}^{a}(\xi)) \! \mp \! 1)
\, \md \xi \! < \! \infty$ (so as to remove secular terms), invoking the
zero-integral conditions, and substituting, finally, into the expression
for $a(\zeta)$ given at the beginning of \textsc{Method}~(ii), one shows
that, for $\zeta \! \approx \! \pm 1$, $\pm 2 (\zeta \! \mp \! 1)a(\zeta)
\! = \! \exp (\mathcal{K}^{\pm} \! + \! \mathcal{O}(\zeta \! \mp \! 1))
(\underbrace{\me^{\frac{\mi \theta}{2}} \! - \! \me^{\frac{\mi \theta}{
2}}}_{0} \! + \mathcal{O}(\zeta \! \mp \! 1))$, with $\mathcal{K}^{\pm} \!
\in \! \mathbb{C} \setminus \{0\}$ and $\mathcal{O}(1)$; hence, $a(\zeta)
\! =_{\zeta \to \pm 1} \! \widehat{a}_{o}^{\pm} \! + \! \mathcal{O}(\zeta
\! \mp \! 1)$, where $\widehat{a}_{o}^{\pm}$ are some $\mathbb{C}$-valued
$\mathcal{O}(1)$ constants (whose explicit expressions are not given here).
\hfill $\square$
\begin{eeeee}
The analysis for the singular limit $\vert r(\pm 1) \vert \! = \! 1$ is
not addressed in this work: it is assumed throughout that $\vert r(\pm
1) \vert \! \not= \! 1$.
\end{eeeee}

The $t$-dependence is re-introduced into the analysis by studying
the $\partial_{t} \varPsi (x,t;\zeta) \! = \! \mathcal{V}(x,t;\zeta)
\linebreak[4]
\cdot \varPsi (x,t;\zeta)$ component of system~(2). It is shown in
\cite{a9} that the scattering map $(S)$ $u_{o}(\cdot) \! \mapsto \!
r(\zeta) \! = \! \mathscr{R}(u_{o}(\cdot))$, which is a bijection for
$u_{o}(x)$ and $r(\zeta)$ belonging to the spaces defined heretofore,
linearises the D${}_{f}$NLSE flow. This leads to the following
\begin{bbbbb}[$\cite{a9,a30}$]
Let $u(x,t)$ be the solution of the Cauchy problem for the
{\rm D${}_{f}$N\-L\-S\-E} with finite-density initial data and $\Psi
(x,t;\zeta)$ the corresponding solution of system~{\rm (2)} defined in
Proposition~{\rm 2.2}. Then $a(\zeta,t) \! = \! a(\zeta)$ and $b(\zeta,t)
\! = \! b(\zeta) \exp (4 \mi k(\zeta) \lambda (\zeta)t)$.
\end{bbbbb}
{}From Proposition~2.5, it follows that, since $a(\zeta,t)$ is independent
of $t$, $a(\zeta)$ is the ``generator'' of the integrals of motion of the
model, and, {}from the definition $r(\zeta,t) \! := \! \tfrac{b(\zeta,t)}
{a(\zeta,t)}$, it follows that $r(\zeta,t)$ evolves in the scattering data
phase space according to the rule $r(\zeta,t) \! = \! r(\zeta) \exp (4 \mi
k(\zeta) \lambda (\zeta)t)$, with $r(\zeta) \! \in \! \mathcal{S}_{\mathbb{
C}}^{1}(\mathbb{R})$. Assembling the above, one arrives at the following
(normalised at $\infty)$ RHP for the $\mathrm{M}(2,\mathbb{C})$-valued
function $m(x,t;\zeta)$.
\begin{ccccc}
Let $u(x,t)$ be the solution of the Cauchy problem for the
{\rm D${}_{f}$NLSE} with finite-density initial data $u(x,0) \! := \! u_{o}
(x) \! =_{x \to \pm \infty} \! u_{o}(\pm \infty)(1 \! + \! o(1))$, where
$u_{o}(\pm \infty) \! := \! \exp (\tfrac{\mi (1 \mp 1) \theta}{2})$, $\theta$
is given in Lemma~{\rm 2.2}, $u_{o}(x) \! \in \! \mathbf{C}^{\infty}(\mathbb{
R})$, and $u_{o}(x) \! - \! u_{o}(\pm \infty) \! \in \! \mathcal{S}_{\mathbb{
C}}(\mathbb{R}_{\pm})$. For $\zeta \! \in \! \mathbb{C}_{+}$, set $\Phi (x,t;
\zeta) \! := \!
\left(
\begin{smallmatrix}
\frac{\Psi^{-}_{11}(x,t;\zeta)}{a(\zeta)} & \, \Psi^{+}_{12}(x,t;\zeta) \\
\frac{\Psi^{-}_{21}(x,t;\zeta)}{a(\zeta)} & \, \Psi^{+}_{22}(x,t;\zeta)
\end{smallmatrix}
\right)$, and, for $\zeta \! \in \! \mathbb{C}_{-}$, set $\Phi (x,t;
\zeta) \! := \!
\left(
\begin{smallmatrix}
\Psi^{+}_{11}(x,t;\zeta) & \, \frac{\Psi^{-}_{12}(x,t;\zeta)}{\overline{
a(\overline{\zeta})}} \\
\Psi^{+}_{21}(x,t;\zeta) & \, \frac{\Psi^{-}_{22}(x,t;\zeta)}{\overline{
a(\overline{\zeta})}}
\end{smallmatrix}
\right)$, where $a(\zeta)$ is given in Lemma~{\rm 2.2}, $\Psi^{\pm}
(x,t;\zeta)$ are the solutions of system~{\rm (2)} defined in
Proposition~{\rm 2.2}, and $\Psi^{\pm}(x,0;\zeta)$ solve $\mathcal{
O}^{\mathcal{D}} \Psi^{\pm}(x,0;\zeta) \! = \!
\left(
\begin{smallmatrix}
0 & 0 \\
0 & 0
\end{smallmatrix}
\right)$ with the asymptotics stated in Definition~{\rm 2.1}. Set $m(x,t;
\zeta) \! := \! \Phi (x,t;\zeta) \exp (\mi k(\zeta)(x \! + \! 2 \lambda
(\zeta) t) \sigma_{3})$. Then: (1) the bounded discrete set $\sigma_{d}$ is
finite; (2) the poles of $m(x,t;\zeta)$ are simple; (3) the first
(respectively~second) column of $m(x,t;\zeta)$ has poles in $\mathbb{C}_{+}$
(respectively~$\mathbb{C}_{-})$ at $\{\varsigma_{n}\}_{n=1}^{N}$
(respectively~$\{\overline{\varsigma_{n}}\}_{n=1}^{N})$, $\varsigma_{n} \! :=
\! \me^{\mi \phi_{n}}$, $\phi_{n} \! \in \! (0,\pi);$ and (4) $m(x,t;\zeta)
\colon \mathbb{C} \setminus (\sigma_{d} \cup \sigma_{c}) \! \to \! \mathrm{
M}(2,\mathbb{C})$ solves the following {\rm RHP:}
\begin{enumerate}
\item[(i)] $m(x,t;\zeta)$ is piecewise meromorphic $\forall \, \zeta \!
\in \! \mathbb{C} \setminus \sigma_{c};$
\item[(ii)] $m_{\pm}(x,t;\zeta) \! := \! \lim_{\varepsilon \downarrow 0}
m(x,t;\zeta \! \pm \! \mi \varepsilon)$ satisfy the jump condition
\begin{equation*}
m_{+}(x,t;\zeta) \! = \! m_{-}(x,t;\zeta) \mathcal{G}(x,t;\zeta), \quad
\zeta \! \in \! \mathbb{R},
\end{equation*}
where $\mathcal{G}(x,t;\zeta) \! := \! \exp (-\mi k(\zeta)(x \! + \! 2 \lambda
(\zeta)t) \mathrm{ad}(\sigma_{3})) \!
\left(
\begin{smallmatrix}
1+r(\zeta)r(\tfrac{1}{\zeta}) & \, \, r(\tfrac{1}{\zeta}) \\
r(\zeta) & \, \, 1
\end{smallmatrix}
\right)$, and $r(\zeta)$, the reflection coefficient of the direct
scattering problem for $\mathcal{O}^{\mathcal{D}}$, satisfies $r(\zeta) \!
=_{\zeta \to 0} \! \mathcal{O}(\zeta)$, $r(\zeta) \! =_{\zeta \to \infty}
\! \mathcal{O}(\zeta^{-1})$, $r(\tfrac{1}{\zeta}) \! = \! -\overline{r
(\overline{\zeta})}$, and $r(\zeta) \! \in \! \mathcal{S}_{\mathbb{C}}^{1}
(\mathbb{R});$
\item[(iii)] for the simple poles of $m(x,t;\zeta)$ at $\{\varsigma_{n}\}_{
n=1}^{N}$ and $\{\overline{\varsigma_{n}}\}_{n=1}^{N}$, there exist
nilpotent matrices, with degree of nilpotency 2, such that the residues of
$m(x,t;\zeta)$ satisfy the polar conditions
\begin{align}
\mathrm{res}(m(x,t;\zeta);\varsigma_{n}) &= \! \lim_{\zeta \, \to \,
\varsigma_{n}}m(x,t;\zeta)g_{n}(x,t) \sigma_{-}, \quad n \! \in \! \{
1,2,\ldots,N\}, \nonumber \\
\mathrm{res}(m(x,t;\zeta);\overline{\varsigma_{n}}) &= \! \sigma_{1}
\overline{\mathrm{res}(m(x,t;\zeta);\varsigma_{n})} \, \sigma_{1}, \quad \,
\, n \! \in \! \{1,2,\ldots,N\}, \nonumber
\end{align}
where $g_{n}(x,t) \! := \! g_{n} \me^{2 \mi k(\varsigma_{n})(x+2 \lambda
(\varsigma_{n})t)}$, with $g_{n} \! = \! b_{n}(\varsigma_{n} \! - \!
\overline{\varsigma_{n}}) \me^{-\mi \widehat{\theta}(\varsigma_{n})} \prod_{
\genfrac{}{}{0pt}{2}{k=1}{k \not= n}}^{N} \! \left( \tfrac{\varsigma_{n}-
\overline{\varsigma_{k}}}{\varsigma_{n}-\varsigma_{k}} \right)$ and
$\widehat{\theta}(z) \! := \! \tfrac{\theta}{2} \! + \! \int_{-\infty}^{+
\infty} \tfrac{\ln (1-\vert r(\mu) \vert^{2})}{(\mu-z)} \, \tfrac{\md \mu}
{2 \pi}$, and $b_{n} \! := \! \vert b_{n} \vert \me^{\mi \theta_{b_{n}}}$,
$\theta_{b_{n}} \! \in \! \{\pm \tfrac{\pi}{2}\}$, are (pure imaginary)
constants associated with the direct scattering problem for the operator
$\mathcal{O}^{\mathcal{D}};$
\item[(iv)] $\det (m(x,t;\zeta)) \vert_{\zeta = \pm 1} \! = \! 0;$
\item[(v)] $m(x,t;\zeta) \! \underset{\zeta \, \to \, 0}{=} \! \tfrac{1}{
\zeta} \sigma_{2} \! + \! \mathcal{O}(1);$
\item[(vi)] as $\zeta \! \to \! \infty$, $\zeta \! \in \! \mathbb{C}
\setminus (\sigma_{d} \cup \sigma_{c})$, $m(x,t;\zeta) \! = \! \mathrm{I}
\! + \! \mathcal{O}(\zeta^{-1});$
\item[(vii)] $m(x,t;\zeta)$ possesses the following symmetry reductions,
$m(x,t;\zeta) \! = \! \sigma_{1} \overline{m(x,t;\overline{\zeta})} \,
\sigma_{1}$ and $m(x,t;\tfrac{1}{\zeta}) \! = \! \zeta m(x,t;\zeta)
\sigma_{2}$.
\end{enumerate}
For $r(\zeta) \! \in \! \mathcal{S}_{\mathbb{C}}^{1}(\mathbb{R})$: (I) the
{\rm RHP} for $m(x,t;\zeta)$ formulated above is uniquely (asymptotically)
solvable; and (II) $\Phi (x,t;\zeta) \! := \! m(x,t;\zeta) \exp (-\mi k(\zeta)
(x \! + \! 2 \lambda (\zeta)t) \sigma_{3})$ solves system~{\rm (2)}, with
\begin{equation}
u(x,t) \! := \! \mi \lim_{\genfrac{}{}{0pt}{2}{\zeta \, \to \, \infty}
{\zeta \, \in \, \mathbb{C} \, \setminus \, (\sigma_{d} \cup \sigma_{
c})}}(\zeta (m(x,t;\zeta) \! - \! \mathrm{I}))_{12}
\end{equation}
the solution of the Cauchy problem for the {\rm D${}_{f}$NLSE},
and
\begin{equation}
\int\nolimits_{+\infty}^{x}(\vert u(\xi,t) \vert^{2} \! - \! 1) \, \md
\xi \! := \! -\mi \lim_{\genfrac{}{}{0pt}{2}{\zeta \, \to \, \infty}
{\zeta \, \in \, \mathbb{C} \, \setminus \, (\sigma_{d} \cup \sigma_{
c})}}(\zeta (m(x,t;\zeta) \! - \! \mathrm{I}))_{11}.
\end{equation}
\end{ccccc}

\emph{Sketch of Proof.} {}From the definition of $\Phi (x,t;\zeta)$, $\zeta
\! \in \! \mathbb{C} \setminus \mathbb{R}$, given in the Lemma, the $\zeta
\! \to \! \infty$ and $\zeta \! \to \! 0$ asymptotics for $\Psi^{\pm}(x,0;
\zeta)$ (respectively~$a(\zeta))$ given in Lemma~2.1
(respectively~Corollary~2.3), the (evolution) rule $r(\zeta) \! \to \!
r(\zeta) \exp (4 \mi k(\zeta) \lambda (\zeta)t)$, and the fact that
$a(\zeta)$ is independent of $t$, one shows that
\begin{gather}
\Phi (x,t;\zeta) \! \underset{\zeta \, \to \, \infty}{=} \! \left(
\mathrm{I} \! + \! \tfrac{1}{\zeta} \!
\left(
\begin{smallmatrix}
\mi \int_{+\infty}^{x}(\vert u(\xi,t) \vert^{2}-1) \, \md \xi & -\mi
u(x,t) \\
\mi \overline{u(x,t)} & -\mi \int_{+\infty}^{x}(\vert u(\xi,t) \vert^{
2}-1) \, \md \xi
\end{smallmatrix}
\right) \! + \! \mathcal{O}(\tfrac{1}{\zeta^{2}}) \right) \! \me^{-\mi
k(\zeta)(x+2 \lambda (\zeta)t) \sigma_{3}}, \nonumber \\
\Phi (x,t;\zeta) \! \underset{\zeta \, \to \, 0}{=} \! \left(
\tfrac{1}{\zeta} \sigma_{2} \! + \! \mathcal{O}(1) \right) \! \me^{-\mi
k(\zeta)(x+2 \lambda (\zeta)t) \sigma_{3}}: \nonumber
\end{gather}
this can also be obtained by mimicking calculations similar to those given
in the proof of Lemma~2.1. {}From the monodromy and unimodularity relations
given in Proposition~2.4, one shows that $\Phi (x,t;\zeta)$, with $\Phi_{
\pm}(x,t;\zeta) \! := \! \lim_{\varepsilon \downarrow 0} \Phi (x,t;\zeta
\! \pm \! \mi \varepsilon)$, satisfies the ``jump'' relation $\Phi_{+}(x,
t;\zeta) \! = \! \Phi_{-}(x,t;\zeta) \!
\left(
\begin{smallmatrix}
1-r(\zeta) \overline{r(\overline{\zeta})} & \, -\overline{r(\overline{
\zeta})} \\
r(\zeta) & \, 1
\end{smallmatrix}
\right)$, $\zeta \! \in \! \mathbb{R}$, where, {}from Corollaries~2.2
and~2.3, and Remark~2.2, $r(\tfrac{1}{\zeta}) \! = \! -\overline{r
(\overline{\zeta})}$, $r(\zeta) \! =_{\zeta \to \infty} \! \mathcal{O}
(\zeta^{-1})$, $r(\zeta) \! =_{\zeta \to 0} \! \mathcal{O}(\zeta)$, and
$r(\zeta) \! \in \! \mathcal{S}_{\mathbb{C}}(\mathbb{R})$. {}From the
definition of $\Phi (x,t;\zeta)$, $\zeta \! \in \! \mathbb{C} \setminus
\mathbb{R}$, and the representation for $a(\zeta)$ given in Lemma~2.2,
one shows that, for $\zeta \! \in \! \mathbb{C}_{+}$, $\Phi (x,t;\zeta)
\! = \!
\left(
\begin{smallmatrix}
\me^{-\mi \widehat{\theta}(\zeta)} \prod_{n=1}^{N} \tfrac{(\zeta-\overline{
\varsigma_{n}})}{(\zeta-\varsigma_{n})} \Psi^{-}_{11}(x,t;\zeta) & \, \, \,
\Psi^{+}_{12}(x,t;\zeta) \\
\me^{-\mi \widehat{\theta}(\zeta)} \prod_{n=1}^{N} \tfrac{(\zeta-\overline{
\varsigma_{n}})}{(\zeta-\varsigma_{n})} \Psi^{-}_{21}(x,t;\zeta) & \, \, \,
\Psi^{+}_{22}(x,t;\zeta)
\end{smallmatrix}
\right)$, and, for $\zeta \! \in \! \mathbb{C}_{-}$, $\Phi (x,t;\zeta) \!
= \!
\left(
\begin{smallmatrix}
\Psi^{+}_{11}(x,t;\zeta) & \, \, \, \me^{\mi \widehat{\theta}(\zeta)} \prod_{
n=1}^{N} \tfrac{(\zeta-\varsigma_{n})}{(\zeta-\overline{\varsigma_{n}})}
\Psi^{-}_{12}(x,t;\zeta) \\
\Psi^{+}_{21}(x,t;\zeta) & \, \, \, \me^{\mi \widehat{\theta}(\zeta)} \prod_{
n=1}^{N} \tfrac{(\zeta-\varsigma_{n})}{(\zeta-\overline{\varsigma_{n}})}
\Psi^{-}_{22}(x,t;\zeta)
\end{smallmatrix}
\right)$, with $\widehat{\theta}(z)$ as defined in the Lemma; hence, it
follows that the polar structure of $\Phi (x,t;\zeta)$ is such that, for
$\zeta \! \in \! \mathbb{C}_{+}$ (respectively~$\zeta \! \in \! \mathbb{
C}_{-})$, the first (respectively~second) column of $\Phi (x,t;\zeta)$ has
(a finitely denumerable number of) simple poles at $\{\varsigma_{n}\}_{n
=1}^{N}$ (respectively~$\{\overline{\varsigma_{n}}\}_{n=1}^{N})$, where
$\varsigma_{n} \! := \! \me^{\mi \phi_{n}}$, $\phi_{n} \! \in \! (0,\pi)$.
Since $a(\zeta) \vert_{\zeta =\varsigma_{n}} \! = \! 0$, $n \! \in \!
\{1,2,\ldots,N\}$, it follows {}from the monodromy relation given in
Proposition~2.4 that $\Psi^{-}(x,t;\varsigma_{n}) \! = \! \Psi^{+}(x,t;
\varsigma_{n}) \!
\left(
\begin{smallmatrix}
0 & \overline{b(\overline{\varsigma_{n}})} \\
b(\varsigma_{n}) & 0
\end{smallmatrix}
\right)$; hence, {}from this relation and the definition of $\Phi (x,t;
\zeta)$ for $\zeta \! \in \! \mathbb{C}_{+}$, it follows that, with the
choice $g_{n} \! = \! b_{n}(\varsigma_{n} \! - \! \overline{\varsigma_{n}})
\me^{-\mi \widehat{\theta}(\varsigma_{n})} \prod_{\genfrac{}{}{0pt}{2}{k
=1}{k \not= n}}^{N} \tfrac{(\varsigma_{n}-\overline{\varsigma_{k}})}{
(\varsigma_{n}-\varsigma_{k})}$, $\mathrm{res}(\Phi (x,t;\zeta);\varsigma_{
n}) \! = \! \lim_{\zeta \to \varsigma_{n}} \Phi (x,t;\zeta)g_{n} \sigma_{-}$,
$n \! \in \! \{1,2,\ldots,N\}$: using the fact that $\widehat{\theta}
(\overline{\varsigma_{n}}) \! = \! \overline{\widehat{\theta}(\varsigma_{
n})}$, a similar argument shows that $\mathrm{res}(\Phi (x,t;\zeta);
\overline{\varsigma_{n}}) \! = \! \sigma_{1} \overline{\mathrm{res}(\Phi
(x,t;\zeta);\varsigma_{n})} \, \sigma_{1}$, $n \! \in \! \{1,2,\ldots,N\}$.
{}From the definition of $\Phi (x,t;\zeta)$, $\zeta \! \in \! \mathbb{C}
\setminus \mathbb{R}$, the jump condition for $\Phi (x,t;\zeta)$, and the
fact that $\Psi (x,t;\zeta)$ satisfies the $\sigma_{1}$ symmetry reduction
$\Psi (x,t;\zeta) \! = \! \sigma_{1} \overline{\Psi (x,t;\overline{\zeta})}
\, \sigma_{1}$, it follows that $\sigma_{1} \overline{\Phi_{\pm}(x,t;
\overline{\zeta})} \, \sigma_{1} \! = \! \Phi_{\mp}(x,t;\zeta)$, namely,
$\Phi (x,t;\zeta) \! = \! \sigma_{1} \overline{\Phi (x,t;\overline{\zeta})}
\, \sigma_{1}$: a similar argument, along with the fact that $r(\tfrac{1}
{\zeta}) \! = \! -\overline{r(\overline{\zeta})}$, shows that $\Phi (x,t;
\tfrac{1}{\zeta}) \! = \! \zeta \Phi (x,t;\zeta) \sigma_{2}$. The fact that
$\det (\Phi (x,t;\zeta)) \vert_{\zeta =\pm 1} \! = \! 0$ is a consequence
of the definition of $\Phi (x,t;\zeta)$, the asymptotics (proof of Lemma~2.3)
$a(\zeta) \! =_{\zeta \to \pm 1} \! \widehat{a}_{o}^{\pm} \! + \! \mathcal{O}
(\zeta \! \mp \! 1)$, with $\widehat{a}_{o}^{\pm} \! \not= \! 0$, and the
degeneracy condition $\det (\Psi (x,t;\zeta)) \vert_{\zeta = \pm 1} \! = \!
0$ (Proposition~2.4). Now, setting $m(x,t;\zeta) \! := \! \Phi (x,t;\zeta)
\exp (\mi k(\zeta)(x \! + \! 2 \lambda (\zeta)t) \sigma_{3})$, one arrives
at items~\emph{(ii)}--\emph{(vii)} of the Lemma: item~\emph{(i)} is an
immediate consequence of the polar structure of
$m(x,t;\zeta)$ and the fact that $m(x,t;\zeta)$ has a
jump discontinuity across $\mathbb{R}$ (oriented {}from $-\infty$ to
$+\infty)$. Since the solution of the RHP for $m(x,t;\zeta)$ can be
written as the ordered factorisation $m(x,t;\zeta) \! = \! (\mathrm{I} \!
+ \! \Delta_{o}(x,t) \zeta^{-1}) \mathscr{P}(x,t;\zeta) m_{d}(x,t;\zeta)
m^{c}(x,t;\zeta)$, where $m^{c}(x,t;\zeta)$ $(= \! \sigma_{1}
\overline{m^{c}(x,t;\overline{\zeta})} \, \sigma_{1})$ $\in \! \mathrm{
SL}(2,\mathbb{C})$, $m_{d}(x,t;\zeta)$ $(= \! \sigma_{1} \overline{
m_{d}(x,t;\overline{\zeta})} \, \sigma_{1})$ $\in \! \mathrm{SL}(2,
\mathbb{C})$ has the representation $m_{d}(x,t;\zeta) \! = \! \mathrm{
I} \! + \! \sum_{n=1}^{N} \! \left( \tfrac{\mathrm{res}(m(x,t;\zeta);
\varsigma_{n})}{(\zeta-\varsigma_{n})} \! + \! \tfrac{\sigma_{1}
\overline{\mathrm{res}(m(x,t;\zeta);\varsigma_{n})} \, \sigma_{
1}}{(\zeta -\overline{\varsigma_{n}})} \right)$, $\mathrm{I} \! + \!
\Delta_{o}(x,t) \zeta^{-1} \! \in \! \mathrm{M}(2,\mathbb{C})$, with
$\Delta_{o}(x,t)$ $(= \! \sigma_{1} \overline{\Delta_{o}(x,t)} \,
\sigma_{1})$ $\in \! \mathrm{GL}(2,\mathbb{C})$, is analytic in a
punctured neighbourhood of the origin, and $\mathscr{P}(x,t;
\zeta)$ $(= \! \sigma_{1} \overline{\mathscr{P}(x,t;\overline{
\zeta})} \, \sigma_{1})$ $\in \! \mathrm{GL}(2,\mathbb{C})$ is
chosen so that $\Delta_{o}(x,t)$ $(= \! \mathscr{P}(x,t;0)m_{d}
(x,t;0)m^{c}(x,t;0) \sigma_{2})$ is idempotent, and, for $\vert \vert
r(\cdot) \vert \vert_{\mathcal{L}^{\infty}(\mathbb{R})} \! := \! \sup_{
\zeta \in \mathbb{R}} \vert r(\zeta) \vert \! < \! 1$, $\tfrac{1}{2}
(\mathcal{G}^{\dagger}(x,t;\overline{\zeta}) \! + \! \mathcal{G}(x,t;
\zeta))$, with ${}^{\dagger}$ denoting Hermitian conjugation, is positive
definite, it follows that, either due to a classical result of Gohberg
and Krein (see \cite{a23} for details) or Zhou's skew Schwarz
reflection invariant symmetry principle (see Theorem~9.3 in
\cite{a31} and Theorem~2.2.10 in \cite{a28}: see, also,
\cite{a32}), and the fact that $\sum_{j=1}^{2} \mathrm{k}_{j} \!
= \! \tfrac{1}{2 \pi} \int_{\mathbb{R}} \md (\arg (\det (\mathcal{
G}(x,t;\xi)))) \! = \! 0$, where $\mathrm{k}_{j}$, $j \! \in \! \{1,2\}$,
denote the partial indices of the RHP factorisation for $m(x,t;
\zeta)$, and the winding number of $(1 \! - \! r(\zeta) \overline{
r(\overline{\zeta})})$ vanishes (see Proposition~1.38 in
\cite{a33}), that is, $\mathrm{W}_{\mathbb{R}}(1 \! - \! r(\zeta)
\overline{r(\overline{\zeta})}) \! = \! \sum_{l \in \{\pm\}} \!
\mathrm{s}(l) \mathrm{n}(l) \! = \! 0$, where $\mathrm{s}(+) \! = \!
-\mathrm{s}(-) \! = \! 1$ and $\mathrm{n}(\pm) \! := \! \mathrm{card}
\{\mathstrut \varsigma_{n}; \, \mp \Im (\varsigma_{n}) \! > \! 0\}$, the
RHP for $m(x,t;\zeta)$ is asymptotically solvable: uniqueness follows
{}from a standard argument (see, for example, Chapter~7 of \cite{a25}).
The fact that $u(x,t)$, defined by Eq.~(3), satisfies the D${}_{f}$NLSE
follows {}from the $\zeta \! \to \! \infty$ asymptotics of $m(x,t;\zeta)$
(see the $\zeta \! \to \! \infty$ asymptotics for $\Phi (x,t;\zeta)$ given
at the beginning of the proof and recall the definition of $m(x,t;\zeta)$
in terms of $\Phi (x,t;\zeta))$. The fact that $\Phi (x,t;\zeta) \! = \!
m(x,t;\zeta) \exp (-\mi k(\zeta)(x \! + \! 2 \lambda (\zeta)t) \sigma_{3})$
satisfies system~(2) is a straightforward calculation. \hfill $\square$
\begin{eeeee}
The precise sense in which the limits for $m(x,t;\zeta)$ are taken is
explained in detail in Chapter~7 of \cite{a25}. It is convenient to define
the formal (space of) scattering data as $\mathscr{SD} \! := \! \left\{
\mathstrut \!
\left(
\begin{smallmatrix}
1-r(\zeta) \overline{r(\overline{\zeta})} & -\overline{r(\overline{\zeta})} \\
r(\zeta) & 1
\end{smallmatrix}
\right), \, \zeta \! \in \! \mathbb{R} \right\} \cup (\cup_{n=1}^{N}\{
\mathstrut \phi_{n}, \, g_{n} \! := \! \vert g_{n} \vert \me^{\mi \phi_{
g_{n}}}\})$ $(\subset \mathcal{M}_{\infty} \times \mathbb{R}^{3n}$, where
$\mathcal{M}_{\infty}$ is some, unspecified for the time being,
infinite-dimensional space): setting $\mathscr{SD}_{o} \! := \! \left\{
\mathstrut \! \left(
\begin{smallmatrix}
1-r(\zeta) \overline{r(\overline{\zeta})} & -\overline{r(\overline{\zeta})} \\
r(\zeta) & 1
\end{smallmatrix}
\right), \, \zeta \! \in \! \mathbb{R} \right\} \cup (\cup_{n=1}^{N}\{
\mathstrut \phi_{n}\})$ $(\subset \mathscr{SD})$, one notes {}from the
expression for $\theta$ given in Lemma~2.2 that $\theta \colon \mathscr{SD}_{
o} \! \to \! [0,2 \pi)$, namely, $\{\mathstrut r(\zeta), \, \zeta \! \in \!
\mathbb{R}\} \cup (\cup_{n=1}^{N}\{\mathstrut \phi_{n}\}) \! \mapsto \! -2
\sum_{n=1}^{N} \phi_{n} \! - \! \int_{-\infty}^{+\infty} \tfrac{\ln (1-\vert
r(\mu) \vert^{2})}{\mu} \, \tfrac{\md \mu}{2 \pi}$.
\end{eeeee}
Since $a(\zeta)$ is independent of $t$, it follows {}from a well-known
result \cite{a9} that $\ln (a(\zeta))$ is the generating function for
the integrals of motion, namely, $\ln (a(\zeta)) \! =_{\genfrac{}{}{0pt}
{2}{\zeta \to \infty}{\zeta \in \mathbb{C}_{+}}} \! \sum_{n=0}^{\infty}
\mathrm{I}_{n}[u,\overline{u}] \zeta^{-n} \! + \! \mathcal{O}(\vert \zeta
\vert^{-\infty})$, where $\mathrm{I}_{n}[u,\overline{u}]$ are local
functionals of $u(x,t)$ and $\overline{u(x,t)}$. This leads to the
following (trace identity)
\begin{bbbbb}
\begin{equation*}
\int\nolimits_{-\infty}^{+\infty}(\vert u(\xi,t) \vert^{2} \! - \! 1) \,
\md \xi \! = \! -2 \sum_{n=1}^{N} \sin (\phi_{n}) \! - \!
\int\nolimits_{-\infty}^{+\infty} \ln (1 \! - \! \vert r(\mu) \vert^{2}) \,
\tfrac{\md \mu}{2 \pi}.
\end{equation*}
\end{bbbbb}

\emph{Proof.} {}From Corollary~2.3, it follows that, for $\zeta \! \in
\! \mathbb{C}_{+}$, $\ln (a(\zeta)) \! =_{\zeta \to \infty} \! \tfrac{
\mi \theta}{2} \! + \! (\mi \int_{-\infty}^{+\infty}(\vert u(\xi,t)
\vert^{2} \! - \! 1) \, \md \xi) \zeta^{-1} \! + \! \mathcal{O}(\zeta^{
-2})$. Using the representation for $a(\zeta)$ given in Lemma~2.2 and
the fact that $r(\zeta) \! \in \! \mathcal{S}_{\mathbb{C}}(\mathbb{R})$,
one deduces that, for $\zeta \! \in \! \mathbb{C}_{+}$, $\ln (a(\zeta))
\! =_{\zeta \to \infty} \! \tfrac{\mi \theta}{2} \! + \! (\sum_{n=1}^{N}
(\overline{\varsigma_{n}} \! - \! \varsigma_{n}) \! + \! \int_{-\infty}^{
+\infty} \ln (1 \! - \! \vert r(\mu) \vert^{2}) \, \tfrac{\md \mu}{2 \pi
\mi}) \zeta^{-1} \! + \! \mathcal{O}(\zeta^{-2})$: noting that $\sum_{n=
1}^{N}(\overline{\varsigma_{n}} \! - \! \varsigma_{n}) \! = \! -2 \mi
\sum_{n=1}^{N} \sin (\phi_{n})$, with $\vert \sum_{n=1}^{N}(\overline{
\varsigma_{n}} \! - \! \varsigma_{n}) \vert \! \leqslant \! 2N$, equating
the above (two) $\zeta \! \to \! \infty$ asymptotics for $\ln (a(\zeta))$,
one obtains the result stated in the Proposition. \hfill $\square$

Heretofore, the discussion has focused on the general case when $\sigma_{
\mathcal{O}^{\mathcal{D}}} \! := \! \mathrm{spec}(\mathcal{O}^{\mathcal{D}})
\! = \! \sigma_{d} \cup \sigma_{c}$, with $\sigma_{d} \cap \sigma_{c} \! =
\! \emptyset$, where $\sigma_{d}$ and $\sigma_{c}$ subsume, respectively,
the ``$N$-dark soliton'' (discrete spectrum) and ``continuum'' (continuous
spectrum) contributions to the solution of the Cauchy problem for the
D${}_{f}$NLSE: in this work, only the asymptotics of solutions to the Cauchy
problem for the D${}_{f}$NLSE in the so-called \emph{solitonless sector},
that is, $\sigma_{d} \! \equiv \! \emptyset$ and $\sigma_{c} \! \not\equiv \!
\emptyset$, are studied (the asymptotic analysis for the general case $\{
\sigma_{d},\sigma_{c}\} \! \not\equiv \! \{\emptyset,\emptyset\}$ is in
progress). Essentially, in the construction of the asymptotic solution of
the RHP for the $\mathrm{M}(2,\mathbb{C})$-valued function $m(x,t;\zeta)$
stated in Lemma~2.4, $\sigma_{d} \! \equiv \! \emptyset$ means dropping all
reference to the polar structure of $m(x,t;\zeta)$ at $\{\varsigma_{n},
\overline{\varsigma_{n}}\}_{n=1}^{N}$: all terms depending, in any
way, on $\{\varsigma_{n},\overline{\varsigma_{n}}\}_{n=1}^{N}$ are
absent {}from the specification of the RHP for $m(x,t;\zeta)$; in
particular, $a(\zeta) \! = \! \me^{\frac{\mi \theta}{2}} \exp \! \left(\!
-\int_{-\infty}^{+\infty} \tfrac{\ln (1-\vert r(\mu) \vert^{2})}{(\mu
-\zeta)} \, \tfrac{\md \mu}{2 \pi \mi} \right)$, $\zeta \! \in \! \mathbb{
C}_{+}$, with $0 \! \leqslant \! \theta \! = \! -\int_{-\infty}^{+\infty}
\tfrac{\ln (1-\vert r(\mu) \vert^{2})}{\mu} \, \tfrac{\md \mu}{2 \pi} \! <
\! 2 \pi$, $m_{d}(x,t;\zeta) \! \equiv \! \mathrm{I}$, and it is sufficient
to take $\mathscr{P}(x,t;\zeta) \! \equiv \! \mathrm{I}$. One is lead to
the following (normalised at $\infty)$ RHP for $m(x,t;\zeta)$:
\begin{ccccc}
Let $u(x,t)$ be the solution of the Cauchy problem for the
{\rm D${}_{f}$NLSE} with finite-density initial data $u(x,0) \! := \!
u_{o}(x) \! =_{x \to \pm \infty} \! u_{o}(\pm \infty)(1 \! + \! o(1))$,
where $u_{o}(\pm \infty) \! := \! \exp (\tfrac{\mi (1 \mp 1) \theta}{2})$,
$0 \! \leqslant \! \theta \! = \! -\int_{-\infty}^{+\infty} \tfrac{\ln
(1-\vert r(\mu) \vert^{2})}{\mu} \, \tfrac{\md \mu}{2 \pi} \! < \! 2 \pi$,
$u_{o}(x) \! \in \! \mathbf{C}^{\infty}(\mathbb{R})$, and $u_{o}(x) \! -
\! u_{o}(\pm \infty) \! \in \! \mathcal{S}_{\mathbb{C}}(\mathbb{R}_{\pm})$.
Then $m(x,t;\zeta) \colon \mathbb{C} \setminus \sigma_{c} \! \to \!
\mathrm{M}(2,\mathbb{C})$ solves the following {\rm RHP:}
\begin{enumerate}
\item[(i)] $m(x,t;\zeta)$ is piecewise holomorphic $\forall \, \zeta \!
\in \! \mathbb{C} \setminus \sigma_{c};$
\item[(ii)] $m_{\pm}(x,t;\zeta) \! := \! \lim_{\varepsilon \downarrow 0}
m(x,t;\zeta \! \pm \! \mi \varepsilon)$ satisfy the jump condition
\begin{equation*}
m_{+}(x,t;\zeta) \! = \! m_{-}(x,t;\zeta) \mathcal{G}(x,t;\zeta), \quad
\zeta \! \in \! \mathbb{R},
\end{equation*}
where $\mathcal{G}(x,t;\zeta) \! := \! \exp (-\mi k(\zeta)(x \! + \! 2
\lambda (\zeta)t) \mathrm{ad}(\sigma_{3})) \!
\left(
\begin{smallmatrix}
1-r(\zeta) \overline{r(\overline{\zeta})} & \, \, -\overline{r(\overline{
\zeta})} \\
r(\zeta) & \, \, 1
\end{smallmatrix}
\right)$, and $r(\zeta)$, the reflection coefficient of the direct
scattering problem for $\mathcal{O}^{\mathcal{D}}$, satisfies $r(\zeta)
\! =_{\zeta \to 0} \! \mathcal{O}(\zeta)$, $r(\zeta) \! =_{\zeta \to
\infty} \! \mathcal{O}(\zeta^{-1})$, $r(\tfrac{1}{\zeta}) \! = \! -
\overline{r(\overline{\zeta})}$, and $r(\zeta) \! \in \! \mathcal{S}_{
\mathbb{C}}^{1}(\mathbb{R});$
\item[(iii)] $\det (m(x,t;\zeta)) \vert_{\zeta = \pm 1} \! = \! 0;$
\item[(iv)] $m(x,t;\zeta) \! \underset{\zeta \, \to \, 0}{=} \! \tfrac{1}
{\zeta} \sigma_{2} \! + \! \mathcal{O}(1);$
\item[(v)] as $\zeta \! \to \! \infty$, $\zeta \! \in \! \mathbb{C}
\setminus \sigma_{c}$, $m(x,t;\zeta) \! = \! \mathrm{I} \! + \! \mathcal{
O}(\zeta^{-1});$
\item[(vi)] $m(x,t;\zeta)$ possesses the following symmetry reductions,
$m(x,t;\zeta) \! = \! \sigma_{1} \overline{m(x,t;\overline{\zeta})} \,
\sigma_{1}$ and $m(x,t;\tfrac{1}{\zeta}) \! = \! \zeta m(x,t;\zeta)
\sigma_{2}$.
\end{enumerate}
For $r(\zeta) \! \in \! \mathcal{S}_{\mathbb{C}}^{1}(\mathbb{R})$: (I) the
{\rm RHP} for $m(x,t;\zeta)$ formulated above is uniquely (asymptotically)
solvable; and (II) $\Phi (x,t;\zeta) \! := \! m(x,t;\zeta) \exp (-\mi k
(\zeta)(x \! + \! 2 \lambda (\zeta) t) \sigma_{3})$, with $\Phi (x,t;\zeta)$
in terms of $\Psi^{\pm}(x,t;\zeta)$ as defined in Lemma~{\rm 2.4} for
$a(\zeta) \! = \! \exp (\tfrac{\mi \theta}{2}) \exp \! \left(\! -\int_{-
\infty}^{+\infty} \tfrac{\ln (1-\vert r(\mu) \vert^{2})}{(\mu -\zeta)} \,
\tfrac{\md \mu}{2 \pi \mi} \right)$, $\zeta \! \in \! \mathbb{C}_{+}$,
solves system~{\rm (2)},
\begin{equation}
u(x,t) \! := \! \mi \lim_{\genfrac{}{}{0pt}{2}{\zeta \, \to \, \infty}
{\zeta \, \in \, \mathbb{C} \, \setminus \, \sigma_{c}}}(\zeta (m(x,t;
\zeta) \! - \! \mathrm{I}))_{12}
\end{equation}
is the solution of the Cauchy problem for the {\rm D${}_{f}$NLSE},
and
\begin{equation}
\int\nolimits_{+\infty}^{x}(\vert u(\xi,t) \vert^{2} \! - \! 1) \, \md
\xi \! := \! -\mi \lim_{\genfrac{}{}{0pt}{2}{\zeta \, \to \, \infty}
{\zeta \, \in \, \mathbb{C} \, \setminus \, \sigma_{c}}}(\zeta (m(x,t;
\zeta) \! - \! \mathrm{I}))_{11}.
\end{equation}
\end{ccccc}
\begin{fffff}
In the solitonless sector $(\sigma_{d} \! \equiv \! \emptyset)$,
\begin{equation*}
\int\nolimits_{-\infty}^{+\infty}(\vert u(\xi,t) \vert^{2} \! - \! 1) \,
\md \xi \! = \! -\int\nolimits_{-\infty}^{+\infty} \ln (1 \! - \! \vert
r(\mu) \vert^{2}) \, \tfrac{\md \mu}{2 \pi}.
\end{equation*}
\end{fffff}

\emph{Proof.} Since, in the solitonless sector, $\Delta_{a} \! \equiv \!
\emptyset$ (Lemma~2.2), it follows that $\sum_{n=1}^{N} \sin (\phi_{
n}) \! = \! 0$: the result stated in the Corollary is now a consequence
of Proposition~2.6. \hfill $\square$
\begin{ccccc}
The solution of the {\rm RHP} for $m(x,t;\zeta) \colon \mathbb{C} \setminus
\sigma_{c} \! \to \! \mathrm{M}(2,\mathbb{C})$ formulated in Lemma
{\rm 2.5} can be written as the following ordered product,
\begin{equation}
m(x,t;\zeta) \! = \! (\mathrm{I} \! + \! \Delta_{o}(x,t) \zeta^{-1})m^{c}
(x,t;\zeta),
\end{equation}
where $\Delta_{o}(x,t) \! \in \! \mathrm{GL}(2,\mathbb{C})$, $\det (\mathrm{
I} \! + \! \Delta_{o}(x,t) \zeta^{-1}) \vert_{\zeta = \pm 1} \! = \! 0$,
$\sigma_{1} \overline{\Delta_{o}(x,t)} \, \sigma_{1} \! = \! \Delta_{o}(x,
t)$, with finite, order 2, matrix involutive structure,
\begin{equation*}
\Delta_{o}(x,t) \! = \!
\left( \begin{array}{cc}
\Delta_{o}^{11}(x,t) \me^{\mi (k+1/2) \pi} & \sqrt{1 \! + \! (\Delta_{o}^{11}
(x,t))^{2}} \, \me^{-\mi \vartheta (x,t)} \\
\sqrt{1 \! + \! (\Delta_{o}^{11}(x,t))^{2}} \, \me^{\mi \vartheta (x,t)} &
\quad \Delta_{o}^{11}(x,t) \me^{-\mi (k+1/2) \pi}
\end{array} \right) \! , \quad k \! \in \! \mathbb{Z},
\end{equation*}
and $\Delta_{o}^{11}(x,t) \colon \mathbb{R} \times \mathbb{R} \! \to
\! \mathbb{R}$ and $\vartheta (x,t) \colon \mathbb{R} \times
\mathbb{R} \! \to \! \mathbb{R} \setminus \{0\}$ obtained {}from the
determining relation $\Delta_{o}(x,t)m^{c}(x,t;0) \! = \! \sigma_{2}$
$(\det (\Delta_{o}(x,t)) \! = \! \det (\sigma_{2}) \! = \! -1)$, and
$m^{c}(x,t;\zeta) \colon \mathbb{C} \setminus \sigma_{c} \! \to \!
\mathrm{SL}(2,\mathbb{C})$ solves the following {\rm RHP:} (1) $m^{c}(x,t;
\zeta)$ is piecewise holomorphic $\forall \, \zeta \! \in \! \mathbb{C}
\setminus \sigma_{c};$ (2) $m^{c}_{+}(x,t;\zeta) \! = \! m^{c}_{-}(x,t;
\zeta) \mathcal{G}(x,t;\zeta)$, $\zeta \! \in \! \mathbb{R}$, where
$\mathcal{G}(x,t;\zeta)$ is defined in Lemma~{\rm 2.5;} (3) as $\zeta \!
\to \! \infty$, $\zeta \! \in \! \mathbb{C} \setminus \sigma_{c}$, $m^{c}
(x,t;\zeta) \! = \! \mathrm{I} \! + \! \mathcal{O}(\zeta^{-1});$ and (4)
$m^{c}(x,t;\zeta)$ satisfies the symmetry reduction $m^{c}(x,t;\zeta) \! =
\! \sigma_{1} \overline{m^{c}(x,t;\overline{\zeta})} \, \sigma_{1}$ and the
condition $(m^{c}(x,t;0) \sigma_{2})^{2} \! = \! \mathrm{I}$.
\end{ccccc}

\emph{Proof.} Follows {}from the ordered (matrix) product given in Eq.~(7)
and Lemma~2.5. \hfill $\square$
\begin{eeeee}
The $\mathrm{M}(2,\mathbb{C})$-valued function $\mathrm{I} \! + \! \Delta_{
o}(x,t) \zeta^{-1}$, with $\Delta_{o}(x,t) \! \in \! \mathrm{GL}(2,\mathbb{
C})$, which is analytic in a punctured neighbourhood of the origin, plays
a role somewhat analogous to the (analytic) matrix-valued function $E(z)$
in \cite{a34}. One can take a Laurent expansion of the form $\mathrm{I} \!
+ \! \sum_{j \in \mathbb{Z}} \widetilde{\Delta}_{j}(x,t) \zeta^{-j}$, with
$\sigma_{1} \overline{\widetilde{\Delta}_{j}(x,t)} \, \sigma_{1} \! = \!
\widetilde{\Delta}_{j}(x,t)$ and $\widetilde{\Delta}_{j}(x,t) \! \in \!
\mathrm{GL}(2,\mathbb{C})$, for the first factor on the right-hand side of
the ordered product in Eq.~(7); however, in order to satisfy the conditions
imposed by Lemma~2.5 on the solution of the RHP for $m(x,t;\zeta)$, one
shows that $\widetilde{\Delta}_{j}(x,t) \! = \!
\left(
\begin{smallmatrix}
0 & 0 \\
0 & 0
\end{smallmatrix}
\right)$, $j \! \in \! \mathbb{Z} \! \setminus \! \{1\}$, and only the
$\widetilde{\Delta}_{1}(x,t) \! := \! \Delta_{o}(x,t)$ term is non-zero.
\end{eeeee}

Hence, the main technical thrust revolves around the explicit (asymptotic)
construction of $m^{c}(x,t;\zeta)$, and the determination of $\Delta_{o}
(x,t)$ is relegated to a purely algebraic computation, namely, $\Delta_{o}
(x,t) \! = \! \sigma_{2}(m^{c}(x,t;0))^{-1}$.
\section{The Beals-Coifman Construction, the Deift-Zhou
N\-o\-n-L\-i\-n\-e\-a\-r Steepest-Descent Method, and Summary of Results}
In this section, the Beals-Coifman (BC) construction \cite{a24} for the
representation of the solution of the (matrix) RHP for $m^{c}(x,t;\zeta)$
on $\mathbb{R}$ stated in Lemma~2.6 is succinctly recapitulated, the
Deift-Zhou (DZ) non-linear steepest-descent method \cite{a27} for the
asymptotic analysis of the RHP for $m^{c}(x,t;\zeta)$ is discussed in
detail, and the results of this work are summarised in Theorems~3.1--3.3.

The solution framework for matrix RHPs of the type stated in Section~2
(for $m^{c}(x,t;\zeta))$ is constructed according to the BC formulation
\cite{a24}, a self-contained synopsis of which, with some requisite
preamble, follows (explicit $x,t$ dependences are temporarily suppressed).
One agrees to call a contour $\Gamma^{\sharp}$ oriented if: (1) $\mathbb{C}
\setminus \Gamma^{\sharp}$ has finitely many open connected components;
(2) $\mathbb{C} \setminus \Gamma^{\sharp}$ is the disjoint union of two,
possibly disconnected, open regions, denoted by $\boldsymbol{\mho}^{
+}$ and $\boldsymbol{\mho}^{-}$; and (3) $\Gamma^{\sharp}$ may be
viewed as either the positively oriented boundary for $\boldsymbol{
\mho}^{+}$ or the negatively oriented boundary for $\boldsymbol{
\mho}^{-}$ ($\mathbb{C} \setminus \Gamma^{\sharp}$ is coloured by
two colours, namely, $\pm)$. Let $\Gamma^{\sharp}$, as a closed
set, be the union of finitely many oriented simple piecewise-smooth
arcs. Denote the set of all self-intersections of $\Gamma^{\sharp}$ by
$\widehat{\Gamma}^{\sharp}$ (with $\mathrm{card}(\widehat{\Gamma}^{
\sharp}) \! < \! \infty$ assumed throughout). Set $\widetilde{\Gamma}^{
\sharp} \! := \! \Gamma^{\sharp} \setminus \widehat{\Gamma}^{\sharp}$.
The BC construction for the solution of a (matrix) RHP, in the absence of a
discrete spectrum and spectral singularities \cite{a28}, on an oriented
contour $\Gamma^{\sharp}$ consists of finding an $\mathrm{M}(2,
\mathbb{C})$-valued function $\mathcal{X}(\lambda)$ such that: (1)
$\mathcal{X}(\lambda)$ is piecewise holomorphic $\forall \, \lambda
\! \in \! \mathbb{C} \setminus \Gamma^{\sharp}$, $\mathcal{X}(\lambda)
\! \! \upharpoonright_{\mathbb{C} \setminus \Gamma^{\sharp}}$ has
continuous extension to $\widetilde{\Gamma}^{\sharp}$, and
$\lim_{\genfrac{}{}{0pt}{2}{\lambda^{\prime} \, \to \, \lambda}{
\lambda^{\prime} \, \in \, \pm \, \mathrm{side} \, \mathrm{of} \,
\widetilde{\Gamma}^{\sharp}}} \int_{\widetilde{\Gamma}^{\sharp}}
\vert \mathcal{X}(\lambda^{\prime}) \! - \! \mathcal{X}_{\pm}(\lambda)
\vert^{2} \, \vert \md \lambda \vert \! = \! 0$; (2) $\mathcal{X}_{+}(\lambda)
\! = \! \mathcal{X}_{-}(\lambda) \upsilon (\lambda)$, $\lambda \! \in \!
\widetilde{\Gamma}^{\sharp}$, for some ``jump'' matrix $\upsilon (\lambda)
\colon \widetilde{\Gamma}^{\sharp} \! \to \! \mathrm{GL}(2,\mathbb{C})$;
and (3) uniformly as $\lambda \! \to \! \infty$, $\lambda \! \in \!
\mathbb{C} \setminus \Gamma^{\sharp}$, $\mathcal{X}(\lambda) \! = \!
\mathrm{I} \! + \! \mathcal{O}(\lambda^{-1})$. Let $\upsilon (\lambda)
\! := \! (\mathrm{I} \! - \! w_{-}(\lambda))^{-1}(\mathrm{I} \! + \!
w_{+}(\lambda))$, $\lambda \! \in \! \widetilde{\Gamma}^{\sharp}$, be a
factorisation for $\upsilon (\lambda)$, where $w_{\pm}(\lambda)$ are some
upper/lower or lower/upper triangular (depending on the orientation of
$\Gamma^{\sharp})$ nilpotent matrices, with degree of nilpotency 2, and
$w_{\pm}(\lambda) \! \in \! \cap_{p \in \{2,\infty\}} \mathcal{L}^{p}_{
\mathrm{M}_{2}(\mathbb{C})}(\widetilde{\Gamma}^{\sharp})$ (if $\widetilde{
\Gamma}^{\sharp}$ is unbounded, one requires that $w_{\pm}(\lambda) \!
=_{\genfrac{}{}{0pt}{2}{\lambda \to \infty}{\lambda \in \widetilde{\Gamma}^{
\sharp}}} \!
\left(
\begin{smallmatrix}
0 & 0 \\
0 & 0
\end{smallmatrix}
\right))$. Define $w(\lambda) \! := \! w_{+}(\lambda) \! + \! w_{-}
(\lambda)$, and introduce the Cauchy operators on $\mathcal{L}^{2}_{
\mathrm{M}_{2}(\mathbb{C})}(\Gamma^{\sharp})$, $(C_{\pm}f)(\lambda)
\! := \! \lim_{\genfrac{}{}{0pt}{2}{\lambda^{\prime} \, \to \, \lambda}
{\lambda^{\prime} \, \in \, \pm \, \mathrm{side} \, \mathrm{of}
\, \Gamma^{\sharp}}} \int_{\Gamma^{\sharp}} \tfrac{f(z)}
{(z-\lambda^{\prime})} \, \tfrac{\md z}{2 \pi \mi}$, where $\lambda^{
\prime} \! \to \! \lambda$ $(\in \! \Gamma^{\sharp})$, $\lambda^{\prime}
\! \in \! \pm$ side of $\Gamma^{\sharp}$, denotes the non-tangential
limits {}from the $\pm$--sides of $\Gamma^{\sharp}$ \textbf{at} $\lambda
\! \in \! \Gamma^{\sharp}$, $f(\cdot) \! \in \! \mathcal{L}^{2}_{
\mathrm{M}_{2}(\mathbb{C})}(\Gamma^{\sharp})$, with $C_{\pm} \colon
\mathcal{L}^{2}_{\mathrm{M}_{2}(\mathbb{C})}(\Gamma^{\sharp}) \!
\to \! \mathcal{L}^{2}_{\mathrm{M}_{2}(\mathbb{C})}(\Gamma^{\sharp})$
bounded in operator norm, namely, $\vert \vert C_{\pm} \vert \vert_{
\mathscr{N}(\Gamma^{\sharp})} \! < \! \infty$, where $\mathscr{N}(\ast)$
denotes the space of all bounded linear operators acting {}from $\mathcal{
L}^{2}_{\mathrm{M}_{2}(\mathbb{C})}(\ast)$ into $\mathcal{L}^{2}_{
\mathrm{M}_{2}(\mathbb{C})}(\ast)$, and $\vert \vert (C_{\pm}f)(\cdot)
\vert \vert_{\mathcal{L}^{2}_{\mathrm{M}_{2}(\mathbb{C})}(\ast)} \!
\leqslant \! \mathrm{const.} \vert \vert f(\cdot) \vert \vert_{\mathcal{L}^{
2}_{\mathrm{M}_{2}(\mathbb{C})}(\ast)}$. Introduce the BC operator on
$\mathcal{L}^{2}_{\mathrm{M}_{2}(\mathbb{C})}(\ast)$, $C_{w}f \! := \!
C_{+}(fw_{-}) \! + \! C_{-}(fw_{+})$; moreover, since $\mathbb{C} \setminus
\Gamma^{\sharp}$ can be coloured by two colours $\pm$, $C_{\pm}$
are complementary projections \cite{a28}, $C_{+}^{2} \! = \! C_{+}$,
$C_{-}^{2} \! = \! -C_{-}$, $C_{+}C_{-} \! = \! C_{-}C_{+} \! = \!
\mathbf{0}$, and $C_{+} \! - \! C_{-} \! = \! \mathbf{1}$: in the case that
$C_{+}$ and $-C_{-}$ are complementary, the contour $\Gamma^{\sharp}$ can
always be oriented in such a way that the $\pm$ regions lie on the $\pm$
sides of the contour, respectively. Re-introducing $x,t$ dependences,
specialising the BC formulation to the solution of the RHP for $m^{c}(x,t;
\zeta)$ on $\sigma_{c} \! = \! \mathbb{R}$ (oriented {}from $-\infty$ to
$+\infty)$, and defining $\mathcal{G}(x,t;\zeta) \! := \! (\mathrm{I} \! -
\! w_{-}^{\mathcal{G}}(x,t;\zeta))^{-1}(\mathrm{I} \! + \! w_{+}^{\mathcal{
G}}(x,t;\zeta))$, the integral representation for $m^{c}(x,t;\zeta)$ is
given by the following
\begin{ccccc}[Beals and Coifman $\cite{a24}$]
Let
\begin{equation*}
\mu^{\mathcal{G}}(x,t;\zeta) \! = \! m^{c}_{+}(x,t;\zeta)(\mathrm{I} \!
+ \! w_{+}^{\mathcal{G}}(x,t;\zeta))^{-1} \! = \! m^{c}_{-}(x,t;\zeta)
(\mathrm{I} \! - \! w_{-}^{\mathcal{G}}(x,t;\zeta))^{-1}.
\end{equation*}
If $\mu^{\mathcal{G}}(x,t;\zeta) \! \in \! \mathrm{I} \! + \! \mathcal{L}^{
2}_{\mathrm{M}_{2}(\mathbb{C})}(\sigma_{c})\footnote{For $f(\zeta) \! \in
\! \mathrm{I} \! + \! \mathcal{L}^{2}_{\mathrm{M}_{2}(\mathbb{C})}(\ast)$,
$\vert \vert f(\cdot) \vert \vert_{\mathrm{I}+\mathcal{L}^{2}_{\mathrm{M}_{
2}(\mathbb{C})}(\ast)} \! := \! (\vert f(\infty) \vert^{2} \! + \! \vert
\vert f(\cdot) \! - \! f(\infty) \vert \vert_{\mathcal{L}^{2}_{\mathrm{M}_{
2}(\mathbb{C})}(\ast)}^{2})^{1/2}$ \cite{a31}.} \! := \! \{\mathstrut
\mathrm{I} \! + \! h(\cdot); \, h(\cdot) \! \in \! \mathcal{L}^{2}_{\mathrm{
M}_{2}(\mathbb{C})}(\sigma_{c})\}$ solves the linear singular integral
equation
\begin{equation*}
(\mathbf{1} \! - \! C_{w^{\mathcal{G}}})(\mu^{\mathcal{G}}(x,t;\zeta) \! -
\! \mathrm{I}) \! = \! C_{w^{\mathcal{G}}} \mathrm{I} \! = \! C_{+}(w_{-}
^{\mathcal{G}}(x,t;\zeta)) \! + \! C_{-}(w_{+}^{\mathcal{G}}(x,t;\zeta)),
\quad \zeta \! \in \! \sigma_{c},
\end{equation*}
where $\mathbf{1}$ is the identity operator on $\mathrm{I} \! + \!
\mathcal{L}^{2}_{\mathrm{M}_{2}(\mathbb{C})}(\sigma_{c})$, then
the solution of the {\rm RHP} for $m^{c}(x,t;\zeta)$ is
\begin{equation*}
m^{c}(x,t;\zeta) \! = \! \mathrm{I} \! + \! \int\nolimits_{\sigma_{c}}
\dfrac{\mu^{\mathcal{G}}(x,t;z)w^{\mathcal{G}}(x,t;z)}{(z \! - \! \zeta)}
\, \dfrac{\md z}{2 \pi \mi}, \quad \zeta \! \in \! \mathbb{C} \setminus
\sigma_{c},
\end{equation*}
where $\mu^{\mathcal{G}}(x,t;z) \! := \! ((\mathbf{1} \! - \! C_{w^{
\mathcal{G}}})^{-1} \mathrm{I})(x,t;z)$, and $w^{\mathcal{G}}(x,t;z)
\! := \! \sum_{l \in \{\pm\}} \! w_{l}^{\mathcal{G}}(x,t;z)$.
\end{ccccc}
\begin{eeeee}
The central difficulty subsumed in the analysis is concerned with the
existence and invertibility of the operator $\mathbf{1} \! - \! C_{w^{
\mathcal{G}}}$, in particular $(\mathbf{1} \! - \! C_{w^{\mathcal{G}}})^{
-1}$, as an operator {}from $\mathcal{L}^{2}_{\mathrm{M}_{2}(\mathbb{C})}
(\sigma_{c}) \! \to \! \mathcal{L}^{2}_{\mathrm{M}_{2}(\mathbb{C})}(\sigma_{
c})$. In Sections~4 and~5, it is shown by explicit construction that,
asymptotically, $\ker (\mathbf{1} \! - \! C_{w^{\mathcal{G}}}) \! \!
\upharpoonright_{\mathcal{L}^{2}_{\mathrm{M}_{2}(\mathbb{C})}(\sigma_{c})}
= \! \emptyset$, or, alternatively, the Fredholm index of $\mathbf{1} \! -
\! C_{w^{\mathcal{G}}}$ on $\mathcal{L}^{2}_{\mathrm{M}_{2}(\mathbb{C})}
(\sigma_{c})$ is zero, that is,
\begin{equation*}
i(\mathbf{1} \! - \! C_{w^{\mathcal{G}}}) \! \! \upharpoonright_{\mathcal{
L}^{2}_{\mathrm{M}_{2}(\mathbb{C})}(\sigma_{c})} \, := \! \left( \dim \ker
(\mathbf{1} \! - \! C_{w^{\mathcal{G}}}) \! - \! \dim \mathrm{coker}
(\mathbf{1} \! - \! C_{w^{\mathcal{G}}}) \right) \! \! \upharpoonright_{
\mathcal{L}^{2}_{\mathrm{M}_{2}(\mathbb{C})}(\sigma_{c})} \, =0.
\end{equation*}
\end{eeeee}
{}From Lemma~2.6, the ordered factorisation of Eq.~(7), and Eq.~(5),
one shows that
\begin{equation}
\begin{split}
u(x,t) \! = \! \mi (\Delta_{o}(x,t)&)_{12}+\int\nolimits_{\sigma_{c}}
(\mu^{\mathcal{G}}(x,t;z))_{11} \, \overline{r(\overline{z})} \, \exp
\! \left(-2 \mi t \theta^{u}(z) \right) \tfrac{\md z}{2 \pi}, \\
\theta^{u}(\zeta) \! &:= \! \tfrac{1}{2}(\zeta \! - \! \tfrac{1}{\zeta})
(z_{o} \! + \! \zeta \! + \! \tfrac{1}{\zeta}), \quad z_{o} \! := \!
\tfrac{x}{t}
\end{split}
\end{equation}
(with an analogous expression for $\int_{+\infty}^{x}(\vert u(\xi,t)
\vert^{2} \! - \! 1) \, \md \xi)$, where $(\star)_{ij}$ denotes the
$(i \, j)$-element of $\star$. It is clear {}from Eq.~(8) that the
determination of $u(x,t)$ comes down to an asymptotic analysis of the
resolvent kernel, $\mu^{\mathcal{G}}(x,t;z)$, for which no \emph{a priori}
explicit information regarding its analytical properties is available:
the remedy to this is achieved via the application of the DZ (non-linear
steepest-descent) method \cite{a27}, complemented by the explicit asymptotic
solution of an auxiliary system of linear singular integral equations.
In analogy with the method of steepest descents, the DZ method begins
by examining the saddle point(s) of the phase function, $\theta^{u}
(\zeta)$. The following cases evince themselves: (i)
$t \! \to \! \pm \infty$ and $x \! \to \! \mp \infty$ such
that $z_{o} \! < \! -2$, $\partial_{\zeta} \theta^{u}(\zeta) \! = \!
\zeta^{-3}(\zeta \! - \! \zeta_{1})(\zeta \! - \! \zeta_{2})(\zeta \! - \!
\zeta_{3})(\zeta \! - \! \zeta_{4})$, where $\{\zeta_{i}\}_{i=1}^{4}$ are
defined in Theorem~3.1, Eqs.~(16) and~(17); (ii) $t \! \to \! \pm \infty$
and $x \! \to \! \pm \infty$ such that $z_{o} \! > \! 2$, $\partial_{\zeta}
\theta^{u}(\zeta) \! = \! \zeta^{-3} (\zeta \! - \! \aleph_{1})(\zeta \! -
\! \aleph_{2})(\zeta \! - \! \aleph_{3})(\zeta \! - \! \aleph_{4})$, where
$\{\aleph_{i}\}_{i=1}^{4}$ are defined in Theorem~3.1, Eq.~(24); (iii) $t
\! \to \! \pm \infty$ and $x \! \to \! \mp \infty$ (respectively~$x \! \to
\! \pm \infty)$ such that $z_{o} \! \in \! (-2,0)$ (respectively~$z_{o} \!
\in \! (0,2))$, $\partial_{\zeta} \theta^{u}(\zeta) \! = \! \zeta^{-3}(\zeta
\! - \! \zeta_{1}^{\sharp})(\zeta \! - \! \overline{\zeta_{1}^{\sharp}})
(\zeta \! - \! \zeta_{3}^{\sharp})(\zeta \! - \! \overline{\zeta_{3}^{
\sharp}})$, where $\zeta_{n}^{\sharp}$, $n \! \in \! \{1,3\}$, are defined
in Theorem~3.2, Eqs.~(36) and~(37); (iv) $t \! \to \! \pm \infty$ and
$x \! \to \! \mp \infty$ (respectively~$x \! \to \! \pm \infty)$ such
that $z_{o} \! \to \! 0^{-}$ (respectively~$z_{o} \! \to \! 0^{+})$,
$\partial_{\zeta} \theta^{u}(\zeta) \! = \! \zeta^{-3}(\zeta \! - \!
\me^{\frac{\mi \pi}{4}})(\zeta \! - \! \me^{-\frac{\mi \pi}{4}})(\zeta
\! - \! \me^{\frac{3 \pi \mi}{4}})(\zeta \! - \! \me^{-\frac{3 \pi \mi}
{4}})$; (v) $t \! \to \! \pm \infty$ and $x \! \to \! \mp \infty$ such
that $z_{o} \! = \! -2$, $\partial_{\zeta} \theta^{u}(\zeta) \! = \!
\zeta^{-3}(\zeta \! - \! 1)^{2}(\zeta \! - \! \me^{\frac{2 \pi \mi}
{3}})(\zeta \! - \! \me^{-\frac{2 \pi \mi}{3}})$; and (vi) $t \! \to
\! \pm \infty$ and $x \! \to \! \pm \infty$ such that $z_{o} \! = \!
2$, $\partial_{\zeta} \theta^{u}(\zeta) \! = \! \zeta^{-3}(\zeta \!
+ \! 1)^{2}(\zeta \! - \! \me^{\frac{\mi \pi}{3}})(\zeta \! - \! \me^{
-\frac{\mi \pi}{3}})$. Cases (i) and (ii) correspond to oscillatory
asymptotics, cases (iii) and (iv) give rise to exponentially decaying
asymptotics, and cases (v) and (vi) give rise to asymptotics which are
related to those of the transcendent of the Painlev\'{e} II equation
(PII) \cite{a35,a36,a37,a38,a39}. In this work, only cases~(i)--(iv)
are considered, and cases~(v) and~(vi) will be studied elsewhere. Hereafter,
the discussion will focus exclusively on how the DZ method is applied to
the subcase $t \! \to \! +\infty$ and $x \! \to \! -\infty$ such that $z_{o}
\! < \! -2$ of case (i): analogous statements/arguments, with corresponding
modifications, apply to cases (ii)--(vi). Succinctly, the DZ method for
(oscillatory) RHPs is based on a succession of transformations (involving
judicious factorisations of the jump matrix), and deformations of $\sigma_{
c}$ (orientations, too), which, as $t \! \to \! +\infty$, convert the
original RHP into an equivalent RHP (in the sense that a solution
of the equivalent RHP gives a solution of the original RHP and
\emph{vice versa}\footnote{In particular, if there are two RHPs, $(\mathcal{
X}_{1}(\lambda),\upsilon_{1}(\lambda),\Gamma_{1})$ and $(\mathcal{X}_{2}
(\lambda),\upsilon_{2}(\lambda),\Gamma_{2})$, say, with $\Gamma_{2} \subset
\Gamma_{1}$ and $\upsilon_{1}(\lambda) \! \! \upharpoonright_{\Gamma_{1}
\setminus \Gamma_{2}} \! =_{t \to +\infty} \! \mathrm{I} \! + \! o(1)$,
then, within the BC framework (Lemma~3.1), and modulo $o(1)$ estimates,
their solutions, $\mathcal{X}_{1}(\lambda)$ and $\mathcal{X}_{2}(\lambda)$,
respectively, are (asymptotically) equal.}) with jump matrix $\mathcal{G}_{
\mathrm{equiv}}(x,t;\zeta)$ of the form $\mathcal{G}_{\mathrm{equiv}}(x,t;
\zeta) \! = \! \mathcal{G}_{\mathrm{model}}(x,t;\zeta) \! + \! \mathcal{
G}_{\mathrm{error}}(x,t;\zeta)$, where $\mathcal{G}_{\mathrm{model}}(x,t;
\zeta)$ is the jump matrix for an explicitly solvable model RHP, and
$\mathcal{G}_{\mathrm{error}}(x,t;\zeta) \! =_{t \to +\infty} \! \mathrm{I}
\! + \! o(1)$. Modulo error estimates, the solution of the original RHP
``tends to'' the solution of the model RHP. The delineation of the DZ
method is as follows:
\begin{enumerate}
\item[(i)] decompose the complex plane of the spectral parameter
$\zeta$ according to the signature of $\Re (\mi t \theta^{u}(\zeta))$
(see Section~4, Figure~1), where $\pm \! \leftrightarrow \! \Re (\mi t
\theta^{u}(\zeta)) \! \gtrless \! 0$, and $\{\zeta_{2},\zeta_{1}\}$ and
$\{\zeta_{3},\overline{\zeta_{3}}\}$ are, respectively, the real and complex
first-order saddle points (note that $\mathbb{C} \setminus \sigma_{c}$ is
partitioned into the disjoint union of two disconnected domains coloured,
respectively, by $\pm$);
\item[(ii)] reorient $\sigma_{c} \! = \! \mathbb{R}$ (oriented {}from
$-\infty$ to $+\infty)$ according to, and consistent with, the signature
of $\Re (\mi t \theta^{u}(\zeta))$, denoted by $\sigma_{c}^{\prime}$
(see Section~4, Figure~2), ``conjugate'' the RHP for $m^{c}(x,t;\zeta)$
according to $\widehat{m}^{c}(x,t;\zeta) \! := \! m^{c}(x,t;\zeta)
(\delta (\zeta))^{-\sigma_{3}}$, where $\delta (\zeta)$ solves the scalar
discontinuous RHP
\begin{align}
\delta_{+}(\zeta) &= \!
\begin{cases}
\delta_{-}(\zeta)(1 \! - \! r(\zeta) \overline{r(\overline{\zeta})}),
&\text{$\Re (\zeta) \! \in \! (-\infty,0) \cup (\zeta_{2},\zeta_{1})$,} \\
\delta_{-}(\zeta) \! = \! \delta (\zeta), &\text{$\Re (\zeta) \! \in \! (0,
\zeta_{2}) \cup (\zeta_{1},+\infty)$,}
\end{cases} \nonumber \\
\delta (\zeta) &\underset{\zeta \, \to \, \infty}{=} \! 1 \! + \! \mathcal{
O}(\zeta^{-1}), \nonumber
\end{align}
with solution $\delta (\zeta) \! = \! \left( \tfrac{\zeta-\zeta_{1}}{\zeta
-\zeta_{2}} \right)^{\mi \nu} \exp \! \left( \int_{-\infty}^{0} \! \tfrac{
\ln (1-\vert r(\mu) \vert^{2})}{(\mu-\zeta)} \, \tfrac{\md \mu}{2 \pi \mi}
\! + \! \int_{\zeta_{2}}^{\zeta_{1}} \ln \! \left( \tfrac{1-\vert r(\mu)
\vert^{2}}{1-\vert r (\zeta_{1}) \vert^{2}} \right) \! \tfrac{1}{(\mu -
\zeta)} \, \tfrac{\md \mu}{2 \pi \mi} \right)$, w\-h\-e\-r\-e $\nu \! :=
\! -\tfrac{1}{2 \pi} \ln (1 \! - \! \vert r(\zeta_{1}) \vert^{2})$, $\delta
(\zeta) \overline{\delta (\overline{\zeta})} \! = \! 1$, and $\delta (\zeta)
\delta (\tfrac{1}{\zeta}) \! = \! \delta (0)$, and derive the following RHP
for $\widehat{m}^{c}(x,t;\zeta) \colon \mathbb{C} \setminus \sigma_{c}^{
\prime} \! \to \! \mathrm{SL}(2,\mathbb{C})$,
\begin{equation*}
\widehat{m}^{c}_{+}(x,t;\zeta) \! = \! \widehat{m}^{c}_{-}(x,t;\zeta) \!
\left(
\begin{smallmatrix}
1 & 0 \\
-\overline{\rho (\overline{\zeta})}(\delta_{-}(\zeta))^{-2} \, \me^{2
\mi t \theta^{u}(\zeta)} & 1
\end{smallmatrix}
\right) \!
\left(
\begin{smallmatrix}
1 & \rho (\zeta) (\delta_{+}(\zeta))^{2} \me^{-2 \mi t \theta^{u}(\zeta)
} \\
0 & 1
\end{smallmatrix}
\right), \quad \zeta \! \in \! \sigma_{c}^{\prime},
\end{equation*}
$\widehat{m}^{c}(x,t;\zeta) \! =_{\genfrac{}{}{0pt}{2}{\zeta \to \infty}
{\zeta \in \mathbb{C} \setminus \sigma_{c}^{\prime}}} \! \mathrm{I} \!
+ \! \mathcal{O}(\zeta^{-1})$, $\widehat{m}^{c}(x,t;\zeta) \! = \! \sigma_{
1} \overline{\widehat{m}^{c}(x,t;\overline{\zeta})} \, \sigma_{1}$, and
$(\widehat{m}^{c}(x,t;0)(\delta (0))^{\sigma_{3}} \sigma_{2})^{2} \! = \!
\mathrm{I}$, where $\rho (\zeta) \! := \!
\begin{cases}
\, \overline{r(\overline{\zeta})}, &\text{$\Re (\zeta) \! \in \! (0,
\zeta_{2}) \cup (\zeta_{1},+\infty)$,} \\
\, -\overline{r(\overline{\zeta})}(1 \! - \! r(\zeta) \overline{r
(\overline{\zeta})})^{-1}, &\text{$\Re (\zeta) \! \in \! (-\infty,0)
\cup (\zeta_{2},\zeta_{1})$;}
\end{cases}$
\item[(iii)] on each interval $(-\infty,0)$, $(0,\zeta_{2})$, $(\zeta_{
2},\zeta_{1})$, and $(\zeta_{1},+\infty)$, replace $r(\zeta)$ by rational
functions, and deform and augment $\sigma_{c}^{\prime}$ to the (oriented)
contour $\Sigma^{\prime}$ (see Section~4, Figure~3) such that the respective
jump matrices on $\sigma_{c}^{\prime} \subset \Sigma^{\prime}$ and the
finite triangular humps (respectively~linear segments) of $\Sigma^{\prime}
\setminus \sigma_{c}^{\prime}$, away {}from the union of the neighbourhoods
of the real first-order saddle points $\zeta_{2}$ and $\zeta_{1}$
(respectively~complex first-order saddle points $\zeta_{3}$ and $\overline{
\zeta_{3}}$), tend to $\mathrm{I}$ as $t \! \to \! +\infty$, and rewrite
the RHP for $\widehat{m}^{c}(x,t;\zeta)$ on $\sigma_{c}^{\prime}$ as an
equivalent RHP for an $\mathrm{SL}(2,\mathbb{C})$-valued function $m^{\sharp}
(x,t;\zeta)$ (see Section~4, Lemma~4.3 for the explicit transformation
{}from $\widehat{m}^{c}(x,t;\zeta)$ to $m^{\sharp}(x,t;\zeta))$ on $\Sigma^{
\prime}$ (this is possible due to the fact that the index of the RHP is a
topological invariant \cite{a23});
\item[(iv)] showing that the contribution of $\zeta_{3}$
(respectively~$\overline{\zeta_{3}})$ to the leading-order asymptotics of
$m^{\sharp}(x,t;\zeta)$ is $\mathcal{O}(\exp (-\widehat{g}(z_{o})t))$,
where $\widehat{g}(z_{o}) \! > \! 0$, truncate $\Sigma^{\prime}$ to the
(oriented) contour $\Sigma^{\sharp}$ (see Section~4, Figure~4) partitioned
such that $\Sigma^{\sharp} \! = \! \Sigma_{A^{\prime}} \cup \Sigma_{B^{
\prime}}$ and $\Sigma_{A^{\prime}} \cap \Sigma_{B^{\prime}} \! = \!
\emptyset$, with $\mathrm{dist}(\Sigma_{A^{\prime}},\Sigma_{B^{\prime}}) \!
=_{t \to +\infty} \! \mathcal{O}(1)$;\item[(v)] localising the jump matrices
of the most rapidly descented RHPs on the union of the truncated disjoint
crosses to the open neighbourhoods of $\zeta_{1}$ and $\zeta_{2}$,
introducing the scaling-shifting operators (see Section~5, Eqs.~(105)
and~(106)) $\mathcal{N}_{A}$, $f(\zeta) \! \mapsto \! (\mathcal{N}_{A}f)
(\widetilde{w}) \! = \! f \! \left( \! \zeta_{2} \! + \! \widetilde{w} /
\tfrac{\vert \zeta_{2}-\zeta_{3} \vert}{\zeta_{2}} \sqrt{\tfrac{2t(\zeta_{
1}-\zeta_{2})}{\zeta_{2}}} \right)$, and $\mathcal{N}_{B}$, $g(\zeta) \!
\mapsto \! (\mathcal{N}_{B}g)(\widetilde{w}) \! = \! g \! \left( \! \zeta_{
1} \! + \! \widetilde{w} / \tfrac{\vert \zeta_{1}-\zeta_{3} \vert}{\zeta_{
1}} \sqrt{\tfrac{2 t (\zeta_{1}-\zeta_{2})}{\zeta_{1}}} \right)$, which
scale and shift (centre) $\Sigma_{A^{\prime}}$ and $\Sigma_{B^{\prime}}$,
respectively, to $\widetilde{w} \! = \! 0$, deducing that the leading-order
asymptotics are $\mathcal{O}(t^{-1/2})$, and showing that the higher order
interaction between the crosses is $\mathcal{O}(t^{-1/2} \ln t)$, one
separates out the contributions of the crosses and shows that $u(x,t)$ can
be written as the linear superposition of the contributions of the various
(disjoint) crosses, and, with additional transformations and scalings,
reduce the RHPs on the crosses to model RHPs (on $\mathbb{R})$ which can
be solved in closed form in terms of parabolic cylinder functions.
\end{enumerate}
\begin{eeeee}
Throughout this work, $M \! \in \! \mathbb{R}_{>1}$ denotes a fixed, bounded
constant, and the ``symbols'' $c^{\mathcal{S}}(\diamondsuit)$, $\underline{
c}(\flat,\natural,\sharp)$, $\underline{c}(z_{1},z_{2},z_{3},z_{4})$,
$\underline{c}(\bullet)$, and $\underline{c}$ appearing in the various error
estimates are to be understood as follows: (1) for $\pm \diamondsuit \! >
\! 0$, $c^{\mathcal{S}}(\diamondsuit) \! \in \! \mathcal{S}_{\mathbb{C}}
(\mathbb{R}_{\pm})$; (2) for $\pm \flat \! > \! 0$, $\underline{c}(\flat,
\natural,\sharp) \! \in \! \mathcal{L}^{\infty}_{\mathbb{C}}(\mathbb{R}_{\pm}
\! \times \! \mathbb{C}^{\ast} \! \times \! \overline{\mathbb{C}^{\ast}})$,
where $\mathbb{C}^{\ast} \! := \! \mathbb{C} \setminus \{0\}$ (and bounded);
(3) for $(z_{1},z_{2}) \! \in \! \mathbb{R}_{\pm} \times \mathbb{R}_{\pm}$,
$\underline{c}(z_{1},z_{2},z_{3},z_{4}) \! \in \! \mathcal{L}^{\infty}_{
\mathbb{C}}(\mathbb{R}^{2}_{\pm} \! \times \! \mathbb{C}^{\ast} \! \times \!
\overline{\mathbb{C}^{\ast}})$; (4) for $\pm \bullet \! > \! 0$, $\underline{
c}(\bullet) \! \in \! \mathcal{L}^{\infty}_{\mathbb{C}}(\mathrm{D}_{\pm})$,
where $\mathrm{D}_{+} \! := \! (0,2)$ and $\mathrm{D}_{-} \! := \! (-2,0)$;
and (5) $\underline{c} \! \in \! \mathbb{C}^{\ast}$. Even though the symbols
$c^{\mathcal{S}}(\diamondsuit)$, $\underline{c}(\flat,\natural,\sharp)$,
$\underline{c}(z_{1},z_{2},z_{3},z_{4})$, $\underline{c}(\bullet)$, and
$\underline{c}$ appearing in the error estimations are not, in general,
equal, and should properly be denoted as $c_{1}(\cdot)$, $c_{2}(\cdot)$,
etc., the simplified notations $c^{\mathcal{S}}(\diamondsuit)$, $\underline{
c}(\flat,\natural,\sharp)$, $\underline{c}(z_{1},z_{2},z_{3},z_{4})$,
$\underline{c}(\bullet)$, and $\underline{c}$ are retained throughout in
order to eschew a flood of superfluous notation, as well as to maintain
consistency with the main theme of this work, namely, to derive explicitly
the leading order asymptotics and the classes to which the errors belong
without regard to their precise $z_{o}$-dependence.
\end{eeeee}
\begin{eeeee}
In Eqs.~(9) and~(18) below, one should keep, everywhere, the upper
(respectively~lower) signs as $t \! \to \! +\infty$ (respectively~$t \! \to
 \! -\infty)$.
\end{eeeee}
\begin{ddddd}
For $r(\zeta) \! \in \! \mathcal{S}_{\mathbb{C}}(\mathbb{R}) \cap \{
\mathstrut h(z); \, \vert \vert h(\cdot) \vert \vert_{\mathcal{L}^{\infty}
(\mathbb{R})} \! := \! \sup_{z \in \mathbb{R}} \vert h(z) \vert \! < \! 1\}$,
let $m(x,t;\zeta)$ be the solution of the Riemann-Hilbert problem formulated
in Lemma~{\rm 2.5}. Let $u(x,t)$, the solution of the Cauchy problem for the
{\rm D${}_{f}$NLSE} with finite-density initial data $u(x,0) \! := \! u_{o}
(x) \! =_{x \to \pm \infty} \! u_{o}(\pm \infty)(1 \! + \! o(1))$, where
$u_{o}(\pm \infty) \! := \! \exp (\tfrac{\mi (1 \mp 1) \theta}{2})$, $0 \!
\leqslant \! \theta \! = \! -\int_{-\infty}^{+\infty} \tfrac{\ln (1-\vert
r(\mu) \vert^{2})}{\mu} \, \tfrac{\md \mu}{2 \pi} \! < \! 2 \pi$, $u_{o}(x)
\! \in \! \mathbf{C}^{\infty}(\mathbb{R})$, and $u_{o}(x) \! - \! u_{o}(\pm
\infty) \! \in \! \mathcal{S}_{\mathbb{C}}(\mathbb{R}_{\pm})$, be defined
by Eq.~{\rm (5)}. Then as $t \! \to \! \pm \infty$ and $x \! \to \! \mp
\infty$ such that $z_{o} \! := \! x/t \! < \! -2$,
\begin{align}
u(x,t) &= \! \me^{-\mi \theta^{\pm}(1)} \! \left( 1 \! + \! \dfrac{\mi \sqrt{
\nu(\zeta_{1})}}{\sqrt{\vert t \vert (\zeta_{1} \! - \! \zeta_{2})} \, (z_{
o}^{2} \! + \! 32)^{1/4}} \! \left( \zeta_{1} \me^{\mp \mi (\Theta^{\pm}(z_{
o},t) \pm (2 \mp 1) \frac{\pi}{4})} \! + \! \zeta_{2} \me^{\pm \mi (\Theta^{
\pm}(z_{o},t) \pm (2 \mp 1) \frac{\pi}{4})} \right) \! \right. \nonumber \\
       &+ \! \left. \mathcal{O} \! \left( \! \left( \dfrac{c^{\mathcal{
S}}(\zeta_{1}) \underline{c}(\zeta_{2},\zeta_{3},\zeta_{4})}{\sqrt{\zeta_{
1}(z_{o}^{2} \! + \! 32)}} \! + \! \dfrac{c^{\mathcal{S}}(\zeta_{2})
\underline{c}(\zeta_{1},\zeta_{3},\zeta_{4})}{\sqrt{\zeta_{2}(z_{o}^{2} \!
+ \! 32)}} \right) \! \dfrac{\ln \vert t \vert}{(\zeta_{1} \! - \! \zeta_{
2}) t} \right) \! \right),
\end{align}
where
\begin{gather}
\theta^{+}(j) \! := \! \int_{-\infty}^{0} \dfrac{\ln (1 \! - \! \vert r(\mu)
\vert^{2})}{\mu^{j}} \, \dfrac{\md \mu}{2 \pi} + \int_{\zeta_{2}}^{\zeta_{
1}} \dfrac{\ln (1 \! - \! \vert r(\mu) \vert^{2})}{\mu^{j}} \, \dfrac{\md
\mu}{2 \pi}, \quad j \! \in \! \{0,1\}, \\
\theta^{-}(l) \! := \! \int_{0}^{\zeta_{2}} \dfrac{\ln (1 \! - \! \vert r(\mu)
\vert^{2})}{\mu^{l}} \, \dfrac{\md \mu}{2 \pi} + \int_{\zeta_{1}}^{+\infty}
\dfrac{\ln (1 \! - \! \vert r(\mu) \vert^{2})}{\mu^{l}} \, \dfrac{\md \mu}
{2 \pi}, \quad l \! \in \! \{0,1\}, \\
\nu(z) \! := \! -\tfrac{1}{2 \pi} \ln (1 \! - \! \vert r(z) \vert^{2}), \\
\Theta^{\pm}(z_{o},t) \! := \! \pm \arg r(\zeta_{1}) \! - \! \arg \Gamma
(\mi \nu (\zeta_{1})) \! \pm \! t(\zeta_{1} \! - \! \zeta_{2})(z_{o} \!
+ \! \zeta_{1} \! + \! \zeta_{2}) \! + \! \nu (\zeta_{1}) \ln \vert t \vert
\nonumber \\
\qquad + \, 3 \nu (\zeta_{1}) \ln (\zeta_{1} \! - \! \zeta_{2}) \! + \!
\tfrac{1}{2} \nu (\zeta_{1}) \ln (z_{o}^{2} \! + \! 32) \! \mp \! \Omega^{
\pm}(\zeta_{1}) \! \pm \! \tfrac{1}{2} \Omega^{\pm}(0), \\
\Omega^{+}(z) \! = \! \frac{1}{\pi} \int_{-\infty}^{0} \ln \! \vert \mu
\! - \! z \vert \md \ln (1 \! - \! \vert r(\mu) \vert^{2}) + \frac{1}{\pi}
\int_{\zeta_{2}}^{\zeta_{1}} \ln \! \vert \mu \! - \! z \vert \md \ln
(1 \! - \! \vert r(\mu) \vert^{2}), \\
\Omega^{-}(z) \! = \! \frac{1}{\pi} \int_{0}^{\zeta_{2}} \ln \! \vert
\mu \! - \! z \vert \md \ln (1 \! - \! \vert r(\mu) \vert^{2}) + \frac{1}
{\pi} \int_{\zeta_{1}}^{+\infty} \ln \! \vert \mu \! - \! z \vert \md \ln
(1 \! - \! \vert r(\mu) \vert^{2}), \\
\zeta_{1} \! := \! \frac{1}{2} \! \left( \! -a_{1} \! + \! \sqrt{a_{1}^{2}
\! - \! 4} \, \right) \! , \quad \, \zeta_{2} \! = \! \frac{1}{\zeta_{1}},
\quad \, \zeta_{3} \! := \! \frac{1}{2} \! \left( \! -a_{2} \! + \! \mi
\sqrt{4 \! - \! a_{2}^{2}} \, \right) \! , \quad \, \zeta_{4} \! = \!
\overline{\zeta_{3}}, \\
a_{1} \! = \! \frac{1}{4} \! \left(z_{o} \! - \! \sqrt{z_{o}^{2} \! + \! 32}
\, \right) \! , \qquad \, a_{2} \! = \! \frac{1}{4} \! \left(z_{o} \! + \!
\sqrt{z_{o}^{2} \! + \! 32} \, \right) \! ,
\end{gather}
$0 \! < \! \zeta_{2} \! < \! \zeta_{1}$, $\vert \zeta_{3} \vert^{2} \! = \!
1$, $a_{1}a_{2} \! = \! -2$, and $\Gamma (\cdot)$ is the gamma function
{\rm \cite{a40}}, and as $t \! \to \! \pm \infty$ and $x \! \to \! \pm
\infty$ such that $z_{o} \! > \! 2$,
\begin{align}
u(x,t) &= \! -\me^{-\mi \phi^{\pm}(1)} \! \left( 1 \! + \! \dfrac{\mi \sqrt{
\nu(\aleph_{4})}}{\sqrt{\vert t \vert (\aleph_{3} \! - \! \aleph_{4})} \,
(z_{o}^{2} \! + \! 32)^{1/4}} \! \left( \aleph_{4} \me^{\mp \mi (\Phi^{\pm}
(z_{o},t) \pm (2 \mp 1) \frac{\pi}{4})} \! + \! \aleph_{3} \me^{\pm \mi
(\Phi^{\pm}(z_{o},t) \pm (2 \mp 1) \frac{\pi}{4})} \right) \! \right.
\nonumber \\
       &+ \! \left. \mathcal{O} \! \left( \! \left( \dfrac{c^{\mathcal{
S}}(\aleph_{3}) \underline{c}(\aleph_{4},\aleph_{1},\aleph_{
2})}{\sqrt{\vert \aleph_{3} \vert (z_{o}^{2} \! + \! 32)}} \!+ \! \dfrac{
c^{\mathcal{S}}(\aleph_{4}) \underline{c}(\aleph_{3},\aleph_{
1},\aleph_{2})}{\sqrt{\vert \aleph_{4} \vert (z_{o}^{2} \! + \! 32)}}
\right) \! \dfrac{\ln \vert t \vert}{(\aleph_{3} \! - \! \aleph_{4}) t}
\right) \! \right),
\end{align}
where
\begin{gather}
\phi^{+}(j) \! := \! \int_{-\infty}^{\aleph_{4}} \dfrac{\ln (1 \! - \! \vert
r(\mu) \vert^{2})}{\mu^{j}} \, \dfrac{\md \mu}{2 \pi} + \int_{\aleph_{3}}
^{0} \dfrac{\ln (1 \! - \! \vert r(\mu) \vert^{2})}{\mu^{j}} \,
\dfrac{\md \mu}{2 \pi}, \quad j \! \in \! \{0,1\}, \\
\phi^{-}(l) \! := \! \int_{\aleph_{4}}^{\aleph_{3}} \dfrac{\ln (1 \! - \!
\vert r(\mu) \vert^{2})}{\mu^{l}} \, \dfrac{\md \mu}{2 \pi} + \int_{0}^{+
\infty} \dfrac{\ln (1 \! - \! \vert r(\mu) \vert^{2})}{\mu^{l}} \, \dfrac{
\md \mu}{2 \pi}, \quad l \! \in \! \{0,1\}, \\
\Phi^{\pm}(z_{o},t) \! := \! \pm \arg r(\aleph_{4}) \! - \! \arg \Gamma (\mi
\nu (\aleph_{4})) \! \pm \! t(\aleph_{4} \! - \! \aleph_{3})(z_{o} \! + \!
\aleph_{3} \! + \! \aleph_{4}) \! + \! \nu (\aleph_{4}) \ln \vert t \vert
\nonumber \\
\quad \, \, + \, 3 \nu (\aleph_{4}) \ln (\aleph_{3} \! - \! \aleph_{4}) \!
+ \! \tfrac{1}{2} \nu (\aleph_{4}) \ln (z_{o}^{2} \! + \! 32) \! \mp \!
\Lambda^{\pm}(\aleph_{4}) \! \pm \! \tfrac{1}{2} \Lambda^{\pm}(0), \\
\Lambda^{+}(z) \! = \! \frac{1}{\pi} \int_{-\infty}^{\aleph_{4}} \ln \!
\vert \mu \! - \! z \vert \md \ln (1 \! - \! \vert r(\mu) \vert^{2})
+ \frac{1}{\pi} \int_{\aleph_{3}}^{0} \ln \! \vert \mu \! - \! z \vert
\md \ln (1 \! - \! \vert r(\mu) \vert^{2}), \\
\Lambda^{-}(z) \! = \! \frac{1}{\pi} \int_{\aleph_{4}}^{\aleph_{3}}
\ln \! \vert \mu \! - \! z \vert \md \ln (1 \! - \! \vert r(\mu) \vert^{2})
+ \frac{1}{\pi} \int_{0}^{+\infty} \ln \! \vert \mu \! - \! z \vert \md
\ln (1 \! - \! \vert r(\mu) \vert^{2}),
\end{gather}
\begin{gather}
\aleph_{1} \! := \! \frac{1}{2} \! \left( \! -a_{1} \! + \! \mi \sqrt{4
\! - \! a_{1}^{2}} \, \right) \! , \quad \, \aleph_{2} \! = \! \overline{
\aleph_{1}}, \quad \aleph_{3} \! := \! \frac{1}{2} \! \left( \! -a_{
2} \! + \! \sqrt{a_{2}^{2} \! - \! 4} \, \right) \! , \quad \, \aleph_{4}
\! = \! \frac{1}{\aleph_{3}},
\end{gather}
$\aleph_{4} \! < \! \aleph_{3} \! < \! 0$, and $\vert \aleph_{1} \vert^{2}
\! = \! 1$. For $u(x,t)$ as defined and given above, let $\int_{+\infty}^{x}
(\vert u(\xi,t) \vert^{2} \! - \! 1) \, \md \xi$ be defined by Eq.~{\rm (6)}.
Then: (i) as $t \! \to \! +\infty$ and $x \! \to \! -\infty$ such that $z_{
o} \! < \! -2$,
\begin{align}
\int_{+\infty}^{x} \! \left( \vert u(\xi,t) \vert^{2} \! - \! 1 \right)
\md \xi &= \theta^{+}(0) - \dfrac{2 \sqrt{\nu(\zeta_{1})} \, \cos (\Theta
^{+}(z_{o},t)+\frac{\pi}{4})}{\sqrt{t(\zeta_{1} \! - \! \zeta_{2})} \,
(z_{o}^{2} \! + \! 32)^{1/4}} \nonumber \\
 &+ \mathcal{O} \! \left( \! \left( \dfrac{c^{\mathcal{S}}(\zeta_{1})
\underline{c}(\zeta_{2},\zeta_{3},\zeta_{4})}{\sqrt{\zeta_{1}(z_{o}^{
2} \! + \! 32)}} \! + \! \dfrac{c^{\mathcal{S}}(\zeta_{2}) \underline{c}
(\zeta_{1},\zeta_{3},\zeta_{4})}{\sqrt{\zeta_{2}(z_{o}^{2} \! + \! 32)}}
\right) \! \dfrac{\ln t}{(\zeta_{1} \! - \! \zeta_{2}) t} \right), \\
\int_{-\infty}^{x} \! \left( \vert u(\xi,t) \vert^{2} \! - \! 1 \right)
\md \xi &= -\theta^{-}(0)-\dfrac{2 \sqrt{\nu(\zeta_{1})} \, \cos (\Theta
^{+}(z_{o},t)+\frac{\pi}{4})}{\sqrt{t(\zeta_{1} \! - \! \zeta_{2})} \,
(z_{o}^{2} \! + \! 32)^{1/4}} \nonumber \\
 &+ \mathcal{O} \! \left( \! \left( \dfrac{c^{\mathcal{S}}(\zeta_{1})
\underline{c}(\zeta_{2},\zeta_{3},\zeta_{4})}{\sqrt{\zeta_{1}(z_{o}^{
2} \! + \! 32)}} \! + \! \dfrac{c^{\mathcal{S}}(\zeta_{2}) \underline{c}
(\zeta_{1},\zeta_{3},\zeta_{4})}{\sqrt{\zeta_{2}(z_{o}^{2} \! + \! 32)}
} \right) \! \dfrac{\ln t}{(\zeta_{1} \! - \! \zeta_{2}) t} \right);
\end{align}
(ii) as $t \! \to \! -\infty$ and $x \! \to \! +\infty$ such that $z_{
o} \! < \! -2$,
\begin{align}
\int_{+\infty}^{x} \! \left( \vert u(\xi,t) \vert^{2} \! - \! 1 \right)
\md \xi &= \theta^{-}(0)-\dfrac{2 \sqrt{\nu(\zeta_{1})} \, \cos (\Theta
^{-}(z_{o},t)-\frac{3 \pi}{4})}{\sqrt{\vert t \vert (\zeta_{1} \! - \!
\zeta_{2})} \, (z_{o}^{2} \! + \! 32)^{1/4}} \nonumber \\
 &+ \mathcal{O} \! \left( \! \left( \dfrac{c^{\mathcal{S}}(\zeta_{1})
\underline{c}(\zeta_{2},\zeta_{3},\zeta_{4})}{\sqrt{\zeta_{1}(z_{o}^{
2} \! + \! 32)}} \! + \! \dfrac{c^{\mathcal{S}}(\zeta_{2}) \underline{c}
(\zeta_{1},\zeta_{3},\zeta_{4})}{\sqrt{\zeta_{2}(z_{o}^{2} \! + \! 32)}
} \right) \! \dfrac{\ln \vert t \vert}{(\zeta_{1} \! - \! \zeta_{2}) t}
\right), \\
\int_{-\infty}^{x} \! \left( \vert u(\xi,t) \vert^{2} \! - \! 1 \right)
\md \xi &= -\theta^{+}(0)-\dfrac{2 \sqrt{\nu(\zeta_{1})} \, \cos (\Theta
^{-}(z_{o},t)-\frac{3 \pi}{4})}{\sqrt{\vert t \vert (\zeta_{1} \! - \!
\zeta_{2})} \, (z_{o}^{2} \! + \! 32)^{1/4}} \nonumber \\
 &+ \mathcal{O} \! \left( \! \left( \dfrac{c^{\mathcal{S}}(\zeta_{1})
\underline{c}(\zeta_{2},\zeta_{3},\zeta_{4})}{\sqrt{\zeta_{1}(z_{o}^{
2} \! + \! 32)}} \! + \! \dfrac{c^{\mathcal{S}}(\zeta_{2}) \underline{c}
(\zeta_{1},\zeta_{3},\zeta_{4})}{\sqrt{\zeta_{2}(z_{o}^{2} \! + \! 32)}
} \right) \! \dfrac{\ln \vert t \vert}{(\zeta_{1} \! - \! \zeta_{2})
t} \right);
\end{align}
(iii) as $t \! \to \! +\infty$ and $x \! \to \! +\infty$ such that
$z_{o} \! > \! 2$,
\begin{align}
\int_{+\infty}^{x} \! \left( \vert u(\xi,t) \vert^{2} \! - \! 1 \right)
\md \xi &= \phi^{-}(0)+\dfrac{2 \sqrt{\nu(\aleph_{4})} \, \cos (\Phi^{
+}(z_{o},t)+\frac{\pi}{4})}{\sqrt{t(\aleph_{3} \! - \! \aleph_{4})} \,
(z_{o}^{2} \! + \! 32)^{1/4}} \nonumber \\
 &+ \mathcal{O} \! \left( \! \left( \dfrac{c^{\mathcal{S}}(\aleph_{3})
\underline{c}(\aleph_{4},\aleph_{1},\aleph_{2})}{\sqrt{\vert \aleph_{
3} \vert (z_{o}^{2} \! + \! 32)}} \! + \! \dfrac{c^{\mathcal{S}}(\aleph_{4})
\underline{c}(\aleph_{3},\aleph_{1},\aleph_{2})}{\sqrt{\vert \aleph_{
4} \vert (z_{o}^{2} \! + \! 32)}} \right) \! \dfrac{\ln t}{(\aleph_{3} \!
- \! \aleph_{4}) t} \right), \\
\int_{-\infty}^{x} \! \left( \vert u(\xi,t) \vert^{2} \! - \! 1 \right)
\md \xi &= -\phi^{+}(0)+\dfrac{2 \sqrt{\nu(\aleph_{4})} \, \cos (\Phi^{
+}(z_{o},t)+\frac{\pi}{4})}{\sqrt{t(\aleph_{3} \! - \! \aleph_{4})} \,
(z_{o}^{2} \! + \! 32)^{1/4}} \nonumber \\
 &+ \mathcal{O} \! \left( \! \left( \dfrac{c^{\mathcal{S}}(\aleph_{3})
\underline{c}(\aleph_{4},\aleph_{1},\aleph_{2})}{\sqrt{\vert \aleph_{
3} \vert (z_{o}^{2} \! + \! 32)}} \! + \! \dfrac{c^{\mathcal{S}}(\aleph_{4})
 \underline{c}(\aleph_{3},\aleph_{1},\aleph_{2})}{\sqrt{\vert \aleph_{
4} \vert (z_{o}^{2} \! + \! 32)}} \right) \! \dfrac{\ln t}{(\aleph_{3} \!
- \! \aleph_{4}) t} \right);
\end{align}
and (iv) as $t \! \to \! -\infty$ and $x \! \to \! -\infty$ such that
$z_{o} \! > \! 2$,
\begin{align}
\int_{+\infty}^{x} \! \left( \vert u(\xi,t) \vert^{2} \! - \! 1 \right)
\md \xi &= \phi^{+}(0) + \dfrac{2 \sqrt{\nu(\aleph_{4})} \, \cos (\Phi
^{-}(z_{o},t)-\frac{3 \pi}{4})}{\sqrt{\vert t \vert (\aleph_{3} \! - \!
\aleph_{4})} \, (z_{o}^{2} \! + \! 32)^{1/4}} \nonumber \\
 &+ \mathcal{O} \! \left( \! \left( \dfrac{c^{\mathcal{S}}(\aleph_{3})
\underline{c}(\aleph_{4},\aleph_{1},\aleph_{2})}{\sqrt{\vert \aleph_{
3} \vert (z_{o}^{2} \! + \! 32)}} \! + \! \dfrac{c^{\mathcal{S}}(\aleph_{4})
\underline{c}(\aleph_{3},\aleph_{1},\aleph_{2})}{\sqrt{\vert \aleph_{
4} \vert (z_{o}^{2} \! + \! 32)}} \right) \! \dfrac{\ln \vert t \vert}{
(\aleph_{3} \! - \! \aleph_{4}) t} \right), \\
\int_{-\infty}^{x} \! \left( \vert u(\xi,t) \vert^{2} \! - \! 1 \right)
\md \xi &= -\phi^{-}(0) + \dfrac{2 \sqrt{\nu(\aleph_{4})} \, \cos (\Phi
^{-}(z_{o},t)-\frac{3 \pi}{4})}{\sqrt{\vert t \vert (\aleph_{3} \! - \!
\aleph_{4})} \, (z_{o}^{2} \! + \! 32)^{1/4}} \nonumber
\end{align}
\begin{align}
 &+ \mathcal{O} \! \left( \! \left( \dfrac{c^{\mathcal{S}}(\aleph_{3})
\underline{c}(\aleph_{4},\aleph_{1},\aleph_{2})}{\sqrt{\vert \aleph_{
3} \vert (z_{o}^{2} \! + \! 32)}} \! + \! \dfrac{c^{\mathcal{S}}(\aleph_{
4}) \underline{c}(\aleph_{3},\aleph_{1},\aleph_{2})}{\sqrt{\vert \aleph_{
4} \vert (z_{o}^{2} \! + \! 32)}} \right) \! \dfrac{\ln \vert t \vert}{
(\aleph_{3} \! - \! \aleph_{4}) t} \right).
\end{align}
\end{ddddd}
\begin{ddddd}
For $r(\zeta) \! \in \! \mathcal{S}_{\mathbb{C}}(\mathbb{R}) \cap \{
\mathstrut h(z); \, \vert \vert h(\cdot) \vert \vert_{\mathcal{L}^{\infty}
(\mathbb{R})} \! := \! \sup_{z \in \mathbb{R}} \vert h(z) \vert \! < \!
1\}$, let $m(x,t;\zeta)$ be the solution of the Riemann-Hilbert problem
formulated in Lemma~{\rm 2.5}. Let $u(x,t)$, the solution of the Cauchy
problem for the {\rm D${}_{f}$NLSE} with finite-density initial data
$u(x,0) \! := \! u_{o}(x) \! =_{x \to \pm \infty} \! u_{o}(\pm \infty)
(1 \! + \! o(1))$, where $u_{o}(\pm \infty) \! := \! \exp (\tfrac{\mi (1
\mp 1) \theta}{2})$, $0 \! \leqslant \! \theta \! = \! -\int_{-\infty}^{+
\infty} \tfrac{\ln (1-\vert r(\mu) \vert^{2})}{\mu} \, \tfrac{\md \mu}{2
\pi} \! < \! 2 \pi$, $u_{o}(x) \! \in \! \mathbf{C}^{\infty}(\mathbb{R})$,
and $u_{o}(x) \! - \! u_{o}(\pm \infty) \! \in \! \mathcal{S}_{\mathbb{C}}
(\mathbb{R}_{\pm})$, be defined by Eq.~{\rm (5)}. Set $s_{1} \! := \! \zeta_{
1}^{\sharp} \! = \! \exp (\mi \widehat{\varphi}_{1})$ and $s_{2} \! := \!
\zeta_{3}^{\sharp} \! = \! \exp (\mi \widehat{\varphi}_{3})$, where $\zeta_{
n}^{\sharp}$ and $\widehat{\varphi}_{n}$, $n \! \in \! \{1,3\}$, are defined
in Eqs.~{\rm (36)} and~{\rm (37)}. Then: (i) for $r(s_{1}) \! = \! \exp (-
\tfrac{\mi \varepsilon_{1} \pi}{2}) \vert r(s_{1}) \vert$, $\varepsilon_{
1} \! \in \! \{\pm 1\}$, $r(\overline{s_{2}}) \! = \! \exp (\tfrac{\mi
\varepsilon_{2} \pi}{2}) \vert r(\overline{s_{2}}) \vert$, $\varepsilon_{2}
\! \in \! \{\pm 1\}$, $0 \! < \! r(s_{2}) \overline{r(\overline{s_{2}})}
\! < \! 1$, and $\varepsilon_{1} \! = \! \varepsilon_{2}$, as $t \! \to \!
+\infty$ and $x \! \to \! -\infty$ such that $z_{o} \! := \! x/t \! \in \!
(-2,0)$,
\begin{align}
u(x,t) \! &= \! \me^{-\mi \psi^{+}(1)} \! \left(1 \! + \! \dfrac{\me^{\frac{
\mi}{2}(\varepsilon_{1} \pi -(\widehat{\varphi}_{1}-\widehat{\varphi}_{
3}))} \me^{-(\widetilde{a}_{+}t+\widetilde{c}_{+})} \beth_{+}}{\widetilde{b}
\sqrt{t}} \! \left( \mi \sin (\tfrac{1}{2}(\widehat{\varphi}_{1} \! + \!
\widehat{\varphi}_{3})) \sinh \! \left(\widetilde{a}_{-}t \! + \! \widetilde{
c}_{-} \right. \right. \right. \nonumber \\
 &\left. \left. \left. \! +\tfrac{1}{8} \ln \! \left( \tfrac{4-a_{1}^{2}}{4
-a_{2}^{2}} \right) \! - \! \ln \beth_{-} \right) \! + \! \cos (\tfrac{1}{2}
(\widehat{\varphi}_{1} \! + \! \widehat{\varphi}_{3})) \cosh \! \left(
\widetilde{a}_{-}t \! + \! \widetilde{c}_{-} \! + \! \tfrac{1}{8} \ln \!
\left( \tfrac{4-a_{1}^{2}}{4-a_{2}^{2}} \right) \! - \! \ln \beth_{-} \right)
\right) \right. \nonumber \\
 &\left. + \, \mathcal{O} \! \left( \dfrac{\underline{c}(z_{o}) \me^{-4
\alpha t}}{\beta t} \right) \right),
\end{align}
and, for $\varepsilon_{1} \! = \! -\varepsilon_{2}$,
\begin{align}
u(x,t) \! &= \! \me^{-\mi \psi^{+}(1)} \! \left(1 \! - \! \dfrac{\me^{\frac{
\mi}{2}(\varepsilon_{1} \pi -(\widehat{\varphi}_{1}-\widehat{\varphi}_{
3}))} \me^{-(\widetilde{a}_{+}t+\widetilde{c}_{+})} \beth_{+}}{
\widetilde{b} \sqrt{t}} \! \left(\mi \sin (\tfrac{1}{2}(\widehat{\varphi}_{
1} \! + \! \widehat{\varphi}_{3})) \cosh \! \left(\widetilde{a}_{-}t \! + \!
\widetilde{c}_{-} \right. \right. \right. \nonumber \\
 &\left. \left. \left. \! +\tfrac{1}{8} \ln \! \left( \tfrac{4-a_{1}^{2}}{4
-a_{2}^{2}} \right) \! - \! \ln \beth_{-} \right) \! + \! \cos (\tfrac{1}{2}
(\widehat{\varphi}_{1} \! + \! \widehat{\varphi}_{3})) \sinh \! \left(
\widetilde{a}_{-}t \! + \! \widetilde{c}_{-} \! + \! \tfrac{1}{8} \ln \!
\left( \tfrac{4-a_{1}^{2}}{4-a_{2}^{2}} \right) \! - \! \ln \beth_{-} \right)
\right) \right. \nonumber \\
 &\left. + \, \mathcal{O} \! \left( \dfrac{\underline{c}(z_{o}) \me^{-4
\alpha t}}{\beta t} \right) \right),
\end{align}
where
\begin{gather}
\psi^{+}(l) \! := \! \int_{-\infty}^{0} \dfrac{\ln (1 \! - \! \vert r
(\mu) \vert^{2})}{\mu^{l}} \, \dfrac{\md \mu}{2 \pi}, \quad l \! \in
\! \{0,1\}, \\
\zeta_{1}^{\sharp} \! := \! -\tfrac{1}{2}(a_{1} \! - \! \mi (4 \! - \!
a_{1}^{2})^{1/2}) \! = \! \me^{\mi \widehat{\varphi}_{1}}, \, \, \, \,
\widehat{\varphi}_{1} \! = \! \arctan \! \left( \tfrac{(4-a_{1}^{2})^{
1/2}}{\vert a_{1} \vert} \right) \! \in \! (0,\tfrac{\pi}{2}), \, \, \, \,
a_{1} \! < \! 0, \, \, \, \, \vert a_{1} \vert \! < \! 2, \\
\zeta_{3}^{\sharp} \! := \! -\tfrac{1}{2}(a_{2} \! - \! \mi (4 \! - \!
a_{2}^{2})^{1/2}) \! = \! \me^{\mi \widehat{\varphi}_{3}}, \, \, \, \,
\widehat{\varphi}_{3} \! = \! -\arctan \! \left( \tfrac{(4-a_{2}^{2})^{
1/2}}{\vert a_{2} \vert} \right) \! \in \! (\tfrac{\pi}{2},\pi), \, \, \,
\, a_{2} \! > \! 0, \, \, \, \, \vert a_{2} \vert \! < \! 2, \\
\widetilde{a}_{\pm} \! := \! \tfrac{1}{2}z_{o}((4 \! - \! a_{1}^{2})^{
1/2} \! \mp \! (4 \! - \! a_{2}^{2})^{1/2}) \! - \! \tfrac{1}{2}(a_{1}(4
\! - \! a_{1}^{2})^{1/2} \! \mp \! a_{2}(4 \! - \! a_{2}^{2})^{1/2}),
\quad \widetilde{a}_{+} \! > \! 0, \\
\widetilde{c}_{\pm} \! := \! \sin (\widehat{\varphi}_{1}) \int\nolimits_{
-\infty}^{0} \dfrac{\ln (1 \! - \! \vert r(\mu) \vert^{2})}{(\mu \! - \!
\cos \widehat{\varphi}_{1})^{2} \! + \! \sin^{2} \widehat{\varphi}_{1}}
\, \dfrac{\md \mu}{2 \pi} \mp \sin (\widehat{\varphi}_{3}) \int\nolimits_{
-\infty}^{0} \dfrac{\ln (1 \! - \! \vert r(\mu) \vert^{2})}{(\mu \! - \!
\cos \widehat{\varphi}_{3})^{2} \! + \! \sin^{2} \widehat{\varphi}_{3}}
\, \dfrac{\md \mu}{2 \pi}, \\
\widetilde{b} \! := \! 2^{-1/2}(z_{o}^{2} \! + \! 32)^{1/4}(4 \! - \!
a_{1}^{2})^{1/8}(4 \! - \! a_{2}^{2})^{1/8}, \\
\beth_{+} \! := \! \left(\dfrac{\vert r(s_{1}) \vert \vert r(\overline{s_{2}})
\vert}{(1 \! - \! r(s_{2}) \overline{r(\overline{s_{2}})})} \right)^{1/2},
\qquad \, \, \beth_{-} \! := \! \left( \dfrac{\vert r(s_{1}) \vert (1 \! - \!
r(s_{2}) \overline{r(\overline{s_{2}})})}{\vert r(\overline{s_{2}}) \vert}
\right)^{1/2}, \\
\alpha \! := \! \min \{\tfrac{1}{2}(z_{o} \! - \! a_{1})(4 \! - \! a_{1}^{
2})^{1/2},-\tfrac{1}{2}(z_{o} \! - \! a_{2})(4 \! - \! a_{2}^{2})^{1/2}\} \,
\, \, (> \! 0), \\
\beta \! := \! \min \{\tfrac{1}{2}(z_{o}^{2} \! + \! 32)^{
1/2}(4 \! - \! a_{1}^{2})^{1/2},\tfrac{1}{2}(z_{o}^{2} \! + \! 32)^{1/2}(4
\! - \! a_{2}^{2})^{1/2}\} \, \, \, (> \! 0),
\end{gather}
and $a_{1}$ and $a_{2}$ are given in Theorem~{\rm 3.1},
Eq.~{\rm (17);} (ii) for $r(\overline{s_{1}}) \! = \! \exp (\tfrac{\mi
\varepsilon_{1} \pi}{2}) \vert r(\overline{s_{1}}) \vert$, $\varepsilon_{
1} \! \in \! \{\pm 1\}$, $r(s_{2}) \! = \! \exp (-\tfrac{\mi \varepsilon_{
2} \pi}{2}) \vert r(s_{2}) \vert$, $\varepsilon_{2} \! \in \! \{\pm 1\}$,
$0 \! < \! r(s_{1}) \overline{r(\overline{s_{1}})} \! < \! 1$, and
$\varepsilon_{1} \! = \! \varepsilon_{2}$, as $t \! \to \! -\infty$ and
$x \! \to \! +\infty$ such that $z_{o} \! \in \! (-2,0)$,
\begin{align}
u(x,t) \! &= \! \me^{-\mi \psi^{-}(1)} \! \left(1 \! + \! \dfrac{\me^{\frac{
\mi}{2}(\varepsilon_{1} \pi +(\widehat{\varphi}_{1}-\widehat{\varphi}_{3}))}
\me^{-(\widetilde{a}_{+} \vert t \vert -\widehat{c}_{+})} \daleth_{+}}{
\widetilde{b} \sqrt{\vert t \vert}} \! \left(\mi \sin (\tfrac{1}{2}(\widehat{
\varphi}_{1} \! + \! \widehat{\varphi}_{3})) \sinh \! \left(\widetilde{a}_{
-} \vert t \vert \! - \! \widehat{c}_{-} \right. \right. \right. \nonumber
\end{align}
\begin{align}
 &\left. \left. \left. \! + \tfrac{1}{8} \ln \! \left( \tfrac{4-a_{1}^{2}}{4-
a_{2}^{2}} \right) \! + \ln \daleth_{-} \right) \! - \! \cos (\tfrac{1}{2}
(\widehat{\varphi}_{1} \! + \! \widehat{\varphi}_{3})) \cosh
\! \left(\widetilde{a}_{-} \vert t \vert \! - \! \widehat{c}_{-} \! + \!
\tfrac{1}{8} \ln \! \left(\tfrac{4-a_{1}^{2}}{4-a_{2}^{2}} \right) \!
+ \! \ln \daleth_{-} \right) \right) \right. \nonumber \\
 &\left. + \, \mathcal{O} \! \left( \dfrac{\underline{c}(z_{o}) \me^{
-4 \alpha \vert t \vert}}{\beta t} \right) \right),
\end{align}
and, for $\varepsilon_{1} \! = \! -\varepsilon_{2}$,
\begin{align}
u(x,t) \! &= \! \me^{-\mi \psi^{-}(1)} \! \left(1 \! - \! \dfrac{\me^{\frac{
\mi}{2}(\varepsilon_{1} \pi +(\widehat{\varphi}_{1}-\widehat{\varphi}_{3}))}
\me^{-(\widetilde{a}_{+} \vert t \vert -\widehat{c}_{+})} \daleth_{+}}{
\widetilde{b} \sqrt{\vert t \vert}} \! \left( \mi \sin (\tfrac{1}{2}
(\widehat{\varphi}_{1} \! + \! \widehat{\varphi}_{3})) \cosh \! \left(
\widetilde{a}_{-} \vert t \vert \! - \! \widehat{c}_{-} \right. \right.
\right. \nonumber \\
 &\left. \left. \left. \! +\tfrac{1}{8} \ln \! \left( \tfrac{4-a_{1}^{2}}{4
-a_{2}^{2}} \right) \! + \ln \daleth_{-} \right) \! - \! \cos (\tfrac{1}{2}
(\widehat{\varphi}_{1} \! + \! \widehat{\varphi}_{3})) \sinh \! \left(
\widetilde{a}_{-} \vert t \vert \! - \! \widehat{c}_{-} \! + \! \tfrac{1}{8}
\ln \! \left(\tfrac{4-a_{1}^{2}}{4-a_{2}^{2}} \right) \! + \! \ln \daleth_{
-} \right) \right) \right. \nonumber \\
 &\left. + \, \mathcal{O} \! \left( \dfrac{\underline{c}(z_{o}) \me^{
-4 \alpha \vert t \vert}}{\beta t} \right) \right),
\end{align}
where
\begin{gather}
\psi^{-}(l) \! := \! \int_{0}^{+\infty} \dfrac{\ln (1 \! - \! \vert r
(\mu) \vert^{2})}{\mu^{l}} \, \dfrac{\md \mu}{2 \pi}, \quad l \! \in
\! \{0,1\}, \\
\widehat{c}_{\pm} \! := \! \sin (\widehat{\varphi}_{1}) \int\nolimits_{
0}^{+\infty} \dfrac{\ln (1 \! - \! \vert r(\mu) \vert^{2})}{(\mu \! - \!
\cos \widehat{\varphi}_{1})^{2} \! + \! \sin^{2} \widehat{\varphi}_{1}}
\, \dfrac{\md \mu}{2 \pi} \mp \sin (\widehat{\varphi}_{3}) \int\nolimits_{
0}^{+\infty} \dfrac{\ln (1 \! - \! \vert r(\mu) \vert^{2})}{(\mu \! - \!
\cos \widehat{\varphi}_{3})^{2} \! + \! \sin^{2} \widehat{\varphi}_{
3}} \, \dfrac{\md \mu}{2 \pi}, \\
\daleth_{+} \! := \! \left(\dfrac{\vert r(\overline{s_{1}}) \vert \vert r(s_{
2}) \vert}{(1 \! - \! r(s_{1}) \overline{r(\overline{s_{1}})})} \right)^{1/2},
\qquad \, \, \daleth_{-} \! := \! \left( \dfrac{\vert r(s_{2}) \vert (1 \!
- \! r(s_{1}) \overline{r(\overline{s_{1}})})}{\vert r(\overline{s_{1}})
\vert} \right)^{1/2};
\end{gather}
(iii) for $r(\overline{s_{2}}) \! = \! \exp(-\tfrac{\mi \varepsilon_{1} \pi}{
2}) \vert r(\overline{s_{2}}) \vert$, $\varepsilon_{1} \! \in \! \{\pm 1\}$,
$r(s_{1}) \! = \! \exp (\tfrac{\mi \varepsilon_{2} \pi}{2}) \vert r(s_{1})
\vert$, $\varepsilon_{2} \! \in \! \{\pm 1\}$, $0 \! < \! r(s_{1}) \overline{
r(\overline{s_{1}})} \! < \! 1$, and $\varepsilon_{1} \! = \! \varepsilon_{
2}$, as $t \! \to \! +\infty$ and $x \! \to \! +\infty$ such that $z_{o}
\! \in \! (0,2)$,
\begin{align}
u(x,t) \! &= \! -\me^{-\mi \psi^{+}(1)} \! \left(1 \! + \! \dfrac{\me^{
\frac{\mi}{2}(\varepsilon_{1} \pi -(\widehat{\varphi}_{1}-\widehat{
\varphi}_{3}))} \me^{-(\widetilde{a}_{+}t+\widetilde{c}_{+})}
\widetilde{\gimel}_{+}}{\widetilde{b} \sqrt{t}} \! \left(\mi \sin (\tfrac{
1}{2}(\widehat{\varphi}_{1} \! + \! \widehat{\varphi}_{3})) \sinh \! \left(
\widetilde{a}_{-}t \! + \! \widetilde{c}_{-} \right. \right. \right.
\nonumber \\
 &\left. \left. \left. \! +\tfrac{1}{8} \ln \! \left( \tfrac{4-a_{1}^{
2}}{4-a_{2}^{2}} \right) \! + \! \ln \widetilde{\gimel}_{-} \right)
\! + \! \cos (\tfrac{1}{2}(\widehat{\varphi}_{1} \! + \! \widehat{
\varphi}_{3})) \cosh \! \left( \widetilde{a}_{-}t \! + \! \widetilde{
c}_{-} \! + \! \tfrac{1}{8} \ln \! \left( \tfrac{4-a_{1}^{2}}{4-a_{
2}^{2}} \right) \! + \! \ln \widetilde{\gimel}_{-} \right) \right)
\right. \nonumber \\
 &\left. + \, \mathcal{O} \! \left( \dfrac{\underline{c}(z_{o})
\me^{-4 \alpha t}}{\beta t} \right) \right),
\end{align}
and, for $\varepsilon_{1} \! = \! -\varepsilon_{2}$,
\begin{align}
u(x,t) \! &= \! -\me^{-\mi \psi^{+}(1)} \! \left(1 \! + \! \dfrac{\me^{
\frac{\mi}{2}(\varepsilon_{1} \pi -(\widehat{\varphi}_{1}-\widehat{
\varphi}_{3}))} \me^{-(\widetilde{a}_{+}t+\widetilde{c}_{+})}
\widetilde{\gimel}_{+}}{\widetilde{b} \sqrt{t}} \! \left(\mi \sin (\tfrac{
1}{2}(\widehat{\varphi}_{1} \! + \! \widehat{\varphi}_{3})) \cosh \! \left(
\widetilde{a}_{-}t \! + \! \widetilde{c}_{-} \right. \right. \right.
\nonumber \\
 &\left. \left. \left. \! +\tfrac{1}{8} \ln \! \left( \tfrac{4-a_{1}^{2}}{
4-a_{2}^{2}} \right) \! + \! \ln \widetilde{\gimel}_{-} \right) \! + \! \cos
(\tfrac{1}{2}(\widehat{\varphi}_{1} \! + \! \widehat{\varphi}_{3})) \sinh
\! \left( \widetilde{a}_{-}t \! + \! \widetilde{c}_{-} \! + \! \tfrac{1}{8}
\ln \! \left( \tfrac{4-a_{1}^{2}}{4-a_{2}^{2}} \right) \! + \! \ln
\widetilde{\gimel}_{-} \right) \right) \right. \nonumber \\
 &\left. + \, \mathcal{O} \! \left( \dfrac{\underline{c}(z_{o})
\me^{-4 \alpha t}}{\beta t} \right) \right),
\end{align}
where
\begin{gather}
\widetilde{\gimel}_{+} \! := \! \left( \dfrac{\vert r(\overline{s_{2}}) \vert
\vert r(s_{1}) \vert}{(1 \! - \! r(s_{1}) \overline{r(\overline{s_{1}})})}
\right)^{1/2}, \qquad \, \, \widetilde{\gimel}_{-} \! := \! \left(\dfrac{
\vert r(\overline{s_{2}}) \vert (1 \! - \! r(s_{1}) \overline{r(\overline{
s_{1}})})}{\vert r(s_{1}) \vert} \right)^{1/2};
\end{gather}
and (iv) for $r(s_{2}) \! = \! \exp (\tfrac{\mi \varepsilon_{1} \pi}{2}) \vert
r(s_{2}) \vert$, $\varepsilon_{1} \! \in \! \{\pm 1\}$, $r(\overline{s_{1}})
\! = \! \exp (-\tfrac{\mi \varepsilon_{2} \pi}{2}) \vert r(\overline{s_{1}})
\vert$, $\varepsilon_{2} \! \in \! \{\pm 1\}$, $0 \! < \! r(s_{2}) \overline{r
(\overline{s_{2}})} \! < \! 1$, and $\varepsilon_{1} \! = \! \varepsilon_{
2}$, as $t \! \to \! -\infty$ and $x \! \to \! -\infty$ such that $z_{o} \!
\in \! (0,2)$,
\begin{align}
u(x,t) \! &= \! -\me^{-\mi \psi^{-}(1)} \! \left( \! 1 \! + \! \dfrac{\me^{
\frac{\mi}{2}(\varepsilon_{1} \pi +(\widehat{\varphi}_{1}-\widehat{\varphi}_{
3}))} \me^{-(\widetilde{a}_{+} \vert t \vert -\widehat{c}_{+})} \widehat{
\gimel}_{+}}{\widetilde{b} \sqrt{\vert t \vert}} \! \left(\mi \sin (\tfrac{
1}{2}(\widehat{\varphi}_{1} \! + \! \widehat{\varphi}_{3})) \sinh \! \left(
\widetilde{a}_{-} \vert t \vert \! - \! \widehat{c}_{-} \right. \right.
\right. \nonumber \\
 &\left. \left. \left. \! +\tfrac{1}{8} \ln \! \left( \tfrac{4-a_{1}^{2}}{4-
a_{2}^{2}} \right) \! - \! \ln \widehat{\gimel}_{-} \right) \! - \! \cos
(\tfrac{1}{2}(\widehat{\varphi}_{1} \! + \! \widehat{\varphi}_{3}))
\cosh \! \left(\widetilde{a}_{-} \vert t \vert \! - \! \widehat{c}_{-}
\! + \! \tfrac{1}{8} \ln \! \left( \tfrac{4-a_{1}^{2}}{4-a_{2}^{2}}
\right) \! - \! \ln \widehat{\gimel}_{-} \right) \right) \right. \nonumber
\\
 &\left. + \, \mathcal{O} \! \left( \dfrac{\underline{c}(z_{o})
\me^{-4 \alpha \vert t \vert}}{\beta t} \right) \right),
\end{align}
and, for $\varepsilon_{1} \! = \! -\varepsilon_{2}$,
\begin{align}
u(x,t) \! &= \! -\me^{-\mi \psi^{-}(1)} \! \left( \! 1 \! + \! \dfrac{\me^{
\frac{\mi}{2}(\varepsilon_{1} \pi +(\widehat{\varphi}_{1}-\widehat{
\varphi}_{3}))} \me^{-(\widetilde{a}_{+} \vert t \vert -\widehat{
c}_{+})} \widehat{\gimel}_{+}}{\widetilde{b} \sqrt{\vert t \vert}} \!
\left(\mi \sin (\tfrac{1}{2}(\widehat{\varphi}_{1} \! + \! \widehat{
\varphi}_{3})) \cosh \! \left(\widetilde{a}_{-} \vert t \vert \! - \!
\widehat{c}_{-} \right. \right. \right. \nonumber \\
 &\left. \left. \left. \! +\tfrac{1}{8} \ln \! \left( \tfrac{4-a_{1}^{2}}{4-
a_{2}^{2}} \right) \! - \! \ln \widehat{\gimel}_{-} \right) \! - \! \cos
(\tfrac{1}{2}(\widehat{\varphi}_{1} \! + \! \widehat{\varphi}_{3}))
\sinh \! \left(\widetilde{a}_{-} \vert t \vert \! - \! \widehat{c}_{-} \!
+ \! \tfrac{1}{8} \ln \! \left( \tfrac{4-a_{1}^{2}}{4-a_{2}^{2}} \right)
\! - \! \ln \widehat{\gimel}_{-} \right) \right) \right. \nonumber \\
 &\left. + \, \mathcal{O} \! \left( \dfrac{\underline{c}(z_{o})
\me^{-4 \alpha \vert t \vert}}{\beta t} \right) \right),
\end{align}
where
\begin{gather}
\widehat{\gimel}_{+} \! := \! \left(\dfrac{\vert r(s_{2}) \vert \vert
r(\overline{s_{1}}) \vert}{(1 \! - \! r(s_{2}) \overline{r(\overline{
s_{2}})})} \right)^{1/2}, \qquad \, \, \widehat{\gimel}_{-} \! := \!
\left(\dfrac{\vert r(\overline{s_{1}}) \vert (1 \! - \! r(s_{2}) \overline{
r(\overline{s_{2}})})}{\vert r(s_{2}) \vert} \right)^{1/2}.
\end{gather}
For $u(x,t)$ as defined and given above, let $\int_{+\infty}^{x}(\vert u(\xi,
t) \vert^{2} \! - \! 1) \, \md \xi$ be defined by Eq.~{\rm (6)}. Then: (i)
for $r(s_{1}) \! = \! \exp (-\tfrac{\mi \varepsilon_{1} \pi}{2}) \vert r(s_{
1}) \vert$, $\varepsilon_{1} \! \in \! \{\pm 1\}$, $r(\overline{s_{2}}) \! =
\! \exp (\tfrac{\mi \varepsilon_{2} \pi}{2}) \vert r(\overline{s_{2}})
\vert$, $\varepsilon_{2} \! \in \! \{\pm 1\}$, $0 \! < \! r(s_{2}) \overline{
r(\overline{s_{2}})} \! < \! 1$, and $\varepsilon_{1} \! = \! \varepsilon_{
2}$, as $t \! \to \! +\infty$ and $x \! \to \! -\infty$ such that $z_{o} \!
\in \! (-2,0)$,
\begin{align}
\int\nolimits_{+\infty}^{x}(\vert u(\xi,t) \vert^{2} \! - \! 1) \, \md
\xi \! &= \! \psi^{+}(0) \! - \! \dfrac{\mathrm{sgn}(\varepsilon_{
1}) \me^{-(\widetilde{a}_{+}t+\widetilde{c}_{+})} \beth_{+}}{
\widetilde{b} \sqrt{t}} \cosh \! \left( \widetilde{a}_{-}t \! + \!
\widetilde{c}_{-} \! + \! \tfrac{1}{8} \ln \! \left( \tfrac{4-a_{1}^{
2}}{4-a_{2}^{2}} \right) \! - \! \ln \beth_{-} \right) \nonumber \\
 &+\mathcal{O} \! \left( \dfrac{\underline{c}(z_{o}) \me^{-4
\alpha t}}{\beta t} \right), \\
\int\nolimits_{-\infty}^{x}(\vert u(\xi,t) \vert^{2} \! - \! 1) \, \md
\xi \! &= \! -\psi^{-}(0) \! - \! \dfrac{\mathrm{sgn}(\varepsilon_{
1}) \me^{-(\widetilde{a}_{+}t+\widetilde{c}_{+})} \beth_{+}}{
\widetilde{b} \sqrt{t}} \cosh \! \left( \widetilde{a}_{-}t \! + \!
\widetilde{c}_{-} \! + \! \tfrac{1}{8} \ln \! \left( \tfrac{4-a_{1}^{
2}}{4-a_{2}^{2}} \right) \! - \! \ln \beth_{-} \right) \nonumber \\
 &+\mathcal{O} \! \left( \dfrac{\underline{c}(z_{o}) \me^{-4
\alpha t}}{\beta t} \right),
\end{align}
and, for $\varepsilon_{1} \! = \! -\varepsilon_{2}$,
\begin{align}
\int\nolimits_{+\infty}^{x}(\vert u(\xi,t) \vert^{2} \! - \! 1) \, \md
\xi \! &= \! \psi^{+}(0) \! + \! \dfrac{\mathrm{sgn}(\varepsilon_{
1}) \me^{-(\widetilde{a}_{+}t+\widetilde{c}_{+})} \beth_{+}}{
\widetilde{b} \sqrt{t}} \sinh \! \left( \widetilde{a}_{-}t \! + \!
\widetilde{c}_{-} \! + \! \tfrac{1}{8} \ln \! \left( \tfrac{4-a_{1}^{
2}}{4-a_{2}^{2}} \right) \! - \! \ln \beth_{-} \right) \nonumber \\
 &+\mathcal{O} \! \left( \dfrac{\underline{c}(z_{o}) \me^{-4
\alpha t}}{\beta t} \right), \\
\int\nolimits_{-\infty}^{x}(\vert u(\xi,t) \vert^{2} \! - \! 1) \, \md
\xi \! &= \! -\psi^{-}(0) \! + \! \dfrac{\mathrm{sgn}(\varepsilon_{
1}) \me^{-(\widetilde{a}_{+}t+\widetilde{c}_{+})} \beth_{+}}{
\widetilde{b} \sqrt{t}} \sinh \! \left( \widetilde{a}_{-}t \! + \!
\widetilde{c}_{-} \! + \! \tfrac{1}{8} \ln \! \left( \tfrac{4-a_{1}^{
2}}{4-a_{2}^{2}} \right) \! - \! \ln \beth_{-} \right) \nonumber \\
 &+\mathcal{O} \! \left( \dfrac{\underline{c}(z_{o}) \me^{-4
\alpha t}}{\beta t} \right);
\end{align}
(ii) for $r(\overline{s_{1}}) \! = \! \exp (\tfrac{\mi \varepsilon_{1} \pi}
{2}) \vert r(\overline{s_{1}}) \vert$, $\varepsilon_{1} \! \in \! \{\pm 1\}$,
$r(s_{2}) \! = \! \exp (-\tfrac{\mi \varepsilon_{2} \pi}{2}) \vert r(s_{2})
\vert$, $\varepsilon_{2} \! \in \! \{\pm 1\}$, $0 \! < \! r(s_{1}) \overline{r
(\overline{s_{1}})} \! < \! 1$, and $\varepsilon_{1} \! = \! \varepsilon_{
2}$, as $t \! \to \! -\infty$ and $x \! \to \! +\infty$ such that $z_{o} \!
\in \! (-2,0)$,
\begin{align}
\int\nolimits_{+\infty}^{x}(\vert u(\xi,t) \vert^{2} \! - \! 1) \, \md
\xi \! &= \! \psi^{-}(0) \! + \! \dfrac{\mathrm{sgn}(\varepsilon_{
1}) \me^{-(\widetilde{a}_{+} \vert t \vert -\widehat{c}_{+})}
\daleth_{+}}{\widetilde{b} \sqrt{\vert t \vert}} \cosh \! \left(
\widetilde{a}_{-} \vert t \vert \! - \! \widehat{c}_{-} \! + \! \tfrac{
1}{8} \ln \! \left( \tfrac{4-a_{1}^{2}}{4-a_{2}^{2}} \right) \! + \! \ln
\daleth_{-} \right) \nonumber \\
 &+\mathcal{O} \! \left( \dfrac{\underline{c}(z_{o}) \me^{-4
\alpha \vert t \vert}}{\beta t} \right), \\
\int\nolimits_{-\infty}^{x}(\vert u(\xi,t) \vert^{2} \! - \! 1) \, \md
\xi \! &= \! -\psi^{+}(0) \! + \! \dfrac{\mathrm{sgn}(\varepsilon_{
1}) \me^{-(\widetilde{a}_{+} \vert t \vert -\widehat{c}_{+})}
\daleth_{+}}{\widetilde{b} \sqrt{ \vert t \vert}} \cosh \! \left(
\widetilde{a}_{-} \vert t \vert \! - \! \widehat{c}_{-} \! + \! \tfrac{
1}{8} \ln \! \left( \tfrac{4-a_{1}^{2}}{4-a_{2}^{2}} \right) \! + \!
\ln \daleth_{-} \right) \nonumber \\
 &+\mathcal{O} \! \left( \dfrac{\underline{c}(z_{o}) \me^{-4
\alpha \vert t \vert}}{\beta t} \right),
\end{align}
and, for $\varepsilon_{1} \! = \! -\varepsilon_{2}$,
\begin{align}
\int\nolimits_{+\infty}^{x}(\vert u(\xi,t) \vert^{2} \! - \! 1) \, \md
\xi \! &= \! \psi^{-}(0) \! - \! \dfrac{\mathrm{sgn}(\varepsilon_{
1}) \me^{-(\widetilde{a}_{+} \vert t \vert -\widehat{c}_{+})}
\daleth_{+}}{\widetilde{b} \sqrt{\vert t \vert}} \sinh \! \left(
\widetilde{a}_{-} \vert t \vert \! - \! \widehat{c}_{-} \! + \! \tfrac{
1}{8} \ln \! \left( \tfrac{4-a_{1}^{2}}{4-a_{2}^{2}} \right) \! + \!
\ln \daleth_{-} \right) \nonumber \\
 &+\mathcal{O} \! \left( \dfrac{\underline{c}(z_{o}) \me^{-4
\alpha \vert t \vert}}{\beta t} \right), \\
\int\nolimits_{-\infty}^{x}(\vert u(\xi,t) \vert^{2} \! - \! 1) \, \md
\xi \! &= \! -\psi^{+}(0) \! - \! \dfrac{\mathrm{sgn}(\varepsilon_{
1}) \me^{-(\widetilde{a}_{+} \vert t \vert -\widehat{c}_{+})}
\daleth_{+}}{\widetilde{b} \sqrt{ \vert t \vert}} \sinh \! \left(
\widetilde{a}_{-} \vert t \vert \! - \! \widehat{c}_{-} \! + \! \tfrac{
1}{8} \ln \! \left( \tfrac{4-a_{1}^{2}}{4-a_{2}^{2}} \right) \! + \!
\ln \daleth_{-} \right) \nonumber \\
 &+\mathcal{O} \! \left( \dfrac{\underline{c}(z_{o}) \me^{-4
\alpha \vert t \vert}}{\beta t} \right);
\end{align}
(iii) for $r(\overline{s_{2}}) \! = \! \exp (-\tfrac{\mi \varepsilon_{1} \pi}
{2}) \vert r(\overline{s_{2}}) \vert$, $\varepsilon_{1} \! \in \! \{\pm 1\}$,
$r(s_{1}) \! = \! \exp (\tfrac{\mi \varepsilon_{2} \pi}{2}) \vert r(s_{1})
\vert$, $\varepsilon_{2} \! \in \! \{\pm 1\}$, $0 \! < \! r(s_{1}) \overline{
r(\overline{s_{1}})} \! < \! 1$, and $\varepsilon_{1} \! = \! \varepsilon_{
2}$, as $t \! \to \! +\infty$ and $x \! \to \! +\infty$ such that $z_{o} \!
\in \! (0,2)$,
\begin{align}
\int\nolimits_{+\infty}^{x}(\vert u(\xi,t) \vert^{2} \! - \! 1) \, \md
\xi \! &= \! \psi^{-}(0) \! + \! \dfrac{\mathrm{sgn}(\varepsilon_{
1}) \me^{-(\widetilde{a}_{+}t+\widetilde{c}_{+})} \widetilde{
\gimel}_{+}}{\widetilde{b} \sqrt{t}} \cosh \! \left(\widetilde{
a}_{-}t \! + \! \widetilde{c}_{-} \! + \! \tfrac{1}{8} \ln \! \left(\tfrac{
4-a_{1}^{2}}{4-a_{2}^{2}} \right) \! + \! \ln \widetilde{\gimel}_{
-} \right) \nonumber \\
 &+\mathcal{O} \! \left( \dfrac{\underline{c}(z_{o}) \me^{-4
\alpha t}}{\beta t} \right), \\
\int\nolimits_{-\infty}^{x}(\vert u(\xi,t) \vert^{2} \! - \! 1) \, \md
\xi \! &= \! -\psi^{+}(0)\! + \! \dfrac{\mathrm{sgn}(\varepsilon_{
1}) \me^{-(\widetilde{a}_{+}t+\widetilde{c}_{+})} \widetilde{
\gimel}_{+}}{\widetilde{b} \sqrt{t}} \cosh \! \left(\widetilde{
a}_{-}t \! + \! \widetilde{c}_{-} \! + \! \tfrac{1}{8} \ln \! \left(\tfrac{
4-a_{1}^{2}}{4-a_{2}^{2}} \right) \! + \! \ln \widetilde{\gimel}_{
-} \right) \nonumber \\
 &+\mathcal{O} \! \left( \dfrac{\underline{c}(z_{o}) \me^{-4
\alpha t}}{\beta t} \right),
\end{align}
and, for $\varepsilon_{1} \! = \! -\varepsilon_{2}$,
\begin{align}
\int\nolimits_{+\infty}^{x}(\vert u(\xi,t) \vert^{2} \! - \! 1) \, \md
\xi \! &= \! \psi^{-}(0) \! + \! \dfrac{\mathrm{sgn}(\varepsilon_{
1}) \me^{-(\widetilde{a}_{+}t+\widetilde{c}_{+})} \widetilde{
\gimel}_{+}}{\widetilde{b} \sqrt{t}} \sinh \! \left(\widetilde{
a}_{-}t \! + \! \widetilde{c}_{-} \! + \! \tfrac{1}{8} \ln \! \left(\tfrac{
4-a_{1}^{2}}{4-a_{2}^{2}} \right) \! + \! \ln \widetilde{\gimel}_{
-} \right) \nonumber \\
 &+\mathcal{O} \! \left( \dfrac{\underline{c}(z_{o}) \me^{-4
\alpha t}}{\beta t} \right), \\
\int\nolimits_{-\infty}^{x}(\vert u(\xi,t) \vert^{2} \! - \! 1) \, \md
\xi \! &= \! -\psi^{+}(0)\! + \! \dfrac{\mathrm{sgn}(\varepsilon_{
1}) \me^{-(\widetilde{a}_{+}t+\widetilde{c}_{+})} \widetilde{
\gimel}_{+}}{\widetilde{b} \sqrt{t}} \sinh \! \left(\widetilde{
a}_{-}t \! + \! \widetilde{c}_{-} \! + \! \tfrac{1}{8} \ln \! \left(\tfrac{
4-a_{1}^{2}}{4-a_{2}^{2}} \right) \! + \! \ln \widetilde{\gimel}_{
-} \right) \nonumber \\
 &+\mathcal{O} \! \left( \dfrac{\underline{c}(z_{o}) \me^{-4 \alpha
t}}{\beta t} \right);
\end{align}
and (iv) for $r(s_{2}) \! = \! \exp (\tfrac{\mi \varepsilon_{1} \pi}{2}) \vert
r(s_{2}) \vert$, $\varepsilon_{1} \! \in \! \{\pm 1\}$, $r(\overline{s_{1}})
\! = \! \exp (-\tfrac{\mi \varepsilon_{2} \pi}{2}) \vert r(\overline{s_{1}})
\vert$, $\varepsilon_{2} \! \in \! \{\pm 1\}$, $0 \! < \! r(s_{2}) \overline{r
(\overline{s_{2}})} \! < \! 1$, and $\varepsilon_{1} \! = \! \varepsilon_{
2}$, as $t \! \to \! -\infty$ and $x \! \to \! -\infty$ such that $z_{o} \!
\in \! (0,2)$,
\begin{align}
\int\nolimits_{+\infty}^{x}(\vert u(\xi,t) \vert^{2} \! - \! 1) \, \md \xi \!
&= \! \psi^{+}(0) \! - \! \dfrac{\mathrm{sgn}(\varepsilon_{1}) \me^{
-(\widetilde{a}_{+} \vert t \vert -\widehat{c}_{+})} \widehat{\gimel}_{
+}}{\widetilde{b} \sqrt{\vert t \vert}} \cosh \! \left(\widetilde{a}_{-}
\vert t \vert \! - \! \widehat{c}_{-} \! + \! \tfrac{1}{8} \ln \! \left(
\tfrac{4-a_{1}^{2}}{4-a_{2}^{2}} \right) \! - \! \ln \widehat{\gimel}_{-}
\right) \nonumber \\
 &+\mathcal{O} \! \left( \dfrac{\underline{c}(z_{o}) \me^{-4
\alpha \vert t \vert}}{\beta t} \right), \\
\int\nolimits_{-\infty}^{x}(\vert u(\xi,t) \vert^{2} \! - \! 1) \, \md
\xi \! &= \! -\psi^{-}(0) \! - \! \dfrac{\mathrm{sgn}(\varepsilon_{1})
\me^{-(\widetilde{a}_{+} \vert t \vert -\widehat{c}_{+})} \widehat{
\gimel}_{+}}{\widetilde{b} \sqrt{\vert t \vert}} \cosh \! \left(
\widetilde{a}_{-} \vert t \vert \! - \! \widehat{c}_{-} \! + \! \tfrac{
1}{8} \ln \! \left( \tfrac{4-a_{1}^{2}}{4-a_{2}^{2}} \right) \! - \! \ln
\widehat{\gimel}_{-} \right) \nonumber \\
 &+\mathcal{O} \! \left( \dfrac{\underline{c}(z_{o}) \me^{-4
\alpha \vert t \vert}}{\beta t} \right),
\end{align}
and, for $\varepsilon_{1} \! = \! -\varepsilon_{2}$,
\begin{align}
\int\nolimits_{+\infty}^{x}(\vert u(\xi,t) \vert^{2} \! - \! 1) \, \md \xi
\! &= \! \psi^{+}(0) \! - \! \dfrac{\mathrm{sgn}(\varepsilon_{1}) \me^{
-(\widetilde{a}_{+} \vert t \vert -\widehat{c}_{+})} \widehat{\gimel}_{
+}}{\widetilde{b} \sqrt{\vert t \vert}} \sinh \! \left(\widetilde{a}_{-}
\vert t \vert \! - \! \widehat{c}_{-} \! + \! \tfrac{1}{8} \ln \! \left(
\tfrac{4-a_{1}^{2}}{4-a_{2}^{2}} \right) \! - \! \ln \widehat{\gimel}_{-}
\right) \nonumber \\
 &+\mathcal{O} \! \left( \dfrac{\underline{c}(z_{o}) \me^{-4
\alpha \vert t \vert}}{\beta t} \right), \\
\int\nolimits_{-\infty}^{x}(\vert u(\xi,t) \vert^{2} \! - \! 1) \, \md
\xi \! &= \! -\psi^{-}(0) \! - \! \dfrac{\mathrm{sgn}(\varepsilon_{1})
\me^{-(\widetilde{a}_{+} \vert t \vert -\widehat{c}_{+})} \widehat{
\gimel}_{+}}{\widetilde{b} \sqrt{\vert t \vert}} \sinh \! \left(
\widetilde{a}_{-} \vert t \vert \! - \! \widehat{c}_{-} \! + \! \tfrac{
1}{8} \ln \! \left( \tfrac{4-a_{1}^{2}}{4-a_{2}^{2}} \right) \! - \! \ln
\widehat{\gimel}_{-} \right) \nonumber \\
 &+\mathcal{O} \! \left( \dfrac{\underline{c}(z_{o}) \me^{-4
\alpha \vert t \vert}}{\beta t} \right).
\end{align}
\end{ddddd}
\begin{ddddd}
For $r(\zeta) \! \in \! \mathcal{S}_{\mathbb{C}}(\mathbb{R}) \cap
\{\mathstrut h(z); \, \vert \vert h(\cdot) \vert \vert_{\mathcal{L}^{
\infty}(\mathbb{R})} \! := \! \sup_{z \in \mathbb{R}} \vert h(z) \vert \! <
\! 1\}$, let $m(x,t;\zeta)$ be the solution of the Riemann-Hilbert problem
formulated in Lemma~{\rm 2.5}. Let $u(x,t)$, the solution of the Cauchy
problem for the {\rm D${}_{f}$NLSE} with finite-density initial data $u(x,0)
\! := \! u_{o}(x) \! =_{x \to \pm \infty} \! u_{o}(\pm \infty)(1 \! + \!
o(1))$, where $u_{o}(\pm \infty) \! := \! \exp (\tfrac{\mi (1 \mp 1) \theta}
{2})$, $0 \! \leqslant \! \theta \! = \! -\int_{-\infty}^{+\infty} \tfrac{
\ln (1-\vert r(\mu) \vert^{2})}{\mu} \, \tfrac{\md \mu}{2 \pi} \! < \! 2
\pi$, $u_{o}(x) \! \in \! \mathbf{C}^{\infty}(\mathbb{R})$, and $u_{o}(x)
\! - \! u_{o}(\pm \infty) \! \in \! \mathcal{S}_{\mathbb{C}}(\mathbb{R}_{
\pm})$, be defined by Eq.~{\rm (5)}. Set $s_{1} \! := \! \exp (\tfrac{\mi
\pi}{4})$ and $s_{2} \! := \! \exp (\tfrac{3 \pi \mi}{4})$. Then: (i) for
$r(s_{1}) \! = \! \exp (-\tfrac{\mi \varepsilon_{1} \pi}{2}) \vert r(s_{1})
\vert$, $\varepsilon_{1} \! \in \! \{\pm 1\}$, $r(\overline{s_{2}}) \! = \!
\exp (\tfrac{\mi \varepsilon_{2} \pi}{2}) \vert r(\overline{s_{2}}) \vert$,
$\varepsilon_{2} \! \in \! \{\pm 1\}$, $0 \! < \! r(s_{2}) \overline{r
(\overline{s_{2}})} \! < \! 1$, and $\varepsilon_{1} \! = \! \varepsilon_{
2}$, as $t \! \to \! +\infty$ and $x \! \to \! -\infty$ such that $z_{o}
\! := \! x/t \! \to \! 0^{-}$,
\begin{gather}
u(x,t) \! = \! \me^{-\mi \psi^{+}(1)} \! \left(1 \! + \! \dfrac{\me^{\frac{
\mi \pi}{2}(\varepsilon_{1}+\frac{3}{2})} \me^{-(2t+\widetilde{
\mathfrak{c}}_{+})} \mathfrak{b}_{+}}{2 \sqrt{t}} \sinh \! \left(
\widetilde{\mathfrak{c}}_{-} \! - \! \ln \mathfrak{b}_{-} \right) \! +
\! \mathcal{O} \! \left( \dfrac{\underline{c} \, \me^{-4t}}{t} \right)
\right),
\end{gather}
and, for $\varepsilon_{1} \! = \! -\varepsilon_{2}$,
\begin{gather}
u(x,t) \! = \! \me^{-\mi \psi^{+}(1)} \! \left(1 \! - \! \dfrac{\me^{
\frac{\mi \pi}{2}(\varepsilon_{1}+\frac{3}{2})} \me^{-(2t+
\widetilde{\mathfrak{c}}_{+})} \mathfrak{b}_{+}}{2 \sqrt{t}}
\cosh \! \left(\widetilde{\mathfrak{c}}_{-} \! - \! \ln \mathfrak{
b}_{-} \right) \! + \! \mathcal{O} \! \left( \dfrac{\underline{c} \,
\me^{-4t}}{t} \right) \right),
\end{gather}
where
\begin{gather}
\widetilde{\mathfrak{c}}_{\pm} \! := \! \dfrac{1}{\sqrt{2}} \int\nolimits_{
-\infty}^{0} \dfrac{\ln (1 \! - \! \vert r(\mu) \vert^{2})}{(\sqrt{2} \,
\mu \! - \! 1)^{2} \! + \! 1} \, \dfrac{\md \mu}{\pi} \mp \dfrac{1}{\sqrt{
2}} \int\nolimits_{-\infty}^{0} \dfrac{\ln (1 \! - \! \vert r(\mu) \vert^{
2})}{(\sqrt{2} \, \mu \! + \! 1)^{2} \! + \! 1} \, \dfrac{\md \mu}{\pi}, \\
\mathfrak{b}_{+} \! := \! \left( \dfrac{\vert r(s_{1}) \vert \vert r(\overline{
s_{2}}) \vert}{(1 \! - \! r(s_{2}) \overline{r(\overline{s_{2}})})} \right)^{
1/2}, \qquad \, \, \mathfrak{b}_{-} \! := \! \left(\dfrac{\vert r(s_{1}) \vert
(1 \! - \! r(s_{2}) \overline{r(\overline{s_{2}})})}{\vert r(\overline{s_{
2}}) \vert} \right)^{1/2},
\end{gather}
and $\psi^{+}(\cdot)$ is defined in Theorem~{\rm 3.2},
Eq.~{\rm (35);} (ii) for $r(\overline{s_{1}}) \! = \! \exp (\tfrac{\mi
\varepsilon_{1} \pi}{2}) \vert r(\overline{s_{1}}) \vert$, $\varepsilon_{
1} \! \in \! \{\pm 1\}$, $r(s_{2}) \! = \! \exp (-\tfrac{\mi \varepsilon_{2}
\pi}{2}) \vert r(s_{2}) \vert$, $\varepsilon_{2} \! \in \! \{\pm 1\}$, $0 \!
< \! r(s_{1}) \overline{r(\overline{s_{1}})} \! < \! 1$, and $\varepsilon_{1}
\! = \! \varepsilon_{2}$, as $t \! \to \! -\infty$ and $x \! \to \! +\infty$
such that $z_{o} \! \to \! 0^{-}$,
\begin{gather}
u(x,t) \! = \! \me^{-\mi \psi^{-}(1)} \! \left(1 \! - \! \dfrac{\me^{\frac{
\mi \pi}{2}(\varepsilon_{1}+\frac{1}{2})} \me^{-(2 \vert t \vert
-\widehat{\mathfrak{c}}_{+})} \mathfrak{d}_{+}}{2 \sqrt{\vert t
\vert}} \sinh \! \left( \widehat{\mathfrak{c}}_{-} \! - \! \ln \mathfrak{
d}_{-} \right) \! + \! \mathcal{O} \! \left( \dfrac{\underline{c} \,
\me^{-4 \vert t \vert}}{t} \right) \right),
\end{gather}
and, for $\varepsilon_{1} \! = \! -\varepsilon_{2}$,
\begin{gather}
u(x,t) \! = \! \me^{-\mi \psi^{-}(1)} \! \left(1 \! + \! \dfrac{\me^{
\frac{\mi \pi}{2}(\varepsilon_{1}-\frac{3}{2})} \me^{-(2 \vert t
\vert -\widehat{\mathfrak{c}}_{+})} \mathfrak{d}_{+}}{2 \sqrt{
\vert t \vert}} \cosh \! \left( \widehat{\mathfrak{c}}_{-} \! - \!
\ln \mathfrak{d}_{-} \right) \! + \! \mathcal{O} \! \left( \dfrac{
\underline{c} \, \me^{-4 \vert t \vert}}{t} \right) \right),
\end{gather}
where
\begin{gather}
\widehat{\mathfrak{c}}_{\pm} \! := \! \dfrac{1}{\sqrt{2}} \int\nolimits_{
0}^{+\infty} \dfrac{\ln (1 \! - \! \vert r(\mu) \vert^{2})}{(\sqrt{2} \, \mu
\! - \! 1)^{2} \! + \! 1} \, \dfrac{\md \mu}{\pi} \mp \dfrac{1}{\sqrt{2}}
\int\nolimits_{0}^{+\infty} \dfrac{\ln (1 \! - \! \vert r(\mu) \vert^{2})}{
(\sqrt{2} \, \mu \! + \! 1)^{2} \! + \! 1} \, \dfrac{\md \mu}{\pi}, \\
\mathfrak{d}_{+} \! := \! \left( \dfrac{\vert r(\overline{s_{1}}) \vert
\vert r(s_{2}) \vert}{(1 \! - \! r(s_{1}) \overline{r(\overline{s_{1}})})}
\right)^{1/2}, \qquad \, \, \mathfrak{d}_{-} \! := \! \left(\dfrac{\vert
r(s_{2}) \vert (1 \! - \! r(s_{1}) \overline{r(\overline{s_{1}})})}{\vert
r(\overline{s_{1}}) \vert} \right)^{1/2},
\end{gather}
and $\psi^{-}(\cdot)$ is defined in Theorem~{\rm 3.2},
Eq.~{\rm (46);} (iii) for $r(\overline{s_{2}}) \! = \! \exp (-\tfrac{\mi
\varepsilon_{1} \pi}{2}) \vert r(\overline{s_{2}}) \vert$, $\varepsilon_{
1} \! \in \! \{\pm 1\}$, $r(s_{1}) \! = \! \exp (\tfrac{\mi \varepsilon_{2}
\pi}{2}) \vert r(s_{1}) \vert$, $\varepsilon_{2} \! \in \! \{\pm 1\}$, $0 \!
< \! r(s_{1}) \overline{r(\overline{s_{1}})} \! < \! 1$, and $\varepsilon_{
1} \! = \! \varepsilon_{2}$, as $t \! \to \! +\infty$ and $x \! \to \!
+\infty$ such that $z_{o} \! \to \! 0^{+}$,
\begin{gather}
u(x,t) \! = \! -\me^{-\mi \psi^{+}(1)} \! \left(1 \! + \! \dfrac{\me^{
\frac{\mi \pi}{2}(\varepsilon_{1}+\frac{3}{2})} \me^{-(2t+\widetilde{
\mathfrak{c}}_{+})} \widetilde{\mathfrak{g}}_{+}}{2 \sqrt{t}}
\sinh \! \left( \widetilde{\mathfrak{c}}_{-} \! + \! \ln \widetilde{
\mathfrak{g}}_{-} \right) \! + \! \mathcal{O} \! \left( \dfrac{
\underline{c} \, \me^{-4t}}{t} \right) \right),
\end{gather}
and, for $\varepsilon_{1} \! = \! -\varepsilon_{2}$,
\begin{gather}
u(x,t) \! = \! -\me^{-\mi \psi^{+}(1)} \! \left(1 \! - \! \dfrac{\me^{
\frac{\mi \pi}{2}(\varepsilon_{1}-\frac{1}{2})} \me^{-(2t+\widetilde{
\mathfrak{c}}_{+})} \widetilde{\mathfrak{g}}_{+}}{2 \sqrt{t}}
\cosh \! \left(\widetilde{\mathfrak{c}}_{-} \! + \! \ln \widetilde{
\mathfrak{g}}_{-} \right) \! + \! \mathcal{O} \! \left( \dfrac{
\underline{c} \, \me^{-4t}}{t} \right) \right),
\end{gather}
where
\begin{gather}
\widetilde{\mathfrak{g}}_{+} \! := \! \left(\dfrac{\vert r(\overline{s_{
2}}) \vert \vert r(s_{1}) \vert}{(1 \! - \! r(s_{1}) \overline{r(\overline{
s_{1}})})} \right)^{1/2}, \qquad \, \, \widetilde{\mathfrak{g}}_{-} \! :=
\! \left(\dfrac{\vert r(\overline{s_{2}}) \vert (1 \! - \! r(s_{1}) \overline{
r(\overline{s_{1}})})}{\vert r(s_{1}) \vert} \right)^{1/2};
\end{gather}
and (iv) for $r(s_{2}) \! = \! \exp (\tfrac{\mi \varepsilon_{1} \pi}{2})
\vert r(s_{2}) \vert$, $\varepsilon_{1} \! \in \! \{\pm 1\}$, $r(\overline{
s_{1}}) \! = \! \exp (-\tfrac{\mi \varepsilon_{2} \pi}{2}) \vert r(\overline{
s_{1}}) \vert$, $\varepsilon_{2} \! \in \! \{\pm 1\}$, $0 \! < \! r(s_{2})
\overline{r(\overline{s_{2}})} \! < \! 1$, and $\varepsilon_{1} \! = \!
\varepsilon_{2}$, as $t \! \to \! -\infty$ and $x \! \to \! -\infty$ such that
$z_{o} \! \to \! 0^{+}$,
\begin{gather}
u(x,t) \! = \! -\me^{-\mi \psi^{-}(1)} \! \left(1 \! - \! \dfrac{\me^{
\frac{\mi \pi}{2}(\varepsilon_{1}+\frac{1}{2})} \me^{-(2 \vert t
\vert -\widehat{\mathfrak{c}}_{+})} \widehat{\mathfrak{g}}_{+}}
{2 \sqrt{\vert t \vert}} \sinh \! \left( \widehat{\mathfrak{c}}_{-} \!
+ \! \ln \widehat{\mathfrak{g}}_{-} \right) \! + \! \mathcal{O}
\! \left( \dfrac{\underline{c} \, \me^{-4 \vert t \vert}}{t} \right)
\right),
\end{gather}
and, for $\varepsilon_{1} \! = \! -\varepsilon_{2}$,
\begin{gather}
u(x,t) \! = \! -\me^{-\mi \psi^{-}(1)} \! \left(1 \! + \! \dfrac{\me^{
\frac{\mi \pi}{2}(\varepsilon_{1}+\frac{1}{2})} \me^{-(2 \vert t
\vert -\widehat{\mathfrak{c}}_{+})} \widehat{\mathfrak{g}}_{
+}}{2 \sqrt{\vert t \vert}} \cosh \! \left(\widehat{\mathfrak{c}}_{
-} \! + \! \ln \widehat{\mathfrak{g}}_{-} \right) \! + \! \mathcal{O}
\! \left( \dfrac{\underline{c} \, \me^{-4 \vert t \vert}}{t} \right)
\right),
\end{gather}
where
\begin{gather}
\widehat{\mathfrak{g}}_{+} \! := \! \left(\dfrac{\vert r(s_{2}) \vert \vert
r(\overline{s_{1}}) \vert}{(1 \! - \! r(s_{2}) \overline{r(\overline{s_{
2}})})} \right)^{1/2}, \qquad \, \, \widehat{\mathfrak{g}}_{-} \! := \!
\left(\dfrac{\vert r(\overline{s_{1}}) \vert (1 \! - \! r(s_{2}) \overline{
r(\overline{s_{2}})})}{\vert r(s_{2}) \vert} \right)^{1/2}.
\end{gather}
For $u(x,t)$ as defined and given above, let $\int_{+\infty}^{x}
(\vert u(\xi,t) \vert^{2} \! - \! 1) \, \md \xi$ be defined by Eq.~{\rm (6)}.
Then: (i) for $r(s_{1}) \! = \! \exp (-\tfrac{\mi \varepsilon_{1} \pi}{2})
\vert r(s_{1}) \vert$, $\varepsilon_{1} \! \in \{\pm 1\}$, $r(\overline{s_{
2}}) \! = \! \exp (\tfrac{\mi \varepsilon_{2} \pi}{2}) \vert r(\overline{s_{
2}}) \vert$, $\varepsilon_{2} \! \in \! \{\pm 1\}$, $0 \! < \! r(s_{2})
\overline{r(\overline{s_{2}})} \! < \! 1$, and $\varepsilon_{1} \! = \!
\varepsilon_{2}$, as $t \! \to \! +\infty$ and $x \! \to \! -\infty$ such
that $z_{o} \! \to \! 0^{-}$,
\begin{align}
\int\nolimits_{+\infty}^{x}(\vert u(\xi,t) \vert^{2} \! - \! 1) \, \md
\xi \! &= \! \psi^{+}(0) \!  - \! \dfrac{\mathrm{sgn}(\varepsilon_{
1}) \me^{-(2t+\widetilde{\mathfrak{c}}_{+})} \mathfrak{b}_{+
}}{2 \sqrt{t}} \cosh \! \left( \widetilde{\mathfrak{c}}_{-} \! - \!
\ln \mathfrak{b}_{-} \right) \! + \! \mathcal{O} \! \left( \dfrac{
\underline{c} \, \me^{-4t}}{t} \right), \\
\int\nolimits_{-\infty}^{x}(\vert u(\xi,t) \vert^{2} \! - \! 1) \, \md
\xi \! &= \! -\psi^{-}(0) \! - \! \dfrac{\mathrm{sgn}(\varepsilon_{
1}) \me^{-(2t+\widetilde{\mathfrak{c}}_{+})} \mathfrak{b}_{+
}}{2 \sqrt{t}} \cosh \! \left( \widetilde{\mathfrak{c}}_{-} \! - \!
\ln \mathfrak{b}_{-} \right) \! + \! \mathcal{O} \! \left( \dfrac{
\underline{c} \, \me^{-4t}}{t} \right),
\end{align}
and, for $\varepsilon_{1} \! = \! -\varepsilon_{2}$,
\begin{align}
\int\nolimits_{+\infty}^{x}(\vert u(\xi,t) \vert^{2} \! - \! 1) \, \md
\xi \! &= \! \psi^{+}(0) \!  + \! \dfrac{\mathrm{sgn}(\varepsilon_{
1}) \me^{-(2t+\widetilde{\mathfrak{c}}_{+})} \mathfrak{b}_{+
}}{2 \sqrt{t}} \sinh \! \left( \widetilde{\mathfrak{c}}_{-} \! - \!
\ln \mathfrak{b}_{-} \right) \! + \! \mathcal{O} \! \left( \dfrac{
\underline{c} \, \me^{-4t}}{t} \right), \\
\int\nolimits_{-\infty}^{x}(\vert u(\xi,t) \vert^{2} \! - \! 1) \, \md
\xi \! &= \! -\psi^{-}(0) \! + \! \dfrac{\mathrm{sgn}(\varepsilon_{
1}) \me^{-(2t+\widetilde{\mathfrak{c}}_{+})} \mathfrak{b}_{+
}}{2 \sqrt{t}} \sinh \! \left( \widetilde{\mathfrak{c}}_{-} \! - \!
\ln \mathfrak{b}_{-} \right) \! + \! \mathcal{O} \! \left( \dfrac{
\underline{c} \, \me^{-4t}}{t} \right);
\end{align}
(ii) for $r(\overline{s_{1}}) \! = \! \exp (\tfrac{\mi \varepsilon_{1}
\pi}{2}) \vert r(\overline{s_{1}}) \vert$, $\varepsilon_{1} \! \in \!
\{\pm 1\}$, $r(s_{2}) \! = \! \exp (-\tfrac{\mi \varepsilon_{2} \pi}{2})
\vert r(s_{2}) \vert$, $\varepsilon_{2} \! \in \! \{\pm 1\}$, $0 \! < \!
r(s_{1}) \overline{r(\overline{s_{1}})} \! < \! 1$, and $\varepsilon_{
1} \! = \! \varepsilon_{2}$, as $t \! \to \! -\infty$ and $x \! \to \!
+\infty$ such that $z_{o} \! \to \! 0^{-}$,
\begin{align}
\int\nolimits_{+\infty}^{x}(\vert u(\xi,t) \vert^{2} \! - \! 1) \, \md
\xi \! &= \! \psi^{-}(0) \! + \! \dfrac{\mathrm{sgn}(\varepsilon_{
1}) \me^{-(2 \vert t \vert-\widehat{\mathfrak{c}}_{+})} \mathfrak{
d}_{+}}{2 \sqrt{\vert t \vert}} \cosh \! \left( \widehat{\mathfrak{
c}}_{-} \! - \! \ln \mathfrak{d}_{-} \right) \! + \! \mathcal{O} \!
\left( \dfrac{\underline{c} \, \me^{-4 \vert t \vert}}{t} \right), \\
\int\nolimits_{-\infty}^{x}(\vert u(\xi,t) \vert^{2} \! - \! 1) \, \md
\xi \! &= \! -\psi^{+}(0) \! + \! \dfrac{\mathrm{sgn}(\varepsilon_{
1}) \me^{-(2 \vert t \vert -\widehat{\mathfrak{c}}_{+})} \mathfrak{
d}_{+}}{2 \sqrt{\vert t \vert}} \cosh \! \left( \widehat{\mathfrak{
c}}_{-} \! - \! \ln \mathfrak{d}_{-} \right) \! + \! \mathcal{O} \!
\left( \dfrac{\underline{c} \, \me^{-4 \vert t \vert}}{t} \right),
\end{align}
and, for $\varepsilon_{1} \! = \! -\varepsilon_{2}$,
\begin{align}
\int\nolimits_{+\infty}^{x}(\vert u(\xi,t) \vert^{2} \! - \! 1) \, \md
\xi \! &= \! \psi^{-}(0) \! + \! \dfrac{\mathrm{sgn}(\varepsilon_{
1}) \me^{-(2 \vert t \vert-\widehat{\mathfrak{c}}_{+})} \mathfrak{
d}_{+}}{2 \sqrt{\vert t \vert}} \sinh \! \left( \widehat{\mathfrak{
c}}_{-} \! - \! \ln \mathfrak{d}_{-} \right) \! + \! \mathcal{O} \!
\left( \dfrac{\underline{c} \, \me^{-4 \vert t \vert}}{t} \right), \\
\int\nolimits_{-\infty}^{x}(\vert u(\xi,t) \vert^{2} \! - \! 1) \, \md
\xi \! &= \! -\psi^{+}(0) \! + \! \dfrac{\mathrm{sgn}(\varepsilon_{
1}) \me^{-(2 \vert t \vert -\widehat{\mathfrak{c}}_{+})} \mathfrak{
d}_{+}}{2 \sqrt{\vert t \vert}} \sinh \! \left( \widehat{\mathfrak{
c}}_{-} \! - \! \ln \mathfrak{d}_{-} \right) \! + \! \mathcal{O} \!
\left( \dfrac{\underline{c} \, \me^{-4 \vert t \vert}}{t} \right);
\end{align}
(iii) for $r(\overline{s_{2}}) \! = \! \exp (-\tfrac{\mi \varepsilon_{1}
\pi}{2}) \vert r(\overline{s_{2}}) \vert$, $\varepsilon_{1} \! \in \! \{
\pm 1\}$, $r(s_{1}) \! = \! \exp (\tfrac{\mi \varepsilon_{2} \pi}{2})
\vert r(s_{1}) \vert$, $\varepsilon_{2} \! \in \! \{\pm 1\}$, $0 \! < \!
r(s_{1}) \overline{r(\overline{s_{1}})} \! < \! 1$, and $\varepsilon_{
1} \! = \! \varepsilon_{2}$, as $t \! \to \! +\infty$ and $x \! \to \!
+\infty$ such that $z_{o} \! \to \! 0^{+}$,
\begin{align}
\int\nolimits_{+\infty}^{x}(\vert u(\xi,t) \vert^{2} \! - \! 1) \, \md
\xi \! &= \! \psi^{-}(0) \! + \! \dfrac{\mathrm{sgn}(\varepsilon_{
1}) \me^{-(2t+\widetilde{\mathfrak{c}}_{+})} \widetilde{
\mathfrak{g}}_{+}}{2 \sqrt{t}} \cosh \! \left( \widetilde{
\mathfrak{c}}_{-} \! + \! \ln \widetilde{\mathfrak{g}}_{-}
\right) \! + \! \mathcal{O} \! \left( \dfrac{\underline{c} \,
\me^{-4t}}{t} \right), \\
\int\nolimits_{-\infty}^{x}(\vert u(\xi,t) \vert^{2} \! - \! 1) \, \md
\xi \! &= \! -\psi^{+}(0) \! + \! \dfrac{\mathrm{sgn}(\varepsilon_{
1}) \me^{-(2t+\widetilde{\mathfrak{c}}_{+})} \widetilde{
\mathfrak{g}}_{+}}{2 \sqrt{t}} \cosh \! \left( \widetilde{
\mathfrak{c}}_{-} \! + \! \ln \widetilde{\mathfrak{g}}_{-}
\right) \! + \! \mathcal{O} \! \left( \dfrac{\underline{c} \,
\me^{-4t}}{t} \right),
\end{align}
and, for $\varepsilon_{1} \! = \! -\varepsilon_{2}$,
\begin{align}
\int\nolimits_{+\infty}^{x}(\vert u(\xi,t) \vert^{2} \! - \! 1) \, \md
\xi \! &= \! \psi^{-}(0) \! + \! \dfrac{\mathrm{sgn}(\varepsilon_{
1}) \me^{-(2t+\widetilde{\mathfrak{c}}_{+})} \widetilde{
\mathfrak{g}}_{+}}{2 \sqrt{t}} \sinh \! \left( \widetilde{
\mathfrak{c}}_{-} \! + \! \ln \widetilde{\mathfrak{g}}_{-}
\right) \! + \! \mathcal{O} \! \left( \dfrac{\underline{c} \,
\me^{-4t}}{t} \right), \\
\int\nolimits_{-\infty}^{x}(\vert u(\xi,t) \vert^{2} \! - \! 1) \, \md
\xi \! &= \! -\psi^{+}(0) \! + \! \dfrac{\mathrm{sgn}(\varepsilon_{
1}) \me^{-(2t+\widetilde{\mathfrak{c}}_{+})} \widetilde{
\mathfrak{g}}_{+}}{2 \sqrt{t}} \sinh \! \left( \widetilde{
\mathfrak{c}}_{-} \! + \! \ln \widetilde{\mathfrak{g}}_{-}
\right) \! + \! \mathcal{O} \! \left( \dfrac{\underline{c} \,
\me^{-4t}}{t} \right);
\end{align}
and (iv) for $r(s_{2}) \! = \! \exp (\tfrac{\mi \varepsilon_{1} \pi}
{2}) \vert r(s_{2}) \vert$, $\varepsilon_{1} \! \in \! \{\pm 1\}$,
$r(\overline{s_{1}}) \! = \! \exp (-\tfrac{\mi \varepsilon_{2} \pi}
{2}) \vert r(\overline{s_{1}}) \vert$, $\varepsilon_{2} \! \in \!
\{\pm 1\}$, $0 \! < \! r(s_{2}) \overline{r(\overline{s_{2}})} \! <
\! 1$, and $\varepsilon_{1} \! = \! \varepsilon_{2}$, as $t \! \to
\! -\infty$ and $x \! \to \! -\infty$ such that $z_{o} \! \to \! 0^{+}$,
\begin{align}
\int\nolimits_{+\infty}^{x}(\vert u(\xi,t) \vert^{2} \! - \! 1) \, \md
\xi \! &= \! \psi^{+}(0) \! - \! \dfrac{\mathrm{sgn}(\varepsilon_{
1}) \me^{-(2 \vert t \vert -\widehat{\mathfrak{c}}_{+})}
\widehat{\mathfrak{g}}_{+}}{2 \sqrt{\vert t \vert}} \cosh \!
\left( \widehat{\mathfrak{c}}_{-} \! + \! \ln \widehat{\mathfrak{
g}}_{-} \right) \! + \! \mathcal{O} \! \left( \dfrac{\underline{c}
\, \me^{-4 \vert t \vert}}{t} \right), \\
\int\nolimits_{-\infty}^{x}(\vert u(\xi,t) \vert^{2} \! - \! 1) \, \md
\xi \! &= \! -\psi^{-}(0) \! - \! \dfrac{\mathrm{sgn}(\varepsilon_{
1}) \me^{-(2 \vert t \vert -\widehat{\mathfrak{c}}_{+})}
\widehat{\mathfrak{g}}_{+}}{2 \sqrt{\vert t \vert}} \cosh \!
\left( \widehat{\mathfrak{c}}_{-} \! + \! \ln \widehat{\mathfrak{
g}}_{-} \right) \! + \! \mathcal{O} \! \left( \dfrac{\underline{c}
\, \me^{-4 \vert t \vert}}{t} \right),
\end{align}
and, for $\varepsilon_{1} \! = \! -\varepsilon_{2}$,
\begin{align}
\int\nolimits_{+\infty}^{x}(\vert u(\xi,t) \vert^{2} \! - \! 1) \, \md
\xi \! &= \! \psi^{+}(0) \! + \! \dfrac{\mathrm{sgn}(\varepsilon_{
1}) \me^{-(2 \vert t \vert -\widehat{\mathfrak{c}}_{+})}
\widehat{\mathfrak{g}}_{+}}{2 \sqrt{\vert t \vert}} \sinh \!
\left( \widehat{\mathfrak{c}}_{-} \! + \! \ln \widehat{\mathfrak{
g}}_{-} \right) \! + \! \mathcal{O} \! \left( \dfrac{\underline{
c} \, \me^{-4 \vert t \vert}}{t} \right), \\
\int\nolimits_{-\infty}^{x}(\vert u(\xi,t) \vert^{2} \! - \! 1) \, \md
\xi \! &= \! -\psi^{-}(0) \! + \! \dfrac{\mathrm{sgn}(\varepsilon_{
1}) \me^{-(2 \vert t \vert -\widehat{\mathfrak{c}}_{+})}
\widehat{\mathfrak{g}}_{+}}{2 \sqrt{\vert t \vert}} \sinh \!
\left(\widehat{\mathfrak{c}}_{-} \! + \! \ln \widehat{\mathfrak{
g}}_{-} \right) \! + \! \mathcal{O} \! \left( \dfrac{\underline{
c} \, \me^{-4 \vert t \vert}}{t} \right).
\end{align}
\end{ddddd}
\begin{eeeee}
In this work, the complete details of the analysis are presented for the
case $t \! \to \! +\infty$ and $x \! \to \! -\infty$ such that $z_{o} \! :=
\! x/t \! < \! -2$, and the case $t \! \to \! -\infty$ and $x \! \to \!
+\infty$ such that $z_{o} \! < \! -2$ is succinctly treated in Section 7: the
remaining cases are analogous (see the Appendix). The DZ method has recently
been extended to tackle asymptotic problems arising in the theory of random
permutations \cite{a41}, orthogonal polynomials and random matrix theory
\cite{a34,a42}, and perturbation theory for integrable NLEEs \cite{a43}
(see, also, the recent extension by Kamvissis~\emph{et al.} \cite{a44}).
\end{eeeee}
\section{The Auxiliary and Truncated RHPs}
In this section, as $t \! \to \! +\infty$ and $x \! \to \! -\infty$ such that
$z_{o} \! := \! x/t \! < \! -2$, the RHP formulated in Lemma~2.6 for $m^{c}
(x,t;\zeta)$ on $\sigma_{c}$ (oriented {}from $-\infty$ to $+\infty)$ is
reformulated as an auxiliary RHP on the augmented contour $\Sigma^{\prime}$
(see Figure~3), which is then dissected to produce an equivalent RHP on the
truncated contour $\Sigma^{\sharp}$ (see Figure~4).
\begin{eeeee}
For notational convenience, except where absolutely necessary and/or where
confusion may arise, explicit $x,t$ dependences are suppressed.
\end{eeeee}

As per the DZ method \cite{a27}, one begins by decomposing the complex plane
of the spectral parameter $\zeta$ according to the signature of $\Re (\mi t
\theta^{u}(\zeta))$ (see Figure~1), where, {}from Eq.~(8), $\theta^{u}(\zeta)
\! = \! \tfrac{1}{2}(\zeta \! - \! \tfrac{1}{\zeta})(z_{o} \! + \! \zeta \! +
\! \tfrac{1}{\zeta})$, with $\{\zeta_{i}\}_{i=1}^{4}$ defined in Theorem~3.1,
Eqs.~(16) and~(17), $0 \! < \! \zeta_{2} \! < \! \zeta_{1}$, $\vert \zeta_{3}
\vert^{2} \! = \! 1$, and $\pm \! \leftrightarrow \! \Re (\mi t \theta^{u}
(\zeta)) \! \gtrless \! 0$. One now reorients $\sigma_{c}$ (oriented {}from
$-\infty$ to $+\infty)$ according to, and consistent with, the signature of
$\Re (\mi t \theta^{u}(\zeta))$, leading to the reoriented contour $\sigma_{
c}^{\prime}$ (see Figure~2). Denoting $m^{c}(\zeta)$ on $\sigma_{c}^{\prime}$
by $\widetilde{m}^{c}(\zeta)$, one shows that $\widetilde{m}^{c}(\zeta)
\colon \mathbb{C} \setminus \sigma_{c}^{\prime} \! \to \! \mathrm{SL}(2,
\mathbb{C})$ solves the following (normalised at $\infty)$ RHP: (1)
$\widetilde{m}^{c}(\zeta)$ is piecewise holomorphic $\forall \, \zeta \! \in
\! \mathbb{C} \setminus \sigma_{c}^{\prime}$; (2) $\widetilde{m}^{c}_{\pm}
(\zeta) \! := \! \lim_{\genfrac{}{}{0pt}{2}{\zeta^{\prime} \, \to \, \zeta}
{\zeta^{\prime} \, \in \, \pm \, \mathrm{side} \, \mathrm{of} \, \sigma_{c}^{
\prime}}} \widetilde{m}^{c}(\zeta^{\prime})$ satisfy the jump condition
$\widetilde{m}_{+}^{c}(\zeta) \! = \! \widetilde{m}_{-}^{c}(\zeta)
\widetilde{\mathcal{G}}^{c}(\zeta)$, $\zeta \! \in \! \mathbb{R}$, where
\begin{equation*}
\widetilde{\mathcal{G}}^{c}(\zeta) \! := \!
\begin{cases}
(\mathrm{I} \! - \! r(\zeta) \me^{2 \mi t \theta^{u}(\zeta)} \sigma_{-})
(\mathrm{I} \! + \! \overline{r(\overline{\zeta})} \, \me^{-2 \mi t \theta^{
u}(\zeta)} \sigma_{+}), &\text{$\zeta \! \in \! (0,\zeta_{2}) \cup (\zeta_{
1},+\infty)$,} \\
(\mathrm{I} \! - \! \overline{r(\overline{\zeta})} \, \me^{-2 \mi t \theta^{
u}(\zeta)} \sigma_{+})(\mathrm{I} \! + \! r(\zeta) \me^{2 \mi t \theta^{u}
(\zeta)} \sigma_{-}), &\text{$\zeta \! \in \! (-\infty,0) \cup (\zeta_{2},
\zeta_{1})$;}
\end{cases}
\end{equation*}
(3) as $\zeta \! \to \! \infty$, $\zeta \! \in \! \mathbb{C} \setminus
\sigma_{c}^{\prime}$, $\widetilde{m}^{c}(\zeta) \! = \! \mathrm{I} \! +
\! \mathcal{O}(\zeta^{-1})$; and (4) $\widetilde{m}^{c}(\zeta)$ satisfies
the symmetry reduction $\widetilde{m}^{c}(\zeta) \! = \! \sigma_{1}
\overline{\widetilde{m}^{c}(\overline{\zeta})} \, \sigma_{1}$ and the
condition $(\widetilde{m}^{c}(0) \sigma_{2})^{2} \! = \! \mathrm{I}$. One
notes {}from the definition of $\widetilde{\mathcal{G}}^{c}(\zeta)$ that,
as $t \! \to \! +\infty$ and $x \! \to \! -\infty$ such that $z_{o}\! <
\! -2$, for $\zeta \! \in \! \mathbb{C}_{\mp} \cap \{\mathstrut \zeta;
\, \Re (\zeta) \! \in \! (0,\zeta_{2}) \cup (\zeta_{1},+\infty)\} \cap
(\cup_{\star \in \{0,\zeta_{2},\zeta_{1}\}} \{\mathstrut \zeta; \, \vert
\zeta \! - \! \star \vert \! > \! \varepsilon \})$, where $\varepsilon$
is an arbitrarily fixed, sufficiently small positive real number, since
$\exp (\mp 2 \mi t \theta^{u}(\zeta))$ have analytic continuation to
$\zeta \! \mp \! \mi 0$, $\vert \exp (\mp 2 \mi t \theta^{u}(\zeta)) \vert
\! \to \! 0$, but, for $\zeta \! \in \! \mathbb{C}_{\pm} \cap \{\mathstrut
\zeta; \, \Re (\zeta) \! \in \! (-\infty,0) \cup (\zeta_{2},\zeta_{1})\}
\cap (\cup_{\star \in \{0,\zeta_{2},\zeta_{1}\}} \{\mathstrut \zeta; \,
\vert \zeta \! - \! \star \vert \! > \! \varepsilon\})$, since $\exp (\pm
2 \mi t \theta^{u}(\zeta))$ have analytic continuation to $\zeta \!
\pm \! \mi 0$, $\vert \exp (\pm 2 \mi t \theta^{u}(\zeta)) \vert \! \to
\! \infty$. In order to control the latter exponential growths, the
triangular factorisation of $\widetilde{\mathcal{G}}^{c}(\zeta)$ for
$\Re (\zeta) \! \in \! (-\infty,0) \cup (\zeta_{2},\zeta_{1})$ must be
changed {}from
$\left(
\begin{smallmatrix}
1 & \blacktriangle \\
0 & 1
\end{smallmatrix}
\right) \! \left(
\begin{smallmatrix}
1 & 0 \\
\blacktriangledown & 1
\end{smallmatrix}
\right)$ to $\left(
\begin{smallmatrix}
1 & 0 \\
\triangledown & 1
\end{smallmatrix}
\right) \! \left(
\begin{smallmatrix}
1 & \vartriangle \\
0 & 1
\end{smallmatrix}
\right)$, that is, upper-lower triangular to lower-upper triangular
refactorisation: to accomplish this, one introduces the ``$\delta
(\cdot)$-function'' \cite{a27}.
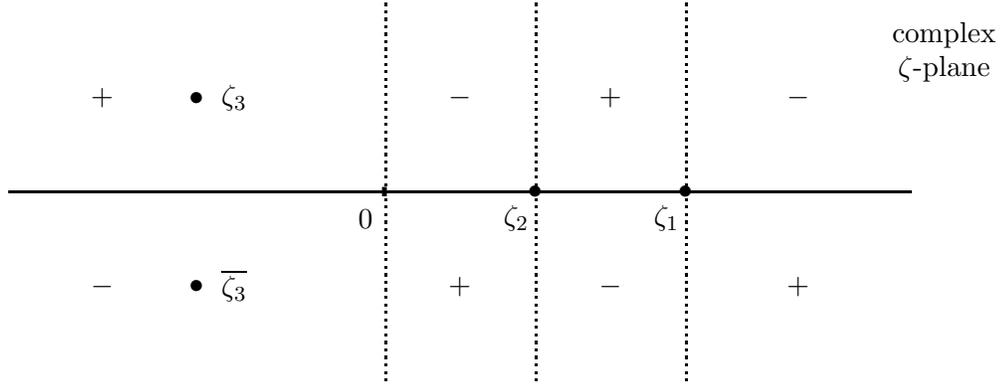
\begin{figure}[tbh]
\begin{center}
\unitlength=1cm
\vspace{0.65cm}
\begin{picture}(12,5)(0,0)
\thicklines
\put(5,2.55){\makebox(0,0){$\centerdot$}}
\put(5,2.60){\makebox(0,0){$\centerdot$}}
\put(7,2.5){\makebox(0,0){$\bullet$}}
\put(9,2.5){\makebox(0,0){$\bullet$}}
\put(2.5,1.25){\makebox(0,0){$\bullet$}}
\put(2.5,3.75){\makebox(0,0){$\bullet$}}
\put(4.75,2.15){\makebox(0,0){$0$}}
\put(6.75,2.15){\makebox(0,0){$\zeta_{2}$}}
\put(8.75,2.15){\makebox(0,0){$\zeta_{1}$}}
\put(3,1.25){\makebox(0,0){$\overline{\zeta_{3}}$}}
\put(3,3.75){\makebox(0,0){$\zeta_{3}$}}
\put(1.25,3.75){\makebox(0,0){$+$}}
\put(1.25,1.25){\makebox(0,0){$-$}}
\put(6,3.75){\makebox(0,0){$-$}}
\put(6,1.25){\makebox(0,0){$+$}}
\put(8,3.75){\makebox(0,0){$+$}}
\put(8,1.25){\makebox(0,0){$-$}}
\put(10.5,3.75){\makebox(0,0){$-$}}
\put(10.5,1.25){\makebox(0,0){$+$}}
\put(11.75,4.05){\shortstack[c]{complex\\$\zeta$-plane}}
\put(0,2.5){\line(1,0){12}}
\qbezier[50](5,0)(5,2.5)(5,5)
\qbezier[50](7,0)(7,2.5)(7,5)
\qbezier[50](9,0)(9,2.5)(9,5)
\end{picture}
\end{center}
\caption{Signature graph of $\Re (\mi t \theta^{u}(\zeta))$ as
$t \! \to \! +\infty$}
\end{figure}
\begin{figure}[htb]
\begin{center}
\unitlength=1cm
\begin{picture}(12,4)(0,0)
\thicklines
\put(5,2.05){\makebox(0,0){$\centerdot$}}
\put(5,2.10){\makebox(0,0){$\centerdot$}}
\put(7,2){\makebox(0,0){$\bullet$}}
\put(9,2){\makebox(0,0){$\bullet$}}
\put(5,1.65){\makebox(0,0){$0$}}
\put(7,1.65){\makebox(0,0){$\zeta_{2}$}}
\put(9,1.65){\makebox(0,0){$\zeta_{1}$}}
\put(0,2){\line(1,0){12}}
\put(0,2){\vector(1,0){2.5}}
\put(8,2){\vector(-1,0){2.0}}
\put(6,2){\vector(1,0){2.0}}
\put(12,2){\vector(-1,0){1.5}}
\put(11.5,2.70){\shortstack[c]{complex\\$\zeta$-plane}}
\end{picture}
\vspace{-1.10cm}
\end{center}
\caption{Reoriented contour $\sigma_{c}^{\prime}$}
\end{figure}
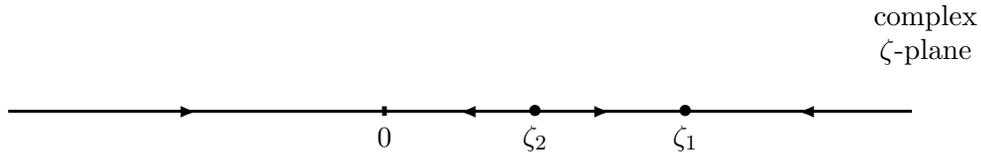
\begin{bbbbb}
Let $\delta (\zeta)$ solve the following scalar discontinuous {\rm RHP:}
\begin{align}
\delta_{+}(\zeta) &= \!
\begin{cases}
\delta_{-}(\zeta)(1 \! - \! r(\zeta) \overline{r(\overline{\zeta})}),
&\text{$\Re (\zeta) \! \in \! (-\infty,0) \cup (\zeta_{2},\zeta_{1})$,}
\\\delta_{-}(\zeta) \! = \! \delta (\zeta), &\text{$\Re (\zeta) \! \in \!
(0,\zeta_{2}) \cup (\zeta_{1},+\infty)$,}
\end{cases} \nonumber \\
\delta (\zeta) &\underset{\zeta \to \infty}{=} \! 1 \! + \! \mathcal{O}
(\zeta^{-1}). \nonumber
\end{align}
The unique solution of this {\rm RHP} can be written as
\begin{equation*}
\delta (\zeta) \! = \! \left( \dfrac{\zeta \! - \! \zeta_{1}}{\zeta \! - \!
\zeta_{2}} \right)^{\mi \nu} \exp \! \left(\int\nolimits_{-\infty}^{0}
\dfrac{\ln (1 \! - \! \vert r(\mu) \vert^{2})}{(\mu \! - \! \zeta)} \,
\dfrac{\md \mu}{2 \pi \mi} \! + \! \int\nolimits_{\zeta_{2}}^{\zeta_{1}} \ln
\! \left( \dfrac{1 \! - \! \vert r(\mu) \vert^{2}}{1 \! - \! \vert r (\zeta_{
1}) \vert^{2}} \right) \! \dfrac{1}{(\mu \! - \! \zeta)} \, \dfrac{\md \mu}
{2 \pi \mi} \right),
\end{equation*}
where $\{\zeta_{i}\}_{i=1}^{2}$ are defined in Theorem~{\rm 3.1},
Eqs.~{\rm (16)} and~{\rm (17)}, and $\nu \! := \! \nu (\zeta_{1}) \! = \!
-\tfrac{1}{2 \pi} \ln (1 \! - \! \vert r(\zeta_{1}) \vert^{2}) \! \in \!
\mathbb{R}_{+}$. Furthermore, the function $\delta (\zeta)$ possesses the
following properties, $\delta (\zeta) \overline{\delta (\overline{\zeta})}
\! = \! 1$, $\delta (\zeta) \delta (\tfrac{1}{\zeta}) \! = \! \delta (0)$,
$\vert \delta_{+}(\zeta) \vert^{2} \! \leqslant \! 1$ and $\vert \delta_{-}
(\zeta) \vert^{2} \! \leqslant \! (1 \! - \! \sup_{z \in \mathbb{R}} \vert
r(z) \vert^{2})^{-1} \! < \! \infty$ $\forall \, \zeta \! \in \! \mathbb{R}$,
and $\vert \vert (\delta (\cdot))^{\pm 1} \vert \vert_{\mathcal{L}^{\infty}
(\mathbb{C})} \! := \! \sup_{\zeta \in \mathbb{C}} \vert (\delta (\zeta))^{
\pm 1} \vert \! < \! \infty$.
\end{bbbbb}

\emph{Proof.} Noting that the index, $\kappa$, associated with the RHP for
$\delta (\zeta)$ stated in the Proposition is zero, namely, $\kappa \! := \!
\tfrac{1}{2 \pi} \! \left[ \arg (1 \! - \! \vert r(\zeta) \vert^{2}) \right]_{
-\infty}^{+\infty} \! = \! 0$, it follows {}from a well-known result
\cite{a45} (see, also, Theorem~A2 in \cite{a24}) that it can be solved
explicitly (and uniquely) to yield $\delta (\zeta) \! = \! \exp \! \left( \!
\left(\int_{-\infty}^{0} \! + \! \int_{\zeta_{2}}^{\zeta_{1}} \right) \!
\tfrac{\ln (1-\vert r(\mu) \vert^{2})}{(\mu-\zeta)} \, \tfrac{\md \mu}{2 \pi
\mi} \right)$: choosing the principal branch of $\ln (\cdot)$ as per item~(9)
of \textsc{Notational Conventions}, one arrives at the expression for $\delta
(\zeta)$ stated in the Proposition. Using the fact that $r(\zeta) \! \in
\! \mathcal{S}_{\mathbb{C}}^{1}(\mathbb{R})$, one shows {}from the
representation of $\delta (\zeta)$ that $\delta (\zeta) \! =_{\zeta \to
\infty} \! 1 \! + \! \mathcal{O}(\zeta^{-1})$; moreover, one deduces that
$\delta (\zeta)$ satisfies the symmetry reduction $\delta (\zeta) \overline{
\delta (\overline{\zeta})} \! = \! 1$. Using the fact that, for $\zeta \!
\in \! \mathbb{R}$, $r(\tfrac{1}{\zeta}) \! = \! -\overline{r(\overline{
\zeta})} \! = \! -\overline{r(\zeta)}$, it follows by a change-of-variable
argument that $\delta (\zeta) \delta (\tfrac{1}{\zeta}) \! = \! \delta (0)$,
where $\delta (0) \! = \! \exp \! \left( \! \left( \int_{-\infty}^{0} \! +
\! \int_{\zeta_{2}}^{\zeta_{1}} \right) \! \tfrac{\ln (1-\vert r(\mu) \vert^{
2})}{\mu} \, \tfrac{\md \mu}{2 \pi \mi} \right)$. Letting $\zeta \! \to \!
\zeta \! \pm \mi 0$, one notes {}from the above symmetry reduction for
$\delta (\zeta)$ that $\delta_{\pm} \overline{\delta_{\mp}(\zeta)} \! = \!
1$; hence, using the relation $\delta_{+}(\zeta) \! = \! \delta_{-}(\zeta)
(1 \! - \! r(\zeta) \overline{r(\overline{\zeta})})$, one deduces that
$\vert \delta_{-}(\zeta) \vert^{2} \! \leqslant \! (1 \! - \! \sup_{z
\in \mathbb{R}} \vert r(z) \vert^{2})^{-1} \! < \! \infty$ $\forall \,
\zeta \! \in \! \mathbb{R}$, {}from which one also shows that $\vert
\delta_{+}(\zeta) \vert^{2} \! \leqslant \! 1$. Setting $\zeta \! := \!
\xi \! + \! \mi \eta$, with $\eta \! \not= \! 0$, one shows {}from the
representation for $\delta (\zeta)$ that $\vert \delta (\zeta) \vert \! =
\! \exp \! \left( \! \left( \int_{-\infty}^{0} \! + \! \int_{\zeta_{2}}^{
\zeta_{1}} \right) \! \tfrac{\eta \ln (1-\vert r(\mu) \vert^{2})}{(\mu -
\xi)^{2}+ \eta^{2}} \, \tfrac{\md \mu}{2 \pi} \right)$, and, using the fact
that $r(\zeta) \! \in \! \mathcal{S}_{\mathbb{C}}^{1}(\mathbb{R})$, one shows
that $\vert \delta (\zeta) \vert \! \leqslant \! \exp \! \left( \tfrac{
\eta \sup_{z \in \mathbb{R}} \vert \ln (1-\vert r(z) \vert^{2}) \vert}{2
\pi} \int_{-\infty}^{+\infty} \tfrac{\md \mu}{(\mu -\xi)^{2}+ \eta^{2}}
\right)$: now, recalling that $\int \! \tfrac{\md x}{x^{2}+p^{2}} \! = \!
\tfrac{1}{p} \arctan (x/p)$, and choosing the principal branch of $\arctan
(\cdot)$, one shows that, $\exists \, \widetilde{M} \! \in \! \mathbb{R}_{
+}$ and bounded such that $\forall \, \zeta \! \in \! \mathbb{C} \setminus
\mathbb{R}$, $\vert \delta (\zeta) \vert \! \leqslant \! \widetilde{M}$ (a
similar argument shows that $\vert (\delta (\zeta))^{-1} \vert \! \leqslant
\! (\widetilde{M})^{-1})$; hence, with the estimates for $\vert \delta_{\pm}
(\zeta) \vert^{2}$, the maximum modulus principle, and the fact that $\{
\mathstrut z \! \in \! \mathbb{C}; \, \delta (z) \! = \! 0\} \! = \!
\emptyset$, one shows that $\vert \vert (\delta (\cdot))^{\pm 1} \vert
\vert_{\mathcal{L}^{\infty}(\mathbb{C})} \! < \! \infty$. \hfill $\square$

Making use of Proposition~4.1, one changes the triangular factorisation
of $\widetilde{\mathcal{G}}^{c}(\zeta)$, with exponential decay of
elements like $\exp (\pm 2 \mi t \theta^{u}(\zeta))$ in their respective
domains of analyticity.
\begin{ccccc}
Define $\widehat{m}^{c}(\zeta) \! := \! \widetilde{m}^{c}(\zeta)(\delta
(\zeta))^{-\sigma_{3}}$, where $\delta (\zeta)$ is given in
Proposition~{\rm 4.1}. Then $\widehat{m}^{c}(\zeta) \colon \mathbb{C}
\setminus \sigma_{c}^{\prime} \! \to \! \mathrm{SL}(2,\mathbb{C})$ solves
the following {\rm RHP:} (1) $\widehat{m}^{c}(\zeta)$ is piecewise holomorphic
$\forall \, \zeta \! \in \! \mathbb{C} \setminus \sigma_{c}^{\prime};$ (2)
$\widehat{m}^{c}_{\pm}(\zeta) \! := \! \lim_{\genfrac{}{}{0pt}{2}{\zeta^{
\prime} \, \to \, \zeta}{\zeta^{\prime} \, \in \, \pm \, \mathrm{side} \,
\mathrm{of} \, \sigma_{c}^{\prime}}} \widehat{m}^{c}(\zeta^{\prime})$ satisfy
the jump condition $\widehat{m}^{c}_{+}(\zeta) \! = \! \widehat{m}^{c}_{-}
(\zeta) \widehat{\mathcal{G}}^{c}(\zeta)$, $\zeta \! \in \! \mathbb{R}$, where
\begin{equation*}
\widehat{\mathcal{G}}^{c}(\zeta) \! := \!
\begin{cases}
(\mathrm{I} \! - \! \overline{\rho (\overline{\zeta})}(\delta (\zeta))
^{-2} \me^{2 \mi t \theta^{u}(\zeta)} \sigma_{-})(\mathrm{I} \! + \! \rho
(\zeta)(\delta (\zeta))^{2} \me^{-2 \mi t \theta^{u}(\zeta)} \sigma_{+}),
&\text{$\zeta \! \in \! (0,\zeta_{2}) \cup (\zeta_{1},+\infty)$,} \\
(\mathrm{I} \! - \! \overline{\rho (\overline{\zeta})}(\delta_{-}(\zeta)
)^{-2} \me^{2 \mi t \theta^{u}(\zeta)} \sigma_{-})(\mathrm{I} \! + \! \rho
(\zeta)(\delta_{+}(\zeta))^{2} \me^{-2 \mi t \theta^{u}(\zeta)} \sigma_{
+}), &\text{$\zeta \! \in \! (-\infty,0) \cup (\zeta_{2},\zeta_{1})$,}
\end{cases}
\end{equation*}
with
\begin{equation*}
\rho (\zeta) \! := \!
\begin{cases}
\, \overline{r(\overline{\zeta})}, &\text{$\zeta \! \in \! (0,\zeta_{
2}) \cup (\zeta_{1},+\infty)$,} \\
\, -\overline{r(\overline{\zeta})}(1 \! - \! r(\zeta) \overline{r
(\overline{\zeta})})^{-1}, &\text{$\zeta \! \in \! (-\infty,0) \cup
(\zeta_{2},\zeta_{1});$}
\end{cases}
\end{equation*}
(3) as $\zeta \! \to \! \infty$, $\zeta \! \in \! \mathbb{C} \setminus
\sigma_{c}^{\prime}$, $\widehat{m}^{c}(\zeta) \! = \! \mathrm{I} \! +
\! \mathcal{O}(\zeta^{-1});$ and (4) $\widehat{m}^{c}(\zeta)$ satisfies
the symmetry reduction $\widehat{m}^{c}(\zeta) \! = \! \sigma_{1}
\overline{\widehat{m}^{c}(\overline{\zeta})} \, \sigma_{1}$ and the
condition $(\widehat{m}^{c}(0)(\delta (0))^{\sigma_{3}} \sigma_{2})^{2}
\! = \! \mathrm{I}$.
\end{ccccc}

\emph{Proof.} Follows {}from the RHP for $\widetilde{m}^{c}(\zeta)$
stated at the beginning of Section~4, the definition $\widehat{m}^{c}
(\zeta) \! := \! \widetilde{m}^{c}(\zeta)(\delta (\zeta))^{-\sigma_{3}}$,
with $\delta (\zeta)$ given in Proposition~4.1, and the identity $\sigma_{1}
\sigma_{1} \! = \! \mathrm{I}$. \hfill $\square$

The first main objective of this section is to reformulate the RHP for
$\widehat{m}^{c}(\zeta)$ as an auxiliary (equivalent) RHP on the
augmented (and oriented) contour $\Sigma^{\prime}$ (see Figure~3).
\begin{eeeee}
The augmented contour $\Sigma^{\prime}$ can be chosen with some
degree of flexibility, not necessarily consisting of straight line segments:
its crucial characteristic is the position of the rays relative to the
lines $\Re (\mi t \theta^{u}(\zeta)) \! = \! 0$.
\end{eeeee}
Since the jump matrices of the RHP on $\Sigma^{\prime}$ must be written
in terms of the jump matrices of the RHP for $\widehat{m}^{c}(\zeta)$ on
$\sigma_{c}^{\prime}$, and since, in general, the reflection coefficient,
$r(\zeta)$, does not have an analytic continuation to $\mathbb{C} \setminus
\mathbb{R}$, one must, as per the DZ method \cite{a27}, decompose,
or split, $r(\zeta)$ into an analytically continuable part and a negligible
non-analytic ``remainder''.
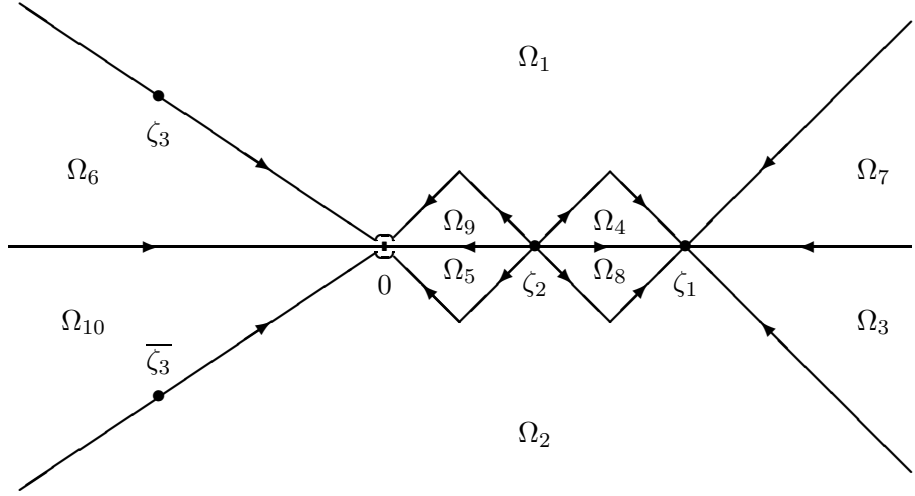
\begin{figure}[htb]
\begin{center}
\unitlength=1.0cm
\begin{picture}(12,8)(0,0.35)
\thicklines
\put(5,4.05){\makebox(0,0){$\centerdot$}}
\put(5,4.10){\makebox(0,0){$\centerdot$}}
\put(7,4){\makebox(0,0){$\bullet$}}
\put(9,4){\makebox(0,0){$\bullet$}}
\put(2,6){\makebox(0,0){$\bullet$}}
\put(2,2){\makebox(0,0){$\bullet$}}
\put(5,3.5){\makebox(0,0){$0$}}
\put(7,3.5){\makebox(0,0){$\zeta_{2}$}}
\put(9,3.5){\makebox(0,0){$\zeta_{1}$}}
\put(2,5.5){\makebox(0,0){$\zeta_{3}$}}
\put(2,2.5){\makebox(0,0){$\overline{\zeta_{3}}$}}
\put(7,6.5){\makebox(0,0){$\Omega_{1}$}}
\put(7,1.5){\makebox(0,0){$\Omega_{2}$}}
\put(1,5){\makebox(0,0){$\Omega_{6}$}}
\put(1,3){\makebox(0,0){$\Omega_{10}$}}
\put(6,4.325){\makebox(0,0){$\Omega_{9}$}}
\put(6,3.675){\makebox(0,0){$\Omega_{5}$}}
\put(8,4.325){\makebox(0,0){$\Omega_{4}$}}
\put(8,3.675){\makebox(0,0){$\Omega_{8}$}}
\put(11.5,5){\makebox(0,0){$\Omega_{7}$}}
\put(11.5,3){\makebox(0,0){$\Omega_{3}$}}
\put(12,4){\vector(-1,0){1.5}}
\put(10.5,4){\line(-1,0){1.5}}
\put(7,4){\vector(1,0){1}}
\put(8,4){\line(1,0){1}}
\put(7,4){\vector(-1,0){1}}
\put(5,4){\line(1,0){1}}
\put(0,4){\vector(1,0){2}}
\put(2,4){\line(1,0){3}}
\put(12,7){\vector(-1,-1){2}}
\put(12,1){\vector(-1,1){2}}
\put(10,5){\line(-1,-1){1}}
\put(10,3){\line(-1,1){1}}
\put(8,5){\vector(1,-1){0.5}}
\put(8,3){\vector(1,1){0.5}}
\put(8.5,4.5){\line(1,-1){0.5}}
\put(8.5,3.5){\line(1,1){0.5}}
\put(7,4){\vector(1,1){0.5}}
\put(7,4){\vector(1,-1){0.5}}
\put(7.5,4.5){\line(1,1){0.5}}
\put(7.5,3.5){\line(1,-1){0.5}}
\put(7,4){\vector(-1,1){0.5}}
\put(7,4){\vector(-1,-1){0.5}}
\put(6.5,4.5){\line(-1,1){0.5}}
\put(6.5,3.5){\line(-1,-1){0.5}}
\put(6,5){\vector(-1,-1){0.50}}
\put(6,3){\vector(-1,1){0.50}}
\put(6,5){\line(-1,-1){0.87}}
\put(6,3){\line(-1,1){0.87}}
\put(2,6){\vector(3,-2){1.5}}
\put(2,2){\vector(3,2){1.5}}
\put(3.5,5){\line(3,-2){1.37}}
\put(3.5,3){\line(3,2){1.37}}
\put(2,6){\line(-3,2){1.85}}
\put(2,2){\line(-3,-2){1.85}}
\put(5,4.075){\oval(0.23,0.15)[t]}
\put(5,3.925){\oval(0.23,0.15)[b]}
\end{picture}
\end{center}
\caption{Augmented contour $\Sigma^{\prime}$}
\end{figure}
\begin{ccccc}
Set $\Sigma^{\prime} \! := \! \mathrm{L} \cup \overline{\mathrm{
L}} \cup \mathbb{R}$, where $\mathrm{L} \! = \! \{\mathstrut \zeta;
\, \zeta \! = \! \zeta_{1} \! + \! \tfrac{v}{\sqrt{2}}(\zeta_{1} \! -
\! \zeta_{2}) \me^{\frac{3 \pi \mi}{4}}, \, -\infty \! < \! v \!
\leqslant \! 1\} \cup \{\mathstrut \zeta; \, \zeta \! = \! \zeta_{2}
\! + \! \tfrac{v}{\sqrt{2}}(\zeta_{1} \! - \! \zeta_{2}) \me^{\frac{
\mi \pi}{4}}, \, v \! \in \! [0,1]\} \cup \{\mathstrut \zeta; \, \zeta
\! = \! \zeta_{2} \! + \! \tfrac{v}{\sqrt{2}} \zeta_{2} \me^{-\frac{3
\pi \mi}{4}}, \, v \! \in \! [0,1]\} \cup \mathrm{L}_{>} \cup \mathrm{
L}_{<}$, with $\mathrm{L}_{>} \! := \! \{\mathstrut \zeta; \, \zeta
\! = \! \tfrac{v}{\sqrt{2}} \zeta_{2} \me^{-\frac{\mi \pi}{4}}, \,
v \! \in \! (0,1]\}$, and $\mathrm{L}_{<} \! := \! \{\mathstrut \zeta;
\, \zeta \! = \! v \me^{\mi \widetilde{\varphi}_{3}}, \, \widetilde{
\varphi}_{3} \! := \! \arg (\zeta_{3}) \! \in \! (\tfrac{\pi}{2},\pi),
\, v \! \in \! \mathbb{R}_{+}\}$. Set $\mathrm{L}_{\varepsilon} \!
:= \! \{\mathstrut \zeta; \, \zeta \! = \! \zeta_{1} \! + \! \tfrac{
v}{\sqrt{2}}(\zeta_{1} \! - \! \zeta_{2}) \me^{\frac{3 \pi \mi}{4}},
\, v \! \in \! (\varepsilon,1]\} \cup \{\mathstrut \zeta; \, \zeta \!
= \! \zeta_{2} \! + \! \tfrac{v}{\sqrt{2}}(\zeta_{1} \! - \! \zeta_{
2}) \me^{\frac{\mi \pi}{4}}, \, v \! \in \! (\varepsilon,1]\} \cup \{
\mathstrut \zeta; \, \zeta \! = \! \zeta_{2} \! + \! \tfrac{v}{\sqrt{
2}} \zeta_{2} \me^{-\frac{3 \pi \mi}{4}}, \, v \! \in \! (\varepsilon,
1]\} \cup \mathrm{L}_{>}$. Let $l \! \in \! \mathbb{Z}_{\geqslant 1}$ be
arbitrary, and choose $k \! = \! 4q \! + \! 1$, with $q \! \in \! \mathbb{
Z}_{\geqslant 1}$ and arbitrarily fixed, so large that $\tfrac{3q+2}{2} \!
- \! \tfrac{1}{2} \! > \! \tfrac{q}{2} \! > \! l$. For each $l \! \in \!
\mathbb{Z}_{\geqslant 1}$, there exists a decomposition of the function
(cf.~Lemma~{\rm 4.1)} $\rho (\zeta)$,
\begin{equation*}
\rho (\zeta) \! = \! h_{I}(\zeta) \! + \! (h_{II}(\zeta) \! + \!
\mathcal{R}(\zeta)), \quad \zeta \! \in \! \mathbb{R},
\end{equation*}
such that $h_{I}(\zeta)$ is analytic on $\mathbb{R}$ (generally, it
has no analytic continuation to $\mathbb{C} \setminus \mathbb{
R})$, $h_{II}(\zeta)$ has an analytic continuation to $\mathrm{L}$,
and $\mathcal{R}(\zeta)$, with $\mathcal{R}(\zeta) \! \equiv \! 0$
$\forall \, \, \zeta \! < \! 0$, is a piecewise-rational function with the
property that $(\tfrac{\md}{\md \zeta})^{j} \rho (\zeta) \vert_{\zeta
\in \{0,\zeta_{2},\zeta_{1}\}} \! = \! (\tfrac{\md}{\md \zeta})^{j}
\mathcal{R}(\zeta) \vert_{\zeta \in \{0,\zeta_{2},\zeta_{1}\}}$,
$j \! \in \! \{0,1,\ldots,k\}$. Then, as $t \! \to \! +\infty$ such that $0 \!
< \! \zeta_{2} \! < \! \tfrac{1}{M} \! < \! M \! < \! \zeta_{1}$ and $\vert
\zeta_{3} \vert^{2} \! = \! 1$, with $M \! \in \! \mathbb{R}_{>1}$ and
bounded, the following estimates are valid:
\begin{align*}
\vert \me^{-2 \mi t \theta^{u}(\zeta)} h_{I}(\zeta) \vert &\leqslant
\tfrac{\vert \underline{c}(\zeta_{1},\zeta_{2},\zeta_{3},\overline{
\zeta_{3}}) \vert}{\vert \zeta_{1} +\zeta_{2} \vert^{3}(\vert \zeta
\vert^{2}+1)t^{l}}, \quad \zeta \! \in \! \mathbb{R}, \\
\left\vert \me^{-2 \mi t \theta^{u}(\zeta)}(\tfrac{h_{I}(\zeta)}{\zeta})
\right\vert &\leqslant \tfrac{\vert \underline{c}(\zeta_{1},
\zeta_{2},\zeta_{3},\overline{\zeta_{3}}) \vert}{\vert \zeta_{1}+
\zeta_{2} \vert^{3}(\vert \zeta \vert^{2}+1)t^{l}}, \quad \zeta \! \in
\! (\cos \widetilde{\varphi}_{3},0) \cup (0,\tfrac{1}{2} \zeta_{2}), \\
\vert \me^{-2 \mi t \theta^{u}(\zeta)}h_{II}(\zeta) \vert &\leqslant
\tfrac{\vert \underline{c}(\zeta_{1},\zeta_{2},\zeta_{3},\overline{
\zeta_{3}}) \vert}{\vert z_{o}+\zeta_{1}+\zeta_{2} \vert^{l}(\vert
\zeta \vert^{2}+1)t^{l}}, \quad \zeta \! \in \! \mathrm{L} \setminus
(\mathrm{L}_{>} \cup \mathrm{L}_{<}), \\
\vert \me^{-2 \mi t \theta^{u}(\zeta)}h_{II}(\zeta) \vert &\leqslant
\tfrac{\vert \underline{c}(\zeta_{1},\zeta_{2},\zeta_{3},\overline{
\zeta_{3}}) \vert}{(\vert \zeta \vert^{2}+1)} \me^{-\gamma_{II}^{0}t},
\quad \zeta \! \in \! \mathrm{L}_{>} \cup \mathrm{L}_{<}, \\
\! \left\vert \me^{-2 \mi t \theta^{u}(\zeta)}(\tfrac{h_{II}(\zeta)}
{\zeta}) \right\vert &\leqslant \! \tfrac{\vert \underline{c}(\zeta_{
1},\zeta_{2},\zeta_{3},\overline{\zeta_{3}}) \vert}{(\vert \zeta
\vert^{2}+1)} \me^{-\gamma_{II}^{1}t}, \quad \zeta \! \in \! (\mathrm{
L}_{>} \cup \mathrm{L}_{<}) \! \setminus \! \{\mathstrut \zeta; \, \zeta
\! = \! v \me^{\mi \widetilde{\varphi}_{3}}, \, v \! \in \! \mathbb{R}_{
>1}\}, \\
\vert \me^{-2 \mi t \theta^{u}(\zeta)} \mathcal{R}(\zeta) \vert
&\leqslant \vert \underline{c}(\zeta_{1},\zeta_{2},\zeta_{3},\overline{
\zeta_{3}}) \vert \me^{-\varepsilon^{2} \gamma^{0}_{\mathcal{R}}t},
\quad \zeta \! \in \! \mathrm{L}_{\varepsilon} \setminus \mathrm{L}_{
>}, \\
\vert \me^{-2 \mi t \theta^{u}(\zeta)} \mathcal{R}(\zeta) \vert
&\leqslant \tfrac{\vert \underline{c}(\zeta_{1},\zeta_{2},\zeta_{3},
\overline{\zeta_{3}}) \vert}{(\vert \zeta \vert^{2}+1)} \me^{-\gamma_{
\mathcal{R}}^{1}t}, \quad \zeta \! \in \! \mathrm{L}_{>}, \\
\left\vert \me^{-2 \mi t \theta^{u}(\zeta)}(\tfrac{\mathcal{R}(\zeta)}
{\zeta}) \right\vert &\leqslant \tfrac{\vert \underline{c}(\zeta_{1},
\zeta_{2},\zeta_{3},\overline{\zeta_{3}}) \vert}{(\vert \zeta \vert^{2}
+1)} \me^{-\gamma_{\mathcal{R}}^{1}t}, \quad \zeta \! \in \! \mathrm{
L}_{>},
\end{align*}
where $\gamma_{II}^{0} \! := \! \min \{\tfrac{1}{2}(a_{2} \! - \! z_{
o}) \sqrt{4 \! - \! a_{2}^{2}},\tfrac{1}{2} \vert z_{o} \vert \zeta_{1}
(2 \! - \! \zeta_{2}^{2}),(\tfrac{1}{2} \vert z_{o} \vert \! - \! \cos
\! \widetilde{\varphi}_{3}) \sin \! \widetilde{\varphi}_{3}\}$ $(\in \!
\mathbb{R}_{+})$, $\gamma_{II}^{1} \! := \! \min \{\tfrac{1}{2} \vert
\sin \! 2 \widetilde{\varphi}_{3} \vert, \tfrac{1}{2} \vert z_{o} \vert
\zeta_{1}(2 \! - \! \zeta_{2}^{2})\}$ $(\in \! \mathbb{R}_{+})$,
$\gamma^{0}_{\mathcal{R}} \! := \! \min \{\tfrac{1}{2} \zeta_{2} \vert
z_{o} \! + \! \zeta_{2} \vert (1 \! - \! 2 \zeta_{1}^{2})^{2},\tfrac{
1}{2}(\zeta_{1} \! - \! \zeta_{2}) \vert z_{o} \! + \! \zeta_{1} \! +
\! \zeta_{2} \vert (1 \! - \! \zeta_{1}^{2})^{2},\tfrac{1}{2}(\zeta_{
1} \! - \! \zeta_{2}) \vert z_{o} \! + \! \zeta_{1} \! + \! \zeta_{2}
\vert (1 \! - \! \tfrac{2 \zeta_{1}^{2}}{\zeta_{1}^{4}+1})^{2}\}$ $(\in
\! \mathbb{R}_{+})$, and $\gamma^{1}_{\mathcal{R}} \! := \! \min
\{\vert z_{o} \vert \zeta_{1}(2 \! - \! \zeta_{2}^{2}),(\vert z_{o} \vert
\! - \! 2 \cos \! \widetilde{\varphi}_{3}) \sin \! \widetilde{\varphi}_{3}
\}$ $(\in \! \mathbb{R}_{+})$, with $z_{o} \! := \! x/t$, and
$\{\zeta_{i}\}_{i=1}^{3}$ and $a_{2}$ defined in Theorem~{\rm 3.1},
Eqs.~{\rm (16)} and~{\rm (17)}. Furthermore, taking conjugates,
$\overline{\rho (\overline{\zeta})} \! = \! \overline{h_{I}(\zeta)} \! +
\! (\overline{h_{II}(\overline{\zeta})} \! + \! \overline{\mathcal{R}
(\overline{\zeta})})$, gives rise to similar estimates for $\me^{2 \mi t
\theta^{u}(\zeta)} \, \overline{\star_{I}(\zeta)}$, $\star_{I}(\zeta)
\! \in \! \{h_{I}(\zeta),h_{I}(\zeta)/\zeta\}$, $\me^{2 \mi t \theta^{u}
(\zeta)} \, \overline{\star_{II}(\overline{\zeta})}$, $\star_{II}(\zeta) \!
\in \! \{h_{II}(\zeta),h_{II}(\zeta)/\zeta\}$, and $\me^{2 \mi t \theta^{
u}(\zeta)} \, \overline{\star_{III}(\overline{\zeta})}$, $\star_{III}
(\zeta) \! \in \! \{\mathcal{R}(\zeta),\mathcal{R}(\zeta)/\zeta\}$, on
$\overline{\mathrm{L}} \cup \mathbb{R};$ moreover, $h_{I}(\zeta)$
(respectively~$\overline{h_{I}(\zeta)})$ $\in \! \cap_{p \in \{1,2,\infty\}}
\! \mathcal{L}^{p}(\mathbb{R})$ (respectively~$\in \! \cap_{p \in \{1,2,
\infty\}} \! \mathcal{L}^{p}(\mathbb{R}))$, and $h_{II}(\zeta)$
(respectively~$\overline{h_{II}(\overline{\zeta})})$ $\in \! \cap_{p \in
\{1,2,\infty\}} \linebreak[4]
\mathcal{L}^{p}(\mathrm{L})$ (respectively~$\in \! \cap_{p \in \{1,2,
\infty\}} \mathcal{L}^{p}(\overline{\mathrm{L}}))$.
\end{ccccc}

\emph{Proof.} Many of the technical details and arguments associated
with the full proof of this Lemma are identical; hence, only two
representative calculations are presented, and the remaining estimations
follow in an analogous manner. Let $\exists \, M \! \in \! \mathbb{R}_{
>1}$ and bounded such that, as $t \! \to \! +\infty$, $0 \! < \! \zeta_{
2} \! < \! \tfrac{1}{M} \! < \! M \! < \! \zeta_{1}$ and $\vert \zeta_{
3} \vert^{2} \! = \! 1$. One begins by considering the non-empty domain
$\{\mathstrut z; \, \Im (z) \! = \! 0\} \cap \{\mathstrut z; \, \Re (z)
\! \in \! (\tfrac{1}{2}(\zeta_{1} \! + \! \zeta_{2}),\zeta_{1})\}$. Let
$k \! \in \! \mathbb{Z}_{\geqslant 1}$ be fixed and sufficiently large
with representation \cite{a27} $k \! = \! 4q \! + \! 1$, $q \! \in \!
\mathbb{Z}_{\geqslant 1}$. Recalling that, {}from Lemma~4.1, for $\Re
(\zeta) \! \in \! (\zeta_{2},\zeta_{1})$, $\rho (\zeta) \! = \! -
\overline{r(\overline{\zeta})} \, (1 \! - \! r(\zeta) \overline{r
(\overline{\zeta})})^{-1}$, one must estimate, as can be seen {}from
the explicit expression for the jump matrix $\widehat{\mathcal{G}}^{c}
(\zeta)$ given in Lemma~4.1, terms of the type $\rho (\zeta)(\delta_{+}
(\zeta))^{2} \me^{-2 \mi t \theta^{u}(\zeta)}$ (respectively~$\overline{
\rho (\overline{\zeta})} \, (\delta_{-}(\zeta))^{-2} \me^{2 \mi t
\theta^{u}(\zeta)})$: in fact, the former will be considered in detail,
with analytic continuation to the bounded domain $\Omega_{4}$
(see Figure~3) where $\Re (\mi t \theta^{u}(\zeta)) \! > \! 0$, whilst
the latter can be estimated (via a Schwarz symmetry principle argument)
analogously with analytic continuation to the bounded domain $\Omega_{
8}$ (see Figure~3) where $\Re (\mi t \theta^{u}(\zeta)) \! < \! 0$.
With the above-given choice of $k$, and thus $q$, consider the Taylor
expansion with integral remainder term for $\rho (\zeta)$, $\rho (\zeta)
\! = \! \sum_{n=0}^{k} \tfrac{1}{n!} \rho^{(n)}(\zeta_{1})(\zeta \! -
\! \zeta_{1})^{n} \! + \! \tfrac{1}{k!} \! \int_{\zeta_{1}}^{\zeta}
\rho^{(k+1)}(\xi)(\zeta \! - \! \xi)^{k} \, \md \xi$, where $\rho^{
(n)}(\zeta_{1}) \! := \! -(\tfrac{\md}{\md \zeta})^{n}(\overline{r(
\overline{\zeta})} \, (1 \! - \! r(\zeta) \overline{r(\overline{\zeta}
)})^{-1}) \vert_{\zeta=\zeta_{1}}$, $n \! \in \! \{0,1,\ldots,k\}$,
$k \! \in \! \mathbb{Z}_{\geqslant 1}$: note that, if one considers
explicitly, say, the limiting case $\zeta_{1} \! \to \! +\infty$
$(\zeta_{2} \! \to \! 0^{+})$, then, as functions of $\zeta_{1}$,
the Taylor coefficients, $\rho^{(n)}(\zeta_{1})$, $n \! \in \! \{0,
1,\ldots,k\}$, $k \! \in \! \mathbb{Z}_{\geqslant 1}$, are in the
$\mathbb{C}$-valued Schwartz class (since $r(\zeta) \! \in \! \mathcal{
S}_{\mathbb{C}}(\mathbb{R}))$. Let $\mathcal{R}(\zeta)$ denote the
``polynomial part'' of this expansion, and $h(\zeta)$ the ``remainder'',
namely, $\mathcal{R}(\zeta) \! := \! \sum_{n=0}^{k} \tfrac{1}{n!} \rho^{
(n)}(\zeta_{1})(\zeta \! - \! \zeta_{1})^{n}$ and $h(\zeta) \! := \!
\tfrac{1}{k!} \! \int_{\zeta_{1}}^{\zeta} \rho^{(k+1)}(\xi)(\zeta \! -
\! \xi)^{k} \, \md \xi$; hence, $\rho (\zeta) \! = \! \mathcal{R}(\zeta)
\! + \! h(\zeta)$. One notes that, with the above-given choice, $\rho
(\zeta) \! - \! \mathcal{R}(\zeta) \! = \! h(\zeta)$ can be interpreted
as the error incurred in approximating $\rho (\zeta)$ by a polynomial
of degree $k$. Noting that, $\forall \, j \! \in \! \{0,1,\ldots,k\}$,
$k \! \in \! \mathbb{Z}_{\geqslant 1}$, $(\tfrac{\md}{\md \zeta})^{j}
\rho (\zeta) \vert_{\zeta=\zeta_{1}} \! = \! (\tfrac{\md}{\md \zeta})
^{j} \mathcal{R}(\zeta) \vert_{\zeta=\zeta_{1}}$, it follows that
$(\tfrac{\md}{\md \zeta})^{j} h(\zeta) \vert_{\zeta=\zeta_{1}} \! = \!
0$, $j \! \in \! \{0,1,\ldots,k\}$, $k \! \in \! \mathbb{Z}_{\geqslant
1}$; hence, as per the DZ 
method \cite{a27}, one uses this latter
property to split, further, $h(\zeta)$ as $h(\zeta) \! := \! h_{I}(\zeta)
\! + \! h_{II}(\zeta)$, which, when combined with $\mathcal{R}(\zeta)$,
shows that $\rho (\zeta) \! = \! h_{I}(\zeta) \! + \! (h_{II}(\zeta) \!
+ \! \mathcal{R}(\zeta))$, where $h_{I}(\zeta)$ is defined for
$\zeta \! \in \! \{\mathstrut z; \, \Im (z) \! = \! 0\} \cap \{\mathstrut z;
\, \Re (z) \! \in \! (\tfrac{1}{2}(\zeta_{1} \! + \! \zeta_{2}),\zeta_{
1})\}$, where $\Re (\mi t \theta^{u}(\zeta)) \! = \! 0$, and has
negligibly ``small norm'' (see below) as $t \! \to \! +\infty$, $h_{II}
(\zeta)$ has analytic continuation to $\mathbb{C}_{+} \cap \{
\mathstrut z; \, \Re (z) \! \in \! (\tfrac{1}{2}(\zeta_{1} \! + \! \zeta_{2}),
\zeta_{1})\} \cap \Omega_{4}$, where $\Re (\mi t \theta^{u}(\zeta))
\! > \! 0$, and $\mathcal{R}(\zeta)$ has trivial analytic continuation
to $\mathbb{C}_{+} \cap \{\mathstrut z; \, \Re (z) \! \in \! (\tfrac{
1}{2}(\zeta_{1} \! + \! \zeta_{2}),\zeta_{1})\} \cap \Omega_{4}$. One
notes {}from the expression for $\theta^{u}(\zeta)$ that $\theta^{u}
(\zeta_{1}) \! < \! \theta^{u}(\tfrac{1}{2}(\zeta_{1} \! + \! \zeta_{
2})) \! < \! 0$; hence, considering the auxiliary function (which will
be needed below) $\alpha (\zeta) \! := \! \zeta^{-3}(\zeta \! - \!
\zeta_{1})^{q}(\zeta \! - \! \zeta_{2})(\zeta \! - \! \zeta_{3})(\zeta
\! - \! \overline{\zeta_{3}})$, $q \! \in \! \mathbb{Z}_{\geqslant 1}$,
set $(\tfrac{h}{\alpha})(\theta^{u}) \! := \!
\begin{cases}
\tfrac{h(\zeta (\theta^{u}))}{\alpha (\zeta (\theta^{u}))}, &\text{
$\theta^{u}(\zeta_{1}) \! < \! \theta^{u} \! < \! \theta^{u}(\tfrac{
1}{2}(\zeta_{1} \! + \! \zeta_{2}))$,} \\
0, &\text{$\theta^{u} \! \in \! \mathbb{R} \setminus (\theta^{u}
(\zeta_{1}),\theta^{u}(\tfrac{1}{2}(\zeta_{1} \! + \! \zeta_{2})))$,}
\end{cases}$ and consider the Fourier transform with respect to (w.r.t.)
$\theta^{u}(\zeta)$. Before proceeding any further, and for future
reference, one notes that $\tfrac{h(\zeta)}{\alpha (\zeta)} \! = \!
\tfrac{\zeta^{3}(\zeta-\zeta_{1})^{3q+2}}{(\zeta-\zeta_{2})(\zeta-
\zeta_{3})(\zeta-\overline{\zeta_{3}})k!} \! \int_{0}^{1} \rho^{(k+1)}
(\zeta_{1} \! + \! (\zeta \! - \! \zeta_{1}) \tau)(1 \! - \! \tau)^{k}
\, \md \tau$, and, for $0 \! < \! \zeta_{2} \! < \! \tfrac{1}{M} \! <
\! M \! < \! \zeta_{1}$ and $\vert \zeta_{3} \vert^{3} \! = \! 1$, since
$r(\zeta) \! \in \! \mathcal{S}_{\mathbb{C}}(\mathbb{R})$, $\vert \vert
r(\cdot) \vert \vert_{\mathcal{L}^{\infty}(\mathbb{R})} \! < \! 1$, and
$\sup_{(\tau,k) \in [0,1] \times \mathbb{Z}_{\geqslant 1}} \vert
\rho^{(k+1)}(\zeta_{1} \! + \! (\zeta \! - \! \zeta_{1}) \tau) \vert \! <
\! \infty$, with $\theta^{u} \! \uparrow \! \theta^{u}(\tfrac{1}{2}
(\zeta_{1} \! + \! \zeta_{2}))$ and $\theta^{u} \! \downarrow \! \theta^{
u}(\zeta_{1})$, $\tfrac{h(\zeta)}{\alpha (\zeta)} \! =_{\zeta \to \zeta_{
1}} \! \mathcal{O}((\zeta \! - \! \zeta_{1})^{3q+2})$: also, one deduces
that $\tfrac{\md \zeta}{\md \theta^{u}} \! =_{\zeta \to \zeta_{1}} \!
\mathcal{O}((\zeta \! - \! \zeta_{1})^{-1})$. Define the Fourier transform
pair (w.r.t.~$\theta^{u}(\zeta))$: $(\tfrac{h}{\alpha})(\zeta) \! := \!
\int_{-\infty}^{+\infty} \me^{\mi s \theta^{u}(\zeta)} \widetilde{(\tfrac{
h}{\alpha})}(s) \tfrac{\md s}{\sqrt{2 \pi}}$, $\tfrac{1}{2}(\zeta_{1} \!
+ \! \zeta_{2}) \! < \! \Re (\zeta) \! < \! \zeta_{1}$, and $\widetilde{
(\tfrac{h}{\alpha})}(s) \! := \! -\int_{\frac{1}{2}(\zeta_{1}+\zeta_{2})
}^{\zeta_{1}} \me^{-\mi s \theta^{u}(\zeta)}(\tfrac{h}{\alpha})(\zeta)
\tfrac{\md \theta^{u}(\zeta)}{\sqrt{2 \pi}}$, $s \! \in \! \mathbb{R}$.
In order to obtain the necessary estimate for $(h_{I}(\zeta) \! + \!
(h_{II}(\zeta) \! + \! \mathcal{R}(\zeta)))(\delta_{+}(\zeta))^{2} \me^{
-2 \mi t \theta^{u}(\zeta)}$, one needs to show that $(\tfrac{h}{\alpha})
(\theta^{u}) \! := \! \tfrac{h(\zeta (\theta^{u}))}{\alpha (\zeta (\theta
^{u}))} \! \in \! \mathcal{H}^{j}(\mathbb{R})$, $0 \! \leqslant \! j \!
\leqslant \! [\tfrac{3q+2}{2}]$, $q \! \in \! \mathbb{Z}_{\geqslant 1}$,
where $\mathcal{H}^{j}(\mathbb{R})$ denotes the $L^{2}$-Sobolev
space with norm $\vert \vert \! \star \! (\cdot) \vert \vert_{\mathcal{
H}(\mathbb{R})} \! := \! (\sum_{j=0}^{[\frac{3q+2}{2}]} \vert \vert
(\tfrac{\md}{\md \theta^{u}})^{j} \star \! (\theta^{u}) \vert \vert_{
\mathcal{L}^{2}(\mathbb{R})}^{2})^{1/2}$. In this particular case,
one must show that, for $0 \! < \! \zeta_{2} \! < \! \tfrac{1}{M} \! < \!
M \! < \! \zeta_{1}$ and $\vert \zeta_{3} \vert^{3} \! = \! 1$, with
$\Re (\zeta) \! \in \! (\tfrac{1}{2}(\zeta_{1} \! + \! \zeta_{2}),\zeta_{
1})$, $\mathrm{I}_{\mathcal{H}} \! := \! \int_{-\infty}^{+\infty} \vert
(\tfrac{\md}{\md \theta^{u}})^{j} \tfrac{h(\zeta (\theta^{u}))}{\alpha
(\zeta (\theta^{u}))} \vert^{2} \, \md \theta^{u} \! < \! \infty$,
$0 \! \leqslant \! j \! \leqslant \! [\tfrac{3q+2}{2}]$, $q \! \in \!
\mathbb{Z}_{\geqslant 1}$: in fact, since $\tfrac{h(\zeta (\theta^{u}))}
{\alpha (\zeta (\theta^{u}))} \! \equiv \! 0$ $\forall \, \, \theta^{u}
\! \in \! \mathbb{R} \setminus (\theta^{u}(\zeta_{1}),\theta^{u}(\tfrac{
1}{2}(\zeta_{1} \! + \! \zeta_{2})))$, the latter integral reduces to
$\mathrm{I}_{\mathcal{H}} \! = \! \int_{\theta^{u}(\zeta_{1})}^{\theta^{
u}(\frac{1}{2}(\zeta_{1}+\zeta_{2}))} \vert (\tfrac{\md}{\md \theta^{u}}
)^{j} \tfrac{h(\zeta (\theta^{u}))}{\alpha (\zeta (\theta^{u}))} \vert^{
2} \, \md \theta^{u}$, which, via the chain rule and a change-of-variable
argument, is shown to be equal to $\mathrm{I}_{\mathcal{H}} \! = \! \int_{
\frac{1}{2}(\zeta_{1}+\zeta_{2})}^{\zeta_{1}} \! \left\vert \tfrac{(\frac{
\md}{\md \zeta})^{j} \frac{h(\zeta)}{\alpha (\zeta)}}{(\frac{\md \theta^{
u}(\zeta)}{\md \zeta})^{j}} \right\vert^{2}(\tfrac{\md \theta^{u}(\zeta)}
{\md \zeta}) \, \md \zeta$. Now, recalling that $\tfrac{h(\zeta)}{\alpha
(\zeta)} \! = \! \tfrac{\zeta^{3}(\zeta-\zeta_{1})^{3q+2}}{(\zeta-\zeta_{
2})(\zeta-\zeta_{3})(\zeta-\overline{\zeta_{3}})k!} \! \int_{0}^{1} \rho^{
(k+1)}(\zeta_{1} \! + \! (\zeta \! - \! \zeta_{1}) \tau)(1 \! - \! \tau)^{
k} \, \md \tau$ and ({}from Section~3) $\tfrac{\md \theta^{u}(\zeta)}{\md
\zeta} \! = \! \zeta^{-3}(\zeta \! - \! \zeta_{1})(\zeta \! - \! \zeta_{2}
)(\zeta \! - \! \zeta_{3})(\zeta \! - \! \overline{\zeta_{3}})$, one shows
that, for $0 \! \leqslant \! j \! \leqslant \! [\tfrac{3q+2}{2}]$, $q \!
\in \! \mathbb{Z}_{\geqslant 1}$, $\mathrm{I}_{\mathcal{H}} \! \leqslant
\! \vert c^{\mathcal{S}}(\zeta_{1}) \underline{c}(\zeta_{2},\zeta_{3},
\overline{\zeta_{3}}) \vert$, $0 \! < \! \zeta_{2} \! < \! \tfrac{1}{M}
\! < \! M \! < \! \zeta_{1}$ and $\vert \zeta_{3} \vert^{2} \! = \! 1$,
that is, $\vert \vert (\tfrac{h}{\alpha})(\cdot) \vert \vert_{\mathcal{
H}(\mathbb{R})} \! = \! (\sum_{j=0}^{[\frac{3q+2}{2}]} \vert \vert
(\tfrac{\md}{\md \theta^{u}})^{j}(\tfrac{h}{\alpha})(\theta^{u}) \vert
\vert_{\mathcal{L}^{2}(\mathbb{R})}^{2})^{1/2} \! < \! \infty$; hence,
by Parseval's Theorem, for $M \! \in \! \mathbb{R}_{>1}$ and
bounded such that $0 \! < \! \zeta_{2} \! < \! \tfrac{1}{M} \! < \! M \! < \!
\zeta_{1}$ and $\vert \zeta_{3} \vert^{2} \! = \! 1$, $\int_{-\infty}^{
+\infty}(1 \! + \! s^{2})^{j} \vert \widetilde{(\tfrac{h}{\alpha})}(s)
\vert^{2} \, \md s \! \leqslant \! \vert c^{\mathcal{S}}(\zeta_{1})
\underline{c}(\zeta_{2},\zeta_{3},\overline{\zeta_{3}}) \vert$, $0 \!
\leqslant \! j \! \leqslant \! [\tfrac{3q+2}{2}]$, $q \! \in \! \mathbb{
Z}_{\geqslant 1}$. Recalling {}from the Fourier transform pair that,
for $\tfrac{1}{2}(\zeta_{1} \! + \! \zeta_{2}) \! < \! \Re (\zeta) \! < \!
\zeta_{1}$, $(\tfrac{h}{\alpha})(\zeta) \! = \! \int_{-\infty}^{+\infty}
\me^{\mi s \theta^{u}(\zeta)} \widetilde{(\tfrac{h}{\alpha})}(s) \tfrac{
\md s}{\sqrt{2 \pi}}$, it follows, by defining $h_{I}(\zeta) \! := \!
\alpha (\zeta) \! \int_{t}^{+\infty} \me^{\mi s \theta^{u}(\zeta)}
\widetilde{(\tfrac{h}{\alpha})}(s) \tfrac{\md s}{\sqrt{2 \pi}}$ and
$h_{II}(\zeta) \! := \! \alpha (\zeta) \! \int_{-\infty}^{t} \me^{\mi
s \theta^{u}(\zeta)} \widetilde{(\tfrac{h}{\alpha})}(s) \tfrac{\md s}
{\sqrt{2 \pi}}$, that $h(\zeta) \! = \! h_{I}(\zeta) \! + \! h_{II}
(\zeta)$. For $\zeta \! \in \! \{\mathstrut z; \, \Im (z) \! = \! 0\}
\cap \{\mathstrut z; \, \Re (z) \! \in \! (\tfrac{1}{2}(\zeta_{1} \! + \!
\zeta_{2}),\zeta_{1})\}$, recalling the bounds for $\delta_{+}(\zeta)$
given in Proposition~4.1, and using the Cauchy-Schwarz inequality
for integrals, one shows that $\vert (\delta_{+}(\zeta))^{2} \me^{-2 \mi
t \theta^{u}(\zeta)}h_{I}(\zeta) \vert \! = \! \vert (\delta_{+}(\zeta))^{
2} \me^{-2 \mi t \theta^{u}(\zeta)} \alpha (\zeta) \! \int_{t}^{+\infty}
\me^{\mi s \theta^{u}(\zeta)} \widetilde{(\tfrac{h}{\alpha})}(s) \tfrac{
\md s}{\sqrt{2 \pi}} \vert \! \leqslant \! (1 \! - \! \sup_{z \in \mathbb{
R}} \vert r(z) \vert^{2}) \tfrac{\vert \alpha (\zeta) \vert}{\sqrt{2 \pi}}
(\int_{t}^{+\infty}(1 \! + \! s^{2})^{-j} \, \md s)^{1/2}(\int_{t}^{+
\infty}(1 \! + \! s^{2})^{j} \vert \widetilde{(\tfrac{h}{\alpha})}(s)
\vert^{2} \, \md s)^{1/2}$: noting that $\int_{t}^{+\infty}(1 \! + \!
s^{2})^{-j} \, \md s \! \leqslant \! \tfrac{t^{-(2j-1)}}{(2j-1)}$, $2j
\! - \! 1 \! > \! 0$, and recalling the Parseval estimate $\int_{t}^{+
\infty}(1 \! + \! s^{2})^{j} \vert \widetilde{(\tfrac{h}{\alpha})}(s)
\vert^{2} \, \md s \! \leqslant \! \int_{-\infty}^{+\infty}(1 \! + \!
s^{2})^{j} \vert \widetilde{(\tfrac{h}{\alpha})}(s) \vert^{2} \, \md s
\! \leqslant \! \vert c^{\mathcal{S}}(\zeta_{1}) \underline{c}(\zeta_{
2},\zeta_{3},\overline{\zeta_{3}}) \vert$, $0 \! \leqslant \! j \!
\leqslant \! [\tfrac{3q+2}{2}]$, it follows that, with $\alpha (\zeta)
\! = \! \zeta^{-3}(\zeta \! - \! \zeta_{1})^{q}(\zeta \! - \! \zeta_{2})
(\zeta \! - \! \zeta_{3})(\zeta \! - \! \overline{\zeta_{3}})$, $\vert
\me^{-2 \mi t \theta^{u}(\zeta)} h_{I}(\zeta) \vert \! \leqslant \!
\tfrac{\vert c^{\mathcal{S}}(\zeta_{1}) \underline{c}(\zeta_{2},\zeta_{
3},\overline{\zeta_{3}}) \vert}{\vert \zeta_{1}+\zeta_{2} \vert^{3}t^{
(j-1/2)}}$, $1 \! \leqslant \! j \! \leqslant \! [\tfrac{3q+2}{2}]$,
$q \! \in \! \mathbb{Z}_{\geqslant 1}$, $\tfrac{1}{2}(\zeta_{1} \! + \!
\zeta_{2}) \! < \! \Re (\zeta) \! < \! \zeta_{1}$, and $0 \! < \! \zeta_{
2} \! < \! \tfrac{1}{M} \! < \! M \! < \! \zeta_{1}$ and $\vert \zeta_{3}
\vert^{2} \! = \! 1$. Since, for $\zeta \! \in \! \mathbb{C}_{+} \cap
\{\mathstrut z; \, \Re (z) \! \in \! (\tfrac{1}{2}(\zeta_{1} \! + \! \zeta_{
2}),\zeta_{1})\} \cap \Omega_{4}$, $\Re (\mi t \theta^{u}(\zeta))
\! > \! 0$, $h_{II}(\zeta)$ has an analytic continuation to the line
(parametrised by $v)$ $\zeta \! = \! \zeta (v) \! = \! \zeta_{1} \! +
\! \tfrac{v}{\sqrt{2}}(\zeta_{1} \! - \! \zeta_{2}) \me^{\frac{3 \pi
\mi}{4}}$, $v \! \in \! [0,1]$; hence, on this line, and using the
Cauchy-Schwarz inequality for integrals, it follows that $\mathrm{
I}_{h_{II}} \! := \! \vert (\delta_{+}(\zeta))^{2} \me^{-2 \mi t \theta^{
u}(\zeta)}h_{II}(\zeta) \vert \! = \! \vert (\delta_{+}(\zeta))^{2} \me^{
-2 \mi t \theta^{u}(\zeta)} \alpha (\zeta) \! \int_{-\infty}^{t} \me^{
\mi s \theta^{u}(\zeta)} \widetilde{(\tfrac{h}{\alpha})}(s) \tfrac{\md
s}{\sqrt{2 \pi}} \vert \! \leqslant \! (1 \! - \! \sup_{z \in \mathbb{
R}} \vert r(z) \vert^{2}) \tfrac{\vert \alpha (\zeta) \vert}{\sqrt{2
\pi}} \me^{-t \Re (\mi \theta^{u}(\zeta))}(\int_{-\infty}^{t}(1 \! +
\linebreak[4]
s^{2})^{-j} \, \md s)^{1/2}(\int_{-\infty}^{t}(1 \! + \! s^{2})^{j}
\vert \widetilde{(\tfrac{h}{\alpha})}(s) \vert^{2} \, \md s)^{1/2}$.
Noting that $\int_{-\infty}^{t}(1 \! + \! s^{2})^{-j} \, \md s \!
\leqslant \! \int_{-\infty}^{+\infty}(1 \! + \! s^{2})^{-j} \, \md s \!
\leqslant \! \int_{-\infty}^{+\infty}(1 \! + \! s^{2})^{-1} \, \md s \!
= \! \pi$, and, {}from the above Parseval estimate $\int_{-\infty}^{t}
(1 \! + \! s^{2})^{j} \vert \widetilde{(\tfrac{h}{\alpha})}(s) \vert^{
2} \, \md s \! \leqslant \! \int_{-\infty}^{+\infty}(1 \! + \! s^{2})^{j}
\vert \widetilde{(\tfrac{h}{\alpha})}(s) \vert^{2} \, \md s \! \leqslant
\! \vert c^{\mathcal{S}}(\zeta_{1}) \underline{c}(\zeta_{2},\zeta_{
3},\overline{\zeta_{3}}) \vert$, $0 \! \leqslant \! j \! \leqslant \!
[\tfrac{3q+2}{2}]$, $q \! \in \! \mathbb{Z}_{\geqslant 1}$, it follows
{}from the definition of $\alpha (\zeta)$ that $\mathrm{I}_{h_{II}}
\! \leqslant \! \tfrac{\vert c^{\mathcal{S}}(\zeta_{1}) \underline{c}
(\zeta_{2},\zeta_{3},\overline{\zeta_{3}}) \vert v^{q}}{\vert \zeta_{
1}+\zeta_{2} \vert^{3} \vert \zeta_{1}-\zeta_{2} \vert^{-q}} \me^{-t
\Re (\mi \theta^{u}(\zeta))}$, $q \! \in \! \mathbb{Z}_{\geqslant 1}$,
$v \! \in \! [0,1]$. Recalling that $\theta^{u}(\zeta) \! = \! \tfrac{1}
{2}(\zeta \! - \! \tfrac{1}{\zeta})(z_{o} \! + \! \zeta \! + \! \tfrac{1}
{\zeta})$, for $\zeta \! = \! \zeta (v) \! = \! \zeta_{1} \! + \! \tfrac{v}
{\sqrt{2}}(\zeta_{1} \! - \! \zeta_{2}) \me^{\frac{3 \pi \mi}{4}}$, $v \!
\in \! [0,1]$, one shows that $\Im ((\zeta \! - \! \tfrac{1}{\zeta})(z_{o}
\! + \! \zeta \! + \! \tfrac{1}{\zeta})) \! = \! \tfrac{1}{2}v(\zeta_{1}
\! - \! \zeta_{2})(\zeta_{1} \! - \! \tfrac{1}{2}(\zeta_{1} \! - \! \zeta_{
2})v)(1 \! - \! \tfrac{1}{(\zeta_{1}-\frac{1}{2}(\zeta_{1}-\zeta_{2})
v)^{2}+(\frac{1}{2}(\zeta_{1} \! - \! \zeta_{2})v)^{2}})^{2} \! + \!
\tfrac{1}{2}v(\zeta_{1} \! - \! \zeta_{2})(\zeta_{1} \! - \! \tfrac{1}
{2}(\zeta_{1} \! - \! \zeta_{2})v)(1 \! + \! \tfrac{1}{(\zeta_{1}-
\frac{1}{2}(\zeta_{1}-\zeta_{2})v)^{2}+(\frac{1}{2}(\zeta_{1} \! -
\! \zeta_{2})v)^{2}})^{2} \! + \! \tfrac{1}{2}v(\zeta_{1} \! - \! \zeta_{
2})z_{o}(1 \! + \! \tfrac{1}{(\zeta_{1}-\frac{1}{2}(\zeta_{1}-\zeta_{
2})v)^{2}+(\frac{1}{2}(\zeta_{1}-\zeta_{2})v)^{2}})$. Setting
$\widehat{a}(v) \! := \! (\zeta_{1} \! - \! \tfrac{1}{2}(\zeta_{1} \! - \!
\zeta_{2})v)^{2} \! + \! (\tfrac{1}{2}(\zeta_{1} \! - \! \zeta_{2})v)^{2}$,
one deduces that, for $v \! \in \! [0,1]$, $\widehat{a}(1) \! \leqslant
\! \widehat{a}(v) \! \leqslant \! \widehat{a}(0)$; hence, noting that
$(\tfrac{\zeta_{1}^{2}-\zeta_{1}(\zeta_{1}-\zeta_{2})v+\frac{1}{2}
(\zeta_{1}-\zeta_{2})^{2}v^{2}+1}{\zeta_{1}^{2}-\zeta_{1}(\zeta_{1}-
\zeta_{2})v+\frac{1}{2}(\zeta_{1}-\zeta_{2})^{2}v^{2}}) \! \geqslant \!
(\tfrac{\zeta_{1}^{2}-\zeta_{1}(\zeta_{1}-\zeta_{2})v+\frac{1}{2}(\zeta_{
1}-\zeta_{2})^{2}v^{2}-1}{\zeta_{1}^{2}-\zeta_{1}(\zeta_{1}-\zeta_{2})
v+\frac{1}{2}(\zeta_{1}-\zeta_{2})^{2}v^{2}})$, $v \! \in \! [0,1]$, one
deduces that $\Im ((\zeta \! - \! \tfrac{1}{\zeta})(z_{o} \! + \! \zeta
\! + \! \tfrac{1}{\zeta})) \! \geqslant \! v(\zeta_{1} \! - \! \zeta_{2})
(\zeta_{1} \! - \! \tfrac{1}{2}(\zeta_{1} \! - \! \zeta_{2})v)(1 \! - \!
(\widehat{a}(v))^{-1})^{2} \! + \! \tfrac{1}{2}v(\zeta_{1} \! - \! \zeta_{
2})z_{o}(1 \! - \! (\widehat{a}(v))^{-1})$. Noting that, for $v \! \in
\! [0,1]$, $1 \! - \! (\widehat{a}(1))^{-1} \! \leqslant \! 1 \! - \!
(\widehat{a}(v))^{-1} \! \leqslant \! 1 \! - \! (\widehat{a}(0))^{
-1}$, $1 \! - \! (\widehat{a}(1))^{-1} \! \geqslant \! (1 \! - \!
(\widehat{a}(1))^{-1})^{2}$, and $2 \! - \! v \! \geqslant \! v$, it
follows that $\Im ((\zeta \! - \! \tfrac{1}{\zeta})(z_{o} \! + \! \zeta
\! + \! \tfrac{1}{\zeta})) \! \geqslant \! \tfrac{1}{2}(\zeta_{1} \!
- \! \zeta_{2})(z_{o} \! + \! \zeta_{1} \! + \! \zeta_{2})(1 \! - \!
(\widehat{a}(1))^{-1})^{2}v^{2}$; hence, since $(\zeta_{1} \! - \!
\zeta_{2}) \! > \! 0$ and $(z_{o} \! + \! \zeta_{1} \! + \! \zeta_{2}) \!
< \! 0$, it follows that $\mi \tfrac{1}{2} \Im ((\zeta \! - \! \tfrac{1}{
\zeta})(z_{o} \! + \! \zeta \! + \! \tfrac{1}{\zeta})) \! \geqslant
\! \tfrac{1}{4}(\zeta_{1} \! - \! \zeta_{2}) \vert z_{o} \! + \! \zeta_{
1} \! + \! \zeta_{2} \vert (1 \! - \! (\widehat{a}(1))^{-1})^{2}v^{2}$,
whence $-t \Re (\mi \theta^{u}(\zeta)) \! \leqslant \! -\tfrac{1}
{4}t(\zeta_{1} \! - \! \zeta_{2}) \vert z_{o} \! + \! \zeta_{1} \! + \!
\zeta_{2} \vert (1 \! - \! (\widehat{a}(1))^{-1})^{2}v^{2}$, $v \! \in
\! [0,1]$. With this inequality, one deduces that, for $\zeta \! = \!
\zeta (v) \! = \! \zeta_{1} \! + \! \tfrac{v}{\sqrt{2}}(\zeta_{1} \! -
\! \zeta_{2}) \me^{\frac{3 \pi \mi}{4}}$, $v \! \in \! [0,1]$, $\vert
\me^{-2 \mi t \theta^{u}(\zeta)}h_{II}(\zeta) \vert \! \leqslant \!
\tfrac{\vert c^{\mathcal{S}}(\zeta_{1}) \underline{c}(\zeta_{2},\zeta_{
3},\overline{\zeta_{3}}) \vert}{\vert \zeta_{1}+\zeta_{2} \vert^{3}t^{
q/2}}$, $q \! \in \! \mathbb{Z}_{\geqslant 1}$, $0 \! < \! \zeta_{2}
\! < \! \tfrac{1}{M} \! < \! M \! < \! \zeta_{1}$ and $\vert \zeta_{
3} \vert^{2} \! = \! 1$. Let $\varepsilon$ be an arbitrarily fixed,
sufficiently small positive real number such that $\{\mathstrut \zeta;
\, \vert \zeta \! - \! \zeta_{1} \vert \! < \! \varepsilon\} \cap \{
\mathstrut \zeta; \, \zeta \! = \! \zeta_{1} \! + \! \tfrac{v}{\sqrt{
2}}(\zeta_{1} \! - \! \zeta_{2}) \me^{\frac{3 \pi \mi}{4}}, \, v \!
\in \! (\varepsilon,1]\} \! = \! \emptyset$, and recall that $\mathcal{
R}(\zeta) \! := \! \sum_{n=0}^{k} \tfrac{1}{n!} \rho^{(n)}(\zeta_{1})
(\zeta \! - \! \zeta_{1})^{n}$, $k \! \in \! \mathbb{Z}_{\geqslant 1}$.
On the line segment $\zeta \! = \! \zeta (v) \! = \! \zeta_{1} \! + \!
\tfrac{v}{\sqrt{2}}(\zeta_{1} \! - \! \zeta_{2}) \me^{\frac{3 \pi \mi}
{4}}$, $\varepsilon \! < \! v \! \leqslant \! 1$, one shows that $\vert
(\delta_{+}(\zeta))^{2} \me^{-2 \mi t \theta^{u}(\zeta)} \mathcal{R}
(\zeta) \vert \! \leqslant \! (1 \! - \! \sup_{z \in \mathbb{R}} \vert r(z)
\vert^{2}) \me^{-2t \Re (\mi \theta^{u}(\zeta))} \sup_{(\zeta_{1},
n) \in (M,+\infty) \times \{0,1,\ldots,k\}} \vert \rho^{(n)}(\zeta_{1})
\vert \sum_{n=0}^{k} \tfrac{\vert \zeta_{1}-\zeta_{2} \vert^{n}v^{
n}}{2^{n/2}n!}$: now, recalling the above inequality for $-t \Re (\mi
\theta^{u}(\zeta))$, and using the formula $\sum_{n=0}^{k} \! \star^{
n} \! = \! \tfrac{1-\star^{k+1}}{1-\star}$, one shows that, on this line
segment, $\vert \me^{-2 \mi t \theta^{u}(\zeta)} \mathcal{R}(\zeta)
\vert \! \leqslant \! \vert \underline{c}(\zeta_{1},\zeta_{2},\zeta_{
3},\overline{\zeta_{3}}) \vert \exp (-\tfrac{1}{2} t \varepsilon^{2}
(\zeta_{1} \! - \! \zeta_{2}) \vert z_{o} \! + \! \zeta_{1} \! + \! \zeta_{
2} \vert (\tfrac{\zeta_{1}^{4}-2 \zeta_{1}^{2}+1}{\zeta_{1}^{4}+
1})^{2})$.

Without loss of generality, and as the second representative calculation,
one considers, say, the non-empty domain $\{\mathstrut z; \, \Im (z)
\! = \! 0\} \cap \{\mathstrut z; \Re (z) \! \geqslant \! \zeta_{1}\}$.
Once again, let $k \! \in \! \mathbb{Z}_{\geqslant 1}$ be fixed and
sufficiently large with representation $k \! = \! 4q \! + \! 1$, $q
\! \in \! \mathbb{Z}_{\geqslant 1}$. Recalling that, {}from Lemma~4.1,
for $\Re (\zeta) \! \in \! (\zeta_{1},+\infty)$, $\rho (\zeta) \! = \!
\overline{r(\overline{\zeta})}$, one must estimate terms of the type
$\rho (\zeta)(\delta (\zeta))^{2} \me^{-2 \mi t \theta^{u}(\zeta)}$
(respectively~$\overline{r(\overline{\zeta})}(\delta (\zeta))^{-2} \me^{
2 \mi t \theta^{u}(\zeta)})$: in fact, the former will be considered in
detail, with analytic continuation to the unbounded sector $\Omega_{3}$
(see~Figure~3) where $\Re (\mi t \theta^{u}(\zeta)) \! > \! 0$, whilst
the latter can be estimated analogously with analytic continuation to
the unbounded sector $\Omega_{7}$ (see~Figure~3) where $\Re
(\mi t \theta^{u}(\zeta)) \! < \! 0$. For $\Re (\zeta) \! \geqslant \! \zeta_{
1}$, consider the following (rational) Taylor expansion with integral
remainder term for $\rho (\zeta)$, $(\zeta \! - \! \mi)^{k+5} \rho (\zeta)
\! = \! \sum_{n=0}^{k} \tfrac{1}{n!} \mu^{(n)}(\zeta_{1})(\zeta \! - \!
\zeta_{1})^{n} \! + \! \tfrac{1}{k!} \! \int_{\zeta_{1}}^{\zeta}((\cdot
\! - \! \mi)^{k+5} \rho (\cdot))^{(k+1)}(\xi)(\zeta \! - \! \xi)^{k} \,
\md \xi$, where $\mu^{(n)}(\zeta_{1}) \! := \! (\tfrac{\md}{\md \zeta}
)^{n}((\zeta \! - \! \mi)^{k+5} \rho (\zeta)) \vert_{\zeta=\zeta_{1}}$,
$n \! \in \! \{0,1,\ldots,k\}$, $k \! \in \! \mathbb{Z}_{\geqslant 1}$:
also, if one explicitly considers, say, the limiting case when $\zeta_{
1} \! \to \! +\infty$ $(\zeta_{2} \! \to \! 0^{+})$, then, as functions
of $\zeta_{1}$, $\mu^{(n)}(\zeta_{1})$, $n \! \in \! \{0,1,\ldots,k\}$,
$k \! \in \! \mathbb{Z}_{\geqslant 1}$, are in the $\mathbb{C}$-valued
Schwartz class (since $r(\zeta) \! \in \! \mathcal{S}_{\mathbb{C}}
(\mathbb{R}))$. Let $\mathcal{R}(\zeta)$ denote the ``rational part''
of this expansion, and $h(\zeta)$ the ``remainder'', namely, $\mathcal{
R}(\zeta) \! := \! \tfrac{\sum_{n=0}^{k} \frac{1}{n!} \mu^{(n)}(\zeta_{
1})(\zeta-\zeta_{1})^{n}}{(\zeta-\mi)^{k+5}}$ and $h(\zeta) \! := \!
\tfrac{\int_{\zeta_{1}}^{\zeta}((\cdot-\mi)^{k+5} \rho (\cdot))^{(k+1)}
(\xi)(\zeta-\xi)^{k} \, \md \xi}{k!(\zeta-\mi)^{k+5}}$; hence, $\rho
(\zeta) \! = \! \mathcal{R}(\zeta) \! + \! h(\zeta)$, and $\rho (\zeta)
\! - \! \mathcal{R}(\zeta) \! = \! h(\zeta)$ can be interpreted as the
error incurred in approximating $\rho (\zeta)$ by a rational function
of finite degree. Noting that, $\forall \, j \! \in \! \{0,1,\ldots,k\}$,
$k \! \in \! \mathbb{Z}_{\geqslant 1}$, $(\tfrac{\md}{\md \zeta})^{j}
\rho (\zeta) \vert_{\zeta=\zeta_{1}} \! = \! (\tfrac{\md}{\md \zeta}
)^{j} \mathcal{R}(\zeta) \vert_{\zeta=\zeta_{1}}$, it follows that
$(\tfrac{\md}{\md \zeta})^{j}h(\zeta) \vert_{\zeta=\zeta_{1}} \! = \!
0$, $j \! \in \! \{0,1,\ldots,k\}$, $k \! \in \! \mathbb{Z}_{\geqslant
1}$; thence, one splits, further, $h(\zeta)$ as $h(\zeta) \! := \! h_{
I}(\zeta) \! + \! h_{II}(\zeta)$, whence $\rho (\zeta) \! = \! h_{I}
(\zeta) \! + \! (h_{II}(\zeta) \! + \! \mathcal{R}(\zeta))$, where $h_{
I}(\zeta)$ is defined for $\zeta \! \in \! \{\mathstrut z; \, \Im (z) \!
= \! 0\} \cap \{\mathstrut z; \, \Re (z) \! \geqslant \! \zeta_{1}\}$,
where $\Re (\mi t \theta^{u}(\zeta)) \! = \! 0$, and has negligibly
``small norm'' (see~below) as $t \! \to \! +\infty$, and $h_{II}(\zeta)$
has an analytic continuation to $\{\mathstrut z; \Re (z) \! \geqslant
\! \zeta_{1}\} \cap \Omega_{3}$, where $\Re (\mi t \theta^{u}(\zeta))
\! > \! 0$, and $\mathcal{R}(\zeta)$ is a rational function of finite
degree with trivial analytic continuation to $\{\mathstrut z; \, \Re (z)
\! \geqslant \! \zeta_{1}\} \cap \Omega_{3}$. Considering the auxiliary
function (which will be needed below) $\beta (\zeta) \! := \! \tfrac{
(\zeta-\zeta_{1})^{q}}{(\zeta-\mi)^{q+2}}$, $q \! \in \! \mathbb{Z}_{
\geqslant 1}$, define $(\tfrac{h}{\beta})(\theta^{u}) \! := \!
\begin{cases}
\tfrac{h(\zeta (\theta^{u}))}{\beta (\zeta (\theta^{u}))}, &\text{$
\theta^{u} \! \geqslant \! \theta^{u}(\zeta_{1})$,} \\
0, &\text{$\theta^{u} \! < \! \theta^{u}(\zeta_{1})$,}
\end{cases}$ and consider the Fourier transform w.r.t.~$\theta^{u}
(\cdot)$. Via a change of variable argument, one shows that $h(\zeta)
\! = \! \tfrac{(\zeta-\zeta_{1})^{k+1}g_{k}(\zeta,\zeta_{1})}{(\zeta-
\mi)^{k+5}}$, where $g_{k}(\zeta,\zeta_{1}) \! := \! \tfrac{1}{k!} \!
\int_{0}^{1}((\cdot \! - \! \mi)^{k+5} \rho (\cdot))^{(k+1)}(\zeta_{
1} \! + \! (\zeta \! - \! \zeta_{1}) \tau)(1 \! - \! \tau)^{k} \, \md
\tau$: since $\rho (\zeta) \! \in \! \mathcal{S}_{\mathbb{C}}(\mathbb{
R})$ and $\exists \, M \! \in \! \mathbb{R}_{>1}$ and bounded such that
$0 \! < \! \zeta_{2} \! < \! \tfrac{1}{M} \! < \! M \! < \! \zeta_{1}$
and $\vert \zeta_{3} \vert^{2} \! = \! 1$, one shows that, $\forall \,
(\zeta,\zeta_{1}) \! \in \! [\zeta_{1},+\infty) \times (M,+\infty)$,
$\exists \, \widehat{\varkappa} \! \in \! \mathbb{R}_{+}$ and bounded
such that $\vert g_{k}^{(j_{o})}(\zeta,\zeta_{1}) \vert \! \leqslant
\! \widehat{\varkappa}$, $(j_{o},k) \! \in \! \mathbb{Z}_{\geqslant
0} \times \mathbb{Z}_{\geqslant 1}$. With the above choice of $\beta
(\zeta)$ (not the only one possible!), one shows that $\tfrac{h(\zeta)}
{\beta (\zeta)} \! = \! \tfrac{(\zeta-\zeta_{1})^{3q+2}g_{k}(\zeta,
\zeta_{1})}{(\zeta-\mi)^{3q+4}}$, with $\tfrac{h(\zeta)}{\beta (\zeta)}
\! =_{\zeta \to +\infty} \! \mathcal{O}(\zeta^{-2})$; hence, $\vert
\int_{\zeta_{1}}^{+\infty} \tfrac{h(z)}{\beta (z)} \, \md z \vert \!
< \! \infty$. Define the Fourier transform pair (w.r.t.~$\theta^{u}
(\cdot))$: $(\tfrac{h}{\beta})(s) \! := \! \int_{-\infty}^{+\infty}
\me^{\mi s \theta^{u}(\zeta)} \widetilde{(\tfrac{h}{\beta})}(s)
\tfrac{\md s}{\sqrt{2 \pi}}$, $\zeta \! \geqslant \! \zeta_{1}$, and
$\widetilde{(\tfrac{h}{\beta})}(s) \! := \! \int_{\theta^{u}(\zeta_{
1})}^{+\infty} \me^{-\mi s \theta^{u}(\zeta)} (\tfrac{h}{\beta})(\zeta)
\tfrac{\md \theta^{u}(\zeta)}{\sqrt{2 \pi}}$. In order to obtain the
necessary estimates for $(h_{I}(\zeta) \! + \! (h_{II}(\zeta) \! +
\! \mathcal{R}(\zeta)))(\delta (\zeta))^{2} \me^{-2 \mi t \theta^{u}
(\zeta)}$, one needs to show that $(\tfrac{h}{\beta})(\theta^{u}) \!
:= \! \tfrac{h(\zeta (\theta^{u}))}{\beta (\zeta (\theta^{u}))} \!
\in \! \mathcal{H}^{j}(\mathbb{R})$, $0 \! \leqslant \! j \! \leqslant
\! [\tfrac{3q+2}{2}]$, $q \! \in \! \mathbb{Z}_{\geqslant 1}$, where
$\mathcal{H}^{j}(\mathbb{R})$ denotes the $L^{2}$-Sobolev space with
norm $\vert \vert \star (\cdot) \vert \vert_{\mathcal{H}(\mathbb{
R})} \! := \! (\sum_{j=0}^{[\frac{3q+2}{2}]} \vert \vert (\tfrac{\md}{
\md \theta^{u}})^{j} \star \! (\theta^{u}) \vert \vert^{2}_{\mathcal{L}
^{2}(\mathbb{R})})^{1/2}$. In this particular case, one must show that,
for $0 \! < \! \zeta_{2} \! < \! \tfrac{1}{M} \! < \! M \! < \! \zeta_{
1}$ and $\vert \zeta_{3} \vert^{2} \! = \! 1$, with $\zeta \! \geqslant
\! \zeta_{1}$, $\widehat{\mathrm{I}}_{\mathcal{H}} \! := \! \int_{-
\infty}^{+\infty} \vert (\tfrac{\md}{\md \theta^{u}})^{j} \tfrac{h
(\zeta (\theta^{u}))}{\beta (\zeta (\theta^{u}))} \vert^{2} \, \md
\theta^{u} \! < \! \infty$, $0 \! \leqslant \! j \! \leqslant \!
[\tfrac{3q+2}{2}]$, $q \! \in \! \mathbb{Z}_{\geqslant 1}$: in fact,
since $\tfrac{h(\zeta (\theta^{u}))}{\beta (\zeta (\theta^{u}))} \!
\equiv \! 0 \, \, \forall \, \, \theta^{u} \! < \! \theta^{u}(\zeta_{
1})$, the latter integral reduces to $\widehat{\mathrm{I}}_{\mathcal{
H}} \! = \! \int_{\theta^{u}(\zeta_{1})}^{+\infty} \vert (\tfrac{\md}
{\md \theta^{u}})^{j} \tfrac{h(\zeta (\theta^{u}))}{\beta (\zeta
(\theta^{u}))} \vert^{2} \, \md \theta^{u}$, which, via the chain rule
and a change-of-variable argument, is shown to be equal to $\widehat{
\mathrm{I}}_{\mathcal{H}} \! = \! \int_{\zeta_{1}}^{+\infty} \!
\left\vert \tfrac{(\frac{\md}{\md \zeta})^{j} \frac{h(\zeta)}{\beta
(\zeta)}}{(\frac{\md \theta^{u}(\zeta)}{\md \zeta})^{j}} \right\vert^{
2}(\tfrac{\md \theta^{u}(\zeta)}{\md \zeta}) \, \md \zeta$. Now,
recalling that $\tfrac{h(\zeta)}{\beta (\zeta)} \! = \! \tfrac{(\zeta
-\zeta_{1})^{3q+2}}{(\zeta-\mi)^{3q+4}k!} \! \int_{0}^{1}((\cdot \! -
\! \mi)^{k+5} \rho (\cdot))^{(k+1)}(\zeta_{1} \! + \! (\zeta \! - \!
\zeta_{1}) \tau)(1 \! - \! \tau)^{k} \, \md \tau$ and ({}from Section~3)
$\tfrac{\md \theta^{u}(\zeta)}{\md \zeta} \! = \! \zeta^{-3}(\zeta \!
- \! \zeta_{1})(\zeta \! - \! \zeta_{2})(\zeta \! - \! \zeta_{3})(\zeta
\! - \! \overline{\zeta_{3}})$, one shows that, upon deducing $(\tfrac{
h}{\beta})^{(j)}(\zeta) \! =_{\zeta \downarrow \zeta_{1}} \! \mathcal{
O}((\zeta \! - \! \zeta_{1})^{3q+2-j})$, $\widehat{\mathrm{I}}_{\mathcal{
H}} \! \leqslant \! \vert c^{\mathcal{S}}(\zeta_{1}) \underline{c}
(\zeta_{2},\zeta_{3},\overline{\zeta_{3}}) \vert$, $0 \! \leqslant \! j
\! \leqslant \! [\tfrac{3q+2}{2}]$, $q \! \in \! \mathbb{Z}_{\geqslant
1}$, $0 \! < \! \zeta_{2} \! < \! \tfrac{1}{M} \! < \! M \! < \! \zeta_{
1}$ and $\vert \zeta_{3} \vert^{2} \! = \! 1$, that is, $\vert \vert
(\tfrac{h}{\beta})(\cdot) \vert \vert_{\mathcal{H}(\mathbb{R})} \! = \!
(\sum_{j=0}^{[\frac{3q+2}{2}]} \vert \vert (\tfrac{\md}{\md \theta^{u}
})^{j}(\tfrac{h}{\beta})(\theta^{u}) \vert \vert^{2}_{\mathcal{L}^{2}
(\mathbb{R})})^{1/2} \! < \! \infty$; hence, by Parseval's Theorem,
for $0 \! < \! \zeta_{2} \! < \! \tfrac{1}{M} \! < \! M \! < \! \zeta_{
1}$ and $\vert \zeta_{3} \vert^{2} \! = \! 1$, $\int_{-\infty}^{+
\infty}(1 \! + \! s^{2})^{j} \vert \widetilde{(\tfrac{h}{\beta})}(s)
\vert^{2} \, \md s \! \leqslant \! \vert c^{\mathcal{S}}(\zeta_{1})
\underline{c}(\zeta_{2},\zeta_{3},\overline{\zeta_{3}}) \vert$, $0 \!
\leqslant \! j \! \leqslant \! [\tfrac{3q+2}{2}]$, $q \! \in \! \mathbb{
Z}_{\geqslant 1}$. Recalling {}from the Fourier transform pair that,
for $\zeta \! \geqslant \! \zeta_{1}$, $(\tfrac{h}{\beta})(\zeta) \! =
\! \int_{-\infty}^{+\infty} \me^{\mi s \theta^{u}(\zeta)} \widetilde{
(\tfrac{h}{\beta})}(s) \tfrac{\md s}{\sqrt{2 \pi}}$, it follows, by
defining $h_{I}(\zeta) \! := \! \beta (\zeta) \! \int_{t}^{+\infty}
\me^{\mi s \theta^{u}(\zeta)} \widetilde{(\tfrac{h}{\beta})}(s) \tfrac{
\md s}{\sqrt{2 \pi}}$ and $h_{II}(\zeta) \! := \! \beta (\zeta) \!
\int_{-\infty}^{t} \me^{\mi s \theta^{u}(\zeta)} \widetilde{(\tfrac{
h}{\beta})}(s) \tfrac{\md s}{\sqrt{2 \pi}}$, that $h(\zeta) \! = \!
h_{I}(\zeta) \! + \! h_{II}(\zeta)$. For $\zeta \! \in \! \{\mathstrut
z; \, \Im (z) \! = \! 0\} \cap \{\mathstrut z; \, \Re (z) \! \geqslant
\! \zeta_{1}\}$, recalling the estimate for $\delta (\zeta)$ given in
Proposition~4.1, using the Cauchy-Schwarz inequality for integrals, as
well as the inequality $\vert \tfrac{\zeta-\zeta_{1}}{\zeta+\zeta_{1}}
\vert \! \leqslant \! 1$, $\zeta \! \geqslant \! \zeta_{1}$, one shows
that $\vert (\delta (\zeta))^{2} \me^{-2 \mi t \theta^{u}(\zeta)}h_{I}
(\zeta) \vert \! = \! \vert (\delta (\zeta))^{2} \me^{-2 \mi t \theta^{
u}(\zeta)} \beta (\zeta) \! \int_{t}^{+\infty} \me^{\mi s \theta^{u}
(\zeta)} \widetilde{(\tfrac{h}{\beta})}(s) \tfrac{\md s}{\sqrt{2 \pi}}
\vert \! \leqslant \! \vert \vert (\delta (\cdot))^{2} \vert \vert_{
\mathcal{L}^{\infty}(\mathbb{C})} \tfrac{\vert \beta (\zeta) \vert}{
\sqrt{2 \pi}}(\int_{t}^{+\infty}(1 \! + \! s^{2})^{-j} \, \md s)^{1/2}
(\int_{t}^{+\infty}(1 \! + \! s^{2})^{j} \vert \widetilde{(\tfrac{h}{
\beta})}(s) \vert^{2} \, \md s)^{1/2}$: noting that $\int_{t}^{+\infty}
(1 \! + \! s^{2})^{-j} \, \md s \! \leqslant \! \tfrac{t^{-(2j-1)}}{(2
j-1)}$, $2j \! - \! 1 \! > \! 0$, and recalling the Parseval estimate
$\int_{t}^{+\infty}(1 \! + \! s^{2})^{j} \vert \widetilde{(\tfrac{h}{
\beta})}(s) \vert^{2} \, \md s \! \leqslant \! \int_{-\infty}^{+\infty}
(1 \! + \! s^{2})^{j} \vert \widetilde{(\frac{h}{\beta})}(s) \vert^{2}
\, \md s \! \leqslant \! \vert c^{\mathcal{S}}(\zeta_{1}) \underline{
c}(\zeta_{2},\zeta_{3},\overline{\zeta_{3}}) \vert$, $0 \! \leqslant \!
j \! \leqslant \! [\tfrac{3q+2}{2}]$, it follows that, with the choice
$\beta (\zeta) \! = \! \tfrac{(\zeta-\zeta_{1})^{q}}{(\zeta-\mi)^{q+2}}$,
$q \! \in \! \mathbb{Z}_{\geqslant 1}$, $\vert \me^{-2 \mi t \theta^{
u}(\zeta)}h_{I}(\zeta) \vert \! \leqslant \! \tfrac{\vert c^{\mathcal{
S}}(\zeta_{1}) \underline{c}(\zeta_{2},\zeta_{3},\overline{\zeta_{3}})
\vert}{\vert \zeta-\mi \vert^{2}t^{(j-1/2)}}$, $1 \! \leqslant \! j \!
\leqslant \! [\tfrac{3q+2}{2}]$, $q \! \in \! \mathbb{Z}_{\geqslant 1}$,
$\zeta \! \geqslant \! \zeta_{1}$, $0 \! < \! \zeta_{2} \! < \! \tfrac{
1}{M} \! < \! M \! < \! \zeta_{1}$ and $\vert \zeta_{3} \vert^{2} \! =
\! 1$. Since, for $\zeta \! \in \! \{\mathstrut z; \, \Re (z) \!
\geqslant \! \zeta_{1}\} \cap \Omega_{3}$, $\Re (\mi t \theta^{u}
(\zeta)) \! > \! 0$, $h_{II}(\zeta)$ has an analytic continuation to
the ray (parametrised by $v)$ $\zeta \! = \! \zeta (v) \! = \! \zeta_{1}
\! + \! \tfrac{v}{\sqrt{2}}(\zeta_{1} \! - \! \zeta_{2}) \me^{-\frac{\mi
\pi}{4}}$, $v \! \in \! \mathbb{R}_{\geqslant 0}$; hence, on this ray,
it follows that $\widehat{\mathrm{I}}_{h_{II}} \! := \! \vert (\delta
(\zeta))^{2} \me^{-2 \mi t \theta^{u}(\zeta)}h_{II}(\zeta) \vert \! =
\! \vert (\delta (\zeta))^{2} \me^{-2 \mi t \theta^{u}(\zeta)} \beta
(\zeta) \! \int_{-\infty}^{t} \! \me^{\mi s \theta^{u}(\zeta)}
\widetilde{(\tfrac{h}{\beta})}(s) \tfrac{\md s}{\sqrt{2 \pi}} \vert \!
\leqslant \! \vert \vert (\delta (\cdot))^{2} \vert \vert_{\mathcal{
L}^{\infty}(\mathbb{C})} \vert \beta (\zeta) \vert \vert \me^{-\mi t
\theta^{u}(\zeta)} \vert \vert \int_{-\infty}^{t} \! \widetilde{(\tfrac{
h}{\beta})}(s) \tfrac{\md s}{\sqrt{2 \pi}} \vert$, \linebreak[4]
thus, {}from the Cauchy-Schwarz inequality for integrals, $\widehat{
\mathrm{I}}_{h_{II}} \! \leqslant \! \vert \vert (\delta (\cdot))^{
2} \vert \vert_{\mathcal{L}^{\infty}(\mathbb{C})} \tfrac{\vert \beta
(\alpha) \vert}{\sqrt{2 \pi}} \me^{-t \Re (\mi \theta^{u}(\zeta))}
\linebreak[4]
\cdot (\int_{-\infty}^{t}(1 \! + \! s^{2})^{-j} \, \md s)^{1/2}(\int_{-
\infty}^{t}(1 \! + \! s^{2})^{j} \vert \widetilde{(\tfrac{h}{\beta})}
(s) \vert^{2} \, \md s)^{1/2}$. Recalling {}from the previous calculation
that $\int_{-\infty}^{t}(1 \! + \! s^{2})^{-j} \, \md s \! \leqslant \!
\pi$, and deducing, {}from the above Parseval estimate, that $\int_{-
\infty}^{t}(1 \! + \! s^{2})^{j} \vert \widetilde{(\tfrac{h}{\beta})}
(s) \vert^{2} \, \md s \! \leqslant \! \int_{-\infty}^{+\infty}(1 \! +
\! s^{2})^{j} \vert \widetilde{(\tfrac{h}{\beta})}(s) \vert^{2} \, \md
s \! \leqslant \! \vert c^{\mathcal{S}}(\zeta_{1}) \underline{c}(\zeta_{
2},\zeta_{3},\overline{\zeta_{3}}) \vert$, $0 \! \leqslant \! j \!
\leqslant \! [\tfrac{3q+2}{2}]$, $q \! \in \! \mathbb{Z}_{\geqslant 1}$,
it follows {}from the definition of $\beta (\zeta)$ that $\widehat{
\mathrm{I}}_{h_{II}} \! \leqslant \! \tfrac{\vert c^{\mathcal{S}}(\zeta_{
1}) \underline{c}(\zeta_{2},\zeta_{3},\overline{\zeta_{3}}) \vert \vert
\zeta_{1}-\zeta_{2} \vert^{q}v^{q}}{\vert \zeta-\mi \vert^{2}} \me^{-t
\Re (\mi \theta^{u}(\zeta))}$, $q \! \in \! \mathbb{Z}_{\geqslant 1}$,
$v \! \geqslant \! 0$. Recalling that $\theta^{u}(\zeta) \! = \! \tfrac{
1}{2}(\zeta \! - \! \tfrac{1}{\zeta})(z_{o} \! + \! \zeta \! + \! \tfrac{
1}{\zeta})$, for $\zeta \! = \! \zeta (v) \! = \! \zeta_{1} \! + \!
\tfrac{v}{\sqrt{2}}(\zeta_{1} \! - \! \zeta_{2}) \me^{-\frac{\mi \pi}{
4}}$, $v \! \in \! \mathbb{R}_{\geqslant 0}$, one shows that $\Re
(\mi \theta^{u}(\zeta)) \! = \! \tfrac{1}{4}v(\zeta_{1} \! - \! \zeta_{2})
(\zeta_{1} \! + \! \tfrac{1}{2}(\zeta_{1} \! - \! \zeta_{2})v)(\tfrac{
\zeta_{1}^{2}+\zeta_{1}(\zeta_{1}-\zeta_{2})v+\frac{1}{2}(\zeta_{1}-
\zeta_{2})^{2}v^{2}-1}{\zeta_{1}^{2}+\zeta_{1}(\zeta_{1}-\zeta_{2})v
+\frac{1}{2}(\zeta_{1}-\zeta_{2})^{2}v^{2}})^{2} \! + \! \tfrac{1}{4}
v(\zeta_{1} \! - \! \zeta_{2})(\zeta_{1} \! + \! \tfrac{1}{2}(\zeta_{
1} \! - \! \zeta_{2})v)(\tfrac{\zeta_{1}^{2}+\zeta_{1}(\zeta_{1}-\zeta_{
2})v+\frac{1}{2}(\zeta_{1}-\zeta_{2})^{2}v^{2}+1}{\zeta_{1}^{2}+\zeta_{
1}(\zeta_{1}-\zeta_{2})v+\frac{1}{2}(\zeta_{1}-\zeta_{2})^{2}v^{2}})^{2}
\! + \! \tfrac{1}{4}v(\zeta_{1} \! - \! \zeta_{2})z_{o}(\tfrac{\zeta_{1}
^{2}+\zeta_{1}(\zeta_{1}-\zeta_{2})v+\frac{1}{2}(\zeta_{1}-\zeta_{2})^{
2}v^{2}+1}{\zeta_{1}^{2}+\zeta_{1}(\zeta_{1}-\zeta_{2})v+\frac{1}{2}
(\zeta_{1}-\zeta_{2})^{2}v^{2}})$. Setting $\widetilde{a}(v) \! := \!
(\zeta_{1} \! + \! \tfrac{1}{2}(\zeta_{1} \! - \! \zeta_{2})v)^{2} \!
+ \! (\tfrac{1}{2}(\zeta_{1} \! - \! \zeta_{2})v)^{2}$, one shows that,
for $v \! \geqslant \! 0$, $\widetilde{a}(v) \! \geqslant \! \widetilde{
a}(0)$; hence, with $(\tfrac{\zeta_{1}^{2}+\zeta_{1}(\zeta_{1}-\zeta_{2}
)v+\frac{1}{2}(\zeta_{1}-\zeta_{2})^{2}v^{2}+1}{\zeta_{1}^{2}+\zeta_{1}
(\zeta_{1}-\zeta_{2})v+\frac{1}{2}(\zeta_{1}-\zeta_{2})^{2}v^{2}}) \!
\geqslant \! (\tfrac{\zeta_{1}^{2}+\zeta_{1}(\zeta_{1}-\zeta_{2})v+
\frac{1}{2}(\zeta_{1}-\zeta_{2})^{2}v^{2}-1}{\zeta_{1}^{2}+\zeta_{1}
(\zeta_{1}-\zeta_{2})v+\frac{1}{2}(\zeta_{1}-\zeta_{2})^{2}v^{2}})$,
$v \! \geqslant \! 0$, one shows that $\Re (\mi \theta^{u}(\zeta))
\! \geqslant \! \tfrac{1}{2}v(\zeta_{1} \! - \! \zeta_{2})(\zeta_{1}
\! + \! \tfrac{1}{2}(\zeta_{1} \! - \! \zeta_{2})v)(\tfrac{\zeta_{1}^{
2}+\zeta_{1}(\zeta_{1}-\zeta_{2})v+\frac{1}{2}(\zeta_{1}-\zeta_{2})^{
2}v^{2}-1}{\zeta_{1}^{2}+\zeta_{1}(\zeta_{1}-\zeta_{2})v+\frac{1}{2}
(\zeta_{1}-\zeta_{2})^{2}v^{2}})^{2} \! + \! \tfrac{v}{4}(\zeta_{1} \!
- \! \zeta_{2})z_{o}(\tfrac{\zeta_{1}^{2}+\zeta_{1}(\zeta_{1}-\zeta_{2}
)v+\frac{1}{2}(\zeta_{1}-\zeta_{2})^{2}v^{2}-1}{\zeta_{1}^{2}+\zeta_{
1}(\zeta_{1}-\zeta_{2})v+\frac{1}{2}(\zeta_{1}-\zeta_{2})^{2}v^{2}})$.
Noting that $(1 \! - \! (\widetilde{a}(0))^{-1}) \! \leqslant \! (1 \!
- \! (\widetilde{a}(v))^{-1}) \! \leqslant \! 1$, $(1 \! - \! \zeta_{
1}^{-2}) \! \geqslant \! (1 \! - \! \zeta_{1}^{-2})^{2}$, and $2 \! +
\! v \! \geqslant \! v$, $v \! \geqslant \! 0$, one establishes that
$\Re (\mi \theta^{u}(\zeta)) \! \geqslant \! \tfrac{1}{4}v(\zeta_{
1} \! - \! \zeta_{2})(1 \! - \! \zeta_{1}^{-2})^{2}(z_{o} \! + \!
(\zeta_{1} \! - \! \zeta_{2})v)$. {}From the identities (valid for
$v \! \geqslant \! 0)$ $1 \! \geqslant \! (v \! + \! 1)^{-1}$, $1 \!
\geqslant \! v(v \! + \! 1)^{-1}$, and $-1 \! \leqslant \! \tfrac{
v-1}{v+1} \! \leqslant \! 1$, and choosing an arbitrarily fixed,
sufficiently small positive real number $\gamma_{o}$ such that
$\gamma_{o} \! < \! (v \! + \! 1)^{-1} \! \leqslant \! 1$, $v \!
\geqslant \! 0$, one deduces that $\Re (\mi \theta^{u}(\zeta)) \!
\geqslant \! \tfrac{1}{4} \gamma_{o}v^{2}(\zeta_{1} \! - \! \zeta_{2}
)(1 \! - \! \zeta_{1}^{-2})^{2} \vert z_{o} \! + \! \zeta_{1} \! - \!
\zeta_{2} \vert$; hence, $-t \Re (\mi \theta^{u}(\zeta)) \! \leqslant
\! -\tfrac{1}{4}t \gamma_{o} v^{2}(\zeta_{1} \! - \! \zeta_{2})(1 \! -
\! \zeta_{1}^{-2})^{2} \vert z_{o} \! + \! \zeta_{1} \! + \! \zeta_{2}
\vert$, $v \! \in \! \mathbb{R}_{\geqslant 0}$. Recalling that $\widehat{
\mathrm{I}}_{h_{II}} \! \leqslant \! \tfrac{\vert c^{\mathcal{S}}(\zeta_{
1}) \underline{c}(\zeta_{2},\zeta_{3},\overline{\zeta_{3}}) \vert \vert
\zeta_{1}-\zeta_{2} \vert^{q}v^{q}}{\vert \zeta-\mi \vert^{2}} \me^{-t
\Re (\mi \theta^{u}(\zeta))}$, it follows {}from the above inequality
for $-t \Re (\mi \theta^{u}(\zeta))$ (on the ray $\zeta \! = \! \zeta
(v) \! = \! \zeta_{1} \! + \! \tfrac{v}{\sqrt{2}}(\zeta_{1} \! - \!
\zeta_{2}) \me^{-\frac{\mi \pi}{4}}$, $v \! \in \! \mathbb{R}_{\geqslant
0})$ that $\vert \me^{-2 \mi t \theta^{u}(\zeta)}h_{II}(\zeta) \vert
\! \leqslant \! \tfrac{\vert c^{\mathcal{S}}(\zeta_{1}) \underline{c}
(\zeta_{2},\zeta_{3},\overline{\zeta_{3}}) \vert}{\vert z_{o}+\zeta_{
1}-\zeta_{2} \vert^{q/2} \vert \zeta-\mi \vert^{2}t^{q/2}}$, $q \! \in
\! \mathbb{Z}_{\geqslant 1}$, $0 \! < \! \zeta_{2} \! < \! \tfrac{1}{M}
\! < \! M \! < \! \zeta_{1}$ and $\vert \zeta_{3} \vert^{2} \! = \! 1$.

Proceeding in the above-demonstrated manner for the remaining domains and
sectors, and setting $\mathcal{R}(\zeta) \! \equiv \! 0$ for $\zeta \! < \!
0$, one obtains the results stated in the Lemma; however, in order to analyse
the (complex conjugate pair of) first-order saddle points at $\zeta_{3}$ and
$\overline{\zeta_{3}}$, one uses the fact that $r(0) \! = \! 0$ and the
solution of the RHP is bounded outside open neighbourhoods of $\{\zeta_{
i}\}_{i=1}^{4}$, and then proceeds according to the Remark in Section~3.1
of \cite{a46}. \hfill $\square$

{}From Lemmae~4.1 and~4.2, one derives the following (normalised
at $\infty)$ RHP for $m^{\sharp}(\zeta)$ on the augmented contour
$\Sigma^{\prime}$:
\begin{ccccc}
Let $\widehat{m}^{c}(\zeta)$ be the solution of the {\rm RHP}
formulated in Lemma~{\rm 4.1}. As $t \! \to \! +\infty$ such that $0 \!
< \! \zeta_{2} \! < \! \tfrac{1}{M} \! < \! M \! < \! \zeta_{1}$ and $\vert
\zeta_{3} \vert^{2} \! = \! 1$, with $M \! \in \! \mathbb{R}_{>1}$ and
bounded, and for arbitrarily fixed, sufficiently large $l \! \in \! \mathbb{
Z}_{\geqslant 1}$, let the estimates in Lemma~{\rm 4.2} be valid.
Set
\begin{equation*}
m^{\sharp}(\zeta) \! := \!
\begin{cases}
\widehat{m}^{c}(\zeta), &\text{$\zeta \! \in \! \Omega_{1} \cup \Omega_{
2}$,} \\
\widehat{m}^{c}(\zeta)(\mathrm{I} \! + \! \underline{w}^{a}_{+}(\zeta))
^{-1}, &\text{$\zeta \! \in \! \Omega_{3} \cup \Omega_{4} \cup \Omega_{
5} \cup \Omega_{6}$,} \\
\widehat{m}^{c}(\zeta)(\mathrm{I} \! - \! \underline{w}^{a}_{-}(\zeta))
^{-1}, &\text{$\zeta \! \in \! \Omega_{7} \cup \Omega_{8} \cup \Omega_{
9} \cup \Omega_{10}$,}
\end{cases}
\end{equation*}
where $\underline{w}^{a}_{\pm}(\zeta) \! := \! (\delta (\zeta))^{\mathrm{
ad}(\sigma_{3})} \me^{-\mi t \theta^{u}(\zeta) \mathrm{ad}(\sigma_{
3})}w^{a}_{\pm}(\zeta)$, with $w^{a}_{+}(\zeta) \! = \! (h_{II}(\zeta)
\! + \! \mathcal{R}(\zeta)) \sigma_{+}$ and $w^{a}_{-}(\zeta) \! = \!
-(\overline{h_{II}(\overline{\zeta})} \! + \! \overline{\mathcal{R}
(\overline{\zeta})}) \sigma_{-}$. Then $m^{\sharp}(\zeta) \colon \mathbb{
C} \setminus \Sigma^{\prime} \! \to \! \mathrm{SL}(2,\mathbb{C})$ solves
the following {\rm RHP:} (1) $m^{\sharp}(\zeta)$ is piecewise holomorphic
$\forall \, \zeta \! \in \! \mathbb{C} \setminus \Sigma^{\prime};$ (2)
$m^{\sharp}_{\pm}(\zeta) \! := \! \lim_{\genfrac{}{}{0pt}{2}{\zeta^{\prime}
\, \to \, \zeta}{\zeta^{\prime} \, \in \, \pm \, \mathrm{side} \, \mathrm{of}
\, \Sigma^{\prime}}} m^{\sharp}(\zeta^{\prime})$ satisfy the jump condition
$m^{\sharp}_{+}(\zeta) \! = \! m^{\sharp}_{-}(\zeta) \mathcal{G}^{\sharp}
(\zeta)$, $\zeta \! \in \! \Sigma^{\prime}$, where $\mathcal{G}^{\sharp}
(\zeta) \! := \! (\mathrm{I} \! - \! w^{\sharp}_{-}(\zeta))^{-1}(\mathrm{I}
\! + \! w^{\sharp}_{+}(\zeta))$, with
\begin{equation*}
(\mathrm{I} \! - \! w^{\sharp}_{-}(\zeta))^{-1}(\mathrm{I} \! + \! w^{
\sharp}_{+}(\zeta)) \! = \!
\begin{cases}
(\mathrm{I} \! - \! \underline{w}^{o}_{-}(\zeta))(\mathrm{I} \! + \!
\underline{w}^{o}_{+}(\zeta))^{-1}, &\text{$\zeta \! \in \! \mathbb{R}$,} \\
(\mathrm{I} \! + \! \underline{w}^{a}_{+}(\zeta)), &\text{$\zeta \! \in \!
\mathrm{L}$,} \\
(\mathrm{I} \! - \! \underline{w}^{a}_{-}(\zeta))^{-1}, &\text{$\zeta \! \in
\! \overline{\mathrm{L}}$,}
\end{cases}
\end{equation*}
$\underline{w}^{o}_{\pm}(\zeta) \! := \! (\delta_{\pm}(\zeta))^{\mathrm{
ad}(\sigma_{3})} \me^{-\mi t \theta^{u}(\zeta) \mathrm{ad}(\sigma_{3})}
w^{o}_{\pm}(\zeta)$, where $w^{o}_{+}(\zeta) \! = \! h_{I}(\zeta) \sigma_{
+}$ and $w^{o}_{-}(\zeta) \! = \! -\overline{h_{I}(\zeta)} \, \sigma_{-}$,
and $\underline{w}^{a}_{\pm}(\zeta)$ are defined above; (3) as $\zeta \!
\to \! \infty$, $\zeta \! \in \! \mathbb{C} \setminus \Sigma^{\prime}$,
$m^{\sharp}(\zeta) \! = \! \mathrm{I} \! + \! \mathcal{O}(\zeta^{-1});$ and
(4) $m^{\sharp}(\zeta)$ satisfies the symmetry reduction $m^{\sharp}
(\zeta) \! = \! \sigma_{1} \overline{m^{\sharp}(\overline{\zeta})} \,
\sigma_{1}$ and the condition $(m^{\sharp}(0)(\delta (0))^{\sigma_{3}}
\sigma_{2})^{2} \! = \! \mathrm{I}$. Furthermore, $\underline{w}^{o}_{\pm}
(\zeta) \! \in \! \cap_{p \in \{1,2,\infty\}} \mathcal{L}^{p}_{\mathrm{
M}_{2}(\mathbb{C})}(\mathbb{R})$, $\underline{w}^{a}_{+}(\zeta) \! \in \!
\cap_{p \in \{1,2,\infty\}} \mathcal{L}^{p}_{\mathrm{M}_{2}(\mathbb{C})}
(\mathrm{L})$, and $\underline{w}^{a}_{-}(\zeta) \! \in \! \cap_{p \in
\{1,2,\infty\}} \mathcal{L}^{p}_{\mathrm{M}_{2}(\mathbb{C})}(\overline{
\mathrm{L}})$.
\end{ccccc}

\emph{Proof.} {}From Lemma~4.1, one rewrites the jump matrix, $\widehat{
\mathcal{G}}^{c}(\zeta)$, in the BC form, $\widehat{\mathcal{G}}^{c}(\zeta)
\! = \! (\mathrm{I} \! - \! \widehat{w}^{c}_{-}(\zeta))^{-1}(\mathrm{I} \!
+ \! \widehat{w}^{c}_{+}(\zeta))$, where $\widehat{w}^{c}_{\pm}(\zeta) \!
:= \! (\delta_{\pm}(\zeta))^{\mathrm{ad}(\sigma_{3})} \me^{-\mi t \theta
^{u}(\zeta) \mathrm{ad}(\sigma_{3})}w^{c}_{\pm}(\zeta)$, with $w^{c}_{
+}(\zeta) \! = \! \rho (\zeta) \sigma_{+}$ and $w_{-}(\zeta) \! = \! -
\overline{\rho (\overline{\zeta})} \, \sigma_{-}$, and $\rho (\zeta)$ as
defined in Lemma~4.1. Defining $m^{\sharp}(\zeta)$ as in the Lemma, one
arrives, as a consequence of the above BC factorisation for the jump
matrix, the RHP for $\widehat{m}^{c}(\zeta)$ formulated in Lemma~4.1, the
decomposition $\rho (\zeta) \! = \! h_{I}(\zeta) \! + \! (h_{II}(\zeta)
\! + \! \mathcal{R}(\zeta))$, and Lemma~4.2, at the RHP for $m^{\sharp}
(\zeta)$. \hfill $\square$

The second main objective of this section is to reformulate the RHP for
$m^{\sharp}(\zeta)$ on $\Sigma^{\prime}$ as an equivalent RHP for $m^{
\Sigma^{\sharp}}(\zeta)$ (see~Lemma~4.6) on the truncated contour (see
Figure~4)
\begin{equation}
\Sigma^{\sharp} \! = \! \Sigma^{\prime} \setminus (\mathbb{R} \cup
(\mathrm{L}_{\varepsilon} \cup \mathrm{L}_{<}) \cup \overline{(\mathrm{
L}_{\varepsilon} \cup \mathrm{L}_{<})}) \! := \! \Sigma_{A^{\prime}}
\cup \Sigma_{B^{\prime}},
\end{equation}
with $\Sigma_{A^{\prime}} \cap \Sigma_{B^{\prime}} \! = \! \emptyset$.
\begin{figure}[htb]
\begin{center}
\unitlength=1cm
\vspace{0.65cm}
\begin{picture}(10,6)(0,0)
\thicklines
\put(7,3){\makebox(0,0){$\bullet$}}
\put(3,3){\makebox(0,0){$\bullet$}}
\put(7,2.5){\makebox(0,0){$\zeta_{1}$}}
\put(3,2.5){\makebox(0,0){$\zeta_{2}$}}
\put(7,4.25){\makebox(0,0){$\Sigma_{B^{\prime}}$}}
\put(3,4.25){\makebox(0,0){$\Sigma_{A^{\prime}}$}}
\put(10,6){\vector(-1,-1){2}}
\put(10,0){\vector(-1,1){2}}
\put(8,4){\line(-1,-1){1}}
\put(8,2){\line(-1,1){1}}
\put(5.85,4.15){\vector(1,-1){1}}
\put(5.85,1.85){\vector(1,1){1}}
\put(5.85,4.15){\line(1,-1){1.15}}
\put(5.85,1.85){\line(1,1){1.15}}
\put(2,2){\vector(1,1){1.375}}
\put(2,4){\vector(1,-1){1.375}}
\put(4,2){\vector(-1,1){1.375}}
\put(4,4){\vector(-1,-1){1.375}}
\put(3,3){\line(-1,1){1.15}}
\put(3,3){\line(-1,-1){1.15}}
\put(3,3){\line(1,1){1.15}}
\put(3,3){\line(1,-1){1.15}}
\end{picture}
\end{center}
\caption{Truncated contour $\Sigma^{\sharp} \! := \! \Sigma_{A^{\prime}}
\cup \Sigma_{B^{\prime}}$}
\end{figure}
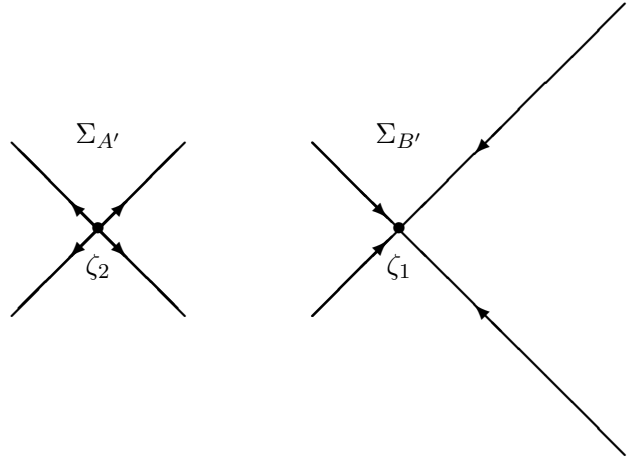
In going {}from the RHP for $m^{\sharp}(\zeta)$ on $\Sigma^{\prime}$ to the
RHP for $m^{\Sigma^{\sharp}}(\zeta)$ on $\Sigma^{\sharp}$, the error
incurred (as a result of the truncation of the integration contour) will be
estimated explicitly, and shown to be, as $t \! \to \! +\infty$, $\mathcal{O}
(\tfrac{\underline{c}(\zeta_{1},\zeta_{2},\zeta_{3},\overline{\zeta_{3}})
\diamondsuit (\zeta)}{\vert z_{o}+\zeta_{1}+\zeta_{2} \vert^{l}t^{l}})$, $l
\! \in \! \mathbb{Z}_{\geqslant 1}$ and arbitrarily large, and $\diamondsuit
(\zeta) \! \in \! \mathcal{L}^{\infty}_{\mathrm{M}_{2}(\mathbb{C})}(\mathbb{
C} \setminus \Sigma^{\sharp})$. In the course of these estimations, it will
be shown that the (asymptotic) contributions of the functions $h_{I}(\zeta)
\colon \mathbb{R} \! \to \! \mathbb{C}$ (respectively~$\overline{h_{I}
(\zeta)} \colon \mathbb{R} \! \to \! \mathbb{C})$ and $h_{II}(\zeta) \colon
\mathrm{L} \! \to \! \mathbb{C}$ (respectively~$\overline{h_{II}(\overline{
\zeta})} \colon \overline{\mathrm{L}} \! \to \! \mathbb{C})$ are ``negligibly
small'' (see~Lemma~4.4), and the contribution to the leading order
asymptotics of the solution of the RHP for $m^{\sharp}(\zeta)$ coming
{}from $\mathbb{R}$, $\mathrm{L}_{\varepsilon} \cup \overline{\mathrm{L}_{
\varepsilon}}$, and $\mathrm{L}_{<} \cup \overline{\mathrm{L}_{<}}$ are
negligible (see~Lemma~4.4). Using Lemma~3.1, the solution of the RHP for
$m^{\sharp}(\zeta)$ on $\Sigma^{\prime}$ has the following integral
representation,
\begin{equation*}
m^{\sharp}(\zeta) \! = \! \mathrm{I} \! + \! \int\nolimits_{\Sigma^{
\prime}} \dfrac{\mu^{\sharp}(z)w^{\sharp}(z)}{(z \! - \! \zeta)} \,
\dfrac{\md z}{2 \pi \mi}, \quad \zeta \! \in \! \mathbb{C} \setminus
\Sigma^{\prime},
\end{equation*}
where $\mu^{\sharp}(\zeta) \! := \! ((\mathbf{1} \! - \! C_{w^{\sharp}})^{
-1} \mathrm{I})(\zeta)$, $C_{w^{\sharp}} \, \star \! := \! C_{+}(\star \,
w^{\sharp}_{-}) \! + \! C_{-}(\star \, w^{\sharp}_{+})$, $\star \! \in \!
\mathcal{L}^{2}_{\mathrm{M}_{2}(\mathbb{C})}(\Sigma^{\prime})$, $(C_{\pm}
\star)(\zeta) \! := \! \lim_{\genfrac{}{}{0pt}{2}{\zeta^{\prime}\, \to \,
\zeta}{\zeta^{\prime} \, \in \, \pm \, \mathrm{side} \, \mathrm{of} \,
\Sigma^{\prime}}} \int_{\Sigma^{\prime}} \tfrac{\star (z)}{z-\zeta} \,
\tfrac{\md z}{2 \pi \mi}$, with $w^{\sharp}_{\pm}(\zeta)$ defined in
Lemma~4.3, and $w^{\sharp}(\zeta) \! := \! \sum_{l \in \{\pm\}} \! w^{
\sharp}_{l}(\zeta)$. One notes {}from Lemma~4.3 that: (1) for $\zeta \! \in
\! \mathbb{R}$, $w^{\sharp}(\zeta) \! = \! w^{\sharp}_{-}(\zeta) \! + \! w^{
\sharp}_{+}(\zeta) \! = \! \underline{w}^{o}_{-}(\zeta) \! + \! \underline{
w}^{o}_{+}(\zeta) \! = \! \left(
\begin{smallmatrix}
0 & -h_{I}(\zeta)(\delta_{+}(\zeta))^{2} \me^{-2 \mi t \theta^{u}(\zeta)}
\\
\overline{h_{I}(\zeta)}(\delta_{-}(\zeta))^{-2} \me^{2 \mi t \theta^{u}
(\zeta)} & 0
\end{smallmatrix}
\right)$; (2) for $\zeta \! \in \! \mathrm{L}$, since $w^{\sharp}_{-}
(\zeta) \! = \! \underline{w}^{a}_{-}(\zeta) \! = \!
\left(
\begin{smallmatrix}
0 & 0 \\
0 & 0
\end{smallmatrix}
\right)$, $w^{\sharp}(\zeta) \! = \! w^{\sharp}_{+}(\zeta) \! = \!
\underline{w}^{a}_{+}(\zeta) \! = \! (h_{II}(\zeta) \! + \! \mathcal{
R}(\zeta))(\delta (\zeta))^{2} \me^{-2 \mi t \theta^{u}(\zeta)} \sigma_{
+}$; and (3) for $\zeta \! \in \! \overline{\mathrm{L}}$, since $w^{
\sharp}_{+}(\zeta) \! = \! \underline{w}^{a}_{+}(\zeta) \! = \!
\left(
\begin{smallmatrix}
0 & 0 \\
0 & 0
\end{smallmatrix}
\right)$, $w^{\sharp}(\zeta) \! = \! w^{\sharp}_{-}(\zeta) \! = \!
\underline{w}^{a}_{-}(\zeta) \! = \! -(\overline{h_{II}(\overline{
\zeta})} \! + \! \overline{\mathcal{R}(\overline{\zeta})})(\delta
(\zeta))^{-2} \me^{2 \mi t \theta^{u}(\zeta)} \sigma_{-}$. To carry
out the second main objective of this section, and guided by the latter
expressions, one decomposes $w^{\sharp}(\zeta) \! = \! \sum_{l \in \{
\pm\}} \! w^{\sharp}_{l}(\zeta)$ as
\begin{equation}
w^{\sharp}(\zeta) \! = \! w^{e}(\zeta) \! + \! w^{\prime}(\zeta),
\qquad w^{e}(\zeta) \! := \! w^{\widehat{a}}(\zeta) \! + \! w^{b}
(\zeta) \! + \! w^{c}(\zeta),
\end{equation}
where: (1) $w^{\widehat{a}}(\zeta) \! := \! w^{\sharp}(\zeta) \! \!
\upharpoonright_{\mathbb{R}}$ has support on $\mathbb{R}$, and
consists of the contribution to $w^{\sharp}(\zeta)$ {}from $h_{I}(\zeta)$
and $\overline{h_{I}(\zeta)}$; (2) $w^{b}(\zeta)$ has support on
$\mathrm{L} \cup \overline{\mathrm{L}}$, and consists of the
contribution to $w^{\sharp}(\zeta)$ {}from $h_{II}(\zeta) \! \!
\upharpoonright_{\mathrm{L}}$ and $\overline{h_{II}(\overline{
\zeta})} \! \! \upharpoonright_{\overline{\mathrm{L}}}$; (3) $w^{c}
(\zeta)$ has support on $\mathrm{L}_{\varepsilon} \cup \overline{
\mathrm{L}_{\varepsilon}}$, and consists of the contribution to
$w^{\sharp}(\zeta)$ {}from $\mathcal{R}(\zeta) \! \! \upharpoonright_{
\mathrm{L}_{\varepsilon}}$ and $\overline{\mathcal{R}(\overline{
\zeta})} \! \! \upharpoonright_{\overline{\mathrm{L}_{\varepsilon}}}$;
and (4) $w^{\prime}(\zeta) \! \! \upharpoonright_{\Sigma^{\prime}
\setminus \Sigma^{\sharp}} \, \equiv
\left(
\begin{smallmatrix}
0 & 0 \\
0 & 0
\end{smallmatrix}
\right)$. It will now be shown that, as $t \! \to \! +\infty$, $w^{e}(\zeta)
\! \to \! \left(
\begin{smallmatrix}
0 & 0 \\
0 & 0
\end{smallmatrix}
\right)$ (see~Lemma~4.4), and the contribution to $w^{\sharp}(\zeta)$
{}from $\mathcal{R}(\zeta)$ and $\overline{\mathcal{R}(\overline{
\zeta})}$ (in some regions, polynomials of degree $k \! \in \! \mathbb{
Z}_{\geqslant 1}$, and in other regions, rational functions of the type
$\tfrac{\mathrm{polynomial} \, \mathrm{of} \, \mathrm{degree} \, k \,
\in \, \mathbb{Z}_{\geqslant 1}}{\vert \zeta \mp \mi \vert^{k+5}})$,
which is lumped into the factor $w^{\prime}(\zeta)$, and has support
on $\Sigma^{\sharp}$, encapsulates the leading-order asymptotics.
\begin{ccccc}
For arbitrarily large $l \! \in \! \mathbb{Z}_{\geqslant 1}$, and
arbitrarily fixed, sufficiently small positive $\varepsilon$, as $t \!
\to \! +\infty$ such that $0 \! < \! \zeta_{2} \! < \! \tfrac{1}{M} \!
< \! M \! < \! \zeta_{1}$ and $\vert \zeta_{3} \vert^{2} \! = \! 1$,
with $M \! \in \! \mathbb{R}_{>1}$ and bounded,
\begin{align*}
\vert \vert w^{\widehat{a}}(\cdot) \vert \vert_{\cap_{p \in \{1,2,
\infty\}} \mathcal{L}^{p}_{\mathrm{M}_{2}(\mathbb{C})}(\mathbb{R})}
&\leqslant \tfrac{\vert \underline{c}(\zeta_{1},\zeta_{2},\zeta_{3},
\overline{\zeta_{3}}) \vert}{\vert \zeta_{1}+\zeta_{2} \vert^{3}t^{
l}}, \\
\vert \vert w^{\widehat{a}}(\cdot)/(\cdot) \vert \vert_{\cap_{p \in
\{1,2,\infty\}} \mathcal{L}^{p}_{\mathrm{M}_{2}(\mathbb{C})}((\cos
(\widetilde{\varphi}_{3}),0) \, \cup \, (0,\frac{1}{2}\zeta_{2}))}
&\leqslant \tfrac{\vert \underline{c}(\zeta_{1},\zeta_{2},\zeta_{3},
\overline{\zeta_{3}}) \vert}{\vert \zeta_{1}+\zeta_{2} \vert^{3}t^{
l}}, \\
\vert \vert w^{b}(\cdot) \vert \vert_{\cap_{p \in \{1,2,\infty\}}
\mathcal{L}^{p}_{\mathrm{M}_{2}(\mathbb{C})}(Q_{0} \, \cup \, \overline{
Q_{0}})} &\leqslant \tfrac{\vert \underline{c}(\zeta_{1},\zeta_{2},
\zeta_{3},\overline{\zeta_{3}}) \vert}{\vert z_{o}+\zeta_{1}+\zeta_{2}
\vert^{l}t^{l}}, \\
\vert \vert w^{b}(\cdot) \vert \vert_{\cap_{p \in \{1,2,\infty\}}
\mathcal{L}^{p}_{\mathrm{M}_{2}(\mathbb{C})}(Q_{1} \, \cup \, \overline{
Q_{1}})} &\leqslant \vert \underline{c}(\zeta_{1},\zeta_{2},\zeta_{3},
\overline{\zeta_{3}}) \vert \exp (-2 \gamma_{\mathcal{R}}^{1}t), \\
\vert \vert w^{b}(\cdot) \vert \vert_{\cap_{p \in \{1,2,\infty\}}
\mathcal{L}^{p}_{\mathrm{M}_{2}(\mathbb{C})}(Q_{2} \, \cup \, \overline{
Q_{2}})} &\leqslant \tfrac{\vert \underline{c}(\zeta_{1},\zeta_{2},
\zeta_{3},\overline{\zeta_{3}}) \vert}{\sqrt{\omega_{1}} \, \sqrt{t}}
\exp (-\omega_{0}t),
\end{align*}
\begin{align*}
\vert \vert w^{b}(\cdot)/(\cdot) \vert \vert_{\cap_{p \in \{1,2,\infty\}}
\mathcal{L}^{p}_{\mathrm{M}_{2}(\mathbb{C})}(Q_{1} \, \cup \, \overline{
Q_{1}})} &\leqslant \vert \underline{c}(\zeta_{1},\zeta_{2},\zeta_{3},
\overline{\zeta_{3}}) \vert \exp (-\gamma_{II}^{1}t), \\
\vert \vert w^{c}(\cdot) \vert \vert_{\cap_{p \in \{1,2,\infty\}}
\mathcal{L}^{p}_{\mathrm{M}_{2}(\mathbb{C})}(Q_{3} \, \cup \, \overline{
Q_{3}})} &\leqslant \vert \underline{c}(\zeta_{1},\zeta_{2},\zeta_{
3},\overline{\zeta_{3}}) \vert \exp (-\varepsilon^{2} \gamma_{\mathcal{
R}}^{0}t), \\
\vert \vert w^{c}(\cdot) \vert \vert_{\cap_{p \in \{1,2,\infty\}}
\mathcal{L}^{p}_{\mathrm{M}_{2}(\mathbb{C})}(\mathrm{L}_{>} \, \cup \,
\overline{\mathrm{L}_{>}})} &\leqslant \vert \underline{c}(\zeta_{1},
\zeta_{2},\zeta_{3},\overline{\zeta_{3}}) \vert \exp (-\zeta_{1} \vert
z_{o} \vert (2 \! - \! \zeta_{2}^{2})t), \\
\vert \vert w^{c}(\cdot)/(\cdot) \vert \vert_{\cap_{p \in \{1,2,\infty\}}
\mathcal{L}^{p}_{\mathrm{M}_{2}(\mathbb{C})}(\mathrm{L}_{>} \, \cup \,
\overline{\mathrm{L}_{>}})} &\leqslant \vert \underline{c}(\zeta_{1},
\zeta_{2},\zeta_{3},\overline{\zeta_{3}}) \vert \exp (-\zeta_{1} \vert
z_{o} \vert (2 \! - \! \zeta_{2}^{2})t), \\
\vert \vert w^{\prime}(\cdot) \vert \vert_{\mathcal{L}^{1}_{\mathrm{
M}_{2}(\mathbb{C})}(\Sigma^{\prime})} \! = \! \vert \vert w^{\prime}
(\cdot) \vert \vert_{\mathcal{L}^{1}_{\mathrm{M}_{2}(\mathbb{C})}
(\Sigma^{\sharp})} &\leqslant \tfrac{\vert \underline{c}(\zeta_{1},
\zeta_{2},\zeta_{3},\overline{\zeta_{3}}) \vert}{\sqrt{(\zeta_{1}-
\zeta_{2}) \vert z_{o}+\zeta_{1}+\zeta_{2} \vert} \, \sqrt{t}}, \\
\vert \vert w^{\prime}(\cdot) \vert \vert_{\mathcal{L}^{2}_{\mathrm{
M}_{2}(\mathbb{C})}(\Sigma^{\prime})} \! = \! \vert \vert w^{\prime}
(\cdot) \vert \vert_{\mathcal{L}^{2}_{\mathrm{M}_{2}(\mathbb{C})}
(\Sigma^{\sharp})} &\leqslant \tfrac{\vert \underline{c}(\zeta_{1},
\zeta_{2},\zeta_{3},\overline{\zeta_{3}}) \vert}{((\zeta_{1}-\zeta_{
2}) \vert z_{o}+\zeta_{1}+\zeta_{2} \vert)^{1/4} \, t^{1/4}},
\end{align*}
where $Q_{0} \! := \! \mathrm{L} \setminus (\mathrm{L}_{>} \cup \mathrm{L}_{
<})$, $Q_{1} \! := \! \mathrm{L}_{>} \cup \{\mathstrut \zeta; \, \zeta \! =
\! v \me^{\mi \widetilde{\varphi}_{3}}, \, \widetilde{\varphi}_{3} \! := \!
\arg (\zeta_{3}) \! \in \! (\tfrac{\pi}{2},\pi), \, v \! \in \! (0,1 \! - \!
\varepsilon)\}$, $Q_{2} \! := \! \{\mathstrut \zeta; \, \zeta \! = \! v \me^{
\mi \widetilde{\varphi}_{3}}, \, \widetilde{\varphi}_{3} \! := \! \arg
(\zeta_{3}) \! \in \! (\tfrac{\pi}{2},\pi), \, v \! \geqslant \! 1 \! - \!
\varepsilon\}$, $Q_{3} \! := \! \mathrm{L}_{\varepsilon} \setminus \mathrm{
L}_{>}$, $\omega_{0} \! := \! \tfrac{1}{2}(a_{2} \! - \! z_{o})(4 \! - \!
a_{2}^{2})^{1/2}$ $(\in \! \mathbb{R}_{+})$, and $\omega_{1} \! := \! \tfrac{
1}{2}(z_{o}^{2} \! + \! 32)^{1/2}(4 \! - \! a_{2}^{2})^{1/2}$ $(\in \!
\mathbb{R}_{+})$, with $\mathrm{L}$, $\mathrm{L}_{>}$, $\mathrm{L}_{<}$,
$\mathrm{L}_{\varepsilon}$, $\gamma_{\mathcal{R}}^{1}$, $\gamma_{II}^{1}$,
and $\gamma_{\mathcal{R}}^{0}$ defined in Lemma~{\rm 4.2}.
\end{ccccc}

\emph{Proof.} Without loss of generality, the bounds for $\vert \vert
w^{b} (\cdot) \vert \vert_{\mathcal{L}^{p}_{\mathrm{M}_{2}(\mathbb{C})
}(\ast)}$ and $\vert \vert w^{b}(\cdot)/(\cdot) \vert \vert_{\mathcal{
L}^{p}_{\mathrm{M}_{2}(\mathbb{C})}(\ast)}$, $p \! \in \! \{1,2,\infty\}$,
and $\vert \vert w^{\prime}(\cdot) \vert \vert_{\mathcal{L}^{p^{\prime}
}_{\mathrm{M}_{2}(\mathbb{C})}(\ast)}$, $p^{\prime} \! \in \! \{1,2\}$,
will be derived: the remaining estimates follow in an analogous manner.
{}From the proof of Lemma~4.2 and the parametrisations of the respective
rays given therein, one shows that, modulo a scalar factor of 2 on the
right-hand side (RHS),
\begin{align*}
\vert \vert w^{b}(\cdot) \vert \vert_{\mathcal{L}^{1}_{\mathrm{M}_{
2}(\mathbb{C})}(Q_{0} \, \cup \, \overline{Q_{0}})} &\leqslant
\int\nolimits_{0}^{+\infty} \tfrac{\vert \underline{c}(\zeta_{1},
\zeta_{2},\zeta_{3},\overline{\zeta_{3}}) \vert v^{q}}{\vert \zeta_{
1}+\frac{v}{\sqrt{2}}(\zeta_{1}-\zeta_{2}) \me^{-\frac{\mi \pi}{4}}-
\mi \vert^{2}} \me^{-\frac{1}{4}t \gamma_{o}(\zeta_{1}-\zeta_{2})(1-
\zeta_{2}^{2})^{2} \vert z_{o}+\zeta_{1}-\zeta_{2} \vert v^{2}} \,
\md v \\
 &+ \int\nolimits_{0}^{1} \vert \underline{c}(\zeta_{1},\zeta_{2},
\zeta_{3},\overline{\zeta_{3}}) \vert v^{q} \me^{-\tfrac{1}{4}t
(\zeta_{1}-\zeta_{2})(1-\tfrac{2 \zeta_{1}^{2}}{\zeta_{1}^{4}+1})^{
2} \vert z_{o}+\zeta_{1}+\zeta_{2} \vert v^{2}} \, \md v \\
 &+ \int\nolimits_{0}^{1} \vert \underline{c}(\zeta_{1},\zeta_{
2},\zeta_{3},\overline{\zeta_{3}}) \vert v^{q} \me^{-\tfrac{1}{4}
t(\zeta_{1}-\zeta_{2})(1-\zeta_{1}^{2})^{2} \vert z_{o}+\zeta_{
1}+\zeta_{2} \vert v^{2}} \, \md v \\
 &+ \int\nolimits_{0}^{1} \vert \underline{c}(\zeta_{1},\zeta_{2},
\zeta_{3},\overline{\zeta_{3}}) \vert v^{q} \me^{-\tfrac{1}{4}t
\zeta_{2}(1-2 \zeta_{1}^{2})^{2} \vert z_{o}+\zeta_{2} \vert v^{2}}
\, \md v,
\end{align*}
$q \! \in \! \mathbb{Z}_{\geqslant 1}$, where $Q_{0}$ is defined in
the Lemma. Since, $\forall \, v \! \geqslant \! 0$, $\vert \zeta_{1}
\! + \! \tfrac{v}{\sqrt{2}}(\zeta_{1} \! - \! \zeta_{2}) \me^{-\frac{
\mi \pi}{4}} \! - \! \mi \vert^{-2} \! \leqslant \! 1$, and the
integrands are positive functions of $v$, it follows, by a
change-of-variable argument and letting the upper limits of integration
tend to $+\infty$, that $\vert \vert w^{b}(\cdot) \vert \vert_{\mathcal{
L}^{1}_{\mathrm{M}_{2}(\mathbb{C})}(Q_{0} \, \cup \, \overline{Q_{0}})}
\! \leqslant \! (\tfrac{\vert \underline{c}(\zeta_{1},\zeta_{2},\zeta_{
3},\overline{\zeta_{3}}) \vert t^{-(q/2+1/2)}}{\vert z_{o}+\zeta_{1}-
\zeta_{2} \vert^{q/2+1/2}} \! + \! \tfrac{\vert \underline{c}(\zeta_{1},
\zeta_{2},\zeta_{3},\overline{\zeta_{3}}) \vert t^{-(q/2+1/2)}}{\vert
z_{o}+\zeta_{1}+\zeta_{2} \vert^{q/2+1/2}} \! + \! \tfrac{\vert
\underline{c}(\zeta_{1},\zeta_{2},\zeta_{3},\overline{\zeta_{3}}) \vert
t^{-(q/2+1/2)}}{\vert z_{o}+\zeta_{2} \vert^{q/2+1/2}}) \! \int_{0}^{+
\infty} \me^{-\xi} \xi^{\frac{q-1}{2}} \, \md \xi$; hence, recalling
that \cite{a40} $\Gamma (z) \! = \! \int_{0}^{+\infty} \me^{-x}x^{z-1}
\, \md x$, $\Re (z) \! > \! 0$, where $\Gamma (\cdot)$ is the gamma
function, and $\int_{0}^{+\infty} \me^{-x^{2}} \, \md x \! = \! \tfrac{
\sqrt{\pi}}{2}$, for $\tfrac{q}{2} \! > \! l$, $l \! \in \! \mathbb{
Z}_{\geqslant 1}$ and arbitrarily large, $\vert \vert w^{b}(\cdot) \vert
\vert_{\mathcal{L}^{1}_{\mathrm{M}_{2}(\mathbb{C})}(Q_{0} \, \cup \,
\overline{Q_{0}})} \! \leqslant \! \tfrac{\vert \underline{c}(\zeta_{
1},\zeta_{2},\zeta_{3},\overline{\zeta_{3}}) \vert}{\vert z_{
o}+\zeta_{1}+\zeta_{2} \vert^{l}t^{l}}$. {}From the estimates given
in Lemma~4.2, it follows that, for $\tfrac{q}{2} \! > \! l$, $\vert
\vert w^{b}(\cdot) \vert \vert_{\mathcal{L}^{\infty}_{\mathrm{M}_{2}
(\mathbb{C})}(Q_{0} \, \cup \, \overline{Q_{0}})} \! \leqslant \!
\tfrac{\vert \underline{c}(\zeta_{1},\zeta_{2},\zeta_{3},\overline{
\zeta_{3}}) \vert}{\vert z_{o}+\zeta_{1}+\zeta_{2} \vert^{l}t^{l}}$,
and, {}from the inequality $\vert \vert w^{b}(\cdot) \vert \vert^{
2}_{\mathcal{L}^{2}_{\mathrm{M}_{2}(\mathbb{C})}(\ast)} \! \leqslant
\! \vert \vert w^{b}(\cdot) \vert \vert_{\mathcal{L}^{\infty}_{
\mathrm{M}_{2}(\mathbb{C})}(\ast)} \vert \vert w^{b}(\cdot) \vert
\vert_{\mathcal{L}^{1}_{\mathrm{M}_{2}(\mathbb{C})}(\ast)}$ and the
latter two estimates, one shows that, for $\tfrac{q}{2} \! > \! l$,
$\vert \vert w^{b}(\cdot) \vert \vert_{\mathcal{L}^{2}_{\mathrm{M}_{
2}(\mathbb{C})}(Q_{0} \, \cup \, \overline{Q_{0}})} \! \leqslant \!
\tfrac{\vert \underline{c}(\zeta_{1},\zeta_{2},\zeta_{3},\overline{
\zeta_{3}}) \vert}{\vert z_{o}+\zeta_{1}+\zeta_{2} \vert^{l}t^{l}}$;
hence, recalling that $\vert \vert w^{b}(\cdot) \vert \vert_{\cap_{p
\in \{1,2,\infty\}} \mathcal{L}^{p}_{\mathrm{M}_{2}(\mathbb{C})}(Q_{
0} \cup \overline{Q_{0}})} \! := \! \sum_{p \in \{1,2,\infty\}} \!
\vert \vert w^{b}(\cdot) \vert \vert_{\mathcal{L}^{p}_{\mathrm{M}_{2}
(\mathbb{C})}(Q_{0} \cup \overline{Q_{0}})}$, one obtains, for $\tfrac{
q}{2} \! > \! l$ and the latter estimates, the result for $\vert \vert
w^{b}(\cdot) \vert \vert_{\cap_{p \in \{1,2,\infty\}} \mathcal{L}^{p}_{
\mathrm{M}_{2}(\mathbb{C})}(Q_{0} \cup \overline{Q_{0}})}$ stated in
the Lemma. {}From the proof of Lemma~4.2, one shows that, modulo
a scalar factor of 2 on the RHS, $\vert \vert w^{b}(\cdot) \vert \vert_{
\mathcal{L}^{1}_{\mathrm{M}_{2}(\mathbb{C})}(Q_{1} \cup \overline{
Q_{1}})} \! \leqslant \! \int_{0}^{1} \! \tfrac{\vert \underline{c}(\zeta_{1},
\zeta_{2},\zeta_{3},\overline{\zeta_{3}}) \vert}{\xi^{2}+1} \exp (-
\tfrac{1}{2}t \vert z_{o} \vert \zeta_{1}(2 \! - \! \zeta_{2}^{2}))
\md \xi \! + \! \int_{0}^{1-\varepsilon} \! \tfrac{\vert \underline{c}
(\zeta_{1},\zeta_{2},\zeta_{3},\overline{\zeta_{3}}) \vert}{\xi^{2}+1}
\exp (-t(\tfrac{1}{2} \vert z_{o} \vert \linebreak[4]
- \! \cos \widetilde{\varphi}_{3}) \sin \widetilde{\varphi}_{3}) \md \xi$,
with $Q_{1}$ as defined in the Lemma: noting that the integrands are
positive functions of $\xi$, letting the upper limits of integration
tend to $+\infty$, and using the fact that, with the principal branch
of $\arctan (\cdot)$ chosen, $\int_{0}^{+\infty} \tfrac{\md \xi}{\xi^{
2}+1} \! = \! \pi/2$, one shows that $\vert \vert w^{b}(\cdot)
\vert \vert_{\mathcal{L}^{1}_{\mathrm{M}_{2}(\mathbb{C})}(Q_{1} \cup
\overline{Q_{1}})} \! \leqslant \! \vert \underline{c}(\zeta_{1},\zeta_{
2},\zeta_{3},\overline{\zeta_{3}}) \vert \me^{-2 \gamma_{\mathcal{R}}^{
1}t}$, with $\gamma_{\mathcal{R}}^{1}$ defined in Lemma~4.2. The estimate
$\vert \vert w^{b}(\cdot) \vert \vert_{\mathcal{L}^{\infty}_{\mathrm{
M}_{2}(\mathbb{C})}(Q_{1} \cup \overline{Q_{1}})} \! \leqslant \! \vert
\underline{c}(\zeta_{1},\zeta_{2},\zeta_{3},\overline{\zeta_{3}}) \vert
\me^{-2 \gamma_{\mathcal{R}}^{1}t}$ is a consequence of Lemma~4.2:
using the latter two estimates and the inequality $\vert \vert w^{b}(\cdot)
\vert \vert^{2}_{\mathcal{L}^{2}_{\mathrm{M}_{2}(\mathbb{C})}(\ast)}
\! \leqslant \! \vert \vert w^{b}(\cdot) \vert \vert_{\mathcal{L}^{
\infty}_{\mathrm{M}_{2}(\mathbb{C})}(\ast)} \vert \vert w^{b}(\cdot)
\vert \vert_{\mathcal{L}^{1}_{\mathrm{M}_{2}(\mathbb{C})}(\ast)}$, one
deduces that $\vert \vert w^{b}(\cdot) \vert \vert_{\mathcal{L}^{2}_{
\mathrm{M}_{2}(\mathbb{C})}(Q_{1} \cup \overline{Q_{1}})} \! \leqslant
\! \vert \underline{c}(\zeta_{1},\zeta_{2},\zeta_{3},\overline{\zeta_{
3}}) \vert \me^{-2 \gamma_{\mathcal{R}}^{1}t}$; hence, one arrives at
the estimate for $\vert \vert w^{b}(\cdot) \vert \vert_{\cap_{p \in \{
1,2,\infty\}} \mathcal{L}^{p}_{\mathrm{M}_{2}(\mathbb{C})}(Q_{1} \cup
\overline{Q_{1}})}$ stated in the Lemma. {}From the proof of Lemma~4.2,
one shows that, modulo a scalar factor of 2 on the RHS, $\vert \vert
w^{b}(\cdot) \vert \vert_{\mathcal{L}^{1}_{\mathrm{M}_{2}(\mathbb{
C})}(Q_{2} \cup \overline{Q_{2}})} \! \leqslant \! \int_{1-\varepsilon}^{
+\infty} \tfrac{\vert \underline{c}(\zeta_{1},\zeta_{2},\zeta_{3},
\overline{\zeta_{3}}) \vert}{(v+1)^{2}+1} \exp (-t \linebreak[4]
\cdot (\omega_{0} \! + \! \tfrac{1}{2} \omega_{1}v^{2})) \md v \!
\leqslant \! \tfrac{\vert \underline{c}(\zeta_{1},\zeta_{2},\zeta_{3},
\overline{\zeta_{3}}) \vert}{\sqrt{\omega_{1}t}} \me^{-\omega_{0}t}
\int_{0}^{+\infty} \me^{-\xi^{2}} \, \md \xi \! \leqslant \! \tfrac{
\vert \underline{c}(\zeta_{1},\zeta_{2},\zeta_{3},\overline{\zeta_{3}})
\vert}{\sqrt{\omega_{1}t}} \me^{-\omega_{0}t}$, with $Q_{2}$, $\omega_{
0}$, and $\omega_{1}$ as defined in the Lemma. The estimate $\vert \vert
w^{b}(\cdot) \vert \vert_{\mathcal{L}^{\infty}_{\mathrm{M}_{2}(\mathbb{
C})}(Q_{2} \cup \overline{Q_{2}})} \! \leqslant \! \tfrac{\vert
\underline{c}(\zeta_{1},\zeta_{2},\zeta_{3},\overline{\zeta_{3}}) \vert}
{\sqrt{\omega_{1}t}} \me^{-\omega_{0}t}$ follows {}from Lemma 4.2: using
the inequality $\vert \vert w^{b}(\cdot) \vert \vert^{2}_{\mathcal{L}^{
2}_{\mathrm{M}_{2}(\mathbb{C})}(\ast)} \! \leqslant \! \vert \vert w^{b}
(\cdot) \vert \vert_{\mathcal{L}^{\infty}_{\mathrm{M}_{2}(\mathbb{C})}
(\ast)} \vert \vert w^{b}(\cdot) \vert \vert_{\mathcal{L}^{1}_{\mathrm{
M}_{2}(\mathbb{C})}(\ast)}$ and the latter two estimates, one deduces
that $\vert \vert w^{b}(\cdot) \vert \vert_{\mathcal{L}^{2}_{\mathrm{
M}_{2}(\mathbb{C})}(Q_{2} \cup \overline{Q_{2}})} \! \leqslant \!
\tfrac{\vert \underline{c}(\zeta_{1},\zeta_{2},\zeta_{3},\overline{
\zeta_{3}}) \vert}{\sqrt{\omega_{1}t}} \me^{-\omega_{0}t}$; hence, one
arrives at the estimate for $\vert \vert w^{b}(\cdot) \vert \vert_{
\cap_{p \in \{1,2,\infty\}} \mathcal{L}^{p}_{\mathrm{M}_{2}(\mathbb{
C})}(Q_{2} \cup \overline{Q_{2}})}$ stated in the Lemma. {}From the
proof of Lemma 4.2, one shows that, modulo a scalar factor of 2 on the
RHS, $\vert \vert w^{b}(\cdot)/(\cdot) \vert \vert_{\mathcal{L}^{1}_{
\mathrm{M}_{2}(\mathbb{C})}(Q_{1} \cup \overline{Q_{1}})} \! \leqslant
\! \int_{0}^{1} \tfrac{\vert \underline{c}(\zeta_{1},\zeta_{2},\zeta_{
3},\overline{\zeta_{3}}) \vert v^{-3q}}{\vert \zeta_{2} \frac{v}{\sqrt{
2}} \me^{-\frac{\mi \pi}{4}} \vert} \exp (-\tfrac{1}{2}t \vert z_{o}
\vert \zeta_{1}(2 \! - \! \zeta_{2}^{2})) \exp (-\tfrac{t}{v^{2}}
\zeta_{1}^{2}(1 \! - \! \tfrac{1}{2} \zeta_{2}^{2})^{2}) \md (\tfrac{
\zeta_{2}v}{\sqrt{2}}) \! + \! \int_{0}^{1-\varepsilon} \tfrac{\vert
\underline{c}(\zeta_{1},\zeta_{2},\zeta_{3},\overline{\zeta_{3}})
\vert}{(\vert \me^{\mi \widetilde{\varphi}_{3}}v \vert^{2}+1)v}
\exp (-\linebreak[4]
\tfrac{t \vert z_{o} \vert}{2v^{2}} \sin \widetilde{\varphi}_{3}) \exp
(-\tfrac{t}{2} \vert \sin 2 \widetilde{\varphi}_{3} \vert) \! \leqslant
\! \tfrac{\vert \underline{c}(\zeta_{1},\zeta_{2},\zeta_{3},\overline{
\zeta_{3}}) \vert}{\zeta_{1}^{3q}(1-\frac{1}{2} \zeta_{2}^{2})^{3q}
t^{3q}} \exp (-\frac{1}{2}t \vert z_{o} \vert \zeta_{1}(2 \! - \! \zeta_{
2}^{2})) \! \int_{0}^{+\infty} \me^{-\xi} \xi^{\frac{3q}{2}-1} \md
\xi \! + \! \vert \underline{c}(\zeta_{1},\zeta_{2},\linebreak[4]
\zeta_{3},\overline{\zeta_{3}}) \vert \me^{-\frac{1}{2} t \vert \sin
2 \widetilde{\varphi}_{3} \vert} \int_{0}^{+\infty} \me^{-\xi} \xi^{-1
} \, \md \xi \! \leqslant \! \vert \underline{c}(\zeta_{1},\zeta_{2},
\zeta_{3},\overline{\zeta_{3}}) \vert \me^{-\gamma_{II}^{1}t}$, $q
\! \in \! \mathbb{Z}_{\geqslant 1}$, with $\gamma_{II}^{1}$ defined
in Lemma 4.2. Also, $\vert \vert w^{b}(\cdot)/(\cdot) \vert \vert_{
\mathcal{L}^{\infty}_{\mathrm{M}_{2}(\mathbb{C})}(Q_{1} \cup \overline{
Q_{1}})} \! \leqslant \! \vert \underline{c}(\zeta_{1},\zeta_{2},\zeta_{
3},\overline{\zeta_{3}}) \vert \me^{-\gamma_{II}^{1}t}$, and, {}from the
latter two estimates and the inequality $\vert \vert w^{b}(\cdot)/(\cdot)
\vert \vert^{2}_{\mathcal{L}^{2}_{\mathrm{M}_{2}(\mathbb{C})}(\ast)} \!
\leqslant \! \vert \vert w^{b}(\cdot)/(\cdot) \vert \vert_{\mathcal{L}^{
\infty}_{\mathrm{M}_{2}(\mathbb{C})}(\ast)} \vert \vert w^{b}(\cdot)/
(\cdot) \vert \vert_{\mathcal{L}^{1}_{\mathrm{M}_{2}(\mathbb{C})}(\ast)}$,
one deduces that $\vert \vert w^{b}(\cdot)/(\cdot) \vert \vert_{\mathcal{
L}^{2}_{\mathrm{M}_{2}(\mathbb{C})}(Q_{1} \cup \overline{Q_{1}})} \!
\leqslant \! \vert \underline{c}(\zeta_{1},\zeta_{2},\zeta_{3},\overline{
\zeta_{3}}) \vert \me^{-\gamma_{II}^{1}t}$; hence, one arrives at the
estimate for $\vert \vert w^{b}(\cdot)/(\cdot) \vert \vert_{\cap_{p \in
\{1,2,\infty\}} \mathcal{L}^{p}_{\mathrm{M}_{2}(\mathbb{C})}(Q_{1} \cup
\overline{Q_{1}})}$ stated in the Lemma. {}From Eq.~(102), the fact that
$w^{\prime}(\zeta) \! \! \upharpoonright_{\Sigma^{\prime} \setminus
\Sigma^{\sharp}} = \!
\left(
\begin{smallmatrix}
0 & 0 \\
0 & 0
\end{smallmatrix}
\right)$, and the proof of Lemma~4.2, one shows that $\vert \vert w^{
\prime}(\cdot) \vert \vert_{\mathcal{L}^{1}_{\mathrm{M}_{2}(\mathbb{
C})}(\Sigma^{\prime})} \! = \! \vert \vert w^{\prime}(\cdot) \vert
\vert_{\mathcal{L}^{1}_{\mathrm{M}_{2}(\mathbb{C})}(\Sigma^{\prime}
\setminus \Sigma^{\sharp} \cup \Sigma^{\sharp})} \! = \! \vert \vert
w^{\prime}(\cdot) \vert \vert_{\mathcal{L}^{1}_{\mathrm{M}_{2}(\mathbb{
C})}(\Sigma^{\prime} \setminus \Sigma^{\sharp})} \! + \! \vert \vert
w^{\prime}(\cdot) \vert \vert_{\mathcal{L}^{1}_{\mathrm{M}_{2}(\mathbb{
C})}(\Sigma^{\sharp})} \! = \! \vert \vert w^{\prime}(\cdot) \vert
\vert_{\mathcal{L}^{1}_{\mathrm{M}_{2}(\mathbb{C})}(\Sigma^{\sharp})}
\! \leqslant \! \int_{0}^{+\infty} \! 2^{-\frac{1}{2}} \linebreak[4]
\cdot \vert \underline{c}(\zeta_{1},\zeta_{2},\zeta_{3},\overline{
\zeta_{3}}) \vert (\zeta_{1} \! - \! \zeta_{2}) \exp (-\tfrac{1}{
2}t(\zeta_{1} \! - \! \zeta_{2})(1 \! - \! \tfrac{2 \zeta_{1}^{2}}
{\zeta_{1}^{4}+1})^{2} \vert z_{o} \! + \! \zeta_{1} \! + \! \zeta_{
2} \vert v^{2}) \md v \! + \! \int_{0}^{+\infty} \! 2^{-\frac{1}
{2}} \vert \underline{c}(\zeta_{1},\zeta_{2},\zeta_{3},\overline{
\zeta_{3}}) \vert (\zeta_{1} \! - \! \zeta_{2}) \exp (-\tfrac{t}{2}
\gamma_{o}(\zeta_{1} \! - \! \zeta_{2})(1 \! - \! \zeta_{2}^{2})^{2}
\vert z_{o} \! + \! \zeta_{1} \! + \! \zeta_{2} \vert v^{2}) \md v
\! \leqslant \! (\vert \underline{c}(\zeta_{1},\zeta_{2},\zeta_{3},
\overline{\zeta_{3}}) \vert (\zeta_{1} \! - \! \zeta_{2})/(\sqrt{t
(\zeta_{1} \! - \! \zeta_{2}) \vert z_{o} \! + \! \zeta_{1} \! + \!
\zeta_{2} \vert} \linebreak[4]
\cdot (1 \! - \! \frac{2 \zeta_{1}^{2}}{\zeta_{1}^{4}+1})) \! + \!
\vert \underline{c}(\zeta_{1},\zeta_{2},\zeta_{3},\overline{\zeta_{
3}}) \vert (\zeta_{1} \! - \! \zeta_{2})/(\sqrt{t \gamma_{o}(\zeta_{
1} \! - \! \zeta_{2}) \vert z_{o} \! + \! \zeta_{1} \! + \! \zeta_{
2} \vert}(1 \! - \! \zeta_{2}^{2}))) \! \int_{0}^{+\infty} \! \me^{
-\xi^{2}} \md \xi \! \leqslant \! \vert \underline{c}(\zeta_{1},
\zeta_{2},\zeta_{3},\linebreak[4]
\overline{\zeta_{3}}) \vert (t(\zeta_{1} \! - \! \zeta_{2}) \vert
z_{o} \! + \! \zeta_{1} \! + \! \zeta_{2} \vert)^{-1/2}$;
hence, $\vert \vert w^{\prime}(\cdot) \vert \vert_{\mathcal{L}^{
2}_{\mathrm{M}_{2}(\mathbb{C})}(\Sigma^{\prime})} \! = \! \vert
\vert w^{\prime}(\cdot) \vert \vert_{\mathcal{L}^{2}_{\mathrm{M}_{
2}(\mathbb{C})}(\Sigma^{\sharp})} \! \leqslant \! \vert \underline{
c}(\zeta_{1},\zeta_{2},\zeta_{3},\overline{\zeta_{3}}) \vert (t(
\zeta_{1} \linebreak[4]
- \! \zeta_{2}) \vert z_{o} \! + \! \zeta_{1} \! + \! \zeta_{2}
\vert)^{-1/4}$. \hfill $\square$
\begin{aaaaa}
Let $\mathscr{N}(\ast)$ denote the space of bounded linear operators
acting {}from $\mathcal{L}^{2}_{\mathrm{M}_{2}(\mathbb{C})}(\ast)$
into $\mathcal{L}^{2}_{\mathrm{M}_{2}(\mathbb{C})}(\ast)$.
\end{aaaaa}

The following Lemma will be proven \emph{a posteriori} (see Section~5,
Lemma~5.4):
\begin{ccccc}
As $t \! \to \! +\infty$ such that $0 \! < \! \zeta_{2} \! < \! \tfrac{
1}{M} \! < \! M \! < \! \zeta_{1}$ and $\vert \zeta_{3} \vert^{2} \!
= \! 1$, with $M \! \in \! \mathbb{R}_{>1}$ and bounded, $(\mathbf{1}
\! - \! C_{w^{\prime}})^{-1} \! \in \! \mathscr{N}(\Sigma^{\prime})$
$(\vert \vert (\mathbf{1} \! - \! C_{w^{\prime}})^{-1} \vert \vert_{
\mathscr{N}(\Sigma^{\prime})} \! < \! \infty)$.
\end{ccccc}
\begin{eeeee}
Actually, the operator $(\mathbf{1} \! - \! C_{w^{\prime}})^{-1}$ acts in
$\mathrm{I} \! + \! \mathcal{L}^{2}_{\mathrm{M}_{2}(\mathbb{C})}(\Sigma^{
\prime})$; however, due to a result of Zhou \cite{a31}, using the Fredholm
alternative, if $(\mathbf{1} \! - \! C_{w^{\prime}})^{-1}$ is invertible on
$\mathcal{L}^{2}_{\mathrm{M}_{2}(\mathbb{C})}(\Sigma^{\prime})$, then it is
invertible on every space set theoretically contained in the span of constant
functions and $\mathcal{L}^{2}_{\mathrm{M}_{2}(\mathbb{C})}(\Sigma^{\prime})$
(which is the case here); hence, one can consider $(\mathbf{1} \! - \! C_{w^{
\prime}})^{-1} \! \! \! \upharpoonright_{\mathcal{L}^{2}_{\mathrm{M}_{2}
(\mathbb{C})}(\Sigma^{\prime})}$. The result stated in Lemma~4.5 should not,
after all, come as a surprise, since the jump matrix of the RHP formulated in
Lemma~2.6, within the framework of the BC formulation introduced in Section~3,
admits a (bounded) algebraic factorisation of the form $\mathcal{G}(\zeta) \!
= \! (\mathrm{I} \! - \! w^{\mathcal{G}}_{-}(\zeta))^{-1}(\mathrm{I} \! + \!
w^{\mathcal{G}}_{+}(\zeta))$, $\zeta \! \in \! \mathbb{R}$ (oriented {}from
$-\infty$ to $+\infty)$, and due to another result of Zhou \cite{a32} (see,
in particular, Proposition~2.16 and the arguments thereafter), one has that,
for $r(\zeta) \! \in \! \mathcal{S}_{\mathbb{C}}^{1}(\mathbb{R})$, $\vert
\vert (\mathbf{1} \! - \! C_{w^{\mathcal{G}}})^{-1} \vert \vert_{\mathscr{N}
(\sigma_{c})} \! \leqslant \! \tfrac{(\lambda_{\mathrm{max}}+1)+\sqrt{
(\lambda_{\mathrm{max}}+1)^{2}-4 \lambda_{\mathrm{min}}}}{2 \lambda_{\mathrm{
min}}} \vert \vert (\mathrm{I} \! - \! w^{\mathcal{G}}_{-}(\cdot))^{-1}
\vert \vert_{\mathcal{L}^{\infty}_{\mathrm{M}_{2}(\mathbb{C})}(\sigma_{c})}$,
with $\lambda_{\mathrm{max}} \! := \! \sup_{\zeta \in \mathbb{R}} \{\mathrm{
maximal} \, \, \mathrm{eigenvalue} \, \, \mathrm{of} \, (\mathcal{G}(\zeta)
\mathcal{G}^{\dag}(\zeta))^{1/2}\}$, where ${}^{\dag}$ denotes Hermitian
conjugation, and $\lambda_{\mathrm{min}} \! := \! \inf_{\zeta \in \mathbb{
R}} \{\mathrm{minimal} \, \, \mathrm{eigenvalue} \, \, \mathrm{of} \,
\tfrac{1}{2}(\mathcal{G}(\zeta) \! + \! \mathcal{G}^{\dag}(\zeta))\}$ (see,
also, Lemma~2.31 in \cite{a33}).
\end{eeeee}
\begin{bbbbb}
As $t \! \to \! +\infty$ such that $0 \! < \! \zeta_{2} \! < \! \tfrac{1}
{M} \! < \! M \! < \! \zeta_{1}$ and $\vert \zeta_{3} \vert^{2} \! = \! 1$,
with $M \! \in \! \mathbb{R}_{>1}$ and bounded, $(\mathbf{1} \! - \! C_{w^{
\sharp}})^{-1} \! \in \! \mathscr{N}(\Sigma^{\prime}) \! \Leftrightarrow \!
(\mathbf{1} \! - \! C_{w^{\prime}})^{-1} \! \in \! \mathscr{N}(\Sigma^{
\prime})$.
\end{bbbbb}

\emph{Proof.} For $(\mathbf{1} \! - \! C_{w^{\sharp}})^{-1} \colon
\mathcal{L}^{2}_{\mathrm{M}_{2}(\mathbb{C})}(\Sigma^{\prime}) \! \to
\! \mathcal{L}^{2}_{\mathrm{M}_{2}(\mathbb{C})}(\Sigma^{\prime})$ and
$(\mathbf{1} \! - \! C_{w^{\prime}})^{-1} \colon \mathcal{L}^{2}_{
\mathrm{M}_{2}(\mathbb{C})}(\Sigma^{\prime}) \! \to \! \mathcal{L}^{
2}_{\mathrm{M}_{2}(\mathbb{C})}(\Sigma^{\prime})$, {}from the second
resolvent identity, $(\mathbf{1} \! - \! C_{w^{\sharp}})^{-1} \! = \!
(\mathbf{1} \! - \! C_{w^{\prime}})^{-1} \! + \! (\mathbf{1} \! - \!
C_{w^{\sharp}})^{-1} (C_{w^{\sharp}} \! - \! C_{w^{\prime}})(\mathbf{
1} \! - \! C_{w^{\prime}})^{-1}$; hence, $\vert \vert (\mathbf{1} \!
- \! C_{w^{\sharp}})^{-1} \vert \vert_{\mathscr{N}(\Sigma^{\prime})}
\! \leqslant \! \vert \vert (\mathbf{1} \! - \! C_{w^{\prime}})^{-1}
\vert \vert_{\mathscr{N}(\Sigma^{\prime})} \! + \! \vert \vert (\mathbf{
1} \! - \! C_{w^{\sharp}})^{-1} \vert \vert_{\mathscr{N}(\Sigma^{\prime}
)} \vert \vert (C_{w^{\sharp}} \! - \! C_{w^{\prime}}) \vert \vert_{
\mathscr{N}(\Sigma^{\prime})} \vert \vert (\mathbf{1} \! - \! C_{w^{
\prime}})^{-1} \vert \vert_{\mathscr{N}(\Sigma^{\prime})}$. Recalling
{}from Section~3 that, for $\star \! \in \! \mathcal{L}^{2}_{\mathrm{
M}_{2}(\mathbb{C})}(\ast)$, $C_{w} \star \! := \! C_{+}(\star w_{-})
\! + \! C_{-}(\star w_{+})$, using the linearity of the Cauchy operators,
$C_{\pm}$, and recalling Eq.~(102), one deduces that $(C_{w^{\sharp}} \!
- \! C_{w^{\prime}}) \star \! = \! C_{+}(\star w^{e}_{-}) \! + \! C_{-}
(\star w^{e}_{+}) \! = \! C_{w^{e}} \star$; hence, $\vert \vert (\mathbf{
1} \! - \! C_{w^{\sharp}})^{-1} \vert \vert_{\mathscr{N}(\Sigma^{\prime})
} \! \leqslant \! \vert \vert (\mathbf{1} \! - \! C_{w^{\prime}})^{-1}
\vert \vert_{\mathscr{N}(\Sigma^{\prime})} \! + \! \vert \vert (\mathbf{
1} \! - \! C_{w^{\sharp}})^{-1} \vert \vert_{\mathscr{N}(\Sigma^{\prime}
)} \vert \vert w^{e}(\cdot) \vert \vert_{\mathcal{L}^{\infty}_{\mathrm{
M}_{2}(\mathbb{C})}(\Sigma^{\prime})} \vert \vert (\mathbf{1} \! - \! C_{
w^{\prime}})^{-1} \vert \vert_{\mathscr{N}(\Sigma^{\prime})}$. Recalling
{}from Eq.~(102) that $w^{e}(\zeta) \! = \! w^{\widehat{a}}(\zeta) \! +
\! w^{b}(\zeta) \! + \! w^{c}(\zeta)$, using the bounds in Lemma~4.4, one
shows that $\vert \vert w^{e}(\cdot) \vert \vert_{\mathcal{L}^{\infty}_{
\mathrm{M}_{2}(\mathbb{C})}(\Sigma^{\prime})} \! \leqslant \! \tfrac{
\vert \underline{c}(\zeta_{1},\zeta_{2},\zeta_{3},\overline{\zeta_{3}})
\vert}{\vert z_{o}+\zeta_{1}+\zeta_{2} \vert^{l}t^{l}} \! + \! \mathcal{
O}(\vert \underline{c}(\zeta_{1},\zeta_{2},\zeta_{3},\overline{\zeta_{
3}}) \vert \exp (-t \min \{2 \gamma_{\mathcal{R}}^{1},\omega_{0},
\varepsilon^{2} \gamma_{\mathcal{R}}^{0},\vert z_{o} \vert \zeta_{1}
(2 \! - \! \zeta_{2}^{2})\}))$, with $l \! \in \! \mathbb{Z}_{\geqslant
1}$ and arbitrarily large; thus, {}from Lemma~4.5 and the estimate for
$\vert \vert w^{e}(\cdot) \vert \vert_{\mathcal{L}^{\infty}_{\mathrm{
M}_{2}(\mathbb{C})}(\Sigma^{\prime})}$, one shows that $\vert \vert
(\mathbf{1} \! - \! C_{w^{\sharp}})^{-1} \vert \vert_{\mathscr{N}
(\Sigma^{\prime})} \! < \! \infty$. \hfill $\square$
\begin{bbbbb}[$\cite{a27}$]
For $\zeta \! \in \! \mathbb{C} \setminus \Sigma^{\prime}$,
\begin{equation*}
\int\nolimits_{\Sigma^{\prime}} \dfrac{((\mathbf{1} \! - \! C_{w^{\sharp}})
^{-1} \mathrm{I})(z)w^{\sharp}(z)}{(z \! - \! \zeta)} \, \dfrac{\md z}{2
\pi \mi} \! = \! \int\nolimits_{\Sigma^{\prime}} \dfrac{((\mathbf{1} \! -
\! C_{w^{\prime}})^{-1} \mathrm{I})(z)w^{\prime}(z)}{(z \! - \! \zeta)} \,
\dfrac{\md z}{2 \pi \mi} \! + \! \mathscr{A} \! + \! \mathscr{B} \! + \!
\mathscr{C} \! + \! \mathscr{D},
\end{equation*}
where
\begin{gather*}
\mathscr{A} \! := \! \int\nolimits_{\Sigma^{\prime}} \dfrac{w^{e}(z)}{(z
\! - \! \zeta)} \, \dfrac{\md z}{2 \pi \mi}, \qquad \, \, \mathscr{B} \! := \!
\int\nolimits_{\Sigma^{\prime}} \dfrac{((\mathbf{1} \! - \! C_{w^{\prime}
})^{-1}(C_{w^{e}} \mathrm{I}))(z)w^{\sharp}(z)}{(z \! - \! \zeta)} \,
\dfrac{\md z}{2 \pi \mi}, \\
\mathscr{C} \! := \! \int\nolimits_{\Sigma^{\prime}} \dfrac{((\mathbf{1}
\! - \! C_{w^{\prime}})^{-1}(C_{w^{\prime}} \mathrm{I}))(z)w^{e}(z)}{(z
\! - \! \zeta)} \, \dfrac{\md z}{2 \pi \mi}, \\
\mathscr{D} \! := \! \int\nolimits_{\Sigma^{\prime}} \dfrac{((\mathbf{1}
\! - \! C_{w^{\prime}})^{-1}C_{w^{e}}(\mathbf{1} \! - \! C_{w^{\sharp}})
^{-1}(C_{w^{\sharp}} \mathrm{I}))(z)w^{\sharp}(z)}{(z \! - \! \zeta)} \,
\dfrac{\md z}{2 \pi \mi}.
\end{gather*}
\end{bbbbb}
\begin{eeeee}
Hereafter, all exponentially small error terms of the type $\mathcal{O}
(\exp (-\diamondsuit t))$, $\diamondsuit \! \in \! \mathbb{R}_{+}$, will be
neglected, and only leading order error terms will be retained.
\end{eeeee}
\begin{bbbbb}
If $(\mathbf{1} \! - \! C_{w^{\prime}})^{-1} \! \in \! \mathscr{N}(\Sigma^{
\prime})$, then, for $\zeta \! \in \! \mathbb{C} \setminus \Sigma^{\prime}$
and arbitrarily large $l \! \in \! \mathbb{Z}_{\geqslant 1}$, as $t \! \to
\! +\infty$ such that $0 \! < \! \zeta_{2} \! < \! \tfrac{1}{M} \! < \! M
\! < \! \zeta_{1}$ and $\vert \zeta_{3} \vert^{2} \! = \! 1$, with $M \!
\in \! \mathbb{R}_{>1}$ and bounded,
\begin{equation*}
\int\nolimits_{\Sigma^{\prime}} \! \dfrac{((\mathbf{1} \! - \! C_{w^{
\sharp}})^{-1} \mathrm{I})(z)w^{\sharp}(z)}{(z \! - \! \zeta)} \dfrac{
\md z}{2 \pi \mi} \! = \! \int\nolimits_{\Sigma^{\prime}} \! \dfrac{((
\mathbf{1} \! - \! C_{w^{\prime}})^{-1} \mathrm{I})(z)w^{\prime}(z)}{(
z \! - \! \zeta)} \dfrac{\md z}{2 \pi \mi} \! + \! \mathcal{O} \! \left(
\dfrac{\underline{c}(\zeta_{1},\zeta_{2},\zeta_{3},\overline{\zeta_{3}
}) f^{\prime}(\zeta)}{\vert z_{o} \! + \! \zeta_{1} \! + \! \zeta_{2}
\vert^{l}t^{l}} \right) \!,
\end{equation*}
with $f^{\prime}(\zeta) \! \in \! \mathcal{L}^{\infty}_{\mathrm{M}_{2}
(\mathbb{C})}(\mathbb{C} \setminus \Sigma^{\prime})$.
\end{bbbbb}

\emph{Proof.} Modulo exponentially small terms (cf.~Remark~4.4),
one must show that $\mathscr{A}$, $\mathscr{B}$, $\mathscr{C}$
and $\mathscr{D}$ have, respectively, for arbitrarily large $l \! \in \!
\mathbb{Z}_{\geqslant 1}$, the estimate $\mathcal{O}(\tfrac{\underline{
c}(\zeta_{1},\zeta_{2},\zeta_{3},\overline{\zeta_{3}}) \diamondsuit
(\zeta)}{\vert z_{o}+\zeta_{1}+\zeta_{2} \vert^{l}t^{l}})$, where
$\vert \vert \diamondsuit (\cdot) \vert \vert_{\mathcal{L}^{\infty}_{
\mathrm{M}_{2}(\mathbb{C})}(\mathbb{C} \setminus \Sigma^{\prime})}
\! < \! \infty$. {}From the definition of $\mathscr{A}$ given in
Proposition~4.3, and Eq.~(102), it follows that $\tfrac{2 \pi \vert
\mathscr{A} \vert}{d_{z,\zeta}} \! \leqslant \! \vert \vert w^{\widehat{
a}}(\cdot) \vert \vert_{\mathcal{L}^{1}_{\mathrm{M}_{2}(\mathbb{C})}
(\mathbb{R})} \! + \! \vert \vert w^{b}(\cdot) \vert \vert_{\mathcal{
L}^{1}_{\mathrm{M}_{2}(\mathbb{C})}(\mathrm{L} \cup \overline{\mathrm{
L}})} \! + \! \vert \vert w^{c}(\cdot) \vert \vert_{\mathcal{L}^{1}_{
\mathrm{M}_{2}(\mathbb{C})}(\mathrm{L}_{\varepsilon} \cup \overline{
\mathrm{L}_{\varepsilon}})} \! \leqslant \! \vert \vert w^{\widehat{
a}}(\cdot) \vert \vert_{\mathcal{L}^{1}_{\mathrm{M}_{2}(\mathbb{C})}
(\mathbb{R})} \linebreak[4]
\! + \! \vert \vert w^{b}(\cdot) \vert \vert_{\mathcal{L}^{1}_{\mathrm{
M}_{2}(\mathbb{C})}(Q_{0} \cup \overline{Q_{0}})} \! + \! \vert \vert
w^{b}(\cdot) \vert \vert_{\mathcal{L}^{1}_{\mathrm{M}_{2}(\mathbb{
C})}(Q_{1} \cup \overline{Q_{1}})} \! + \! \vert \vert w^{b}(\cdot)
\vert \vert_{\mathcal{L}^{1}_{\mathrm{M}_{2}(\mathbb{C})}(Q_{2} \cup
\overline{Q_{2}})} \! + \! \vert \vert w^{c}(\cdot) \vert \vert_{
\mathcal{L}^{1}_{\mathrm{M}_{2}(\mathbb{C})}(Q_{3} \cup \overline{Q_{
3}})} \! + \! \vert \vert w^{c}(\cdot) \vert \vert_{\mathcal{L}^{1}_{
\mathrm{M}_{2}(\mathbb{C})}(\mathrm{L}_{>} \cup \overline{\mathrm{L}_{
>}})}$, where $d_{z,\zeta} \! := \! \sup_{(z,\zeta) \in \Sigma^{\prime}
\times \mathbb{C} \setminus \Sigma^{\prime}} \vert (z \! - \! \zeta)^{-
1} \vert$ $(< \! \infty)$; hence, {}from the estimates given in
Lemma~4.4, modulo exponentially small terms, one deduces that $\vert
\mathscr{A} \vert \! \leqslant \! \tfrac{\vert \underline{c}(\zeta_{
1},\zeta_{2},\zeta_{3},\overline{\zeta_{3}}) \vert d_{z,\zeta}}{\vert
z_{o}+\zeta_{1}+\zeta_{2} \vert^{l}t^{l}}$. {}From the definition of
$\mathscr{B}$ given in Proposition~4.3, Eq.~(102), and Lemma~4.5, $2
\pi \vert \mathscr{B} \vert (d_{z,\zeta})^{-1} \! \leqslant \! \vert
\vert (\mathbf{1} \! - \! C_{w^{\prime}})^{-1} \vert \vert_{\mathscr{
N}(\Sigma^{\prime})} \vert \vert C_{w^{e}} \mathrm{I} \vert \vert_{
\mathcal{L}^{2}_{\mathrm{M}_{2}(\mathbb{C})}(\Sigma^{\prime})} \vert
\vert w^{\sharp}(\cdot) \vert \vert_{\mathcal{L}^{2}_{\mathrm{M}_{2}
(\mathbb{C})}(\Sigma^{\prime})} \! \leqslant \! \vert \underline{
c}(\zeta_{1},\zeta_{2},\zeta_{3},\overline{\zeta_{3}}) \vert (\vert
\vert w^{\widehat{a}}(\cdot) \vert \vert_{\mathcal{L}^{2}_{\mathrm{
M}_{2}(\mathbb{C})}(\mathbb{R})} \! + \! \vert \vert w^{b}(\cdot)
\linebreak[4]
\vert \vert_{\mathcal{L}^{2}_{\mathrm{M}_{2}(\mathbb{C})}(\mathrm{L}
\cup \overline{\mathrm{L}})} \! + \! \vert \vert w^{c}(\cdot) \vert
\vert_{\mathcal{L}^{2}_{\mathrm{M}_{2}(\mathbb{C})}(\mathrm{L}_{
\varepsilon} \cup \overline{\mathrm{L}_{\varepsilon}})})(\vert \vert
w^{\prime}(\cdot) \vert \vert_{\mathcal{L}^{2}_{\mathrm{M}_{2}(\mathbb{
C})}(\Sigma^{\prime})} \! + \! \vert \vert w^{\widehat{a}}(\cdot) \vert
\vert_{\mathcal{L}^{2}_{\mathrm{M}_{2}(\mathbb{C})}(\mathbb{R})} \!
+ \! \vert \vert w^{b}(\cdot) \vert \vert_{\mathcal{L}^{2}_{\mathrm{M}_{
2}(\mathbb{C})}(\mathrm{L} \cup \overline{\mathrm{L}})} + \vert \vert
w^{c}(\cdot) \vert \vert_{\mathcal{L}^{2}_{\mathrm{M}_{2}(\mathbb{C})}
(\mathrm{L}_{\varepsilon} \cup \overline{\mathrm{L}_{\varepsilon}})})$;
using the estimates given in Lemma~4.4, and recalling that $w^{\prime}
(\zeta) \! \! \upharpoonright_{\Sigma^{\prime} \setminus \Sigma^{\sharp}}
= \! \left(
\begin{smallmatrix}
0 & 0 \\
0 & 0
\end{smallmatrix}
\right)$, one shows that $2 \pi \vert \mathscr{B} \vert (d_{z,\zeta})^{
-1} \! \leqslant \! \tfrac{\vert \underline{c}(\zeta_{1},\zeta_{2},
\zeta_{3},\overline{\zeta_{3}}) \vert}{\vert z_{o}+\zeta_{1}+\zeta_{2}
\vert^{l}t^{l}} \! \left(\tfrac{\vert \underline{c}(\zeta_{1},\zeta_{
2},\zeta_{3},\overline{\zeta_{3}}) \vert}{((\zeta_{1}-\zeta_{2}) \vert
z_{o}+\zeta_{1}+\zeta_{2} \vert)^{1/4}t^{1/4}} \! + \! \tfrac{\vert
\underline{c}(\zeta_{1},\zeta_{2},\zeta_{3},\overline{\zeta_{3}}) \vert}
{\vert z_{o}+\zeta_{1}+\zeta_{2} \vert^{l}t^{l}} \right)$, whence $\vert
\mathscr{B} \vert \! \leqslant \! \tfrac{\vert \underline{c}(\zeta_{
1},\zeta_{2},\zeta_{3},\overline{\zeta_{3}}) \vert d_{z,\zeta}}{\vert
z_{o}+\zeta_{1}+\zeta_{2} \vert^{l}t^{l}}$. {}From the definition
of $\mathscr{C}$ given in Proposition~4.3, Eq.~(102), Lemma~4.5,
and recalling that $w^{\prime}(\zeta) \! \! \upharpoonright_{\Sigma^{
\prime} \setminus \Sigma^{\sharp}} = \!
\left(
\begin{smallmatrix}
0 & 0 \\
0 & 0
\end{smallmatrix}
\right)$, $\tfrac{2 \pi \vert \mathscr{C} \vert}{d_{z,\zeta}} \leqslant
\vert \vert (\mathbf{1} - C_{w^{\prime}})^{-1} \vert \vert_{\mathscr{
N}(\Sigma^{\prime})} \vert \vert C_{w^{\prime}} \mathrm{I} \vert
\vert_{\mathcal{L}^{2}_{\mathrm{M}_{2}(\mathbb{C})}(\Sigma^{\prime})}
\linebreak[4]
\cdot \vert \vert w^{e}(\cdot) \vert \vert_{\mathcal{L}^{2}_{\mathrm{M}_{2}
(\mathbb{C})}(\Sigma^{\prime})} \! \leqslant \! \vert \underline{c}(\zeta_{
1},\zeta_{2},\zeta_{3},\overline{\zeta_{3}}) \vert \, \vert \vert w^{\prime}
(\cdot) \vert \vert_{\mathcal{L}^{2}_{\mathrm{M}_{2}(\mathbb{C})}(\Sigma^{
\sharp})}(\vert \vert w^{\widehat{a}}(\cdot) \vert \vert_{\mathcal{L}^{2}_{
\mathrm{M}_{2}(\mathbb{C})}(\mathbb{R})} \! + \! \vert \vert w^{b}(\cdot)
\vert \vert_{\mathcal{L}^{2}_{\mathrm{M}_{2}(\mathbb{C})}(\mathrm{L} \cup
\overline{\mathrm{L}})} \! + \! \vert \vert w^{c}(\cdot) \vert \vert_{
\mathcal{L}^{2}_{\mathrm{M}_{2}(\mathbb{C})}(\mathrm{L}_{\varepsilon} \cup
\overline{\mathrm{L}_{\varepsilon}})})$; thus, using the estimates given in
Lemma~4.4, one shows that $\tfrac{2 \pi \vert \mathscr{C} \vert}{d_{z,\zeta}}
\! \leqslant \! \tfrac{\vert \underline{c}(\zeta_{1},\zeta_{2},\zeta_{3},
\overline{\zeta_{3}}) \vert}{\vert z_{o}+\zeta_{1}+\zeta_{2} \vert^{l}t^{l}}
\tfrac{\vert \underline{c}(\zeta_{1},\zeta_{2},\zeta_{3},\overline{\zeta_{
3}}) \vert}{((\zeta_{1}-\zeta_{2}) \vert z_{o}+\zeta_{1}+\zeta_{2} \vert)^{
1/4}t^{1/4}}$, whence, $\vert \mathscr{C} \vert \! \leqslant \! \tfrac{\vert
\underline{c}(\zeta_{1},\zeta_{2},\zeta_{3},\overline{\zeta_{3}}) \vert d_{
z,\zeta}}{\vert z_{o}+\zeta_{1}+\zeta_{2} \vert^{l}t^{l}}$. {}From the
definition of $\mathscr{D}$ given in Proposition~4.3, Eq.~(102), Lemma~4.5,
and noting that $w^{\prime}(\zeta) \! \! \upharpoonright_{\Sigma^{\prime}
\setminus \Sigma^{\sharp}} = \!
\left(
\begin{smallmatrix}
0 & 0 \\
0 & 0
\end{smallmatrix}
\right)$, it follows that
\begin{align*}
\dfrac{2 \pi \vert \mathscr{D} \vert}{d_{z,\zeta}} \! \leqslant& \, \vert
\vert (\mathbf{1} \! - \! C_{w^{\prime}})^{-1} \vert \vert_{\mathscr{N}
(\Sigma^{\prime})} \vert \vert C_{w^{e}} \vert \vert_{\mathscr{N}(\Sigma^{
\prime})} \vert \vert (\mathbf{1} \! - \! C_{w^{\sharp}})^{-1} \vert \vert_{
\mathscr{N}(\Sigma^{\prime})} \vert \vert C_{w^{\sharp}} \mathrm{I} \vert
\vert_{\mathcal{L}^{2}_{\mathrm{M}_{2}(\mathbb{C})}(\Sigma^{\prime})} \\
\times& \, \vert \vert w^{\sharp}(\cdot) \vert \vert_{\mathcal{L}^{2}_{
\mathrm{M}_{2}(\mathbb{C})}(\Sigma^{\prime})} \! \leqslant \! \vert
\underline{c}(\zeta_{1},\zeta_{2},\zeta_{3},\overline{\zeta_{3}}) \vert
\vert \vert w^{e}(\cdot) \vert \vert_{\mathcal{L}^{\infty}_{\mathrm{M}_{2}
(\mathbb{C})}(\Sigma^{\prime})} \vert \vert w^{\sharp}(\cdot) \vert \vert_{
\mathcal{L}^{2}_{\mathrm{M}_{2}(\mathbb{C})}(\Sigma^{\prime})}^{2} \\
\leqslant& \, \vert \underline{c}(\zeta_{1},\zeta_{2},\zeta_{3},\overline{
\zeta_{3}}) \vert (\vert \vert w^{\widehat{a}}(\cdot) \vert \vert_{\mathcal{
L}^{\infty}_{\mathrm{M}_{2}(\mathbb{C})}(\mathbb{R})} \! + \! \vert \vert
w^{b}(\cdot) \vert \vert_{\mathcal{L}^{\infty}_{\mathrm{M}_{2}(\mathbb{C})}
(\mathrm{L} \cup \overline{\mathrm{L}})} \! + \! \vert \vert w^{c}(\cdot)
\vert \vert_{\mathcal{L}^{\infty}_{\mathrm{M}_{2}(\mathbb{C})}(\mathrm{L}_{
\varepsilon} \cup \overline{\mathrm{L}_{\varepsilon}})}) \\
\times& \, \sum_{\genfrac{}{}{0pt}{2}{n_{1}+n_{2}+n_{3}+n_{4}=2}{0 \leqslant
n_{i} \leqslant 2}} \! \tfrac{2!}{n_{1}!n_{2}!n_{3}!n_{4}!} \vert \vert w^{
\prime}(\cdot) \vert \vert_{\mathcal{L}^{2}_{\mathrm{M}_{2}(\mathbb{C})}
(\Sigma^{\sharp})}^{n_{1}} \vert \vert w^{\widehat{a}}(\cdot) \vert \vert_{
\mathcal{L}^{2}_{\mathrm{M}_{2}(\mathbb{C})}(\mathbb{R})}^{n_{2}} \vert \vert
w^{b}(\cdot) \vert \vert_{\mathcal{L}^{2}_{\mathrm{M}_{2}(\mathbb{C})}
(\mathrm{L} \cup \overline{\mathrm{L}})}^{n_{3}} \\
\times& \, \vert \vert w^{c}(\cdot) \vert \vert_{\mathcal{L}^{2}_{\mathrm{
M}_{2}(\mathbb{C})}(\mathrm{L}_{\varepsilon} \cup \overline{\mathrm{L}_{
\varepsilon}})}^{n_{4}},
\end{align*}
whence, using Chebyshev's inequality\footnote{If $a_{1} \! \geqslant \! a_{2}
\! \geqslant \! \cdots \! \geqslant \! a_{n}$ and $b_{1} \! \geqslant \! b_{
2} \! \geqslant \! \cdots \! \geqslant \! b_{n}$, $(a_{1} \! + \! a_{2} \! +
\! \cdots \! + \! a_{n})(b_{1} \! + \! b_{2} \! + \! \cdots \! + \! b_{n}) \!
\leqslant \! n \sum_{i=1}^{n}a_{i}b_{i}$.}, one shows that
\begin{align*}
\dfrac{\pi \vert \mathscr{D} \vert (2d_{z,\zeta})^{-1}}{\vert \underline{c}
(\zeta_{1},\zeta_{2},\zeta_{3},\overline{\zeta_{3}}) \vert} \! \leqslant& \,
(\vert \vert w^{\widehat{a}}(\cdot) \vert \vert_{\mathcal{L}^{\infty}_{
\mathrm{M}_{2}(\mathbb{C})}(\mathbb{R})} \! + \! \vert \vert w^{b}(\cdot)
\vert \vert_{\mathcal{L}^{\infty}_{\mathrm{M}_{2}(\mathbb{C})}(\mathrm{L}
\cup \overline{\mathrm{L}})} \! + \! \vert \vert w^{c}(\cdot) \vert \vert_{
\mathcal{L}^{\infty}_{\mathrm{M}_{2}(\mathbb{C})}(\mathrm{L}_{\varepsilon}
\cup \overline{\mathrm{L}_{\varepsilon}})}) \\
\times& \, (\vert \vert w^{\prime}(\cdot) \vert \vert_{\mathcal{L}^{2}_{
\mathrm{M}_{2}(\mathbb{C})}(\Sigma^{\sharp})}^{2} \! + \! \vert \vert w^{
\widehat{a}}(\cdot) \vert \vert_{\mathcal{L}^{2}_{\mathrm{M}_{2}(\mathbb{
C})}(\mathbb{R})}^{2} \! + \! \vert \vert w^{b}(\cdot) \vert \vert_{\mathcal{
L}^{2}_{\mathrm{M}_{2}(\mathbb{C})}(\mathrm{L} \cup \overline{\mathrm{L}})}^{
2} \\
+& \, \vert \vert w^{c}(\cdot) \vert \vert_{\mathcal{L}^{2}_{\mathrm{M}_{
2}(\mathbb{C})}(\mathrm{L}_{\varepsilon} \cup \overline{\mathrm{L}_{
\varepsilon}})}^{2});
\end{align*}
hence, {}from the estimates given in Lemma~4.4, $\tfrac{\pi \vert \mathscr{
D} \vert}{2d_{z,\zeta}} \! \leqslant \! \tfrac{\vert \underline{c}(\zeta_{
1},\zeta_{2},\zeta_{3},\overline{\zeta_{3}}) \vert}{\vert z_{o}+\zeta_{1}+
\zeta_{2} \vert^{l}t^{l}} \tfrac{\vert \underline{c}(\zeta_{1},\zeta_{2},
\zeta_{3},\overline{\zeta_{3}}) \vert}{\sqrt{(\zeta_{1}-\zeta_{2}) \vert z_{
o}+\zeta_{1}+\zeta_{2} \vert} \, \sqrt{t}}$, whence $\vert \mathscr{D} \vert
\! \leqslant \! \tfrac{\vert \underline{c}(\zeta_{1},\zeta_{2},\zeta_{3},
\overline{\zeta_{3}}) \vert d_{z,\zeta}}{\vert z_{o}+\zeta_{1}+\zeta_{2}
\vert^{l}t^{l}}$. {}From the above-derived bounds for $\mathscr{A}$,
$\mathscr{B}$, $\mathscr{C}$, and $\mathscr{D}$, one arrives at the result
stated in the Proposition. \hfill $\square$
\begin{aaaaa}
Recall the definition of the {\rm BC} operator: $C_{\star} \diamondsuit \!
:= \! C_{+}(\diamondsuit \star_{-}) \! + \! C_{-}(\diamondsuit \star_{
+})$, $\diamondsuit \! \in \! \mathcal{L}^{2}_{\mathrm{M}_{2}(\mathbb{
C})}(\ast)$, where $C_{\pm} \colon \mathcal{L}^{2}_{\mathrm{M}_{2}
(\mathbb{C})}(\ast) \! \to \! \mathcal{L}^{2}_{\mathrm{M}_{2}(\mathbb{
C})}(\ast)$ are the (bounded) Cauchy operators introduced at the
beginning of Section~{\rm 3}. Let: (1) $C_{w^{\sharp}} \! := \! C_{w^{\sharp}
}^{\Sigma^{\prime}} \colon \mathcal{L}^{2}_{\mathrm{M}_{2}(\mathbb{C})}
(\Sigma^{\prime}) \! \to \! \mathcal{L}^{2}_{\mathrm{M}_{2}(\mathbb{C})}
(\Sigma^{\prime})$ denote the {\rm BC} operator with $\star \! \leftrightarrow
\! w^{\sharp};$ (2) $C_{w^{\prime}} \! := \! C_{w^{\prime}}^{s} \colon
\mathcal{L}^{2}_{\mathrm{M}_{2}(\mathbb{C})}(s) \! \to \! \mathcal{
L}^{2}_{\mathrm{M}_{2}(\mathbb{C})}(s)$, $s \! \in \! \{\Sigma^{
\prime},\Sigma^{\sharp}\}$, denotes the {\rm BC} operator with $\star \!
\leftrightarrow \! w^{\prime};$ (3) $\mathbf{1}_{s}$, $s \! \in \!
\{\Sigma^{\prime},\Sigma^{\sharp}\}$, denotes the identity operator
on $\mathcal{L}^{2}_{\mathrm{M}_{2}(\mathbb{C})}(s);$ and (4) $w^{
\sharp}_{\Sigma^{\prime}}(\zeta) \! := \! w^{\sharp}(\zeta) \! \!
\upharpoonright_{\Sigma^{\prime}}$ and $w^{\Sigma^{\sharp}}(\zeta)
\! := \! w^{\prime}(\zeta) \! \! \upharpoonright_{\Sigma^{\sharp}}$.
\end{aaaaa}

Note also that, since $w^{\prime}(\zeta) \! \! \! \upharpoonright_{
\Sigma^{\prime} \setminus \Sigma^{\sharp}} \, = \left(
\begin{smallmatrix}
0 & 0 \\
0 & 0
\end{smallmatrix}
\right)$, $C_{w^{\prime}}^{\Sigma^{\prime}} \star \! = \! C_{+}^{
\Sigma^{\prime} \setminus \Sigma^{\sharp} \cup \Sigma^{\sharp}}(
\star w^{\prime}_{-}) \! + \! C_{-}^{\Sigma^{\prime} \setminus
\Sigma^{\sharp} \cup \Sigma^{\sharp}}(\star w_{+}^{\prime}) \! = \!
C_{+}^{\Sigma^{\sharp}}(\star w^{\prime}_{-}) \! + \! C_{-}^{\Sigma^{
\sharp}}(\star w_{+}^{\prime}) \! = \! C_{w^{\prime}}^{\Sigma^{\sharp}
} \star$.
\begin{fffff}
If $(\mathbf{1}_{\Sigma^{\prime}} \! - \! C_{w^{\prime}}^{\Sigma^{
\prime}})^{-1} \! \in \! \mathscr{N}(\Sigma^{\prime})$, then, for
$\zeta \! \in \! \mathbb{C} \setminus \Sigma^{\sharp}$ and arbitrarily
large $l \! \in \! \mathbb{Z}_{\geqslant 1}$, as $t \! \to \! +\infty$
such that $0 \! < \! \zeta_{2} \! < \! \tfrac{1}{M} \! < \! M \! < \!
\zeta_{1}$ and $\vert \zeta_{3} \vert^{2} \! = \! 1$, with $M \! \in
\! \mathbb{R}_{>1}$ and bounded,
\begin{align*}
\int\nolimits_{\Sigma^{\prime}} \dfrac{((\mathbf{1}_{\Sigma^{\prime}}
\! - \! C_{w^{\sharp}}^{\Sigma^{\prime}})^{-1} \mathrm{I})(z)w^{\sharp}_{
\Sigma^{\prime}}(z)}{(z \! - \! \zeta)} \, \dfrac{\md z}{2 \pi \mi} &= \!
\int\nolimits_{\Sigma^{\sharp}} \dfrac{((\mathbf{1}_{\Sigma^{\sharp}} \! -
\! C_{w^{\Sigma^{\sharp}}}^{\Sigma^{\sharp}})^{-1} \mathrm{I})(z)w^{
\Sigma^{\sharp}}(z)}{(z \! - \! \zeta)} \, \dfrac{\md z}{2 \pi \mi} \\
 &+ \mathcal{O} \! \left( \dfrac{\underline{c}(\zeta_{1},\zeta_{2},\zeta_{
3},\overline{\zeta_{3}})f^{\sharp}(\zeta)}{\vert z_{o} \! + \! \zeta_{1}
\! + \! \zeta_{2} \vert^{l}t^{l}} \right),
\end{align*}
with $f^{\sharp}(\zeta) \! \in \! \mathcal{L}^{\infty}_{\mathrm{
M}_{2}(\mathbb{C})}(\mathbb{C} \setminus \Sigma^{\sharp})$.
\end{fffff}

\emph{Proof.} Set $\mu^{\Sigma^{\prime}}(\zeta) \! := \! ((\mathbf{
1}_{\Sigma^{\prime}} \! - \! C_{w^{\sharp}}^{\Sigma^{\prime}})^{-1}
\mathrm{I})(\zeta)$ and $\mu^{\Sigma^{\sharp}}(\zeta) \! := \! ((
\mathbf{1}_{\Sigma^{\sharp}} \! - \! C_{w^{\Sigma^{\sharp}}}^{\Sigma^{
\sharp}})^{-1} \mathrm{I})(\zeta)$. Then, {}from Proposition~4.4,
Definition~4.2, and the fact that $w^{\prime}(\zeta) \! \!
\upharpoonright_{\Sigma^{\prime} \setminus \Sigma^{\sharp}} = \!
\left(
\begin{smallmatrix}
0 & 0 \\
0 & 0
\end{smallmatrix}
\right)$,
\begin{gather*}
\int\nolimits_{\Sigma^{\prime}} \tfrac{\mu^{\Sigma^{\prime}}(z)w^{
\sharp}_{\Sigma^{\prime}}(z)}{(z-\zeta)} \tfrac{\md z}{2 \pi \mi}
\! = \! \int\nolimits_{\Sigma^{\prime} \setminus \Sigma^{\sharp}
\cup \Sigma^{\sharp}} \tfrac{((\mathbf{1}_{\Sigma^{\prime}}-C_{w^{
\prime}}^{\Sigma^{\prime}})^{-1} \mathrm{I})(z)w^{\prime}(z)}{(z-
\zeta)} \tfrac{\md z}{2 \pi \mi} \! + \! \mathscr{E} \\
\! = \! \int\nolimits_{\Sigma^{\prime} \setminus \Sigma^{\sharp}}
\! \tfrac{(((\mathbf{1}_{\Sigma^{\prime}}-C_{w^{\prime}}^{\Sigma^{
\prime}})^{-1} \mathrm{I})(z) \upharpoonright_{\Sigma^{\prime}
\setminus \Sigma^{\sharp}})(w^{\prime}(z) \upharpoonright_{\Sigma^{
\prime} \setminus \Sigma^{\sharp}})}{(z-\zeta)} \tfrac{\md z}{2 \pi
\mi} \! + \! \int\nolimits_{\Sigma^{\sharp}} \! \tfrac{(((\mathbf{
1}_{\Sigma^{\prime}}-C_{w^{\prime}}^{\Sigma^{\prime}})^{-1} \mathrm{I}
)(z) \upharpoonright_{\Sigma^{\sharp}})(w^{\prime}(z) \upharpoonright_{
\Sigma^{\sharp}})}{(z-\zeta)} \tfrac{\md z}{2 \pi \mi} \! + \! \mathscr{E} \\
\! = \! \int\nolimits_{\Sigma^{\sharp}} \tfrac{((\mathbf{1}_{\Sigma^{
\sharp}}-C_{w^{\Sigma^{\sharp}}}^{\Sigma^{\sharp}})^{-1} \mathrm{I})
(z)w^{\Sigma^{\sharp}}(z)}{(z-\zeta)} \tfrac{\md z}{2 \pi \mi} \! +
\! \mathscr{E} \! = \! \int\nolimits_{\Sigma^{\sharp}} \tfrac{\mu^{
\Sigma^{\sharp}}(z)w^{\Sigma^{\sharp}}(z)}{(z-\zeta)} \tfrac{\md z}
{2 \pi \mi} \! + \! \mathscr{E},
\end{gather*}
where $\mathscr{E} \! := \! \mathcal{O} \! \left( \tfrac{\underline{
c}(\zeta_{1},\zeta_{2},\zeta_{3},\overline{\zeta_{3}})f^{\sharp}
(\zeta)}{\vert z_{o}+\zeta_{1}+\zeta_{2} \vert^{l}t^{l}} \right)$,
with $\vert \vert f^{\sharp}(\cdot) \vert \vert_{\mathcal{L}^{\infty}_{
\mathrm{M}_{2}(\mathbb{C})}(\mathbb{C} \setminus \Sigma^{
\sharp})} \! < \! \infty$. \hfill $\square$

As a consequence of Corollary~4.1, one can now consider, to $\mathcal{O} \!
\left( \tfrac{\underline{c}(\zeta_{1},\zeta_{2},\zeta_{3},\overline{\zeta_{
3}}) \diamondsuit (\zeta)}{\vert z_{o}+\zeta_{1}+\zeta_{2} \vert^{l}t^{l}}
\right)$, with arbitrarily large $l \! \in \! \mathbb{Z}_{\geqslant 1}$, and
$\vert \vert \diamondsuit (\cdot) \vert \vert_{\mathcal{L}^{\infty}_{\mathrm{
M}_{2}(\mathbb{C})}(\mathbb{C} \setminus \Sigma^{\sharp})} \! < \! \infty$,
the (normalised at $\infty)$ RHP for $m^{\Sigma^{\sharp}}(\zeta):=m^{\sharp}
(\zeta) \! \! \! \! \upharpoonright_{\Sigma^{\sharp}}$ on $\Sigma^{\sharp}$,
thus realising the second main objective of this section, and culminating
in the following
\begin{ccccc}
As $t \! \to \! +\infty$ such that $0 \! < \! \zeta_{2} \! < \! \tfrac{1}{M}
\! < \! M \! < \! \zeta_{1}$ and $\vert \zeta_{3} \vert^{2} \! = \! 1$, with
$M \! \in \! \mathbb{R}_{>1}$ and bounded, $m^{\Sigma^{\sharp}}(\zeta) \! :=
\! m^{\sharp}(\zeta) \! \! \upharpoonright_{\Sigma^{\sharp}}$ solves the
following {\rm RHP:} (1) $m^{\Sigma^{\sharp}}(\zeta)$ is piecewise
holomorphic $\forall \, \zeta \! \in \! \mathbb{C} \setminus \Sigma^{
\sharp};$ (2) $m^{\Sigma^{\sharp}}_{\pm}(\zeta) \! := \! \lim_{\genfrac{}{}
{0pt}{2}{\zeta^{\prime} \, \to \, \zeta}{\zeta^{\prime} \, \in \, \pm \,
\mathrm{side} \, \mathrm{of} \, \Sigma^{\sharp}}}m^{\Sigma^{\sharp}}(\zeta^{
\prime})$ satisfy the jump condition $m^{\Sigma^{\sharp}}_{+}(\zeta) \! = \!
m^{\Sigma^{\sharp}}_{-}(\zeta)(\mathrm{I} \! - \! w^{\Sigma^{\sharp}}_{-}
(\zeta))^{-1}(\mathrm{I} \! + \! w^{\Sigma^{\sharp}}_{+}(\zeta))$, $\zeta \!
\in \! \Sigma^{\sharp}$, where
\begin{gather*}
w^{\Sigma^{\sharp}}_{+}(\zeta) \! = \! (\delta (\zeta))^{\mathrm{ad}
(\sigma_{3})} \exp (-\mi t \theta^{u}(\zeta) \mathrm{ad}(\sigma_{3}))
\mathcal{R}(\zeta) \sigma_{+}, \quad w^{\Sigma^{\sharp}}_{-}(\zeta)
\! = \! \left(
\begin{smallmatrix}
0 & 0 \\
0 & 0
\end{smallmatrix}
\right), \quad \, \, \, \zeta \! \in \! \mathrm{L} \setminus \widetilde{
\mathrm{L}}_{\varepsilon} \subset \Sigma^{\sharp}, \\
w^{\Sigma^{\sharp}}_{+}(\zeta) \! = \!
\left(
\begin{smallmatrix}
0 & 0 \\
0 & 0
\end{smallmatrix}
\right), \quad \, \, w^{\Sigma^{\sharp}}_{-}(\zeta) \! = \! -(\delta (\zeta)
)^{\mathrm{ad}(\sigma_{3})} \exp (-\mi t \theta^{u}(\zeta) \mathrm{ad}
(\sigma_{3})) \overline{\mathcal{R}(\zeta)} \, \sigma_{-}, \quad \zeta
\! \in \! \overline{\mathrm{L}} \setminus \overline{\widetilde{\mathrm{
L}}_{\varepsilon}} \subset \Sigma^{\sharp},
\end{gather*}
with $\widetilde{\mathrm{L}}_{\varepsilon} \! := \! \mathrm{L}_{
\varepsilon} \cup \mathrm{L}_{>} \cup \mathrm{L}_{<};$ (3) as $\zeta \!
\to \! \infty$, $\zeta \! \in \! \mathbb{C} \setminus \Sigma^{\sharp}$,
$m^{\Sigma^{\sharp}}(\zeta) \! = \! \mathrm{I} \! + \! \mathcal{O}
(\zeta^{-1});$ and (4) $m^{\Sigma^{\sharp}}(\zeta)$ satisfies the
symmetry reduction $m^{\Sigma^{\sharp}}(\zeta) \! = \! \sigma_{1}
\overline{m^{\Sigma^{\sharp}}(\overline{\zeta})} \, \sigma_{1}$
and the condition $(m^{\Sigma^{\sharp}}(0)(\delta (0))^{\sigma_{
3}} \sigma_{2})^{2} \! = \! \mathrm{I}$. Furthermore, $w^{\Sigma^{
\sharp}}_{\pm}(\zeta) \! \in \! \cap_{p \in \{1,2,\infty\}} \mathcal{
L}^{p}_{\mathrm{M}_{2}(\mathbb{C})}(\Sigma^{\sharp})$.
\end{ccccc}

\emph{Proof.} Follows {}from Lemma~4.3, Lemma~4.5, Corollary~4.1, and
the definition of $m^{\Sigma^{\sharp}}(\zeta)$ given in the Lemma.
\hfill $\square$
\begin{eeeee}
In Lemma~4.6, $\overline{\mathcal{R}(\zeta)}$ denotes the
piecewise-rational function $\mathcal{R}(\zeta)$ with the complex
conjugated coefficients.
\end{eeeee}

Using Lemma~3.1, the solution of the RHP for $m^{\Sigma^{\sharp}}
(\zeta)$ on $\Sigma^{\sharp}$ stated in Lemma~4.6 has the integral
representation
\begin{equation}
m^{\Sigma^{\sharp}}(\zeta) \! = \! \mathrm{I} \! + \! \int\nolimits_{
\Sigma^{\sharp}} \dfrac{\mu^{\Sigma^{\sharp}}(z)w^{\Sigma^{\sharp}}(z)}
{(z \! - \! \zeta)} \, \dfrac{\md z}{2 \pi \mi}, \quad \zeta \! \in \!
\mathbb{C} \setminus \Sigma^{\sharp},
\end{equation}
where $\mu^{\Sigma^{\sharp}}(\zeta) \! := \! ((\mathbf{1}_{\Sigma^{
\sharp}} \! - \! C^{\Sigma^{\sharp}}_{w^{\Sigma^{\sharp}}})^{-1}
\mathrm{I})(\zeta)$, and $w^{\Sigma^{\sharp}}(\zeta) \! := \! \sum_{
l \in \{\pm\}} \! w_{l}^{\Sigma^{\sharp}}(\zeta)$.
\section{Towards the Model RHP}
In this section, the RHP for $m^{\Sigma^{\sharp}}(\zeta)$ on $\Sigma^{
\sharp}$ stated in Lemma~4.6 is reduced to RHPs on the two disjoint crosses
$\Sigma_{A^{\prime}}$ and $\Sigma_{B^{\prime}}$, and it is shown that, as
$t \! \to \! +\infty$, the leading term of asymptotics of the (singular)
integral representation for $m^{\Sigma^{\sharp}}(\zeta)$ can be written as
the linear superposition of two (singular) integrals corresponding to the
solution of two auxiliary RHPs, each of which is defined on one of the
disjoint crosses. Furthermore, the basic bound on $(\mathbf{1}_{\Sigma^{
\sharp}} \! - \! C^{\Sigma^{\sharp}}_{w^{\Sigma^{\sharp}}})^{-1}$, namely,
$(\mathbf{1}_{\Sigma^{\sharp}} \! - \! C^{\Sigma^{\sharp}}_{w^{\Sigma^{
\sharp}}})^{-1} \! \in \! \mathscr{N}(\Sigma^{\sharp})$, is proved, whence,
as a consequence of Corollary~4.1, Definition~4.2, Proposition~4.4, and
Proposition~4.2, the basic bound on $(\mathbf{1} \! - \! C_{w^{\prime}})^{
-1}$, that is, $(\mathbf{1} \! - \! C_{w^{\prime}})^{-1} \! \in \! \mathscr{
N}(\Sigma^{\prime})$, follows.

To formulate a number of exact results, some notational preamble
is necessary. Recalling that, for $\diamondsuit (\zeta) \! \in \!
\mathcal{L}^{2}_{\mathrm{M}_{2}(\mathbb{C})}(\Sigma^{\sharp})$, the
BC operator is defined as $C^{\Sigma^{\sharp}}_{w^{\Sigma^{\sharp}}}
\diamondsuit \! := \! C_{+}(\diamondsuit w^{\Sigma^{\sharp}}_{-}) \!
+ \! C_{-}(\diamondsuit w^{\Sigma^{\sharp}}_{+})$, where $(C_{\pm}
\diamondsuit)(\zeta) \! := \! \lim_{\genfrac{}{}{0pt}{2}{\zeta^{
\prime} \, \to \, \zeta}{\zeta^{\prime} \, \in \, \pm \, \mathrm{side}
\, \mathrm{of} \, \Sigma^{\sharp}}} \! \int_{\Sigma^{\sharp}} \tfrac{
\diamondsuit (z)}{(z-\zeta^{\prime})} \, \tfrac{\md z}{2 \pi \mi}$,
one shows that, for $w^{\Sigma^{\sharp}}(\zeta) \! := \! \sum_{l
\in \{A^{\prime},B^{\prime}\}} \! w^{\Sigma_{l}}(\zeta)$, where $w^{
\Sigma_{l}}(\zeta) \! := \! w^{\Sigma^{\sharp}}(\zeta) \! \!
\upharpoonright_{\Sigma_{l}}$ and $w^{\Sigma_{j}}(\zeta) \! = \!
\left(
\begin{smallmatrix}
0 & 0 \\
0 & 0
\end{smallmatrix}
\right)$, $\zeta \! \in \! \Sigma_{l}$, $l \! \not= \! j \! \in \!
\{A^{\prime},B^{\prime}\}$, $C^{\Sigma^{\sharp}}_{w^{\Sigma^{\sharp}}}
\diamondsuit \! = \! C^{\Sigma^{\sharp}}_{w^{\Sigma_{A^{\prime}}}}
\diamondsuit \! + \! C^{\Sigma^{\sharp}}_{w^{\Sigma_{B^{\prime}}}}
\diamondsuit$, where $C^{\Sigma^{\sharp}}_{w^{\Sigma_{l}}} \diamondsuit
\! := \! (C_{+}(\diamondsuit w^{\Sigma^{\sharp}}_{-}) \! + \! C_{-}
(\diamondsuit w^{\Sigma^{\sharp}}_{+})) \! \! \! \upharpoonright_{\Sigma_{
l}}$, $l \! \in \! \{A^{\prime},B^{\prime}\}$; hence, as an operator
on $\mathcal{L}^{2}_{\mathrm{M}_{2}(\mathbb{C})}(\Sigma^{\sharp})$,
$C^{\Sigma^{\sharp}}_{w^{\Sigma^{\sharp}}} \! := \! \sum_{l \in \{A^{
\prime},B^{\prime}\}} \! C^{\Sigma_{l}}_{w^{\Sigma_{l}}}$. Writing
$\Sigma^{\sharp} \! = \! \Sigma_{A^{\prime}} \cup \Sigma_{B^{\prime}}$,
with $\Sigma_{A^{\prime}} \cap \Sigma_{B^{\prime}} \! = \! \emptyset$,
extend the (oriented) contours $\Sigma_{A^{\prime}}$ and $\Sigma_{B^{
\prime}}$ (with orientations unchanged), respectively, to the oriented
contours
\begin{align*}
\widehat{\Sigma}_{A^{\prime}} \! :=& \{\mathstrut \zeta (v); \, \zeta (v) \!
= \! \zeta_{2} \! + \! \tfrac{v}{\sqrt{2}}(\zeta_{1} \! - \! \zeta_{2}) \exp
(\pm \tfrac{\mi \pi}{4}), \, v \! \in \! [0,+\infty)\} \\
\cup& \{\mathstrut \zeta (v); \, \zeta (v) \! = \! \zeta_{2} \! + \! \tfrac{
v}{\sqrt{2}} \zeta_{2} \exp (\pm \tfrac{3 \pi \mi}{4}), \, v \! \in \! [0,
+\infty)\}, \\
\widehat{\Sigma}_{B^{\prime}} \! :=& \{\mathstrut \zeta (v); \,
\zeta (v) \! = \! \zeta_{1} \! + \! \tfrac{v}{\sqrt{2}}(\zeta_{1} \!
- \! \zeta_{2}) \exp (\pm \tfrac{3 \pi \mi}{4}), \, v \! \in \! \mathbb{
R}\},
\end{align*}
and denote by $\Sigma_{A}$ and $\Sigma_{B}$, respectively, the
contours $\{\mathstrut \varpi (v); \, \varpi (v) \! = \! v(\zeta_{1} \! - \!
\zeta_{2}) \exp (\pm \tfrac{\mi \pi}{4}), \, v \! \in \! \mathbb{R}\}$
oriented ``outward'', as in $\Sigma_{A^{\prime}}$ and $\widehat{\Sigma}_{
A^{\prime}}$, and ``inward'', as in $\Sigma_{B^{\prime}}$ and $\widehat{
\Sigma}_{B^{\prime}}$. For $l \! \in \! \{A^{\prime},B^{\prime}\}$,
define $\widehat{w}^{\widehat{\Sigma}_{l}}(\zeta) \! = \! \sum_{k \in
\{\pm\}} \! \widehat{w}_{k}^{\widehat{\Sigma}_{l}}(\zeta)$ on $\widehat{
\Sigma}_{l}$ via
\begin{equation}
\widehat{w}^{\widehat{\Sigma}_{l}}(\zeta) \! := \!
\begin{cases}
w^{\Sigma_{l}}(\zeta) \! = \! \sum_{k \in \{\pm\}} \! w^{\Sigma_{l}}_{
k}(\zeta), &\text{$\zeta \! \in \! \Sigma_{l} \subset \widehat{\Sigma}_{
l}$,} \\
\left(
\begin{smallmatrix}
0 & 0 \\
0 & 0
\end{smallmatrix}
\right), &\text{$\zeta \! \in \! \widehat{\Sigma}_{l} \setminus \Sigma_{l}$.}
\end{cases}
\end{equation}
The corresponding BC operators on $\Sigma_{l}$, $l \! \in \! \{A,B\}$, are
denoted by $C^{\Sigma_{l}}_{w^{\Sigma_{l}}}$, and, on $\widehat{\Sigma}_{
l^{\prime}}$, by $C^{\widehat{\Sigma}_{l^{\prime}}}_{\widehat{w}^{\widehat{
\Sigma}_{l^{\prime}}}}$. Introduce the following scaling-shifting operators:
\begin{equation}
\begin{split}
\mathcal{N}_{A} &\colon \mathcal{L}^{2}(\widehat{\Sigma}_{A^{\prime}
}) \! \to \! \mathcal{L}^{2}(\Sigma_{A}), \qquad \, \, f(\zeta) \! \mapsto
\! (\mathcal{N}_{A}f)(\widetilde{w}) \! = \! f(\zeta_{2} \! + \!
\varepsilon_{A}(\widetilde{w})), \\
\mathcal{N}_{B} &\colon \mathcal{L}^{2}(\widehat{\Sigma}_{B^{\prime}
}) \! \to \! \mathcal{L}^{2}(\Sigma_{B}), \qquad \, \, g(\zeta) \! \mapsto
\! (\mathcal{N}_{B}g)(\widetilde{w}) \! = \! g(\zeta_{1} \! + \!
\varepsilon_{B}(\widetilde{w})),
\end{split}
\end{equation}
where
\begin{equation}
\varepsilon_{A}(\widetilde{w}) \! := \! \widetilde{w}/\tfrac{\vert
\zeta_{2}-\zeta_{3} \vert}{\zeta_{2}} \! \left(\tfrac{2t(\zeta_{1}-
\zeta_{2})}{\zeta_{2}} \right)^{1/2}, \qquad \, \, \varepsilon_{B}
(\widetilde{w}) \! := \! \widetilde{w}/\tfrac{\vert \zeta_{1}-\zeta_{
3} \vert}{\zeta_{1}} \! \left( \tfrac{2t(\zeta_{1}-\zeta_{2})}{\zeta_{
1}} \right)^{1/2}.
\end{equation}
Noting {}from the expressions for $w^{\Sigma^{\sharp}}_{\pm}(\zeta)$
given in Lemma~4.6 that, modulo factors like $\mathcal{R}(\zeta)$ and
$\overline{\mathcal{R}(\zeta)}$, the elements of the jump matrix for
$m^{\Sigma^{\sharp}}(\zeta)$ are proportional to $(\delta (\zeta))^{
\pm 2} \exp (\mp 2 \mi t \theta^{u}(\zeta))$, one considers the
``action'' of $\mathcal{N}_{k}$, $k \! \in \! \{A,B\}$, on such terms.
\begin{bbbbb}
Let $\mathcal{N}_{k}$, $k \! \in \! \{A,B\}$, be the operators defined
in Eq.~{\rm (105)}. Then, for $k \! \in \! \{A,B\}$, with $\varepsilon_{k}
(\widetilde{w})$ defined in Eq.~{\rm (106)},
\begin{equation*}
(\mathcal{N}_{k}(\delta^{\pm 2} \me^{\mp 2 \mi t \theta^{u}}))
(\widetilde{w}) \! = \! (\delta_{k}^{0})^{\pm 2}(\delta_{k}^{1}
(\widetilde{w}))^{\pm 2},
\end{equation*}
where
\begin{align*}
\delta_{A}^{0} \! &:= \! \vert \zeta_{2} \! - \! \zeta_{3} \vert^{\mi
\nu}(2t(\zeta_{1} \! - \! \zeta_{2})^{3} \zeta_{2}^{-3})^{\frac{\mi
\nu}{2}} \me^{\chi (\zeta_{2})} \exp (\tfrac{\mi t}{2}(\zeta_{1} \!
- \! \zeta_{2})(z_{o} \! + \! \zeta_{1} \! + \! \zeta_{2})), \\
\delta_{B}^{0} \! &:= \! \vert \zeta_{1} \! - \! \zeta_{3} \vert^{-\mi
\nu}(2t(\zeta_{1} \! - \! \zeta_{2})^{3} \zeta_{1}^{-3})^{-\frac{\mi
\nu}{2}} \me^{\chi (\zeta_{1})} \exp (-\tfrac{\mi t}{2}(\zeta_{1} \! -
\! \zeta_{2})(z_{o} \! + \! \zeta_{1} \! + \! \zeta_{2})), \\
\delta_{A}^{1}(\widetilde{w}) \! &:= \! (-\widetilde{w})^{-\mi \nu} \!
\left( \tfrac{-(\zeta_{1}-\zeta_{2})+\varepsilon_{A}(\widetilde{w})}
{-(\zeta_{1}-\zeta_{2})} \right)^{\mi \nu} \me^{\chi (\zeta_{2}+
\varepsilon_{A}(\widetilde{w}))-\chi (\zeta_{2})} \me^{\frac{\mi
\widetilde{w}^{2}}{4}} \exp \! \left( \tfrac{\mi \theta^{u}_{3}
(\zeta_{2}) \zeta_{2}^{9/2} \widetilde{w}^{3}}{3!(2(\zeta_{1}-\zeta_{
2}))^{3/2} \vert \zeta_{2}-\zeta_{3} \vert^{3} \sqrt{t}} \right), \\
\delta_{B}^{1}(\widetilde{w}) \! &:= \! (\widetilde{w})^{\mi \nu} \!
\left( \tfrac{(\zeta_{1}-\zeta_{2})}{(\zeta_{1}-\zeta_{2})+\varepsilon_{
B}(\widetilde{w})} \right)^{\mi \nu} \me^{\chi (\zeta_{1}+\varepsilon_{
B}(\widetilde{w}))-\chi (\zeta_{1})} \me^{-\frac{\mi \widetilde{w}^{2}}
{4}} \exp \! \left( -\tfrac{\mi \theta^{u}_{3}(\zeta_{1}) \zeta_{1}^{
9/2} \widetilde{w}^{3}}{3! (2(\zeta_{1}-\zeta_{2}))^{3/2} \vert \zeta_{
1}-\zeta_{3} \vert^{3} \sqrt{t}} \right),
\end{align*}
with $\nu$ defined in Proposition~{\rm 4.1},
\begin{equation*}
\chi (\zeta) \! := \! \int\nolimits_{-\infty}^{0} \dfrac{\ln (1 \! - \!
\vert r(\mu) \vert^{2})}{(\mu \! - \! \zeta)} \, \dfrac{\md \mu}{2 \pi
\mi} \! + \! \int\nolimits_{\zeta_{2}}^{\zeta_{1}} \ln \! \left( \dfrac{
1 \! - \! \vert r(\mu) \vert^{2}}{1 \! - \! \vert r(\zeta_{1}) \vert^{
2}} \right) \! \dfrac{1}{(\mu \! - \! \zeta)} \, \dfrac{\md \mu}{2 \pi
\mi},
\end{equation*}
$\theta^{u}_{3}(\zeta_{n}) \! := \! 2 \zeta_{n}^{-3} \! \left(2(\zeta_{
1} \! - \! \zeta_{2})(\zeta_{n} \! - \! \cos \widetilde{\varphi}_{3}) \!
+ \! (-1)^{n+1} \vert \zeta_{n} \! - \! \zeta_{3} \vert^{2}(1 \! + \!
3(-1)^{n} \zeta_{n}^{-1}(\zeta_{1} \! - \! \zeta_{2})) \right)$, $n \!
\in \! \{1,2\}$, $\widetilde{\varphi}_{3} \! := \! \arg (\zeta_{3}) \!
\in \! (\tfrac{\pi}{2},\pi)$, and $(\pm \widetilde{w})^{\pm \mi \nu}
\! := \! \exp \! \left(\pm \mi \nu \ln (\pm \widetilde{w}) \right)$ with
branch cuts along $\mp \mathbb{R}_{+}$.
\end{bbbbb}

\emph{Proof.} Consequence of the expression for $\delta (\zeta)$
given in Proposition~4.1, the formula (Eq.~(8)) $\theta^{u}(\zeta)
\! = \! \tfrac{1}{2}(\zeta \! - \! \tfrac{1}{\zeta})(z_{o} \! + \! \zeta \! +
\! \tfrac{1}{\zeta})$, the definition of the operators $\mathcal{N}_{
k}$, $k \! \in \! \{A,B\}$, given in Eq.~(105), and Eq.~(106). \hfill
$\square$

For $k \! \in \! \{A,B\}$, define
\begin{equation}
\Delta_{k}^{0} \! := \! (\delta_{k}^{0})^{\sigma_{3}},
\end{equation}
and let $\widetilde{\Delta}_{k}^{0}$ denote (the operator of)
right multiplication by $\Delta_{k}^{0}$:
\begin{equation}
\widetilde{\Delta}_{k}^{0} \phi \! := \! \phi \Delta_{k}^{0}.
\end{equation}
\begin{eeeee}
One notes {}from Proposition~5.1 that, for $k \! \in \! \{A,B\}$, since
$\chi (\zeta_{k})$ are pure imaginary, $\vert \delta_{k}^{0} \vert \! = \!
1$; furthermore, {}from Eq.~(108) and the aforementioned, one also shows
that $\widetilde{\Delta}_{k}^{0}$ are unitary operators, namely,
$(\widetilde{\Delta}_{k}^{0})^{\dagger} \! = \! (\widetilde{\Delta}_{k}^{
0})^{-1}$.
\end{eeeee}
\begin{bbbbb}
For $k \! \in \! \{A,B\}$,
\begin{equation*}
C^{\widehat{\Sigma}_{k^{\prime}}}_{\widehat{w}^{\widehat{\Sigma}_{
k^{\prime}}}} \! = \! (\mathcal{N}_{k})^{-1}(\widetilde{\Delta}_{k
}^{0})^{-1}C^{\Sigma_{k}}_{w^{\Sigma_{k}}}(\widetilde{\Delta}_{k}^{
0}) \mathcal{N}_{k}, \qquad
w^{\Sigma_{k}} \! = \! w^{\Sigma_{k}}(\cdot) \! := \! ((\Delta_{k}^{
0})^{-1}(\mathcal{N}_{k} \widehat{w}^{\widehat{\Sigma}_{k^{\prime}}})
(\Delta_{k}^{0}))(\cdot),
\end{equation*}
where
\begin{align*}
C^{\Sigma_{k}}_{w^{\Sigma_{k}}} \! \! \upharpoonright_{\mathcal{L}^{
2}_{\mathrm{M}_{2}(\mathbb{C})}(\widetilde{\mathrm{L}}_{k})} \, = C_{
-}(\cdot ((\delta_{k}^{1}(\widetilde{w}))^{2} \mathcal{R}(\zeta_{s(k)}
\! + \! \varepsilon_{k}(\widetilde{w})) \sigma_{+})), \\
C^{\Sigma_{k}}_{w^{\Sigma_{k}}} \! \! \upharpoonright_{\mathcal{L}^{
2}_{\mathrm{M}_{2}(\mathbb{C})}(\overline{\widetilde{\mathrm{L}}_{
k}})} \, = -C_{+}(\cdot ((\delta_{k}^{1}(\widetilde{w}))^{-2} \,
\overline{\mathcal{R}(\zeta_{s(k)} \! + \! \varepsilon_{k}(\widetilde{
w}))} \sigma_{-})),
\end{align*}
$s(A) \! = \! 2$, $s(B) \!= \! 1$, and the rays $\widetilde{\mathrm{
L}}_{k}$ are defined as
\begin{align*}
\widetilde{\mathrm{L}}_{A} \! :=& \{\mathstrut \widetilde{w}; \,
\widetilde{w} \! = \! v(t(\zeta_{1} \! - \! \zeta_{2})^{3} \zeta_{2}^{-
3})^{1/2} \vert \zeta_{2} \! - \! \zeta_{3} \vert \exp (\tfrac{\mi \pi}
{4}), \, v \! \in \! [0,+\infty)\} \\
 \cup& \{\mathstrut \widetilde{w}; \, \widetilde{w} \! = \!
v(t(\zeta_{1} \! - \! \zeta_{2}) \zeta_{2}^{-1})^{1/2} \vert
\zeta_{2} \! - \! \zeta_{3} \vert \exp (-\tfrac{3 \pi \mi}{4}), \,
v \! \in \! [0,+\infty)\}, \\
\widetilde{\mathrm{L}}_{B} \! :=& \{\mathstrut \widetilde{w}; \,
\widetilde{w} \! = \! v(t(\zeta_{1} \! - \! \zeta_{2})^{3} \zeta_{1}^{-
3})^{1/2} \vert \zeta_{1} \! - \! \zeta_{3} \vert \exp (\tfrac{3 \pi \mi}
{4}), \, v \! \in \! \mathbb{R}\},
\end{align*}
so that $\Sigma_{k} \! = \! \widetilde{\mathrm{L}}_{k} \cup \overline{
\widetilde{\mathrm{L}}_{k}}$.
\end{bbbbb}

\emph{Proof.} The case $k \! = \! B$ is considered: the case $k \! =
\! A$ is analogous. Recalling the definition of the Cauchy operators,
$C_{\pm}$, the BC operator, applying the operator $\mathcal{N}_{B}$
defined in Eq.~(105), using Eq.~(107) (in particular, the $\widetilde{
w}$-independence of $\delta_{B}^{0})$, and the action (Eq.~(108)) and
unitarity (Remark~5.1) of $\widetilde{\Delta}_{B}^{0}$, one obtains,
via a change-of-variable argument, the expression for $C^{\widehat{
\Sigma}_{B^{\prime}}}_{\widehat{w}^{\widehat{\Sigma}_{B^{\prime}}}}$
stated in the Proposition, where $C^{\Sigma_{B}}_{w^{\Sigma_{B}}} \!
= \! C_{(\Delta_{B}^{0})^{-1}(\mathcal{N}_{B} \widehat{w}^{\widehat{
\Sigma}_{B^{\prime}}})(\Delta_{B}^{0})} \! = \! C_{+}(\cdot (\Delta_{
B}^{0})^{-1}(\mathcal{N}_{B} \widehat{w}^{\widehat{\Sigma}_{B^{
\prime}}}_{-})(\Delta_{B}^{0})) \! + \! C_{-}(\cdot (\Delta_{B}^{0})^{-1}
(\mathcal{N}_{B} \widehat{w}^{\widehat{\Sigma}_{B^{\prime}}}_{+})
(\Delta_{B}^{0}))$. {}From the definition of $\widehat{w}^{\widehat{
\Sigma}_{B^{\prime}}}(\zeta)$ given in Eq.~(104) and recalling that
$w^{\Sigma_{B}}(\zeta) \! := \! w^{\Sigma^{\sharp}}(\zeta) \! \!
\upharpoonright_{\Sigma_{B}}$, one shows, {}from the expression
for $w^{\Sigma^{\sharp}}(\zeta) \! = \! \sum_{l \in \{\pm\}} \! w_{l}^{
\Sigma^{\sharp}}(\zeta)$ given in Lemma~4.6, that: (1) $((\Delta_{
B}^{0})^{-1}(\mathcal{N}_{B} \widehat{w}_{-}^{\widehat{\Sigma}_{
B^{\prime}}})(\Delta_{B}^{0}))(\widetilde{w}) \! \! \upharpoonright_{
\widetilde{\mathrm{L}}_{B}}= \! ((\Delta_{B}^{0})^{-1}(\mathcal{N}_{
B} \widehat{w}_{+}^{\widehat{\Sigma}_{B^{\prime}}})(\Delta_{B}^{
0}))(\widetilde{w}) \! \! \upharpoonright_{\overline{\widetilde{\mathrm{
L}}_{B}}} \, = \!
\left(
\begin{smallmatrix}
0 & 0 \\
0 & 0
\end{smallmatrix}
\right)$; (2) for $\widetilde{w} \! \in \! \{\mathstrut z; \, z \! =
\! v(t(\zeta_{1} \! - \! \zeta_{2})^{3} \zeta_{1}^{-3})^{1/2} \vert
\zeta_{1} \! - \! \zeta_{3} \vert \me^{\frac{3 \pi \mi}{4}}, \, -\infty
\linebreak[4]
< \! v \! < \! \varepsilon\} \subset \widetilde{\mathrm{L}}_{B}$, $((
\Delta_{B}^{0})^{-1}(\mathcal{N}_{B} \widehat{w}_{+}^{\widehat{
\Sigma}_{B^{\prime}}})(\Delta_{B}^{0}))(\widetilde{w}) \! = \! (\delta_{
B}^{1}(\widetilde{w}))^{2} \mathcal{R}(\zeta_{1} \! + \! \varepsilon_{
B}(\widetilde{w})) \sigma_{+}$, and, for $\widetilde{w} \! \in \!
\widetilde{\mathrm{L}}_{B} \setminus \{\mathstrut z; \, z \! = \! v(t
(\zeta_{1} \! - \! \zeta_{2})^{3} \zeta_{1}^{-3})^{1/2} \vert \zeta_{
1} \! - \! \zeta_{3} \vert \me^{\frac{3 \pi \mi}{4}}, \, -\infty \! <
\! v \! < \! \varepsilon\}$, $((\Delta_{B}^{0})^{-1}(\mathcal{N}_{B}
\widehat{w}_{+}^{\widehat{\Sigma}_{B^{\prime}}})(\Delta_{B}^{
0}))(\widetilde{w}) \! = \!
\left(
\begin{smallmatrix}
0 & 0 \\
0 & 0
\end{smallmatrix}
\right)$; and (3) for $\widetilde{w} \! \in \! \{\mathstrut z; \, z \!
= \! v(t(\zeta_{1} \! - \! \zeta_{2})^{3} \zeta_{1}^{-3})^{1/2} \vert
\zeta_{1} \! - \! \zeta_{3} \vert \me^{-\frac{3 \pi \mi}{4}}, \, -\infty
\! < \! v \! < \! \varepsilon\} \! \subset \! \overline{\widetilde{
\mathrm{L}}_{B}}$, $((\Delta_{B}^{0})^{-1}(\mathcal{N}_{B} \widehat{
w}_{-}^{\widehat{\Sigma}_{B^{\prime}}})(\Delta_{B}^{0})) \linebreak[4]
(\widetilde{w}) \! = \! -(\delta_{B}^{1}(\widetilde{w}))^{-2} \,
\overline{\mathcal{R}(\zeta_{1} \! + \! \varepsilon_{B}(\widetilde{
w}))} \, \sigma_{-}$, and, for $\widetilde{w} \! \in \! \overline{
\widetilde{\mathrm{L}}_{B}} \setminus \{\mathstrut z; \, z \! = \! v(t
(\zeta_{1} \! - \! \zeta_{2})^{3} \zeta_{1}^{-3})^{1/2} \vert \zeta_{
1} \! - \! \zeta_{3} \vert \me^{-\frac{3 \pi \mi}{4}}, \, -\infty \!
< \! v \! < \! \varepsilon\}$, $((\Delta_{B}^{0})^{-1}(\mathcal{N}_{
B} \widehat{w}_{-}^{\widehat{\Sigma}_{B^{\prime}}})(\Delta_{B}^{0}))
(\widetilde{w}) \! = \!
\left(
\begin{smallmatrix}
0 & 0 \\
0 & 0
\end{smallmatrix}
\right)$. With the expressions for $((\Delta_{B}^{0})^{-1} \linebreak[4]
\cdot (\mathcal{N}_{B} \widehat{w}_{\pm}^{\widehat{\Sigma}_{B^{\prime}}
})(\Delta_{B}^{0}))(\widetilde{w})$, $\widetilde{w} \! \in \! \Sigma_{B
}$ $(= \! \widetilde{\mathrm{L}}_{B} \cup \overline{\widetilde{\mathrm{
L}}_{B}})$, and the formula for $C^{\Sigma_{B}}_{w^{\Sigma_{B}}}$ given
earlier in the proof, one arrives at the expression for $C^{\Sigma_{B}
}_{w^{\Sigma_{B}}} \! \! \upharpoonright_{\mathcal{L}^{2}_{\mathrm{M}_{
2}(\mathbb{C})}(\ast)}$, $\ast \! \in \! \{\widetilde{\mathrm{L}}_{B},
\overline{\widetilde{\mathrm{L}}_{B}}\}$, stated in the Proposition.
\hfill $\square$

{}From the formulae stated in Proposition~5.1 and the definition of the rays
$\widetilde{\mathrm{L}}_{k}$, $k \! \in \! \{A,B\}$, given in
Proposition~5.2, as $t \! \to \! +\infty$: (1) for $\widetilde{w} \! \in \!
\widetilde{\mathrm{L}}_{k}$, $k \! \in \! \{A,B\}$, $(\delta_{k}^{1}
(\widetilde{w}))^{2} \mathcal{R}(\zeta_{s(k)} \! + \! \varepsilon_{k}
(\widetilde{w})) \! \to \! (\mathrm{sgn}(k) \widetilde{w})^{2 \mi \mathrm{
sgn}(k) \nu} \exp (-\tfrac{\mi}{2} \mathrm{sgn}(k) \widetilde{w}^{2})
\mathcal{R}(\zeta_{s(k)}^{\pm})$, with $s(A) \! = \! 2$, $s(B) \! = \! 1$,
and $-\mathrm{sgn}(A) \! = \! \mathrm{sgn}(B) \! = \! 1$; and (2) for
$\widetilde{w} \! \in \! \overline{\widetilde{\mathrm{L}}_{k}}$, $k \! \in
\! \{A,B\}$, $(\delta_{k}^{1}(\widetilde{w}))^{-2} \overline{\mathcal{R}
(\zeta_{s(k)} \! + \! \varepsilon_{k}(\widetilde{w}))} \! \to \! (\mathrm{
sgn}(k) \widetilde{w})^{-2 \mi \mathrm{sgn}(k) \nu} \exp (\tfrac{\mi}{2}
\mathrm{sgn}(k) \linebreak[4]
\cdot \widetilde{w}^{2}) \overline{\mathcal{R}(\zeta_{s(k)}^{\pm})}$ (see
Lemma~5.1 for the definition of $\mathcal{R}
(\zeta_{s(k)}^{\pm}))$.
\begin{ccccc}
Let $\gamma \! \in \! (0,\tfrac{1}{2})$ be an arbitrarily fixed,
sufficiently small real number, $s(A) \! = \! 2$, $s(B) \! = \! 1$,
$\widehat{s}(A) \! = \! 1$, $\widehat{s}(B) \! = \! 2$, $-\mathrm{sgn}
(A) \! = \! \mathrm{sgn}(B) \! = \! 1$, $\widetilde{\mathrm{L}}_{k}$,
$k \! \in \! \{A,B\}$, be the rays defined in Proposition~{\rm 5.2},
and $\varepsilon_{k}(\widetilde{w})$ be defined by Eq.~{\rm (106)}.
Then, for $k \! \in \! \{A,B\}$, as $t \! \to \! +\infty$ such that $0 \!
< \! \zeta_{2} \! < \! \tfrac{1}{M} \! < \! M \! < \! \zeta_{1}$ and
$\vert \zeta_{3} \vert^{2} \! = \! 1$, with $M \! \in \! \mathbb{R}_{
>1}$ and bounded,
\begin{align*}
\vert \vert (\delta_{k}^{1}(\widetilde{w}))^{2} \mathcal{R}(\zeta_{
s(k)} \! + \! \varepsilon_{k}(\widetilde{w})) \! &- \! (\mathrm{sgn}
(k) \widetilde{w})^{2 \mi \mathrm{sgn}(k) \nu} \exp (-\tfrac{\mi}{2}
\mathrm{sgn}(k) \widetilde{w}^{2}) \mathcal{R}(\zeta_{s(k)}^{\pm})
\vert \vert_{\mathcal{L}^{\infty}(\widetilde{\mathrm{L}}_{k})} \\
 &\leqslant \! \tfrac{\vert c^{\mathcal{S}}(\zeta_{s(k)}) \vert
\vert \underline{c}(\zeta_{\widehat{s}(k)},\zeta_{3},\overline{
\zeta_{3}}) \vert}{\vert \zeta_{s(k)}-\zeta_{3} \vert \sqrt{(\zeta_{
1}-\zeta_{2})}} \tfrac{\ln (t)}{\sqrt{t}} \exp \! \left(-\tfrac{1}{2}
\gamma \widetilde{E}(k)v_{k}^{2}t \right), \quad \widetilde{
w} \! \in \! \widetilde{\mathrm{L}}_{k}, \\
\vert \vert (\delta_{k}^{1}(\widetilde{w}))^{-2} \, \overline{\mathcal{
R}(\zeta_{s(k)} \! + \! \varepsilon_{k}(\widetilde{w}))} &- \!
(\mathrm{sgn}(k) \widetilde{w})^{-2 \mi \mathrm{sgn}(k) \nu} \exp
(\tfrac{\mi}{2} \mathrm{sgn}(k) \widetilde{w}^{2}) \overline{\mathcal{
R}(\zeta_{s(k)}^{\pm})} \vert \vert_{\mathcal{L}^{\infty}(\overline{
\widetilde{\mathrm{L}}_{k}})} \\
 &\leqslant \! \tfrac{\vert c^{\mathcal{S}}(\zeta_{s(k)}) \vert \vert
\underline{c}(\zeta_{\widehat{s}(k)},\zeta_{3},\overline{\zeta_{3}}
) \vert}{\vert \zeta_{s(k)}-\zeta_{3} \vert \sqrt{(\zeta_{1}-\zeta_{
2})}} \tfrac{\ln (t)}{\sqrt{t}} \exp \! \left(-\tfrac{1}{2} \gamma
\widetilde{E}(k)v_{k}^{2}t \right), \quad \widetilde{w} \! \in \!
\overline{\widetilde{\mathrm{L}}_{k}},
\end{align*}
where $\mathcal{R}(\zeta_{1}^{+}) \! := \! \lim_{\Re (\zeta) \downarrow
\zeta_{1}} \! \mathcal{R}(\zeta) \! = \! \overline{r(\zeta_{1})}$,
$\mathcal{R}(\zeta_{1}^{-}) \! := \! \lim_{\Re (\zeta) \uparrow \zeta_{1}}
\! \mathcal{R}(\zeta) \! = \! -\overline{r(\zeta_{1})}(1 \! - \! \vert
r(\zeta_{1}) \vert^{2})^{-1}$, $\mathcal{R}(\zeta_{2}^{+}) \! := \! \lim_{
\Re (\zeta) \downarrow \zeta_{2}} \! \mathcal{R}(\zeta) \! = \! r(\zeta_{1})
(1 \! - \! \vert r(\zeta_{1}) \vert^{2})^{-1}$, $\mathcal{R}(\zeta_{2}^{-})
\! := \! \lim_{\Re (\zeta) \uparrow \zeta_{2}} \! \mathcal{R}(\zeta) \! =
\! -r(\zeta_{1})$, $\widetilde{E}(A) \! := \! \zeta_{2}^{-1}(\zeta_{1} \!
- \! \zeta_{2}) \vert \zeta_{2} \! - \! \zeta_{3} \vert^{2} \min \{\zeta_{
2}^{-2}(\zeta_{1} \! - \! \zeta_{2})^{2},1\}$, $\widetilde{E}(B) \! := \!
\zeta_{1}^{-3}(\zeta_{1} \! - \! \zeta_{2})^{3} \vert \zeta_{1} \! - \!
\zeta_{3} \vert^{2}$, $0 \! < \! v_{A} \! < \! \widetilde{\varepsilon}$,
and $-\infty \! < \! v_{B} \! < \! \widetilde{\varepsilon}$, with
$\widetilde{\varepsilon}$ some judiciously fixed small positive real number.
\end{ccccc}

\emph{Proof.} Without loss of generality, the $\mathcal{L}^{\infty}
(\ast)$ bound for the case $k \! = \! B$ and $\widetilde{w} \! \in \!
\widetilde{\mathrm{L}}_{B}$ is considered: the remaining cases
follow in an analogous manner. One begins by writing, for $\gamma
\! \in \! (0,\tfrac{1}{2})$ and $\widetilde{w} \! \in \! \widetilde{
\mathrm{L}}_{B}$,
\begin{gather*}
(\delta_{B}^{1}(\widetilde{w}))^{2} \mathcal{R}(\zeta_{1} \! + \!
\varepsilon_{B}(\widetilde{w})) \! - \! (\widetilde{w})^{2 \mi \nu}
\me^{-\frac{\mi \widetilde{w}^{2}}{2}} \mathcal{R}(\zeta_{1}^{\pm}) \! =
\! \me^{-\frac{\mi \gamma \widetilde{w}^{2}}{2}}(\me^{-\frac{\mi \gamma
\widetilde{w}^{2}}{2}} [\mathcal{R}(\zeta_{1} \! + \! \varepsilon_{B}
(\widetilde{w}))(\tfrac{(\zeta_{1}-\zeta_{2})}{(\zeta_{1}-\zeta_{2})+
\varepsilon_{B}(\widetilde{w})})^{2 \mi \nu}
\end{gather*}
\begin{gather*}
\cdot (\widetilde{w})^{2 \mi \nu} \me^{-\frac{\mi (1-2 \gamma)
\widetilde{w}^{2}}{2}(1+\frac{\theta^{u}_{3}(\zeta_{1}) \zeta_{
1}^{9/2} \widetilde{w}}{3 \sqrt{2} \, (1-2 \gamma) \sqrt{t} \,
(\zeta_{1}-\zeta_{2})^{3/2} \vert \zeta_{1}-\zeta_{3} \vert^{3}
})} \me^{2(\chi (\zeta_{1}+\varepsilon_{B}(\widetilde{w}))-\chi
(\zeta_{1}))} \! - \! \mathcal{R}(\zeta_{1}^{\pm})(\widetilde{
w})^{2 \mi \nu} \me^{-\frac{\mi (1-2 \gamma) \widetilde{w}^{2}}
{2}}]).
\end{gather*}
One notes that $\me^{-\frac{\mi \gamma}{2} \widetilde{w}^{2}} \! = \!
\exp (-\tfrac{1}{2} \gamma \zeta_{1}^{-3}(\zeta_{1} \! - \! \zeta_{
2})^{3} \vert \zeta_{1} \! - \! \zeta_{3} \vert^{2} v_{B}^{2} t)$,
$-\infty \! < \! v_{B} \! < \! \widetilde{\varepsilon}$, which gives
rise to the exponential factor stated in the Lemma, and $\me^{-\mi
(1-2 \gamma) \frac{\widetilde{w}^{2}}{2}} \! = \! \exp (-(\tfrac{1}
{2} \! - \! \gamma) \zeta_{1}^{-3}(\zeta_{1} \! - \! \zeta_{2})^{3}
\vert \zeta_{1} \! - \! \zeta_{3} \vert^{2} v_{B}^{2}t)$. {}From the
definition of $\mathcal{R}(\zeta)$ given in the formulation and proof
of Lemma~4.2, and the fact that $r(\zeta) \! \in \! \mathcal{S}_{
\mathbb{C}}(\mathbb{R}) \cap \{\mathstrut h(z); \, \vert \vert h(\cdot)
\vert \vert_{\mathcal{L}^{\infty}(\mathbb{R})} \! := \! \sup_{z \in
\mathbb{R}} \vert h(z) \vert \! < \! 1\}$, one shows that, for $v_{
B} \! \in \! (-\infty,\widetilde{\varepsilon})$, $\vert \mathcal{
R}(\zeta_{1} \! + \! \varepsilon_{B}(\widetilde{w})) \vert \! = \!
\vert \mathcal{R}(\zeta_{1} \! + \! \tfrac{1}{\sqrt{2}}v_{B}(\zeta_{
1} \! - \! \zeta_{2}) \me^{\frac{3 \pi \mi}{4}}) \vert \! \leqslant
\! \vert \underline{c}(\zeta_{1},\zeta_{2},\zeta_{3},\overline{\zeta_{
3}}) \vert$. One shows that $\sup_{-\infty < v_{B} < \widetilde{
\varepsilon}} \vert (\tfrac{(\zeta_{1}-\zeta_{2})}{(\zeta_{1}-\zeta_{
2})+\varepsilon_{B}(\widetilde{w})})^{2 \mi \nu} \vert \! = \! \sup_{
-\infty < v_{B} < \widetilde{\varepsilon}} \vert \me^{2 \nu \arg (1
+ \frac{1}{\sqrt{2}} v_{B} \me^{\frac{3 \pi \mi}{4}})} \vert \!
\leqslant \! \me^{2 \pi \nu_{m}} \! \leqslant \! \vert \underline{
c}(\zeta_{1},\zeta_{2},\zeta_{3},\overline{\zeta_{3}}) \vert$, with
$0 \! < \! \nu \! := \! \nu (\zeta_{1}) \! \leqslant \! \nu_{m}
\! := \! -\tfrac{1}{2 \pi} \ln (1 \! - \! \sup_{z \in \mathbb{R}}
\vert r(z) \vert^{2})$, since $\arg (1 \! + \! \tfrac{1}{\sqrt{2}}
v_{B} \me^{\frac{3 \pi \mi}{4}}) \! \in \! (-\pi,\pi)$, $-\infty
\! < \! v_{B} \! < \! \widetilde{\varepsilon}$: also, $\vert
(\widetilde{w})^{2 \mi \nu} \vert \! \leqslant \! \me^{2 \pi \nu_{
m}} \! \leqslant \! \vert \underline{c}(\zeta_{1},\zeta_{2},\zeta_{
3},\overline{\zeta_{3}}) \vert$. For the exponential term $E \! :=
\! \exp (-\mi (\tfrac{1}{2} \! - \! \gamma) \widetilde{w}^{2}(1 \!
+ \! \tfrac{\theta^{u}_{3}(\zeta_{1}) \zeta_{1}^{9/2} \widetilde{w}}
{3 \sqrt{2} \, (1-2 \gamma) \sqrt{t} \, (\zeta_{1}-\zeta_{2})^{3/2}
\vert \zeta_{1}-\zeta_{3} \vert^{3}}))$, in light of the estimation
for $\me^{-\mi (\frac{1}{2}-\gamma) \widetilde{w}^{2}}$ given earlier
in the proof, one must study the sign of $\widetilde{\mathscr{R}} \!
:= \! \Re (1 \! + \! \tfrac{\theta^{u}_{3}(\zeta_{1}) \zeta_{1}^{9/2}
\widetilde{w}}{3 \sqrt{2} \, (1-2 \gamma) \sqrt{t} \, (\zeta_{1}-
\zeta_{2})^{3/2} \vert \zeta_{1}-\zeta_{3} \vert^{3}})$. One shows
that, for $-\infty \! < \! v_{B} \! < \! \widetilde{\varepsilon}$,
$\widetilde{\mathscr{R}} \! = \! 1 \! - \! \tfrac{\theta^{u}_{3}
(\zeta_{1}) \zeta_{1}^{3}v_{B}}{6(1-2 \gamma) \vert \zeta_{1}-\zeta_{
3} \vert^{2}}$. {}From Eqs.~(16) and~(17), and the formula for
$\theta^{u}_{3}(\zeta_{1})$ given in Proposition~5.1, one shows that,
for $z_{o} \! < \! -2$, $\tfrac{\theta^{u}_{3}(\zeta_{1}) \zeta_{1}^{
3}}{\vert \zeta_{1}-\zeta_{3} \vert^{2}} \! > \! 0$, whence $\tfrac{
\theta^{u}_{3}(\zeta_{1}) \zeta_{1}^{3}}{6(1-2 \gamma) \vert \zeta_{
1}-\zeta_{3} \vert^{2}} \! > \! 0$; hence, for $-\infty \! < \! v_{B}
\! < \! \widetilde{\varepsilon}$, and choosing $\widetilde{\varepsilon}$
so (small) that $1 \! - \! \tfrac{\theta^{u}_{3}(\zeta_{1}) \zeta_{1}^{
3} \widetilde{\varepsilon}}{6(1-2 \gamma) \vert \zeta_{1}-\zeta_{3}
\vert^{2}} \! > \! 0$, one deduces that $\widetilde{\mathscr{R}} \! >
\! 0$, {}from which it follows that $\vert E \vert \! \leqslant \! \exp
(-(\tfrac{1}{2} \! - \! \gamma) \zeta_{1}^{-3}(\zeta_{1} \! - \!
\zeta_{2})^{3} \vert \zeta_{1} \! - \! \zeta_{3} \vert^{2} \widetilde{
\mathscr{R}}v_{B}^{2}t) \! \leqslant \! \vert \underline{c}(\zeta_{1},
\zeta_{2},\zeta_{3},\overline{\zeta_{3}}) \vert$. The boundedness of
$\me^{2(\chi (\zeta_{1}+\varepsilon_{B}(\widetilde{w}))-\chi (\zeta_{
1}))}$ is a consequence of the inequality $\vert \vert (\delta (\cdot)
)^{\pm 1} \vert \vert_{\mathcal{L}^{\infty}(\mathbb{C})} \! < \!
\infty$ (Proposition~4.1) and the formula for $\chi (\zeta)$ given in
Proposition~5.1. Via a Taylor expansion argument, one shows that $\vert
\me^{-\frac{\mi \gamma}{2} \widetilde{w}^{2}}(\mathcal{R}(\zeta_{1} \!
+ \! \varepsilon_{B}(\widetilde{w})) \! - \! \mathcal{R}(\zeta_{1}))
\vert \! \leqslant \! \tfrac{\vert \underline{c}(\zeta_{1},\zeta_{2},
\zeta_{3},\overline{\zeta_{3}}) \vert \vert \vert \partial_{\bullet}
\mathcal{R}(\bullet) \vert \vert_{\mathcal{L}^{\infty}(\mathbb{R})}}
{\vert \zeta_{1}-\zeta_{3} \vert \sqrt{t(\zeta_{1}-\zeta_{2})^{3}}}$.
For $\me^{-\frac{\mi \gamma}{2} \widetilde{w}^{2}}((\tfrac{(\zeta_{1}
-\zeta_{2})}{(\zeta_{1}-\zeta_{2})+\varepsilon_{B}(\widetilde{w})})^{
2 \mi \nu} \! - \! 1)$, one shows that $\vert \me^{-\frac{\mi \gamma}
{2} \widetilde{w}^{2}}((\tfrac{(\zeta_{1}-\zeta_{2})}{(\zeta_{1}-
\zeta_{2})+\varepsilon_{B}(\widetilde{w})})^{2 \mi \nu} \! - \! 1)
\vert \! \leqslant \! 2 \vert \me^{-\frac{\mi \gamma}{2} \widetilde{
w}^{2}} \vert \vert \nu \! \int_{1}^{\varepsilon_{B}(\widetilde{w})/
(\zeta_{1}-\zeta_{2})} \! \xi^{-2 \mi \nu -1} \md \xi \vert \!
\leqslant \! \sqrt{2} \, \nu_{m} \exp (-\frac{1}{2} \gamma \zeta_{
1}^{-3}(\zeta_{1} \! - \! \zeta_{2})^{3} \vert \zeta_{1} \! - \!
\zeta_{3} \vert^{2}v_{B}^{2}t)v_{B} \sup_{s \in [0,1]} \{\mathstrut
\vert z^{-2 \mi \nu -1} \vert; \, z \! = \! 1 \! + \! \tfrac{1}{\sqrt{
2}}sv_{B} \me^{\frac{3 \pi \mi}{4}}, \, -\infty \! < \! v_{B} \! < \!
\widetilde{\varepsilon}\} \! \leqslant \! \tfrac{\vert \underline{c}
(\zeta_{1},\zeta_{2},\zeta_{3},\overline{\zeta_{3}}) \vert}{\vert
\zeta_{1}-\zeta_{3} \vert \sqrt{t(\zeta_{1}-\zeta_{2})^{3}}}$, since
$\vert z^{-2 \mi \nu -1} \vert \! \leqslant \! \me^{2 \pi \nu_{m}}((1
\! - \! \tfrac{1}{2}sv_{B})^{2} \! + \! (\tfrac{1}{2}sv_{B})^{2})^{-1/2} \!
< \! \infty$, $(s,v_{B}) \! \in \! [0,1] \! \times \! (-\infty,\widetilde{
\varepsilon})$. Using the inequality $\vert \me^{\diamondsuit} \! - \! 1
\vert \! \leqslant \! \sup_{s \in [0,1]} \vert \me^{s \diamondsuit} \vert
\vert \diamondsuit \vert$,
\begin{align*}
\vert \me^{-\frac{\mi \gamma}{2} \widetilde{w}^{2}}(\me^{2(\chi (\zeta_{1}+
\varepsilon_{B}(\widetilde{w}))-\chi (\zeta_{1}))} \! - \! 1 ) \vert \!
\leqslant& \, 2 \exp (-\tfrac{1}{2} \gamma \zeta_{1}^{-3}(\zeta_{1} \! - \!
\zeta_{2})^{3} \vert \zeta_{1} \! - \! \zeta_{3} \vert^{2}v_{B}^{2}t) \\
\times& \, \sup_{s \in [0,1]} \vert \me^{2s(\chi (\zeta_{1}+\varepsilon_{B}
(\widetilde{w}))-\chi (\zeta_{1}))} \vert \vert \chi (\zeta_{1} \! + \!
\varepsilon_{B}(\widetilde{w})) \! - \! \chi (\zeta_{1}) \vert \\
\leqslant& \, \vert \underline{c}(\zeta_{1},\zeta_{2},\zeta_{3},\overline{
\zeta_{3}}) \vert \exp (-\tfrac{1}{2} \gamma \zeta_{1}^{-3}(\zeta_{1} \! -
\! \zeta_{2})^{3} \vert \zeta_{1} \! - \! \zeta_{3} \vert^{2}v_{B}^{2}t) \\
\times& \, \vert \chi (\zeta_{1} \!+ \! \varepsilon_{B}(\widetilde{w})) \!
- \! \chi (\zeta_{1}) \vert.
\end{align*}
Recalling the definition of $\chi (\zeta)$ given in Proposition~5.1, one
writes, via an integration by parts argument, $\chi (\zeta_{1} \! + \!
\varepsilon_{B}(\widetilde{w})) \! - \! \chi (\zeta_{1}) \! = \! \tfrac{\mi}
{2 \pi}(\int_{-\infty}^{-M_{o}} \! + \! \int_{-M_{o}}^{-\delta_{o}} \! + \!
\int_{-\delta_{o}}^{0} \! + \! \int_{\zeta_{2}}^{\zeta_{1}})(\ln (\mu \! -
\! \zeta_{1} \! - \! \varepsilon_{B}(\widetilde{w})) \! - \! \ln (\mu \! -
\! \zeta_{1})) \md \ln \! \left( \tfrac{1-\vert r(\mu) \vert^{2}}{1 \! - \!
\vert r(\zeta_{1}) \vert^{2}} \right) \! := \! \widetilde{\mathscr{I}}_{1} \!
+ \! \widetilde{\mathscr{I}}_{2} \! + \! \widetilde{\mathscr{I}}_{3} \! + \!
\widetilde{\mathscr{I}}_{4}$, with $M_{o}$ (respectively~$\delta_{o})$ an
arbitrarily fixed, finite (respectively~sufficiently small) positive real
number. Using the fact that $r(\zeta) \! \in \! \mathcal{S}_{\mathbb{C}}
(\mathbb{R})$, one shows that $\vert \widetilde{\mathscr{I}}_{j} \vert \!
\leqslant \! \mathcal{O}(\vert \varepsilon_{B}(\widetilde{w}) \vert)$, $j \!
\in \! \{1,2\}$; hence, $\exp (-\tfrac{1}{2} \gamma \zeta_{1}^{-3}(\zeta_{
1} \! - \! \zeta_{2})^{3} \vert \zeta_{1} \!- \! \zeta_{3} \vert^{2}v_{
B}^{2}t)(\vert \widetilde{\mathscr{I}}_{1} \vert \! + \! \vert \widetilde{
\mathscr{I}}_{2} \vert) \! \leqslant \! \tfrac{\vert \underline{c}(\zeta_{1},
\zeta_{2},\zeta_{3},\overline{\zeta_{3}}) \vert}{\vert \zeta_{1}-\zeta_{3}
\vert \sqrt{t(\zeta_{1}-\zeta_{2})}}$. Recalling that $r(0) \! = \! 0$ and
$\vert \vert r(\cdot) \vert \vert_{\mathcal{L}^{\infty}(\mathbb{R})} \! < \!
1$, one shows that $\exp (-\tfrac{1}{2} \gamma \zeta_{1}^{-3}(\zeta_{1} \! -
\! \zeta_{2})^{3} \vert \zeta_{1} \! - \! \zeta_{3} \vert^{2}v_{B}^{2}t)
\vert \widetilde{\mathscr{I}}_{3} \vert \! \leqslant \! \tfrac{\vert
\underline{c}(\zeta_{1},\zeta_{2},\zeta_{3},\overline{\zeta_{3}}) \vert}
{\vert \zeta_{1}-\zeta_{3} \vert \sqrt{t(\zeta_{1}-\zeta_{2})}}$. Using the
Lipschitz property of the reflection coefficient, namely, $\vert r(z_{1}) \!
- \! r(z_{2}) \vert \! \leqslant \! \mathrm{A}_{r} \vert z_{1} \! - \! z_{2}
\vert$, with Lipschitz constant $\mathrm{A}_{r} \! > \! 0$, and the fact
that $r(\zeta) \! \in \! \mathcal{S}_{\mathbb{C}}^{1}(\mathbb{R})$, one
shows that $\exp (-\tfrac{1}{2} \gamma \zeta_{1}^{-3}(\zeta_{1} \! - \!
\zeta_{2})^{3} \vert \zeta_{1} \! - \! \zeta_{3} \vert^{2}v_{B}^{2}t) \vert
\widetilde{\mathscr{I}}_{4} \vert \! \leqslant \! \tfrac{\vert \underline{c}
(\zeta_{1},\zeta_{2},\zeta_{3},\overline{\zeta_{3}}) \vert}{\vert \zeta_{1}-
\zeta_{3} \vert \sqrt{t(\zeta_{1}-\zeta_{2})}} \! + \! \tfrac{\vert c^{
\mathcal{S}}(\zeta_{1}) \vert \vert \underline{c}(\zeta_{2},\zeta_{3},
\overline{\zeta_{3}}) \vert}{\vert \zeta_{1}-\zeta_{3} \vert \sqrt{(\zeta_{
1}-\zeta_{2})}} \tfrac{\ln t}{\sqrt{t}}$; hence,
\begin{align*}
\exp (-\tfrac{1}{2} \gamma \zeta_{1}^{-3}(\zeta_{1} \! - \! \zeta_{2})^{3}
\vert \zeta_{1} \! - \! \zeta_{3} \vert^{2}v_{B}^{2}t) \vert \chi (\zeta_{1}
\! + \! \varepsilon_{B}(\widetilde{w})) \! - \! \chi (\zeta_{1}) \vert \!
\leqslant& \, \tfrac{\vert \underline{c}(\zeta_{1},\zeta_{2},\zeta_{3},
\overline{\zeta_{3}}) \vert}{\vert \zeta_{1}-\zeta_{3} \vert \sqrt{t(\zeta_{
1}-\zeta_{2})}} \\
+& \, \tfrac{\vert c^{\mathcal{S}}(\zeta_{1}) \vert \vert \underline{c}
(\zeta_{2},\zeta_{3},\overline{\zeta_{3}}) \vert}{\vert \zeta_{1}-\zeta_{3}
\vert \sqrt{(\zeta_{1}-\zeta_{2})}} \tfrac{\ln t}{\sqrt{t}}.
\end{align*}
Using the inequality $\vert \me^{\diamondsuit} \! - \! 1 \vert \! \leqslant
\! \sup_{s \in [0,1]} \vert \tfrac{\md}{\md s} \me^{s \diamondsuit} \vert$,
\begin{align*}
\vert \me^{-\frac{\mi \gamma}{2} \widetilde{w}^{2}}(\widetilde{w})^{2 \mi \nu}
\me^{-\mi (\frac{1}{2}-\gamma) \widetilde{w}^{2}}(\me^{-\mi \blacklozenge} \!
- \! 1) \vert \! \leqslant& \, \exp (-\tfrac{1}{2}(1 \! - \! \gamma) \zeta_{
1}^{-3}(\zeta_{1} \! - \! \zeta_{2})^{3} \vert \zeta_{1} \! - \! \zeta_{3}
\vert^{2}v_{B}^{2}t) \\
\times& \, \me^{2 \pi \nu_{m}} \vert \blacklozenge \vert \sup_{s \in [0,1]}
\vert \me^{-\mi s \blacklozenge} \vert \! \leqslant \! \tfrac{\vert
\underline{c}(\zeta_{1},\zeta_{2},\zeta_{3},\overline{\zeta_{3}}) \vert}
{\vert \zeta_{1}-\zeta_{3} \vert^{3} \sqrt{t(\zeta_{1}-\zeta_{2})^{3}}},
\end{align*}
where $\blacklozenge \! := \! \tfrac{\theta^{u}_{3}(\zeta_{1}) \zeta_{1}^{
9/2} \widetilde{w}^{3}}{6 \sqrt{2} \, \sqrt{t} \, (\zeta_{1}-\zeta_{2})^{3/2}
\vert \zeta_{1}-\zeta_{3} \vert^{3}}$. Gathering the above-derived bounds,
one deduces that, for $v_{B} \! \in \! (-\infty,\widetilde{\varepsilon})$,
\begin{align*}
\vert \vert (\delta_{B}^{1}(\widetilde{w}))^{2} \mathcal{R}(\zeta_{1} \! + \!
\varepsilon_{B}(\widetilde{w})) -& (\widetilde{w})^{2 \mi \nu} \exp (-
\tfrac{\mi}{2} \widetilde{w}^{2}) \mathcal{R}(\zeta_{1}^{\pm}) \vert \vert_{
\mathcal{L}^{\infty}(\widetilde{\mathrm{L}}_{B})} \! \leqslant \! \tfrac{
\vert c^{\mathcal{S}}(\zeta_{1}) \vert \vert \underline{c}(\zeta_{2},\zeta_{
3},\overline{\zeta_{3}}) \vert}{\vert \zeta_{1}-\zeta_{3} \vert \sqrt{(\zeta_{
1}-\zeta_{2})}} \\
\times& \exp (-\tfrac{1}{2} \gamma \zeta_{1}^{-3}(\zeta_{1} \! - \! \zeta_{
2})^{3} \vert \zeta_{1} \! - \! \zeta_{3} \vert^{2}v_{B}^{2}t) \, \tfrac{\ln
t}{\sqrt{t}}. \qquad \qquad \qquad \qquad \qquad \quad \, \square
\end{align*}
\begin{bbbbb}
For general operators $C^{\Sigma^{\sharp}}_{w^{\Sigma_{k^{
\prime}}}}$, $k \! \in \! \{1,2,\ldots,N\}$, if $(\mathbf{1}_{\Sigma^{
\sharp}} \! - \! C^{\Sigma^{\sharp}}_{w^{\Sigma_{k^{\prime}}}})^{
-1}$ exists, then
\begin{align*}
(\mathbf{1}_{\Sigma^{\sharp}}&+ \! \sum_{i=1}^{N}C^{\Sigma^{\sharp}
}_{w^{\Sigma_{i^{\prime}}}}(\mathbf{1}_{\Sigma^{\sharp}} \! - \! C^{
\Sigma^{\sharp}}_{w^{\Sigma_{i^{\prime}}}})^{-1})(\mathbf{1}_{\Sigma^{
\sharp}} \! - \! \sum_{j=1}^{N} C^{\Sigma^{\sharp}}_{w^{\Sigma_{j^{
\prime}}}}) \! = \! \mathbf{1}_{\Sigma^{\sharp}} \\
&-\sum_{i=1}^{N} \sum_{j=1}^{N}(1 \! - \! \delta_{ij})(\mathbf{1}_{
\Sigma^{\sharp}} \! - \! C^{\Sigma^{\sharp}}_{w^{\Sigma_{i^{\prime}}
}})^{-1}C^{\Sigma^{\sharp}}_{w^{\Sigma_{i^{\prime}}}} C^{\Sigma^{
\sharp}}_{w^{\Sigma_{j^{\prime}}}}, \\
(\mathbf{1}_{\Sigma^{\sharp}}&- \! \sum_{j=1}^{N} C^{\Sigma^{\sharp}
}_{w^{\Sigma_{j^{\prime}}}})(\mathbf{1}_{\Sigma^{\sharp}} \! + \!
\sum_{i=1}^{N}C^{\Sigma^{\sharp}}_{w^{\Sigma_{i^{\prime}}}}(\mathbf{
1}_{\Sigma^{\sharp}} \! - \! C^{\Sigma^{\sharp}}_{w^{\Sigma_{i^{\prime}
}}})^{-1}) \! = \! \mathbf{1}_{\Sigma^{\sharp}} \\
&-\sum_{i=1}^{N} \sum_{j=1}^{N}(1 \! - \! \delta_{ij})C^{\Sigma^{\sharp}
}_{w^{\Sigma_{i^{\prime}}}}C^{\Sigma^{\sharp}}_{w^{\Sigma_{j^{\prime}}}
}(\mathbf{1}_{\Sigma^{\sharp}} \! - \! C^{\Sigma^{\sharp}}_{w^{\Sigma_{
j^{\prime}}}})^{-1},
\end{align*}
where $\delta_{ij}$ is the Kronecker delta.
\end{bbbbb}

\emph{Proof.} Follows {}from the assumption of the existence of the
general operators $(\mathbf{1}_{\Sigma^{\sharp}} \! - \! C^{\Sigma^{
\sharp}}_{w^{\Sigma_{k^{\prime}}}})^{-1}$, $k \! \in \! \{1,2,\ldots,N\}$,
the second resolvent identity, and an induction argument. \hfill $\square$
\begin{ccccc}
For $\alpha \! \not= \! \beta \! \in \! \{A^{\prime},B^{\prime}\}$, as
$t \! \to \! +\infty$ such that $0 \! < \! \zeta_{2} \! < \! \tfrac{1}
{M} \! < \! M \! < \! \zeta_{1}$ and $\vert \zeta_{3} \vert^{2} \! = \!
1$, with $M \! \in \! \mathbb{R}_{>1}$ and bounded,
\begin{gather*}
\vert \vert C^{\Sigma^{\sharp}}_{w^{\Sigma_{\alpha}}}C^{\Sigma^{\sharp}}_{
w^{\Sigma_{\beta}}} \vert \vert_{\mathscr{N}(\Sigma^{\sharp})} \! \leqslant
\! \dfrac{\vert \underline{c}(\zeta_{1},\zeta_{2},\zeta_{3},\overline{
\zeta_{3}}) \vert}{(\zeta_{1} \! - \! \zeta_{2})^{3/2} \sqrt{\vert z_{o} \!
+ \! \zeta_{1} \! + \! \zeta_{2} \vert t}}, \\
\vert \vert C^{\Sigma^{\sharp}}_{w^{\Sigma_{\alpha}}} C^{\Sigma^{\sharp}}_{
w^{\Sigma_{\beta}}} \vert \vert_{\mathcal{L}^{\infty}_{\mathrm{M}_{2}
(\mathbb{C})}(\Sigma^{\sharp}) \to \mathcal{L}^{2}_{\mathrm{M}_{2}(\mathbb{
C})}(\Sigma^{\sharp})} \! \leqslant \! \dfrac{\vert \underline{c}(\zeta_{1},
\zeta_{2},\zeta_{3},\overline{\zeta_{3}}) \vert}{(\zeta_{1} \! - \! \zeta_{
2})^{7/4}(\vert z_{o} \! + \! \zeta_{1} \! + \! \zeta_{2} \vert t)^{3/4}}.
\end{gather*}
\end{ccccc}

\emph{Proof.} Without loss of generality, the bounds for $C^{\Sigma^{
\sharp}}_{w^{\Sigma_{A^{\prime}}}}C^{\Sigma^{\sharp}}_{w^{\Sigma_{B^{
\prime}}}}$ are proved. For $\phi (z) \! \in \! \mathcal{L}^{2}_{
\mathrm{M}_{2}(\mathbb{C})}(\Sigma^{\sharp})$, {}from the definition
of the BC and Cauchy operators (and their associativity properties),
it follows that
\begin{align*}
\mathcal{X}(z) \! :=& \, (C^{\Sigma^{\sharp}}_{w^{\Sigma_{A^{\prime}}}}
C^{\Sigma^{\sharp}}_{w^{\Sigma_{B^{\prime}}}} \phi)(z) \\
=& \, (C^{\Sigma^{\sharp}}_{-}(C^{\Sigma^{\sharp}}_{+}(\phi w^{\Sigma_{B^{
\prime}}}_{-})w^{\Sigma_{A^{\prime}}}_{+}))(z) \! + \! (C^{\Sigma^{\sharp}
}_{+}(C^{\Sigma^{\sharp}}_{-}(\phi w^{\Sigma_{B^{\prime}}}_{+})w^{\Sigma_{
A^{\prime}}}_{-}))(z) \\
+& \, \lim_{\genfrac{}{}{0pt}{2}{\zeta^{\prime \prime} \to z}{\zeta^{\prime
\prime} \, \in \, + \, \mathrm{side} \, \mathrm{of} \, \Sigma_{A^{\prime}}}}
\! \int_{\Sigma_{A^{\prime}}} \lim_{\genfrac{}{}{0pt}{2}{\zeta^{\prime} \to
\zeta}{\zeta^{\prime} \, \in \, + \, \mathrm{side} \, \mathrm{of} \, \Sigma_{
B^{\prime}}}} \! \int_{\Sigma_{B^{\prime}}} \tfrac{\phi (\xi)w^{\Sigma_{B^{
\prime}}}_{-}(\xi)w^{\Sigma_{A^{\prime}}}_{-}(\zeta)}{(\xi -\zeta^{\prime})
(\zeta -\zeta^{\prime \prime})} \, \tfrac{\md \xi}{2 \pi \mi} \, \tfrac{\md
\zeta}{2 \pi \mi} \\
+& \, \lim_{\genfrac{}{}{0pt}{2}{\zeta^{\prime \prime} \to z}{\zeta^{\prime
\prime} \, \in \, - \, \mathrm{side} \, \mathrm{of} \, \Sigma_{A^{\prime}}}}
\! \int_{\Sigma_{A^{\prime}}} \lim_{\genfrac{}{}{0pt}{2}{\zeta^{\prime} \,
\to \, \zeta}{\zeta^{\prime} \, \in \, - \, \mathrm{side} \, \mathrm{of} \,
\Sigma_{B^{\prime}}}} \! \int_{\Sigma_{B^{\prime}}} \tfrac{\phi (\xi)w^{
\Sigma_{B^{\prime}}}_{+}(\xi)w^{\Sigma_{A^{\prime}}}_{+}(\zeta)}{(\xi -
\zeta^{\prime})(\zeta -\zeta^{\prime \prime})} \, \tfrac{\md \xi}{2 \pi \mi}
\, \tfrac{\md \zeta}{2 \pi \mi};
\end{align*}
but $w^{\Sigma_{B^{\prime}}}_{-}(\xi)w^{\Sigma_{A^{\prime}}}_{-}(\zeta) \!
= \! w^{\Sigma_{B^{\prime}}}_{+}(\xi)w^{\Sigma_{A^{\prime}}}_{+}(\zeta) \!
= \!
\left(
\begin{smallmatrix}
0 & 0 \\
0 & 0
\end{smallmatrix}
\right)$, thus
\begin{equation*}
\mathcal{X}(z) \! = \! (C^{\Sigma^{\sharp}}_{-}(C^{\Sigma^{\sharp}}_{+}
(\phi w^{\Sigma_{B^{\prime}}}_{-})w^{\Sigma_{A^{\prime}}}_{+}))(z) \! + \!
(C^{\Sigma^{\sharp}}_{+}(C^{\Sigma^{\sharp}}_{-}(\phi w^{\Sigma_{B^{\prime}
}}_{+})w^{\Sigma_{A^{\prime}}}_{-}))(z).
\end{equation*}
Now,
\begin{align*}
\vert \vert \mathcal{X}(\bullet) \vert \vert_{\mathcal{L}^{2}_{\mathrm{M}_{2}
(\mathbb{C})}(\Sigma^{\sharp})} \! \leqslant& \, \left\vert \left\vert
\int_{\Sigma_{A^{\prime}}} \! \left(\int_{\Sigma_{B^{\prime}}} \tfrac{\phi
(\xi)w^{\Sigma_{B^{\prime}}}_{-}(\xi)}{(\xi -\zeta_{+})} \, \tfrac{\md \xi}
{2 \pi \mi} \right) \! \tfrac{w^{\Sigma_{A^{\prime}}}_{+}(\zeta)}{(\zeta -
\bullet_{-})} \, \tfrac{\md \zeta}{2 \pi \mi} \right\vert \right\vert_{
\mathcal{L}^{2}_{\mathrm{M}_{2}(\mathbb{C})}(\Sigma^{\sharp})} \\
+& \, \left\vert \left\vert \int_{\Sigma_{A^{\prime}}} \! \left(\int_{\Sigma_{
B^{\prime}}} \! \tfrac{\phi (\xi)w^{\Sigma_{B^{\prime}}}_{+}(\xi)}{(\xi -
\zeta_{-})} \, \tfrac{\md \xi}{2 \pi \mi} \right) \! \tfrac{w^{\Sigma_{A^{
\prime}}}_{-}(\zeta)}{(\zeta -\bullet_{+})} \, \tfrac{\md \zeta}{2 \pi \mi}
\right\vert \right\vert_{\mathcal{L}^{2}_{\mathrm{M}_{2}(\mathbb{C})}
(\Sigma^{\sharp})} \\
\leqslant& \, \vert \underline{c}(\zeta_{1},\zeta_{2},\zeta_{3},\overline{
\zeta_{3}}) \vert \vert \vert w^{\Sigma_{A^{\prime}}}_{+}(\cdot) \vert
\vert_{\mathcal{L}^{2}_{\mathrm{M}_{2}(\mathbb{C})}(\Sigma_{A^{\prime}})} \\
\times& \, \sup_{\zeta \in \Sigma_{A^{\prime}}} \! \left\vert \int_{\Sigma_{
B^{\prime}}} \tfrac{\phi (\xi)w^{\Sigma_{B^{\prime}}}_{-}(\xi)}{(\xi -\zeta)}
\, \tfrac{\md \xi}{2 \pi \mi} \right\vert \! + \! \vert \underline{c}(\zeta_{
1},\zeta_{2},\zeta_{3},\overline{\zeta_{3}}) \vert \\
\times& \, \vert \vert w^{\Sigma_{A^{\prime}}}_{-}(\cdot) \vert \vert_{
\mathcal{L}^{2}_{\mathrm{M}_{2}(\mathbb{C})}(\Sigma_{A^{\prime}})} \sup_{
\zeta \in \Sigma_{A^{\prime}}} \! \left\vert \int_{\Sigma_{B^{\prime}}}
\tfrac{\phi (\xi)w^{\Sigma_{B^{\prime}}}_{+}(\xi)}{(\xi -\zeta)} \, \tfrac{
\md \xi}{2 \pi \mi} \right\vert \\
\leqslant& \, ((\zeta_{1} \! - \! \zeta_{2})^{-1} \vert \underline{c}(\zeta_{
1},\zeta_{2},\zeta_{3},\overline{\zeta_{3}}) \vert \vert \vert w^{\Sigma_{
A^{\prime}}}_{+}(\cdot) \vert \vert_{\mathcal{L}^{2}_{\mathrm{M}_{2}(\mathbb{
C})}(\Sigma_{A^{\prime}})} \\
\times& \, \vert \vert w^{\Sigma_{B^{\prime}}}_{-}(\cdot) \vert \vert_{
\mathcal{L}^{2}_{\mathrm{M}_{2}(\mathbb{C})}(\Sigma_{B^{\prime}})} \! + \!
(\zeta_{1} \! - \! \zeta_{2})^{-1} \vert \underline{c}(\zeta_{1},\zeta_{2},
\zeta_{3},\overline{\zeta_{3}}) \vert \\
\times& \, \vert \vert w^{\Sigma_{A^{\prime}}}_{-}(\cdot) \vert \vert_{
\mathcal{L}^{2}_{\mathrm{M}_{2}(\mathbb{C})}(\Sigma_{A^{\prime}})} \vert
\vert w^{\Sigma_{B^{\prime}}}_{+}(\cdot) \vert \vert_{\mathcal{L}^{2}_{
\mathrm{M}_{2}(\mathbb{C})}(\Sigma_{B^{\prime}})}) \vert \vert \phi (\cdot)
\vert \vert_{\mathcal{L}^{2}_{\mathrm{M}_{2}(\mathbb{C})}(\Sigma^{\sharp})},
\end{align*}
since $\sup_{(\zeta,\xi) \in \Sigma_{A^{\prime}} \times \Sigma_{B^{\prime}}}
\vert (\xi \! - \! \zeta)^{-1} \vert \! \leqslant \! \tfrac{\vert \underline{
c}(\zeta_{1},\zeta_{2},\zeta_{3},\overline{\zeta_{3}}) \vert}{(\zeta_{1}-
\zeta_{2})}$; thus, {}from the bound for $\vert \vert w^{\prime}(\cdot)
\vert \vert_{\mathcal{L}^{2}_{\mathrm{M}_{2}(\mathbb{C})}(\Sigma^{\sharp})}$
given in Lemma~4.4, one arrives at $\vert \vert \mathcal{X}(\cdot) \vert
\vert_{\mathcal{L}^{2}_{\mathrm{M}_{2}(\mathbb{C})}(\Sigma^{\sharp})} \!
\leqslant \! \tfrac{\vert \underline{c}(\zeta_{1},\zeta_{2},\zeta_{3},
\overline{\zeta_{3}}) \vert}{(\zeta_{1}-\zeta_{2})^{3/2} \sqrt{t \vert z_{o}
+\zeta_{1}+\zeta_{2} \vert}} \vert \vert \phi (\cdot) \vert \vert_{\mathcal{
L}^{2}_{\mathrm{M}_{2}(\mathbb{C})}(\Sigma^{\sharp})}$, whence, one deduces
the bound for $\vert \vert C^{\Sigma^{\sharp}}_{w^{\Sigma_{A^{\prime}}}}C^{
\Sigma^{\sharp}}_{w^{\Sigma_{B^{\prime}}}} \vert \vert_{\mathscr{N}(\Sigma^{
\sharp})}$ stated in the Lemma. Proceeding analogously as above, but
considering, instead, the $\mathcal{L}^{\infty}_{\mathrm{M}_{2}(\mathbb{C})}
(\Sigma^{\sharp})$ bound on $\phi (z)$, one shows that
\begin{align*}
\vert \vert \mathcal{X}(\cdot) \vert \vert_{\mathcal{L}^{\infty}_{\mathrm{
M}_{2}(\mathbb{C})}(\Sigma^{\sharp}) \to \mathcal{L}^{2}_{\mathrm{M}_{2}
(\mathbb{C})}(\Sigma^{\sharp})} \! \leqslant& \, ((\zeta_{1} \! - \! \zeta_{
2})^{-1} \vert \underline{c}(\zeta_{1},\zeta_{2},\zeta_{3},\overline{\zeta_{
3}}) \vert \vert \vert w^{\Sigma_{A^{\prime}}}_{+}(\cdot) \vert \vert_{
\mathcal{L}^{2}_{\mathrm{M}_{2}(\mathbb{C})}(\Sigma_{A^{\prime}})} \\
\times& \, \vert \vert w^{\Sigma_{B^{\prime}}}_{-}(\cdot) \vert \vert_{
\mathcal{L}^{1}_{\mathrm{M}_{2}(\mathbb{C})}(\Sigma_{B^{\prime}})} \! + \!
(\zeta_{1} \! - \! \zeta_{2})^{-1} \vert \underline{c}(\zeta_{1},\zeta_{2},
\zeta_{3},\overline{\zeta_{3}}) \vert \\
\times& \, \vert \vert w^{\Sigma_{A^{\prime}}}_{-}(\cdot) \vert \vert_{
\mathcal{L}^{2}_{\mathrm{M}_{2}(\mathbb{C})}(\Sigma_{A^{\prime}})} \vert
\vert w^{\Sigma_{B^{\prime}}}_{+}(\cdot) \vert \vert_{\mathcal{L}^{1}_{
\mathrm{M}_{2}(\mathbb{C})}(\Sigma_{B^{\prime}})}) \vert \vert \phi (\cdot)
\vert \vert_{\mathcal{L}^{\infty}_{\mathrm{M}_{2}(\mathbb{C})}(\Sigma^{
\sharp})}:
\end{align*}
now, using, for $p \! \in \! \{1,2\}$, the bounds for $\vert \vert w^{\prime}
(\cdot) \vert \vert_{\mathcal{L}^{p}_{\mathrm{M}_{2}(\mathbb{C})}(\Sigma^{
\sharp})}$ given in Lemma~4.4, one arrives at $\vert \vert \mathcal{X}(\cdot)
\vert \vert_{\mathcal{L}^{\infty}_{\mathrm{M}_{2}(\mathbb{C})}(\Sigma^{
\sharp}) \to \mathcal{L}^{2}_{\mathrm{M}_{2}(\mathbb{C})}(\Sigma^{\sharp})}
\! \leqslant \! (\zeta_{1} \! - \! \zeta_{2})^{-7/4} \vert \underline{c}
(\zeta_{1},\zeta_{2},\zeta_{3},\overline{\zeta_{3}}) \vert (t \vert z_{o}
\! + \! \zeta_{1} \! + \! \zeta_{2} \vert)^{-3/4} \vert \vert \phi (\cdot)
\vert \vert_{\mathcal{L}^{\infty}_{\mathrm{M}_{2}(\mathbb{C})}(\Sigma^{
\sharp})}$; hence, one deduces the bound for $\vert \vert C^{\Sigma^{
\sharp}}_{w^{\Sigma_{A^{\prime}}}} C^{\Sigma^{\sharp}}_{w^{\Sigma_{B^{
\prime}}}} \vert \vert_{\mathcal{L}^{\infty}_{\mathrm{M}_{2}(\mathbb{C})}
(\Sigma^{\sharp}) \to \mathcal{L}^{2}_{\mathrm{M}_{2}(\mathbb{C})}(\Sigma^{
\sharp})}$. \hfill $\square$
\begin{ccccc}
If, for $k \! \in \! \{A,B\}$, $(\mathbf{1}_{\Sigma_{k^{\prime}}} \!
- \! C^{\Sigma_{k^{\prime}}}_{w^{\Sigma_{k^{\prime}}}})^{-1} \! \in
\! \mathscr{N}(\Sigma_{k^{\prime}})$, then, as $t \! \to \! +\infty$
such that $0 \! < \zeta_{2} \! < \! \tfrac{1}{M} \! < \! M \! < \!
\zeta_{1}$ and $\vert \zeta_{3} \vert^{2} \! = \! 1$, with $M \!
\in \! \mathbb{R}_{>1}$ and bounded, for $\zeta \! \in \! \mathbb{
C} \setminus \Sigma^{\sharp}$,
\begin{align*}
\int\nolimits_{\Sigma^{\sharp}} \dfrac{((\mathbf{1}_{\Sigma^{\sharp}
} \! - \! C^{\Sigma^{\sharp}}_{w^{\Sigma^{\sharp}}})^{-1} \mathrm{I}
)(z) w^{\Sigma^{\sharp}}(z)}{(z \! - \! \zeta)} \, \dfrac{\md z}{2
\pi \mi} &= \sum_{k \in \{A,B\}} \int\nolimits_{\Sigma_{k^{\prime}}
} \dfrac{((\mathbf{1}_{\Sigma_{k^{\prime}}} \! - \! C^{\Sigma_{k^{
\prime}}}_{w^{\Sigma_{k^{\prime}}}})^{-1} \mathrm{I})(z)w^{\Sigma_{
k^{\prime}}}(z)}{(z \! - \! \zeta)} \, \dfrac{\md z}{2 \pi \mi} \\
&+ \mathcal{O} \! \left( \dfrac{\underline{c}(\zeta_{1},\zeta_{2},
\zeta_{3},\overline{\zeta_{3}})f^{\Sigma^{\sharp}}(\zeta)}{(\zeta_{
1} \! - \! \zeta_{2})^{2} \vert z_{o} \! + \! \zeta_{1} \! + \!
\zeta_{2} \vert t} \right),
\end{align*}
with $f^{\Sigma^{\sharp}}(\zeta) \! \in \! \mathcal{L}^{\infty}_{
\mathrm{M}_{2}(\mathbb{C})}(\mathbb{C} \setminus \Sigma^{\sharp})$.
\end{ccccc}

\emph{Proof.} {}From Proposition~5.3 (with associations $N \! = \!
2$, $1 \! \leftrightarrow \! A$, and $2 \! \leftrightarrow \! B)$
and the second resolvent identity, one obtains $(\mathbf{1}_{\Sigma^{
\sharp}} \! - \! \sum_{k \in \{A,B\}} \! C^{\Sigma_{k^{\prime}}}_{
w^{\Sigma_{k^{\prime}}}})^{-1} \! = \! D_{\Sigma^{\sharp}} \! + \!
D_{\Sigma^{\sharp}}(\mathbf{1}_{\Sigma^{\sharp}} \! - \! \mathbf{E
}_{\Sigma^{\sharp}})^{-1} \mathbf{E}_{\Sigma^{\sharp}}$, where $D_{
\Sigma^{\sharp}} \! := \! \mathbf{1}_{\Sigma^{\sharp}} \! + \! \sum_{
k \in \{A,B\}} \! C^{\Sigma_{k^{\prime}}}_{w^{\Sigma_{k^{\prime}}}}
(\mathbf{1}_{\Sigma_{k^{\prime}}} \! - \! C^{\Sigma_{k^{\prime}}}_{
w^{\Sigma_{k^{\prime}}}})^{-1}$, and $\mathbf{E}_{\Sigma^{\sharp}} \!
:= \! \sum_{\alpha,\beta \in \{A,B\}} \! (1 \! - \! \delta_{\alpha
\beta}) C^{\Sigma_{\alpha^{\prime}}}_{w^{\Sigma_{\alpha^{\prime}}}}
C^{\Sigma_{\beta^{\prime}}}_{w^{\Sigma_{\beta^{\prime}}}}(\mathbf{
1}_{\Sigma_{\beta^{\prime}}} \! - \! C^{\Sigma_{\beta^{\prime}}}_{
w^{\Sigma_{\beta^{\prime}}}})^{-1}$, with $\delta_{\alpha \beta}$
the Kronecker delta; hence, for $\zeta \! \in \! \mathbb{C} \setminus
\Sigma^{\sharp}$, $\mathrm{I}^{\sharp} \! := \! \int_{\Sigma^{\sharp}
} \! \tfrac{((\mathbf{1}_{\Sigma^{\sharp}}-C^{\Sigma^{\sharp}}_{w^{
\Sigma^{\sharp}}})^{-1} \mathrm{I})(z)w^{\Sigma^{\sharp}}(z)}{(z-\zeta)
} \linebreak[4]
\cdot \tfrac{\md z}{2 \pi \mi} \! = \! \int_{\Sigma^{\sharp}} \! \tfrac{
(D_{\Sigma^{\sharp}} \mathrm{I})(z)w^{\Sigma^{\sharp}}(z)}{(z-\zeta)} \,
\tfrac{\md z}{2 \pi \mi} \! + \! \int_{\Sigma^{\sharp}} \! \tfrac{((D_{
\Sigma^{\sharp}}(\mathbf{1}_{\Sigma^{\sharp}}-\mathbf{E}_{\Sigma^{\sharp}
})^{-1} \mathbf{E}_{\Sigma^{\sharp}}) \mathrm{I})(z)w^{\Sigma^{\sharp}}
(z)}{(z-\zeta)} \, \tfrac{\md z}{2 \pi \mi}$. Recall that $\Sigma^{\sharp}
\! = \! \Sigma_{A^{\prime}} \cup \Sigma_{B^{\prime}}$. Since, for $k \!
\in \! \{A,B\}$, $\vert \vert C^{\Sigma_{k^{\prime}}}_{w^{\Sigma_{k^{
\prime}}}} \vert \vert_{\mathscr{N}(\Sigma_{k^{\prime}})} \! \leqslant
\! \vert \underline{c}(\zeta_{1},\zeta_{2},\zeta_{3},\overline{\zeta_{
3}}) \vert \vert \vert w^{\Sigma_{k^{\prime}}}(\cdot) \vert \vert_{
\mathcal{L}^{2}_{\mathrm{M}_{2}(\mathbb{C})}(\Sigma_{k^{\prime}})}
\! \leqslant \! \vert \underline{c}(\zeta_{1},\zeta_{2},\zeta_{3},
\overline{\zeta_{3}}) \vert \linebreak[4]
\cdot ((\zeta_{1} \! - \! \zeta_{2}) \vert z_{o} \! + \! \zeta_{1}
\! + \! \zeta_{2} \vert t)^{-1/4} \! \leqslant \! \vert \underline{
c}(\zeta_{1},\zeta_{2},\zeta_{3},\overline{\zeta_{3}}) \vert$
(Lemma~~4.4), and, by assumption, $(\mathbf{1}_{\Sigma_{k^{\prime}}}
\! - \! C^{\Sigma_{k^{\prime}}}_{w^{\Sigma_{k^{\prime}}}})^{-1}
\linebreak[4]
\in \! \mathscr{N}(\Sigma_{k^{\prime}})$, it follows, {}from the bounds
given in Lemma~5.2 and the second resolvent identity, that $\vert
\vert D_{\Sigma^{\sharp}} \vert \vert_{\mathscr{N}(\Sigma^{\sharp})}
\! \leqslant \! \vert \underline{c}(\zeta_{1},\zeta_{2},\zeta_{3},
\overline{\zeta_{3}}) \vert$ and $\vert \vert (\mathbf{1}_{\Sigma^{
\sharp}} \! - \! \mathbf{E}_{\Sigma^{\sharp}})^{-1} \vert \vert_{
\mathscr{N}(\Sigma^{\sharp})} \! \leqslant \! \vert \underline{c}
(\zeta_{1},\zeta_{2},\zeta_{3},\overline{\zeta_{3}}) \vert$. {}From
the second resolvent identity, the definition of $\mathbf{E}_{\Sigma^{
\sharp}}$, the fact that $\vert \vert (C^{\Sigma_{k^{\prime}}}_{w^{
\Sigma_{k^{\prime}}}} \mathrm{I})(\cdot) \vert \vert_{\mathcal{L}^{
2}_{\mathrm{M}_{2}(\mathbb{C})}(\Sigma^{\sharp})} \! \leqslant \!
\vert \underline{c}(\zeta_{1},\zeta_{2},\zeta_{3},\overline{\zeta_{
3}}) \vert \vert \vert w^{\Sigma_{k^{\prime}}}(\cdot) \vert \vert_{
\mathcal{L}^{2}_{\mathrm{M}_{2}(\mathbb{C})}(\Sigma^{\sharp})} \!
\leqslant \! \vert \underline{c}(\zeta_{1},\zeta_{2},\zeta_{3},
\overline{\zeta_{3}}) \vert \vert \vert w^{\Sigma^{\sharp}}(\cdot)
\vert \vert_{\mathcal{L}^{2}_{\mathrm{M}_{2}(\mathbb{C})}(\Sigma^{
\sharp})} \! \leqslant \! \vert \underline{c}(\zeta_{1},\zeta_{2},
\zeta_{3},\overline{\zeta_{3}}) \vert ((\zeta_{1} \! - \! \zeta_{
2}) \vert z_{o} \! + \! \zeta_{1} \! + \! \zeta_{2} \vert t)^{-1/4}$
(Lemma~4.4), $k \! \in \! \{A,B\}$, and the Cauchy-Schwarz inequality
for integrals, one shows that, modulo the term $d_{z}(\zeta) \! :=
\! \sup_{(z,\zeta) \in \Sigma^{\sharp} \times \mathbb{C} \setminus
\Sigma^{\sharp}} \vert (z \! - \! \zeta)^{-1} \vert$, $\vert \vert
\tfrac{(\mathbf{E}_{\Sigma^{\sharp}} \mathrm{I})(\cdot) w^{\Sigma^{
\sharp}}(\cdot)}{(\cdot -\zeta)} \vert \vert_{\mathcal{L}^{2}_{
\mathrm{M}_{2}(\mathbb{C})}(\Sigma^{\sharp})} \linebreak[4]
\leqslant \sum_{\alpha,\beta \in \{A,B\}} \! (1-\delta_{\alpha \beta}
) \vert \vert (C^{\Sigma_{\alpha^{\prime}}}_{w^{\Sigma_{\alpha^{
\prime}}}}C^{\Sigma_{\beta^{\prime}}}_{w^{\Sigma_{\beta^{\prime}}
}} \mathrm{I})(\cdot) \vert \vert_{\mathcal{L}^{2}_{\mathrm{M}_{
2}(\mathbb{C})}(\Sigma^{\sharp})} \vert \vert w^{\Sigma^{\sharp}}
(\cdot) \vert \vert_{\mathcal{L}^{2}_{\mathrm{M}_{2}(\mathbb{C})}
(\Sigma^{\sharp})}+\sum_{\alpha,\beta \in \{A,B\}} \! (1-\delta_{
\alpha \beta}) \linebreak[4]
\cdot \vert \vert C^{\Sigma_{\alpha^{\prime}}}_{w^{\Sigma_{\alpha^{
\prime}}}} C^{\Sigma_{\beta^{\prime}}}_{w^{\Sigma_{\beta^{\prime}}}}
\vert \vert_{\mathscr{N}(\Sigma^{\sharp})} \vert \vert (\mathbf{1
}_{\Sigma_{\beta^{\prime}}} \! - \! C^{\Sigma_{\beta^{\prime}}}_{
w^{\Sigma_{\beta^{\prime}}}})^{-1} \vert \vert_{\mathscr{N}(\Sigma^{
\sharp})} \vert \vert w^{\Sigma^{\sharp}}(\cdot) \vert \vert^{2}_{
\mathcal{L}^{2}_{\mathrm{M}_{2}(\mathbb{C})}(\Sigma^{\sharp})}$: using
the bounds given in L\-e\-m\-m\-ae~4.4 and~5.2, one arrives at $\vert
\vert \tfrac{(\mathbf{E}_{\Sigma^{\sharp}} \mathrm{I})(\cdot)w^{
\Sigma^{\sharp}}(\cdot)}{(\cdot -\zeta)} \vert \vert_{\mathcal{L}^{
2}_{\mathrm{M}_{2}(\mathbb{C})}(\Sigma^{\sharp})} \! \leqslant \!
\tfrac{\vert \underline{c}(\zeta_{1},\zeta_{2},\zeta_{3},\overline{
\zeta_{3}}) \vert d_{z}(\zeta)}{(\zeta_{1}-\zeta_{2})^{2} \vert z_{o}
+\zeta_{1}+\zeta_{2} \vert t}$; hence, $(\mathbf{1}_{\Sigma^{\sharp}}
\! - \! \sum_{k \in \{A,B\}} \! C^{\Sigma_{k^{\prime}}}_{w^{\Sigma_{
k^{\prime}}}})^{-1} \! \in \! \mathscr{N}(\Sigma^{\sharp})$, and
$\mathrm{I}^{\sharp} \! = \! \int_{\Sigma^{\sharp}} \! \tfrac{(D_{
\Sigma^{\sharp}} \mathrm{I})(z)w^{\Sigma^{\sharp}}(z)}{(z-\zeta)} \,
\tfrac{\md z}{2 \pi \mi} \! + \! \mathcal{O}(\tfrac{\underline{c}(
\zeta_{1},\zeta_{2},\zeta_{3},\overline{\zeta_{3}})d_{z}(\zeta)}{(
\zeta_{1}-\zeta_{2})^{2} \vert z_{o}+\zeta_{1}+\zeta_{2} \vert t})$.
{}From the definition of $D_{\Sigma^{\sharp}}$ and an application
of the second resolvent identity, one shows that, for $\zeta \! \in \!
\mathbb{C} \setminus \Sigma^{\sharp}$,
\begin{align*}
\mathrm{I}^{\sharp} &= \! \sum_{k \in \{A,B\}} \int_{\Sigma_{k^{
\prime}}} \dfrac{((\mathbf{1}_{\Sigma_{k^{\prime}}} \! - \! C^{
\Sigma_{k^{\prime}}}_{w^{\Sigma_{k^{\prime}}}})^{-1} \mathrm{I})
(z)w^{\Sigma_{k^{\prime}}}(z)}{(z \! - \! \zeta)} \, \dfrac{\md
z}{2 \pi \mi} + \sum_{\alpha,\beta \in \{A,B\}}(1 \! - \! \delta_{
\alpha \beta}) \\
&\times \int_{\Sigma^{\sharp}} \dfrac{(C^{\Sigma_{\alpha^{\prime}}
}_{w^{\Sigma_{\alpha^{\prime}}}}(\mathbf{1}_{\Sigma_{\alpha^{\prime}
}} \! - \! C^{\Sigma_{\alpha^{\prime}}}_{w^{\Sigma_{\alpha^{\prime}}
}})^{-1} \mathrm{I})(z)w^{\Sigma_{\beta^{\prime}}}(z)}{(z \! - \!
\zeta)} \, \dfrac{\md z}{2 \pi \mi} + \mathcal{O} \! \left(\dfrac{
\underline{c}(\zeta_{1},\zeta_{2},\zeta_{3},\overline{\zeta_{3}})
d_{z}(\zeta)}{(\zeta_{1} \! - \! \zeta_{2})^{2} \vert z_{o} \! +
\! \zeta_{1} \! + \! \zeta_{2} \vert t} \right).
\end{align*}
The latter two integrals are now estimated: applying, again, the
second resolvent identity, one shows that, for $\alpha \! \not= \!
\beta \! \in \! \{A,B\}$ and $\zeta \! \in \! \mathbb{C} \setminus
\Sigma^{\sharp}$,
\begin{align*}
& \quad \left\vert \int\nolimits_{\Sigma^{\sharp}} \tfrac{(C^{\Sigma_{
\alpha^{\prime}}}_{w^{\Sigma_{\alpha^{\prime}}}}(\mathbf{1}_{\Sigma_{
\alpha^{\prime}}} - C^{\Sigma_{\alpha^{\prime}}}_{w^{\Sigma_{\alpha^{
\prime}}}})^{-1} \mathrm{I})(z)w^{\Sigma_{\beta^{\prime}}}(z)}{(z-\zeta)}
\, \tfrac{\md z}{2 \pi \mi} \right\vert \! \leqslant \! \left\vert
\int\nolimits_{\Sigma^{\sharp}} \tfrac{(C^{\Sigma_{\alpha^{\prime}}}_{
w^{\Sigma_{\alpha^{\prime}}}} \mathrm{I})(z)w^{\Sigma_{\beta^{\prime}}}
(z)}{(z-\zeta)} \, \tfrac{\md z}{2 \pi \mi} \right\vert \\
& + \left\vert \int\nolimits_{\Sigma^{\sharp}} \tfrac{((C^{
\Sigma_{\alpha^{\prime}}}_{w^{\Sigma_{\alpha^{\prime}}}}(\mathbf{
1}_{\Sigma_{\alpha^{\prime}}} - C^{\Sigma_{\alpha^{\prime}}}_{w^{
\Sigma_{\alpha^{\prime}}}})^{-1}C^{\Sigma_{\alpha^{\prime}}}_{w^{
\Sigma_{\alpha^{\prime}}}}) \mathrm{I})(z)w^{\Sigma_{\beta^{\prime}
}}(z)}{(z-\zeta)} \, \tfrac{\md z}{2 \pi \mi} \right\vert \\
& \leqslant \left\vert \int\nolimits_{\Sigma_{\beta^{\prime}}} \!
\left( \int\nolimits_{\Sigma_{\alpha^{\prime}}} \tfrac{w^{\Sigma_{
\alpha^{\prime}}}(\mu)}{(\mu -z)} \, \tfrac{\md \mu}{2 \pi \mi}
\right) \tfrac{w^{\Sigma_{\beta^{\prime}}}(z)}{(z-\zeta)} \, \tfrac{
\md z}{2 \pi \mi} \right\vert \\
& + \left\vert \int\nolimits_{\Sigma_{\beta^{\prime}}} \! \left(
\int\nolimits_{\Sigma_{\alpha^{\prime}}} \tfrac{(((\mathbf{1}_{\Sigma_{
\alpha^{\prime}}} - C^{\Sigma_{\alpha^{\prime}}}_{w^{\Sigma_{\alpha^{
\prime}}}})^{-1}C^{\Sigma_{\alpha^{\prime}}}_{w^{\Sigma_{\alpha^{
\prime}}}}) \mathrm{I})(\mu)w^{\Sigma_{\alpha^{\prime}}}(\mu)}{(\mu
-z)} \, \tfrac{\md \mu}{2 \pi \mi} \right) \tfrac{w^{\Sigma_{\beta^{
\prime}}}(z)}{(z-\zeta)} \, \tfrac{\md z}{2 \pi \mi} \right\vert \\
& \leqslant \tfrac{d_{z}(\zeta)}{(2 \pi)^{2}} \sup_{(\mu,z) \in
\Sigma_{\alpha^{\prime}} \times \Sigma_{\beta^{\prime}}} \vert \tfrac{
1}{(\mu -z)} \vert \vert \vert w^{\Sigma_{\alpha^{\prime}}}(\cdot)
\vert \vert_{\mathcal{L}^{1}_{\mathrm{M}_{2}(\mathbb{C})}(\Sigma_{
\alpha^{\prime}})} \vert \vert w^{\Sigma_{\beta^{\prime}}}(\cdot)
\vert \vert_{\mathcal{L}^{1}_{\mathrm{M}_{2}(\mathbb{C})}(\Sigma_{
\beta^{\prime}})} \\
 & + \tfrac{d_{z}(\zeta)}{(2 \pi)^{2}} \sup_{(\mu,z) \in  \Sigma_{
\alpha^{\prime}} \times \Sigma_{\beta^{\prime}}} \vert \tfrac{1}{
(\mu -z)} \vert \vert \vert (((\mathbf{1}_{\Sigma_{\alpha^{\prime}
}} \! - \! C^{\Sigma_{\alpha^{\prime}}}_{w^{\Sigma_{\alpha^{\prime}
}}})^{-1}C^{\Sigma_{\alpha^{\prime}}}_{w^{\Sigma_{\alpha^{\prime}}}
}) \mathrm{I})(\cdot) \vert \vert_{\mathcal{L}^{2}_{\mathrm{M}_{2}
(\mathbb{C})}(\Sigma_{\alpha^{\prime}})} \\
 & \times \vert \vert w^{\Sigma_{\alpha^{\prime}}}(\cdot) \vert \vert_{
\mathcal{L}^{2}_{\mathrm{M}_{2}(\mathbb{C})}(\Sigma_{\alpha^{\prime}})
} \vert \vert w^{\Sigma_{\beta^{\prime}}}(\cdot) \vert \vert_{\mathcal{
L}^{1}_{\mathrm{M}_{2}(\mathbb{C})}(\Sigma_{\beta^{\prime}})} \\
& \leqslant \tfrac{\vert \underline{c}(\zeta_{1},\zeta_{2},\zeta_{3},
\overline{\zeta_{3}}) \vert d_{z}(\zeta)}{(\zeta_{1}-\zeta_{2})} \vert
\vert w^{\Sigma_{\alpha^{\prime}}}(\cdot) \vert \vert_{\mathcal{L}^{
1}_{\mathrm{M}_{2}(\mathbb{C})}(\Sigma_{\alpha^{\prime}})} \vert
\vert w^{\Sigma_{\beta^{\prime}}}(\cdot) \vert \vert_{\mathcal{L}^{
1}_{\mathrm{M}_{2}(\mathbb{C})}(\Sigma_{\beta^{\prime}})} \\
& + \tfrac{\vert \underline{c}(\zeta_{1},\zeta_{2},\zeta_{3},\overline{
\zeta_{3}}) \vert d_{z}(\zeta)}{(\zeta_{1}-\zeta_{2})} \vert \vert
(\mathbf{1}_{\Sigma_{\alpha^{\prime}}} \! - \! C^{\Sigma_{\alpha^{\prime}
}}_{w^{\Sigma_{\alpha^{\prime}}}})^{-1} \vert \vert_{\mathscr{N}(\Sigma_{
\alpha^{\prime}})} \vert \vert (C^{\Sigma_{\alpha^{\prime}}}_{w^{\Sigma_{
\alpha^{\prime}}}} \mathrm{I})(\cdot) \vert \vert_{\mathcal{L}^{2}_{
\mathrm{M}_{2}(\mathbb{C})}(\Sigma_{\alpha^{\prime}})} \\
& \times \vert \vert w^{\Sigma_{\alpha^{\prime}}}(\cdot) \vert \vert_{
\mathcal{L}^{2}_{\mathrm{M}_{2}(\mathbb{C})}(\Sigma_{\alpha^{\prime}})}
\vert \vert w^{\Sigma_{\beta^{\prime}}}(\cdot) \vert \vert_{\mathcal{
L}^{1}_{\mathrm{M}_{2}(\mathbb{C})}(\Sigma_{\beta^{\prime}})} \\
& \leqslant \tfrac{\vert \underline{c}(\zeta_{1},\zeta_{2},\zeta_{3},
\overline{\zeta_{3}}) \vert d_{z}(\zeta)}{(\zeta_{1}-\zeta_{2})} \vert
\vert w^{\Sigma_{\alpha^{\prime}}}(\cdot) \vert \vert_{\mathcal{L}^{
1}_{\mathrm{M}_{2}(\mathbb{C})}(\Sigma_{\alpha^{\prime}})} \vert \vert
w^{\Sigma_{\beta^{\prime}}}(\cdot) \vert \vert_{\mathcal{L}^{1}_{
\mathrm{M}_{2}(\mathbb{C})}(\Sigma_{\beta^{\prime}})} \\
& + \tfrac{\vert \underline{c}(\zeta_{1},\zeta_{2},\zeta_{3},\overline{
\zeta_{3}}) \vert d_{z}(\zeta)}{(\zeta_{1}-\zeta_{2})} \vert \vert w^{
\Sigma_{\alpha^{\prime}}}(\cdot) \vert \vert_{\mathcal{L}^{2}_{\mathrm{
M}_{2}(\mathbb{C})}(\Sigma_{\alpha^{\prime}})}^{2} \vert \vert w^{
\Sigma_{\beta^{\prime}}}(\cdot) \vert \vert_{\mathcal{L}^{1}_{\mathrm{
M}_{2}(\mathbb{C})}(\Sigma_{\beta^{\prime}})},
\end{align*}
since $\sup_{(\mu,z) \in \Sigma_{\alpha^{\prime}} \times \Sigma_{
\beta^{\prime}}} \vert (\mu \! - \! z)^{-1} \vert \! \leqslant \!
\vert \underline{c}(\zeta_{1},\zeta_{2},\zeta_{3},\overline{\zeta_{
3}}) \vert (\zeta_{1} \! - \! \zeta_{2})^{-1}$, $\vert \vert (\mathbf{
1}_{\Sigma_{\alpha^{\prime}}} \! - \! C^{\Sigma_{\alpha^{\prime}}}_{
w^{\Sigma_{\alpha^{\prime}}}})^{-1} \vert \vert_{\mathscr{N}(\Sigma_{
\alpha^{\prime}})} \! \leqslant \! \vert \underline{c}(\zeta_{1},\zeta_{
2},\zeta_{3},\overline{\zeta_{3}}) \vert$, and $\vert \vert (C^{\Sigma_{
\alpha^{\prime}}}_{w^{\Sigma_{\alpha^{\prime}}}} \mathrm{I})(\cdot)
\vert \vert_{\mathcal{L}^{2}_{\mathrm{M}_{2}(\mathbb{C})}(\Sigma_{
\alpha^{\prime}})} \! \leqslant \! \vert \underline{c}(\zeta_{1},
\zeta_{2},\zeta_{3},\overline{\zeta_{3}}) \vert \vert \vert w^{\Sigma_{
\alpha^{\prime}}}(\cdot) \vert \vert_{\mathcal{L}^{2}_{\mathrm{M}_{
2}(\mathbb{C})}(\Sigma_{\alpha^{\prime}})}$; thus, using the fact
that $\vert \vert w^{\Sigma_{q}}(\cdot) \vert \vert_{\mathcal{L}^{
p}_{\mathrm{M}_{2}(\mathbb{C})}(\Sigma_{q})} \! \leqslant \! \vert
\vert w^{\Sigma^{\sharp}}(\cdot) \vert \vert_{\mathcal{L}^{p}_{\mathrm{
M}_{2}(\mathbb{C})}(\Sigma^{\sharp})}$, $q \! \in \! \{\alpha^{\prime},
\beta^{\prime}\}$, $p \! \in \! \{1,2\}$, {}from the bounds given in
Lemma~4.4, one arrives at
\begin{equation*}
\left\vert \int\nolimits_{\Sigma^{\sharp}} \dfrac{(C^{\Sigma_{\alpha^{
\prime}}}_{w^{\Sigma_{\alpha^{\prime}}}}(\mathbf{1}_{\Sigma_{\alpha^{
\prime}}} \! - \! C^{\Sigma_{\alpha^{\prime}}}_{w^{\Sigma_{\alpha^{
\prime}}}})^{-1} \mathrm{I})(z)w^{\Sigma_{\beta^{\prime}}}(z)}{(z \! -
\! \zeta)} \, \dfrac{\md z}{2 \pi \mi} \right\vert \! \leqslant \!
\dfrac{\vert \underline{c}(\zeta_{1},\zeta_{2},\zeta_{3},\overline{
\zeta_{3}}) \vert d_{z}(\zeta)}{(\zeta_{1} \! - \! \zeta_{2})^{2} \vert
z_{o} \! + \! \zeta_{1} \! + \! \zeta_{2} \vert t},
\end{equation*}
whence, recalling the expression for $\mathrm{I}^{\sharp}$ given earlier
in the proof, one obtains, for $\zeta \! \in \! \mathbb{C} \setminus
\Sigma^{\sharp}$, the result stated in the Lemma. \hfill $\square$
\begin{bbbbb}
As $t \! \to \! +\infty$ such that $0 \! < \! \zeta_{2} \! < \! \tfrac{1}{M}
\! < \! M \! < \! \zeta_{1}$ and $\vert \zeta_{3} \vert^{2} \! = \! 1$, with
$M \! \in \! \mathbb{R}_{>1}$ and bounded, the solution of the {\rm RHP}
for $m^{\sharp}(\zeta) \colon \mathbb{C} \setminus \Sigma^{\prime} \!
\to \! \mathrm{SL}(2,\mathbb{C})$ formulated in Lemma~{\rm 4.3} has
the integral representation
\begin{align*}
m^{\sharp}(\zeta) &= \mathrm{I} \! + \! \sum_{k \in \{A,B\}}
\int\nolimits_{\Sigma_{k^{\prime}}} \dfrac{((\mathbf{1}_{\Sigma_{
k^{\prime}}} \! - \! C^{\Sigma_{k^{\prime}}}_{w^{\Sigma_{k^{\prime}}}
})^{-1} \mathrm{I})(z) w^{\Sigma_{k^{\prime}}}(z)}{(z \! - \! \zeta)}
\, \dfrac{\md z}{2 \pi \mi} \\
&+ \mathcal{O} \! \left( \dfrac{\underline{c}(\zeta_{1},\zeta_{
2},\zeta_{3},\overline{\zeta_{3}}) \diamondsuit^{\sharp}(\zeta)}
{(\zeta_{1} \! - \! \zeta_{2})^{2} \vert z_{o} \! + \! \zeta_{1}
\! + \! \zeta_{2} \vert t} \right), \quad \zeta \! \in \! \mathbb{
C} \setminus \Sigma_{A^{\prime}} \cup \Sigma_{B^{\prime}},
\end{align*}
where $w^{\Sigma_{k^{\prime}}}(\zeta) \! = \! \sum_{l \in \{\pm\}}
\! w_{l}^{\Sigma_{k^{\prime}}}(\zeta)$, with $w^{\Sigma_{k^{\prime}
}}_{\pm}(\zeta) \! := \! w^{\Sigma^{\sharp}}_{\pm}(\zeta) \! \!
\upharpoonright_{\Sigma_{k^{\prime}}}$ and $w^{\Sigma^{\sharp}
}_{\pm}(\zeta)$ given in Lemma~{\rm 4.6}, and $\diamondsuit^{
\sharp}(\zeta) \! \in \! \mathcal{L}^{\infty}_{\mathrm{M}_{2}(\mathbb{
C})}(\mathbb{C} \setminus \Sigma_{A^{\prime}} \cup \Sigma_{B^{
\prime}})$.
\end{bbbbb}

\emph{Proof.} {}From Lemma~5.3 and the (singular) integral
representation for the solution of the RHP for $m^{\Sigma^{\sharp}}
(\zeta) \colon \mathbb{C} \setminus \Sigma^{\sharp} \! \to \! \mathrm{
SL}(2,\mathbb{C})$ formulated in Lemma~4.6 (cf.~Eq.~(103)), one shows
that, as $t \to +\infty$ such that $0<\zeta_{2}<\tfrac{1}{M}<M<\zeta_{
1}$ and $\vert \zeta_{3} \vert^{2}=1$, with $M \in \mathbb{R}_{>1}$
and bounded, $m^{\Sigma^{\sharp}}(\zeta)=\mathrm{I}+\int\nolimits_{
\Sigma^{\sharp}} \tfrac{((\mathbf{1}_{\Sigma^{\sharp}}-C^{\Sigma^{
\sharp}}_{w^{\Sigma^{\sharp}}})^{-1} \mathrm{I})(z)w^{\Sigma^{\sharp}
}(z)}{(z-\zeta)} \, \tfrac{\md z}{2 \pi \mi}=\sum_{k \in \{A,B\}}
\int\nolimits_{\Sigma_{k^{\prime}}} \tfrac{((\mathbf{1}_{\Sigma_{k^{
\prime}}}-C^{\Sigma_{k^{\prime}}}_{w^{\Sigma_{k^{\prime}}}})^{-1}
\mathrm{I})(z)w^{\Sigma_{k^{\prime}}}(z)}{(z-\zeta)} \, \tfrac{\md
z}{2 \pi \mi}+\mathcal{O}(\tfrac{\underline{c}(\zeta_{1},\zeta_{2},
\zeta_{3},\overline{\zeta_{3}}) \diamondsuit^{\Sigma^{\sharp}}(\zeta)
}{(\zeta_{1}-\zeta_{2})^{2} \vert z_{o}+\zeta_{1}+\zeta_{2} \vert t})$,
$\zeta \! \in \! \mathbb{C} \setminus \Sigma^{\sharp}$: now, {}from
the RHP for $m^{\sharp}(\zeta) \colon \mathbb{C} \setminus \Sigma^{
\prime} \! \to \! \mathrm{SL}(2,\mathbb{C})$ formulated in Lemma~4.3
and the corresponding integral representation $m^{\sharp}(\zeta) \!
= \! \mathrm{I} \! + \! \int_{\Sigma^{\prime}} \! \tfrac{((\mathbf{
1}-C_{w^{\sharp}})^{-1} \mathrm{I})(z)w^{\sharp}(z)}{(z-\zeta)} \,
\tfrac{\md z}{2 \pi \mi}$, $\zeta \! \in \! \mathbb{C} \setminus
\Sigma^{\prime}$, Proposition~4.4, Corollary~4.1, the definition of
$w^{\Sigma_{k^{\prime}}}(\zeta)$, $k \! \in \! \{A,B\}$, given in
the Proposition, and the fact that $\mathcal{O}(\tfrac{\underline{
c}(\zeta_{1},\zeta_{2},\zeta_{3},\overline{\zeta_{3}})}{\vert z_{
o}+\zeta_{1}+\zeta_{2} \vert^{l}t^{l}}) \! + \! \mathcal{O}(\tfrac{
\underline{c}(\zeta_{1},\zeta_{2},\zeta_{3},\overline{\zeta_{3}})}
{(\zeta_{1}-\zeta_{2})^{2} \vert z_{o}+\zeta_{1}+\zeta_{2} \vert
t}) \! = \! \mathcal{O}(\tfrac{\underline{c}(\zeta_{1},\zeta_{2},
\zeta_{3},\overline{\zeta_{3}})}{(\zeta_{1}-\zeta_{2})^{2} \vert
z_{o}+\zeta_{1}+\zeta_{2} \vert t})$, $l \! \in \! \mathbb{Z}_{
\geqslant 1}$ and arbitrarily large, one arrives at the result
stated in the Proposition. \hfill $\square$

Heretofore, it has been shown that (Lemma 5.3, Proposition 5.2,
Corollary 4.1, Definition 4.2, Proposition 4.4, and Proposition
4.2) $\{(\mathbf{1}_{\Sigma_{k}} \! - \! C^{\Sigma_{k}}_{w^{\Sigma_{
k}}})^{-1} \! \in \! \mathscr{N}(\Sigma_{k})\}_{k \in \{A,B\}} \!
\Rightarrow \! \{(\mathbf{1}_{\widehat{\Sigma}_{k^{\prime}}} \! - \!
C^{\widehat{\Sigma}_{k^{\prime}}}_{\widehat{w}^{\widehat{\Sigma}_{k^{
\prime}}}})^{-1} \! \in \! \mathscr{N}(\widehat{\Sigma}_{k^{\prime}})
\}_{k \in \{A,B\}} \! \Rightarrow \! \{(\mathbf{1}_{\Sigma_{k^{\prime}
}} \! - \! C^{\Sigma_{k^{\prime}}}_{w^{\Sigma_{k^{\prime}}}})^{-1
} \! \in \! \mathscr{N}(\Sigma_{k^{\prime}})\}_{k \in \{A,B\}} \!
\Rightarrow \! (\mathbf{1}_{\Sigma^{\sharp}} \! - \! C^{\Sigma^{
\sharp}}_{w^{\Sigma^{\sharp}}})^{-1} \! \in \! \mathscr{N}(\Sigma^{
\sharp}) \! \Rightarrow \! (\mathbf{1} \! - \! C_{w^{\prime}})^{-1}
\! \in \! \mathscr{N}(\Sigma^{\prime}) \! \Rightarrow \! (\mathbf{1}
\! - \! C_{w^{\sharp}})^{-1} \! \in \! \mathscr{N}(\Sigma^{\prime})$;
thus, in order to prove the basic bound $\vert \vert (\mathbf{1} \!
- \! C_{w^{\prime}})^{-1} \vert \vert_{\mathscr{N}(\Sigma^{\prime})}
\! < \! \infty$, one must show that $\vert \vert (\mathbf{1}_{\Sigma_{k}}
\! - \! C^{\Sigma_{k}}_{w^{\Sigma_{k}}})^{-1} \vert \vert_{\mathscr{N}
(\Sigma_{k})} \! < \! \infty$, $k \! \in \! \{A,B\}$: this is the
programme of the following Lemma.
\begin{eeeee}
Essentially, Lemma~5.4 (see below) was proved in \cite{a27}
(see~Proposition~3.109, pp.~340--346); however, it is succinctly reworked
here because a very important model RHP which arises in it, and which is
solved asymptotically in Section~6 and is essential for the proof of
Lemma~6.1 (see below), is necessary in order to obtain the results stated
in Theorem~3.1.
\end{eeeee}
\begin{ccccc}
As $t \! \to \! +\infty$ such that $0 \! < \! \zeta_{2} \! < \! \tfrac{
1}{M} \! < \! M \! < \! \zeta_{1}$ and $\vert \zeta_{3} \vert^{2} \! =
\! 1$, with $M \! \in \! \mathbb{R}_{>1}$ and bounded, $(\mathbf{1}_{
\Sigma_{k}} \! - \! C^{\Sigma_{k}}_{w^{\Sigma_{k}}})^{-1} \! \in \!
\mathscr{N}(\Sigma_{k})$, $k \! \in \! \{A,B\}$.
\end{ccccc}

\emph{Proof.} Without loss of generality, the case $k \! = \! B$ is
considered: the case $k \! = \! A$ follows in an analogous manner.
{}From the bounds given in Lemma~5.1, it is to be understood (for
the purposes of this proof) that $\Sigma_{B}$-related quantities
encapsulate the $\tfrac{\vert c^{\mathcal{S}}(\zeta_{1}) \vert \vert
\underline{c}(\zeta_{2},\zeta_{3},\overline{\zeta_{3}}) \vert}{\vert
\zeta_{1}-\zeta_{3} \vert \sqrt{(\zeta_{1}-\zeta_{2})}} \tfrac{\ln
t}{\sqrt{t}}$ error terms (since $\exp (-\tfrac{1}{2} \gamma \zeta_{
1}^{-3}(\zeta_{1} \! - \! \zeta_{2})^{3} \vert \zeta_{1} \! - \!
\zeta_{3} \vert^{2}v_{B}^{2}t) \! \leqslant \! 1$, $\gamma \! \in \!
(0,\tfrac{1}{2})$, $v_{B} \! \in \! (-\infty,\widetilde{\varepsilon}
))$; therefore, by $\Sigma_{B^{0}}$-related quantities, one denotes
the ``master'', or leading, terms of the $\Sigma_{B}$-related quantities,
such that, symbolically, $\vert \vert (\Sigma_{B} \! \! - \! \! \mathrm{
related}) \! - \! (\Sigma_{B^{0}} \! \! - \! \! \mathrm{related}) \vert
\vert_{\cap_{p \in \{1,2,\infty\}} \mathcal{L}^{p}(\Sigma_{B})} \!
\leqslant \! \tfrac{\vert c^{\mathcal{S}}(\zeta_{1}) \vert \vert
\underline{c}(\zeta_{2},\zeta_{3},\overline{\zeta_{3}}) \vert}{
\vert \zeta_{1}-\zeta_{3} \vert \sqrt{(\zeta_{1}-\zeta_{2})}} \tfrac{
\ln t}{\sqrt{t}}$, and associated with the master, or $\Sigma_{B^{
0}}$-related, terms, is the master, and normalised at $\infty$ $(m^{
\Sigma_{B^{0}}}(\infty) \! = \! \mathrm{I})$ RHP $m^{\Sigma_{B^{
0}}}_{+}(\widetilde{w}) \! = \! m^{\Sigma_{B^{0}}}_{-}(\widetilde{w
})(\mathrm{I} \! - \! w^{\Sigma_{B^{0}}}_{-}(\widetilde{w}))^{-1}
(\mathrm{I} \! + \! w^{\Sigma_{B^{0}}}_{+}(\widetilde{w}))$,
$\widetilde{w} \! \in \! \Sigma_{B}$ (see Figure~5(a)), where
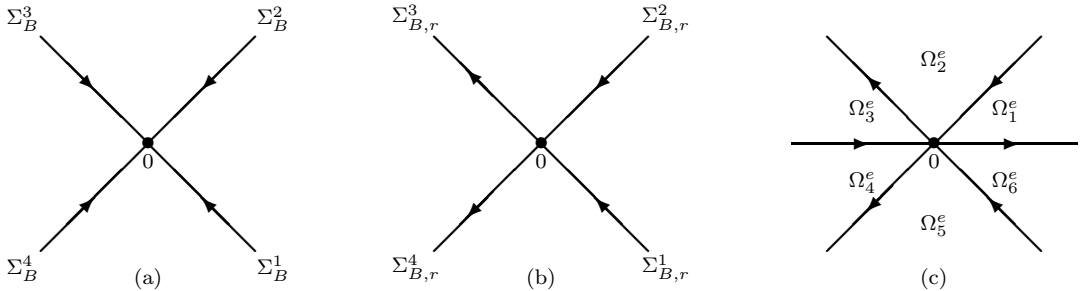
\begin{figure}[bht]
\begin{center}
\unitlength=0.95cm
\begin{picture}(15,4)(-0.325,0)
\thicklines
\put(2,2){\line(1,1){0.75}}
\put(3.5,3.5){\vector(-1,-1){0.75}}
\put(2,2){\line(-1,1){0.75}}
\put(0.5,3.5){\vector(1,-1){0.75}}
\put(2,2){\line(1,-1){0.75}}
\put(3.5,0.5){\vector(-1,1){0.75}}
\put(2,2){\line(-1,-1){0.75}}
\put(0.5,0.5){\vector(1,1){0.75}}
\put(3.75,3.75){\makebox(0,0){$\scriptstyle{}\Sigma_{B}^{2}$}}
\put(3.75,0.25){\makebox(0,0){$\scriptstyle{}\Sigma_{B}^{1}$}}
\put(0.25,3.75){\makebox(0,0){$\scriptstyle{}\Sigma_{B}^{3}$}}
\put(0.25,0.25){\makebox(0,0){$\scriptstyle{}\Sigma_{B}^{4}$}}
\put(2,1.75){\makebox(0,0){$\scriptstyle{}0$}}
\put(2,2){\makebox(0,0){$\bullet$}}
\put(2,0.10){\makebox(0,0){$\scriptstyle{}{\rm (a)}$}}
\put(7.5,2){\line(1,1){0.75}}
\put(9.0,3.5){\vector(-1,-1){0.75}}
\put(7.5,2){\line(1,-1){0.75}}
\put(9.0,0.5){\vector(-1,1){0.75}}
\put(7.5,2){\vector(-1,1){1.0425}}
\put(6,3.5){\line(1,-1){0.75}}
\put(7.5,2){\vector(-1,-1){1.0425}}
\put(6,0.5){\line(1,1){0.75}}
\put(7.5,2.0){\makebox(0,0){$\bullet$}}
\put(9.25,3.75){\makebox(0,0){$\scriptstyle{}\Sigma_{B,r}^{2}$}}
\put(9.25,0.25){\makebox(0,0){$\scriptstyle{}\Sigma_{B,r}^{1}$}}
\put(5.75,3.75){\makebox(0,0){$\scriptstyle{}\Sigma_{B,r}^{3}$}}
\put(5.75,0.25){\makebox(0,0){$\scriptstyle{}\Sigma_{B,r}^{4}$}}
\put(7.5,1.75){\makebox(0,0){$\scriptstyle{}0$}}
\put(7.5,0.10){\makebox(0,0){$\scriptstyle{}{\rm (b)}$}}
\put(11,2){\vector(1,0){1.1}}
\put(12,2){\line(1,0){1.0}}
\put(13,2){\vector(1,0){1.2}}
\put(14,2){\line(1,0){1}}
\put(13,2){\line(1,1){0.75}}
\put(14.5,3.5){\vector(-1,-1){0.75}}
\put(13,2){\line(1,-1){0.75}}
\put(14.5,0.5){\vector(-1,1){0.75}}
\put(13,2){\vector(-1,1){0.951}}
\put(11.5,3.5){\line(1,-1){0.75}}
\put(13,2){\vector(-1,-1){0.951}}
\put(11.5,0.5){\line(1,1){0.75}}
\put(13,1.75){\makebox(0,0){$\scriptstyle{}0$}}
\put(13,2){\makebox(0,0){$\bullet$}}
\put(14,2.45){\makebox(0,0){$\scriptstyle{}\Omega_{1}^{e}$}}
\put(13,3.15){\makebox(0,0){$\scriptstyle{}\Omega_{2}^{e}$}}
\put(12,2.45){\makebox(0,0){$\scriptstyle{}\Omega_{3}^{e}$}}
\put(12,1.45){\makebox(0,0){$\scriptstyle{}\Omega_{4}^{e}$}}
\put(13,0.85){\makebox(0,0){$\scriptstyle{}\Omega_{5}^{e}$}}
\put(14,1.45){\makebox(0,0){$\scriptstyle{}\Omega_{6}^{e}$}}
\put(13,0.10){\makebox(0,0){$\scriptstyle{}{\rm (c)}$}}
\end{picture}
\vspace{-0.65cm}
\end{center}
\caption{(a) $\Sigma_{B}$; (b) $\Sigma_{B,r}$; and (c)
$\Sigma_{e} := \Sigma_{B,r} \cup \mathbb{R}$.}
\end{figure}
\begin{gather*}
w^{\Sigma_{B^{0}}}_{+}(\widetilde{w}) \! = \!
\begin{cases}
(\widetilde{w})^{2 \mi \nu} \me^{-\frac{\mi}{2} \widetilde{w}^{2}}
\overline{r(\zeta_{1})} \, \sigma_{+}, &\text{$\widetilde{w} \! \in
\! \Sigma_{B}^{1}$,} \\
-(\widetilde{w})^{2 \mi \nu} \me^{-\frac{\mi}{2} \widetilde{w}^{2}}
\tfrac{\overline{r(\zeta_{1})}}{(1-\vert r(\zeta_{1}) \vert^{2})}
\sigma_{+}, &\text{$\widetilde{w} \! \in \! \Sigma_{B}^{3}$,}
\end{cases} \quad w^{\Sigma_{B^{0}}}_{-}(\widetilde{w}) \! = \!
\left(
\begin{smallmatrix}
0 & 0 \\
0 & 0
\end{smallmatrix}
\right), \, \, \, \widetilde{w} \! \in \! \Sigma_{B}^{1} \cup
\Sigma_{B}^{3}, \\
w^{\Sigma_{B^{0}}}_{-}(\widetilde{w}) \! = \!
\begin{cases}
-(\widetilde{w})^{-2 \mi \nu} \me^{\frac{\mi}{2} \widetilde{w}^{2}}
r(\zeta_{1}) \sigma_{-}, &\text{$\widetilde{w} \! \in \! \Sigma_{B}^{
2}$,} \\
(\widetilde{w})^{-2 \mi \nu} \me^{\frac{\mi}{2} \widetilde{w}^{2}}
\tfrac{r(\zeta_{1})}{(1-\vert r(\zeta_{1}) \vert^{2})} \sigma_{-},
&\text{$\widetilde{w} \! \in \! \Sigma_{B}^{4}$,}
\end{cases} \quad \, \, w^{\Sigma_{B^{0}}}_{+}(\widetilde{w}) \! = \!
\left(
\begin{smallmatrix}
0 & 0 \\
0 & 0
\end{smallmatrix}
\right), \, \, \, \widetilde{w} \! \in \! \Sigma_{B}^{2} \cup
\Sigma_{B}^{4}.
\end{gather*}
Define the functions (Proposition~5.2) $w^{\Sigma_{B}}_{\pm} \!
(\widetilde{w}) \! := \! ((\Delta_{B}^{0})^{-1}(\mathcal{N}_{B}
\widehat{w}_{\pm}^{\widehat{\Sigma}_{B^{\prime}}})(\Delta_{B}^{0})
)(\widetilde{w})$ and $w^{\Sigma_{B^{0}}}_{\pm} \! (\widetilde{w})
\! := \! ((\Delta_{B}^{0})^{-1}(\mathcal{N}_{B} \widehat{w}_{\pm}^{
\widehat{\Sigma}_{B^{0}}})(\Delta_{B}^{0}))(\widetilde{w})$, with
$\Delta_{B}^{0}$ defined by Eq.~(107) and~Proposition~5.1, and
$\mathcal{N}_{B}$ defined by Eq.~(105). Recall that $C^{\Sigma_{
B}}_{w^{\Sigma_{B}}} \! := \! C_{+}(\cdot w^{\Sigma_{B}}_{-}) \! +
\! C_{-}(\cdot w^{\Sigma_{B}}_{+})$, and set $C^{\Sigma_{B}}_{w^{
\Sigma_{B^{0}}}} \! := \! C_{+}(\cdot w^{\Sigma_{B^{0}}}_{-}) \ +
\! C_{-}(\cdot w^{\Sigma_{B^{0}}}_{+})$. Since $w^{\Sigma_{q}}(
\widetilde{w}) \! = \! \sum_{l \in \{\pm\}} \! w^{\Sigma_{q}}_{l}
(\widetilde{w})$, $q \! \in \! \{B,B^{0}\}$, it follows {}from the
above expressions for $w^{\Sigma_{B^{0}}}_{\pm}(\widetilde{w})$,
Lemma~5.1, and the linearity of $\mathcal{N}_{B}$ that, as $t \!
\to \! +\infty$ such that $0 \! < \! \zeta_{2} \! < \! \tfrac{1}
{M} \! < \! M \! < \! \zeta_{1}$ and $\vert \zeta_{3} \vert^{2}
\! = \! 1$, with $M \! \in \! \mathbb{R}_{>1}$ and bounded, $\vert
\vert (w^{\Sigma_{B}} \! - \! w^{\Sigma_{B^{0}}})(\cdot) \vert
\vert_{\cap_{p \in \{1,2,\infty\}} \mathcal{L}^{p}_{\mathrm{M}_{2}
(\mathbb{C})}(\Sigma_{B})} \! = \! \vert \vert ((\Delta_{B}^{0})^{
-1}(\mathcal{N}_{B} \widehat{w}^{\widehat{\Sigma}_{B^{\prime}}})
(\Delta_{B}^{0}) \! - \! (\Delta_{B}^{0})^{-1}(\mathcal{N}_{B}
\widehat{w}^{\widehat{\Sigma}_{B^{0}}})(\Delta_{B}^{0}))(\cdot)
\vert \vert_{\cap_{p \in \{1,2,\infty\}} \mathcal{L}^{p}_{\mathrm{
M}_{2}(\mathbb{C})}(\Sigma_{B})} \! \leqslant \! \tfrac{\vert c^{
\mathcal{S}}(\zeta_{1}) \vert \vert \underline{c}(\zeta_{2},\zeta_{
3},\overline{\zeta_{3}}) \vert}{\vert \zeta_{1}-\zeta_{3} \vert
\sqrt{(\zeta_{1}-\zeta_{2})}}
\linebreak[4]
\cdot \tfrac{\ln t}{\sqrt{t}}$; hence, {}from the linearity of the
Cauchy operators, $\vert \vert C^{\Sigma_{B}}_{w^{\Sigma_{B}}} \!
- \! C^{\Sigma_{B}}_{w^{\Sigma_{B^{0}}}} \vert \vert_{\mathscr{N}
(\Sigma_{B})} \! = \! \vert \vert C_{+}(\cdot (w^{\Sigma_{B}}_{-}
\! - \! w^{\Sigma_{B^{0}}}_{-})) \! + \! C_{-}(\cdot (w^{\Sigma_{
B}}_{+} \! - \! w^{\Sigma_{B^{0}}}_{+})) \vert \vert_{\mathscr{N}
(\Sigma_{B})} \! \leqslant \! \tfrac{\vert c^{\mathcal{S}}(\zeta_{
1}) \vert \vert \underline{c}(\zeta_{2},\zeta_{3},\overline{\zeta_{
3}}) \vert}{\vert \zeta_{1}-\zeta_{3} \vert \sqrt{(\zeta_{1}-\zeta_{
2})}} \tfrac{\ln t}{\sqrt{t}}$, and, consequently, via the second
resolvent identity, one deduces that $(\mathbf{1}_{\Sigma_{B}} \! -
\! C^{\Sigma_{B}}_{w^{\Sigma_{B^{0}}}})^{-1} \! \in \! \mathscr{N}
(\Sigma_{B}) \! \Rightarrow \! (\mathbf{1}_{\Sigma_{B}} \! - \! C^{
\Sigma_{B}}_{w^{\Sigma_{B}}})^{-1} \! \in \! \mathscr{N}(\Sigma_{B})$.
Reorient $\Sigma_{B}$ as in Figure~5(b), and denote the reoriented
contour as $\Sigma_{B,r}$. Define $w^{\Sigma_{B,r}}(\widetilde{w}) \!
:= \! \sum_{l \in \{\pm\}} \! w^{\Sigma_{B,r}}_{l}(\widetilde{w})$,
where $w^{\Sigma_{B,r}}_{\pm}(\widetilde{w}) \! = \!
\begin{cases}
w^{\Sigma_{B^{0}}}_{\pm}(\widetilde{w}), &\text{$\widetilde{w} \!
\in \! \Sigma_{B,r}^{1} \cup \Sigma_{B,r}^{2}$,} \\
-w^{\Sigma_{B^{0}}}_{\mp}(\widetilde{w}), &\text{$\widetilde{w} \!
\in \! \Sigma_{B,r}^{3} \cup \Sigma_{B,r}^{4}$,}
\end{cases}$ and consider the operator $C^{\Sigma_{B,r}}_{w^{\Sigma_{
B,r}}} \! := \! C_{+}(\cdot w^{\Sigma_{B,r}}_{-}) \! + \! C_{-}(\cdot
w^{\Sigma_{B,r}}_{+})$ $(= \! C^{\Sigma_{B}}_{w^{\Sigma_{B^{0}}}})$,
where the Cauchy operators are now taken with respect to the oriented
contour $\Sigma_{B,r}$; hence, {}from the second resolvent identity,
$(\mathbf{1}_{\Sigma_{B,r}} \! - \! C^{\Sigma_{B,r}}_{w^{\Sigma_{
B,r}}})^{-1} \! \in \! \mathscr{N}(\Sigma_{B,r}) \! \Rightarrow \!
(\mathbf{1}_{\Sigma_{B}} \! - \! C^{\Sigma_{B}}_{w^{\Sigma_{B^{0}}}
})^{-1} \! \in \! \mathscr{N}(\Sigma_{B})$. Extend $\Sigma_{B,r} \!
\to \! \Sigma_{e} \! := \! \Sigma_{B,r} \cup \mathbb{R}$, with the
orientation in Figure~5(c), and set $w^{\Sigma_{e}}_{\pm}
(\widetilde{w}) \! := \!
\begin{cases}
w^{\Sigma_{B,r}}_{\pm}(\widetilde{w}), &\text{$\widetilde{w} \!
\in \! \Sigma_{B,r} \subset \Sigma_{e}$,} \\
\left(
\begin{smallmatrix}
0 & 0 \\
0 & 0
\end{smallmatrix}
\right), &\text{$\widetilde{w} \! \in \! \Sigma_{e} \setminus
\Sigma_{B,r}$.}
\end{cases}$ Define $C^{\Sigma_{e}}_{w^{\Sigma_{e}}} \! := \! C_{+}
(\cdot w^{\Sigma_{e}}_{-}) \! + \! C_{-}(\cdot w^{\Sigma_{e}}_{+})$,
with the Cauchy operators now taken with respect to the oriented
contour $\Sigma_{e}$; hence, via the second resolvent identity, it
follows that $(\mathbf{1}_{\Sigma_{e}} \! - \! C^{\Sigma_{e}}_{w^{
\Sigma_{e}}})^{-1} \! \in \! \mathscr{N}(\Sigma_{e}) \! \Rightarrow
\! (\mathbf{1}_{\Sigma_{B,r}} \! - \! C^{\Sigma_{B,r}}_{w^{\Sigma_{
B,r}}})^{-1} \! \in \! \mathscr{N}(\Sigma_{B,r})$. Define the following
piecewise-analytic $2 \! \times \! 2$ matrix-valued function:
\begin{equation*}
\phi^{\Sigma_{B^{0}}}(\widetilde{w}) \! := \!
\begin{cases}
(\widetilde{w})^{-\mi \nu \sigma_{3}}, &\text{$\widetilde{w} \!
\in \! \Omega^{e}_{2} \cup \Omega^{e}_{5}$,} \\
(\widetilde{w})^{-\mi \nu \sigma_{3}}v^{e}_{2}(\widetilde{w}),
&\text{$\widetilde{w} \! \in \! \Omega^{e}_{1}$,} \\
(\widetilde{w})^{-\mi \nu \sigma_{3}}v^{e}_{3}(\widetilde{w}),
&\text{$\widetilde{w} \! \in \! \Omega^{e}_{3}$,} \\
(\widetilde{w})^{-\mi \nu \sigma_{3}}(v^{e}_{1}(\widetilde{w}))^{-1},
&\text{$\widetilde{w} \! \in \! \Omega^{e}_{6}$,} \\
(\widetilde{w})^{-\mi \nu \sigma_{3}}(v^{e}_{4}(\widetilde{w}))^{-1},
&\text{$\widetilde{w} \! \in \! \Omega^{e}_{4}$,}
\end{cases}
\end{equation*}
where
\begin{align*}
v^{e}_{1}(\widetilde{w}) \! &= \! (\widetilde{w})^{\mi \nu \mathrm{
ad}(\sigma_{3})} \exp (-\tfrac{\mi}{4} \widetilde{w}^{2} \mathrm{ad}
(\sigma_{3}))(\mathrm{I} \! + \! \overline{r(\zeta_{1})} \, \sigma_{+}), \\
v^{e}_{2}(\widetilde{w}) \! &= \! (\widetilde{w})^{\mi \nu \mathrm{
ad}(\sigma_{3})} \exp (-\tfrac{\mi}{4} \widetilde{w}^{2} \mathrm{ad}
(\sigma_{3}))(\mathrm{I} \! - \! r(\zeta_{1}) \sigma_{-}), \\
v^{e}_{3}(\widetilde{w}) \! &= \! (\widetilde{w})^{\mi \nu \mathrm{
ad} (\sigma_{3})} \exp (-\tfrac{\mi}{4} \widetilde{w}^{2} \mathrm{ad}
(\sigma_{3}))(\mathrm{I} \! + \! \tfrac{\overline{r(\zeta_{1})}}{(1-\vert
r(\zeta_{1}) \vert^{2})} \sigma_{+}), \\
v^{e}_{4}(\widetilde{w}) \! &= \! (\widetilde{w})^{\mi \nu \mathrm{
ad}(\sigma_{3})} \exp (-\tfrac{\mi}{4} \widetilde{w}^{2} \mathrm{ad}
(\sigma_{3}))(\mathrm{I} \! - \! \tfrac{r(\zeta_{1})}{(1-\vert r(\zeta_{1})
\vert^{2})} \sigma_{-}),
\end{align*}
with
\begin{equation*}
v^{e}_{1}(\widetilde{w}) \! - \! \mathrm{I} \! \underset{\underset{
\widetilde{w} \in \Omega^{e}_{6}}{\widetilde{w} \to \infty}}{=} \!
o(1), \quad \, v^{e}_{2}(\widetilde{w}) \! - \! \mathrm{I} \! \underset{
\underset{\widetilde{w} \in \Omega^{e}_{1}}{\widetilde{w} \to
\infty}}{=} \! o(1), \quad \, v^{e}_{3}(\widetilde{w}) \! - \! \mathrm{
I} \! \underset{\underset{\widetilde{w} \in \Omega^{e}_{3}}{
\widetilde{w} \to \infty}}{=} \! o(1), \quad \, v^{e}_{4}(\widetilde{
w}) \! - \! \mathrm{I} \! \underset{\underset{\widetilde{w} \in
\Omega^{e}_{4}}{\widetilde{w} \to \infty}}{=} \! o(1).
\end{equation*}
For $\widetilde{w} \! \in \! \Sigma_{e}$, set $\mathscr{V}^{e,\phi}
(\widetilde{w}) \! := \! \phi^{\Sigma_{B^{0}}}_{-}(\widetilde{w})
(\mathrm{I} \! - \! w^{\Sigma_{e}}_{-}(\widetilde{w}))^{-1}(\mathrm{
I} \! + \! w^{\Sigma_{e}}_{+}(\widetilde{w}))(\phi^{\Sigma_{B^{
0}}}_{+}(\widetilde{w}))^{-1}$, where $\phi^{\Sigma_{B^{0}}}_{
\pm}(\widetilde{w}) \! := \! \lim_{\genfrac{}{}{0pt}{2}{\widetilde{
w}^{\prime} \, \to \, \widetilde{w}}{\widetilde{w}^{\prime} \, \in \,
\pm \, \mathrm{side} \, \mathrm{of} \, \Sigma_{e}}}\phi^{\Sigma_{
B^{0}}}(\widetilde{w}^{\prime})$ denote the non-tangential limits of
$\phi^{\Sigma_{B^{0}}}(\widetilde{w})$ as $\widetilde{w}$ approaches
$\Sigma_{e}$ {}from the ``$\pm$'' sides, respectively. {}From the definition
of $w^{\Sigma_{e}}_{\pm}(\widetilde{w})$ given above, for $\widetilde{
w} \! \in \! \Sigma_{B,r}$ $(\subset \Sigma_{e})$, $\mathscr{V}^{e,
\phi}(\widetilde{w}) \! = \! \mathrm{I}$, and, for $\widetilde{w} \!
\in \! \Sigma_{e} \setminus \Sigma_{B,r}$ $(= \! \mathbb{R})$, $\mathscr{
V}^{e,\phi}(\widetilde{w}) \! = \! \phi^{\Sigma_{B^{0}}}_{-}(\widetilde{
w})(\phi^{\Sigma_{B^{0}}}_{+}(\widetilde{w}))^{-1}$; hence, {}from the
definition of $\phi^{\Sigma_{B^{0}}}(\widetilde{w})$ and the formulae
for $v^{e}_{k}(\widetilde{w})$, $k \! \in \! \{1,2,3,4\}$, given above,
taking the principal branch of the logarithm for $\widetilde{w} \! < \!
0$, one shows that $\mathscr{V}^{e,\phi}(\widetilde{w}) \! = \!
\begin{cases}
\me^{-\frac{\mi}{4} \widetilde{w}^{2} \mathrm{ad}(\sigma_{3})}V
(\zeta_{1}), &\text{$\widetilde{w} \! \in \! \mathbb{R}_{+}$,} \\
\me^{-\frac{\mi}{4} \widetilde{w}^{2} \mathrm{ad}(\sigma_{3})}V
(\zeta_{1}), &\text{$-\widetilde{w} \! \in \! \mathbb{R}_{+}$,}
\end{cases}$ where $V(\zeta_{1}) \! := \!
\left(
\begin{smallmatrix}
1-\vert r(\zeta_{1}) \vert^{2} & -\overline{r(\zeta_{1})} \\
r(\zeta_{1}) & 1
\end{smallmatrix}
\right)$, with $\det (V(\zeta_{1})) \! = \! 1$ and $\mathrm{tr}(V
(\zeta_{1})) \! = \! 2 \! - \! \vert r(\zeta_{1}) \vert^{2} \! > \!
0$ (since $\vert \vert r(\cdot) \vert \vert_{\mathcal{L}^{\infty}
(\mathbb{R})} \! < \! 1)$. Hence, the jump matrix, $\mathscr{V}^{e,
\phi}(\widetilde{w})$, is characterised as follows: $\mathscr{V}^{
e,\phi}(\widetilde{w}) \! = \!
\begin{cases}
\mathrm{I}, &\text{$\widetilde{w} \! \in \! \Sigma_{B,r}$,} \\
\me^{-\frac{\mi}{4} \widetilde{w}^{2} \mathrm{ad}(\sigma_{3})}
\! \left(
\begin{smallmatrix}
1-\vert r(\zeta_{1}) \vert^{2} & -\overline{r(\zeta_{1})} \\
r(\zeta_{1}) & 1
\end{smallmatrix}
\right), &\text{$\widetilde{w} \! \in \! \Sigma_{e} \setminus
\Sigma_{B,r}$.}
\end{cases}$ On $\mathbb{R}$, one has that $\mathscr{V}^{e,\phi}
(\widetilde{w}) \! = \! (\mathrm{I}-\overline{r(\zeta_{1})} \,
\me^{-\frac{\mi}{2} \widetilde{w}^{2}} \linebreak[4]
\cdot \sigma_{+})(\mathrm{I} \! + \! r(\zeta_{1}) \me^{\frac{\mi}
{2} \widetilde{w}^{2}} \sigma_{-}) \! := \! (\mathrm{I} \! - \!
w^{e,\phi}_{-}(\widetilde{w}))^{-1}(\mathrm{I} \! + \! w^{e,\phi}_{
+}(\widetilde{w}))$. Let $C^{\Sigma_{e}}_{w^{e,\phi}} \! := \! C_{+
}(\cdot w^{e,\phi}_{-}) \! + \! C_{-}(\cdot w^{e,\phi}_{+})$ be the
associated operator on $\Sigma_{e}$, with $w^{e,\phi}(\widetilde{w})
\! := \! \sum_{l \in \{\pm\}} \! w^{e,\phi}_{l}(\widetilde{w})$, and
the orientation is that of $\Sigma_{e}$: the boundedness of $\vert
\vert (\mathrm{1}_{\Sigma_{e}} \! - \! C^{\Sigma_{e}}_{w^{\Sigma_{e}
}})^{-1} \vert \vert_{\mathscr{N}(\Sigma_{e})}$ follows {}from the
boundedness of $\vert \vert (\mathbf{1}_{\Sigma_{e} \upharpoonright
\mathbb{R}} \! - \! C^{\Sigma_{e} \upharpoonright \mathbb{R}}_{w^{
e,\phi} \upharpoonright \mathbb{R}})^{-1} \vert \vert_{\mathscr{N}
(\Sigma_{e} \upharpoonright \mathbb{R})}$, that is, $(\mathbf{
1}_{\Sigma_{e} \upharpoonright \mathbb{R}} \! - \! C^{\Sigma_{
e} \upharpoonright \mathbb{R}}_{w^{e,\phi} \upharpoonright
\mathbb{R}})^{-1} \! \in \! \mathscr{N}(\Sigma_{e} \! \!
\upharpoonright \mathbb{R}) \! \Rightarrow \! (\mathrm{1}_{
\Sigma_{e}} \! - \! C^{\Sigma_{e}}_{w^{\Sigma_{e}}})^{-1} \! \in
\! \mathscr{N}(\Sigma_{e})$, where $C^{\Sigma_{e} \upharpoonright
\mathbb{R}}_{w^{e,\phi} \upharpoonright \mathbb{R}} \colon \mathcal{
L}^{2}_{\mathrm{M}_{2}(\mathbb{C})}(\mathbb{R}) \! \to \! \mathcal{
L}^{2}_{\mathrm{M}_{2}(\mathbb{C})}(\mathbb{R})$ is the operator
associated with the restriction of $w^{e,\phi}(\widetilde{w})$
to $\mathbb{R}$. Now, $\vert \vert C^{\Sigma_{e} \upharpoonright
\mathbb{R}}_{w^{e,\phi} \upharpoonright \mathbb{R}} \vert \vert_{
\mathscr{N}(\mathbb{R})} \! \leqslant \! \max_{1 \leqslant i,j
\leqslant 2} \sup_{\widetilde{w} \in \mathbb{R}} \vert w^{e,
\phi}_{+}(\widetilde{w}) \! + \! w^{e,\phi}_{-}(\widetilde{w})
\vert \! \leqslant \! \sup_{\widetilde{w} \in \mathbb{R}} \vert
\me^{-\frac{\mi}{2} \widetilde{w}^{2}} \overline{r(\zeta_{1})}
\vert$ $(= \! \sup_{\widetilde{w} \in \mathbb{R}} \vert \me^{
\frac{\mi}{2} \widetilde{w}^{2}}r(\zeta_{1}) \vert)$ $\leqslant
\! \vert \vert r(\cdot) \vert \vert_{\mathcal{L}^{\infty}(\mathbb{
R})} \! < \! 1$; hence, by the second resolvent identity, $\vert
\vert (\mathbf{1}_{\Sigma_{e} \upharpoonright \mathbb{R}} \! - \!
C^{\Sigma_{e} \upharpoonright \mathbb{R}}_{w^{e,\phi} \upharpoonright
\mathbb{R}})^{-1} \vert \vert_{\mathscr{N}(\Sigma_{e} \upharpoonright
\mathbb{R})} \! \leqslant \! \vert \underline{c}(\zeta_{1},\zeta_{2},
\zeta_{3},\overline{\zeta_{3}}) \vert (1 \! - \! \vert \vert r(\cdot)
\vert \vert_{\mathcal{L}^{\infty}(\mathbb{R})})^{-1} \! < \! \infty$,
whence, $(\mathbf{1}_{\Sigma_{e}} \! - \! C^{\Sigma_{e}}_{w^{\Sigma_{
e}}})^{-1} \! \in \! \mathscr{N}(\Sigma_{e})$: this completes the
proof. \hfill $\square$
\section{Asymptotic Solution of the Model RHP}
In this section the $\mathcal{O}(1)$ asymptotics of $\mu^{\Sigma^{\sharp}}
(\zeta) \! := \! ((\mathbf{1}_{\Sigma^{\sharp}} \! - \! C^{\Sigma^{\sharp}
}_{w^{\Sigma^{\sharp}}})^{-1} \mathrm{I})(\zeta)$ (Eq.~(103)), having an
explicit representation in terms of parabolic cylinder functions, is
obtained, and the RHP for $m^{c}(\zeta)$ formulated in Lemma~2.6 is solved
asymptotically, whence, the leading-order asymptotics of $u(x,t)$ and related
integrals are derived.
\begin{ccccc}
Let $\varepsilon$ be an arbitrarily fixed, sufficiently small positive
real number, and, for $\lambda \! \in \! \{\zeta_{2},\zeta_{1}\}$, set
$\mathbb{U}(\lambda;\varepsilon) \! := \! \{ \mathstrut z; \, \vert
z \! - \! \lambda \vert \! < \! \varepsilon\}$. Then, as $t \! \to
\! +\infty$ such that $0 \! < \! \zeta_{2} \! < \! \tfrac{1}{M} \!
< \! M \! < \! \zeta_{1}$ and $\vert \zeta_{3} \vert^{2} \! = \! 1$,
with $M \! \in \! \mathbb{R}_{>1}$ and bounded, for $\zeta \! \in \!
\mathbb{C} \setminus \cup_{\lambda \, \in \, \{\zeta_{2},\zeta_{1}\}
} \mathbb{U}(\lambda;\varepsilon)$, $m^{c}(\zeta)$ has the following
asymptotics,
\begin{align*}
m^{c}_{11}(\zeta) \! &= \! \delta (\zeta) \! \left( 1 \! + \! \mathcal{O} \!
\left( \! \left( \dfrac{c^{\mathcal{S}}(\zeta_{1}) \underline{c}(\zeta_{
2},\zeta_{3},\overline{\zeta_{3}})}{\sqrt{\zeta_{2}(z_{o}^{2} \! + \!
32)} \, (\zeta \! - \! \zeta_{1})} \! + \! \dfrac{c^{\mathcal{S}}(\zeta_{
2}) \underline{c}(\zeta_{1},\zeta_{3},\overline{\zeta_{3}})}{\sqrt{
\zeta_{1}(z_{o}^{2} \! + \! 32)} \, (\zeta \! - \! \zeta_{2})} \right) \!
\dfrac{\ln t}{(\zeta_{1} \! - \! \zeta_{2}) t} \right) \! \right), \\
m^{c}_{12}(\zeta) \! &= \! \dfrac{\me^{\frac{\mi \Omega^{+}(0)}{
2}}}{\delta (\zeta)} \! \left( \dfrac{\sqrt{\nu (\zeta_{1})} \, \zeta_{1
}^{2 \mi \nu (\zeta_{1})}}{\sqrt{t(\zeta_{1} \! - \! \zeta_{2})} \,
(z_{o}^{2} \! + \! 32)^{1/4}} \! \left( \dfrac{\zeta_{1} \me^{-\mi
(\Theta^{+}(z_{o},t)+\frac{\pi}{4})}}{(\zeta \! - \! \zeta_{1})} \! + \!
\dfrac{\zeta_{2} \me^{\mi (\Theta^{+}(z_{o},t)+\frac{\pi}{4})}}{
(\zeta \! - \! \zeta_{2})} \! \right) \right. \\
 &+ \left. \! \mathcal{O} \! \left( \! \left( \dfrac{c^{\mathcal{S}}
(\zeta_{1}) \underline{c}(\zeta_{2},\zeta_{3},\overline{\zeta_{
3}})}{\sqrt{\zeta_{2}(z_{o}^{2} \! + \! 32)} \, (\zeta \! - \! \zeta_{1})}
\! + \! \dfrac{c^{\mathcal{S}}(\zeta_{2}) \underline{c}(\zeta_{1},
\zeta_{3},\overline{\zeta_{3}})}{\sqrt{\zeta_{1}(z_{o}^{2} \! + \!
32)} \, (\zeta \! - \! \zeta_{2})} \right) \! \dfrac{\ln t}{(\zeta_{1} \! - \!
\zeta_{2}) t} \right) \! \right), \\
m^{c}_{21}(\zeta) \! &= \! \dfrac{\delta (\zeta)}{\me^{\frac{\mi
\Omega^{+}(0)}{2}}} \! \left( \dfrac{\sqrt{\nu (\zeta_{1})} \, \zeta_{
1}^{-2 \mi \nu (\zeta_{1})}}{\sqrt{t(\zeta_{1} \! - \! \zeta_{2})} \, (z_{
o}^{2} \! + \! 32)^{1/4}} \! \left( \dfrac{\zeta_{1} \me^{\mi (\Theta^{+}
(z_{o},t)+\frac{\pi}{4})}}{(\zeta \! - \! \zeta_{1})} \! + \! \dfrac{\zeta_{2}
\me^{-\mi (\Theta^{+}(z_{o},t)+\frac{\pi}{4})}}{(\zeta \! - \! \zeta_{2})}
\! \right) \right. \\
 &+ \left. \! \mathcal{O} \! \left( \! \left( \dfrac{c^{\mathcal{S}}
(\zeta_{1}) \underline{c}(\zeta_{2},\zeta_{3},\overline{\zeta_{
3}})}{\sqrt{\zeta_{2}(z_{o}^{2} \! + \! 32)} \, (\zeta \! - \! \zeta_{1})} \!
+ \! \dfrac{c^{\mathcal{S}}(\zeta_{2}) \underline{c}(\zeta_{1},\zeta_{
3},\overline{\zeta_{3}})}{\sqrt{\zeta_{1}(z_{o}^{2} \! + \! 32)} \, (\zeta
\! - \! \zeta_{2})} \right) \! \dfrac{\ln t}{(\zeta_{1} \! - \! \zeta_{2}) t}
\right) \! \right),
\end{align*}
\begin{align*}
m^{c}_{22}(\zeta) \! &= \! \dfrac{1}{\delta (\zeta)} \! \left( 1 \! + \!
\mathcal{O} \! \left( \! \left( \dfrac{c^{\mathcal{S}}(\zeta_{1})
\underline{c}(\zeta_{2},\zeta_{3},\overline{\zeta_{3}})}{\sqrt{
\zeta_{2}(z_{o}^{2} \! + \! 32)} \, (\zeta \! - \! \zeta_{1})} \! + \! \dfrac{
c^{\mathcal{S}}(\zeta_{2}) \underline{c}(\zeta_{1},\zeta_{3},
\overline{\zeta_{3}})}{\sqrt{\zeta_{1}(z_{o}^{2} \! + \! 32)} \, (\zeta
\! - \! \zeta_{2})} \right) \! \dfrac{\ln t}{(\zeta_{1} \! - \! \zeta_{2}) t}
\right) \! \right),
\end{align*}
where $\delta (\zeta)$ is given in Proposition~{\rm 4.1}, $\nu (z)$, $\Theta^{
+}(z_{o},t)$, $\Omega^{+}(z)$, and $\{\zeta_{i}\}_{i=1}^{3}$ are defined in
Theorem~{\rm 3.1}, Eqs.~{\rm (12)}--{\rm (14)}, {\rm (16)}, and~{\rm (17)},
$\sup_{\zeta \in \mathbb{C} \, \setminus \cup_{\lambda \in \{\zeta_{2},\zeta_{
1}\}} \mathbb{U}(\lambda;\varepsilon)} \vert (\zeta \! - \! \zeta_{k})^{-1}
\vert \! \leqslant \! M^{c}$, with $M^{c} \! \in \! \mathbb{R}_{+}$ (and
bounded), $k \! \in \! \{1,2\}$, $m^{c}(\zeta) \! = \! \sigma_{1} \overline{
m^{c}(\overline{\zeta})} \, \sigma_{1}$, and $(m^{c}(0) \sigma_{2})^{2} \! =
\! \mathrm{I}$.
\end{ccccc}

\emph{Proof}. {}From Lemma~5.3, the fact that $w^{\Sigma_{k^{\prime}}}_{\pm}
(\zeta) \! := \! w^{\Sigma^{\sharp}}_{\pm}(\zeta) \! \! \! \upharpoonright_{
\Sigma_{k^{\prime}}}$, $k \! \in \! \{A,B\}$, $C^{\Sigma^{\sharp}}_{w^{
\Sigma^{\sharp}}} \! := \! \sum_{k \in \{A,B\}} \! C^{\Sigma_{k^{\prime}}}_{
w^{\Sigma_{k^{\prime}}}}$, and Lemma~4.6, one shows that, as $t \! \to \!
+\infty$ such that $0 \! < \! \zeta_{2} \! < \! \tfrac{1}{M} \! < \! M \! <
\! \zeta_{1}$ and $\vert \zeta_{3} \vert^{2} \! = \! 1$, with $M \! \in \!
\mathbb{R}_{>1}$ and bounded, for $\zeta \! \in \! \mathbb{C} \setminus
\Sigma_{A^{\prime}} \cup \Sigma_{B^{\prime}}$,
\begin{align*}
m^{\Sigma^{\sharp}}_{11}(\zeta) &= \! 1 \! - \! \left( \int\nolimits_{
\zeta_{1}+\varepsilon \me^{-\frac{3 \pi \mi}{4}}}^{\zeta_{1}+0^{-} \me^{-
\frac{3 \pi \mi}{4}}}+\int\nolimits_{\zeta_{1}+\infty \me^{\frac{\mi \pi}
{4}}}^{\zeta_{1}+0^{+} \me^{\frac{\mi \pi}{4}}} \right) \dfrac{\mu^{\Sigma_{
B^{\prime}}}_{12}(\xi) \overline{\mathcal{R}(\xi)}(\delta (\xi))^{-2} \me^{
2 \mi t \theta^{u}(\xi)}}{(\xi \! - \! \zeta)} \, \dfrac{\md \xi}{2 \pi \mi}
\\
&- \left(\int\nolimits_{\zeta_{2}+0^{+} \me^{-\frac{\mi \pi}{4}}}^{\zeta_{2}
+\varepsilon \me^{-\frac{\mi \pi}{4}}}+\int\nolimits_{\zeta_{2}+0^{-} \me^{
\frac{3 \pi \mi}{4}}}^{\zeta_{2}+\varepsilon \me^{\frac{3 \pi \mi}{4}}}
\right) \dfrac{\mu^{\Sigma_{A^{\prime}}}_{12}(\xi) \overline{\mathcal{R}
(\xi)}(\delta (\xi))^{-2} \me^{2 \mi t \theta^{u}(\xi)}}{(\xi \! - \! \zeta)}
\, \dfrac{\md \xi}{2 \pi \mi} + \mathcal{E}^{\Sigma^{\sharp}}_{11}(\zeta),
\\
m^{\Sigma^{\sharp}}_{12}(\zeta) &= \! \left( \int\nolimits_{\zeta_{1}+\infty
\me^{-\frac{\mi \pi}{4}}}^{\zeta_{1}+0^{+} \me^{-\frac{\mi \pi}{4}}}+
\int\nolimits_{\zeta_{1}+\varepsilon \me^{\frac{3 \pi \mi}{4}}}^{\zeta_{1}+
0^{-} \me^{\frac{3 \pi \mi}{4}}} \right) \dfrac{\mu^{\Sigma_{B^{\prime}}}_{
11}(\xi) \mathcal{R}(\xi)(\delta (\xi))^{2} \me^{-2 \mi t \theta^{u}(\xi)}}
{(\xi \! - \! \zeta)} \, \dfrac{\md \xi}{2 \pi \mi} \\
 &+ \left(\int\nolimits_{\zeta_{2}+0^{-} \me^{-\frac{3 \pi \mi}{4}}}^{\zeta_{
2}+\varepsilon \me^{-\frac{3 \pi \mi}{4}}}+\int\nolimits_{\zeta_{2}+0^{+}
\me^{\frac{\mi \pi}{4}}}^{\zeta_{2}+\varepsilon \me^{\frac{\mi \pi}{4}}}
\right) \dfrac{\mu^{\Sigma_{A^{\prime}}}_{11}(\xi) \mathcal{R}(\xi)(\delta
(\xi))^{2} \me^{-2 \mi t \theta^{u}(\xi)}}{(\xi \! - \! \zeta)} \, \dfrac{
\md \xi}{2 \pi \mi}+\mathcal{E}^{\Sigma^{\sharp}}_{12}(\zeta), \\
m^{\Sigma^{\sharp}}_{21}(\zeta) &= \! - \left(\int\nolimits_{\zeta_{1}+
\varepsilon \me^{-\frac{3 \pi \mi}{4}}}^{\zeta_{1}+0^{-} \me^{-\frac{3 \pi
\mi}{4}}}+\int\nolimits_{\zeta_{1}+\infty \me^{\frac{\mi \pi}{4}}}^{\zeta_{
1}+0^{+} \me^{\frac{\mi \pi}{4}}} \right) \dfrac{\mu^{\Sigma_{B^{\prime}}}_{
22}(\xi) \overline{\mathcal{R}(\xi)}(\delta (\xi))^{-2} \me^{2 \mi t
\theta^{u}(\xi)}}{(\xi \! - \! \zeta)} \, \dfrac{\md \xi}{2 \pi \mi} \\
&- \left(\int\nolimits_{\zeta_{2}+0^{+} \me^{-\frac{\mi \pi}{4}}}^{\zeta_{2}
+\varepsilon \me^{-\frac{\mi \pi}{4}}}+\int\nolimits_{\zeta_{2}+0^{-} \me^{
\frac{3 \pi \mi}{4}}}^{\zeta_{2}+\varepsilon \me^{\frac{3 \pi \mi}{4}}}
\right) \dfrac{\mu^{\Sigma_{A^{\prime}}}_{22}(\xi) \overline{\mathcal{R}
(\xi)}(\delta (\xi))^{-2} \me^{2 \mi t \theta^{u}(\xi)}}{(\xi \! - \!
\zeta)} \, \dfrac{\md \xi}{2 \pi \mi}+\mathcal{E}^{\Sigma^{\sharp}}_{21}
(\zeta), \\
m^{\Sigma^{\sharp}}_{22}(\zeta) &= \! 1 \! + \! \left( \int\nolimits_{\zeta_{
1}+\infty \me^{-\frac{\mi \pi}{4}}}^{\zeta_{1}+0^{+} \me^{-\frac{\mi \pi}{4}}}
+\int\nolimits_{\zeta_{1}+\varepsilon \me^{\frac{3 \pi \mi}{4}}}^{\zeta_{1}+
0^{-} \me^{\frac{3 \pi \mi}{4}}} \right) \dfrac{\mu^{\Sigma_{B^{\prime}}}_{
21}(\xi) \mathcal{R}(\xi)(\delta (\xi))^{2} \me^{-2 \mi t \theta^{u}
(\xi)}}{(\xi \! - \! \zeta)} \, \dfrac{\md \xi}{2 \pi \mi} \\
&+ \left( \int\nolimits_{\zeta_{2}+0^{-} \me^{-\frac{3 \pi \mi}{4}}}^{\zeta_{
2}+\varepsilon \me^{-\frac{3 \pi \mi}{4}}}+\int\nolimits_{\zeta_{2}+0^{+}
\me^{\frac{\mi \pi}{4}}}^{\zeta_{2}+\varepsilon \me^{\frac{\mi \pi}{4}}}
\right) \dfrac{\mu^{\Sigma_{A^{\prime}}}_{21}(\xi) \mathcal{R}(\xi)(\delta
(\xi))^{2} \me^{-2 \mi t \theta^{u}(\xi)}}{(\xi \! - \! \zeta)} \, \dfrac{
\md \xi}{2 \pi \mi}+\mathcal{E}^{\Sigma^{\sharp}}_{22}(\zeta),
\end{align*}
where $\mu^{\Sigma_{l}}(\zeta) \! := \! ((\mathbf{1}_{\Sigma_{l}} \! - \! C^{
\Sigma_{l}}_{w^{\Sigma_{l}}})^{-1} \mathrm{I})(\zeta)$, $l \! \in \! \{A,
B\}$, $\varepsilon$ is an arbitrarily fixed, sufficiently small positive real
number, and $\mathcal{E}_{ij}^{\Sigma^{\sharp}}(\zeta) \! := \! \mathcal{O}
\! \left(\tfrac{\underline{c}(\zeta_{1},\zeta_{2},\zeta_{3},\overline{\zeta_{
3}})f_{ij}^{\Sigma^{\sharp}}(\zeta)}{(\zeta_{1}-\zeta_{2})^{2} \vert z_{o}+
\zeta_{1}+\zeta_{2} \vert t} \right)$, with $\vert \vert f_{ij}^{\Sigma^{
\sharp}}(\cdot) \vert \vert_{\mathcal{L}^{\infty}(\mathbb{C} \setminus
\Sigma_{A^{\prime}} \cup \Sigma_{B^{\prime}})} \! < \! \infty$. In order to
proceed, explicit expressions for the $\mathcal{O}(1)$ asymptotics of $\mu^{
\Sigma_{k^{\prime}}}(\zeta)$, $k \! \in \! \{A,B\}$, are needed: without loss
of generality, $\mu^{\Sigma_{B^{\prime}}}(\zeta)$, associated with $\zeta_{
1}$, is considered in detail (an analogous argument applies for $\mu^{\Sigma_{
A^{\prime}}}(\zeta)$, associated with $\zeta_{2})$. {}From the proof of
Lemma~5.4, introduce the $2 \times 2$ matrix-valued function $\mathcal{D}^{
\Sigma_{B^{0}}}(\widetilde{w}) \! := \! m^{\Sigma_{B^{0}}}(\widetilde{w})
(\phi^{\Sigma_{B^{0}}}(\widetilde{w}))^{-1} \exp (-\frac{\mi}{4} \widetilde{
w}^{2} \sigma_{3})$, $\widetilde{w} \! \in \! \mathbb{C} \setminus \Sigma_{
e}$, and note that, for $\widetilde{w} \! \in \! \Sigma_{B}$ $(= \! \Sigma_{
e} \setminus \mathbb{R}$), $\mathcal{D}^{\Sigma_{B^{0}}}_{+}(\widetilde{w})
\! = \! \mathcal{D}^{\Sigma_{B^{0}}}_{-}(\widetilde{w})$, and, for
$\widetilde{w} \! \in \! \mathbb{R}$ $(= \! \Sigma_{e} \setminus \Sigma_{
B})$, oriented {}from $-\infty$ to $+\infty$, $\mathcal{D}^{\Sigma_{B^{0}}}_{
+}(\widetilde{w}) \! = \! \mathcal{D}^{\Sigma_{B^{0}}}_{-}(\widetilde{w})
V(\zeta_{1})$, where $V(\zeta_{1}) \! := \!
\left(
\begin{smallmatrix}
1-\vert r(\zeta_{1}) \vert^{2} & -\overline{r(\zeta_{1})} \\
r(\zeta_{1}) & 1
\end{smallmatrix}
\right)$, with (cf.~the asymptotics for $v^{e}_{k}(\widetilde{w})$, $k
\! \in \! \{1,2,3,4\}$, given in the proof of Lemma~5.4) asymptotics
$\mathcal{D}^{\Sigma_{B^{0}}}(\widetilde{w}) \! =_{\genfrac{}{}
{0pt}{2}{\widetilde{w} \to \infty}{\widetilde{w} \in \mathbb{C}
\setminus \mathbb{R}}} \! (\mathrm{I} \! - \! \tfrac{1}{\widetilde{w}}
m^{\Sigma_{B^{0}}}_{1} \! + \! \mathcal{O}(\tfrac{1}{\widetilde{
w}^{2}}))(\widetilde{w})^{\mi \nu \sigma_{3}} \exp (-\tfrac{\mi}{4}
\widetilde{w}^{2} \sigma_{3})$, and $\mathcal{D}^{\Sigma_{B^{0}}
}(\widetilde{w}) \! = \! \sigma_{1} \overline{\mathcal{D}^{\Sigma_{
B^{0}}}(\overline{\widetilde{w}})} \, \sigma_{1}$; hence, $\mathcal{
D}^{\Sigma_{B^{0}}}(\widetilde{w})$ is analytic (and bounded) $\forall
\, \widetilde{w} \! \in \! \mathbb{C} \setminus \mathbb{R}$, and solves
the latter RHP on $\mathbb{R}$. Since $\det (\mathcal{D}^{\Sigma_{B^{
0}}}(\widetilde{w}))$ is an analytic and bounded $\mathbb{C}$-valued
function, it is, by Liouville's Theorem, a constant: in this case,
since $\det (V(\zeta_{1})) \! = \! 1$, $\det (\mathcal{D}^{\Sigma_{
B^{0}}}(\widetilde{w})) \! = \! 1$. By (partial) differentiation,
it follows that $\partial_{\widetilde{w}} \mathcal{D}^{\Sigma_{B^{
0}}}_{+}(\widetilde{w}) \! = \! \partial_{\widetilde{w}} \mathcal{
D}^{\Sigma_{B^{0}}}_{-}(\widetilde{w})V(\zeta_{1})$, $\widetilde{w}
\! \in \! \mathbb{R}$; hence, $\partial_{\widetilde{w}} \mathcal{D}^{
\Sigma_{B^{0}}}(\widetilde{w})(\mathcal{D}^{\Sigma_{B^{0}}}(\widetilde{
w}))^{-1}$ has no jumps across $\mathbb{R}$, and is an entire function
of $\widetilde{w}$. Recalling the definition of $\mathcal{D}^{\Sigma_{
B^{0}}}(\widetilde{w})$ given above, and its asymptotics, one shows that
$(\partial_{\widetilde{w}} \mathcal{D}^{\Sigma_{B^{0}}}(\widetilde{w}) \!
+ \! \tfrac{\mi}{2} \widetilde{w} \sigma_{3} \mathcal{D}^{\Sigma_{B^{0}}}
(\widetilde{w}))(\mathcal{D}^{\Sigma_{B^{0}}}(\widetilde{w}))^{-1} \!
=_{\genfrac{}{}{0pt}{2}{\widetilde{w} \to \infty}{\widetilde{w} \in
\mathbb{C} \setminus \mathbb{R}}} \! -\tfrac{\mi}{2}[\sigma_{3},m^{
\Sigma_{B^{0}}}_{1}] \! + \! \mathcal{O}(\widetilde{w}^{-1})$: applying,
now, a generalisation of Liouville's Theorem to the left-hand side of
the latter asymptotics, one arrives at the following (linear) matrix ODE
for $\mathcal{D}^{\Sigma_{B^{0}}}(\widetilde{w})$,
\begin{equation*}
\partial_{\widetilde{w}} \mathcal{D}^{\Sigma_{B^{0}}}(\widetilde{w})
\! + \! \tfrac{\mi}{2} \widetilde{w} \sigma_{3} \mathcal{D}^{\Sigma_{
B^{0}}}(\widetilde{w}) \! = \! \beta^{\Sigma_{B^{0}}} \mathcal{D}^{
\Sigma_{B^{0}}}(\widetilde{w}),
\end{equation*}
where $\beta^{\Sigma_{B^{0}}} \! := \! -\tfrac{\mi}{2}[\sigma_{3},m^{
\Sigma_{B^{0}}}_{1}] \! = \! \beta^{\Sigma_{B^{0}}}_{21} \sigma_{-} \! +
\! \beta^{\Sigma_{B^{0}}}_{12} \sigma_{+}$, whence, since $(m^{\Sigma_{
B^{0}}}_{1})_{12} \! = \! \overline{(m^{\Sigma_{B^{0}}}_{1})_{21}}$,
$\beta^{\Sigma_{B^{0}}}_{12} \! = \! \overline{\beta^{\Sigma_{B^{0}}}_{
21}}$. The method of solution for such matrix ODEs is well known (see, for
example, \cite{a6,a27,a47}): following \cite{a27} (see~Section~4,
pp.~350--353), and recalling that $\mathcal{D}^{\Sigma_{B^{0}}}_{+}
(\widetilde{w}) \! = \! \mathcal{D}^{\Sigma_{B^{0}}}_{-}(\widetilde{w}) \!
\left(
\begin{smallmatrix}
1-\vert r(\zeta_{1}) \vert^{2} & -\overline{r(\zeta_{1})} \\
r(\zeta_{1}) & 1
\end{smallmatrix}
\right)$, $\widetilde{w} \! \in \! \mathbb{R}$, one shows that the solution
of the above matrix ODE for $\mathcal{D}^{\Sigma_{B^{0}}}(\widetilde{w})$
is $(\pm \! \leftrightarrow \! \widetilde{w} \! \in \! \mathbb{C}_{\pm})$
\begin{gather*}
(\mathcal{D}^{\Sigma_{B^{0}}}_{+}(\widetilde{w}))_{11} \! = \! \me^{
-\frac{3 \pi \nu}{4}} \mathbf{D}_{\mi \nu}(\me^{-\frac{3 \pi \mi}{4}
} \widetilde{w}), \qquad (\mathcal{D}^{\Sigma_{B^{0}}}_{+}(\widetilde{
w}))_{22} \! = \! \me^{\frac{\pi \nu}{4}} \mathbf{D}_{-\mi \nu}(\me^{-
\frac{\mi \pi}{4}} \widetilde{w}), \\
(\mathcal{D}^{\Sigma_{B^{0}}}_{+}(\widetilde{w}))_{12} \! = \!
(\beta^{\Sigma_{B^{0}}}_{21})^{-1} \me^{\frac{\pi \nu}{4}}
(\partial_{\widetilde{w}} \mathbf{D}_{-\mi \nu}(\me^{-\frac{
\mi \pi}{4}} \widetilde{w}) \! - \! \tfrac{\mi}{2} \widetilde{w}
\mathbf{D}_{-\mi \nu}(\me^{-\frac{\mi \pi}{4}} \widetilde{w})), \\
(\mathcal{D}^{\Sigma_{B^{0}}}_{+}(\widetilde{w}))_{21} \! = \!
(\beta^{\Sigma_{B^{0}}}_{12})^{-1} \me^{-\frac{3 \pi \nu}{4}}
(\partial_{\widetilde{w}} \mathbf{D}_{\mi \nu}(\me^{-\frac{3
\pi \mi}{4}} \widetilde{w}) \! + \! \tfrac{\mi}{2} \widetilde{
w} \mathbf{D}_{\mi \nu}(\me^{-\frac{3 \pi \mi}{4}} \widetilde{w})), \\
(\mathcal{D}^{\Sigma_{B^{0}}}_{-}(\widetilde{w}))_{11} \! = \! \me^{
\frac{\pi \nu}{4}} \mathbf{D}_{\mi \nu}(\me^{\frac{\mi \pi}{4}}
\widetilde{w}), \qquad (\mathcal{D}^{\Sigma_{B^{0}}}_{-}(\widetilde{
w}))_{22} \! = \! \me^{-\frac{3 \pi \nu}{4}} \mathbf{D}_{-\mi \nu}
(\me^{\frac{3 \pi \mi}{4}} \widetilde{w}), \\
(\mathcal{D}^{\Sigma_{B^{0}}}_{-}(\widetilde{w}))_{12} \! = \!
(\beta^{\Sigma_{B^{0}}}_{21})^{-1} \me^{-\frac{3 \pi \nu}{4}}
(\partial_{\widetilde{w}} \mathbf{D}_{-\mi \nu}(\me^{\frac{3
\pi \mi}{4}} \widetilde{w}) \! - \! \tfrac{\mi}{2} \widetilde{w}
\mathbf{D}_{-\mi \nu}(\me^{\frac{3 \pi \mi}{4}} \widetilde{w})), \\
(\mathcal{D}^{\Sigma_{B^{0}}}_{-}(\widetilde{w}))_{21} \! = \!
(\beta^{\Sigma_{B^{0}}}_{12})^{-1} \me^{\frac{\pi \nu}{4}}
(\partial_{\widetilde{w}} \mathbf{D}_{\mi \nu}(\me^{\frac{
\mi \pi}{4}} \widetilde{w}) \! + \! \tfrac{\mi}{2} \widetilde{w}
\mathbf{D}_{\mi \nu}(\me^{\frac{\mi \pi}{4}} \widetilde{w})),
\end{gather*}
where $\mathbf{D}_{\ast}(\cdot)$ is the parabolic cylinder function
\cite{a40}, and $\beta^{\Sigma_{B^{0}}}_{12} \! = \! \overline{\beta^{
\Sigma_{B^{0}}}_{21}} \! = \! \tfrac{\sqrt{2 \pi} \, \me^{-\frac{\pi \nu}
{2}} \me^{\frac{\mi \pi}{4}}}{r(\zeta_{1}) \overline{\Gamma (\mi \nu)}}$;
using the identity \cite{a40} $\vert \Gamma (\mi \nu) \vert^{2} \! = \!
\tfrac{\pi}{\nu \sinh (\pi \nu)}$, and recalling that $\nu \! := \! \nu
(\zeta_{1}) \! = \! -\tfrac{1}{2 \pi} \ln (1 \! - \! \vert r(\zeta_{
1}) \vert^{2})$ $(> \! 0)$, one deduces that $\vert \beta^{\Sigma_{B^{
0}}}_{12} \vert^{2} \! = \! \vert \beta^{\Sigma_{B^{0}}}_{21} \vert^{2}
\! = \! \nu$. {}From the latter expressions, one obtains explicit formulae
for $m^{\Sigma_{B^{0}}}_{\pm}(\widetilde{w})$ $(=\mathcal{D}^{\Sigma_{B^{
0}}}_{\pm}(\widetilde{w}) \me^{\frac{\mi}{4} \widetilde{w}^{2} \sigma_{3}}
\phi^{\Sigma_{B^{0}}}_{\pm}(\widetilde{w}))$. Letting, for $k \! = \! A$,
$m^{\Sigma_{A^{0}}}(\widetilde{w})$ and $\phi^{\Sigma_{A^{0}}}(\widetilde{
w})$ be the analogues of $m^{\Sigma_{B^{0}}}(\widetilde{w})$ and $\phi^{
\Sigma_{B^{0}}}(\widetilde{w})$, and carrying through with an analysis
analogous to that presented in the proof of Lemma~5.4 and above, one shows
that, for $\mathcal{D}^{\Sigma_{A^{0}}}(\widetilde{w}) \! := \! m^{\Sigma_{
A^{0}}}(\widetilde{w})(\phi^{\Sigma_{A^{0}}}(\widetilde{w}))^{-1} \exp
(\tfrac{\mi}{4} \widetilde{w}^{2} \sigma_{3})$: (1) $\mathcal{D}^{\Sigma_{
A^{0}}}_{+}(\widetilde{w}) \! = \! \mathcal{D}^{\Sigma_{A^{0}}}_{-}
(\widetilde{w}) \!
\left(
\begin{smallmatrix}
1-\vert r(\zeta_{2}) \vert^{2} & -\overline{r(\zeta_{2})} \\
r(\zeta_{2}) & 1
\end{smallmatrix}
\right)$, $\widetilde{w} \! \in \! \mathbb{R}$, with asymptotics
$\mathcal{D}^{\Sigma_{A^{0}}}(\widetilde{w}) \! =_{\genfrac{}{}
{0pt}{2}{\widetilde{w} \to \infty}{\widetilde{w} \in \mathbb{C}
\setminus \mathbb{R}}} \! (\mathrm{I} \! - \! \tfrac{1}{\widetilde{
w}}m^{\Sigma_{A^{0}}}_{1} \! + \! \mathcal{O}(\tfrac{1}{\widetilde{
w}^{2}}))(-\widetilde{w})^{-\mi \nu \sigma_{3}} \exp (\tfrac{\mi}{4}
\widetilde{w}^{2} \sigma_{3})$, and $\mathcal{D}^{\Sigma_{A^{0}}
}(\widetilde{w}) \! = \! \sigma_{1} \overline{\mathcal{D}^{\Sigma_{
A^{0}}}(\overline{\widetilde{w}})} \, \sigma_{1}$; and (2)
\begin{equation*}
\partial_{\widetilde{w}} \mathcal{D}^{\Sigma_{A^{0}}}(\widetilde{w})
\! - \! \tfrac{\mi}{2} \widetilde{w} \sigma_{3} \mathcal{D}^{\Sigma_{
A^{0}}}(\widetilde{w}) \! = \! \beta^{\Sigma_{A^{0}}} \mathcal{D}^{
\Sigma_{A^{0}}}(\widetilde{w}),
\end{equation*}
where $\beta^{\Sigma_{A^{0}}} \! := \! \tfrac{\mi}{2}[\sigma_{3},m^{\Sigma_{
A^{0}}}_{1}] \! = \! \beta^{\Sigma_{A^{0}}}_{21} \sigma_{-} \! + \! \beta^{
\Sigma_{A^{0}}}_{12} \sigma_{+}$, whence, since $(m^{\Sigma_{A^{0}}}_{1})_{
12} \! = \! \overline{(m^{\Sigma_{A^{0}}}_{1})_{21}}$, $\beta^{\Sigma_{A^{
0}}}_{12} \! = \! \overline{\beta^{\Sigma_{A^{0}}}_{21}}$, with solution
given by $(\pm \! \leftrightarrow \! \widetilde{w} \! \in \! \mathbb{C}_{
\pm})$
\begin{gather*}
(\mathcal{D}^{\Sigma_{A^{0}}}_{+}(\widetilde{w}))_{11} \! = \! \me^{-\frac{
3 \pi \nu}{4}} \mathbf{D}_{-\mi \nu}(\me^{-\frac{\mi \pi}{4}} \widetilde{w}),
\qquad (\mathcal{D}^{\Sigma_{A^{0}}}_{+}(\widetilde{w}))_{22} \! = \! \me^{
\frac{\pi \nu}{4}} \mathbf{D}_{\mi \nu}(\me^{-\frac{3 \pi \mi}{4}}
\widetilde{w}), \\
(\mathcal{D}^{\Sigma_{A^{0}}}_{+}(\widetilde{w}))_{12} \! = \! (\beta^{
\Sigma_{A^{0}}}_{21})^{-1} \me^{\frac{\pi \nu}{4}}(\partial_{\widetilde{w}}
\mathbf{D}_{\mi \nu}(\me^{-\frac{3 \pi \mi}{4}} \widetilde{w}) \! + \!
\tfrac{\mi}{2} \widetilde{w} \mathbf{D}_{\mi \nu}(\me^{-\frac{3 \pi \mi}{4}}
\widetilde{w})), \\
(\mathcal{D}^{\Sigma_{A^{0}}}_{+}(\widetilde{w}))_{21} \! = \! (\beta^{
\Sigma_{A^{0}}}_{12})^{-1} \me^{-\frac{3 \pi \nu}{4}}(\partial_{\widetilde{
w}} \mathbf{D}_{-\mi \nu}(\me^{-\frac{\mi \pi}{4}} \widetilde{w}) \! - \!
\tfrac{\mi}{2} \widetilde{w} \mathbf{D}_{-\mi \nu}(\me^{-\frac{\mi \pi}{4}}
\widetilde{w})), \\
(\mathcal{D}^{\Sigma_{A^{0}}}_{-}(\widetilde{w}))_{11} \! = \! \me^{\frac{
\pi \nu}{4}} \mathbf{D}_{-\mi \nu}(\me^{\frac{3 \pi \mi}{4}} \widetilde{w}),
\qquad (\mathcal{D}^{\Sigma_{A^{0}}}_{-}(\widetilde{w}))_{22} \! = \! \me^{
-\frac{3 \pi \nu}{4}} \mathbf{D}_{\mi \nu}(\me^{\frac{\mi \pi}{4}}
\widetilde{w}),
\end{gather*}
\begin{gather*}
(\mathcal{D}^{\Sigma_{A^{0}}}_{-}(\widetilde{w}))_{12} \! = \! (\beta^{
\Sigma_{A^{0}}}_{21})^{-1} \me^{-\frac{3 \pi \nu}{4}}(\partial_{\widetilde{
w}} \mathbf{D}_{\mi \nu}(\me^{\frac{\mi \pi}{4}} \widetilde{w}) \! + \!
\tfrac{\mi}{2} \widetilde{w} \mathbf{D}_{\mi \nu}(\me^{\frac{\mi \pi}{4}}
\widetilde{w})), \\
(\mathcal{D}^{\Sigma_{A^{0}}}_{-}(\widetilde{w}))_{21} \! = \! (\beta^{
\Sigma_{A^{0}}}_{12})^{-1} \me^{\frac{\pi \nu}{4}}(\partial_{\widetilde{w}}
\mathbf{D}_{-\mi \nu}(\me^{\frac{3 \pi \mi}{4}} \widetilde{w}) \! - \!
\tfrac{\mi}{2} \widetilde{w} \mathbf{D}_{-\mi \nu}(\me^{\frac{3 \pi \mi}{4}}
\widetilde{w})),
\end{gather*}
where $\beta^{\Sigma_{A^{0}}}_{12} \! = \! \overline{\beta^{\Sigma_{
A^{0}}}_{21}} \! = \! \tfrac{\sqrt{2 \pi} \, \me^{-\frac{\pi \nu}{2}}
\me^{-\frac{\mi \pi}{4}}}{\overline{r(\zeta_{1})} \, \Gamma (\mi \nu)
}$, and $\vert \beta^{\Sigma_{A^{0}}}_{12} \vert^{2} \! = \! \vert
\beta^{\Sigma_{A^{0}}}_{21} \vert^{2} \! = \! \nu$. {}From the latter
results, one obtains explicit formulae for $m^{\Sigma_{A^{0}}}_{\pm}
(\widetilde{w})$ $(=\mathcal{D}^{\Sigma_{A^{0}}}_{\pm}(\widetilde{w})
\me^{-\frac{\mi}{4} \widetilde{w}^{2} \sigma_{3}} \phi^{\Sigma_{A^{0
}}}_{\pm}(\widetilde{w}))$. Thus, {}from the above analysis and the
proof of Lemma~5.4, one collects the following leading-order asymptotic
results (as $t \! \to \! +\infty$ such that $0 \! < \! \zeta_{2} \!
< \! \tfrac{1}{M} \! < \! M \! < \! \zeta_{1}$ and $\vert \zeta_{3}
\vert^{2} \! = \! 1$, with $M \! \in \! \mathbb{R}_{>1}$ and bounded)
for $\mu^{\Sigma_{k^{\prime}}}(\cdot)$, $k \! \in \! \{A,B\}$: (1)
$\mu^{\Sigma_{B^{\prime}}}(\widetilde{w}) \! := \! ((\mathbf{1}_{
\Sigma_{B^{\prime}}} \! - \! C^{\Sigma_{B^{\prime}}}_{w^{\Sigma_{B^{
\prime}}}})^{-1} \mathrm{I})(\widetilde{w}) \! = \! m^{\Sigma_{B^{
0}}}_{+}(\widetilde{w})(\mathrm{I} \! + \! w^{\Sigma_{B^{0}}}_{+}
(\widetilde{w}))^{-1}(1 \! + \! o(1)) \! = \! \mathcal{D}^{\Sigma_{
B^{0}}}_{+}(\widetilde{w}) \me^{\frac{\mi}{4} \widetilde{w}^{2}
\sigma_{3}} \phi^{\Sigma_{B^{0}}}_{+}(\widetilde{w})(\mathrm{I} \!
+ \! w^{\Sigma_{B^{0}}}_{+}(\widetilde{w}))^{-1}(\mathrm{I} \! + \!
o(1)) \! = \! m^{\Sigma_{B^{0}}}_{-}(\widetilde{w})(\mathrm{I} \! -
\! w^{\Sigma_{B^{0}}}_{-}(\widetilde{w}))^{-1}(1 \! + \! o(1)) \! =
\! \mathcal{D}^{\Sigma_{B^{0}}}_{-}(\widetilde{w}) \me^{\frac{\mi}{4}
\widetilde{w}^{2} \sigma_{3}} \phi^{\Sigma_{B^{0}}}_{-}(\widetilde{
w})(\mathrm{I} \! - \! w^{\Sigma_{B^{0}}}_{-}(\widetilde{w}))^{-1}
(\mathrm{I} \! + \! o(1))$, $\widetilde{w} \! \in \! \Sigma_{B}$; and
(2) $\mu^{\Sigma_{A^{\prime}}}(\widetilde{w}) \! := \! ((\mathbf{1}_{
\Sigma_{A^{\prime}}} \! - \! C^{\Sigma_{A^{\prime}}}_{w^{\Sigma_{A^{
\prime}}}})^{-1} \linebreak[4]
\cdot \mathrm{I})(\widetilde{w}) \! = \! m^{\Sigma_{A^{0}}}_{+}
(\widetilde{w})(\mathrm{I} \! + \! w^{\Sigma_{A^{0}}}_{+}(\widetilde{
w}))^{-1}(1 \! + \! o(1)) \! = \! \mathcal{D}^{\Sigma_{A^{0}}}_{+}
(\widetilde{w}) \me^{-\frac{\mi}{4} \widetilde{w}^{2} \sigma_{3}}
\phi^{\Sigma_{A^{0}}}_{+}(\widetilde{w})(\mathrm{I} \! + \! w^{
\Sigma_{A^{0}}}_{+}(\widetilde{w}))^{-1}(\mathrm{I} \! + \! o(1))
\! = \! m^{\Sigma_{A^{0}}}_{-}(\widetilde{w})(\mathrm{I} \! - \!
w^{\Sigma_{A^{0}}}_{-}(\widetilde{w}))^{-1}(1 \! + \! o(1)) \! = \!
\mathcal{D}^{\Sigma_{A^{0}}}_{-}(\widetilde{w}) \me^{-\frac{\mi}{4
} \widetilde{w}^{2} \sigma_{3}} \phi^{\Sigma_{A^{0}}}_{-}(\widetilde{
w})(\mathrm{I} \! - \! w^{\Sigma_{A^{0}}}_{-}(\widetilde{w}))^{-1}
(\mathrm{I} \! + \! o(1))$, $\widetilde{w} \! \in \! \Sigma_{A}$.
In fact, using the method of successive approximations and proceeding
as in Section~2 of \cite{a48} (see, also, Section~3 of \cite{a49}, and
the proof of Lemma~4.1 in \cite{a50}), one can expand $\mu^{\Sigma_{
k^{\prime}}}(\widetilde{w})$, $k \! \in \! \{A,B\}$, into Neumann-type
series, and improve the $o(1)$ estimate to $\mathcal{O}(\mu^{\Sigma_{
k^{\prime}}}_{1,1}(\widetilde{w}) \tfrac{\ln t}{\sqrt{t}})$, with $\vert
\vert \mu^{\Sigma_{k^{\prime}}}_{1,1}(\cdot) \vert \vert_{\mathcal{L}^{
2}_{\mathrm{M}_{2}(\mathbb{C})}(\Sigma_{k^{\prime}})} \! < \! \infty$;
however, the $o(1)$ estimates are sufficient for the purposes of obtaining
the leading order asymptotics of this proof. Since the evaluation of the
integrals appearing in the formulae for $m^{\Sigma^{\sharp}}_{ij}(\zeta)$,
$i,j \! \in \! \{1,2\}$, are analogous, consider, say, and without loss
of generality, the following integral (appearing in the expression for
$m^{\Sigma^{\sharp}}_{12}(\zeta))$:
\begin{equation*}
\mathrm{I}^{\Sigma^{\sharp}} := \int_{\zeta_{1}+\infty \me^{-\frac{\mi \pi}
{4}}}^{\zeta_{1}+0^{+} \me^{-\frac{\mi \pi}{4}}} \dfrac{\mu^{\Sigma_{B^{
\prime}}}_{11}(\xi) \mathcal{R}(\xi)(\delta (\xi))^{2} \exp (-2 \mi t
\theta^{u}(\xi))}{(\xi \! - \! \zeta)} \, \dfrac{\md \xi}{2 \pi \mi}.
\end{equation*}
Making the change of variable (Eqs.~(105) and~(106)) $\xi \! = \! \zeta_{1}
\! + \! \varepsilon_{B}(\widetilde{w})$, where $\varepsilon_{B}(\widetilde{
w}) \! := \! \tfrac{\widetilde{w}}{\mathcal{X}_{B} \sqrt{t}}$, with $\mathcal{
X}_{B} \! = \! \left(\tfrac{2(\zeta_{1}-\zeta_{2})}{\zeta_{1}} \right)^{1/2}
\! \tfrac{\vert \zeta_{1}-\zeta_{3} \vert}{\zeta_{1}}$, setting $\mu^{\Sigma_{
B^{\prime}}}_{11}(\zeta_{1} \! + \! \varepsilon_{B}(\widetilde{w})) \! := \!
\mu^{\Sigma_{B^{\prime}}}_{11}(\widetilde{w})$, and recalling the action
of $\mathcal{N}_{B}$ given in Proposition~5.1, namely, $(\mathcal{N}_{B}
(\delta^{2} \me^{-2 \mi t \theta^{u}}))(\widetilde{w}) \! = \! (\delta_{
B}^{0})^{2}(\delta_{B}^{1}(\widetilde{w}))^{2}$, with $\delta_{B}^{0}$ and
$\delta_{B}^{1}(\widetilde{w})$ defined in Proposition~5.1,
\begin{align*}
\mathrm{I}^{\Sigma^{\sharp}} &= \int\nolimits_{\infty \me^{-\frac{
\mi \pi}{4}}}^{0^{+}} \dfrac{\mu^{\Sigma_{B^{\prime}}}_{11}(\zeta_{
1} \! + \! \varepsilon_{B}(\widetilde{w})) \mathcal{R}(\zeta_{1}
\! + \! \varepsilon_{B}(\widetilde{w}))(\delta (\zeta_{1} \! + \!
\varepsilon_{B}(\widetilde{w})))^{2} \me^{-2 \mi t \theta^{u}(\zeta_{
1}+\varepsilon_{B}(\widetilde{w}))}}{(\zeta_{1} \! + \! \varepsilon_{
B}(\widetilde{w}) \! - \! \zeta)} \, \dfrac{\md \varepsilon_{B}
(\widetilde{w})}{2 \pi \mi} \\
 &= \int\nolimits_{0^{+}}^{\infty \me^{-\frac{\mi \pi}{4}}} \dfrac{
\mu^{\Sigma_{B^{\prime}}}_{11}(\widetilde{w}) \mathcal{R}(\zeta_{1}
\! + \! \varepsilon_{B}(\widetilde{w}))(\delta (\zeta_{1} \! + \!
\varepsilon_{B}(\widetilde{w})))^{2} \me^{-2 \mi t \theta^{u}(\zeta_{
1}+\varepsilon_{B}(\widetilde{w}))}}{((\zeta \! - \! \zeta_{1}) \! -
\! \varepsilon_{B}(\widetilde{w}))} \, \dfrac{\md \varepsilon_{B}
(\widetilde{w})}{2 \pi \mi} \\
 &= \dfrac{1}{(\zeta \! - \! \zeta_{1})} \! \int\nolimits_{0^{+}}^{
\infty \me^{-\frac{\mi \pi}{4}}} \dfrac{\mu^{\Sigma_{B^{\prime}}}_{
11}(\widetilde{w}) \mathcal{R}(\zeta_{1} \! + \! \varepsilon_{B}
(\widetilde{w}))(\mathcal{N}_{B}(\delta^{2} \me^{-2 \mi t \theta^{
u}}))(\widetilde{w})}{(1 \! - \! \varepsilon_{B}(\widetilde{w})(\zeta
\! - \! \zeta_{1})^{-1})} \, \dfrac{\md \varepsilon_{B}(\widetilde{w}
)}{2 \pi \mi} \\
 &= \dfrac{1}{(\zeta \! - \! \zeta_{1})} \! \int\nolimits_{0^{+}}^{
\infty \me^{-\frac{\mi \pi}{4}}} \dfrac{\mu^{\Sigma_{B^{\prime}}}_{
11}(\widetilde{w}) \mathcal{R}(\zeta_{1} \! + \! \varepsilon_{B}
(\widetilde{w}))(\delta_{B}^{0})^{2}(\delta_{B}^{1}(\widetilde{w
}))^{2}}{(1 \! - \! \varepsilon_{B}(\widetilde{w})(\zeta \! - \!
\zeta_{1})^{-1})} \, \dfrac{\md \varepsilon_{B}(\widetilde{w})}
{2 \pi \mi} \\
 &= \dfrac{(\delta_{B}^{0})^{2}}{(\zeta \! - \! \zeta_{1})} \!
\int\nolimits_{0^{+}}^{\infty \me^{-\frac{\mi \pi}{4}}} \dfrac{
\mu^{\Sigma_{B^{\prime}}}_{11}(\widetilde{w})(\delta_{B}^{1}
(\widetilde{w}))^{2} \mathcal{R}(\zeta_{1} \! + \! \varepsilon_{
B}(\widetilde{w}))}{(1 \! - \! \varepsilon_{B}(\widetilde{w})
(\zeta \! - \! \zeta_{1})^{-1})} \, \dfrac{\md \varepsilon_{B}
(\widetilde{w})}{2 \pi \mi}.
\end{align*}
Recalling {}from Lemma~5.1 that
\begin{equation*}
(\delta_{B}^{1}(\widetilde{w}))^{2} \mathcal{R}(\zeta_{1} \! + \!
\varepsilon_{B}(\widetilde{w})) \! = \! (\widetilde{w})^{2 \mi \nu}
\exp (-\tfrac{\mi \widetilde{w}^{2}}{2}) \mathcal{R}(\zeta_{1}^{
+}) \! + \! \mathcal{O} \! \left( \tfrac{c^{\mathcal{S}}(\zeta_{1})
\underline{c}(\zeta_{2},\zeta_{3},\overline{\zeta_{3}})}{\vert
\zeta_{1}-\zeta_{3} \vert \sqrt{(\zeta_{1}-\zeta_{2})}} \tfrac{\ln
(t)}{\sqrt{t}} \exp (-\tfrac{\mi \gamma \widetilde{w}^{2}}{2})
\right),
\end{equation*}
with $\mathcal{R}(\zeta_{1}^{+}) \! := \! \lim_{\Re (\zeta) \downarrow
\zeta_{1}} \! \mathcal{R}(\zeta) \! = \! \overline{r(\zeta_{1})}$, and
$\gamma \! \in \! (0,\tfrac{1}{2})$, and noting the exponential decay
of the factors $\me^{-\frac{\mi}{2} \widetilde{w}^{2}}$ and $\me^{-
\frac{\mi \gamma}{2} \widetilde{w}^{2}}$ along the (unbounded) ray
$0^{+}$ to $\infty \me^{-\frac{\mi \pi}{4}}$, one expands the factor
$1/(1 \! - \! \varepsilon_{B}(\widetilde{w})(\zeta \! - \! \zeta_{1}
)^{-1})$ in a geometric progression and, recalling that $\varepsilon_{
B}(\widetilde{w}) \! = \! \tfrac{\widetilde{w}}{\mathcal{X}_{B} \sqrt{
t}}$, using the following rules {}from asymptotic analysis, $\mathcal{
O}(\star_{1})o(\star_{2}) \! = \! o(\star_{1} \star_{2})$ and $\mathcal{
O}(\star) \! + \! o(\star) \! = \! \mathcal{O}(\star)$, along with the
fact that, for $\epsilon$ some arbitrarily fixed, sufficiently small
positive real number, $t^{\epsilon} \ln t \! > \! 1$, one arrives at
the following (limiting) result:
\begin{align*}
\mathrm{I}^{\Sigma^{\sharp}} &= \dfrac{\overline{r(\zeta_{1})}
(\delta_{B}^{0})^{2}}{2 \pi \mi (\zeta \! - \! \zeta_{1}) \mathcal{X}_{
B} \sqrt{t}} \int\nolimits_{0}^{\infty \me^{-\frac{\mi \pi}{4}}} \! \mu^{
\Sigma_{B^{\prime}}}_{11}(\widetilde{w})(\widetilde{w})^{2 \mi
\nu} \exp (-\tfrac{\mi \widetilde{w}^{2}}{2}) \, \md \widetilde{w} \\
 &+ \mathcal{O} \! \left( \tfrac{c^{\mathcal{S}}(\zeta_{1})
\underline{c}(\zeta_{2},\zeta_{3},\overline{\zeta_{3}})(\delta_{
B}^{0})^{2}}{(\zeta -\zeta_{1}) \vert \zeta_{1}-\zeta_{3} \vert
\sqrt{(\zeta_{1}-\zeta_{2})} \, \mathcal{X}_{B}} \, \tfrac{\ln
(t)}{t} \int\nolimits_{0}^{\infty \me^{-\frac{\mi \pi}{4}}} \!
\mu^{\Sigma_{B^{\prime}}}_{11}(\widetilde{w}) \exp (-\tfrac{
\mi \gamma \widetilde{w}^{2}}{2}) \, \md \widetilde{w} \right).
\end{align*}
Recalling the formula for $\mu^{\Sigma_{B^{\prime}}}(\widetilde{w})$
given heretofore in the proof, using the definition of $\phi^{\Sigma_{
B^{0}}}(\widetilde{w})$ given in the proof of Lemma~5.4, and noting
that $w^{\Sigma_{B^{0}}}_{+}(\widetilde{w})$ is nilpotent, with degree
of nilpotency 2, one deduces that $\mu^{\Sigma_{B^{\prime}}}_{11}
(\widetilde{w}) \! = \! (\mathcal{D}^{\Sigma_{B^{0}}}_{-}(\widetilde{
w}))_{11} \me^{\frac{\mi}{4} \widetilde{w}^{2}}(\widetilde{w})^{-\mi
\nu} \! = \! \me^{\frac{\pi \nu}{4}} \mathbf{D}_{\mi \nu}(\me^{\frac{
\mi \pi}{4}} \widetilde{w}) \me^{\frac{\mi}{4} \widetilde{w}^{2}}
(\widetilde{w})^{-\mi \nu}$: making one more change of variable,
namely, $z \! = \! \me^{\frac{\mi \pi}{4}} \widetilde{w}$, one
arrives at
\begin{equation*}
\mathrm{I}^{\Sigma^{\sharp}} \! = \! \dfrac{\overline{r(\zeta_{1})}
(\delta_{B}^{0})^{2} \me^{\frac{\pi \nu}{2}} \me^{-\frac{\mi \pi}{4}}}
{2 \pi \mi (\zeta \! - \! \zeta_{1}) \mathcal{X}_{B} \sqrt{t}}
\int\nolimits_{0}^{+\infty} \mathbf{D}_{\mi \nu}(z) z^{\mi \nu} \me^{
-\frac{z^{2}}{4}} \, \md z \! + \! \mathcal{O} \! \left( \tfrac{c^{
\mathcal{S}}(\zeta_{1}) \underline{c}(\zeta_{2},\zeta_{3},\overline{
\zeta_{3}})(\delta_{B}^{0})^{2}}{(\zeta -\zeta_{1}) \vert \zeta_{1}-
\zeta_{3} \vert \sqrt{(\zeta_{1}-\zeta_{2})} \, \mathcal{X}_{B}} \,
\tfrac{\ln (t)}{t} \, \mathrm{I}_{\gamma}^{\Sigma^{\sharp}} \right),
\end{equation*}
where
\begin{equation*}
\mathrm{I}_{\gamma}^{\Sigma^{\sharp}} := \int\nolimits_{0}^{+\infty}
\mathbf{D}_{\mi \nu}(z) z^{-\mi \nu} \exp \! \left(\tfrac{1}{2}(\tfrac{1}{2}
\! - \! \gamma)z^{2} \right) \md z.
\end{equation*}
Recall {}from the proof of Lemma~5.1 that $0 \! < \! \nu \! := \!
\nu (\zeta_{1}) \! \leqslant \! \nu_{m} \! := \! -\tfrac{1}{2 \pi}
\ln (1 \! - \! \sup_{z \in \mathbb{R}} \vert r(z) \vert^{2})$: let
$\nu_{m}^{\sharp}$ be an arbitrarily fixed, sufficiently large
positive real number with $\nu_{m}^{\sharp} \! \gg \! \nu_{m}$ (in
particular, choose $\nu_{m}^{\sharp} \! \gg \! \sqrt{2})$. Set
\begin{equation*}
\mathrm{I}_{\gamma}^{\Sigma^{\sharp}} \! = \! \left( \int\nolimits_{
0}^{\nu_{m}}+\int\nolimits_{\nu_{m}}^{\nu_{m}^{\sharp}}+\int\nolimits_{
\nu_{m}^{\sharp}}^{+\infty} \right) \mathbf{D}_{\mi \nu}(z) z^{-\mi \nu}
\exp \! \left(\tfrac{1}{2}(\tfrac{1}{2} \! - \! \gamma)z^{2} \right) \md z \!
:= \! \mathrm{I}_{\gamma}^{1} \! + \! \mathrm{I}_{\gamma}^{2} \! + \!
\mathrm{I}_{\gamma}^{3}.
\end{equation*}
Via a change-of-variable argument and the fact that, for $(z,\gamma)
\! \in \! [0,\nu_{m}) \times (0,\tfrac{1}{2})$, $\me^{-\frac{\gamma}
{2}z^{2}} \! \leqslant \! 1$, one deduces that $\vert \mathrm{I}_{
\gamma}^{1} \vert \! \leqslant \! \nu_{m} \sup_{x \in [0,1)} \vert
\mathbf{D}_{\mi \nu}(\nu_{m}x) \vert \int_{0}^{1} \me^{\frac{\nu_{
m}^{2}x^{2}}{4}} \, \md x \! < \! \infty$. For $z \! \in \! [\nu_{m},
\nu_{m}^{\sharp})$, one uses the following representation for the
parabolic cylinder function \cite{a40}, $\mathbf{D}_{\mi \nu}(z)
\! = \! 2^{\frac{\mi \nu}{2}} \me^{-\frac{z^{2}}{4}}(\tfrac{\sqrt{
\pi}}{\Gamma (\frac{1-\mi \nu}{2})} \mathbf{F}(-\tfrac{\mi \nu}{2},
\linebreak[4]
\tfrac{1}{2};\tfrac{z^{2}}{2}) \! - \! \tfrac{\sqrt{2 \pi} \, z}{
\Gamma (-\frac{\mi \nu}{2})} \mathbf{F}(\tfrac{1-\mi \nu}{2},\tfrac{
3}{2};\tfrac{z^{2}}{2}))$, where $\mathbf{F}(a,b;x) \! := \! \sum_{
n=0}^{\infty} \tfrac{(a)_{n}x^{n}}{(b)_{n}n!}$ is the confluent
hypergeometric function, and $(\star)_{n} \! := \! \star (\star \!
+ \! 1) \cdots (\star \! + \! n \! - \! 1) \! = \! \tfrac{\Gamma
(\star +n)}{\Gamma (\star)}$ is the Pochhammer symbol; hence,
$\mathrm{I}_{\gamma}^{2} \! = \! \tfrac{2^{\frac{\mi \nu}{2}}
\sqrt{\pi}}{\Gamma (\frac{1-\mi \nu}{2})} \sum_{n=0}^{\infty}
\tfrac{(-\frac{\mi \nu}{2})_{n}2^{-n}}{(\frac{1}{2})_{n}n!} \!
\int_{\nu_{m}}^{\nu_{m}^{\sharp}}z^{-\mi \nu}z^{2n} \me^{-\frac{
\gamma}{2}z^{2}} \, \md z \! - \! \tfrac{2^{\frac{\mi \nu}{2}}
\sqrt{2 \pi}}{\Gamma (-\frac{\mi \nu}{2})} \sum_{n=0}^{\infty}
\tfrac{(\frac{1-\mi \nu}{2})_{n}2^{-n}}{(\frac{3}{2})_{n}n!}
\! \int_{\nu_{m}}^{\nu_{m}^{\sharp}}z^{-\mi \nu}z^{2n+1} \me^{
-\frac{\gamma}{2}z^{2}} \, \md z$. \linebreak[4]
Recall that $\me^{\star} \! = \! \sum_{n=0}^{\infty} \tfrac{\star^{
n}}{n!}$, where the convergence of the series is absolute and
uniform for $\star$ in compact subsets of $\mathbb{R}$: using the
series expansion for $\me^{\star}$ and integrating term-by-term,
noting that, $\forall \, (m,n) \! \in \! \mathbb{Z}_{\geqslant 0}
\times \mathbb{Z}_{\geqslant 0}$, $((2(n \! + \! m) \! + \! s)^{2}
\! + \! \nu^{2})^{-1/2} \! \leqslant \! (1 \! + \! \nu^{2})^{-1/2}$,
$s \! \in \! \{1,2\}$, and using the identities \cite{a40} $\vert
\Gamma (\tfrac{1-\mi \nu}{2}) \vert \! = \! \sqrt{\tfrac{\pi}{\cosh
(\pi \nu/2)}}$ and $\vert \Gamma (-\tfrac{\mi \nu}{2}) \vert \! =
\! \sqrt{\tfrac{2 \pi}{\nu \sinh (\pi \nu/2)}}$, and the functional
relation $\Gamma (1 \! + \! x) \! = \! x \Gamma (x)$, via the
well-known result $\Gamma (1/2) \! = \! \sqrt{\pi}$ and a
majorisation argument, one shows that
\begin{equation*}
\vert \mathrm{I}_{\gamma}^{2} \vert \leqslant \sqrt{\tfrac{\nu \sinh (\pi
\nu)}{2(1+\nu^{2})}} \sum_{s \in \{\nu_{m},\nu_{m}^{\sharp}\}} \sum_{n=
0}^{\infty} \left( \tfrac{1}{\sqrt{2}} \vert \Gamma (n \! - \! \tfrac{\mi
\nu}{2}) \vert \! + \! s \vert \Gamma (n \! + \! \tfrac{1}{2} \! - \!
\tfrac{\mi \nu}{2}) \vert \right) \! \tfrac{s^{2n+1} \me^{\gamma s^{2}/2}}
{2^{n}n! \Gamma (n+\frac{1}{2})}.
\end{equation*}
{}From the above gamma function identities and repeated application
of the functional relation $\Gamma (1 \! + \! x) \! = \! x \Gamma (x)$,
one shows that, for $n \! \in \! \mathbb{Z}_{\geqslant 1}$, $\tfrac{
\vert \Gamma (n-\frac{\mi \nu}{2}) \vert}{\Gamma (n+\frac{1}{2})} \!
\leqslant \! \tfrac{((n-1)^{2}+(\frac{\nu}{2})^{2})^{\frac{n}{2}}2^{
n+\frac{1}{2}}}{(2n-1)!! \sqrt{\nu \sinh (\pi \nu/2)}}$ and $\tfrac{
\vert \Gamma (n+\frac{1}{2}-\frac{\mi \nu}{2}) \vert}{\Gamma (n+\frac{
1}{2})} \! \leqslant \! \tfrac{((n-\frac{1}{2})^{2}+(\frac{\nu}{2})^{
2})^{\frac{n}{2}}2^{n}}{(2n-1)!! \sqrt{\cosh (\pi \nu/2)}}$, where
$(2n \! - \! 1)!! \! = \! (2n \! - \! 1) \cdots 3 \cdot 1$: using the
latter inequalities, one arrives at
\begin{align*}
\vert \mathrm{I}_{\gamma}^{2} \vert &\leqslant \sqrt{\tfrac{\cosh
(\pi \nu/2)}{1+\nu^{2}}} \, \sum_{s \in \{\nu_{m},\nu_{m}^{\sharp}
\}} \left( 1 \! + \! \sum_{n=1}^{\infty} \tfrac{((n-1)^{2}+(\frac{
\nu}{2})^{2})^{n/2}s^{2n}}{n!(2n-1)!!} \right) \! s \me^{\gamma s^{
2}/2} \\
 &+ \sqrt{\tfrac{\nu \sinh (\pi \nu/2)}{1+\nu^{2}}} \sum_{s \in \{
\nu_{m},\nu_{m}^{\sharp}\}} \left( 1 \! + \! \sum_{n=1}^{\infty}
\tfrac{((n-\frac{1}{2})^{2}+(\frac{\nu}{2})^{2})^{n/2}s^{2n}}{n!
(2n-1)!!} \right) \! s^{2} \me^{\gamma s^{2}/2}.
\end{align*}
A straightforward application of the Ratio Test shows that the
latter two (infinite) series converge absolutely (hence converge);
thus, $\vert \mathrm{I}_{\gamma}^{2} \vert \! < \! \infty$. For $z
\! \in \! [\nu_{m}^{\sharp},+\infty)$, with $\vert \arg z \vert \! < \!
\tfrac{3 \pi}{4}$, one uses the following asymptotic expansion
for $\mathbf{D}_{\mi \nu}(z)$ \cite{a40}:
\begin{equation*}
\mathbf{D}_{\mi \nu}(z) \! \underset{\underset{\vert \arg z \vert <
\frac{3 \pi}{4}}{z \to \infty}}{=} \! z^{\mi \nu} \me^{-\frac{z^{2}}{4}}
\! \left( \, \sum_{n=0}^{N} \tfrac{(-1)^{n} \prod_{k=0}^{2n-1}(\mi \nu -k)}
{(2n)!!z^{2n}}+\mathcal{O} \! \left( \tfrac{\prod_{k=0}^{2N+1}(\mi \nu
-k)}{(2N+2)!!z^{2N+2}} \right) \right),
\end{equation*}
where $N \! \in \! \mathbb{N}$, $(2n)!! \! = \! (2n) \cdots 4 \cdot 2$,
and $\prod_{k=0}^{-1}(\mi \nu \! - \! k) \! := \! 1$. With this asymptotic
representation for $\mathbf{D}_{\mi \nu}(z)$, one presents
$\mathrm{I}_{\gamma}^{3}$ as $\mathrm{I}_{\gamma}^{3} \! = \!
\widetilde{\mathrm{I}}_{\gamma}^{3} \! + \! \widehat{\mathrm{I}}_{
\gamma}^{3}$, where $\widetilde{\mathrm{I}}_{\gamma}^{3} \! := \!
\int_{\nu_{m}^{\sharp}}^{+\infty} \me^{-\frac{\gamma}{2}z^{2}} \,
\md z$, and $\widehat{\mathrm{I}}_{\gamma}^{3} \! := \! \int_{\nu_{
m}^{\sharp}}^{+\infty} \me^{-\frac{\gamma}{2}z^{2}} \! \left( \sum_{
n=1}^{N} \tfrac{(-1)^{n}\prod_{k=0}^{2n-1}(\mi \nu -k)}{(2n)!!z^{2n
}} \! + \! \mathcal{O} \! \left( \tfrac{\prod_{k=0}^{2N+1}(\mi \nu
-k)}{(2N+2)!!z^{2N+2}} \right) \! \right) \! \md z$. Write $\int_{
\nu_{m}^{\sharp}}^{+\infty} \! \me^{-\frac{\gamma}{2}z^{2}} \md z \!
= \! \int_{0}^{+\infty} \! \me^{-\frac{\gamma}{2}z^{2}} \linebreak[4]
\cdot \md z \! - \! \int_{0}^{\nu_{m}^{\sharp}} \me^{-\frac{\gamma}
{2}z^{2}} \, \md z$. Recalling that $\Gamma (1/2) \! = \! \sqrt{\pi}$,
via a change-of-variable argument, one shows that $\int_{0}^{+\infty}
\me^{-\frac{\gamma}{2}z^{2}} \, \md z \! = \! (\pi /2\gamma)^{1/2}$.
Recalling that $\me^{\star} \! = \! \sum_{n=0}^{\infty} \tfrac{
\star^{n}}{n!}$, one integrates (the second integral) term-by-term
and, using the fact that, for $n \! \in \! \mathbb{Z}_{\geqslant 0}$,
$(2n \! + \! 1)^{-1} \! \leqslant \! 1$ and $2^{-n} \! \leqslant \! 1$,
via a change-of-variable and majorisation argument, one shows that $\vert
\int_{0}^{\nu_{m}^{\sharp}} \me^{-\frac{\gamma}{2}z^{2}} \, \md z \vert
\! \leqslant \! \nu_{m}^{\sharp} \me^{\gamma (\nu_{m}^{\sharp})^{2}}$;
hence, $\vert \widetilde{\mathrm{I}}_{\gamma}^{3} \vert \! \leqslant \!
(\pi/2 \gamma)^{1/2} \! + \! \nu_{m}^{\sharp} \me^{\gamma (\nu_{m}^{
\sharp})^{2}} \! < \! \infty$ (one can also use the fact that the second
integral has an explicit representation in terms of the incomplete gamma
function \cite{a40} to obtain a similar estimate). Via a change-of-variable
argument, one arrives at $\widehat{\mathrm{I}}_{\gamma}^{3} \! \sim \!
\tfrac{1}{\sqrt{2 \gamma}} \sum_{n \geqslant 1} \! \tfrac{(-1)^{n}(\prod_{
k=0}^{2n-1}(\mi \nu -k)) \gamma^{n}}{(2n)!!2^{n}} \! \int_{\frac{1}{2}
\gamma (\nu_{m}^{\sharp})^{2}}^{+\infty} x^{-(n+1/2)} \me^{-x} \, \md x$:
using the following definite integral \cite{a40} $\int_{u}^{+\infty}x^{-
\nu^{\prime}} \me^{-x} \, \md x \! = \! u^{-\nu^{\prime}/2} \me^{-u/2}
\mathbf{W}_{-\frac{\nu^{\prime}}{2},\frac{(1-\nu^{\prime})}{2}}(u)$,
$u \! > \! 0$, where $\mathbf{W}_{z_{1},z_{2}}(z)$ is the Whittaker
function, $\mathbf{W}_{z_{1},z_{2}}(z):=\tfrac{\Gamma (-2z_{2})}{\Gamma
(\frac{1}{2}-z_{1}-z_{2})} \me^{-z/2}z^{1/2+z_{2}} \mathbf{F}(z_{2}-z_{1}
+\tfrac{1}{2},2z_{2}+1;z) \! + \! \tfrac{\Gamma (2z_{2})}{\Gamma (\frac{
1}{2}-z_{1}+z_{2})} \me^{-z/2}z^{1/2-z_{2}} \linebreak[4]
\cdot \mathbf{F}(-z_{2} \! - \! z_{1} \! + \! \tfrac{1}{2},-2z_{2} \! + \!
1;z)$, and the gamma function identity \cite{a40} $\Gamma (\tfrac{1}{2} \!
- \! n) \! = \! \tfrac{(-1)^{n}2^{n} \sqrt{\pi}}{(2n-1)!!}$, $n \! \in \!
\mathbb{Z}_{\geqslant 1}$, one shows, upon noting the identity $(2n)!!(2n
\! - \! 1)!! \! = \! (2n)!$, that
\begin{align*}
\vert \widehat{\mathrm{I}}_{\gamma}^{3} \vert \! &\leqslant \! \tfrac{\sqrt{
\pi} \, \nu_{m}^{\sharp} \me^{-\frac{\gamma (\nu_{m}^{\sharp})^{2}}{2}}}{2}
\! \left( \sum_{n=1}^{N} \tfrac{2^{n} \prod_{k=0}^{2n-1}(\nu^{2}+k^{2})^{
\frac{1}{2}}}{(2n)!(\nu_{m}^{\sharp})^{2n}} \right) \! \sum_{m=0}^{\infty}
\tfrac{(\gamma (\nu_{m}^{\sharp})^{2})^{m}}{2^{m} \vert \Gamma (m-n+\frac{3}
{2}) \vert} \! + \! \left(\tfrac{\pi}{2 \gamma} \right)^{1/2} \sum_{n=1}^{N}
\tfrac{\gamma^{n} \prod_{k=0}^{2n-1}(\nu^{2}+k^{2})^{\frac{1}{2}}}{(2n)!} \\
&+\mathcal{O} \! \left( \tfrac{2^{N} \prod_{k=0}^{2N+1}(\nu^{2}+k^{2})^{
\frac{1}{2}}}{(2N+2)!(\nu_{m}^{\sharp})^{2N}} \sum_{m=0}^{\infty} \tfrac{
(\gamma (\nu_{m}^{\sharp})^{2})^{m}}{2^{m} \vert \Gamma (m-N+\frac{1}{2})
\vert} \! + \! \tfrac{\gamma^{N} \prod_{k=0}^{2N+1}(\nu^{2}+k^{2})^{\frac{1}
{2}}}{(2N+2)!} \right).
\end{align*}
Recalling that $\nu_{m}^{\sharp} \! \gg \! \sqrt{2}$ and $\gamma \! \in \!
(0,\tfrac{1}{2})$, a straightforward application of the Ratio Test shows that
the respective series above are absolutely convergent (hence convergent);
thus, via the inequality $\vert \star_{1} \star_{2} \vert \! \leqslant \!
\vert \star_{1} \vert \vert \star_{2} \vert$, one deduces that $\vert
\widehat{\mathrm{I}}_{\gamma}^{3} \vert \! < \! \infty$. Collecting the
above-derived estimates, one arrives at $\mathrm{I}_{\gamma}^{\Sigma^{
\sharp}} \! = \! \sum_{k=1}^{3} \mathrm{I}_{\gamma}^{k} \! \leqslant \!
\underline{c}(\zeta_{1},\zeta_{2},\zeta_{3},\overline{\zeta_{3}})$  $(= \!
\mathcal{O}(1))$; hence,
\begin{equation*}
\mathrm{I}^{\Sigma^{\sharp}} \! = \! \tfrac{\overline{r(\zeta_{1})}
(\delta_{B}^{0})^{2} \me^{\frac{\pi \nu}{2}} \me^{-\frac{\mi \pi}{
4}}}{2 \pi \mi (\zeta-\zeta_{1}) \mathcal{X}_{B} \sqrt{t}} \int_{0
}^{+\infty} \mathbf{D}_{\mi \nu}(z)z^{\mi \nu} \me^{-\frac{z^{2}}{
4}} \, \md z + \mathcal{O} \! \left( \tfrac{c^{\mathcal{S}}(\zeta_{
1}) \underline{c}(\zeta_{2},\zeta_{3},\overline{\zeta_{3}})(\delta_{
B}^{0})^{2}}{(\zeta-\zeta_{1}) \vert \zeta_{1}-\zeta_{3} \vert
\sqrt{(\zeta_{1}-\zeta_{2})} \, \, \mathcal{X}_{B}} \, \tfrac{
\ln t}{t} \right).
\end{equation*}
Proceeding as above for the remaining integrals for $m^{\Sigma^{
\sharp}}_{ij}(\zeta)$, $i,j \! \in \! \{1,2\}$, given at the beginning of
the proof, letting $\varepsilon \! \to \! +\infty$ \textbf{in the limits
of integration}, and neglecting exponentially small terms
(cf.~Remark~4.4), one arrives at
\begin{align*}
m^{\Sigma^{\sharp}}_{11}(\zeta) \! &= \! 1 \! - \! \tfrac{r(\zeta_{1})
(\delta_{B}^{0})^{-2} \me^{\frac{\pi \nu}{2}} \me^{\frac{\mi \pi}{4}
}}{2 \pi \mi (\zeta -\zeta_{1}) \beta^{\Sigma_{B^{0}}}_{21} \mathcal{
X}_{B} \sqrt{t}} \int\nolimits_{0}^{+\infty}(\me^{-\frac{\mi \pi}{4}}
\partial_{z} \mathbf{D}_{-\mi \nu}(z) \! - \! \tfrac{\mi}{2} \me^{
\frac{\mi \pi}{4}}z \mathbf{D}_{-\mi \nu}(z))z^{-\mi \nu} \me^{-
\frac{z^{2}}{4}} \, \md z \\
 &+ \tfrac{r(\zeta_{1})(1-\vert r(\zeta_{1}) \vert^{2})^{-1}
(\delta_{B}^{0})^{-2} \me^{-\frac{3 \pi \mi}{4}}}{2 \pi \mi
(\zeta -\zeta_{1}) \beta^{\Sigma_{B^{0}}}_{21} \me^{\frac{3 \pi
\nu}{2}} \mathcal{X}_{B} \sqrt{t}} \int\nolimits_{0}^{+\infty}
(\me^{\frac{3 \pi \mi}{4}} \partial_{z} \mathbf{D}_{-\mi \nu}(z)
\! - \! \tfrac{\mi}{2} \me^{-\frac{3 \pi \mi}{4}}z \mathbf{D}_{-
\mi \nu}(z))z^{-\mi \nu} \me^{-\frac{z^{2}}{4}} \, \md z \\
 &- \tfrac{\overline{r(\zeta_{1})}(\delta_{A}^{0})^{-2} \me^{-
\frac{\pi \nu}{2}}(-1)^{\mi \nu} \me^{\frac{3 \pi \mi}{4}}}{2 \pi
\mi (\zeta -\zeta_{2}) \beta^{\Sigma_{A^{0}}}_{21} \mathcal{
X}_{A} \sqrt{t}} \int\nolimits_{0}^{+\infty}(\me^{-\frac{3 \pi \mi}{
4}} \partial_{z} \mathbf{D}_{\mi \nu}(z) \! + \! \tfrac{\mi}{2} \me^{
\frac{3 \pi \mi}{4}}z \mathbf{D}_{\mi \nu}(z))z^{\mi \nu} \me^{-
\frac{z^{2}}{4}} \, \md z \\
 &+ \tfrac{\overline{r(\zeta_{1})}(1 -\vert r(\zeta_{1}) \vert^{2}
)^{-1}(\delta_{A}^{0})^{-2}(-1)^{\mi \nu} \me^{-\frac{\mi \pi}{4}
}}{2 \pi \mi (\zeta -\zeta_{2}) \beta^{\Sigma_{A^{0}}}_{21} \me^{
\frac{\pi \nu}{2}} \mathcal{X}_{A} \sqrt{t}} \int\nolimits_{0}^{+
\infty}(\me^{\frac{\mi \pi}{4}} \partial_{z} \mathbf{D}_{\mi \nu}
(z) \! + \! \tfrac{\mi}{2} \me^{-\frac{\mi \pi}{4}}z \mathbf{D}_{
\mi \nu}(z)) z^{\mi \nu} \me^{-\frac{z^{2}}{4}} \, \md z \\
 &+ \mathcal{O} \! \left( \! \left( \tfrac{c^{\mathcal{S}}(\zeta_{
1}) \underline{c}(\zeta_{2},\zeta_{3},\overline{\zeta_{3}})(\delta_{
B}^{0})^{-2}}{(\zeta -\zeta_{1}) \vert \zeta_{1}-\zeta_{3} \vert
\sqrt{(\zeta_{1}-\zeta_{2})} \, \, \mathcal{X}_{B}}+\tfrac{c^{
\mathcal{S}}(\zeta_{2}) \underline{c}(\zeta_{1},\zeta_{3},\overline{
\zeta_{3}})(\delta_{A}^{0})^{-2}}{(\zeta -\zeta_{2}) \vert \zeta_{
2}-\zeta_{3} \vert \sqrt{(\zeta_{1}-\zeta_{2})} \, \, \mathcal{X}_{
A}} \right) \! \tfrac{\ln t}{t} \right), \\
m^{\Sigma^{\sharp}}_{12}(\zeta) \! &= \! \left( \tfrac{\overline{
r(\zeta_{1})}(\delta_{B}^{0})^{2} \me^{\frac{\pi \nu}{2}} \me^{-
\frac{\mi \pi}{4}}}{2 \pi \mi (\zeta -\zeta_{1}) \mathcal{X}_{B}
\sqrt{t}} \! - \! \tfrac{\overline{r(\zeta_{1})}(1-\vert r(\zeta_{
1}) \vert^{2})^{-1}(\delta_{B}^{0})^{2} \me^{\frac{3 \pi \mi}{4}}}
{2 \pi \mi (\zeta -\zeta_{1}) \me^{\frac{3 \pi \nu}{2}} \mathcal{
X}_{B} \sqrt{t}} \right) \! \int\nolimits_{0}^{+\infty} \mathbf{
D}_{\mi \nu}(z)z^{\mi \nu} \me^{-\frac{z^{2}}{4}} \, \md z \\
 &+ \left( \tfrac{r(\zeta_{1})(\delta_{A}^{0})^{2} \me^{-\frac{\pi
\nu}{2}} \me^{-\frac{3 \pi \mi}{4}}}{2 \pi \mi (\zeta -\zeta_{2})
(-1)^{\mi \nu} \mathcal{X}_{A} \sqrt{t}} \! - \! \tfrac{r(\zeta_{
1})(1-\vert r(\zeta_{1}) \vert^{2})^{-1}(\delta_{A}^{0})^{2}
\me^{\frac{\mi \pi}{4}}}{2 \pi \mi (\zeta -\zeta_{2}) \me^{\frac{
\pi \nu}{2}}(-1)^{\mi \nu} \mathcal{X}_{A} \sqrt{t}} \right) \!
\int\nolimits_{0}^{+\infty} \mathbf{D}_{-\mi \nu}(z)z^{-\mi \nu}
\me^{-\frac{z^{2}}{4}} \, \md z \\
 &+ \mathcal{O} \! \left( \! \left( \tfrac{c^{\mathcal{S}}(\zeta_{
1}) \underline{c}(\zeta_{2},\zeta_{3},\overline{\zeta_{3}})(\delta_{
B}^{0})^{2}}{(\zeta -\zeta_{1}) \vert \zeta_{1}-\zeta_{3} \vert
\sqrt{(\zeta_{1}-\zeta_{2})} \, \, \mathcal{X}_{B}} + \tfrac{c^{
\mathcal{S}}(\zeta_{2}) \underline{c}(\zeta_{1},\zeta_{3},\overline{
\zeta_{3}})(\delta_{A}^{0})^{2}}{(\zeta -\zeta_{2}) \vert \zeta_{2}
-\zeta_{3} \vert \sqrt{(\zeta_{1}-\zeta_{2})} \, \, \mathcal{X}_{
A}} \right) \! \tfrac{\ln t}{t} \right), \\
m^{\Sigma^{\sharp}}_{21}(\zeta) \! &= \! -\left( \tfrac{r(\zeta_{1}
)(\delta_{B}^{0})^{-2} \me^{\frac{\pi \nu}{2}} \me^{\frac{\mi \pi}{
4}}}{2 \pi \mi (\zeta -\zeta_{1}) \mathcal{X}_{B} \sqrt{t}} \! - \!
\tfrac{r(\zeta_{1})(1-\vert r(\zeta_{1}) \vert^{2})^{-1}(\delta_{
B}^{0})^{-2} \me^{-\frac{3 \pi \mi}{4}}}{2 \pi \mi (\zeta -\zeta_{
1}) \me^{\frac{3 \pi \nu}{2}} \mathcal{X}_{B} \sqrt{t}} \right) \!
\int\nolimits_{0}^{+\infty} \mathbf{D}_{-\mi \nu}(z)z^{-\mi \nu}
\me^{-\frac{z^{2}}{4}} \, \md z \\
 &- \left( \tfrac{\overline{r(\zeta_{1})}(\delta_{A}^{0})^{-2} \me^{
-\frac{\pi \nu}{2}} \me^{\frac{3 \pi \mi}{4}}}{2 \pi \mi (\zeta -
\zeta_{2})(-1)^{-\mi \nu} \mathcal{X}_{A} \sqrt{t}} \! - \! \tfrac{
\overline{r(\zeta_{1})}(1-\vert r(\zeta_{1}) \vert^{2})^{-1}(\delta_{
A}^{0})^{-2} \me^{-\frac{\mi \pi}{4}}}{2 \pi \mi (\zeta -\zeta_{2}
) \me^{\frac{\pi \nu}{2}}(-1)^{-\mi \nu} \mathcal{X}_{A} \sqrt{t}}
\right) \! \int\nolimits_{0}^{+\infty} \mathbf{D}_{\mi \nu}(z) z^{
\mi \nu} \me^{-\frac{z^{2}}{4}} \, \md z \\
 &+ \mathcal{O} \! \left( \! \left( \tfrac{c^{\mathcal{S}}(\zeta_{
1}) \underline{c}(\zeta_{2},\zeta_{3},\overline{\zeta_{3}})(\delta_{
B}^{0})^{-2}}{(\zeta -\zeta_{1}) \vert \zeta_{1}-\zeta_{3} \vert
\sqrt{(\zeta_{1}-\zeta_{2})} \, \, \mathcal{X}_{B}}+\tfrac{c^{
\mathcal{S}}(\zeta_{2}) \underline{c}(\zeta_{1},\zeta_{3},\overline{
\zeta_{3}})(\delta_{A}^{0})^{-2}}{(\zeta -\zeta_{2}) \vert \zeta_{2}
-\zeta_{3} \vert \sqrt{(\zeta_{1}-\zeta_{2})} \, \, \mathcal{X}_{
A}} \right) \! \tfrac{\ln t}{t} \right), \\
m^{\Sigma^{\sharp}}_{22}(\zeta) \! &= \! 1 \! + \! \tfrac{\overline{
r(\zeta_{1})}(\delta_{B}^{0})^{2} \me^{\frac{\pi \nu}{2}} \me^{-\frac{
\mi \pi}{4}}}{2 \pi \mi (\zeta -\zeta_{1}) \beta^{\Sigma_{B^{0}}}_{12}
\mathcal{X}_{B} \sqrt{t}} \int\nolimits_{0}^{+\infty}(\me^{\frac{\mi
\pi}{4}} \partial_{z} \mathbf{D}_{\mi \nu}(z) \! + \! \tfrac{\mi}{2}
\me^{-\frac{\mi \pi}{4}}z\mathbf{D}_{\mi \nu}(z))z^{\mi \nu} \me^{-
\frac{z^{2}}{4}} \, \md z \\
 &-\tfrac{\overline{r(\zeta_{1})}(1-\vert r(\zeta_{1}) \vert^{2})^{
-1}(\delta_{B}^{0})^{2} \me^{\frac{3 \pi \mi}{4}}}{2 \pi \mi (\zeta
-\zeta_{1}) \beta^{\Sigma_{B^{0}}}_{12} \me^{\frac{3 \pi \nu}{2}}
\mathcal{X}_{B} \sqrt{t}} \int\nolimits_{0}^{+\infty}(\me^{-\frac{3
\pi \mi}{4}} \partial_{z} \mathbf{D}_{\mi \nu}(z) \! + \! \tfrac{\mi}
{2} \me^{\frac{3 \pi \mi}{4}}z \mathbf{D}_{\mi \nu}(z))z^{\mi \nu}
\me^{-\frac{z^{2}}{4}} \, \md z \\
 &+ \tfrac{r(\zeta_{1})(\delta_{A}^{0})^{2} \me^{-\frac{\pi \nu}
{2}} \me^{-\frac{3 \pi \mi}{4}}}{2 \pi \mi (\zeta -\zeta_{2})
\beta^{\Sigma_{A^{0}}}_{12}(-1)^{\mi \nu} \mathcal{X}_{A} \sqrt{t}}
\int\nolimits_{0}^{+\infty}(\me^{\frac{3 \pi \mi}{4}} \partial_{z}
\mathbf{D}_{-\mi \nu}(z) \! - \! \tfrac{\mi}{2} \me^{-\frac{3 \pi
\mi}{4}}z \mathbf{D}_{-\mi \nu}(z))z^{-\mi \nu} \me^{-\frac{z^{2}}
{4}} \, \md z \\
 &- \tfrac{r(\zeta_{1})(1-\vert r(\zeta_{1}) \vert^{2})^{-1}(\delta_{
A}^{0})^{2} \me^{\frac{\mi \pi}{4}}}{2 \pi \mi (\zeta -\zeta_{2})
\beta^{\Sigma_{A^{0}}}_{12} \me^{\frac{\pi \nu}{2}}(-1)^{\mi \nu}
\mathcal{X}_{A} \sqrt{t}} \int\nolimits_{0}^{+\infty}(\me^{-\frac{
\mi \pi}{4}} \partial_{z} \mathbf{D}_{-\mi \nu}(z) \! - \! \tfrac{
\mi}{2} \me^{\frac{\mi \pi}{4}}z \mathbf{D}_{-\mi \nu}(z))z^{-\mi
\nu} \me^{-\frac{z^{2}}{4}} \, \md z \\
 &+ \mathcal{O} \! \left( \! \left( \tfrac{c^{\mathcal{S}}(\zeta_{
1}) \underline{c}(\zeta_{2},\zeta_{3},\overline{\zeta_{3}})(\delta_{
B}^{0})^{2}}{(\zeta -\zeta_{1}) \vert \zeta_{1}-\zeta_{3} \vert
\sqrt{(\zeta_{1}-\zeta_{2})} \, \, \mathcal{X}_{B}}+\tfrac{c^{
\mathcal{S}}(\zeta_{2}) \underline{c}(\zeta_{1},\zeta_{3},\overline{
\zeta_{3}})(\delta_{A}^{0})^{2}}{(\zeta -\zeta_{2}) \vert \zeta_{2}
-\zeta_{3} \vert \sqrt{(\zeta_{1}-\zeta_{2})} \, \, \mathcal{X}_{
A}} \right) \! \tfrac{\ln t}{t} \right),
\end{align*}
where $\delta_{A}^{0}$ and $\delta_{B}^{0}$ are defined in Proposition~5.1,
\begin{gather*}
\chi (\zeta_{1}) \! := \! \dfrac{\mi}{2 \pi} \int_{-\infty}^{0} \ln \vert
\mu \! - \! \zeta_{1} \vert \md \ln (1 \! - \! \vert r(\mu) \vert^{2})
\! + \! \dfrac{\mi}{2 \pi} \int_{\zeta_{2}}^{\zeta_{1}} \ln \vert
\mu \! - \! \zeta_{1} \vert \md \ln (1 \! - \! \vert r(\mu) \vert^{2}), \\
\chi (\zeta_{2}) \! := \! -\chi (\zeta_{1}) \! + \! \dfrac{\mi}{2 \pi}
\int_{-\infty}^{0} \ln \vert \mu \vert \md \ln (1 \! - \! \vert
r(\mu) \vert^{2}) \! + \! \dfrac{\mi}{2 \pi} \int_{\zeta_{2}}^{
\zeta_{1}} \ln \vert \mu \vert \md \ln (1 \! - \! \vert r(\mu)
\vert^{2}), \\
\mathcal{X}_{B} \! = \! \tfrac{\vert \zeta_{1}-\zeta_{3} \vert}{\zeta_{1}}
\sqrt{\tfrac{2(\zeta_{1}-\zeta_{2})}{\zeta_{1}}} \,, \qquad \quad \mathcal{
X}_{A} = \tfrac{\vert \zeta_{2}-\zeta_{3} \vert}{\zeta_{2}} \sqrt{\tfrac{
2(\zeta_{1}-\zeta_{2})}{\zeta_{2}}} \,, \\
\beta^{\Sigma_{B^{0}}}_{12}=\overline{\beta^{\Sigma_{B^{0}}}_{21}}=\tfrac{
\sqrt{2 \pi} \, \me^{-\frac{\pi \nu}{2}} \me^{\frac{\mi \pi}{4}}}{r(\zeta_{
1}) \, \overline{\Gamma (\mi \nu)}}, \qquad \quad \beta^{\Sigma_{A^{0}}}_{
12}=\overline{\beta^{\Sigma_{A^{0}}}_{21}}=\tfrac{\sqrt{2 \pi} \, \me^{-
\frac{\pi \nu}{2}} \me^{-\frac{\mi \pi}{4}}}{\overline{r(\zeta_{1})} \,
\Gamma (\mi \nu)}.
\end{gather*}
{}From the above expressions for $\mathcal{X}_{B}$ and $\mathcal{X}_{A}$,
one notes that the terms $\vert \zeta_{1} \! - \! \zeta_{3} \vert \zeta_{
1}^{-1}$ and $\vert \zeta_{2} \! - \! \zeta_{3} \vert \zeta_{2}^{-1}$ must
be calculated: using the expressions for $\{\zeta_{i}\}_{i=1}^{3}$ defined
in Theorem~3.1, Eqs.~(16) and~(17), along with the fact that $\zeta_{1}
\zeta_{2} \! = \! \zeta_{3} \overline{\zeta_{3}} \! = \! 1$, one shows
that $\vert \zeta_{k} \! - \! \zeta_{3} \vert \zeta_{k}^{-1} \! = \! (2
\zeta_{k})^{-1/2}(z_{o}^{2} \! + \! 32)^{1/4}$, $k \! \in \! \{1,2\}$.
Finally, assembling the above formulae, using the identities \cite{a40}
$\partial_{z} \mathbf{D}_{z_{1}}(z) \! = \! \tfrac{1}{2}(z_{1} \mathbf{
D}_{z_{1}-1}(z) \! - \! \mathbf{D}_{z_{1}+1}(z))$, $z \mathbf{D}_{z_{1}}
(z) \! = \! \mathbf{D}_{z_{1}+1}(z) \! + \! z_{1} \mathbf{D}_{z_{1}-1}
(z)$, and $\vert \Gamma (\mi \nu) \vert^{2} \! = \! \tfrac{\pi}{\nu \sinh
(\pi \nu)}$, and the integral \cite{a40}
\begin{equation*}
\int_{0}^{+\infty} \mathbf{D}_{-z_{1}}(z)z^{z_{2}-1} \exp (-\tfrac{
z^{2}}{4}) \, \md z = \dfrac{\sqrt{\pi} \, \exp (-\frac{1}{2}(z_{1} \!
+ \! z_{2}) \ln 2) \Gamma (z_{2})}{\Gamma (\frac{1}{2}(z_{1} \!
+ \! z_{2}) \! + \! \frac{1}{2})}, \qquad \Re (z_{2}) \! > \! 0,
\end{equation*}
one obtains, upon using (repeatedly) the relation $\vert r(\zeta_{1}) \vert
\vert \Gamma (\mi \nu) \vert \nu \me^{\frac{\pi \nu}{2}} \! = \! \sqrt{2 \pi
\nu}$, and the fact that, for $\zeta \! \in \! (\mathbb{C} \setminus \cup_{
\lambda \in \{\zeta_{2},\zeta_{1}\}} \mathbb{U}(\lambda;\varepsilon)) \cap
(\Omega_{1} \cup \Omega_{2})$, $m^{c}(\zeta) \! = \! m^{\Sigma^{\sharp}}
(\zeta)(\delta (\zeta))^{\sigma_{3}}(\mathrm{I} \! + \! \mathcal{O}(\tfrac{
\underline{c}(\zeta_{1},\zeta_{2},\zeta_{3},\overline{\zeta_{3}}) \diamondsuit
(\zeta)}{\vert z_{o}+\zeta_{1}+\zeta_{2} \vert^{l}t^{l}}))$, with arbitrarily
large $l \! \in \! \mathbb{Z}_{\geqslant 1}$, $\vert \vert \diamondsuit
(\cdot) \vert \vert_{\mathcal{L}^{\infty}_{\mathrm{M}_{2}(\mathbb{C})}
(\mathbb{C} \setminus \cup_{\lambda \in \{\zeta_{2},\zeta_{1}\}} \mathbb{U}
(\lambda;\varepsilon))} \! < \! \infty$, and $\delta (\zeta)$ given in
Proposition~4.1, the result stated in the Lemma; furthermore, one shows that
the symmetry reduction $m^{c}(\zeta) \! = \! \sigma_{1} \overline{m^{c}
(\overline{\zeta})} \, \sigma_{1}$ is satisfied, and verifies that, to
$\mathcal{O}(t^{-1} \ln t)$, $(m^{c}(0) \sigma_{2})^{2} \! = \! \mathrm{
I}$. \hfill $\square$
\begin{bbbbb}
As $t \! \to \! +\infty$ such that $0 \! < \! \zeta_{2} \! < \! \tfrac{1}{M}
\! < \! M \! < \! \zeta_{1}$ and $\vert \zeta_{3} \vert^{2} \! = \! 1$, with
$M \! \in \! \mathbb{R}_{>1}$ and bounded,
\begin{align*}
(\Delta_{o})_{11} \! &= \! -\tfrac{2 \mi \sqrt{\nu (\zeta_{1})} \, \cos
(\Theta^{+}(z_{o},t)+\frac{\pi}{4})}{\sqrt{t(\zeta_{1}-\zeta_{2})} \, (z_{
o}^{2}+32)^{1/4}} \! + \! \mathcal{O} \! \left( \! \left( \tfrac{c^{\mathcal{
S}}(\zeta_{1}) \underline{c}(\zeta_{2},\zeta_{3},\overline{\zeta_{3}})}{\sqrt{
\zeta_{1}(z_{o}^{2}+32)}} \! + \! \tfrac{c^{\mathcal{S}}(\zeta_{2})
\underline{c}(\zeta_{1},\zeta_{3},\overline{\zeta_{3}})}{\sqrt{\zeta_{2}(z_{
o}^{2}+32)}} \right) \! \tfrac{\ln t}{(\zeta_{1}-\zeta_{2})t} \right), \\
(\Delta_{o})_{12} \! &= \! -\mi \exp \! \left(-\mi \left(\int_{-\infty}^{0}
\dfrac{\ln (1 \! - \! \vert r(\mu) \vert^{2})}{\mu} \, \dfrac{\md \mu}{2
\pi} \! + \! \int_{\zeta_{2}}^{\zeta_{1}} \dfrac{\ln (1 \! - \! \vert r(\mu)
\vert^{2})}{\mu} \, \frac{\md \mu}{2 \pi} \right) \right) \\
&\times \left( 1 \! + \! \mathcal{O} \! \left( \! \left( \dfrac{c^{
\mathcal{S}}(\zeta_{1}) \underline{c}(\zeta_{2},\zeta_{3},\overline{
\zeta_{3}})}{\sqrt{\zeta_{1}(z_{o}^{2} \! + \! 32)}} \! + \! \dfrac{
c^{\mathcal{S}}(\zeta_{2}) \underline{c}(\zeta_{1},\zeta_{3},\overline{
\zeta_{3}})}{\sqrt{\zeta_{2}(z_{o}^{2} \! + \! 32)}} \right) \!
\dfrac{\ln t}{(\zeta_{1} \! - \! \zeta_{2})t} \right) \right), \\
(\Delta_{o})_{21} \! &= \! \mi \exp \! \left( \mi \left(\int_{-\infty}^{0}
\dfrac{\ln (1 \! - \! \vert r(\mu) \vert^{2})}{\mu} \, \dfrac{\md \mu}{2
\pi} \! + \! \int_{\zeta_{2}}^{\zeta_{1}} \dfrac{\ln (1 \! - \! \vert r(\mu)
\vert^{2})}{\mu} \, \frac{\md \mu}{2 \pi} \right) \right) \\
 &\times \left( 1 \! + \! \mathcal{O} \! \left( \! \left( \dfrac{
c^{\mathcal{S}}(\zeta_{1}) \underline{c}(\zeta_{2},\zeta_{3},
\overline{\zeta_{3}})}{\sqrt{\zeta_{1}(z_{o}^{2} \! + \! 32)}} \!
+ \! \dfrac{c^{\mathcal{S}}(\zeta_{2}) \underline{c}(\zeta_{1},
\zeta_{3},\overline{\zeta_{3}})}{\sqrt{\zeta_{2}(z_{o}^{2} \! +
\! 32)}} \right) \! \dfrac{\ln t}{(\zeta_{1} \! - \! \zeta_{2})t} \right)
\right), \\
(\Delta_{o})_{22} \! &= \! \tfrac{2 \mi \sqrt{\nu (\zeta_{1})} \, \cos
(\Theta^{+}(z_{o},t)+\frac{\pi}{4})}{\sqrt{t(\zeta_{1}-\zeta_{2})} \, (z_{
o}^{2}+32)^{1/4}} \! + \! \mathcal{O} \! \left( \! \left( \tfrac{c^{\mathcal{
S}}(\zeta_{1}) \underline{c}(\zeta_{2},\zeta_{3},\overline{\zeta_{3}})}{\sqrt{
\zeta_{1}(z_{o}^{2}+32)}} \! + \! \tfrac{c^{\mathcal{S}}(\zeta_{2}) \underline{
c}(\zeta_{1},\zeta_{3},\overline{\zeta_{3}})}{\sqrt{\zeta_{2}(z_{o}^{2}+32)}}
\right) \! \tfrac{\ln t}{(\zeta_{1}-\zeta_{2})t} \right),
\end{align*}
where $\nu (z)$ and $\Theta^{+}(z_{o},t)$ are given in Theorem~{\rm 3.1},
Eqs.~{\rm (12)}--{\rm (14)}.
\end{bbbbb}

\emph{Proof.} One recalls {}from Lemma~2.6 that $(x,t$ dependences
suppressed) $\Delta_{o}m^{c}(0) \! = \! \sigma_{2}$: one deduces {}from this
that $(\Delta_{o})_{11} \! = \! \mi m^{c}_{21}(0)$, $(\Delta_{o})_{12} \! =
\! -\mi m^{c}_{11}(0)$, $(\Delta_{o})_{21} \! = \! \mi m^{c}_{22}(0)$, and
$(\Delta_{o})_{22} \! = \! -\mi m^{c}_{12}(0)$. Using the formulae for $m^{
c}_{ij}(\zeta)$, $i,j \! \in \! \{1,2\}$, given in Lemma~6.1, and the fact
that, via an integration by parts argument, $\delta (0) \zeta_{1}^{-2 \mi
\nu (\zeta_{1})} \exp (-\tfrac{\mi \Omega^{+}(0)}{2}) \! = \! 1$, one
obtains the result stated in the Proposition; furthermore, one also verifies
that, to $\mathcal{O}(t^{-1} \ln t)$, $\Delta_{o}$ has the order 2 matrix
involutive structure stated in Lemma~2.6, and $\det (\Delta_{o}) \! = \!
-1$. \hfill $\square$
\begin{ccccc}
As $t \! \to \! +\infty$ such that $0 \! < \! \zeta_{2} \! < \! \tfrac{1}{M}
\! < \! M \! < \! \zeta_{1}$ and $\vert \zeta_{3} \vert^{2} \! = \! 1$, with
$M \! \in \! \mathbb{R}_{>1}$ and bounded, $u(x,t)$, the solution of the
Cauchy problem for the {\rm D${}_{f}$NLSE}, has the asymptotics stated in
Theorem~{\rm 3.1}, Eqs.~{\rm (9)}, {\rm (10)}, {\rm (12)}, {\rm (13)},
{\rm (14)}, {\rm (16)}, and~{\rm (17)}, and $\int_{\pm \infty}^{x}(\vert
u(\xi,t) \vert^{2} \! - \! 1) \, \md \xi$ have the asymptotics stated in
Theorem~{\rm 3.1}, Eqs.~{\rm (25)} and~{\rm (26)}, with $\theta^{+}(z)$
(respectively~$\theta^{-}(z))$ defined in Theorem~{\rm 3.1}, Eq.~{\rm (10)}
(respectively~Eq.~{\rm (11))}.
\end{ccccc}

\emph{Proof.} {}From the proof of Lemma~2.4, one recalls that $(x,t$
dependences suppressed) $m(\zeta) \! =_{\genfrac{}{}{0pt}{2}{\zeta \to \infty}
{\zeta \in \mathbb{C} \setminus \mathbb{R}}} \! \mathrm{I} \! + \! \tfrac{1}
{\zeta} \!
\left(
\begin{smallmatrix}
\mi \int_{+\infty}^{x}(\vert u(\xi,t) \vert^{2}-1) \, \md \xi &
-\mi u(x,t) \\
\mi \overline{u(x,t)} & -\mi \int_{+\infty}^{x}(\vert u(\xi,t)
\vert^{2}-1) \, \md \xi
\end{smallmatrix}
\right) \! + \! \mathcal{O}(\zeta^{-2})$; in particular, and without loss of
generality, the $\zeta \! \to \! \infty$ asymptotics are taken in the domain
$\Omega_{1} \cup \Omega_{2} \subset \mathbb{C} \setminus \mathbb{R}$
(Figure~3). One recalls the ordered factorisation for $m(\zeta)$ given in
Lemma~2.6, Eq.~(7), namely, $m(\zeta) \! = \! (\mathrm{I} \! + \! \Delta_{o}
\zeta^{-1})m^{c}(\zeta)$, with $m^{c}(\zeta)$ (respectively~$\Delta_{o})$
given in Lemma~6.1 (respectively~Proposition~6.1). {}From the $\zeta \! \to
\! \infty$, $\zeta \! \in \! \Omega_{1} \cup \Omega_{2}$, asymptotics of
$m(\zeta)$ and its ordered factorisation (both given above), one arrives at
(and the complex conjugates of) $-\mi u(x,t) \! = \! \lim_{\genfrac{}{}{0pt}
{2}{\zeta \to \infty}{\zeta \in \Omega_{1} \cup \Omega_{2}}}(\zeta (m(\zeta)
\! - \! \mathrm{I}))_{12} \! = \! (\Delta_{o})_{12} \! + \! \lim_{\genfrac{}
{}{0pt}{2}{\zeta \to \infty}{\zeta \in \Omega_{1} \cup \Omega_{2}}}(\zeta
(m^{c}(\zeta) \! - \! \mathrm{I}))_{12}$ and $\mi \int_{+\infty}^{x}(\vert
u(\xi,t) \vert^{2} \! - \! 1) \, \md \xi \! = \! \lim_{\genfrac{}{}{0pt}{2}
{\zeta \to \infty}{\zeta \in \Omega_{1} \cup \Omega_{2}}}(\zeta (m(\zeta) \!
- \! \mathrm{I}))_{11} \! = \! (\Delta_{o})_{11} \! + \! \lim_{\genfrac{}{}
{0pt}{2}{\zeta \to \infty}{\zeta \in \Omega_{1} \cup \Omega_{2}}}(\zeta (m^{
c}(\zeta) \! - \! \mathrm{I}))_{11}$: using the expressions for $m^{c}_{ij}
(\zeta)$ and $(\Delta_{o})_{ij}$, $i,j \! \in \! \{1,2\}$, given in Lemma~6.1
and Proposition~6.1, respectively, the asymptotics $(\delta (\zeta))^{\pm 1}
\! =_{\genfrac{}{}{0pt}{2}{\zeta \to \infty}{\zeta \in \Omega_{1} \cup
\Omega_{2}}} \! 1 \! \pm \! \mi \! \left(\int_{-\infty}^{0} \ln (1 \! - \!
\vert r(\mu) \vert^{2}) \, \tfrac{\md \mu}{2 \pi} \! + \! \int_{\zeta_{2}}^{
\zeta_{1}} \ln (1 \! - \! \vert r(\mu) \vert^{2}) \, \tfrac{\md \mu}{2 \pi}
\right) \! \zeta^{-1} \! + \! \mathcal{O}(\zeta^{-2})$, and the trace
identity (cf.~Corollary~2.4) $\int_{-\infty}^{+\infty}(\vert u(\xi,t) \vert^{
2} \! - \! 1) \, \md \xi \! = \! -\int_{-\infty}^{+\infty} \ln (1 \! - \!
\vert r(\mu) \vert^{2}) \, \tfrac{\md \mu}{2 \pi}$, one obtains the result
stated in the Lemma. \hfill $\square$
\section{Asymptotics as $t \! \to \! -\infty$}
In this section, as $t \! \to \! -\infty$ and $x \! \to \! +\infty$ such that
$z_{o} \! := \! x/t \! < \! -2$, the RHP for $m^{c}(\zeta)$ on $\sigma_{c}$
formulated in Lemma~2.6 is discussed succinctly, and the asymptotics of $u
(x,t)$ and related integrals are obtained. As the calculations subsumed in
this section are analogous to those presented in Sections~4--6, only final
results, with in one instance a sketch of a proof, are given.

As $t \! \to \! -\infty$ and $x \! \to \! +\infty$ such that $z_{o} \! < \!
-2$, one begins by decomposing the complex plane of the spectral parameter
$\zeta$ according to the signature of $\Re (\mi t \theta^{u}(\zeta))$ (see
Figure~6), where, {}from Eq.~(8), $\theta^{u}(\zeta) \! = \! \tfrac{1}{2}
(\zeta \! - \! \tfrac{1}{\zeta})(z_{o} \! + \! \zeta \! + \! \tfrac{1}{
\zeta})$, with $\{\zeta_{i}\}_{i=1}^{4}$ defined in Theorem~3.1, Eqs.~(16)
and~(17), $0 \! < \! \zeta_{2} \! < \! \zeta_{1}$, $\vert \zeta_{3} \vert^{
2} \! = \! 1$, and $\pm \! \leftrightarrow \! \Re (\mi t \theta^{u}(\zeta))
\! \gtrless \! 0$.
\begin{figure}[tbh]
\begin{center}
\unitlength=1cm
\vspace{0.70cm}
\begin{picture}(12,5)(0,0)
\thicklines
\put(5,2.55){\makebox(0,0){$\centerdot$}}
\put(5,2.60){\makebox(0,0){$\centerdot$}}
\put(7,2.5){\makebox(0,0){$\bullet$}}
\put(9,2.5){\makebox(0,0){$\bullet$}}
\put(2.5,1.25){\makebox(0,0){$\bullet$}}
\put(2.5,3.75){\makebox(0,0){$\bullet$}}
\put(4.75,2.15){\makebox(0,0){$0$}}
\put(6.75,2.15){\makebox(0,0){$\zeta_{2}$}}
\put(8.75,2.15){\makebox(0,0){$\zeta_{1}$}}
\put(3,1.25){\makebox(0,0){$\overline{\zeta_{3}}$}}
\put(3,3.75){\makebox(0,0){$\zeta_{3}$}}
\put(1.25,3.75){\makebox(0,0){$-$}}
\put(1.25,1.25){\makebox(0,0){$+$}}
\put(6,3.75){\makebox(0,0){$+$}}
\put(6,1.25){\makebox(0,0){$-$}}
\put(8,3.75){\makebox(0,0){$-$}}
\put(8,1.25){\makebox(0,0){$+$}}
\put(10.5,3.75){\makebox(0,0){$+$}}
\put(10.5,1.25){\makebox(0,0){$-$}}
\put(11.75,4.05){\shortstack[c]{complex\\$\zeta$-plane}}
\put(0,2.5){\line(1,0){12}}
\qbezier[50](5,0)(5,2.5)(5,5)
\qbezier[50](7,0)(7,2.5)(7,5)
\qbezier[50](9,0)(9,2.5)(9,5)
\end{picture}
\end{center}
\caption{Signature graph of $\Re (\mi t \theta^{u}(\zeta))$ as
$t \! \to \! -\infty$}
\end{figure}
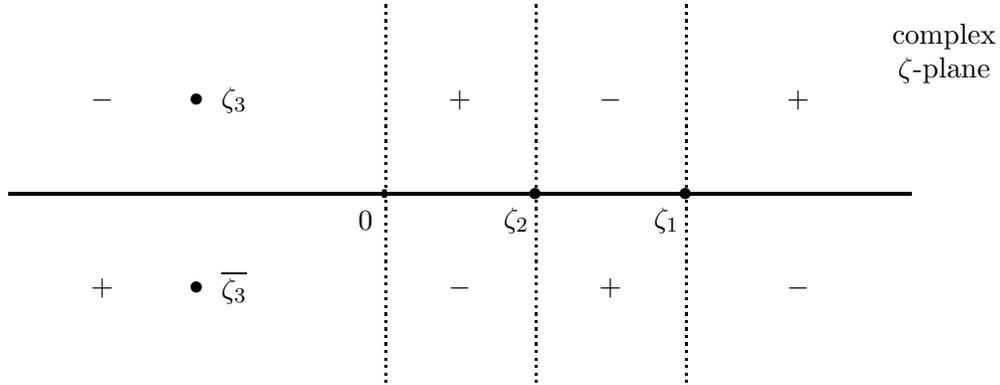
One now reorients $\sigma_{c}$, oriented {}from $-\infty$ to $+\infty$,
according to, and consistent with, the signature graph of $\Re (\mi t
\theta^{u}(\zeta))$, leading to the reoriented contour $\sigma_{c}^{
\prime \prime}$ (see Figure~7).
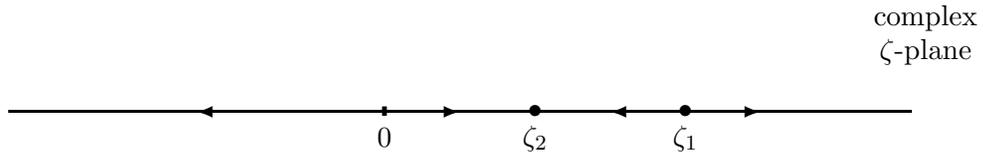
\begin{figure}[tbh]
\begin{center}
\unitlength=1cm
\begin{picture}(12,4)(0,0)
\thicklines
\put(5,2.05){\makebox(0,0){$\centerdot$}}
\put(5,2.10){\makebox(0,0){$\centerdot$}}
\put(0,2){\line(1,0){12}}
\put(7,2){\makebox(0,0){$\bullet$}}
\put(9,2){\makebox(0,0){$\bullet$}}
\put(5,1.65){\makebox(0,0){$0$}}
\put(7,1.65){\makebox(0,0){$\zeta_{2}$}}
\put(9,1.65){\makebox(0,0){$\zeta_{1}$}}
\put(5,2){\vector(-1,0){2.5}}
\put(5,2){\vector(1,0){1.0}}
\put(9,2){\vector(-1,0){1.0}}
\put(9,2){\vector(1,0){1.0}}
\put(11.5,2.70){\shortstack[c]{complex\\$\zeta$-plane}}
\end{picture}
\vspace{-1.20cm}
\end{center}
\caption{Reoriented contour $\sigma_{c}^{\prime \prime}$}
\end{figure}
Denoting $m^{c}(\zeta)$ on
$\sigma_{c}^{\prime \prime}$ by $\mathscr{M}^{c}(\zeta)$, one shows that
(recalling Lemma~2.6) $\mathscr{M}^{c}(\zeta) \colon \mathbb{C} \setminus
\sigma_{c}^{\prime \prime} \! \to \! \mathrm{SL}(2,\mathbb{C})$ solves the
following (normalised at $\infty)$ RHP: (1) $\mathscr{M}^{c}(\zeta)$ is
piecewise holomorphic $\forall \, \zeta \! \in \! \mathbb{C} \setminus
\sigma_{c}^{\prime \prime}$; (2) $\mathscr{M}^{c}_{\pm}(\zeta) \! := \!
\lim_{\genfrac{}{}{0pt}{2}{\zeta^{\prime} \, \to \, \zeta}{\zeta^{\prime} \,
\in \, \pm \, \mathrm{side} \, \mathrm{of} \, \sigma_{c}^{\prime \prime}}}
\! \mathscr{M}^{c}(\zeta^{\prime})$ satisfy the jump condition $\mathscr{M}^{
c}_{+}(\zeta) \! = \! \mathscr{M}^{c}_{-}(\zeta) \mathscr{G}^{c}(\zeta)$,
$\zeta \! \in \! \sigma_{c}^{\prime \prime}$, where
\begin{equation*}
\mathscr{G}^{c}(\zeta) \! := \!
\begin{cases}
(\mathrm{I} \! - \! \overline{r(\overline{\zeta})} \, \me^{-2 \mi
t \theta^{u}(\zeta)} \sigma_{+})(\mathrm{I} \! + \! r(\zeta) \me^{
2 \mi t \theta^{u}(\zeta)} \sigma_{-}), &\text{$\zeta \! \in \! (0,
\zeta_{2}) \cup (\zeta_{1},+\infty),$} \\
(\mathrm{I} \! - \! r(\zeta) \me^{2 \mi t \theta^{u}(\zeta)} \sigma_{
-})(\mathrm{I} \! + \! \overline{r(\overline{\zeta})} \, \me^{-2 \mi
t \theta^{u}(\zeta)} \sigma_{+}), &\text{$\zeta \! \in \! (-\infty,0)
\cup (\zeta_{2},\zeta_{1});$}
\end{cases}
\end{equation*}
(3) as $\zeta \! \to \! \infty$, $\zeta \! \in \! \mathbb{C} \setminus
\sigma_{c}^{\prime \prime}$, $\mathscr{M}^{c}(\zeta) \! = \! \mathrm{
I} \! + \! \mathcal{O}(\zeta^{-1})$; and (4) $\mathscr{M}^{c}(\zeta)$
satisfies the symmetry reduction $\mathscr{M}^{c}(\zeta) \! = \!
\sigma_{1} \overline{\mathscr{M}^{c}(\overline{\zeta})} \, \sigma_{1}$
and the condition $(\mathscr{M}^{c}(0) \sigma_{2})^{2} \! = \! \mathrm{
I}$. The analogue, therefore, of Proposition~4.1 and Lemma~4.1 is the
following
\begin{ccccc}
Let $\widetilde{\delta}(\zeta)$ solve the following scalar discontinuous
{\rm RHP:}
\begin{align*}
\widetilde{\delta}_{+}(\zeta) \! &= \!
\begin{cases}
\widetilde{\delta}_{-}(\zeta)(1 \! - \! r(\zeta) \overline{r(\overline{
\zeta})}), &\text{$\Re (\zeta) \! \in \! (0,\zeta_{2}) \cup (\zeta_{1},
+\infty),$} \\
\widetilde{\delta}_{-}(\zeta) \! = \! \widetilde{\delta}(\zeta), &\text{
$\Re (\zeta) \! \in \! (-\infty,0) \cup (\zeta_{2},\zeta_{1}),$}
\end{cases} \\
\widetilde{\delta}(\zeta) \! &\underset{\zeta \to \infty}{=} \! 1 \! +
\! \mathcal{O}(\zeta^{-1}),
\end{align*}
with index $\kappa \! := \! \tfrac{1}{2 \pi}[\arg (1 \! - \! r(\zeta)
\overline{r(\overline{\zeta})})]_{-\infty}^{+\infty} \! = \! 0$. The
unique solution of this {\rm RHP} can be written as
\begin{equation*}
\widetilde{\delta}(\zeta) \! = \! \left( \dfrac{\zeta \! - \! \zeta_{2}
}{\zeta \! - \! \zeta_{1}} \right)^{\mi \nu} \exp \! \left( \dfrac{\mi}
{2 \pi} \! \left( \int\nolimits_{0}^{\zeta_{2}}+\int\nolimits_{\zeta_{
1}}^{+\infty} \right) \ln (\mu \! - \! \zeta) \md \ln (1 \! - \! \vert
r(\mu) \vert^{2}) \right),
\end{equation*}
where $\{\zeta_{i}\}_{i=1}^{2}$ are defined in Theorem~{\rm 3.1},
Eqs.~{\rm (16)} and~{\rm (17)}, $\nu \! := \! \nu (\zeta_{1}) \! = \!
-\tfrac{1}{2 \pi} \ln (1 \! - \! \vert r(\zeta_{1}) \vert^{2}) \! \in \!
\mathbb{R}_{+}$, $\widetilde{\delta}(\zeta) \overline{\widetilde{\delta}
(\overline{\zeta})} \! = \! 1$, $\widetilde{\delta}(\zeta) \widetilde{\delta}
(\tfrac{1}{\zeta}) \! = \! \widetilde{\delta}(0) \! = \! \exp \! \left( \!
\left( \int_{0}^{\zeta_{2}} \! + \! \int_{\zeta_{1}}^{+\infty} \right) \!
\tfrac{\ln (1-\vert r(\mu) \vert^{2})}{\mu} \, \tfrac{\md \mu}{2 \pi}
\right)$, $\vert \widetilde{\delta}_{+}(\zeta) \vert^{2} \! \leqslant \! 1$
and $\vert \widetilde{\delta}_{-}(\zeta) \vert^{2} \! \leqslant \! (1 \! -
\sup_{z \in \mathbb{R}} \vert r(z) \vert^{2})^{-1} \! < \! \infty \, \forall
\, \, \zeta \! \in \! \mathbb{R}$, and $\vert \vert (\widetilde{\delta}
(\cdot))^{\pm 1} \vert \vert_{\mathcal{L}^{\infty}(\mathbb{C})} \! := \!
\sup_{\zeta \in \mathbb{C}} \vert (\widetilde{\delta}(\zeta))^{\pm 1} \vert
\! < \! \infty$. Set $\widetilde{\mathscr{M}}^{c}(\zeta) \! := \! \mathscr{
M}^{c}(\zeta)(\widetilde{\delta}(\zeta))^{-\sigma_{3}}$. Then $\widetilde{
\mathscr{M}}^{c}(\zeta) \colon \mathbb{C} \setminus \sigma_{c}^{\prime
\prime} \! \to \! \mathrm{SL}(2,\mathbb{C})$ solves the following {\rm RHP:}
(1) $\widetilde{\mathscr{M}}^{c}(\zeta)$ is piecewise holomorphic $\forall
\, \zeta \! \in \! \mathbb{C} \setminus \sigma_{c}^{\prime \prime};$ (2)
$\widetilde{\mathscr{M}}^{c}_{\pm}(\zeta) \! := \! \lim_{\genfrac{}{}{0pt}
{2}{\zeta^{\prime} \, \to \, \zeta}{\zeta^{\prime} \, \in \, \pm \, \mathrm{
side} \, \mathrm{of} \, \sigma_{c}^{\prime \prime}}} \! \widetilde{\mathscr{
M}}^{c}(\zeta^{\prime})$ satisfy the jump condition $\widetilde{\mathscr{
M}}^{c}_{+}(\zeta) \! = \! \widetilde{\mathscr{M}}^{c}_{-}(\zeta)
\widetilde{\mathscr{G}}^{c}(\zeta)$, $\zeta \! \in \! \sigma_{c}^{\prime
\prime}$, where
\begin{equation*}
\widetilde{\mathscr{G}}^{c}(\zeta) \! := \!
\begin{cases}
(\mathrm{I} \! - \! \overline{\widetilde{\rho}(\overline{\zeta})}
(\widetilde{\delta}_{-}(\zeta))^{-2} \me^{2 \mi t \theta^{u}(\zeta)}
\sigma_{-})(\mathrm{I} \! + \! \widetilde{\rho}(\zeta)(\widetilde{
\delta}_{+}(\zeta))^{2} \me^{-2 \mi t \theta^{u}(\zeta)} \sigma_{+}),
&\text{$\zeta \! \in \! (0,\zeta_{2}) \cup (\zeta_{1},+\infty),$} \\
(\mathrm{I} \! - \! \overline{\widetilde{\rho}(\overline{\zeta})}
(\widetilde{\delta}(\zeta))^{-2} \me^{2 \mi t \theta^{u}(\zeta)}
\sigma_{-})(\mathrm{I} \! + \! \widetilde{\rho}(\zeta)(\widetilde{
\delta}(\zeta))^{2} \me^{-2 \mi t \theta^{u}(\zeta)} \sigma_{+}),
&\text{$\zeta \! \in \! (-\infty,0) \cup (\zeta_{2},\zeta_{1}),$}
\end{cases}
\end{equation*}
with
\begin{equation*}
\widetilde{\rho}(\zeta) \! := \!
\begin{cases}
\, -\overline{r(\overline{\zeta})}(1 \! - \! r(\zeta) \overline{r
(\overline{\zeta})})^{-1}, &\text{$\zeta \! \in \! (0,\zeta_{2})
\cup (\zeta_{1},+\infty),$} \\
\, \overline{r(\overline{\zeta})}, &\text{$\zeta \! \in \! (-\infty,
0) \cup (\zeta_{2},\zeta_{1});$}
\end{cases}
\end{equation*}
(3) as $\zeta \! \to \! \infty$, $\zeta \! \in \! \mathbb{C} \setminus
\sigma_{c}^{\prime \prime}$, $\widetilde{\mathscr{M}}^{c}(\zeta) \! = \!
\mathrm{I} \! + \! \mathcal{O}(\zeta^{-1});$ and (4) $\widetilde{\mathscr{
M}}^{c}(\zeta)$ satisfies the symmetry reduction $\widetilde{\mathscr{
M}}^{c}(\zeta) \! = \! \sigma_{1} \overline{\widetilde{\mathscr{M}}^{
c}(\overline{\zeta})} \, \sigma_{1}$ and the condition $(\widetilde{
\mathscr{M}}^{c}(0)(\widetilde{\delta}(0))^{\sigma_{3}} \sigma_{2})^{2}
\! = \! \mathrm{I}$.
\end{ccccc}
Now, proceeding according to an analysis analogous to that presented in
Section~4, one arrives at the following ``model'' RHP on $\widetilde{
\Sigma}$ (see Figure~8) for $\widetilde{\mathscr{M}}^{\widetilde{\Sigma}}
(\zeta)$ (analogue of Lemma~4.6):
\begin{figure}[htb]
\begin{center}
\unitlength=1cm
\vspace{0.60cm}
\begin{picture}(10,6)(0,0)
\thicklines
\put(7,3){\makebox(0,0){$\bullet$}}
\put(3,3){\makebox(0,0){$\bullet$}}
\put(7,2.5){\makebox(0,0){$\zeta_{1}$}}
\put(3,2.5){\makebox(0,0){$\zeta_{2}$}}
\put(7,4.25){\makebox(0,0){$\widetilde{\Sigma}_{B}$}}
\put(3,4.25){\makebox(0,0){$\widetilde{\Sigma}_{A}$}}
\put(6,2){\vector(1,1){2.5}}
\put(6,4){\vector(1,-1){2.5}}
\put(6,2){\line(1,1){3.5}}
\put(6,4){\line(1,-1){3.5}}
\put(8,2){\vector(-1,1){1.5}}
\put(8,4){\vector(-1,-1){1.5}}
\put(5.85,4.15){\line(1,-1){1.15}}
\put(5.85,1.85){\line(1,1){1.15}}
\put(4.2,4.2){\vector(-1,-1){1}}
\put(4.2,1.8){\vector(-1,1){1}}
\put(1.8,4.2){\vector(1,-1){1}}
\put(1.8,1.8){\vector(1,1){1}}
\put(3,3){\line(-1,1){1.20}}
\put(3,3){\line(-1,-1){1.20}}
\put(3,3){\line(1,1){1.20}}
\put(3,3){\line(1,-1){1.20}}
\end{picture}
\vspace{-0.35cm}
\end{center}
\caption{Truncated contour $\widetilde{\Sigma} \! := \!
\widetilde{\Sigma}_{A} \cup \widetilde{\Sigma}_{B}$}
\end{figure}
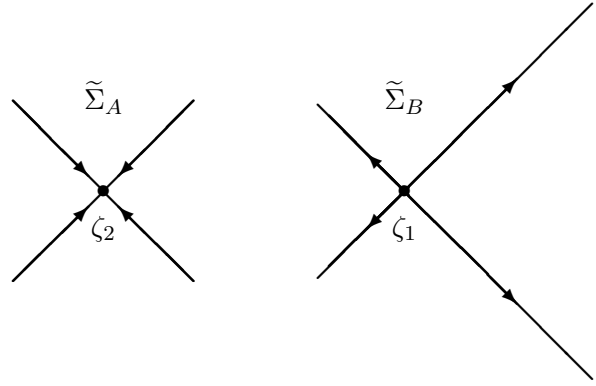
\begin{ccccc}
Set $\widetilde{\Sigma} \! := \! \widehat{\mathscr{L}} \cup
\overline{\widehat{\mathscr{L}}}$, where $\widehat{\mathscr{L}} \!
= \! \{\mathstrut \zeta; \, \zeta \! = \! \zeta_{1} \! + \! \tfrac{
v}{\sqrt{2}}(\zeta_{1} \! - \! \zeta_{2}) \me^{\frac{3 \pi \mi}{4}},
\, -\infty \! < \! v \! < \! \varepsilon\} \cup \{\mathstrut \zeta;
\, \zeta \! = \! \zeta_{2} \! + \! \tfrac{v}{\sqrt{2}}(\zeta_{1} \!
- \! \zeta_{2}) \me^{\frac{\mi \pi}{4}}, \, 0 \! \leqslant \! v \!
< \! \varepsilon\} \cup \{\mathstrut \zeta; \, \zeta \! = \! \zeta_{
2} \! + \! \tfrac{v}{\sqrt{2}} \zeta_{2} \me^{-\frac{3 \pi \mi}{4}},
\, 0 \! \leqslant \! v \! < \! \varepsilon\}$, and $\varepsilon$ is
an arbitrarily fixed, sufficiently small positive real number. As
$t \! \to \! -\infty$ such that $0 \! < \! \zeta_{2} \! < \! \tfrac{1}{
M} \! < \! M \! < \! \zeta_{1}$ and $\vert \zeta_{3} \vert^{2} \!
= \! 1$, with $M \! \in \! \mathbb{R}_{>1}$ and bounded, $\widetilde{
\mathscr{M}}^{\widetilde{\Sigma}}(\zeta) \colon \mathbb{C} \setminus
\widetilde{\Sigma} \! \to \! \mathrm{SL}(2,\mathbb{C})$ solves the
following {\rm RHP:} (1) $\widetilde{\mathscr{M}}^{\widetilde{\Sigma}}
(\zeta)$ is piecewise holomorphic $\forall \, \zeta \! \in \! \mathbb{C}
\setminus \widetilde{\Sigma};$ (2) $\widetilde{\mathscr{M}}^{\widetilde{
\Sigma}}_{\pm}(\zeta) \! := \! \lim_{\genfrac{}{}{0pt}{2}{\zeta^{\prime}
\, \to \, \zeta}{\zeta^{\prime} \, \in \, \pm \, \mathrm{side} \,
\mathrm{of} \, \widetilde{\Sigma}}} \widetilde{\mathscr{M}}^{\widetilde{
\Sigma}}(\zeta^{\prime})$ satisfy the jump condition $\widetilde{\mathscr{
M}}^{\widetilde{\Sigma}}_{+}(\zeta) \! = \! \widetilde{\mathscr{M}}^{
\widetilde{\Sigma}}_{-}(\zeta)(\mathrm{I} \! - \! \widetilde{w}_{-}^{
\widetilde{\Sigma}}(\zeta))^{-1}(\mathrm{I} \! + \! \widetilde{w}_{+}^{
\widetilde{\Sigma}}(\zeta))$, $\zeta \! \in \! \widetilde{\Sigma}$,
where
\begin{align*}
\widetilde{w}_{+}^{\widetilde{\Sigma}}(\zeta) \! &= \!
\left(
\begin{smallmatrix}
0 & 0 \\
0 & 0
\end{smallmatrix}
\right), \quad \, \widetilde{w}_{-}^{\widetilde{\Sigma}}(\zeta) \! =
\! -(\widetilde{\delta}(\zeta))^{\mathrm{ad}(\sigma_{3})} \exp (-\mi
t \theta^{u}(\zeta) \mathrm{ad}(\sigma_{3})) \overline{\widetilde{
\mathcal{R}}(\zeta)} \, \sigma_{-}, \quad \zeta \! \in \! \widehat{
\mathscr{L}} \subset \widetilde{\Sigma}, \\
\widetilde{w}_{+}^{\widetilde{\Sigma}}(\zeta) \! &= \! (\widetilde{
\delta}(\zeta))^{\mathrm{ad}(\sigma_{3})} \exp (-\mi t \theta^{u}
(\zeta) \mathrm{ad}(\sigma_{3})) \widetilde{\mathcal{R}}(\zeta)
\sigma_{+}, \quad \, \widetilde{w}_{-}^{\widetilde{\Sigma}}(\zeta)
\! = \! \left(
\begin{smallmatrix}
0 & 0 \\
0 & 0
\end{smallmatrix}
\right), \quad \, \, \, \zeta \! \in \! \overline{\widehat{\mathscr{L}}}
\subset \widetilde{\Sigma},
\end{align*}
with $\widetilde{\mathcal{R}}(\zeta)$ the analogue of $\mathcal{
R}(\zeta)$ defined in Lemma~{\rm 4.2;} (3) as $\zeta \! \to \! \infty$,
$\zeta \! \in \! \mathbb{C} \setminus \widetilde{\Sigma}$,
$\widetilde{\mathscr{M}}^{\widetilde{\Sigma}}(\zeta) \! = \! \mathrm{
I} \! + \! \mathcal{O}(\zeta^{-1});$ and (4) $\widetilde{\mathscr{
M}}^{\widetilde{\Sigma}}(\zeta)$ satisfies the symmetry reduction
$\widetilde{\mathscr{M}}^{\widetilde{\Sigma}}(\zeta) \! = \! \sigma_{
1} \overline{\widetilde{\mathscr{M}}^{\widetilde{\Sigma}}(\overline{
\zeta})} \, \sigma_{1}$ and the condition $(\widetilde{\mathscr{M}}^{
\widetilde{\Sigma}}(0)(\widetilde{\delta}(0))^{\sigma_{3}} \sigma_{
2})^{2} \! = \! \mathrm{I}$. Furthermore, $\widetilde{w}_{\pm}^{
\widetilde{\Sigma}}(\zeta) \! \in \! \cap_{p \in \{1,2,\infty\}} \mathcal{
L}^{p}_{\mathrm{M}_{2}(\mathbb{C})}(\widetilde{\Sigma})$.
\end{ccccc}
\begin{eeeee}
As per the analysis of Section~4, one shows that the relation between
$\widetilde{\mathscr{M}}^{\widetilde{\Sigma}}(\zeta)$ and $\widetilde{
\mathscr{M}}^{c}(\zeta)$ is $\widetilde{\mathscr{M}}^{c}(\zeta) \! =
\! \widetilde{\mathscr{M}}^{\widetilde{\Sigma}}(\zeta)(\mathrm{I} \!
+ \! \mathcal{O}(\tfrac{\underline{c}(\zeta_{1},\zeta_{2},\zeta_{3},
\overline{\zeta_{3}}) \widetilde{\diamondsuit}(\zeta)}{\vert z_{o}+
\zeta_{1}+\zeta_{2} \vert^{l}t^{l}}))$, with arbitrarily large $l
\! \in \! \mathbb{Z}_{\geqslant 1}$, and $\widetilde{\diamondsuit}
(\zeta) \! \in \! \mathcal{L}^{\infty}_{\mathrm{M}_{2}(\mathbb{C})}
(\mathbb{C} \setminus \widetilde{\Sigma})$.
\end{eeeee}
Using Lemma~3.1, the solution of the RHP for $\widetilde{\mathscr{M}}^{
\widetilde{\Sigma}}(\zeta)$ on $\widetilde{\Sigma}$ stated in Lemma~7.2
has the integral representation
\begin{equation*}
\widetilde{\mathscr{M}}^{\widetilde{\Sigma}}(\zeta) \! = \! \mathrm{
I} \! + \! \int\nolimits_{\widetilde{\Sigma}} \dfrac{\widetilde{
\mu}^{\widetilde{\Sigma}}(z) \widetilde{w}^{\widetilde{\Sigma}}(z)}
{(z \! - \! \zeta)} \, \dfrac{\md z}{2 \pi \mi}, \quad \zeta \! \in
\! \mathbb{C} \setminus \widetilde{\Sigma},
\end{equation*}
where $\widetilde{\mu}^{\widetilde{\Sigma}}(\zeta) \! := \! ((\mathbf{
1}_{\widetilde{\Sigma}} \! - \! C^{\widetilde{\Sigma}}_{\widetilde{w}^{
\widetilde{\Sigma}}})^{-1} \mathrm{I})(\zeta)$, and $\widetilde{w}^{
\widetilde{\Sigma}}(\zeta) \! := \! \sum_{l \in \{\pm\}} \! \widetilde{
w}_{l}^{\widetilde{\Sigma}}(\zeta)$.

Now, proceeding as per the analysis of Section~5, one solves, asymptotically,
the model RHP for $\widetilde{\mathscr{M}}^{\widetilde{\Sigma}}(\zeta)$ on
$\widetilde{\Sigma}$ formulated in Lemma~7.2, and arrives at the following
(analogue of Lemma~6.1)
\begin{ccccc}
Let $\varepsilon$ be an arbitrarily fixed, sufficiently small positive
real number, and, for $\lambda \! \in \! \{\zeta_{2},\zeta_{1}\}$, set
$\mathbb{U}(\lambda;\varepsilon) \! := \! \{ \mathstrut z; \, \vert
z \! - \! \lambda \vert \! < \! \varepsilon\}$. Then, as $t \! \to \!
-\infty$ such that $0 \! < \! \zeta_{2} \! < \! \tfrac{1}{M} \! < \! M
\! < \! \zeta_{1}$ and $\vert \zeta_{3} \vert^{2} \! = \! 1$, with $M
\! \in \! \mathbb{R}_{>1}$ and bounded, for $\zeta \! \in \! \mathbb{
C} \setminus \cup_{\lambda \, \in \, \{\zeta_{2},\zeta_{1}\}} \mathbb{
U}(\lambda;\varepsilon)$, $m^{c}(\zeta)$ has the following asymptotics,
\begin{align*}
m^{c}_{11}(\zeta) \! &= \! \widetilde{\delta}(\zeta) \! \left( 1 \! + \!
\mathcal{O} \! \left( \! \left( \dfrac{c^{\mathcal{S}}(\zeta_{1})
\underline{c}(\zeta_{2},\zeta_{3},\overline{\zeta_{3}})}{\sqrt{\zeta_{2}
(z_{o}^{2} \! + \! 32)} \, (\zeta \! - \! \zeta_{1})} \! + \! \dfrac{c^{
\mathcal{S}}(\zeta_{2}) \underline{c}(\zeta_{1},\zeta_{3},\overline{
\zeta_{3}})}{\sqrt{\zeta_{1}(z_{o}^{2} \! + \! 32)} \, (\zeta \! - \!
\zeta_{2})} \right) \! \dfrac{\ln \vert t \vert}{(\zeta_{1} \! - \!
\zeta_{2}) t} \right) \! \right), \\
m^{c}_{12}(\zeta) \! &= \! \dfrac{\me^{\frac{\mi \Omega^{-}(0)}{2}}}
{\widetilde{\delta}(\zeta)} \! \left( \dfrac{\sqrt{\nu (\zeta_{1})} \,
\zeta_{1}^{-2 \mi \nu (\zeta_{1})}}{\sqrt{\vert t \vert (\zeta_{1} \!
- \! \zeta_{2})} \, (z_{o}^{2} \! + \! 32)^{1/4}} \! \left( \dfrac{\zeta_{
1} \me^{\mi (\Theta^{-}(z_{o},t)-\frac{3 \pi}{4})}}{(\zeta \! - \!
\zeta_{1})} \! + \! \dfrac{\zeta_{2} \me^{-\mi (\Theta^{-}(z_{o},t)
-\frac{3 \pi}{4})}}{(\zeta \! - \! \zeta_{2})} \! \right) \right.
\end{align*}
\begin{align*}
 &+ \left. \! \mathcal{O} \! \left( \! \left( \dfrac{c^{\mathcal{S}}
(\zeta_{1}) \underline{c}(\zeta_{2},\zeta_{3},\overline{\zeta_{
3}})}{\sqrt{\zeta_{2}(z_{o}^{2} \! + \! 32)} \, (\zeta \! - \! \zeta_{1})}
\! + \! \dfrac{c^{\mathcal{S}}(\zeta_{2}) \underline{c}(\zeta_{1},
\zeta_{3},\overline{\zeta_{3}})}{\sqrt{\zeta_{1}(z_{o}^{2} \! + \!
32)} \, (\zeta \! - \! \zeta_{2})} \right) \! \dfrac{\ln \vert t \vert}{
(\zeta_{1} \! - \! \zeta_{2}) t} \right) \! \right), \\
m^{c}_{21}(\zeta) \! &= \! \dfrac{\widetilde{\delta}(\zeta)}{\me^{
\frac{\mi \Omega^{-}(0)}{2}}} \! \left( \dfrac{\sqrt{\nu (\zeta_{1})} \,
\zeta_{1}^{2 \mi \nu (\zeta_{1})}}{\sqrt{\vert t \vert (\zeta_{1} \! - \!
\zeta_{2})} \, (z_{o}^{2} \! + \! 32)^{1/4}} \! \left( \dfrac{\zeta_{
1} \me^{-\mi (\Theta^{-}(z_{o},t)-\frac{3 \pi}{4})}}{(\zeta \! - \!
\zeta_{1})} \! + \! \dfrac{\zeta_{2} \me^{\mi (\Theta^{-}(z_{o},
t)-\frac{3 \pi}{4})}}{(\zeta \! - \! \zeta_{2})} \! \right) \right. \\
 &+ \left. \! \mathcal{O} \! \left( \! \left( \dfrac{c^{\mathcal{S}}
(\zeta_{1}) \underline{c}(\zeta_{2},\zeta_{3},\overline{\zeta_{
3}})}{\sqrt{\zeta_{2}(z_{o}^{2} \! + \! 32)} \, (\zeta \! - \! \zeta_{1})}
\! + \! \dfrac{c^{\mathcal{S}}(\zeta_{2}) \underline{c}(\zeta_{1},
\zeta_{3},\overline{\zeta_{3}})}{\sqrt{\zeta_{1}(z_{o}^{2} \! + \!
32)} \, (\zeta \! - \! \zeta_{2})} \right) \! \dfrac{\ln \vert t \vert}{
(\zeta_{1} \! - \! \zeta_{2}) t} \right) \! \right), \\
m^{c}_{22}(\zeta) \! &= \! \dfrac{1}{\widetilde{\delta}(\zeta)} \!
\left( 1 \! + \! \mathcal{O} \! \left( \! \left( \dfrac{c^{\mathcal{S}}
(\zeta_{1}) \underline{c}(\zeta_{2},\zeta_{3},\overline{\zeta_{3}
})}{\sqrt{\zeta_{2}(z_{o}^{2} \! + \! 32)} \, (\zeta \! - \! \zeta_{1})}
\! + \! \dfrac{c^{\mathcal{S}}(\zeta_{2}) \underline{c}(\zeta_{1},
\zeta_{3},\overline{\zeta_{3}})}{\sqrt{\zeta_{1}(z_{o}^{2} \! +
\! 32)} \, (\zeta \! - \! \zeta_{2})} \right) \! \dfrac{\ln \vert t \vert}
{(\zeta_{1} \! - \! \zeta_{2}) t} \right) \! \right),
\end{align*}
where $\widetilde{\delta}(\zeta)$ is given in Lemma~{\rm 7.1}, $\nu (z)$,
$\Theta^{-}(z_{o},t)$, $\Omega^{-}(z)$, and $\{\zeta_{i}\}_{i=1}^{3}$ are
defined in Theorem~{\rm 3.1}, Eqs.~{\rm (12)}, {\rm (13)}, {\rm (15)},
{\rm (16)}, and {\rm (17)}, $\sup_{\zeta \in \mathbb{C} \, \setminus \cup_{
\lambda \in \{\zeta_{2},\zeta_{1}\}} \mathbb{U}(\lambda;\varepsilon)} \vert
(\zeta \!- \! \zeta_{k})^{-1} \vert \! \leqslant \! M^{\widetilde{c}}$, with
$M^{\widetilde{c}} \! \in \! \mathbb{R}_{+}$ (and bounded), $k \! \in \! \{
1,2\}$, $m^{c}(\zeta) \! = \! \sigma_{1} \overline{m^{c}(\overline{\zeta})}
\, \sigma_{1}$, and $(m^{c}(0) \sigma_{2})^{2} \! = \! \mathrm{I}$.
\end{ccccc}

\emph{Sketch of Proof.} Proceeding as in the proof of Lemma~6.1, as
$t \! \to \! -\infty$ such that $0 \! < \! \zeta_{2} \! < \! \tfrac{
1}{M} \! < \! M \! < \! \zeta_{1}$ and $\vert \zeta_{3} \vert^{2} \!
= \! 1$, with $M \! \in \! \mathbb{R}_{>1}$ and bounded, for $\zeta
\! \in \! \mathbb{C} \setminus \cup_{\lambda \, \in \, \{\zeta_{2},
\zeta_{1}\}} \mathbb{U}(\lambda;\varepsilon)$, one arrives at
\begin{align*}
\widetilde{\mathscr{M}}^{\widetilde{\Sigma}}_{11}(\zeta) \! &= \! 1
\! + \! \tfrac{r(\zeta_{1})(\widetilde{\delta}_{B}^{0})^{-2} \me^{-
\frac{3 \pi \nu}{2}} \me^{\frac{3 \pi \mi}{4}}}{2 \pi \mi (\zeta
-\zeta_{1}) \widetilde{\beta}^{\widetilde{\Sigma}_{B^{0}}}_{21}
\widetilde{\mathcal{X}}_{B} \sqrt{\vert t \vert}} \int\nolimits_{
0}^{+\infty}(\me^{-\frac{3 \pi \mi}{4}} \partial_{z} \mathbf{D}_{
\mi \nu}(z) \! + \! \tfrac{\mi}{2} \me^{\frac{3 \pi \mi}{4}}z
\mathbf{D}_{\mi \nu}(z))z^{\mi \nu} \me^{-\frac{z^{2}}{4}} \,
\md z \\
 &- \tfrac{r(\zeta_{1})(1-\vert r(\zeta_{1}) \vert^{2})^{-1}
(\widetilde{\delta}_{B}^{0})^{-2} \me^{-\frac{\mi \pi}{4}}}{2 \pi
\mi (\zeta -\zeta_{1}) \widetilde{\beta}^{\widetilde{\Sigma}_{B^{
0}}}_{21} \me^{-\frac{\pi \nu}{2}} \widetilde{\mathcal{X}}_{B}
\sqrt{\vert t \vert}} \int\nolimits_{0}^{+\infty}(\me^{\frac{\mi
\pi}{4}} \partial_{z} \mathbf{D}_{\mi \nu}(z) \! + \! \tfrac{\mi}
{2} \me^{-\frac{\mi \pi}{4}}z \mathbf{D}_{\mi \nu}(z))z^{\mi \nu}
\me^{-\frac{z^{2}}{4}} \, \md z \\
 &+ \tfrac{\overline{r(\zeta_{1})}(\widetilde{\delta}_{A}^{0})^{-2}
\me^{-\frac{\pi \nu}{2}}(-1)^{-\mi \nu} \me^{\frac{\mi \pi}{4}}}{2
\pi \mi (\zeta -\zeta_{2}) \widetilde{\beta}^{\widetilde{\Sigma}_{
A^{0}}}_{21} \widetilde{\mathcal{X}}_{A} \sqrt{\vert t \vert}}
\int\nolimits_{0}^{+\infty}(\me^{-\frac{\mi \pi}{4}} \partial_{z}
\mathbf{D}_{-\mi \nu}(z) \! - \! \tfrac{\mi}{2} \me^{\frac{\mi \pi}
{4}}z \mathbf{D}_{-\mi \nu}(z))z^{-\mi \nu} \me^{-\frac{z^{2}}{4}}
\, \md z \\
 &- \tfrac{\overline{r(\zeta_{1})}(1 -\vert r(\zeta_{1}) \vert^{
2})^{-1}(\widetilde{\delta}_{A}^{0})^{-2} \me^{-\frac{3 \pi \mi}
{4}}}{2 \pi \mi (\zeta -\zeta_{2}) \widetilde{\beta}^{\widetilde{
\Sigma}_{A^{0}}}_{21} \me^{\frac{\pi \nu}{2}}(-1)^{\mi \nu} \widetilde{
\mathcal{X}}_{A} \sqrt{\vert t \vert}} \int\nolimits_{0}^{+\infty}
(\me^{\frac{3 \pi \mi}{4}} \partial_{z} \mathbf{D}_{-\mi \nu}(z)
\! - \! \tfrac{\mi}{2} \me^{-\frac{3 \pi \mi}{4}}z \mathbf{D}_{-
\mi \nu}(z))z^{-\mi \nu} \me^{-\frac{z^{2}}{4}} \, \md z \\
 &+ \mathcal{O} \! \left( \! \left( \tfrac{c^{\mathcal{S}}
(\zeta_{1}) \underline{c}(\zeta_{2},\zeta_{3},\overline{\zeta_{3}})
(\widetilde{\delta}_{B}^{0})^{-2}}{(\zeta -\zeta_{1}) \vert \zeta_{
1}-\zeta_{3} \vert \sqrt{(\zeta_{1}-\zeta_{2})} \, \, \widetilde{
\mathcal{X}}_{B}} + \tfrac{c^{\mathcal{S}}(\zeta_{2}) \underline{
c}(\zeta_{1},\zeta_{3},\overline{\zeta_{3}})(\widetilde{\delta}_{
A}^{0})^{-2}}{(\zeta -\zeta_{2}) \vert \zeta_{2}-\zeta_{3} \vert
\sqrt{(\zeta_{1}-\zeta_{2})} \, \, \widetilde{\mathcal{X}}_{A}}
\right) \! \tfrac{\ln \vert t \vert}{t} \right), \\
\widetilde{\mathscr{M}}^{\widetilde{\Sigma}}_{12}(\zeta) \! &= \!
\left( \tfrac{\overline{r(\zeta_{1})}(1-\vert r(\zeta_{1}) \vert^{
2})^{-1}(\widetilde{\delta}_{B}^{0})^{2} \me^{\frac{\mi \pi}{4}}}
{2 \pi \mi (\zeta -\zeta_{1}) \me^{-\frac{\pi \nu}{2}} \widetilde{
\mathcal{X}}_{B} \sqrt{\vert t \vert}} \! - \! \tfrac{\overline{r
(\zeta_{1})}(\widetilde{\delta}_{B}^{0})^{2} \me^{-\frac{3 \pi
\nu}{2}} \me^{-\frac{3 \pi \mi}{4}}}{2 \pi \mi (\zeta -\zeta_{
1}) \widetilde{\mathcal{X}}_{B} \sqrt{\vert t \vert}} \right) \!
\int\nolimits_{0}^{+\infty} \mathbf{D}_{-\mi \nu}(z)z^{-\mi \nu}
\me^{-\frac{z^{2}}{4}} \, \md z \\
 &+ \left( \tfrac{r(\zeta_{1})(1-\vert r(\zeta_{1}) \vert^{2})^{
-1}(\widetilde{\delta}_{A}^{0})^{2} \me^{\frac{3 \pi \mi}{4}}}{2
\pi \mi (\zeta -\zeta_{2}) \me^{\frac{\pi \nu}{2}}(-1)^{-\mi \nu}
\widetilde{\mathcal{X}}_{A} \sqrt{\vert t \vert}} \! - \! \tfrac{
r(\zeta_{1})(\widetilde{\delta}_{A}^{0})^{2} \me^{-\frac{\pi \nu}
{2}} \me^{-\frac{\mi \pi}{4}}}{2 \pi \mi (\zeta -\zeta_{2})(-1)^{
-\mi \nu} \widetilde{\mathcal{X}}_{A} \sqrt{\vert t \vert}} \right)
\! \int\nolimits_{0}^{+\infty} \mathbf{D}_{\mi \nu}(z)z^{\mi \nu}
\me^{-\frac{z^{2}}{4}} \, \md z \\
 &+ \mathcal{O} \! \left( \! \left( \tfrac{c^{\mathcal{S}}
(\zeta_{1}) \underline{c}(\zeta_{2},\zeta_{3},\overline{\zeta_{
3}})(\widetilde{\delta}_{B}^{0})^{2}}{(\zeta -\zeta_{1}) \vert
\zeta_{1}-\zeta_{3} \vert \sqrt{(\zeta_{1}-\zeta_{2})} \, \,
\widetilde{\mathcal{X}}_{B}}+\tfrac{c^{\mathcal{S}}(\zeta_{2})
\underline{c}(\zeta_{1},\zeta_{3},\overline{\zeta_{3}})(\widetilde{
\delta}_{A}^{0})^{2}}{(\zeta -\zeta_{2}) \vert \zeta_{2}-\zeta_{
3} \vert \sqrt{(\zeta_{1}-\zeta_{2})} \, \, \widetilde{\mathcal{
X}}_{A}} \right) \! \tfrac{\ln \vert t \vert}{t} \right), \\
\widetilde{\mathscr{M}}^{\widetilde{\Sigma}}_{21}(\zeta) \! &= \!
-\left( \tfrac{r(\zeta_{1})(1-\vert r(\zeta_{1}) \vert^{2})^{-1}
(\widetilde{\delta}_{B}^{0})^{-2} \me^{-\frac{\mi \pi}{4}}}{2 \pi
\mi (\zeta -\zeta_{1}) \me^{-\frac{\pi \nu}{2}} \widetilde{\mathcal{
X}}_{B} \sqrt{\vert t \vert}} \! - \! \tfrac{r(\zeta_{1})(\widetilde{
\delta}_{B}^{0})^{-2} \me^{-\frac{3 \pi \nu}{2}} \me^{\frac{3 \pi
\mi}{4}}}{2 \pi \mi (\zeta -\zeta_{1}) \widetilde{\mathcal{X}}_{B}
\sqrt{\vert t \vert}} \right) \! \int\nolimits_{0}^{+\infty} \mathbf{
D}_{\mi \nu}(z)z^{\mi \nu} \me^{-\frac{z^{2}}{4}} \, \md z \\
 &- \left( \tfrac{\overline{r(\zeta_{1})}(1-\vert r(\zeta_{1})
\vert^{2})^{-1}(\widetilde{\delta}_{A}^{0})^{-2} \me^{-\frac{3
\pi \mi}{4}}}{2 \pi \mi (\zeta -\zeta_{2}) \me^{\frac{\pi \nu}
{2}}(-1)^{\mi \nu} \widetilde{\mathcal{X}}_{A} \sqrt{\vert t \vert}}
\! - \! \tfrac{\overline{r(\zeta_{1})}(\widetilde{\delta}_{A}^{0})^{
-2} \me^{-\frac{\pi \nu}{2}} \me^{\frac{\mi \pi}{4}}}{2 \pi \mi (\zeta
-\zeta_{2})(-1)^{\mi \nu} \widetilde{\mathcal{X}}_{A} \sqrt{\vert t
\vert}} \right) \! \int\nolimits_{0}^{+\infty} \mathbf{D}_{-\mi \nu}
(z)z^{-\mi \nu} \me^{-\frac{z^{2}}{4}} \, \md z \\
 &+ \mathcal{O} \! \left( \! \left( \tfrac{c^{\mathcal{S}}(\zeta_{1})
\underline{c}(\zeta_{2},\zeta_{3},\overline{\zeta_{3}})(\widetilde{
\delta}_{B}^{0})^{-2}}{(\zeta -\zeta_{1}) \vert \zeta_{1}-\zeta_{3}
\vert \sqrt{(\zeta_{1}-\zeta_{2})} \, \, \widetilde{\mathcal{X}}_{B}}
+\tfrac{c^{\mathcal{S}}(\zeta_{2}) \underline{c}(\zeta_{1},\zeta_{3},
\overline{\zeta_{3}})(\widetilde{\delta}_{A}^{0})^{-2}}{(\zeta -\zeta_{
2}) \vert \zeta_{2}-\zeta_{3} \vert \sqrt{(\zeta_{1}-\zeta_{2})} \, \,
\widetilde{\mathcal{X}}_{A}} \right) \! \tfrac{\ln \vert t \vert}{t}
\right), \\
\widetilde{\mathscr{M}}^{\widetilde{\Sigma}}_{22}(\zeta) \! &= \! 1
\! - \! \tfrac{\overline{r(\zeta_{1})}(\widetilde{\delta}_{B}^{0})^{
2} \me^{-\frac{3 \pi \nu}{2}} \me^{-\frac{3 \pi \mi}{4}}}{2 \pi \mi
(\zeta -\zeta_{1}) \widetilde{\beta}^{\widetilde{\Sigma}_{B^{0}}}_{
12} \widetilde{\mathcal{X}}_{B} \sqrt{\vert t \vert}} \int\nolimits_{
0}^{+\infty}(\me^{\frac{3 \pi \mi}{4}} \partial_{z} \mathbf{D}_{-\mi
\nu}(z) \! - \! \tfrac{\mi}{2} \me^{-\frac{3 \pi \mi}{4}}z \mathbf{
D}_{-\mi \nu}(z))z^{-\mi \nu} \me^{-\frac{z^{2}}{4}} \, \md z \\
 &+ \tfrac{\overline{r(\zeta_{1})}(1-\vert r(\zeta_{1}) \vert^{2})^{
-1}(\widetilde{\delta}_{B}^{0})^{2} \me^{\frac{\mi \pi}{4}}}{2 \pi
\mi (\zeta -\zeta_{1}) \widetilde{\beta}^{\widetilde{\Sigma}_{B^{0}
}}_{12} \me^{-\frac{\pi \nu}{2}} \widetilde{\mathcal{X}}_{B} \sqrt{
\vert t \vert}} \int\nolimits_{0}^{+\infty}(\me^{-\frac{\mi \pi}{4}
} \partial_{z} \mathbf{D}_{-\mi \nu}(z) \! - \! \tfrac{\mi}{2} \me^{
\frac{\mi \pi}{4}}z \mathbf{D}_{-\mi \nu}(z))z^{-\mi \nu} \me^{-
\frac{z^{2}}{4}} \, \md z \\
 &- \tfrac{r(\zeta_{1})(\widetilde{\delta}_{A}^{0})^{2} \me^{-\frac{
\pi \nu}{2}}(-1)^{\mi \nu} \me^{-\frac{\mi \pi}{4}}}{2 \pi \mi (\zeta
-\zeta_{2}) \widetilde{\beta}^{\widetilde{\Sigma}_{A^{0}}}_{12}
\widetilde{\mathcal{X}}_{A} \sqrt{\vert t \vert}} \int\nolimits_{
0}^{+\infty}(\me^{\frac{\mi \pi}{4}} \partial_{z} \mathbf{D}_{\mi
\nu}(z) \! + \! \tfrac{\mi}{2} \me^{-\frac{\mi \pi}{4}}z \mathbf{
D}_{\mi \nu}(z))z^{\mi \nu} \me^{-\frac{z^{2}}{4}} \, \md z
\end{align*}
\begin{align*}
 &+ \tfrac{r(\zeta_{1})(1-\vert r(\zeta_{1}) \vert^{2})^{-1}
(\widetilde{\delta}_{A}^{0})^{2}(-1)^{\mi \nu} \me^{\frac{3
\pi \mi}{4}}}{2 \pi \mi (\zeta -\zeta_{2}) \widetilde{\beta}^{
\widetilde{\Sigma}_{A^{0}}}_{12} \me^{\frac{\pi \nu}{2}} \widetilde{
\mathcal{X}}_{A} \sqrt{\vert t \vert}} \int\nolimits_{0}^{+\infty}
(\me^{-\frac{3 \pi \mi}{4}} \partial_{z} \mathbf{D}_{\mi \nu}(z) \!
+ \! \tfrac{\mi}{2} \me^{\frac{3 \pi \mi}{4}}z \mathbf{D}_{\mi \nu}
(z))z^{\mi \nu} \me^{-\frac{z^{2}}{4}} \, \md z \\
 &+ \mathcal{O} \! \left( \! \left( \tfrac{c^{\mathcal{S}}
(\zeta_{1}) \underline{c}(\zeta_{2},\zeta_{3},\overline{\zeta_{3}})
(\widetilde{\delta}_{B}^{0})^{2}}{(\zeta -\zeta_{1}) \vert \zeta_{
1}-\zeta_{3} \vert \sqrt{(\zeta_{1}-\zeta_{2})} \, \, \widetilde{
\mathcal{X}}_{B}}+\tfrac{c^{\mathcal{S}}(\zeta_{2}) \underline{
c}(\zeta_{1},\zeta_{3},\overline{\zeta_{3}})(\widetilde{\delta}_{
A}^{0})^{2}}{(\zeta -\zeta_{2}) \vert \zeta_{2}-\zeta_{3} \vert \sqrt{
(\zeta_{1}-\zeta_{2})} \, \, \widetilde{\mathcal{X}}_{A}} \right) \!
\tfrac{\ln \vert t \vert}{t} \right),
\end{align*}
where
\begin{gather*}
\widetilde{\delta}_{B}^{0} \! := \! \vert \zeta_{1} \! - \! \zeta_{3}
\vert^{\mi \nu}(2 \vert t \vert (\zeta_{1} \! - \! \zeta_{2})^{3} \zeta_{
1}^{-3})^{\frac{\mi \nu}{2}} \me^{\widetilde{\chi}(\zeta_{1})} \exp
(-\tfrac{\mi t}{2}(\zeta_{1} \! - \! \zeta_{2})(z_{o} \! + \! \zeta_{1}
\! + \! \zeta_{2})), \\
\widetilde{\delta}_{A}^{0} \! := \! \vert \zeta_{2} \! - \! \zeta_{3}
\vert^{-\mi \nu}(2 \vert t \vert (\zeta_{1} \! - \! \zeta_{2})^{3} \zeta_{
2}^{-3})^{-\frac{\mi \nu}{2}} \me^{\widetilde{\chi}(\zeta_{2})} \exp
(\tfrac{\mi t}{2}(\zeta_{1} \! - \! \zeta_{2})(z_{o} \! + \! \zeta_{1} \!
+ \! \zeta_{2})), \\
\widetilde{\chi}(\zeta_{1}) \! := \! \dfrac{\mi}{2 \pi} \int_{0}^{\zeta_{2}}
\ln \vert \mu \! - \! \zeta_{1} \vert \md \ln (1 \! - \! \vert r(\mu)
\vert^{2}) \! + \! \dfrac{\mi}{2 \pi} \int_{\zeta_{1}}^{+\infty} \ln \vert
\mu \! - \! \zeta_{1} \vert \md \ln (1 \! - \! \vert r(\mu) \vert^{2}), \\
\widetilde{\chi}(\zeta_{2}) \! := \! -\widetilde{\chi}(\zeta_{1}) \! + \!
\dfrac{\mi}{2 \pi} \int_{0}^{\zeta_{2}} \ln \vert \mu \vert \md \ln (1 \!
- \! \vert r(\mu) \vert^{2}) \! + \! \dfrac{\mi}{2 \pi} \int_{\zeta_{1}}^{
+\infty} \ln \vert \mu \vert \md \ln (1 \! - \! \vert r(\mu) \vert^{2}), \\
\widetilde{\mathcal{X}}_{B} \! = \! \mathcal{X}_{B}, \quad \widetilde{
\mathcal{X}}_{A} \! = \! \mathcal{X}_{A}, \quad \widetilde{\beta}^{
\widetilde{\Sigma}_{B^{0}}}_{12}=\overline{\widetilde{\beta}^{\widetilde{
\Sigma}_{B^{0}}}_{21}} \! := \! \tfrac{\sqrt{2 \pi} \, \me^{-\frac{\pi \nu}
{2}} \me^{\frac{3 \pi \mi}{4}}}{r(\zeta_{1}) \Gamma (\mi \nu)}, \quad
\widetilde{\beta}^{\widetilde{\Sigma}_{A^{0}}}_{12}=\overline{\widetilde{
\beta}^{\widetilde{\Sigma}_{A^{0}}}_{21}} \! := \! \tfrac{\sqrt{2 \pi} \,
\me^{-\frac{\pi \nu}{2}} \me^{-\frac{3 \pi \mi}{4}}}{\overline{r(\zeta_{1})}
\, \overline{\Gamma (\mi \nu)}},
\end{gather*}
and $\mathbf{D}_{\ast}(\cdot)$ is the parabolic cylinder function \cite{a40}.
Using the relations $\vert \zeta_{k} \! - \! \zeta_{3} \vert \zeta_{k}^{-1}
\! = \! (2 \zeta_{k})^{-1/2}(z_{o}^{2} \! + \! 32)^{1/4}$, $k \! \in \! \{1,
2\}$, the identities \cite{a40} $\partial_{z} \mathbf{D}_{z_{1}}(z) \! = \!
\tfrac{1}{2}(z_{1} \mathbf{D}_{z_{1}-1}(z) \! - \! \mathbf{D}_{z_{1}+1}(z))$,
$z \mathbf{D}_{z_{1}}(z) \! = \! \mathbf{D}_{z_{1}+1}(z) \linebreak[4]+ \!
z_{1} \mathbf{D}_{z_{1}-1}(z)$, and $\vert \Gamma (\mi \nu) \vert^{2} \! =
\! \tfrac{\pi}{\nu \sinh (\pi \nu)}$, the integral \cite{a40} $\int_{0}^{+
\infty} \! \mathbf{D}_{-z_{1}}(z)z^{z_{2}-1} \me^{-\frac{z^{2}}{4}} \, \md z
\! = \! \sqrt{\pi} \, \Gamma (z_{2}) \linebreak[4]
\cdot 2^{-\frac{1}{2}(z_{1}+z_{2})}(\Gamma (\tfrac{1}{2}(z_{1} \! + \! z_{2})
\! + \! \tfrac{1}{2}))^{-1}$, $\Re (z_{2}) \! > \! 0$, the relation $\vert r
(\zeta_{1}) \vert \vert \Gamma (\mi \nu) \vert \nu \me^{\frac{\pi \nu}{2}} \!
= \! \sqrt{2 \pi \nu}$, and the fact that (Lemma~7.1 and~Remark~7.1), for
$\zeta \! \in \! (\mathbb{C} \setminus \cup_{\lambda \in \{\zeta_{2},\zeta_{
1}\}} \mathbb{U}(\lambda;\varepsilon)) \cap (\Omega_{1} \cup \Omega_{2})$,
$m^{c}(\zeta) \! = \! \widetilde{\mathscr{M}}^{\widetilde{\Sigma}}(\zeta)
(\widetilde{\delta}(\zeta))^{\sigma_{3}}(\mathrm{I} \! + \! \mathcal{O}
(\tfrac{\underline{c}(\zeta_{1},\zeta_{2},\zeta_{3},\overline{\zeta_{3}})
\widehat{\diamondsuit}(\zeta)}{\vert z_{o}+\zeta_{1}+\zeta_{2} \vert^{l}
t^{l}}))$, with arbitrarily large $l \! \in \! \mathbb{Z}_{\geqslant 1}$,
$\widehat{\diamondsuit}(\zeta) \! \in \! \mathcal{L}^{\infty}_{\mathrm{M}_{2}
(\mathbb{C})}(\mathbb{C} \setminus \cup_{\lambda \in \{\zeta_{2},\zeta_{1}\}}
\mathbb{U}(\lambda;\varepsilon))$, and $\widetilde{\delta}(\zeta)$ given in
Lemma~7.1, one obtains the result stated in the Lemma; furthermore, one shows
that the symmetry reduction $m^{c}(\zeta) \! = \! \sigma_{1} \overline{m^{c}
(\overline{\zeta})} \, \sigma_{1}$ is satisfied, and verifies that, to
$\mathcal{O}(t^{-1} \ln \vert t \vert)$, $(m^{c}(0) \sigma_{2})^{2} \! = \!
\mathrm{I}$. \hfill $\square$

{}From the relation $\Delta_{o}m^{c}(0) \! = \! \sigma_{2}$ and Lemma~7.3,
one obtains the following (analogue of Proposition~6.1)
\begin{bbbbb}
As $t \! \to \! -\infty$ such that $0 \! < \! \zeta_{2} \! < \!
\tfrac{1}{M} \! < \! M \! < \! \zeta_{1}$ and $\vert \zeta_{3}
\vert^{2} \! = \! 1$, with $M \! \in \! \mathbb{R}_{>1}$ and
bounded,
\begin{align*}
(\Delta_{o})_{11} \! &= \! -\tfrac{2 \mi \sqrt{\nu (\zeta_{1})} \, \cos
(\Theta^{-}(z_{o},t)-\frac{3 \pi}{4})}{\sqrt{\vert t \vert (\zeta_{1}-\zeta_{
2})} \, (z_{o}^{2}+32)^{1/4}} \! + \! \mathcal{O} \! \left( \! \left( \tfrac{
c^{\mathcal{S}}(\zeta_{1}) \underline{c}(\zeta_{2},\zeta_{3},\overline{\zeta_{
3}})}{\sqrt{\zeta_{1}(z_{o}^{2}+32)}} \! + \! \tfrac{c^{\mathcal{S}}(\zeta_{
2}) \underline{c}(\zeta_{1},\zeta_{3},\overline{\zeta_{3}})}{\sqrt{\zeta_{2}
(z_{o}^{2}+32)}} \right) \! \tfrac{\ln \vert t \vert}{(\zeta_{1}-\zeta_{2})t}
\right), \\
(\Delta_{o})_{12} \! &= \! -\mi \exp \! \left(-\mi \left(\int_{0}^{\zeta_{
2}} \dfrac{\ln (1 \! - \! \vert r(\mu) \vert^{2})}{\mu} \, \dfrac{\md \mu}
{2 \pi} \! + \! \int_{\zeta_{1}}^{+\infty} \dfrac{\ln (1 \! - \! \vert
r(\mu) \vert^{2})}{\mu} \, \frac{\md \mu}{2 \pi} \right) \right) \\
 &\times \left( 1 \! + \! \mathcal{O} \! \left( \! \left( \dfrac{c^{\mathcal{
S}}(\zeta_{1}) \underline{c}(\zeta_{2},\zeta_{3},\overline{\zeta_{3}})}{
\sqrt{\zeta_{1}(z_{o}^{2} \! + \! 32)}} \! + \! \dfrac{c^{\mathcal{S}}
(\zeta_{2}) \underline{c}(\zeta_{1},\zeta_{3},\overline{\zeta_{3}})}{\sqrt{
\zeta_{2}(z_{o}^{2} \! + \! 32)}} \right) \! \dfrac{\ln \vert t \vert}{
(\zeta_{1} \! - \! \zeta_{2})t} \right) \right), \\
(\Delta_{o})_{21} \! &= \! \mi \exp \! \left(\mi \left(\int_{0}^{\zeta_{2}}
\dfrac{\ln (1 \! - \! \vert r(\mu) \vert^{2})}{\mu} \, \dfrac{\md \mu}{2
\pi} \! + \! \int_{\zeta_{1}}^{+\infty} \dfrac{\ln (1 \! - \! \vert r(\mu)
\vert^{2})}{\mu} \, \frac{\md \mu}{2 \pi} \right) \right) \\
 &\times \left( 1 \! + \! \mathcal{O} \! \left( \! \left( \dfrac{c^{\mathcal{
S}}(\zeta_{1}) \underline{c}(\zeta_{2},\zeta_{3},\overline{\zeta_{3}})}{\sqrt{
\zeta_{1}(z_{o}^{2} \! + \! 32)}} \! + \! \dfrac{c^{\mathcal{S}}(\zeta_{2})
\underline{c}(\zeta_{1},\zeta_{3},\overline{\zeta_{3}})}{\sqrt{\zeta_{2}(z_{
o}^{2} \! + \! 32)}} \right) \! \dfrac{\ln \vert t \vert}{(\zeta_{1} \! - \!
\zeta_{2})t} \right) \right), \\
(\Delta_{o})_{22} \! &= \! \tfrac{2 \mi \sqrt{\nu (\zeta_{1})} \, \cos
(\Theta^{-}(z_{o},t)-\frac{3 \pi}{4})}{\sqrt{\vert t \vert (\zeta_{1}-\zeta_{
2})} \, (z_{o}^{2}+32)^{1/4}} \! + \! \mathcal{O} \! \left( \! \left( \tfrac{
c^{\mathcal{S}}(\zeta_{1}) \underline{c}(\zeta_{2},\zeta_{3},\overline{\zeta_{
3}})}{\sqrt{\zeta_{1}(z_{o}^{2}+32)}} \! + \! \tfrac{c^{\mathcal{S}}(\zeta_{
2}) \underline{c}(\zeta_{1},\zeta_{3},\overline{\zeta_{3}})}{\sqrt{\zeta_{2}
(z_{o}^{2}+32)}} \right) \! \tfrac{\ln \vert t \vert}{(\zeta_{1}-\zeta_{2})
t} \right),
\end{align*}
where $\nu (z)$ and $\Theta^{-}(z_{o},t)$ are given in Theorem~{\rm 3.1},
Eqs.~{\rm (12)}, {\rm (13)}, and {\rm (15)}.
\end{bbbbb}

Finally, the analogue of Lemma~6.2 is
\begin{ccccc}
As $t \! \to \! -\infty$ such that $0 \! < \! \zeta_{2} \! < \! \tfrac{1}{M}
\! < \! M \! < \! \zeta_{1}$ and $\vert \zeta_{3} \vert^{2} \! = \! 1$, with
$M \! \in \! \mathbb{R}_{>1}$ and bounded, $u(x,t)$, the solution of the
Cauchy problem for the {\rm D${}_{f}$NLSE}, has the asymptotics stated in
Theorem~{\rm 3.1}, Eqs.~{\rm (9)}, {\rm (11)}, {\rm (12)}, {\rm (13)},
{\rm (15)}, {\rm (16)}, and~{\rm (17)}, and $\int_{\pm \infty}^{x}(\vert
u(\xi,t) \vert^{2} \! - \! 1) \, \md \xi$ have the asymptotics stated in
Theorem~{\rm 3.2}, Eqs.~{\rm (27)} and~{\rm (28)}, with $\theta^{+}(z)$
(respectively~$\theta^{-}(z))$ given in Theorem~{\rm 3.1}, Eq.~{\rm (10)}
(respectively~Eq.~{\rm (11))}.
\end{ccccc}

\textbf{\Large{Acknowledgements}}

The author is very grateful to K.~T.-R.~McLaughlin and X.~Zhou for the
invitation to the AMS Special Session on Integrable Systems and
Riemann-Hilbert Problems (Birmingham, Alabama, U.~S.~A., Nov.~10--12, 2000),
where partial results of this work were presented, A.~V.~Kitaev for fruitful
discussions, X.~Zhou for a copy of \cite{a43}, and M.~Kovalyov and R.~Saxton
for encouragement. The author is especially grateful to K.~T.-R.~McLaughlin,
S.~Shipman, A.~Tovbis, S.~Venakides and X.~Zhou for many helpful comments
and suggestions. The author is also very grateful to the referee for a
meticulous reading of the manuscript and for many helpful suggestions. The
author dedicates this work to C.~Roth.
\clearpage
\section*{Appendix}
\setcounter{section}{1}
\setcounter{z0}{1}
\setcounter{z1}{1}
In order to obtain the results stated in Theorems~3.2 and~3.3, the following
Lemmae and Propositions, which are the analogues of Lemmae~6.1 and~7.3
and Propositions~6.1 and~7.1, are necessary.
\begin{ay}
Let $\varepsilon$ be an arbitrarily fixed, sufficiently small positive
real number, and, for $\lambda \! \in \! \widetilde{\mathfrak{J}} \! := \!
\{(s_{1})^{\pm 1},(s_{2})^{\pm 1}\}$, where $s_{1} \! := \! \zeta_{1}^{
\sharp} \! = \! \exp (\mi \widehat{\varphi}_{1})$ and $s_{2} \! := \! \zeta_{
3}^{\sharp} \! = \! \exp (\mi \widehat{\varphi}_{3})$, with $\zeta_{n}^{
\sharp}$ and $\widehat{\varphi}_{n}$, $n \! \in \! \{1,3\}$, defined in
Theorem~{\rm 3.2}, Eqs.~{\rm (36)} and~{\rm (37)}, set $\mathbb{
U}(\lambda;\varepsilon) \! := \! \{\mathstrut z; \, \vert z \! - \! \lambda
\vert \! < \! \varepsilon\}$. Then, for $r(s_{1}) \! = \! \exp (-\tfrac{\mi
\varepsilon_{1} \pi}{2}) \vert r(s_{1}) \vert$, $\varepsilon_{1} \! \in \!
\{\pm 1\}$, $r(\overline{s_{2}}) \! = \! \exp (\tfrac{\mi \varepsilon_{2}
\pi}{2}) \vert r(\overline{s_{2}}) \vert$, $\varepsilon_{2} \! \in \! \{\pm
1\}$, $0 \! < \! r(s_{2}) \overline{r(\overline{s_{2}})} \! < \! 1$, and
$\zeta \! \in \! \mathbb{C} \setminus \cup_{\lambda \, \in \, \widetilde{
\mathfrak{J}}} \, \mathbb{U}(\lambda;\varepsilon)$, as $t \! \to \! +\infty$
and $x \! \to \! -\infty$ such that $z_{o} \! := \! x/t \! \in \! (-2,0)$,
$m^{c}(\zeta)$ has the following asymptotics,
\begin{align*}
m^{c}_{11}(\zeta) \! &= \! \delta (\zeta) \! \left( 1 \! + \! \mathcal{O}
\! \left( \! \left( \dfrac{\underline{c}(z_{o})}{(\zeta \! - \! s_{1})} \!
+ \! \dfrac{\underline{c}(z_{o})}{(\zeta \! - \! \overline{s_{2}})} \right)
\! \dfrac{\me^{-4 \alpha t}}{\beta t} \right) \right), \\
m^{c}_{12}(\zeta) \! &= \! \dfrac{1}{\delta (\zeta)} \! \left( \dfrac{
n_{1} \me^{-(2a_{0}t+ \sin (\widehat{\varphi}_{1}) \int_{-\infty}^{0}
\! \frac{\ln (1-\vert r(\mu) \vert^{2})}{(\mu - \cos \widehat{\varphi}_{
1})^{2}+\sin^{2} \widehat{\varphi}_{1}} \frac{\md \mu}{\pi})} \me^{
-\mi (\widehat{\varphi}_{1}+\int_{-\infty}^{0} \! \frac{(\mu - \cos
\widehat{\varphi}_{1}) \ln (1-\vert r(\mu) \vert^{2})}{(\mu - \cos
\widehat{\varphi}_{1})^{2}+\sin^{2} \widehat{\varphi}_{1}} \frac{
\md \mu}{\pi})}}{2 (\vert r(s_{1}) \vert)^{-1}(b_{0}t)^{1/2}(\zeta \! -
\! \overline{s_{1}})} \right. \\
 &+\left. \dfrac{n_{2} \me^{-(2\widehat{a}_{0}t-\sin (\widehat{\varphi}_{
3}) \int_{-\infty}^{0} \! \frac{\ln (1-\vert r(\mu) \vert^{2})}{(\mu - \cos
\widehat{\varphi}_{3})^{2}+\sin^{2} \widehat{\varphi}_{3}} \frac{\md
\mu}{\pi})} \me^{\mi (\widehat{\varphi}_{3}-\int_{-\infty}^{0} \! \frac{
(\mu - \cos \widehat{\varphi}_{3}) \ln (1-\vert r(\mu) \vert^{2})}{(\mu -
\cos \widehat{\varphi}_{3})^{2}+\sin^{2} \widehat{\varphi}_{3}} \frac{
\md \mu}{\pi})}}{2 (\vert r(\overline{s_{2}}) \vert)^{-1}(1 \! - \! r(s_{2})
\overline{r(\overline{s_{2}})})(\widehat{b}_{0}t)^{1/2}(\zeta \! - \! s_{
2})} \right. \\
 &+\left. \mathcal{O} \! \left( \! \left( \dfrac{\underline{c}(z_{o})}
{(\zeta \! - \! \overline{s_{1}})} \! + \! \dfrac{\underline{c}(z_{o})}{
(\zeta \! - \! s_{2})} \right) \! \dfrac{\me^{-4 \alpha t}}{\beta t} \right)
\right), \\
m^{c}_{21}(\zeta) \! &= \! \delta (\zeta) \! \left( \dfrac{n_{1} \me^{
-(2a_{0}t+\sin (\widehat{\varphi}_{1}) \int_{-\infty}^{0} \! \frac{\ln
(1-\vert r(\mu) \vert^{2})}{(\mu - \cos \widehat{\varphi}_{1})^{2}+
\sin^{2} \widehat{\varphi}_{1}} \frac{\md \mu}{\pi})} \me^{\mi
(\widehat{\varphi}_{1}+\int_{-\infty}^{0} \! \frac{(\mu - \cos \widehat{
\varphi}_{1}) \ln (1-\vert r(\mu) \vert^{2})}{(\mu - \cos \widehat{
\varphi}_{1})^{2}+\sin^{2} \widehat{\varphi}_{1}} \frac{\md \mu}
{\pi})}}{2 (\vert r(s_{1}) \vert)^{-1}(b_{0}t)^{1/2}(\zeta \! - \! s_{1})}
\right. \\
 &+\left. \dfrac{n_{2} \me^{-(2\widehat{a}_{0}t- \sin (\widehat{
\varphi}_{3}) \int_{-\infty}^{0} \! \frac{\ln (1-\vert r(\mu) \vert^{2})}{
(\mu - \cos \widehat{\varphi}_{3})^{2}+\sin^{2} \widehat{\varphi}_{
3}} \frac{\md \mu}{\pi})} \me^{\mi (-\widehat{\varphi}_{3}+\int_{-
\infty}^{0} \! \frac{(\mu - \cos \widehat{\varphi}_{3}) \ln (1-\vert
r(\mu) \vert^{2})}{(\mu - \cos \widehat{\varphi}_{3})^{2}+\sin^{2}
\widehat{\varphi}_{3}} \frac{\md \mu}{\pi})}}{2(\vert r(\overline{
s_{2}}) \vert)^{-1}(1 \! - \! r(s_{2}) \overline{r(\overline{s_{2}})})
(\widehat{b}_{0}t)^{1/2}(\zeta \! - \! \overline{s_{2}})} \right. \\
 &+\left. \mathcal{O} \! \left( \! \left( \dfrac{\underline{c}(z_{o})}
{(\zeta \! - \! s_{1})} \! + \! \dfrac{\underline{c}(z_{o})}{(\zeta \! - \!
\overline{s_{2}})} \right) \! \dfrac{\me^{-4 \alpha t}}{\beta t} \right)
\right), \\
m^{c}_{22}(\zeta) \! &= \! \dfrac{1}{\delta (\zeta)} \! \left( 1 \! + \!
\mathcal{O} \! \left( \! \left( \dfrac{\underline{c}(z_{o})}{(\zeta \!
- \! \overline{s_{1}})} \! + \! \dfrac{\underline{c}(z_{o})}{(\zeta \! -
\! s_{2})} \right) \! \dfrac{\me^{-4 \alpha t}}{\beta t} \right) \right),
\end{align*}
where $n_{1} \! := \! \mathrm{sgn}(\varepsilon_{1})$, $n_{2} \! :=
\! \mathrm{sgn}(\varepsilon_{2})$,
\begin{gather*}
\delta (\zeta) \! := \! \exp \! \left( \int\nolimits_{-\infty}^{0} \! \dfrac{
\ln (1 \! - \! \vert r(\mu) \vert^{2})}{(\mu \! - \! \zeta)} \, \dfrac{\md
\mu}{2 \pi \mi} \right), \qquad a_{0} \! := \! \tfrac{1}{2}(z_{o} \! - \!
a_{1})(4 \! - \! a_{1}^{2})^{1/2} \, \, \, (>0), \\
\widehat{a}_{0} \! := \! -\tfrac{1}{2}(z_{o} \! - \! a_{2})(4 \! - \! a_{
2}^{2})^{1/2} \, \, \, (>0), \qquad b_{0} \! := \! \tfrac{1}{2}(z_{o}^{2}
\! + \! 32)^{1/2}(4 \! - \! a_{1}^{2})^{1/2} \, \, \, (>0), \\
\widehat{b}_{0} \! := \! \tfrac{1}{2}(z_{o}^{2} \! + \! 32)^{1/2}(4 \! -
\! a_{2}^{2})^{1/2} \, \, \, (>0), \qquad \alpha \! := \! \min \{a_{0},
\widehat{a}_{0}\}, \quad  \beta \! := \! \min\{b_{0},\widehat{b}_{
0}\},
\end{gather*}
with $a_{1}$ and $a_{2}$ given in Theorem~{\rm 3.1}, Eq.~{\rm (17)}, $\sup_{
\zeta \in \mathbb{C} \, \setminus \cup_{\lambda \in \widetilde{\mathfrak{J}}}
\mathbb{U}(\lambda;\varepsilon)} \vert (\zeta \! - \! (\zeta_{n}^{\sharp})^{
l})^{-1} \vert \! \leqslant \! \widetilde{\mathfrak{M}}$, with $\widetilde{
\mathfrak{M}} \in \mathbb{R}_{+}$ (and bounded), $n \! \in \! \{1,3\}$, $l
\! \in \! \{\pm 1\}$, $m^{c}(\zeta) \! = \! \sigma_{1} \overline{m^{c}
(\overline{\zeta})} \, \sigma_{1}$, and $(m^{c}(0) \sigma_{2})^{2} \! = \!
\mathrm{I}$.
\end{ay}
\begin{by}
Let $s_{1} \! := \! \zeta_{1}^{\sharp} \! = \! \exp (\mi \widehat{\varphi}_{
1})$ and $s_{2} \! := \! \zeta_{3}^{\sharp} \! = \! \exp (\mi \widehat{
\varphi}_{3})$, where $\zeta_{n}^{\sharp}$ and $\widehat{\varphi}_{n}$, $n
\! \in \! \{1,3\}$, are defined in Theorem~{\rm 3.2}, Eqs.~{\rm (36)}
and~{\rm (37)}. For $r(s_{1}) \! = \! \exp (-\tfrac{\mi \varepsilon_{1}
\pi}{2}) \vert r(s_{1}) \vert$, $\varepsilon_{1} \! \in \! \{\pm 1\}$,
$r(\overline{s_{2}}) \! = \! \exp (\tfrac{\mi \varepsilon_{2} \pi}{2}) \vert
r(\overline{s_{2}}) \vert$, $\varepsilon_{2} \! \in \! \{\pm 1\}$, $0 \! <
\! r(s_{2}) \overline{r(\overline{s_{2}})} \! < \! 1$, and $\varepsilon_{1}
\! = \! \varepsilon_{2}$, as $t \! \to \! +\infty$ and $x \! \to \! -\infty$
such that $z_{o} \! := \! x/t \! \in \! (-2,0)$,
\begin{gather*}
(\Delta_{o})_{11} \! = \! -\dfrac{n_{1} \mi \me^{-(\widetilde{a}_{+}t+
\widetilde{c}_{+})} \beth_{+}}{\widetilde{b} \sqrt{t}} \cosh \! \left(
\widetilde{a}_{-}t \! + \! \widetilde{c}_{-} \! + \! \tfrac{1}{8} \ln \!
\left( \tfrac{4-a_{1}^{2}}{4-a_{2}^{2}} \right) \! - \! \ln \beth_{-}
\right) \! + \! \mathcal{O} \! \left( \dfrac{\underline{c}(z_{o}) \me^{-4
\alpha t}}{\beta t} \right),
\end{gather*}
\begin{gather*}
(\Delta_{o})_{12} \! = \! -\mi \me^{-\mi \psi^{+}(1)} \! \left(1 \! + \!
\mathcal{O} \! \left(\dfrac{\underline{c}(z_{o}) \me^{-4 \alpha t}}{\beta t}
\right) \right), \quad (\Delta_{o})_{21} \! = \! \mi \me^{\mi \psi^{+}(1)}
\! \left(1 \! + \! \mathcal{O} \! \left(\dfrac{\underline{c}(z_{o}) \me^{
-4 \alpha t}}{\beta t} \right) \right), \\
(\Delta_{o})_{22} \! = \! \dfrac{n_{1} \mi \me^{-(\widetilde{a}_{+}t+
\widetilde{c}_{+})} \beth_{+}}{\widetilde{b} \sqrt{t}} \cosh \! \left(
\widetilde{a}_{-}t \! + \! \widetilde{c}_{-} \! + \! \tfrac{1}{8} \ln \!
\left( \tfrac{4-a_{1}^{2}}{4-a_{2}^{2}} \right) \! - \! \ln \beth_{-}
\right) \! + \! \mathcal{O} \! \left( \dfrac{\underline{c}(z_{o}) \me^{
-4 \alpha t}}{\beta t} \right),
\end{gather*}
and, for $\varepsilon_{1} \! = \! -\varepsilon_{2}$,
\begin{gather*}
(\Delta_{o})_{11} \! = \! \dfrac{n_{1} \mi \me^{-(\widetilde{a}_{+}t+
\widetilde{c}_{+})} \beth_{+}}{\widetilde{b} \sqrt{t}} \sinh \! \left(
\widetilde{a}_{-}t \! + \! \widetilde{c}_{-} \! + \! \tfrac{1}{8} \ln \!
\left( \tfrac{4-a_{1}^{2}}{4-a_{2}^{2}} \right) \! - \! \ln \beth_{-}
\right) \! + \! \mathcal{O} \! \left( \dfrac{\underline{c}(z_{o}) \me^{-4
\alpha t}}{\beta t} \right), \\
(\Delta_{o})_{12} \! = \! -\mi \me^{-\mi \psi^{+}(1)} \! \left(1 \! + \!
\mathcal{O} \! \left( \dfrac{\underline{c}(z_{o}) \me^{-4 \alpha t}}{\beta
t} \right) \right), \quad (\Delta_{o})_{21} \! = \! \mi \me^{\mi \psi^{+}
(1)} \! \left(1 \! + \! \mathcal{O} \! \left(\dfrac{\underline{c}(z_{o})
\me^{-4 \alpha t}}{\beta t} \right) \right), \\
(\Delta_{o})_{22} \! = \! -\dfrac{n_{1} \mi \me^{-(\widetilde{a}_{+}t+
\widetilde{c}_{+})} \beth_{+}}{\widetilde{b} \sqrt{t}} \sinh \! \left(
\widetilde{a}_{-}t \! + \! \widetilde{c}_{-} \! + \! \tfrac{1}{8} \ln \!
\left( \tfrac{4-a_{1}^{2}}{4-a_{2}^{2}} \right) \! - \! \ln \beth_{-}
\right) \! + \! \mathcal{O} \! \left( \dfrac{\underline{c}(z_{o}) \me^{-4
\alpha t}}{\beta t} \right),
\end{gather*}
where $n_{1} \! := \! \mathrm{sgn}(\varepsilon_{1})$, $\psi^{+}(\cdot)$,
$\widetilde{a}_{\pm}$, $\widetilde{c}_{\pm}$, $\widetilde{b}$, $\beth_{
\pm}$, $\alpha$, and $\beta$ are defined in Theorem~{\rm 3.2},
Eqs.~{\rm (35)}, {\rm (38)}, {\rm (39)}, {\rm (40)}, {\rm (41)}, {\rm (42)},
and~{\rm (43)}, and $a_{1}$ and $a_{2}$ are given in Theorem~{\rm 3.1},
Eq.~{\rm (17)}.
\end{by}
\stepcounter{z0}
\stepcounter{z1}
\begin{ay}
Let $\varepsilon$ be an arbitrarily fixed, sufficiently small positive
real number, and, for $\lambda \! \in \! \widetilde{\mathfrak{J}} \! := \!
\{(s_{1})^{\pm 1},(s_{2})^{\pm 1}\}$, where $s_{1} \! := \! \zeta_{1}^{
\sharp} \! = \! \exp (\mi \widehat{\varphi}_{1})$ and $s_{2} \! := \!
\zeta_{3}^{\sharp} \! = \! \exp (\mi \widehat{\varphi}_{3})$, with
$\zeta_{n}^{\sharp}$ and $\widehat{\varphi}_{n}$, $n \! \in \! \{1,3\}$,
defined in Theorem~{\rm 3.2}, Eqs.~{\rm (36)} and~{\rm (37)}, set
$\mathbb{U}(\lambda;\varepsilon) \! := \! \{ \mathstrut z; \, \vert z \! - \!
\lambda \vert \! < \! \varepsilon\}$. Then, for $r(\overline{s_{1}}) \! = \!
\exp (\tfrac{\mi \varepsilon_{1} \pi}{2}) \vert r(\overline{s_{1}}) \vert$,
$\varepsilon_{1} \! \in \! \{\pm 1\}$, $r(s_{2}) \! = \! \exp (-\tfrac{\mi
\varepsilon_{2} \pi}{2}) \vert r(s_{2}) \vert$, $\varepsilon_{2} \! \in \!
\{\pm 1\}$, $0 \! < \! r(s_{1}) \overline{r(\overline{s_{1}})} \! < \! 1$, and
$\zeta \! \in \! \mathbb{C} \setminus \cup_{\lambda \, \in \, \widetilde{
\mathfrak{J}}} \, \mathbb{U}(\lambda;\varepsilon)$, as $t \! \to \! -\infty$
and $x \! \to \! +\infty$ such that $z_{o} \! := \! x/t \! \in \! (-2,0)$,
$m^{c}(\zeta)$ has the following asymptotics,
\begin{align*}
m^{c}_{11}(\zeta) \! &= \! \widetilde{\delta}(\zeta) \! \left( 1 \! + \!
\mathcal{O} \! \left( \! \left( \dfrac{\underline{c}(z_{o})}{(\zeta \! - \!
\overline{s_{1}})} \! + \! \dfrac{\underline{c}(z_{o})}{(\zeta \! - \! s_{2})}
\right) \! \dfrac{\me^{-4 \alpha \vert t \vert}}{\beta t} \right) \right), \\
m^{c}_{12}(\zeta) \! &= \! -\dfrac{1}{\widetilde{\delta}(\zeta)} \! \left(
\dfrac{n_{1} \me^{-(2a_{0} \vert t \vert -\sin (\widehat{\varphi}_{1})
\int_{0}^{+\infty} \! \frac{\ln (1-\vert r(\mu) \vert^{2})}{(\mu - \cos
\widehat{\varphi}_{1})^{2}+\sin^{2} \widehat{\varphi}_{1}} \frac{
\md \mu}{\pi})} \me^{\mi (\widehat{\varphi}_{1}-\int_{0}^{+\infty} \!
\frac{(\mu - \cos \widehat{\varphi}_{1}) \ln (1-\vert r(\mu) \vert^{2})}
{(\mu - \cos \widehat{\varphi}_{1})^{2}+\sin^{2} \widehat{\varphi}_{
1}} \frac{\md \mu}{\pi})}}{2(\vert r(\overline{s_{1}}) \vert)^{-1}(1 \! - \!
r(s_{1}) \overline{r(\overline{s_{1}})})(b_{0} \vert t \vert)^{1/2}(\zeta
\! - \! s_{1})} \right. \\
 &+\left. \dfrac{n_{2} \me^{-(2\widehat{a}_{0} \vert t \vert +\sin
(\widehat{\varphi}_{3}) \int_{0}^{+\infty} \! \frac{\ln (1-\vert r(\mu)
\vert^{2})}{(\mu - \cos \widehat{\varphi}_{3})^{2}+\sin^{2} \widehat{
\varphi}_{3}} \frac{\md \mu}{\pi})} \me^{-\mi (\widehat{\varphi}_{3}+\int_{
0}^{+\infty} \! \frac{(\mu - \cos \widehat{\varphi}_{3}) \ln (1-\vert r(\mu)
\vert^{2})}{(\mu - \cos \widehat{\varphi}_{3})^{2}+\sin^{2} \widehat{
\varphi}_{3}} \frac{\md \mu}{\pi})}}{2(\vert r(s_{2}) \vert)^{-1}(\widehat{
b}_{0} \vert t \vert)^{1/2}(\zeta \! - \! \overline{s_{2}})} \right. \\
 &+\left. \mathcal{O} \! \left( \! \left( \dfrac{\underline{c}(z_{o})}
{(\zeta \! - \! s_{1})} \! + \! \dfrac{\underline{c}(z_{o})}{(\zeta \! - \!
\overline{s_{2}})} \right) \! \dfrac{\me^{-4 \alpha \vert t \vert}}{\beta t}
\right) \right), \\
m^{c}_{21}(\zeta) \! &= \! -\widetilde{\delta}(\zeta) \! \left( \dfrac{
n_{1} \me^{-(2a_{0} \vert t \vert -\sin (\widehat{\varphi}_{1}) \int_{
0}^{+\infty} \! \frac{\ln (1-\vert r(\mu) \vert^{2})}{(\mu - \cos \widehat{
\varphi}_{1})^{2}+\sin^{2} \widehat{\varphi}_{1}} \frac{\md \mu}{
\pi})} \me^{-\mi (\widehat{\varphi}_{1}-\int_{0}^{+\infty} \! \frac{
(\mu - \cos \widehat{\varphi}_{1}) \ln (1-\vert r(\mu) \vert^{2})}{(
\mu - \cos \widehat{\varphi}_{1})^{2}+\sin^{2} \widehat{\varphi}_{
1}} \frac{\md \mu}{\pi})}}{2(\vert r(\overline{s_{1}}) \vert)^{-1}(1 \!
- \! r(s_{1}) \overline{r(\overline{s_{1}})})(b_{0} \vert t \vert)^{1/2}
(\zeta \! - \! \overline{s_{1}})} \right. \\
 &+\left. \dfrac{n_{2} \me^{-(2\widehat{a}_{0} \vert t \vert +\sin
(\widehat{\varphi}_{3}) \int_{0}^{+\infty} \! \frac{\ln (1-\vert r(\mu)
\vert^{2})}{(\mu - \cos \widehat{\varphi}_{3})^{2}+\sin^{2} \widehat{
\varphi}_{3}} \frac{\md \mu}{\pi})} \me^{\mi (\widehat{\varphi}_{3}+
\int_{0}^{+\infty} \! \frac{(\mu - \cos \widehat{\varphi}_{3}) \ln (1-\vert
r(\mu) \vert^{2})}{(\mu - \cos \widehat{\varphi}_{3})^{2}+\sin^{2}
\widehat{\varphi}_{3}} \frac{\md \mu}{\pi})}}{2(\vert r(s_{2}) \vert)^{
-1}(\widehat{b}_{0} \vert t \vert)^{1/2}(\zeta \! - \! s_{2})} \right. \\
 &+\left. \mathcal{O} \! \left( \! \left( \dfrac{\underline{c}(z_{o})}
{(\zeta \! - \! \overline{s_{1}})} \! + \! \dfrac{\underline{c}(z_{o})}
{(\zeta \! - \! s_{2})} \right) \! \dfrac{\me^{-4 \alpha \vert t \vert}}
{\beta t} \right) \right), \\
m^{c}_{22}(\zeta) \! &= \! \dfrac{1}{\widetilde{\delta}(\zeta)} \!
\left(1 \! + \! \mathcal{O} \! \left( \! \left( \dfrac{\underline{c}(z_{o})}
{(\zeta \! - \! s_{1})} \! + \! \dfrac{\underline{c}(z_{o})}{(\zeta \! - \!
\overline{s_{2}})} \right) \! \dfrac{\me^{-4 \alpha \vert t \vert}}{\beta
t} \right) \right),
\end{align*}
where $n_{1} \! := \! \mathrm{sgn}(\varepsilon_{1})$, $n_{2} \! := \!
\mathrm{sgn}(\varepsilon_{2})$,
\begin{gather*}
\widetilde{\delta}(\zeta) \! := \! \exp \! \left(\int\nolimits_{0}^{+\infty}
\! \dfrac{\ln (1 \! - \! \vert r(\mu) \vert^{2})}{(\mu \! - \! \zeta)} \,
\dfrac{\md \mu}{2 \pi \mi} \right),
\end{gather*}
$a_{0}$, $\widehat{a}_{0}$, $b_{0}$, $\widehat{b}_{0}$, $\alpha$, and $\beta$
are defined in Lemma~{\rm A.1}, $\sup_{\zeta \in \mathbb{C} \, \setminus
\cup_{\lambda \in \widetilde{\mathfrak{J}}} \mathbb{U}(\lambda;\varepsilon)}
\vert (\zeta \! - \! (\zeta_{n}^{\sharp})^{l})^{-1} \vert \! \leqslant \!
\widehat{\mathfrak{M}}$, with $\widehat{\mathfrak{M}} \! \in \! \mathbb{R}_{
+}$ (and bounded), $n \! \in \! \{1,3\}$, $l \! \in \! \{\pm 1\}$, $m^{c}
(\zeta) \! = \! \sigma_{1} \overline{m^{c}(\overline{\zeta})} \, \sigma_{
1}$, and $(m^{c}(0) \sigma_{2})^{2} \! = \! \mathrm{I}$.
\end{ay}
\begin{by}
Let $s_{1} \! := \! \zeta_{1}^{\sharp} \! = \! \exp (\mi \widehat{\varphi}_{
1})$ and $s_{2} \! := \! \zeta_{3}^{\sharp} \! = \! \exp (\mi \widehat{
\varphi}_{3})$, where $\zeta_{n}^{\sharp}$ and $\widehat{\varphi}_{n}$,
$n \! \in \! \{1,3\}$, are defined in Theorem~{\rm 3.2}, Eqs.~{\rm (36)}
and~{\rm (37)}. For $r(\overline{s_{1}}) \! = \! \exp (\tfrac{\mi
\varepsilon_{1} \pi}{2}) \vert r(\overline{s_{1}}) \vert$, $\varepsilon_{1}
\! \in \! \{\pm 1\}$, $r(s_{2}) \! = \! \exp (-\tfrac{\mi \varepsilon_{2}
\pi}{2}) \vert r(s_{2}) \vert$, $\varepsilon_{2} \! \in \! \{\pm 1\}$, $0 \!
< \! r(s_{1}) \overline{r(\overline{s_{1}})} \! < \! 1$, and $\varepsilon_{
1} \! = \! \varepsilon_{2}$, as $t \! \to \! -\infty$ and $x \! \to \!
+\infty$ such that $z_{o} \! := \! x/t \! \in \! (-2,0)$,
\begin{gather*}
(\Delta_{o})_{11} \! = \! \dfrac{n_{1} \mi \me^{-(\widetilde{a}_{+} \vert t
\vert -\widehat{c}_{+})} \daleth_{+}}{\widetilde{b} \sqrt{\vert t \vert}}
\cosh \! \left(\widetilde{a}_{-} \vert t \vert \! - \! \widehat{c}_{-} \! +
\! \tfrac{1}{8} \ln \! \left(\tfrac{4-a_{1}^{2}}{4-a_{2}^{2}} \right) \! +
\! \ln \daleth_{-} \right) \! + \! \mathcal{O} \! \left(\dfrac{\underline{c}
(z_{o}) \me^{-4 \alpha \vert t \vert}}{\beta t} \right), \\
(\Delta_{o})_{12} \! = \! -\mi \me^{-\mi \psi^{-}(1)} \! \left( 1 \! + \!
\mathcal{O} \! \left(\dfrac{\underline{c}(z_{o}) \me^{-4 \alpha \vert t
\vert}}{\beta t} \right) \right), \quad (\Delta_{o})_{21} \! = \! \mi \me^{
\mi \psi^{-}(1)} \! \left(1 \! + \! \mathcal{O} \! \left(\dfrac{\underline{
c}(z_{o}) \me^{-4 \alpha \vert t \vert}}{\beta t} \right) \right), \\
(\Delta_{o})_{22} \! = \! -\dfrac{n_{1} \mi \me^{-(\widetilde{a}_{+} \vert t
\vert -\widehat{c}_{+})} \daleth_{+}}{\widetilde{b} \sqrt{\vert t \vert}}
\cosh \! \left( \widetilde{a}_{-} \vert t \vert \! - \! \widehat{c}_{-} \! +
\! \tfrac{1}{8} \ln \! \left(\tfrac{4-a_{1}^{2}}{4-a_{2}^{2}} \right) \! + \!
\ln \daleth_{-} \right) \! + \! \mathcal{O} \! \left( \dfrac{\underline{c}
(z_{o}) \me^{-4 \alpha \vert t \vert}}{\beta t} \right),
\end{gather*}
and, for $\varepsilon_{1} \! = \! -\varepsilon_{2}$,
\begin{gather*}
(\Delta_{o})_{11} \! = \! -\dfrac{n_{1} \mi \me^{-(\widetilde{a}_{+} \vert t
\vert -\widehat{c}_{+})} \daleth_{+}}{\widetilde{b} \sqrt{\vert t \vert}}
\sinh \! \left(\widetilde{a}_{-} \vert t \vert \! - \! \widehat{c}_{-} \! +
\! \tfrac{1}{8} \ln \! \left(\tfrac{4-a_{1}^{2}}{4-a_{2}^{2}} \right) \! +
\! \ln \daleth_{-} \right) \! + \! \mathcal{O} \! \left(\dfrac{\underline{c}
(z_{o}) \me^{-4 \alpha \vert t \vert}}{\beta t} \right), \\
(\Delta_{o})_{12} \! = \! -\mi \me^{-\mi \psi^{-}(1)} \! \left( 1 \! + \!
\mathcal{O} \! \left(\dfrac{\underline{c}(z_{o}) \me^{-4 \alpha \vert t
\vert}}{\beta t} \right) \right), \quad (\Delta_{o})_{21} \! = \! \mi \me^{
\mi \psi^{-}(1)} \! \left(1 \! + \! \mathcal{O} \! \left(\dfrac{\underline{
c}(z_{o}) \me^{-4 \alpha \vert t \vert}}{\beta t} \right) \right), \\
(\Delta_{o})_{22} \! = \! \dfrac{n_{1} \mi \me^{-(\widetilde{a}_{+} \vert t
\vert -\widehat{c}_{+})} \daleth_{+}}{\widetilde{b} \sqrt{\vert t \vert}}
\sinh \! \left( \widetilde{a}_{-} \vert t \vert \! - \! \widehat{c}_{-} \! +
\! \tfrac{1}{8} \ln \! \left(\tfrac{4-a_{1}^{2}}{4-a_{2}^{2}} \right) \! +
\! \ln \daleth_{-} \right) \! + \! \mathcal{O} \! \left( \dfrac{\underline{
c}(z_{o}) \me^{-4 \alpha \vert t \vert}}{\beta t} \right),
\end{gather*}
where $n_{1} \! := \! \mathrm{sgn}(\varepsilon_{1})$, $\widetilde{a}_{\pm}$,
$\widetilde{b}$, $\alpha$, $\beta$, $\psi^{-}(\cdot)$, $\widehat{c}_{\pm}$,
and $\daleth_{\pm}$ are defined in Theorem~{\rm 3.2}, Eqs.~{\rm (38)},
{\rm (40)}, {\rm (42)}, {\rm (43)}, {\rm (46)}, {\rm (47)}, and~{\rm (48)},
and $a_{1}$ and $a_{2}$ are given in Theorem~{\rm 3.1}, Eq.~{\rm (17)}.
\end{by}
\stepcounter{z0}
\stepcounter{z1}
\begin{ay}
Let $\varepsilon$ be an arbitrarily fixed, sufficiently small positive real
number, and, for $\lambda \! \in \! \widehat{\mathfrak{J}} \! := \! \{(s_{
1})^{\pm 1},(s_{2})^{\pm 1}\}$, where $s_{1} \! := \! \exp (\tfrac{\mi \pi}
{4})$ and $s_{2} \! := \! \exp (\tfrac{3 \pi \mi}{4})$, set $\mathbb{U}
(\lambda;\varepsilon) \! := \! \{ \mathstrut z; \, \vert z \! - \! \lambda
\vert \! < \! \varepsilon\}$. Then, for $r(s_{1}) \! = \! \exp (-\tfrac{\mi
\varepsilon_{1} \pi}{2}) \vert r(s_{1}) \vert$, $\varepsilon_{1} \! \in \!
\{\pm 1\}$, $r(\overline{s_{2}}) \! = \! \exp (\tfrac{\mi \varepsilon_{2}
\pi}{2}) \vert r(\overline{s_{2}}) \vert$, $\varepsilon_{2} \! \in \! \{\pm
1\}$, $0 \! < \! r(s_{2}) \overline{r(\overline{s_{2}})} \! < \! 1$, and
$\zeta \! \in \! \mathbb{C} \setminus \cup_{\lambda \, \in \, \widehat{
\mathfrak{J}}} \, \mathbb{U}(\lambda;\varepsilon)$, as $t \! \to \! +\infty$
and $x \! \to \! -\infty$ such that $z_{o} \! := \! x/t \! \to \! 0^{-}$,
$m^{c}(\zeta)$ has the following asymptotics,
\begin{align*}
m^{c}_{11}(\zeta) \! &= \! \delta (\zeta) \! \left( 1 \! + \! \mathcal{
O} \! \left( \! \left( \dfrac{\underline{c}}{(\zeta \! - \! s_{1})} \! + \!
\dfrac{\underline{c}}{(\zeta \! - \! \overline{s_{2}})} \right) \!
\dfrac{\me^{-4t}}{t} \right) \right), \\
m^{c}_{12}(\zeta) \! &= \! \dfrac{1}{\delta (\zeta)} \! \left(\dfrac{
\mathrm{sgn}(\varepsilon_{1}) \me^{-(2t+\sqrt{2} \int_{-\infty}^{0}
\! \frac{\ln (1-\vert r(\mu) \vert^{2})}{(\sqrt{2} \, \mu - 1)^{2}+1} \frac{
\md \mu}{\pi})} \me^{-\mi (\frac{\pi}{4}+\sqrt{2} \int_{-\infty}^{0} \!
\frac{(\sqrt{2} \, \mu -1) \ln (1-\vert r(\mu) \vert^{2})}{(\sqrt{2} \, \mu
-1)^{2}+1} \frac{\md \mu}{\pi})}}{4(\vert r(s_{1}) \vert)^{-1} \sqrt{t}
\, (\zeta \! - \! \overline{s_{1}})} \right. \\
 &+\left. \dfrac{\mathrm{sgn}(\varepsilon_{2}) \me^{-(2t-\sqrt{2}
\int_{-\infty}^{0} \! \frac{\ln (1-\vert r(\mu) \vert^{2})}{(\sqrt{2} \, \mu
+1)^{2}+1} \frac{\md \mu}{\pi})} \me^{\mi (\frac{3 \pi}{4}-\sqrt{2}
\int_{-\infty}^{0} \! \frac{(\sqrt{2} \, \mu +1) \ln (1-\vert r(\mu) \vert^{
2})}{(\sqrt{2} \, \mu +1)^{2}+1} \frac{\md \mu}{\pi})}}{4(\vert
r(\overline{s_{2}}) \vert)^{-1}(1 \! - \! r(s_{2}) \overline{r(\overline{
s_{2}})}) \sqrt{t} \, (\zeta \! - \! s_{2})} \right. \\
 &+\left. \mathcal{O} \! \left( \! \left( \dfrac{\underline{c}}{(\zeta \!
- \! \overline{s_{1}})} \! + \! \dfrac{\underline{c}}{(\zeta \! - \! s_{
2})} \right) \! \dfrac{\me^{-4t}}{t} \right) \right), \\
m^{c}_{21}(\zeta) \! &= \! \delta (\zeta) \! \left( \dfrac{\mathrm{sgn}
(\varepsilon_{1}) \me^{-(2t+\sqrt{2} \int_{-\infty}^{0} \! \frac{\ln (1-
\vert r(\mu) \vert^{2})}{(\sqrt{2} \, \mu -1)^{2}+1} \frac{\md \mu}{\pi})}
\me^{\mi (\frac{\pi}{4}+\sqrt{2} \int_{-\infty}^{0} \! \frac{(\sqrt{2} \,
\mu -1) \ln (1-\vert r(\mu) \vert^{2})}{(\sqrt{2} \, \mu -1)^{2}+1} \frac{
\md \mu}{\pi})}}{4(\vert r(s_{1}) \vert)^{-1} \sqrt{t} \, (\zeta \! - \!
s_{1})} \right. \\
&+\left. \dfrac{\mathrm{sgn}(\varepsilon_{2}) \me^{-(2t-\sqrt{2} \int_{
-\infty}^{0} \! \frac{\ln (1-\vert r(\mu) \vert^{2})}{(\sqrt{2} \, \mu +1)^{
2}+1} \frac{\md \mu}{\pi})} \me^{\mi (-\frac{3 \pi}{4}+\sqrt{2} \int_{-
\infty}^{0} \! \frac{(\sqrt{2} \, \mu +1) \ln (1-\vert r(\mu) \vert^{2})}
{(\sqrt{2} \, \mu +1)^{2}+1} \frac{\md \mu}{\pi})}}{4(\vert r(\overline{s_{
2}}) \vert)^{-1}(1 \! - \! r(s_{2}) \overline{r(\overline{s_{2}})}) \sqrt{
t} \, (\zeta \! - \! \overline{s_{2}})} \right.
\end{align*}
\begin{align*}
 &+\left. \mathcal{O} \! \left( \! \left( \dfrac{\underline{c}}{(\zeta \!
- \! s_{1})} \! + \! \dfrac{\underline{c}}{(\zeta \! - \! \overline{s_{2}})}
\right) \! \dfrac{\me^{-4t}}{t} \right) \right), \\
m^{c}_{22}(\zeta) \! &= \! \dfrac{1}{\delta (\zeta)} \! \left( 1 \! +
\! \mathcal{O} \! \left( \! \left( \dfrac{\underline{c}}{(\zeta \! - \!
\overline{s_{1}})} \! + \! \dfrac{\underline{c}}{(\zeta \! - \! s_{2})}
\right) \! \dfrac{\me^{-4t}}{t} \right) \right),
\end{align*}
where $\delta (\zeta)$ is defined in Lemma~{\rm A.1}, $\sup_{\zeta \in
\mathbb{C} \, \setminus \cup_{\lambda \in \widehat{\mathfrak{J}}} \mathbb{U}
(\lambda;\varepsilon)} \vert (\zeta \! - \! \underline{\widehat{\zeta}})^{
-1} \vert \! \leqslant \! \mathfrak{M}^{\natural}$, with $\mathfrak{M}^{
\natural} \! \in \! \mathbb{R}_{+}$ (and bounded), $\underline{\widehat{
\zeta}} \! \in \! \widehat{\mathfrak{J}}$, $m^{c}(\zeta) \! = \! \sigma_{1}
\overline{m^{c}(\overline{\zeta})} \, \sigma_{1}$, and $(m^{c}(0) \sigma_{
2})^{2} \! = \! \mathrm{I}$.
\end{ay}
\begin{by}
Let $s_{1} \! := \! \exp (\tfrac{\mi \pi}{4})$ and $s_{2} \! := \! \exp
(\tfrac{3 \pi \mi}{4})$. For $r(s_{1}) \! = \! \exp (-\tfrac{\mi \varepsilon_{
1} \pi}{2}) \vert r(s_{1}) \vert$, $\varepsilon_{1} \! \in \! \{\pm 1\}$,
$r(\overline{s_{2}}) \! = \! \exp (\tfrac{\mi \varepsilon_{2} \pi}{2}) \vert
r(\overline{s_{2}}) \vert$, $\varepsilon_{2} \! \in \! \{\pm 1\}$, $0 \! < \!
r(s_{2}) \overline{r(\overline{s_{2}})} \! < \! 1$, and $\varepsilon_{1} \!
= \! \varepsilon_{2}$, as $t \! \to \! +\infty$ and $x \! \to \! -\infty$
such that $z_{o} \! := \! x/t \! \to \! 0^{-}$,
\begin{gather*}
(\Delta_{o})_{11} \! = \! -\dfrac{\mathrm{sgn}(\varepsilon_{1}) \mi \me^{-
(2t+\widetilde{\mathfrak{c}}_{+})} \mathfrak{b}_{+}}{2 \sqrt{t}} \cosh \!
\left(\widetilde{\mathfrak{c}}_{-} \! - \! \ln \mathfrak{b}_{-} \right) \! +
\! \mathcal{O} \! \left( \dfrac{\underline{c} \, \me^{-4t}}{t} \right), \\
(\Delta_{o})_{12} \! = \! -\mi \me^{-\mi \psi^{+}(1)} \! \left(1 \! + \!
\mathcal{O} \! \left(\dfrac{\underline{c} \, \me^{-4t}}{t} \right) \right),
\quad (\Delta_{o})_{21} \! = \! \mi \me^{\mi \psi^{+}(1)} \! \left(1 \! + \!
\mathcal{O} \! \left(\dfrac{\underline{c} \, \me^{-4t}}{t} \right) \right), \\
(\Delta_{o})_{22} \! = \! \dfrac{\mathrm{sgn}(\varepsilon_{1}) \mi \me^{-
(2t+\widetilde{\mathfrak{c}}_{+})} \mathfrak{b}_{+}}{2 \sqrt{t}} \cosh \!
\left( \widetilde{\mathfrak{c}}_{-} \! - \! \ln \mathfrak{b}_{-} \right) \!
+ \! \mathcal{O} \! \left( \dfrac{\underline{c} \, \me^{-4t}}{t} \right),
\end{gather*}
and, for $\varepsilon_{1} \! = \! -\varepsilon_{2}$,
\begin{gather*}
(\Delta_{o})_{11} \! = \! \dfrac{\mathrm{sgn}(\varepsilon_{1}) \mi \me^{-
(2t+\widetilde{\mathfrak{c}}_{+})} \mathfrak{b}_{+}}{2 \sqrt{t}} \sinh \!
\left(\widetilde{\mathfrak{c}}_{-} \! - \! \ln \mathfrak{b}_{-} \right) \! +
\! \mathcal{O} \! \left( \dfrac{\underline{c} \, \me^{-4t}}{t} \right), \\
(\Delta_{o})_{12} \! = \! -\mi \me^{-\mi \psi^{+}(1)} \! \left(1 \! + \!
\mathcal{O} \! \left(\dfrac{\underline{c} \, \me^{-4t}}{t} \right) \right),
\quad (\Delta_{o})_{21} \! = \! \mi \me^{\mi \psi^{+}(1)} \! \left(1 \! + \!
\mathcal{O} \! \left(\dfrac{\underline{c} \, \me^{-4t}}{t} \right) \right),
\\
(\Delta_{o})_{22} \! = \! -\dfrac{\mathrm{sgn}(\varepsilon_{1}) \mi \me^{
-(2t+\widetilde{\mathfrak{c}}_{+})} \mathfrak{b}_{+}}{2 \sqrt{t}} \sinh \!
\left(\widetilde{\mathfrak{c}}_{-} \! - \! \ln \mathfrak{b}_{-} \right) \!
+ \! \mathcal{O} \! \left( \dfrac{\underline{c} \, \me^{-4t}}{t} \right),
\end{gather*}
where $\psi^{+}(\cdot)$ is defined in Theorem~{\rm 3.2}, Eq.~{\rm (35)},
and $\widetilde{\mathfrak{c}}_{\pm}$ and $\mathfrak{b}_{\pm}$ are defined
in Theorem~{\rm 3.3}, Eqs.~{\rm (73)} and~{\rm (74)}.
\end{by}
\stepcounter{z0}
\stepcounter{z1}
\begin{ay}
Let $\varepsilon$ be an arbitrarily fixed, sufficiently small positive real
number, and, for $\lambda \! \in \! \widehat{\mathfrak{J}} \! := \! \{(s_{
1})^{\pm 1},(s_{2})^{\pm 1}\}$, where $s_{1} \! := \! \exp (\tfrac{\mi \pi}
{4})$ and $s_{2} \! := \! \exp (\tfrac{3 \pi \mi}{4})$, set $\mathbb{U}
(\lambda;\varepsilon) \! := \! \{\mathstrut z; \, \vert z \! - \! \lambda
\vert \! < \! \varepsilon\}$. Then, for $r(\overline{s_{1}}) \! = \! \exp
(\tfrac{\mi \varepsilon_{1} \pi}{2}) \vert r(\overline{s_{1}}) \vert$,
$\varepsilon_{1} \! \in \! \{\pm 1\}$, $r(s_{2}) \! = \! \exp (-\tfrac{\mi
\varepsilon_{2} \pi}{2}) \vert r(s_{2}) \vert$, $\varepsilon_{2} \! \in \!
\{\pm 1\}$, $0 \! < \! r(s_{1}) \overline{r(\overline{s_{1}})} \! < \! 1$,
and $\zeta \! \in \! \mathbb{C} \setminus \cup_{\lambda \, \in \, \widehat{
\mathfrak{J}}} \, \mathbb{U}(\lambda;\varepsilon)$, as $t \! \to \! -\infty$
and $x \! \to \! +\infty$ such that $z_{o} \! := \! x/t \! \to \! 0^{-}$,
$m^{c}(\zeta)$ has the following asymptotics,
\begin{align*}
m^{c}_{11}(\zeta) \! &= \! \widetilde{\delta}(\zeta) \! \left( 1 \! +
\! \mathcal{O} \! \left( \! \left( \dfrac{\underline{c}}{(\zeta \! - \!
\overline{s_{1}})} \! + \! \dfrac{\underline{c}}{(\zeta \! - \! s_{
2})} \right) \! \dfrac{\me^{-4 \vert t \vert}}{t} \right) \right), \\
m^{c}_{12}(\zeta) \! &= \! -\dfrac{1}{\widetilde{\delta}(\zeta)} \! \left(
\dfrac{\mathrm{sgn}(\varepsilon_{1}) \me^{-(2 \vert t \vert -\sqrt{2}
\int_{0}^{+\infty} \! \frac{\ln (1-\vert r(\mu) \vert^{2})}{(\sqrt{2} \, \mu
-1)^{2}+1} \frac{\md \mu}{\pi})} \me^{\mi (\frac{\pi}{4}-\sqrt{2} \int_{
0}^{+\infty} \! \frac{(\sqrt{2} \, \mu -1) \ln (1-\vert r(\mu) \vert^{2})}
{(\sqrt{2} \, \mu -1)^{2}+1} \frac{\md \mu}{\pi})}}{4(\vert r(\overline{
s_{1}}) \vert)^{-1}(1 \! - \! r(s_{1}) \overline{r(\overline{s_{1}})})
\sqrt{\vert t \vert} \, (\zeta \! - \! s_{1})} \right. \\
 &+\left. \dfrac{\mathrm{sgn}(\varepsilon_{2}) \me^{-(2 \vert t \vert
+\sqrt{2} \int_{0}^{+\infty} \! \frac{\ln (1-\vert r(\mu) \vert^{2})}{(\sqrt{
2} \, \mu +1)^{2}+1} \frac{\md \mu}{\pi})} \me^{-\mi (\frac{3 \pi}{4}+
\sqrt{2} \int_{0}^{+\infty} \! \frac{(\sqrt{2} \, \mu +1) \ln (1-\vert r(\mu)
\vert^{2})}{(\sqrt{2} \, \mu +1)^{2}+1} \frac{\md \mu}{\pi})}}{4(\vert
r(s_{2}) \vert)^{-1} \sqrt{\vert t \vert} \, (\zeta \! - \! \overline{s_{2}})}
\right. \\
 &+\left. \mathcal{O} \! \left( \! \left( \dfrac{\underline{c}}{(\zeta \!
- \! s_{1})} \! + \! \dfrac{\underline{c}}{(\zeta \! - \! \overline{s_{2}})}
\right) \! \dfrac{\me^{-4 \vert t \vert}}{t} \right) \right), \\
m^{c}_{21}(\zeta) \! &= \! -\widetilde{\delta}(\zeta) \! \left(\dfrac{
\mathrm{sgn}(\varepsilon_{1}) \me^{-(2 \vert t \vert -\sqrt{2} \int_{
0}^{+\infty} \! \frac{\ln (1-\vert r(\mu) \vert^{2})}{(\sqrt{2} \, \mu
-1)^{2}+1} \frac{\md \mu}{\pi})} \me^{-\mi (\frac{\pi}{4}-\sqrt{2} \int_{
0}^{+\infty} \! \frac{(\sqrt{2} \, \mu -1) \ln (1-\vert r(\mu) \vert^{2})}
{(\sqrt{2} \, \mu -1)^{2}+1} \frac{\md \mu}{\pi})}}{4(\vert r(\overline{
s_{1}}) \vert)^{-1}(1 \! - \! r(s_{1}) \overline{r(\overline{s_{1}})})
\sqrt{\vert t \vert} \, (\zeta \! - \! \overline{s_{1}})} \right.
\end{align*}
\begin{align*}
 &+\left. \dfrac{\mathrm{sgn}(\varepsilon_{2}) \me^{-(2 \vert t \vert +
\sqrt{2} \int_{0}^{+\infty} \! \frac{\ln (1-\vert r(\mu) \vert^{2})}
{(\sqrt{2} \, \mu +1)^{2}+1} \frac{\md \mu}{\pi})} \me^{\mi (\frac{3
\pi}{4}+\sqrt{2} \int_{0}^{+\infty} \! \frac{(\sqrt{2} \, \mu +1) \ln
(1-\vert r(\mu) \vert^{2})}{(\sqrt{2} \, \mu +1)^{2}+1} \frac{\md \mu}
{\pi})}}{4(\vert r(s_{2}) \vert)^{-1} \sqrt{\vert t \vert} \, (\zeta \!
- \! s_{2})} \right. \\
 &+\left. \mathcal{O} \! \left( \! \left( \dfrac{\underline{c}}{(\zeta \!
- \! \overline{s_{1}})} \! + \! \dfrac{\underline{c}}{(\zeta \! - \! s_{
2})} \right) \! \dfrac{\me^{-4 \vert t \vert}}{t} \right) \right), \\
m^{c}_{22}(\zeta) \! &= \! \dfrac{1}{\widetilde{\delta}(\zeta)} \! \left(
1 \! + \! \mathcal{O} \! \left( \! \left( \dfrac{\underline{c}}{(\zeta \! -
\! s_{1})} \! + \! \dfrac{\underline{c}}{(\zeta \! - \! \overline{s_{2}})}
\right) \! \dfrac{\me^{-4 \vert t \vert}}{t} \right) \right),
\end{align*}
where $\widetilde{\delta}(\zeta)$ is defined in Lemma~{\rm A.2}, $\sup_{
\zeta \in \mathbb{C} \, \setminus \cup_{\lambda \in \widehat{\mathfrak{J}}}
\mathbb{U}(\lambda;\varepsilon)} \vert (\zeta \! - \! \underline{\widehat{
\zeta}})^{-1} \vert \! \leqslant \! \mathfrak{M}^{\sharp}$, with $\mathfrak{
M}^{\sharp} \! \in \! \mathbb{R}_{+}$ (and bounded), $\underline{\widehat{
\zeta}} \! \in \! \widehat{\mathfrak{J}}$, $m^{c}(\zeta) \! = \! \sigma_{1}
\overline{m^{c}(\overline{\zeta})} \, \sigma_{1}$, and $(m^{c}(0) \sigma_{
2})^{2} \! = \! \mathrm{I}$.
\end{ay}
\begin{by}
Let $s_{1} \! := \! \exp (\tfrac{\mi \pi}{4})$ and $s_{2} \! := \! \exp
(\tfrac{3 \pi \mi}{4})$. For $r(\overline{s_{1}}) \! = \! \exp (\tfrac{\mi
\varepsilon_{1} \pi}{2}) \vert r(\overline{s_{1}}) \vert$, $\varepsilon_{1}
\! \in \! \{\pm 1\}$, $r(s_{2}) \! = \! \exp (-\tfrac{\mi \varepsilon_{2}
\pi}{2}) \vert r(s_{2}) \vert$, $\varepsilon_{2} \! \in \{\pm 1\}$, $0 \!
< \! r(s_{1}) \overline{r(\overline{s_{1}})} \! < \! 1$, and $\varepsilon_{
1} \! = \! \varepsilon_{2}$, as $t \! \to \! -\infty$ and $x \! \to \!
+\infty$ such that $z_{o} \! := \! x/t \! \to \! 0^{-}$,
\begin{gather*}
(\Delta_{o})_{11} \! = \! \dfrac{\mathrm{sgn}(\varepsilon_{1}) \mi \me^{-(2
\vert t \vert -\widehat{\mathfrak{c}}_{+})} \mathfrak{d}_{+}}{2 \sqrt{\vert
t \vert}} \cosh \! \left(\widehat{\mathfrak{c}}_{-} \! - \! \ln \mathfrak{
d}_{-} \right) \! + \! \mathcal{O} \! \left(\dfrac{\underline{c} \, \me^{-4
\vert t \vert}}{t} \right), \\
(\Delta_{o})_{12} \! = \! -\mi \me^{-\mi \psi^{-}(1)} \! \left(1 \! + \!
\mathcal{O} \! \left(\dfrac{\underline{c} \, \me^{-4 \vert t \vert}}{t}
\right) \right), \quad (\Delta_{o})_{21} \! = \! \mi \me^{\mi \psi^{-}(1)}
\! \left(1 \! + \! \mathcal{O} \! \left(\dfrac{\underline{c} \, \me^{-4
\vert t \vert}}{t} \right) \right), \\
(\Delta_{o})_{22} \! = \! -\dfrac{\mathrm{sgn}(\varepsilon_{1}) \mi \me^{-
(2 \vert t \vert -\widehat{\mathfrak{c}}_{+})} \mathfrak{d}_{+}}{2 \sqrt{
\vert t \vert}} \cosh \! \left( \widehat{\mathfrak{c}}_{-} \! - \! \ln
\mathfrak{d}_{-} \right) \! + \! \mathcal{O} \! \left( \dfrac{\underline{c}
\, \me^{-4 \vert t \vert}}{t} \right),
\end{gather*}
and, for $\varepsilon_{1} \! = \! -\varepsilon_{2}$,
\begin{gather*}
(\Delta_{o})_{11} \! = \! \dfrac{\mathrm{sgn}(\varepsilon_{1}) \mi \me^{-
(2 \vert t \vert -\widehat{\mathfrak{c}}_{+})} \mathfrak{d}_{+}}{2 \sqrt{
\vert t \vert}} \sinh \! \left( \widehat{\mathfrak{c}}_{-} \! - \! \ln
\mathfrak{d}_{-} \right) \! + \! \mathcal{O} \! \left(\dfrac{\underline{c}
\, \me^{-4 \vert t \vert}}{t} \right), \\
(\Delta_{o})_{12} \! = \! -\mi \me^{-\mi \psi^{-}(1)} \! \left(1 \! + \!
\mathcal{O} \! \left(\dfrac{\underline{c} \, \me^{-4 \vert t \vert}}{t}
\right) \right), \quad (\Delta_{o})_{21} \! = \! \mi \me^{\mi \psi^{-}(1)}
\! \left(1 \! + \! \mathcal{O} \! \left(\dfrac{\underline{c} \, \me^{-4
\vert t \vert}}{t} \right) \right), \\
(\Delta_{o})_{22} \! = \! -\dfrac{\mathrm{sgn}(\varepsilon_{1}) \mi \me^{-
(2 \vert t \vert -\widehat{\mathfrak{c}}_{+})} \mathfrak{d}_{+}}{2 \sqrt{
\vert t \vert}} \sinh \! \left( \widehat{\mathfrak{c}}_{-} \! - \! \ln
\mathfrak{d}_{-} \right) \! + \! \mathcal{O} \! \left( \dfrac{\underline{c}
\, \me^{-4 \vert t \vert}}{t} \right),
\end{gather*}
where $\psi^{-}(\cdot)$ is defined in Theorem~{\rm 3.2}, Eq.~{\rm (46)},
and $\widehat{\mathfrak{c}}_{\pm}$ and $\mathfrak{d}_{\pm}$ are defined in
Theorem~{\rm 3.3}, Eqs.~{\rm (77)} and~{\rm (78)}.
\end{by}


\clearpage
\begin{thebibliography}{111}
\bibitem{a1} G.~P.~Agrawal, 2nd edn., \emph{Nonlinear Fiber Optics},
Academic Press, San Diego, 1995.
\bibitem{a2} A.~M.~Weiner, ``Dark optical solitons'', pp.~378--408,
in J.~R.~Taylor, ed., \emph{Optical Solitons - Theory and Experiment},
Cambridge Studies in Modern Optics, Vol.~10, Cambridge University Press,
Cambridge, 1992.
\bibitem{a3} Y.~Kodama, ``The Whitham Equations for Optical
Communications: Mathematical Theory of NRZ'', SIAM J. Appl. Math.,
Vol.~59, No.~6, pp.~2162--2192, 1999.
\bibitem{a4} D.~Sh.~Lundina and V.~A.~Marchenko, ``Compactness of
the Set of Multisoliton Solutions of the Nonlinear Schr\"{o}dinger
Equation'', Russian Acad. Sci. Sb. Math., Vol.~75, No.~2,
pp.~429--443, 1993.
\bibitem{a5} P.~D.~Miller, ``Zero-crosstalk junctions made {}from dark
solitons'', Phys. Rev. E, Vol.~53, No.~4, pp.~4137--4142, 1996. P.~D.~Miller
and N.~N.~Akhmediev, ``Transfer matrices for multiport devices made {}from
solitons'', Phys. Rev. E, Vol.~53, No.~4, pp.~4098--4106, 1996.
\bibitem{a6} S.~P.~Novikov, S.~V.~Manakov, L.~P.~Pitaevskii, and
V.~E.~Zakharov, \emph{Theory of Solitons: The Inverse Scattering
Method}, Plenum, New York, 1984.
\bibitem{a7} M.~J.~Ablowitz and P.~A.~Clarkson, \emph{Solitons,
Nonlinear Evolution Equations and Inverse Scattering}, LMS~149,
Cambridge University Press, Cambridge, 1991.
\bibitem{a8} R.~Beals, P.~Deift, and X.~Zhou, ``The Inverse Scattering
Transform on the Line'', pp.~7--32, in A.~S.~Fokas and V.~E.~Zakharov,
eds., \emph{Important Developments in Soliton Theory}, Springer Series
in Nonlinear Dynamics, Springer-Verlag, New York, 1993.
\bibitem{a9} L.~D.~Faddeev and L.~A.~Takhtajan, \emph{Hamiltonian
Methods in the Theory of Solitons}, Springer-Verlag, Berlin, 1987.
\bibitem{a10} V.~E.~Zakharov and A.~B.~Shabat, ``Interaction between
solitons in a stable medium'', Sov. Phys.~JETP, Vol.~37, No.~5,
pp.~823--828, 1973.
\bibitem{a11} T.~Kawata and H.~Inoue, ``Inverse Scattering Method for
the Nonlinear Evolution Equations under Nonvanishing Conditions'', J.
Phys. Soc. Japan, Vol.~44, No.~5, pp.~1722--1729, 1978.
\bibitem{a12} N.~Asano and Y.~Kato, ``Non-self-adjoint Zakharov-Shabat
operator with a potential of the finite asymptotic values. II. Inverse
problem'', J. Math. Phys., Vol.~25, No.~3, pp.~570--588, 1984. N.~Asano
and Y.~Kato, ``Non-self-adjoint Zakharov-Shabat operator with a potential
of the finite asymptotic values. I. Direct spectral and scattering
problems'', J. Math. Phys., Vol.~22, No.~12, pp.~2780--2793, 1981.
\bibitem{a13} A.~Boutet~de~Monvel and V.~Marchenko, ``The Cauchy problem
for nonlinear Schr\"{o}dinger equation with bounded initial data'', Mat.
Fiz. Anal. Geom., Vol.~4, No.~1/2, pp.~3--45, 2000.
\bibitem{a14} M.~Boiti and F.~Pempinelli, ``The Spectral Transform
for the NLS Equation with Left-Right Asymmetric Boundary Conditions'',
Il Nuovo Cimento, Vol.~69B, No.~2, pp.~213--227, 1982.
\bibitem{a15} T.~Kawata, J.~Sakai, and N.~Kobayashi, ``Inverse Method for
the Mixed Nonlinear Schr\"{o}dinger Equation and Soliton Solutions'', J.
Phys. Soc. Japan, Vol.~48, No.~4, pp.~1371--1379, 1980. T.~Kawata and
H.~Inoue, ``Exact Solutions of the Derivative Nonlinear Schr\"{o}dinger
Equation under the Nonvanishing Conditions'', J. Phys. Soc. Japan, Vol.~44,
No.~6, pp.~1968--1976, 1978.
\bibitem{a16} A.~Cohen and T.~ Kappeler, ``Scattering and Inverse
Scattering for Steplike Potentials in the Schr\"{o}dinger Equation'',
Indiana Univ. Math. J., Vol.~34, No.~1, pp.~127--180, 1985.
\bibitem{a17} V.~A.~Marchenko, ``The Cauchy Problem for the KdV
Equation with Non-Decreasing Initial Data'', pp.~273--318, in
\emph{What is Integrability?}, V.~E.~Zakharov, ed., Springer Series
in Nonlinear Dynamics, Springer-Verlag, Berlin, 1991.
\bibitem{a18} A.~Boutet~de~Monvel, E.~Ya.~Khruslov, and V.~P.~Kotlyarov,
``The Cauchy problem for the sinh-Gordon equation and regular solitons'',
Inverse Problems, Vol.~14, No.~6, pp.~1403--1427, 1998. A.~B.~Borisov and
V.~V.~Kiseliev, ``Inverse problem for an elliptic sine-Gordon equation
with an asymptotic behaviour of the cnoidal-wave type'', Inverse Problems,
Vol.~5, No.~6, pp.~959--982, 1989.
\bibitem{a19} A.~R.~Its and A.~F.~Ustinov, ``The time asymptotics of the
solution of the Cauchy problem for the nonlinear Schr\"{o}dinger equation
with finite density boundary conditions'', Dokl. Akad. Nauk SSSR, Vol.~291,
No.~1, pp.~91--95, 1986 (in Russian).
\bibitem{a20} A.~R.~Its and A.~F.~Ustinov, ``Formulation of the Scattering
Theory for the Nonlinear Schr\"{o}dinger Equation with Boundary
Conditions of the Finite Density Type in a Soliton-Free Sector'', J. Sov.
Math., Vol.~54, No.~3, pp.~900--905, 1991.
\bibitem{a21} V.~E.~Zakharov and A.~B.~Shabat, ``Integration of the
non-linear equations of mathematical physics by the method of the
inverse scattering transform. II'', Funct. Anal. Appl., Vol.~13, No.~3,
pp.~166--173, 1980.
\bibitem{a22} P.~A.~Deift, S.~Kamvissis, T.~Kriecherbauer, and X.~Zhou,
``The Toda Rarefaction Problem'', Comm. Pure Appl. Math., Vol.~49, No.~1,
pp.~35--83, 1996.
\bibitem{a23} K.~Clancey and I.~Gohberg, \emph{Factorization of Matrix
Functions and Singular Integral Operators}, Operator Theory: Advances and
Applications, Vol.~3, Birkh\"{a}user, Basel, 1981.
\bibitem{a24} R.~Beals and R.~R.~Coifman, ``Scattering and Inverse
Scattering for First Order Systems'', Comm. Pure Appl. Math., Vol.~37,
No.~1, pp.~39--90, 1984.
\bibitem{a25} P.~Deift, \emph{Orthogonal Polynomials and Random Matrices:
A Riemann-Hilbert Approach}, Courant Lecture Notes in Mathematics, Vol.~3,
CIMS, New York, 1999.
\bibitem{a26} A.~S.~Fokas, ``On the integrability of linear and nonlinear
partial differential equations'', J. Math. Phys., Vol.~41, No.~6,
pp.~4188--4237, 2000.
\bibitem{a27} P.~Deift and X.~Zhou, ``A Steepest descent method for
oscillatory Riemann-Hilbert problems. Asymptotics for the MKdV equation'',
Ann. of Math., Vol.~137, No.~2, pp.~295--368, 1993.
\bibitem{a28} X.~Zhou, ``Direct and Inverse Scattering Transforms with
Arbitrary Spectral Singularities'', Comm. Pure Appl. Math., Vol.~42,
No.~7, pp.~ 895--938, 1989.
\bibitem{a29} X.~Zhou, ``Inverse Scattering Transform for Systems with
Rational Spectral Dependence'', J. Differential Equations, Vol.~115,
No.~2, pp.~277--303, 1995.
\bibitem{a30} N.-N.~Huang and Z.-Y.~Chen, ``Zakharov-Shabat Equations for
Dark Solitons to the NLS Equation'', Commun. Theor. Phys., Vol.~20, No.~2,
pp.~187--194, 1993.
\bibitem{a31} X.~Zhou, ``The Riemann-Hilbert Problem and Inverse
Scattering'', SIAM J. Math. Anal., Vol.~20, No.~4, pp.~966--986, 1989.
\bibitem{a32} X.~Zhou, ``Strong Regularizing Effect of Integrable
Systems'', Comm. PDE, Vol.~22, Nos.~3 \& 4, pp.~503--526, 1997.
\bibitem{a33} X.~Zhou, ``$L^{2}$-Sobolev Space Bijectivity of the
Scattering and Inverse Scattering Transforms'', Comm. Pure Appl. Math.,
Vol.~51, No.~7, pp.~697--731, 1998.
\bibitem{a34} P.~Deift, T.~Kriecherbauer, K.~T.-R.~McLaughlin,
S.~Venakides, and X.~Zhou, ``Strong Asymptotics of Orthogonal
Polynomials with Respect to Exponential Weights'', Comm. Pure Appl.
Math., Vol.~52, No.~12, pp.~1491--1552, 1999.
\bibitem{a35} H.~Flaschka and A.~C.~Newell, ``Monodromy- and
Spectrum-Preserving Deformations I'', Comm. Math. Phys., Vol.~76,
No.~1, pp.~65--116, 1980.
\bibitem{a36} A.~S.~Fokas and M.~J.~Ablowitz, ``On the Initial Value
Problem of the Second Painlev\'{e} Transcendent'', Comm. Math. Phys.,
Vol.~91, No.~3, pp.~381--403, 1983.
\bibitem{a37} A.~R.~Its and V.~Yu.~Novokshenov, \emph{The
Isomonodromy Deformation Method in the Theory of Painlev\'{e}
Equations}, LNM~1191, Springer-Verlag, Berlin, 1986.
\bibitem{a38} A.~R.~Its and A.~A.~Kapaev, ``The Method of Isomonodromy
Deformations and Connection Formulas for the Second Painlev\'{e}
Transcendent'', Math. USSR Izvestiya, Vol.~31, No.~1, pp.~193--207, 1988.
\bibitem{a39} P.~Deift and X.~Zhou, ``Asymptotics for the Painlev\'{e}
II Equation'', Comm. Pure Appl. Math., Vol.~48, No.~3, pp.~277--337,
1995.
\bibitem{a40} I.~S.~Gradshteyn and I.~M.~Ryzhik, \emph{Tables of
Integrals, Series, and Products}, 5th edn., A.~Jeffrey, ed., Academic
Press, San Diego, 1994.
\bibitem{a41} J.~Baik, P.~Deift, and K.~Johansson, ``On the Distribution
of the Length of the Longest Increasing Subsequence of Random
Permutations'', J. Amer. Math. Soc., Vol.~12, No.~4, pp.~1119--1178,
1999.
\bibitem{a42} P.~Deift, T.~Kriecherbauer, K.~T.-R.~McLaughlin,
S.~Venakides, and X.~Zhou, ``Uniform Asymptotics for Polynomials
Orthogonal with Respect to Varying Exponential Weights and
Applications to Universality Questions in Random Matrix Theory'',
Comm. Pure Appl. Math., Vol.~52, No.~11, pp.~1335--1425, 1999.
\bibitem{a43} P.~Deift and X.~Zhou, ``Perturbation theory for infinite
dimensional integrable systems on the line. A case study'', Preprint,
2000.
\bibitem{a44} S.~Kamvissis, K.~T.-R.~McLaughlin, and P.~D.~Miller,
``Semiclassical Soliton Ensembles for the Focusing Nonlinear
Schr\"{o}dinger Equation'', \texttt{arXiv:nlin.SI/0012034}.
\bibitem{a45} F.~D.~Gakhov, \emph{Boundary Value Problems}, Dover,
New York, 1990.
\bibitem{a46} P.-J.~Cheng, S.~Venakides, and X.~Zhou, ``Long-Time
Asymptotics for the Pure Radiation Solution of the Sine-Gordon Equation'',
Comm. PDE, Vol.~24, Nos.~7 \& 8, pp.~1195--1262, 1999.
\bibitem{a47} A.~R.~Its, ``Asymptotics of Solutions of the Nonlinear
Schr\"{o}dinger Equation and Isomonodromic Deformations of Systems of
Linear Differential Equations'', Soviet Math. Dokl., Vol.~24, No.~3,
pp.~452--456, 1982.
\bibitem{a48} P.~Deift and X.~Zhou, ``Long-time Asymptotics for
Integrable Systems. Higher Order Theory'', Comm. Math. Phys.,
Vol.~165, No.~1, pp.~175--191, 1994.
\bibitem{a49} A.~H.~Vartanian, ``Higher Order Asymptotics of the
Modified Non-Linear Schr\"{o}dinger Equation'', Comm. PDE, Vol.~25,
Nos.~5 \& 6, pp.~1043--1098, 2000.
\bibitem{a50} A.~V.~Kitaev and A.~H.~Vartanian, ``Asymptotics of Solutions
to the Modified Nonlinear Schr\"{o}dinger Equation: Solitons on a
Nonvanishing Continuous Background'', SIAM J. Math. Anal., Vol.~30, No.~4,
pp.~787--832, 1999.
\end{thebibliography}
\end{document}